%% file: main.tex
\documentclass[a4paper,12pt]{book}
\usepackage[utf8]{inputenc}
\usepackage[T1]{fontenc}
\usepackage[english]{babel}
\usepackage[left=2cm,right=2cm,top=2.5cm,bottom=2.5cm]{geometry}
\usepackage{amsmath}
\usepackage{amsfonts}
\usepackage{amssymb}
\usepackage{graphicx}
\usepackage{color} 
\definecolor{Prune}{RGB}{99,0,60}
\usepackage{mdframed}
\usepackage[absolute]{textpos}
\usepackage{scrextend}
\usepackage{colortbl}
\usepackage{multirow}
\usepackage{multicol} 
\usepackage{pstricks}
\usepackage{verbatim}
\usepackage{slashed}
\usepackage{hyperref}
\usepackage{braket}
\usepackage[square,sort,comma,numbers]{natbib}
\usepackage{dsfont}
\usepackage{bm}
\usepackage{fancyhdr}
\usepackage{mathrsfs}
\usepackage{pifont}
\usepackage[titletoc]{appendix}
\usepackage{multirow}
\usepackage{bbm}
\usepackage{subcaption}
\usepackage{float}
\usepackage[many]{tcolorbox}
\usepackage{empheq}
\usetikzlibrary{shadows}

\tcbset{
  highlight math style={
    enhanced,
    colframe=red!60!black,
    colback=yellow!50,
    arc=4pt,
    boxrule=1pt,
    drop fuzzy shadow
  }
}
\newcommand\bbone{\ensuremath{\mathbbm{1}}}
\def\dif{\textrm{d}}
\def\th{\theta}
\def\ptjet{p_{T,\textrm{jet}}}
\def\Njet{N_{\textrm{jets}}}
\def\ktmin{k_{\perp,\textrm{min}}}
\def\as{\alpha_s}
\def\om{\omega}
\def\abar{\bar{\alpha}_s}
\def\zc{z_{\textrm{cut}}}
\def\thetacut{\theta_{\textrm{cut}}}
\newcommand{\rmd}{{\rm d}}
\newcommand{\erf}{\textrm{Erf}}
\newcommand{\nSD}{n_{\textrm{SD}}}
\newcommand{\Lc}{L_{\textrm{cut}}}
\newcommand{\Nmie}{N_{\textrm{mie}}}
\newcommand{\omBH}{\omega_{\textrm{BH}}}
\newcommand{\ombr}{\omega_{\textrm{br}}}
\newcommand{\omc}{\omega_c}
\def\obr{\omega_{\rm br}}
\def\oc{\omega_c}
\def\amed{\alpha_{s,\rm med}}
\newcommand{\qhat}{\hat{q}}
\newcommand{\tcoh}{t_{\textrm{coh}}}
\newcommand{\tvac}{t_{i,\textrm{vac}}}
\newcommand{\tint}{t_{i,\textrm{med}}}
\newcommand{\tbr}{t_{\textrm{br}}}
\newcommand{\Zmie}{Z^{\textrm{mie}}}
\newcommand{\mie}{\textrm{mie}}
\newcommand{\tprop}{t_{\textrm{prop}}}
\newcommand{\Tr}{\textrm{Tr}}
\newcommand{\LQCD}{\Lambda_{\textrm{QCD}}}
\newcommand{\thqq}{\theta_{q\bar{q}}}

\newcommand{\vac}{\textrm{vac}}
\newcommand{\med}{\textrm{med}}
\newcommand{\brem}{\textrm{brem}}
\newcommand{\tf}{t_f}

\newcommand{\kt}{k_{\perp}}
\newcommand{\xmark}{\text{\ding{55}}}
\newcommand{\cut}{\textrm{cut}}
\newcommand{\LL}{\textrm{LL}}
\newcommand{\NLL}{\textrm{NLL}}
\newcommand{\sym}{\textrm{sym}}
\newcommand{\njets}{$N_\text{jets}$}
\newcommand{\Njets}{N_\text{jets}}
\newcommand{\sub}{\textrm{sub}}

\newcommand{\eqn}[1]{Eq.~\eqref{#1}}
\newcommand{\beq}{\begin{equation}}
\newcommand{\eeq}{\end{equation}}
\newcommand{\minus}{\!-\!}
\newcommand{\del}{\partial}
\newcommand{\rme}{{\rm e}}
\newcommand{\nn}{\nonumber\\}
\newcommand{\lab}{\textrm{lab}}

\begin{document}

\begin{titlepage}

\newgeometry{left=7.5cm,bottom=2cm, top=1cm, right=1cm}

\tikz[remember picture,overlay] \node[opacity=1,inner sep=0pt] at (-28mm,-135mm){\includegraphics{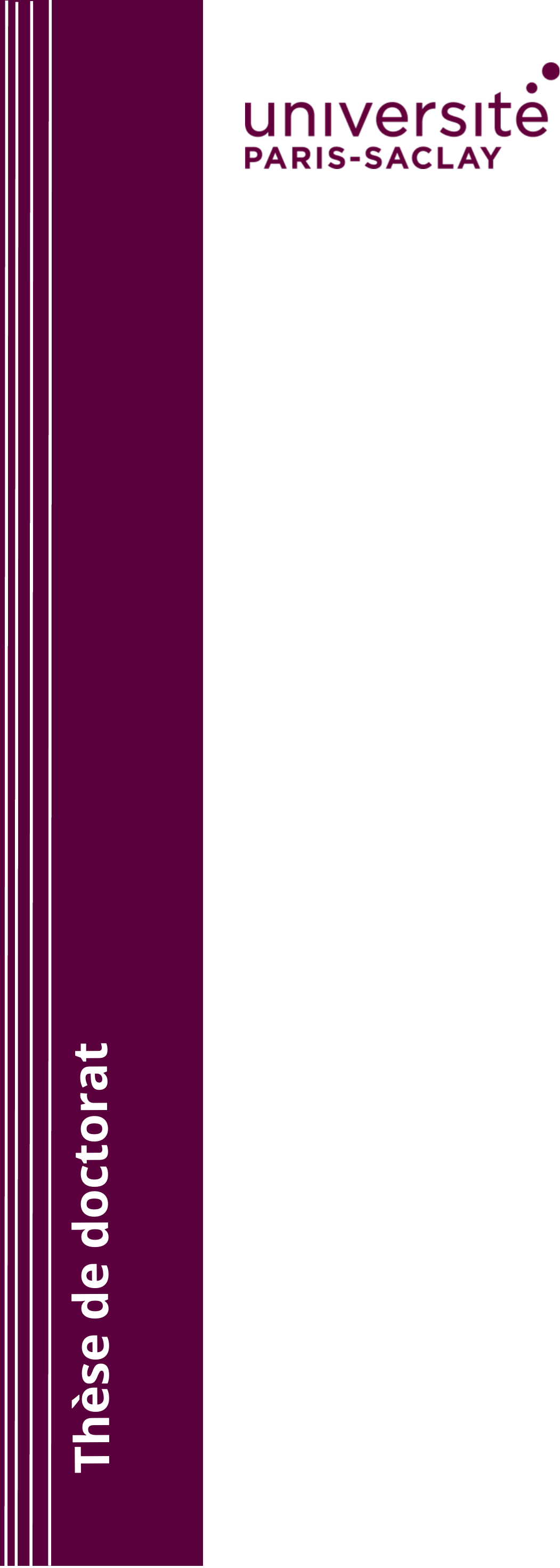}};

\fontfamily{fvs}\fontseries{m}\selectfont


\color{white}

\begin{picture}(0,0)

\put(-150,-735){\rotatebox{90}{NNT: 2020UPASP002}}
\end{picture}
 

\flushright
\vspace{10mm} 
\color{Prune}
\fontfamily{fvs}\fontseries{m}\fontsize{22}{26}\selectfont
  Jet evolution in a dense QCD medium


\normalsize
\vspace{1.5cm}

\color{black}
\textbf{Thèse de doctorat de l'Université Paris-Saclay}

\vspace{15mm}

École doctorale n$^{\circ}$ 564 : physique en l'\^{I}le-de-France (PIF)\\
\small Spécialité de doctorat: physique\\
\footnotesize Unité de recherche: Université Paris-Saclay, CNRS, CEA, Institut de physique théorique, 91191, Gif-sur-Yvette, France.\\
\footnotesize Référent: Faculté des sciences d’Orsay
\vspace{15mm}

\textbf{Thèse présentée et soutenue à Gif-sur-Yvette, le 4 septembre 2020, par}\\
\bigskip
\Large {\color{Prune} \textbf{Paul Caucal}}

\vspace{\fill} 

\flushleft \small \textbf{Composition du jury:}
\bigskip

\scriptsize
\begin{tabular}{|p{8cm}l}
\arrayrulecolor{Prune}
\textbf{Samuel Wallon} &   Président\\ 
Professeur des universités, Université Paris-Saclay, \\ IJCLab & \\
\textbf{Carlos Salgado} &  Rapporteur \& Examinateur  \\ 
Professeur des universités, Galician Institute of \\ High Energy Physics   &   \\ 
\textbf{Konrad Tywoniuk} &  Rapporteur \& Examinateur \\ 
Directeur de recherche, Université de Bergen  &   \\ 
\textbf{Matteo Cacciari} &  Examinateur \\ 
Professeur des universités, Université Paris Diderot, LPTHE   &   \\ 
\textbf{Leticia Cunqueiro} &  Examinatrice \\ 
Chargée de recherche, Oak Ridge National Laboratory, CERN   &   \\ 


\end{tabular} 

\medskip
\begin{tabular}{|p{8cm}l}\arrayrulecolor{white}
\textbf{Edmond Iancu} &   Directeur\\ 
Directeur de recherche, Institut de Physique Théorique & \\
\textbf{Gregory Soyez} &   Codirecteur\\ 
Directeur de recherche, Institut de Physique Théorique  &   \\ 

\end{tabular}

\end{titlepage}

\include{abstract}

\include{remerciements}

\tableofcontents

\include{intro}

\part{Theory: jets in a dense QCD medium}
\label{part:theory}

\include{intro-part1}

\include{chapter2}

\include{chapter3}

\include{chapter4}

\include{chapter5}

\part{Phenomenology: jet quenching at the LHC}
\label{part:phenomenology}

\include{intro-part2}

\include{chapter6}

\include{chapter7}

\include{chapter8}

\include{conclusion}

\phantomsection
\appendix
\include{appendix}

\phantomsection
\addcontentsline{toc}{part}{Bibliography}
\bibliographystyle{utcaps}
\bibliography{biblio/biblio-full}
\phantomsection
\newpage
\thispagestyle{empty}
\mbox{}
\newpage
\ifthispageodd{\newpage\thispagestyle{empty}\null\newpage}{}
\thispagestyle{empty}
\newgeometry{top=1.5cm, bottom=1.25cm, left=2cm, right=2cm}
\fontfamily{rm}\selectfont

\lhead{}
\rhead{}
\rfoot{}
\cfoot{}
\lfoot{}

\noindent 
\includegraphics[height=2.45cm]{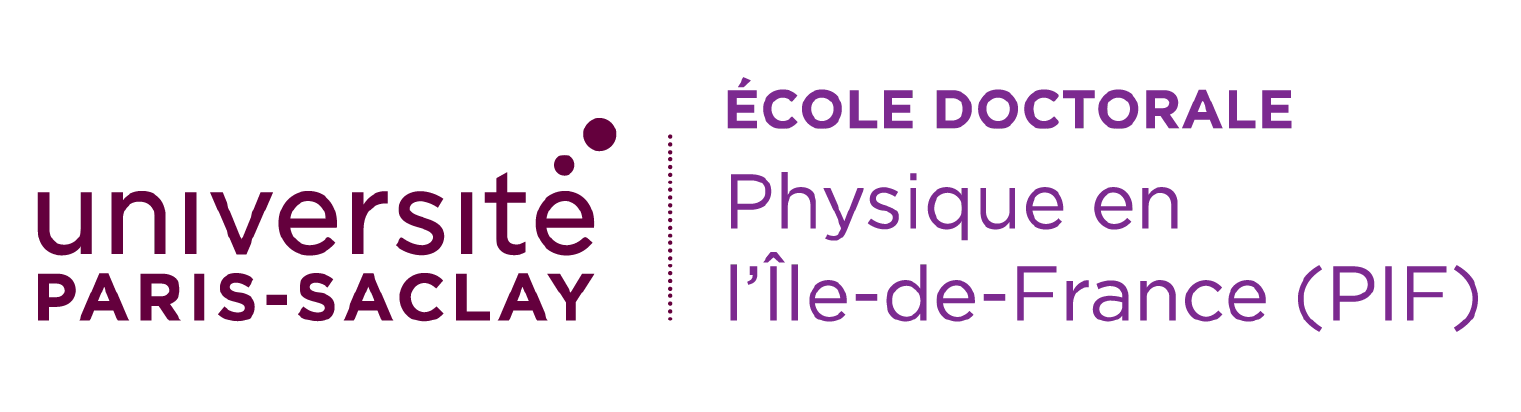}
\vspace{1cm}

\begin{mdframed}[linecolor=Prune,linewidth=1]
\vspace{-.25cm}
\paragraph*{Titre:} Evolution des jets dans un plasma quarks-gluons dense

\begin{small}
\vspace{-.25cm}
\paragraph*{Mots clés:} Chromodynamique quantique, Plasma quarks-gluons, Jet

\vspace{-.5cm}
\begin{multicols}{2}
\paragraph*{Résumé:} Afin de sonder les propriétés du plasma quarks-gluons créé dans les collisions d'ions lourds, on mesure des observables associées à la propagation de jets en son sein. Un jet est une gerbe collimatée de hadrons de haute énergie générée par des émissions successives de partons à partir d'un quark ou d'un gluon virtuel produit par la collision. Quand de telles gerbes se propagent dans le milieu dense créé par la collision des noyaux, leurs interactions avec ce milieu entraînent une modification dans leurs propriétés, phénomène appelé "réduction des jets". Dans cette thèse, nous développons une nouvelle théorie permettant de comprendre la réduction des jets. Nous calculons pour la première fois les effets du milieu dense sur les émissions de type vide dans les jets, c'est à dire les émissions déclenchées par la virtualité initiale du parton source. Une nouvelle image physique pour l'évolution des jets est présentée, dans laquelle les émissions de type vide sont factorisées en temps par rapport à celles induites par le milieu. Cette image est markovienne, donc adaptée pour une implémentation Monte-Carlo des cascades de partons que nous développons dans le programme {\tt JetMed}. Nous nous intéressons ensuite aux prédictions de notre théorie sur des observables de jets, et en particulier le facteur de modification nucléaire des jets $R_{AA}$, la distribution Soft Drop $z_g$ et la fonction de fragmentation. Ces prédictions se révèlent être en bon accord avec les mesures du LHC.
\end{multicols}
\end{small}
\end{mdframed}

\begin{mdframed}[linecolor=Prune,linewidth=1]
\vspace{-.25cm}
\paragraph*{Title:} Jet evolution in a dense QCD medium

\begin{small}
\vspace{-.25cm}

\vspace{-.5cm}
\begin{multicols}{2}
\paragraph*{Abstract:} To probe the properties of the quark-gluon plasma created in heavy-ion collisions, a very useful class of observables refers to the propagation of energetic jets. A jet is a collimated spray of particles generated via successive parton branchings, starting with a virtual quark or gluon produced by the collision. When such a jet is produced in the dense environment of a nucleus-nucleus collision, its interactions with the surrounding medium lead to a modification of its properties, phenomenon known as jet quenching. In this thesis, we develop a new theory to describe jet quenching. We compute for the first time the effects of the medium on
multiple vacuum-like emissions, that is emissions triggered by the virtuality of the initial parton. We present a new physical picture for jet evolution, with notably a factorisation in time between vacuum-like emissions and medium-induced emissions.
This picture is Markovian, hence well suited for a Monte Carlo implementation that we develop in the parton shower {\tt JetMed}. We then investigate the phenomenological consequences of our new picture on jet observables and especially the jet nuclear modification factor $R_{AA}$, the Soft Drop $z_g$ distribution and the jet fragmentation function. Our Monte Carlo results prove to be in good agreement with the LHC measurements.
\end{multicols}
\end{small}
\end{mdframed}

\vspace{2cm} 
\fontfamily{fvs}\fontseries{m}\selectfont
\begin{tabular}{p{14cm}r}
\multirow{3}{16cm}[+0mm]{{\color{Prune} Université Paris-Saclay\\
Espace Technologique / Immeuble Discovery\\
Route de l’Orme aux Merisiers RD 128 / 91190 Saint-Aubin, France}} & \multirow{3}{2.19cm}[+6mm]{\includegraphics[height=2.19cm]{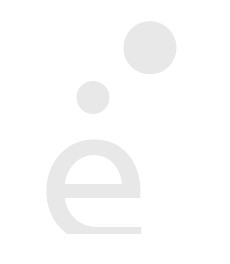}}\\
\end{tabular}

\end{document}

%% file: abstract.tex
\chapter*{Abstract}

Besides the emblematic studies of the Higgs boson and the search of new physics beyond the Standard Model, another goal of the LHC experimental program is the study of the quark-gluon plasma (QGP), a phase of nuclear matter that exists at high temperature or density, and in which the quarks and gluons are deconfined. This state of matter is now re-created in the laboratory in high-energy nucleus-nucleus collisions.
To probe the properties of the QGP, a very useful class of observables refers to the propagation of energetic jets. A jet is a collimated spray of hadrons generated via successive parton branchings, starting with a highly energetic and highly virtual parton (quark or gluon) produced by the collision. When such a jet is produced in the dense environment of a nucleus-nucleus collision, its interactions with the surrounding medium lead to a modification of its physical properties, phenomenon known as jet quenching.

In this thesis, we develop a new theory to describe jet quenching phenomena. Using a leading, double logarithmic approximation in perturbative QCD, we compute for the first time the effects of the medium on
multiple vacuum-like emissions, that is emissions triggered by the virtuality of the initial parton. We show that, due to the scatterings off the plasma, the in-medium parton showers differ from the vacuum ones in two crucial aspects: their phase-space is reduced and the first emission outside the medium can violate angular ordering. A new physical picture emerges from these observations, with notably a factorization in time between vacuum-like emissions and medium-induced parton branchings, the former constrained by the presence of the medium.
This picture is Markovian, hence well suited for a Monte Carlo implementation. We develop then a Monte Carlo parton shower called {\tt JetMed} which combines consistently both the vacuum-like shower and the medium-induced emissions. 

With this numerical tool at our disposal, we investigate the phenomenological consequences of our new picture on jet observables and especially the jet nuclear modification factor $R_{AA}$, the Soft Drop $z_g$ distribution and the jet fragmentation function. Our Monte Carlo results are in good agreement with the LHC measurements. We find that the energy loss by the jet is increasing with the jet transverse momentum, due to a rise in the number of partonic sources via vacuum-like emissions. This is a key element in our description of both $R_{AA}$ and the $z_g$ distribution. For the latter, we identify two main nuclear effects: incoherent jet energy loss and hard medium-induced emissions.
Regarding the fragmentation function, the qualitative behaviour that we find is in agreement with the experimental observations at the LHC: a pronounced nuclear enhancement at both ends of the spectrum. While the enhancement of hard-fragmenting jets happens to be strongly correlated with $R_{AA}$, hence controlled by jet energy loss, the enhancement of soft fragments is driven by the violation of angular ordering mechanism and the hard medium-induced emissions. 
We finally propose a new observable, which describes the jet fragmentation into subjets and is infrared-and-collinear safe by construction (therefore less sensitive to hadronisation effects) and we present Monte Carlo predictions for the associated nuclear modification factor.

\chapter*{Résumé}

Outre les tests du Modèle Standard des particules, un autre objectif du programme expérimental du grand collisionneur de hadrons (LHC) est l'étude du plasma de quarks et de gluons, une phase de la matière qui existe à haute température ou densité, et dans laquelle les quarks et les gluons sont déconfinés. Ce plasma est aujourd'hui recréé en laboratoire dans des collisions d'ions lourds de haute énergie. Pour sonder les propriétés du plasma, on mesure des observables associées à la propagation de ``jets'' en son sein. Un jet est une gerbe collimatée de hadrons tr\`{e}s énergétiques générée par des émissions successives de partons à partir d'un quark ou d'un gluon virtuel produit par la collision. Quand de telles gerbes se propagent dans le milieu dense créé par la collision de noyaux lourds, leurs interactions avec ce milieu entraînent une modification de leurs propriétés physiques, phénomène que l'on appelle "réduction des jets".

Dans cette thèse, nous développons une nouvelle théorie permettant de décrire la réduction des jets. En utilisant l'approximation double logarithmique en chromodynamique quantique perturbative, nous calculons pour la première fois les effets du milieu dense sur les émissions de type vide dans les jets. Ces émissions sont précisément celles déclenchées par la virtualité initiale du parton source. Nous montrons que les cascades de partons dans le milieu diffèrent de celles qui se développent dans le vide à cause des diffusions multiples par les constituants du milieu: l'espace de phase pour les émissions de type vide est réduit et la première émission à l'extérieur du milieu peut violer l'ordonnancement angulaire. Une nouvelle image physique émerge alors de ces observations, dans laquelle les émissions de type vide sont factorisées en temps par rapport à celles induites par le milieu. Cette image a l'avantage d'être markovienne, et donc adaptée pour une implémentation Monte-Carlo des cascades de partons, que nous développons dans le programme {\tt JetMed}. Ce programme combine donc de fa\c{c}on coh\'{e}rente les cascades de type vide et les cascades induites par le milieu.

Grâce à cet outil numérique, nous nous intéressons ensuite aux prédictions de notre théorie sur des observables de jets, et en particulier le facteur de modification nucléaire des jets $R_{AA}$, la distribution Soft Drop $z_g$ et la fonction de fragmentation. Ces prédictions se révèlent être en bon accord avec les mesures du LHC. Nous trouvons que la perte d'énergie des jets augmente avec leur impulsion transverse à cause d'une augmentation du nombre de sources partoniques produites par les émissions de type vide dans le milieu. C'est un élément essentiel dans notre description de $R_{AA}$ et de $z_g$. Pour cette dernière observable, nous identifions deux principaux effets nucléaires: la perte d'énergie incohérente des jets et les émissions induites par le milieu relativement dures. Le comportement de la fonction de fragmentation que nous obtenons est en accord avec les mesures du LHC ; en particulier nous observons une augmentation prononcée du nombre de fragments durs et mous. Dans la partie dure, cette augmentation est corrélée avec $R_{AA}$ et donc controllée par la perte d'énergie des jets. Au contraire, l'augmentation de fragments mous est due à la violation de l'ordonnancement angulaire et aux émissions dures à petit angle induites par le milieu. Nous proposons finalement une nouvelle observable qui décrit la fragmentation des jets en termes de sous-jets, donc moins sensible aux effets d'hadronisation, et nous la calculons en collisions d'ions lourds.

%% file: remerciements.tex
\chapter*{Remerciements - Acknowledgements}

\hspace{\parindent} Cette th\`{e}se est le fruit de trois ans de travail, dont trois mois et demi d'\'{e}criture entre avril et juillet 2020. Je tiens \`{a} remercier ici les personnes qui ont rendu possible l'\'{e}mergence de ce manuscrit. En premier lieu, il y a \'{e}videmment mes directeurs, Edmond et Gregory. Ils m'ont fait d\'{e}couvrir leur domaine de recherche et leur passion avec une g\'{e}n\'{e}rosit\'{e} rare. Je mesure de plus en plus la chance que j'ai eu de travailler avec eux pendant ces trois ann\'{e}es, et je pense qu'une th\`{e}se ne suffirait pas pour \'{e}crire tout ce qu'ils m'ont apport\'{e}.

\vspace{0.4cm}

Je souhaite ensuite remercier Al Mueller qui est aussi \`{a} l'origine des travaux pr\'{e}sent\'{e}s dans la suite de ce document. Les discussions que nous avons pu avoir ont \'{e}t\'{e} riches d'enseignement. Ces moments dans ma vie de physicien balbutiant me sont pr\'{e}cieux.

\vspace{0.4cm}

Merci aux membres de l'\'{e}quipe QCD de l'Institut de Physique Th\'{e}orique, permanents ou seulement de passage, en particulier Jean-Paul Blaizot, Fran\c{c}ois Gelis, Giuliano Giacalone, Davide Napoletano, Jean-Yves Ollitrault et Vincent Theeuwes dont la pr\'{e}sence au laboratoire a donn\'{e} \`{a} mes journ\'{e}es de travail cette dimension humaine essentielle. Les \'{e}changes scientifiques sur des sujets connexes au mien m'ont beaucoup aid\'{e} \`{a} me faire une id\'{e}e d'ensemble du domaine dans lequel s'inscrit cette th\`{e}se.

\vspace{0.4cm}

I thank the referees Carlos Salgado and Konrad Tywoniuk for accepting to read and comment this way too long manuscript. I thank also Matteo Cacciari, Leticia Cunqueiro and Samuel Wallon for accepting the invitation to be part of my jury, and for having physically been at my PhD defence in spite of the complications caused by the Covid19.

\vspace{0.4cm}

I would like to thank all the researchers of the nuclear theory group at the Brookhaven National Laboratory for their hospitality during my stay in July 2019. Je voudrais en particulier remercier Yacine Mehtar-Tani pour cette invitation. Ce s\'{e}jour a \'{e}t\'{e} tr\`{e}s enrichissant et stimulant pour la suite.

\vspace{0.4cm}

Merci aux membres de la SFDJAP pour ces r\'{e}unions fort instructives qui m'ont permis de prendre du recul sur mon propre travail. Mes charges d'enseignement en physique et les \'{e}tudiants de premi\`{e}re ann\'{e}e de licence en physique et biologie de l'universit\'{e} d'Orsay ont aussi beaucoup contribu\'{e} \`{a} cette prise de recul indispensable. Je remercie mes \'{e}l\`{e}ves et les responsables des unit\'{e}s d'enseignement, en particulier Brigitte Pansu et Cyprien Morize.

\vspace{0.4cm}

Pour conclure, et non sans \'{e}motions, merci \`{a} mes amis, \`{a} ma famille, et \`{a} Manon.

%% file: intro.tex
\chapter{General introduction}
\label{chapter:intro}

\section{Quantum chromodynamics}

\subsection{Quarks and gluons}

Quantum chromodynamics (QCD) is the quantum theory of the strong interaction. The strong interaction binds together the protons and neutrons in atomic nuclei. However, protons and neutrons and the other bound states of the strong force called \textit{hadrons}, such as the pions, kaons, etc, which are detected in high-energy experiments are not the elementary degrees of freedom of the theory. These elementary degrees of freedom from which quantum chromodynamics is formulated are called \textit{quarks} and \textit{gluons}: they are the fundamental constituents of the hadronic matter, the matter sensitive to the strong interaction. 

The existence of ``point-like'' constituents inside hadrons is revealed in high energy scattering experiments, such as the deeply inelastic electron-proton scattering: at asymptotically high momentum transferred (squared) $Q^2$ during the collision process, 
the scattering cross-section behaves as if the proton was made of free elementary particles. This phenomenon was one of the first experimental evidence for quarks and gluons within hadrons \cite{Bjorken:1968dy,Bjorken:1969ja,Feynman:1969ej}.

In high energy experiments like deep inelastic electron-proton scattering or those described in Section~\ref{sec:HIC} of this introduction, special relativity effects cannot be neglected. The theoretical framework to deal with quarks and gluons is thus quantum field theory. Quarks and gluons are the elementary degrees of freedom in the sense that they appear as \textit{elementary fields} in the Lagrangian $\mathcal{L}_{\rm QCD}(\psi_f,A_\mu)$ of quantum chromodynamics. $\psi_f(x)$ are the quark Dirac fields labelled by their flavour indices $f=u,d,s,c,b,t$ for the six quark flavours of the Standard Model and $A_\mu(x)$ is the gluon vector field.

\subsection{Non abelian local gauge symmetry and asymptotic freedom}

\paragraph{Local $SU(3)$ gauge symmetry.} This Lagrangian is strongly constrained by the \textit{local} gauge invariance principle. More precisely, on imposes that $\mathcal{L}_{\rm QCD}$ is invariant under $SU(N_c=3)$ local transformations of the quark fields:
\begin{equation}\label{gauge-trans-q}
 \psi'_f(x)=U(x)\psi_f(x)\,,\qquad U(x)=\exp(i\epsilon_a(x)t^a)
\end{equation}
with $\{t^a\}_{1\le a\le N_c^2-1}$ a set of generators in the fundamental representation of the $\mathfrak{su}(N_c)$ Lie algebra and $\epsilon_a(x)$ arbitrary functions of the space-time coordinates. The index $a$ is called a colour index. Since the group $SU(N_c)$ is non commutative for $N_c>1$, the $SU(N_c)$ gauge symmetry is said non-abelian \cite{Yang:1954ek}. As part of the fundamental representation of $SU(N_c)$, quarks carry a colour charge, pretty much like electrons and positrons carry an electric charge in quantum electrodynamics (QED). The number $N_c$ of possible ``colours'' is $3$ in the physical world.

A system of \textit{free} quarks with (bare) masses $m_f$ is simply described by the free Lagrangian\footnote{We adopt standard particle physics units with $\hbar=c=1$.}:
\begin{equation}\label{free-quarks}
 \mathcal{L}(\psi_f)=\sum\limits_{f=1}^{n_f}\,\bar{\psi}_f\big(i\gamma^\mu \partial_\mu-m_{f}\big)\psi_f
\end{equation}
but is not invariant under \eqref{gauge-trans-q}.
In order for this Lagrangian to be invariant under the transformation \eqref{gauge-trans-q}, a gluon field $A_\mu$ belonging to the adjoint representation of $SU(N_c)$ must be included, with the following transformation rule:
\begin{equation}\label{gauge-trans-g}
 A'_\mu(x)=U(x)A_\mu U^{-1}(x)-\frac{i}{g_s}\partial_\mu U(x)U^{-1}(x)\,,\qquad A_\mu(x)=A_\mu^a(x)t^a,
\end{equation}
and the quark fields must couple with $A_\mu$ via a covariant derivative $\partial_{\mu}\rightarrow D_\mu=\partial_\mu-ig_sA_\mu$ in \eqref{free-quarks}. $g_s$ is the (bare) strong coupling constant that quantifies the strength of the strong interaction. The fact that gluons carry also a colour charge, as member of a non-trivial representation of the gauge group, leads to a rich set of phenomena specific to the strong interaction. 

One can then check that the following Lagrangian is invariant under the transformations \eqref{gauge-trans-q}-\eqref{gauge-trans-g} and is unique if renormalizability and parity invariance are also enforced:
\begin{equation}\label{QCD-Lagrangian}
 \mathcal{L}_{\rm QCD}(\psi_f,A)=-\frac{1}{4}\Tr F_{\mu\nu}F^{\mu\nu}+\sum\limits_{f=1}^{n_f}\,\bar{\psi}_f\big(i\gamma^\mu D_\mu-m_{f}\big)\psi_f
\end{equation}
 The gauge field tensor $F_{\mu\nu}$ depends on the gluon gauge field $A_{\mu}$ as:
\begin{align}
 F_{\mu\nu}&=\partial_\mu A_\nu-\partial_\nu A_\mu+g_s[A_\mu, A_\nu]
\end{align}
This gluon field is the analogous of the photon field in quantum electrodynamics: gluons are said to mediate the strong interaction. This mediation is realized by means of the covariant derivative $D_\mu$ in the second term of the Lagrangian. The QCD Lagrangian is remarkable in several respects.

\paragraph{Gluon self-interaction.} The gauge field $A_\mu$ is self interacting at tree level because of the non-abelian term $g_s[A_\mu,A_\nu]$. In comparison, the gauge field of quantum electrodynamics is the photon field, and photons do not interact between themselves at the lowest non trivial order in the electromagnetic coupling $e$. This has important phenomenological consequences, especially for QCD jets, as we shall see.

\paragraph{Asymptotic freedom.} As it is, the Lagrangian \eqref{QCD-Lagrangian} has no predictive power beyond tree-level in perturbation theory because of the UV divergences appearing from the one-loop order. These divergences are cured thanks to the renormalization procedure: the bare coupling $g_s$ is discarded for the benefit of the \textit{renormalized} strong coupling constant $g$. A by-product of this procedure is that $g$ acquires a scale dependence, $g(\mu)$, with $\mu$ a sliding energy scale one chooses to define the renormalized coupling. Measurable quantities are invariant with respect to the choice of $\mu$. To satisfy this requirement, the variations of the function $g(\mu)$ are constrained by the $\beta$-function of QCD which is calculable order by order in perturbation theory (hence assuming that $g(\mu)$ is small)\footnote{In this thesis, the natural logarithm of $x$ is noted $\log(x)$.}:
\begin{equation}
 \frac{\partial g(\mu)}{\partial \log(\mu)}=\beta(g(\mu))=-\beta_0\frac{g^3}{4\pi}+\mathcal{O}(g^5)\,,\qquad\beta_0=\frac{11C_A-2n_f}{12\pi}>0
\end{equation}
$C_A=(N_c^2-1)/(2N_c)$ is the Casimir factor of the adjoint representation of $\mathfrak{su}(N_c)$.
At one loop order in perturbation theory, that is using only the first term in the Taylor expansion of $\beta(g)$, one gets:
 \begin{equation}
  \alpha_s(\mu)=\frac{1}{\beta_0\log\left(\frac{\mu^2}{\LQCD^2}\right)}\,,\qquad \alpha_s=\frac{g^2}{4\pi}
 \end{equation}
where $\LQCD\sim 0.2$ GeV\footnote{1 GeV $\simeq1.6\times 10^{-19}$ J and 1 fm = $1\times 10^{-15}$ m.} is the confinement or QCD scale discussed in the next point. A striking feature of the strong running coupling is that $\alpha_s(\mu)\rightarrow 0$ as $\mu\rightarrow \infty$. This property is known as \textit{asymptotic freedom} \cite{Gross:1973id,Politzer:1973fx,tHooft:1973mfk,tHooft:1985mkt}. For a physical process governed by the strong interaction and involving a typical momentum transfer of order $Q$, the more $Q$ is large: $Q\gg \LQCD$, the more $\alpha_s(Q)$ is small and hence the more the process becomes similar to what one would expect of a non interacting theory of quarks and gluons. In this regime, perturbative techniques, generically gathered under the term ``perturbative QCD'' (pQCD) are the standard tools to predict the behaviour of such physical processes. This is the realm of high energy collision experiments such as those performed at the Large Hadron Collider (LHC).

\paragraph{Confinement.} On the other hand, when $\mu\rightarrow \Lambda_{\rm QCD}\sim0.2$ GeV, the strong coupling becomes very large suggesting that perturbative techniques and perturbative degrees of freedom (quarks and gluons) become irrelevant. This is closely related to another property of QCD, \textit{confinement}, even if it is still not clear how confinement emerges from the QCD Lagrangian. This property states that in normal conditions of temperature and pressure, free quarks and gluons cannot exist and are always confined within colour neutral bound states called hadrons. At least, this is an experimental fact up to now: there are no free quarks and gluons, nor coloured particles, measured in low temperature/pressure experiments. At low energy scales, below or around $\LQCD$, the QCD Lagrangian is ``solved'' using lattice field theory. The quark and gluon fields are anchored on a finite lattice and the equations are studied numerically on a computer. Lattice QCD is very powerful to compute the mass spectrum of hadrons for instance.

\subsection{Phases of hadronic matter}

In our discussion of confinement, we have avoided to precise what we meant by ``normal conditions of temperature and pressure''. This paragraph intends to specify these conditions. Let us enclose some hadronic matter in a finite box of volume $V$, in thermodynamical equilibrium with a heat bath of temperature $T$ and a particle reservoir with baryon chemical potential $\mu_B$. The baryon chemical potential is the chemical potential associated with the baryon number $ B$:
\begin{equation}
 B=\frac{1}{3}(N_q-N_{\bar{q}})
\end{equation}
for a system with $N_q$ quark and $N_{\bar{q}}$ antiquark. How does this matter behave as the temperature  of the heat bath, the volume of the box  or the chemical potential $\mu_B$ are varied? This section is a short review of what is known on this subject. 

\begin{figure}[t]
   \centering
      \includegraphics[width=0.6\textwidth]{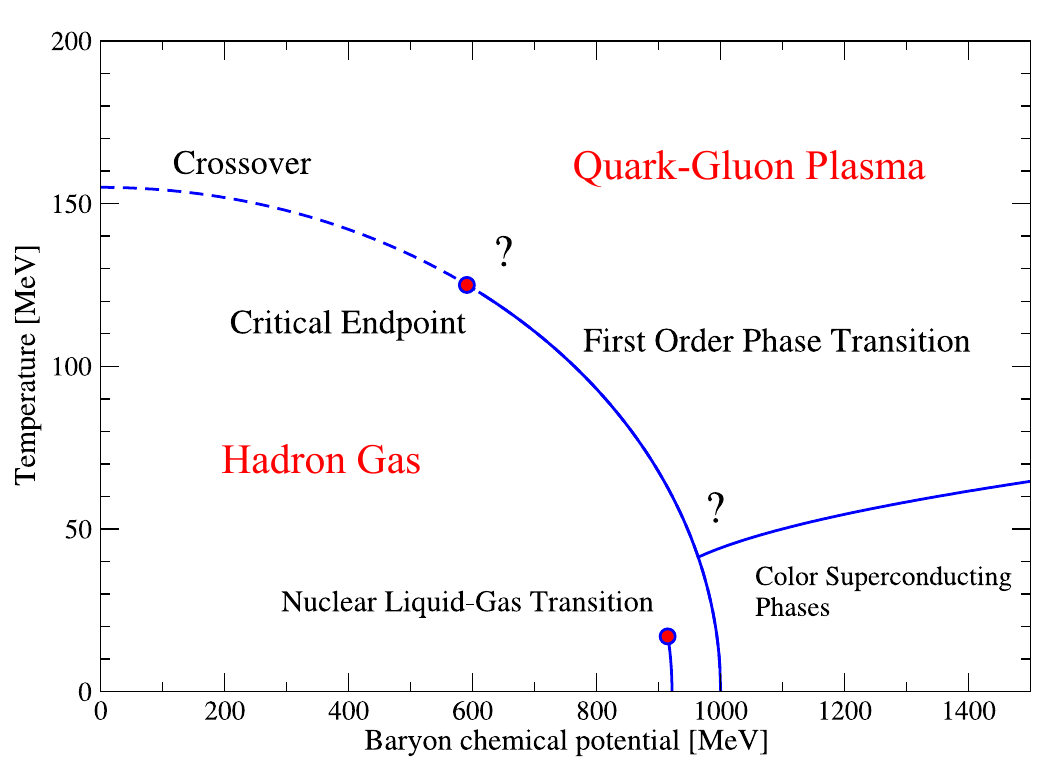}
    \caption{\small The conjectured phase diagram of QCD matter (extracted from \cite{Fischer:2018sdj}) projected on the plane $(\mu_B,T)$. The Boltzmann constant is set to one $k_B=1$ for the units of the horizontal and vertical axis.  The nuclear liquid-gas phase transition is the small line at the bottom of the diagram. The low temperature but high $\mu_B$ part of the diagram with the ``colour superconducting'' phase is not discussed here, but is relevant for neutron star physics.}
    \label{Fig:QCD-ps}
\end{figure}

\paragraph{Basic features.} Some features can be guessed with very few physical inputs. For instance, as the radius of hadrons is of order $R_h\sim1$ fm, there should exist a critical volume for our box, of order $V_c=4\pi R_h^3/3$. Indeed, for a volume smaller than $V_c$, hadrons cannot fit in the box any more. 
Another interesting regime is the very high temperature regime. We have seen that pQCD has a self-emergent scale characteristic of the strong interaction, $\Lambda_{\rm QCD}$: one expects that quarks and gluons are the relevant degrees of freedom for $T\gg \LQCD/k_B$, because the strong coupling constant $\alpha_s(k_BT)$ --- naturally evaluated at a scale of the order of the thermal energy of the system --- becomes very small from the asymptotic freedom property of QCD. Thus, in the high temperature phase $T\gg\LQCD/k_B$, the system behaves like an ideal gas of (free) quarks and gluons. This phase is known as quark-gluon plasma. In the present thesis, there is almost no need to give further details on this phase as all the knowledge required to understand these lines is encoded in this definition. For completeness, we provide now some additional materials from hadronic physics and lattice QCD on the conjectured phase diagram of QCD shown Fig.~\ref{Fig:QCD-ps}.

\paragraph{Liquid-gas phase transition of nuclear matter.}  We start from the ordinary temperature $T\sim 270$-290 K and baryon chemical potential $0.9\lesssim\mu_b\lesssim 1$ GeV$/k_B$. Under these conditions, the hadronic matter in our box is a mixed phase between vacuum and droplets of Fermi liquid formed by the nucleons --- protons and neutrons --- inside nuclei. The interaction between the nucleons looks very much like the Van der Waals interaction between electrostatic dipoles: this is confirmed by lattice QCD results on the nucleon-nucleon force and by calculation in the chiral field theory which is an effective field theory of QCD at low energy (see \cite{Fukushima:2013rx} for a review). Thus, in the same way as water undergoes a liquid-gas transition, we expect a first order phase transition towards a Fermi gas of free nucleons when the temperature increases. The critical point of the transition is located around $T'_c\sim 10$-20 MeV$/k_B$ and $\mu'_{Bc}\sim900$-$920$ MeV$/k_B$.

\paragraph{Hadron resonance gas.} Now, let us heat even more this system of free nucleons (or pions at lower chemical potential). This corresponds to the phase diagram of hadronic matter for $\mu_B\lesssim 1$ GeV and $T\gtrsim 10$ MeV. If we heat this system, we know from experiments that more and more hadronic resonances are produced: the number density $\rho(m)$ of hadronic states with mass between $m$ and $m+\dif m$ grows exponentially $\rho\sim\exp(c m)$. This hadronic phase is called ``hadron resonance gas'' (the degrees of freedom are still hadrons). The exponential growth cannot go forever, otherwise, the partition function of the system diverges at a temperature known as the Hagedorn temperature $T_h=1/c$ \cite{Hagedorn:1965st}. From experimental data, $T_h\approx 150$ MeV$/k_B$, value close to $\LQCD$. This suggests that $T_h$ should be interpreted as an upper limit for the hadronic phase and should be close to the transition temperature between this phase and a deconfined phase where quarks and gluons are liberated \cite{Cabibbo:1975ig}. The exact nature of this phase transition is still unknown.

\begin{figure}[t] 
  \centering
  \begin{subfigure}[t]{0.52\textwidth}
    \includegraphics[width=\textwidth]{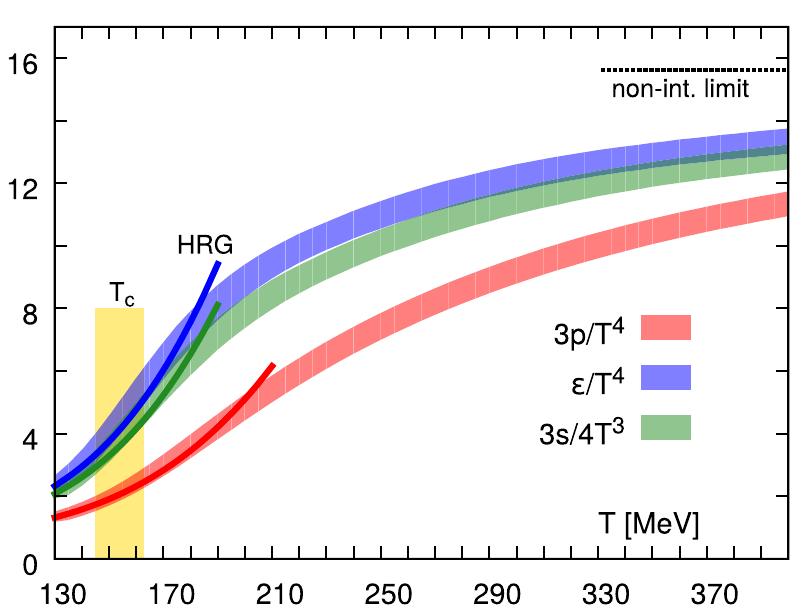}
  \end{subfigure}
  \hfill
  \begin{subfigure}[t]{0.43\textwidth}
    \includegraphics[width=\textwidth]{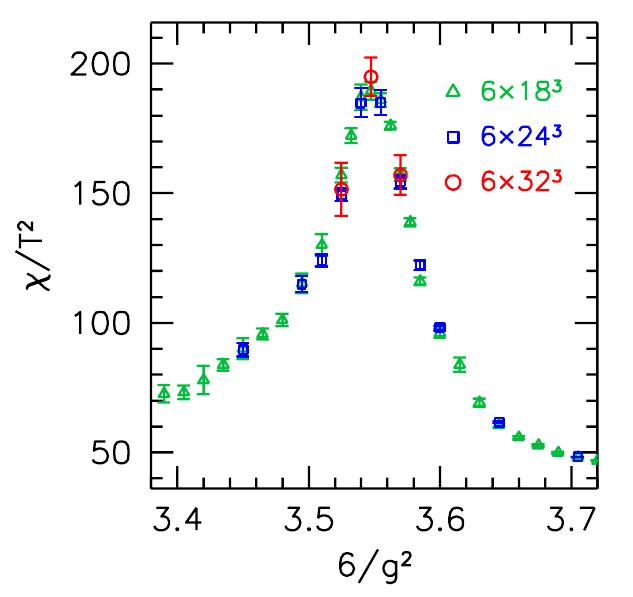}
  \end{subfigure}
  \caption{\small (Left) QCD equation of state at zero chemical potential $\mu_B=0$ obtained from lattice calculations by the HotQCD collaboration \cite{Bazavov:2014pvz}. The curve labelled ``HRG'' corresponds to the prediction from the hadron resonance gas model, whereas the dotted line at the top is the non-interacting (free) gas limit. (Right) Variation of the chiral susceptibility defined by \eqref{chiral-su-def} with $6/g^2$ ($g$ is the gauge coupling, and $T$ grows with $1/g^2$) for several lattice volumes $N_t\times N_s^3\times a^4$ with $a$ the lattice spacing. The figure is taken from \cite{Aoki:2006we}.} \label{Fig:QCD-eos}
\end{figure}

\paragraph{Lattice QCD results for $\mu_B=0$.} As explained in the first part of this introduction, lattice QCD is extremely useful to understand QCD in the strong coupling regime. In thermal QCD at finite temperature, the partition function is evaluated numerically on a finite lattice. Unfortunately, because of the so-called \textit{sign problem}, lattice simulations cannot access non-zero baryon chemical potential (at least until now) \cite{deForcrand:2010ys}. Hence, lattice QCD provides reliable informations on the QCD phase space restricted to the line $\mu_B=0$. Results from the HotQCD collaboration \cite{Bazavov:2014pvz} are shown in Fig.~\ref{Fig:QCD-eos}-left for the energy density $\epsilon$, the pressure $p$ and the entropy density $s=(\epsilon+p)/T$ as a function of the temperature $T$ of the bath. At low temperature, the predictions from the hadron resonance gas description are in good agreement  with the lattice QCD results. At high temperature, the curves converge towards the non-interacting limit of a free gas of massless quarks and gluons following the Stefan-Boltzmann equation of state:
\begin{equation}
 \frac{3p}{T^4}=\frac{\epsilon}{T^4}=n\frac{\pi^2}{30}
\end{equation}
with $n=2(N_c^2-1)+7N_cn_f/2$ the total number of internal degrees of freedom. However, even at $T\sim 400$ MeV$/k_B$, the free limit is not reached suggesting that residual interactions have still an important role, in agreement with perturbative results in QCD at finite temperature. To sum up, at zero chemical potential, lattice QCD predicts a gradual transition between the hadron resonance gas and the deconfined phase.

There is another transition related to the deconfinement transition known as the chiral phase transition: at high temperature $T\gg\LQCD/k_B$, the light quarks (up and down) can be considered as massless and the QCD Lagrangian \eqref{QCD-Lagrangian} is symmetric under global chiral transformations of the light quark fields. At low temperature $T\ll\LQCD/k_B$, chiral symmetry is spontaneously broken, with the pions being the Nambu-Goldstone bosons associated with this symmetry breaking. Lattice QCD enables to have a more quantitative insight on the location of the transition. The order parameter is the chiral susceptibility defined by:
\begin{equation}\label{chiral-su-def}
 \chi =\frac{T}{V}\frac{\partial \log(Z_{\rm QCD})}{\partial m_l}
\end{equation}
where $Z_{\rm QCD}$ is the QCD partition function in the grand canonical ensemble at $\mu_B=0$ and $m_l$ is the mass of the light quarks. A plot of the chiral susceptibility for several values of the volume lattice is shown in Fig.~\ref{Fig:QCD-eos}-right from \cite{Aoki:2006we}.  A clear peak is visible as a function of $6/g^2$ ($6/g^2$ scales like the temperature $T$ of the heat bath). However, this peak is independent of the volume of the lattice, showing that the chiral transition at $\mu_B=0$ is not first nor second order but rather an analytic cross-over. The location of the peak defines the pseudo-critical temperature $T_c$ of the transition, whose value is around 150 MeV$/k_B$.

For non vanishing chemical potential the nature of the chiral transition is still unknown, even if there is a hint for a first order phase transition as shown Fig.~\ref{Fig:QCD-ps}. The determination of the critical point is an active field of research both on the theoretical and experimental sides. Regarding the link between the chiral and the deconfinement transition, they should occur at about the same temperature. Therefore, understanding the chiral transition is crucial to precise the shape of the phase diagram of hadronic matter.

\section{Heavy-ion collisions}
\label{sec:HIC}

How do physicists probe the QCD phase diagram experimentally? The very early Universe was probably a hot soup of free quarks and gluons, so physicists could look for relics of this epoch to infer some properties of the quark-gluon plasma. Also, the heart of neutron stars might be an interesting natural laboratory for studying QCD at high density. Besides these two occurrences of QCD matter at extreme temperature or density in Nature, this matter is studied on Earth in high energy experiments by colliding together heavy nuclei with a large number of nucleons. The two main accelerators used for this purpose are the Relativistic Heavy Ion Collider (RHIC) based at the Brookhaven National Laboratory on Long Island (US) and the Large Hadron Collider (LHC) based at the border between France and Switzerland, near Geneva.

\subsection{Experimental aspects}

We start by describing briefly some experimental aspects relating to nucleus-nucleus collisions relevant for this thesis, and especially the second Part. At RHIC, the heavy nuclei are mainly atoms of copper (Cu), gold (Au) or uranium (U). At the LHC, most of the data discussed in Part~\ref{part:phenomenology} come from lead-lead (Pb) collisions analysis. A run of xenon-xenon collisions (Xe) also took place at the LHC during fall 2017.

\begin{figure}[t] 
  \centering
  \begin{subfigure}[t]{0.68\textwidth}
    \includegraphics[width=\textwidth]{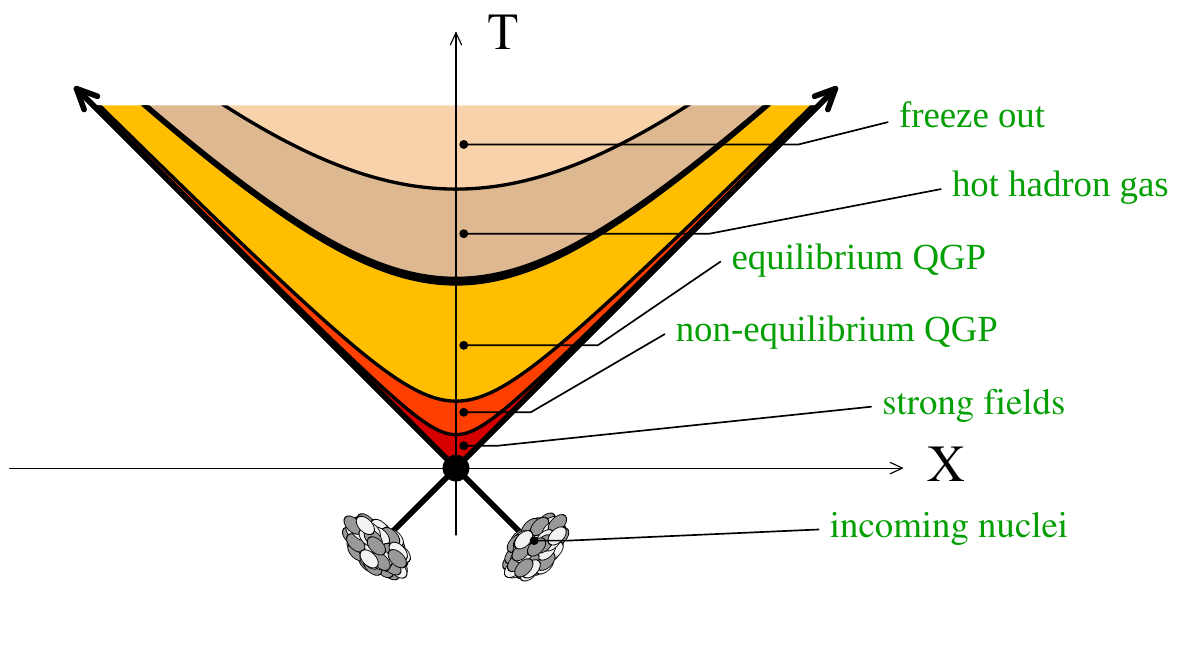}
  \end{subfigure}
  \hfill
  \begin{subfigure}[t]{0.3\textwidth}
    \includegraphics[width=\textwidth]{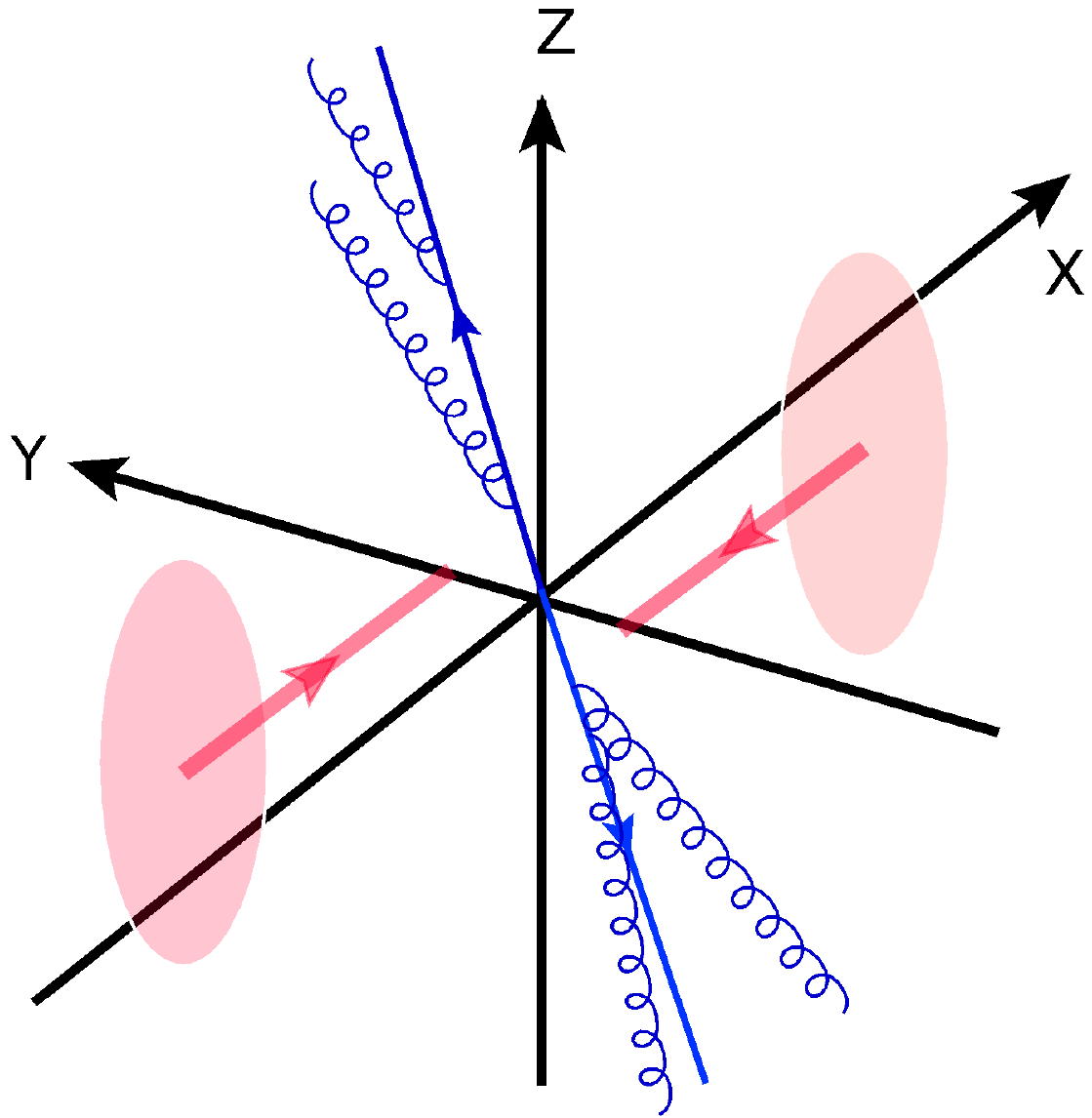}
  \end{subfigure}
  \caption{\small (Left) Space-time diagram of a nucleus-nucleus collision (figure from \cite{Iancu:2012xa}). The horizontal axis is the beam axis, and the vertical axis is the proper time of an observer in the laboratory frame. As the nuclei are ultra-relativistic, their worldlines lie on the light cone with apex at the collision event. By Lorentz contraction, the nuclei appear as thin sheets of matter. The hyperboles are lines with constant fluid proper time $\tau=\sqrt{T^2-X^2}$: it is the proper time of an observer comoving with a fluid element in the Bjorken expansion model. The several stages of the collision between two values of $\tau$ are described. The first stage, during which the partons (mostly gluons) are freed by the collision, lasts $\sim 0.2$ fm. In the second stage, with $0.2\lesssim\tau\lesssim1$ fm, the partonic matter rapidly approaches thermal equilibrium. The longest stage in the yellow region corresponds to a QGP in thermal equilibrium, up to $\tau\sim10$ fm at the LHC. The cooling of the QGP leads to a hot system of hadron resonances which finally becomes a system of free hadrons at the freeze-out proper time $\tau\sim 10-20$ fm. (Right) The coordinate system in the laboratory frame adopted in this thesis. The two incoming nuclei are shown \textit{before} the collision, together with their trajectories. A dijet event is illustrated as well: two cascades of energetic partons develop back to back in the transverse plane with respect to the beam axis.}\label{Fig:cartoon}
\end{figure}

In such high-energy collisions, two beams of heavy and entirely ionised nuclei accelerated at nearly the speed of the light circulate in opposite directions in a large synchrotron. The Lorentz factor of the nuclei in larger than $2500$ at the LHC! These two beams meet in specific locations of the ring where particle detectors are built. When the two beams meet, two nuclei belonging to their respective beam collide and create a myriad of particles ($\sim35000$ on average in frontal collisions at the LHC)  which are measured in the detectors. Understanding what happens between the collision and the particle detection is the main goal of these experiments. The current ``standard model'' of a high energy nucleus-nucleus collision is pictured in Fig.~\ref{Fig:cartoon}-left. Shortly after the collision, a plasma of free quarks and gluons is created that lives during approximatively 10 fm ($\sim10^{-23}$ seconds). After being formed, it expands essentially along the beam direction, cools down and hadronises into final-state hadrons.

To understand the experimental data shown in Chapter~\ref{chapter:intro-part2}, a bit of vocabulary and kinematic notations are useful. 

\paragraph{Laboratory frame.} The laboratory frame is an inertial frame such that the center of mass frame of the two nuclei is fixed at the origin $O$ of the coordinate system. In this frame, the total $3$-momenta of the colliding nuclei vanishes and $O$ is the location of the collision event. To each event in Minkowksi space, one associates a four vector $X^\mu\bold{e}_\mu$, with $\{\bold{e}_\mu\}$ the basis vectors of the laboratory coordinate system and $X^\mu$ the coordinates of the event. This basis is chosen so that $\bold{e}_1=\bold{e}_{\textsc{x}}$ is parallel to the beam axis. The vector $\bold{e}_2=\bold{e}_{\textsc{y}}$ is parallel to the impact vector. The impact vector is a vector in the plane transverse to the collision axis connecting the centers of the two incoming nuclei. The vector $\bold{e}_3=\mathbf{e}_{\textsc{z}}$ lies in the transverse plane and is orthogonal to $\bold{e}_2$. Finally, $\bold{e}_0=\mathbf{e}_{\textsc{t}}=u$ where $u$ is the 4-velocity of an observer fixed in the laboratory frame. This coordinate system is pictured Fig.~\ref{Fig:cartoon}-right. For collisions with exactly zero impact parameter, the choice of the vectors $\bold{e}_2$ and $\bold{e}_3$ is arbitrary as long as the basis $\{\bold{e}_\mu\}$ is orthonormal and direct (in the sense of the metric tensor $g_{\mu\nu}$).

In this frame and coordinate system, one defines the following quantities related to a generic $4$-momentum $p=p^\mu\bold{e}_\mu$ of a particle:
\begin{align}
\mbox{Energy: }&\qquad& E=&p^0\\
 \mbox{Transverse momentum: }&\qquad& p_T=&\sqrt{p_2^2+p_3^2}\\
 \mbox{Rapidity: }&\qquad& y=&\frac{1}{2}\log\left(\frac{p^0+p^1}{p^0-p^1}\right)\\
  \mbox{Pseudo-rapidity: }&\qquad& \eta=&\frac{1}{2}\log\left(\frac{\sqrt{p_1^2+p_2^2+p_3^2}+p^1}{\sqrt{p_1^2+p_2^2+p_3^2}-p^1}\right)
\end{align}
For massless particles ($p^2=p^\mu p_\mu=0$), one has $y=\eta$, and in the mid-rapidity region ($y=0$), one has $E\simeq p_T$. In this thesis, the geometrical aspects of high-energy massless particle production are simplified: they are considered as produced in the transverse plane with $p^1=0$. Energy and transverse momentum are therefore synonyms. The pseudo-rapidity is related to the angle $\th$ in the spherical coordinate with axis $\bold{e}_1$ by $\eta=-\log(\tan(\th/2))$. Finally the azimuthal angle in these spherical coordinates is defined as $p_T\cos(\phi)=p^{\textsc{y}}$.

\paragraph{Collision energy.} An important quantity is the center of mass energy of the collision, noted $\sqrt{s}$ defined by the first Mandelstam variable:
\begin{equation}
 s=(p_1+p_2)^2
\end{equation}
where $p_1$ and $p_2$ are the 4-momenta of the two colliding particles. Experimentalists control this quantity by tuning the beam energy $E_{\rm beam}$ in the laboratory frame, since $\sqrt{s}=2E_{\rm beam}$. In nucleus-nucleus collision, this quantity is averaged per nucleons, so that the common quantity is the $\sqrt{s}$ per nucleon pair:
\begin{equation}
 \sqrt{s_{\rm NN}}=\frac{Z}{A}\sqrt{s}
\end{equation}
for a generic nucleus ${}^A_Z X$. At the LHC, most of the data related to jets in heavy-ion collisions discussed in this thesis have $\sqrt{s_{\rm NN}}=2.76$ TeV or $\sqrt{s_{\rm NN}}=5.02$ TeV.

\paragraph{Centrality.} The impact vector is difficult to measure experimentally and especially its length, the impact parameter. Instead, experimentalists use the concept of centrality. Centrality can be obtained directly from data. There are several ways of defining a centrality. Starting from a set of events called minimum-biased (meaning there is no further requirement on how these data should be, besides the registration trigger), these events are sorted according to the value of an observable $N$ which is a monotonic function of the impact parameter. For instance, one can bin the events according to the total particle multiplicity or the total energy deposited in the detectors. One easily understands that the more the impact parameter is small, the more the total multiplicity or energy is large. Another possible observable is the energy deposited in the ``zero-degree'' calorimeter, located close to the beam direction. This energy is related to the number of spectator nucleons, and therefore increases as the impact parameter increases.

A centrality class $0-c(n)$ is then a subset of the data such that a fraction $c(n)$ satisfies $N\ge n$. For instance, in Fig.~\ref{fig:ATLAS-RAA}, $0-10$\% centrality means that the events selected for the measurement belong to the 10\% events with highest particle multiplicity.

To relate quantitatively the centrality to the impact parameter $b$, one uses the following geometric relation (see e.g. \cite{Florkowski:2010zz}):
\begin{equation}
 c\simeq \frac{\pi b^2}{\sigma^{AA}_{\rm in}}
\end{equation}
This formula neglects the fluctuations in the probability distribution of measuring a value $n$ of $N$ given the impact parameter $b$, and assumes that the nucleus $A$ behaves like a ``black disk'' for $b\lesssim R$, with $R$ the nucleus radius.

\begin{figure}[t] 
\centering
  \begin{subfigure}[t]{0.57\textwidth}
    \centering
    \includegraphics[width=\textwidth]{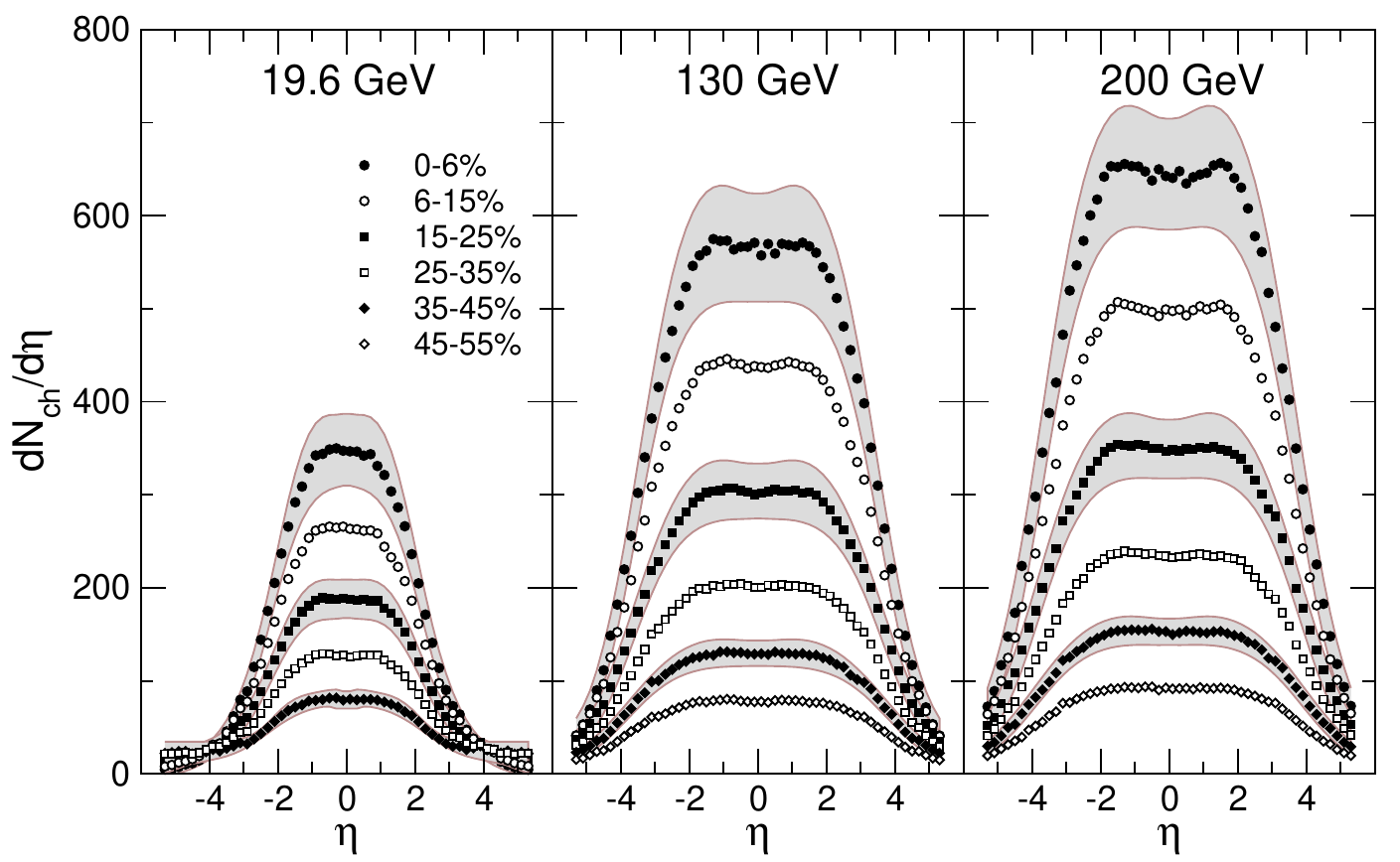}
  \end{subfigure}
  \hfill
  \begin{subfigure}[t]{0.38\textwidth}
  \centering
    \includegraphics[width=\textwidth]{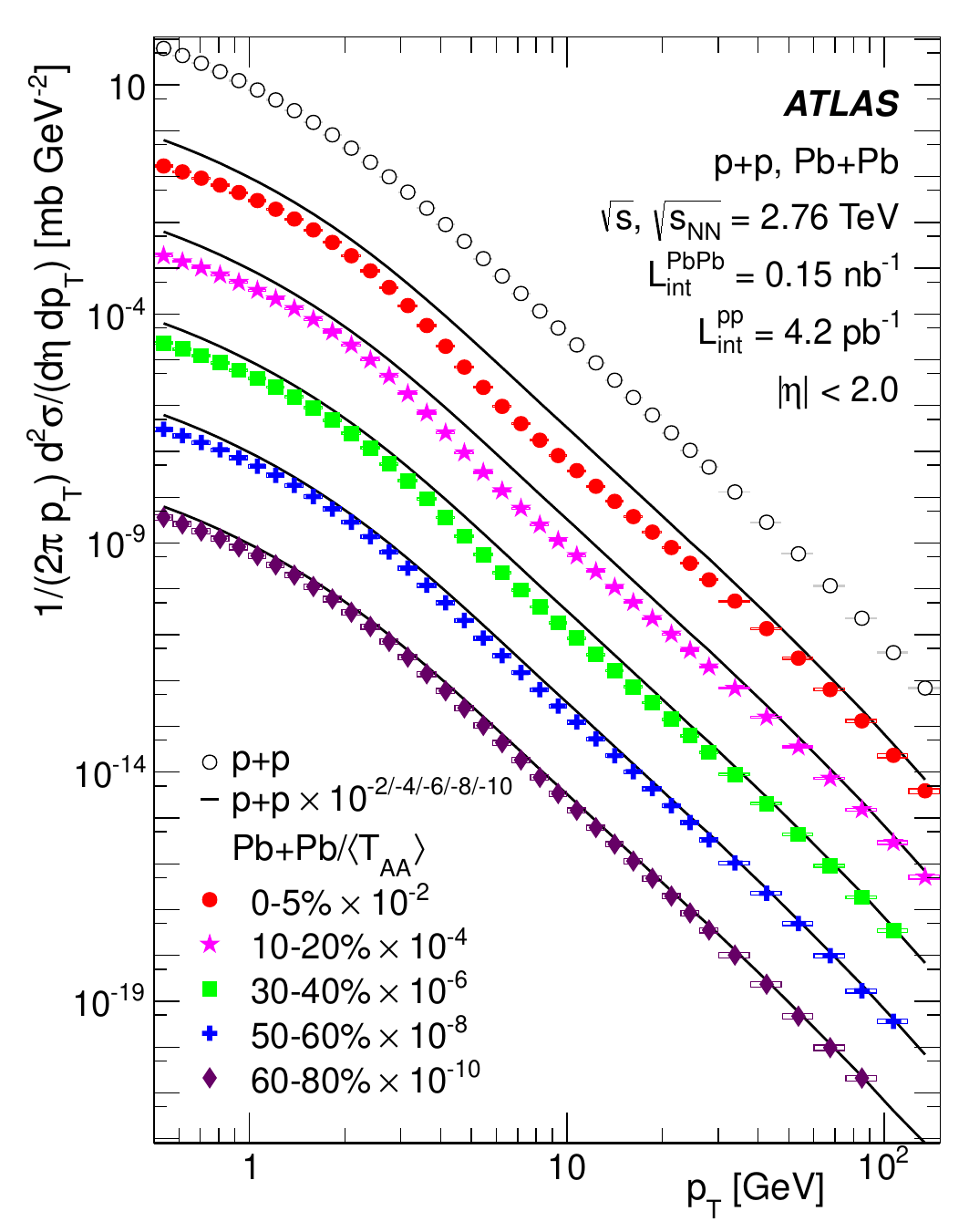}
  \end{subfigure}
  \caption{\small(Left) Pseudo-rapidity distribution of charged particles measured by the PHOBOS detector at RHIC in gold-gold collisions for several centrality classes \cite{Back:2002wb}. The bands correspond to systematic uncertainties. From the left to the right panels, $\sqrt{s_{NN}}$ increases from 19.6 GeV up to 200 GeV. The more the centrality is small, the more the rapidity plateau is high. (Right) $p_T$ spectra of charged particles measured by the ATLAS Collaboration \cite{Aad:2015wga} in proton-proton and lead-lead collisions for several centrality classes. Note that the Pb-Pb spectra are rescaled by the thickness function $\langle T_{AA}\rangle$ to account for the higher hard scattering rate in Pb-Pb collisions (w.r.t\ $pp$) as a single nucleus contains several nucleons.} \label{Fig:single-spec}
\end{figure}

\paragraph{Single particle spectra.} As emphasized, nucleus-nucleus collisions at high energy produce a myriad of particles. To organise the data and get an intuition about the event shape, one first looks at particle spectra. Once a centrality class is selected, experimentalists measure in this class the multiplicity of particles more differentially with respect to their kinematics or their species. There are two basics but important features regarding the particle spectra as a function of the rapidity (or pseudo-rapidity) and transverse momentum:
\begin{itemize}
 \item At mid-rapidity, for $|y|\simeq|\eta|\lesssim 2$, the particle multiplicity (averaged over all events in a centrality class) exhibits a flat plateau (see Fig.~\ref{Fig:single-spec}-left), meaning that the distribution is approximatively boost-invariant along the beam direction. This boost invariance is a natural consequence of the initial symmetry of the collision: at very high center of mass energy, all inertial observers sightly boosted with respect to the beam axis in the laboratory frame see the same collision of two ultra-relativistic nuclei, and thus the same final particle distribution. As a boost along the beam is equivalent to a shift in the rapidity $y$, this explains the central rapidity plateau. The boost invariance with respect to the beam axis is often called ``Bjorken hypothesis'' or ``Bjorken model'' \cite{Bjorken:1982qr}.
 
 \item The transverse momentum spectra of detected particles is steeply falling with $p_T$, as shown Fig.~\ref{Fig:single-spec}-right. The majority of particles produced in heavy-ion collisions are ``soft'', with $p_T\lesssim\mathcal{O}(1)$ GeV at the LHC. It has recently been observed that the mean $p_T$ of charged hadrons is proportional to the temperature of the plasma \cite{Gardim:2019xjs}. The shape of the spectrum shown in Fig.~\ref{Fig:single-spec}-right naturally divides the possible measurements into two categories: those related to the ``bulk'' where soft particles dominate, and those related to the hard tail of the $p_T$ spectrum, called ``hard probes''. As we shall see, the physical mechanisms at play are very different, although each kind provides valuable informations on the nature of the matter produced in heavy-ion collisions.
\end{itemize}

\subsection{Flow in heavy-ion collisions}

\begin{figure}[t] 
\centering
  \begin{subfigure}[t]{0.48\textwidth}
    \centering
    \includegraphics[width=\textwidth]{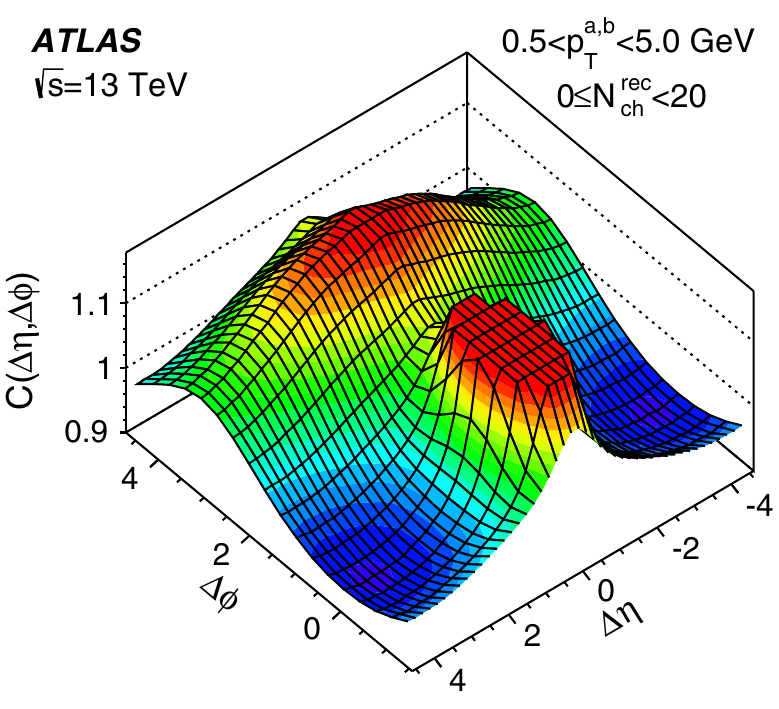}
  \end{subfigure}
  \hfill
  \begin{subfigure}[t]{0.48\textwidth}
  \centering
    \includegraphics[width=\textwidth]{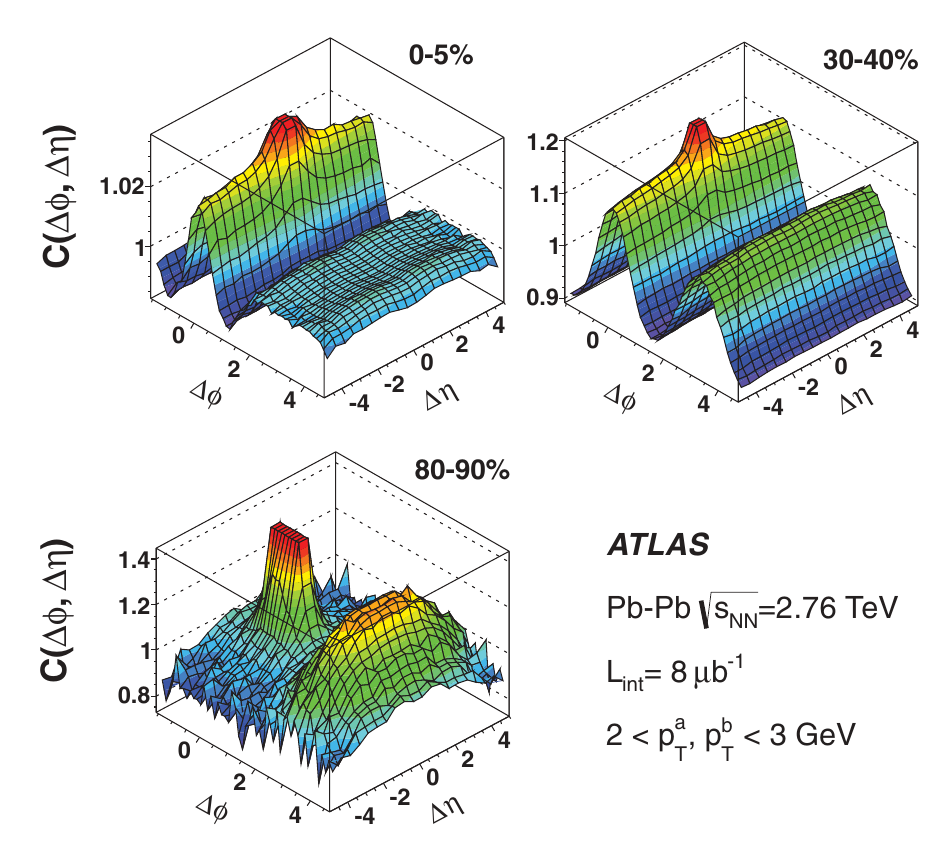}
  \end{subfigure}
  \caption{\small (Left) Two-particle correlation measured by the ATLAS Collaboration in $pp$ collisions with $\sqrt{s}=13$ TeV  and small total multiplicity (less than 20 reconstructed charged particles in the final state)  \cite{Aad:2015gqa}. The peak at $\Delta\phi=\Delta\eta=0$ due to QCD jets has been truncated. The other local maximum at $\Delta\eta=0$ and $\Delta\phi=\pi$ is caused primarily by back-to-back dijet events (i.e. momentum conservation). (Right) Same quantity measured by the ATLAS Collaboration in lead-lead collisions with $\sqrt{s_{NN}}=2.76$ TeV \cite{ATLAS:2012at}. The panels display three different centrality classes. For peripheral collisions (80-90\% centrality), the 2-particle correlation distribution looks like the one in proton-proton collisions on the left. In most central collisions (0-10\%) or a mid-centrality (30-40\%), the boost invariance is manifest, one observes a $\Delta\phi$ wave, and strong correlations at $\Delta\phi=0$ even for large values of $\Delta\eta$.} \label{Fig:flow}
\end{figure}

To understand more precisely the pattern of the bulk particles, one can perform a more refined measurement than the simple rapidity or $p_T$ distribution we have just discussed: measuring pair correlations. More precisely, one can measure the distribution of particle pairs separated by an azimuthal angle $\Delta\phi$ and (pseudo)-rapidity $\Delta \eta$ in a given centrality class. This distribution $C(\Delta\phi,\Delta\eta)$ is shown Fig.~\ref{Fig:flow}-right, as measured by the ATLAS detector at the LHC in PbPb collisions. On the left figure, the same distribution is shown in $pp$ collisions for comparisons. They are dramatically different, enlightening that new physics is involved in heavy-ion collisions with respect to $pp$. Indeed, if a heavy-ion collision was only an incoherent ``sum'' of elementary $pp$ collisions, the distributions should look the same.

For central and no too peripheral collisions, the pair correlation is almost independent of $\eta$, the large peak at $\Delta \phi=\Delta \eta =0$ excepted. As a function of $\Delta\phi$, one observes by eye a cosine modulation of the correlation, maximal at $\Delta\phi=0$ and $\pi$. The picture in $pp$ is very different even if the peak at $\Delta\phi=\Delta\eta =0$ is present. The correlation depends strongly on $\eta$ and there is no cosine-like modulation. We leave the discussion of the peak at $\Delta\phi=\Delta\eta =0$ for the next section: we will see that QCD predicts strong correlations between almost collinear particles (thus with $\Delta \phi=\Delta\eta=0$) due to the existence of high $p_T$ QCD \textit{jets}.

On the contrary, what explains the specific pattern of the two-particle correlations in heavy-ion collisions is generically called ``flow''. The flow picture for multi-particle correlations relies on the following hypothesis (see e.g.~\cite{Luzum:2011mm}): particles are emitted independently in a single event so that all multi-particle correlations in an event can be obtained from the single particle probability distribution. This probability distribution $P_1(\phi,y,p_T)$ determines the probability of having one particle with a given rapidity, azimuth and transverse momentum. This probability density is itself a random variable, as it fluctuates from one event to another. Non trivial correlations among pairs (as well as higher numbers of particles) are generated by these fluctuations. 

In the Bjorken model, $P_1(\phi,y,p_T)$ does not depend upon rapidity, but has a non trivial dependence with the azimuthal angle. It is usually expanded as a Fourier series:
\begin{equation}\label{flow-single}
  P_1(\phi,y,p_T)=\frac{1}{2\pi}\sum_{n=-\infty}^{+\infty}V_n(p_T)\exp(-in\phi)
\end{equation}
and $v_n=V_n/V_0$ is called $n^{\rm th}$ anisotropic flow.

What motivates the flow hypothesis? If the medium created after the collision thermalizes, the flow hypothesis is naturally justified, since thermalization destroys initial correlations so that particles are finally emitted independently. If thermalization is a sufficient condition for flow, one should keep in mind that it is not a necessary condition \cite{Dusling:2017aot}. The great success of the flow hypothesis combined with hydrodynamical predictions for $P_1$ in explaining the experimental data is in itself the best argument in its favour.

From a given model (usually an hydrodynamic picture for the medium evolution), theorists can predict the single particle distribution $P_1$ and its fluctuations. How is this compared to data? One cannot measure directly this probability distribution because it is defined event-by-event: the particle spectrum in one event has intrinsic large statistical fluctuations (essentially because the number of particles is not that large once kinematic cuts are applied). Instead, by measuring the pair correlation as in Fig.~\ref{Fig:flow}, which is averaged of many events, one can infer the value of the underlying theoretical single particle probability density. Indeed, the flow hypothesis leads to the following probability distribution for the event-by-event pair correlation (neglecting for simplicity the $p_T$ dependence of $V_n$):
\begin{align}
C(\Delta\phi,\Delta\eta)&=\frac{1}{2\pi}\sum_{n=-\infty}^{+\infty}|V_n|^2\exp(n\Delta\phi)\\
&=\frac{V_0^2}{2\pi}\Big(1+2\sum_{n=1}^{+\infty}|v_n|^2\cos(n\Delta\phi)\Big)
\end{align}
so that the $n^{\rm th}$ Fourier harmonics of $C(\Delta\phi)$, traditionally called $v_n\{2\}$ \cite{Borghini:2001vi} reads
\begin{equation}
v_n\{2\}\equiv\langle\langle \exp(n\Delta\phi)\rangle\rangle= \frac{\langle|V_n|^2\rangle}{\langle V_0^2\rangle}
\end{equation}
where the double bracket on the left is an average over pairs in a single event followed by an average over all events in a centrality class. The average on the right is an event average since $P_1$ and therefore $V_n$ are also random variables.

In the hydrodynamical picture of the plasma evolution, $P_1$ is dictated by the anisotropy of the initial energy density profile. This holds for the values of $V_n$ in a single event, but also for their fluctuations (meaning that the fluctuations of the initial geometry translate into fluctuations of $P_1$). In non-central collisions, the medium created after the collision is strongly deformed in the transverse plane (it has an ``almond shape''): pressure gradients push more particles in the direction parallel to the impact parameter than in its perpendicular direction. This means that one expects higher pair correlations for $\Delta\phi=0$ and $\Delta\phi=\pi$. This explains the strong cosine modulation, or mathematically the large $v_2$ (elliptic flow \cite{Ollitrault:1997vz}), observed in Fig.~\ref{Fig:flow}-right at mid-centrality where the initial collision anisotropy is larger. Higher harmonics in the modulation are driven by $v_3$, $v_4$ etc, \cite{Alver:2010dn,Gardim:2011xv} themselves related to the anisotropy of the initial energy density profile and its fluctuations.

Measurements of $v_n\{2\}$ are confronted to theoretical predictions of $v_n$. This constrains the parameters entering into the hydrodynamical calculation of the single particle distribution $P_1$ such as the viscosity of the plasma \cite{Romatschke:2007mq} (see \cite{Heinz:2013th} for a review).

%

\subsection{Other probes of the quark-gluon plasma}

It is not our intention to give an exhaustive discussion of the experimental probes of the quark-gluon plasma. Beside flow measurements, other informations about the QCD matter produced in nucleus-nucleus collisions are inferred from  electromagnetic probes: for instance if the medium thermalizes, thermal photons are emitted and their distribution can provide interesting informations about the thermodynamics of the plasma (see \cite{Alam:1996fd} for a review). There are also a lot of theoretical and experimental efforts to measure the production of bound states of heavy quark-antiquark, called quarkonia, in heavy-ion collisions (see \cite{Andronic:2015wma} for a recent review). The hot surrounding medium can destroy the binding between the quark and the antiquark. The suppression of quarkonia was therefore proposed as a signature of quark-gluon plasma formation \cite{Matsui:1986dk}.

\section{QCD jets and jet quenching}

Jets are collimated spray of energetic hadrons. The formation of such structure at high energy is predicted by perturbative QCD: the enhancement of soft and collinear radiative processes is a common feature of (unbroken) gauge theories with a massless gauge boson such as QED or QCD. In QCD, as gluons can easily radiate soft and collinear gluons because of the self interaction highlighted in the first section, the development of soft/collinear cascades is accentuated with respect to QED. Thus, the formation of jets is a direct consequence of the non-abelian nature of the gauge symmetry. Nowadays, jets are used on a daily basis in collider experiments and precise tests of the Standard Model often rely on precise predictions regarding QCD jet production. For a theoretical introduction to jets in $pp$ collisions, we refer the reader to the first part of Chapter~\ref{chapter:jet}.

In heavy-ion collisions, jets are also produced. This shows that even if new physical phenomena such as flow are at stake in nucleus-nucleus collisions, the standard features of the strong interaction like jet production have not disappeared.
In Fig.~\ref{Fig:flow}-left, the peak at $\Delta\phi=\Delta\eta=0$ is caused by QCD jets, since a jet is essentially a collection of collinear particles. In flow physics, QCD jets are considered as ``non-flow'' correlations which are removed by implementing a rapidity gap $\Delta\eta>\Delta\eta_{\rm cut}$ when measuring $v_n\{2\}$. On the contrary, for jet physics in heavy-ion collisions, the main topic of this thesis, bulk particles and flow are treated as part of the background which must be removed. 

To reduce the effect of this background, it is advantageous to focus on high $p_T$ jets, as those produced at the LHC. Jet observables belong then to the set of ``hard probes'' of the quark-gluon plasma.  By the Heisenberg uncertainty principle, high-$p_T$ jets are formed over a very short time, of order $1/p_T$, after the collision and thus before the medium formation. For such short time scales, it is allowed to deal with the perturbative degrees of freedom of QCD: the quarks and gluons. The general physical picture for the jet evolution in proton-proton collision is the following: a hard scattering between two partons inside the incoming protons typically produces a pair of high $p_T$ partons that propagate back-to-back in the transverse plane. By hard scattering, we mean that the momentum transfer is large, of order $p_T$. What happens next to one of these partons? As the parton is not yet on its mass-shell (strictly speaking, on-shell quarks or gluons do not exist), it can further radiate, predominantly soft and collinear gluons, which further radiate and so on (as illustrated Fig~\ref{Fig:cartoon}-left). This partonic cascade develops over a large time scale, of order $1/\LQCD\gg 1/p_T$ since for times larger than $1/\LQCD$ the partonic picture breaks down.

In heavy-ion collisions, this cascade interacts with the surrounding medium created after the collision. We shall discuss extensively these interactions in this thesis (e.g. the Introduction~\ref{chapter:intro-part1} of Part~\ref{part:theory} for a brief overview of the plasma-parton interactions). These latter have a significant effect on jet measurements in heavy-ion collisions with respect to $pp$ collisions. The modification of jet properties in heavy-ion collisions is commonly named jet quenching. We refer the reader to Chapter~\ref{chapter:intro-part2} for an introduction about jet quenching measurements at the LHC. The most salient one is the strong suppression of high $p_T$ jets in lead-lead collisions at the LHC. 

This thesis deals with jet quenching physics. Starting from pQCD and the parton language, we develop a new picture for the parton shower in the presence of a quark-gluon plasma (Part~\ref{part:theory}). This picture includes so far only the expected dominant effects of the plasma-jet interaction in the so-called leading logarithmic approximation of pQCD. We tried to circumscribe the validity of our approximations and acknowledge the limitations of this picture as best as possible. The good qualitative agreements between our calculations and the measurements done at the LHC for several observables (Part~\ref{part:phenomenology}) are encouraging, and we hope for a better understanding of jet quenching phenomena from this new foundation in the future.

\section{Reading guide}

To conclude this introduction, we would like to provide a short ``reading guide'' to precise the internal logic of this thesis and highlight the new contributions to the field. This thesis is divided into two parts. The first part is theoretical, while the second part is dedicated to the phenomenology of jet quenching at the LHC. 

The main purpose of Part~\ref{part:theory} is to explain in details the new factorised picture of jet evolution in a dense QCD medium which is the heart of this work: this is done in Chapter~\ref{chapter:DLApic} and the description of the Monte Carlo parton shower based on these ideas is the subject of Chapter~\ref{chapter:MC}. 

In that respect, the interest of the two opening chapters, Chapter~\ref{chapter:emissions} and Chapter~\ref{chapter:jet} is twofold. On one hand, it enables to provide the material necessary to understand the arguments presented in Chapter~\ref{chapter:DLApic} and the concepts used in Part \ref{part:phenomenology}. On the other hand, even though no significant breakthrough is presented in these two introductory chapters, there are some new developments:
\begin{enumerate}

 \item We give new (as far as we know) simple analytic formulas for the on-shell and off-shell gluon emission spectra in a dense medium (Section \ref{subsub:TMdep-BDMPS}, Eq.~\eqref{onshell-final}-\eqref{offshell-final}) with full dependence on transverse momenta and energy, valid in an expanding medium. We use these formulas to justify the factorisation between virtuality driven processes and medium-induced processes detailed in Chapter \ref{chapter:DLApic}.
 
 \item Still in an expanding medium, the  medium-induced spectrum  integrated over transverse momenta from a colour singlet dipole is derived in Section \ref{sub:mie-antenna}, Eq.~\eqref{final-bdmpsz-inter}.
 
 \item In the same spirit, the effect of decoherence on in-medium vacuum-like emissions with \textit{short formation times} studied in Section \ref{sub:decoherence} is very important for the resummation scheme presented in Chapter~\ref{chapter:DLApic}, Section~\ref{sec:DLresum}. This calculation is an important step to prove that angular ordering holds inside the medium.
 
 \item  In Chapter~\ref{chapter:jet}, Section~\ref{subsub:frag-ISD} (and Appendix~\ref{app:ISD}), we present and calculate a new jet substructure observable relevant in heavy-ion collisions and studied in this context in Chapter~\ref{chapter:FF}: the fragmentation function from subjets \cite{Caucal:2020xad,Caucal:2020vvb}.
 
 \item In Chapter~\ref{chapter:jet}, Section~\ref{sub:med-expansion} we give a 
criterion for the medium dilution so that medium-induced jet fragmentation satisfies scaling properties.
\end{enumerate}

Then, Chapter \ref{chapter:DLApic} is essentially based on the work published in \cite{Caucal:2018dla}. We give further arguments to justify the physics exposed in this short Letter, and we clearly delineate the approximations behind the resummation scheme. The extension of \cite{Caucal:2018dla} to Bjorken expanding media is also extensively discussed. 

In Chapter~\ref{chapter:MC}, the Monte Carlo parton shower {\tt JetMed} \cite{Caucal:2018ofz} is presented in details, and compared to other existing in-medium Monte Carlo parton showers.

The phenomenological study of Part~\ref{part:phenomenology} is entirely based on the theoretical picture for jet fragmentation and its implementation as a Monte Carlo parton shower described in details in the first part. All the phenomenological results on the jet nuclear modification factor and jet substructure observables in PbPb collisions presented in Part~\ref{part:phenomenology} are new. We point out that, Section \ref{sub:RAA-R} excepted, Chapter \ref{chapter:RAA}, Chapter~\ref{chapter:jet-sub} and Chapter~\ref{chapter:FF} are essentially extracted from the following papers \cite{Caucal:2019uvr,Caucal:2020lro,Caucal:2020xad}.

%% file: intro-part1.tex
\chapter{Introduction: jet quenching theory}
\label{chapter:intro-part1}

The next chapter develops in detail the formalism for calculating emission processes in a dense medium. We will make of course several assumptions, both in the modelling of the medium and in the dominant effects at play. On the contrary, with the present chapter, we would like to provide a broad overview of the field in order for the reader to get an intuition about the context of this thesis.

We have divided this introduction into two sections. The first section focuses on the energy loss of a test particle while propagating in a quark-gluon plasma. This problem is very well documented in the literature since the first measurement of the suppression of highly energetic hadrons in gold-gold collisions at RHIC.

The second section deals with the interplay between the medium effects (and in particular the energy loss covered in the first section) and the virtuality of the leading parton produced in the hard partonic scattering. This virtuality is the driven mechanism of jet fragmentation in the vacuum. There are many models for describing this interplay, but a first principle derivation in QCD is often missing.

\section{Parton energy loss in media}

There are two approaches for the problem of parton energy loss in a quark-gluon plasma depending on the modelling of the medium. For a weakly coupled medium, the medium can be described as a collection of scattering centers, and one can calculate the interaction between the test parton and the scattering centers using weak coupling techniques. In this section, we review the two main mechanisms for energy loss at weak coupling: the collisional and the radiative energy loss. 

I will not discuss AdS/CFT calculations in the strong coupling regime, as this thesis relies on perturbative methods. We refer to \cite{Shuryak:2008eq} for a review on this topic. Nevertheless, we point out that weak coupling methods in QCD at finite temperature work very well even when the strong coupling constant evaluated at the plasma thermal energy $\alpha_s(k_BT_{\rm p})$ is not that small ($\sim 0.4$) \cite{Blaizot:2003tw}.

\subsection{Collisional or elastic energy loss}

\begin{figure}
 \centering
 \includegraphics[width=0.48\textwidth]{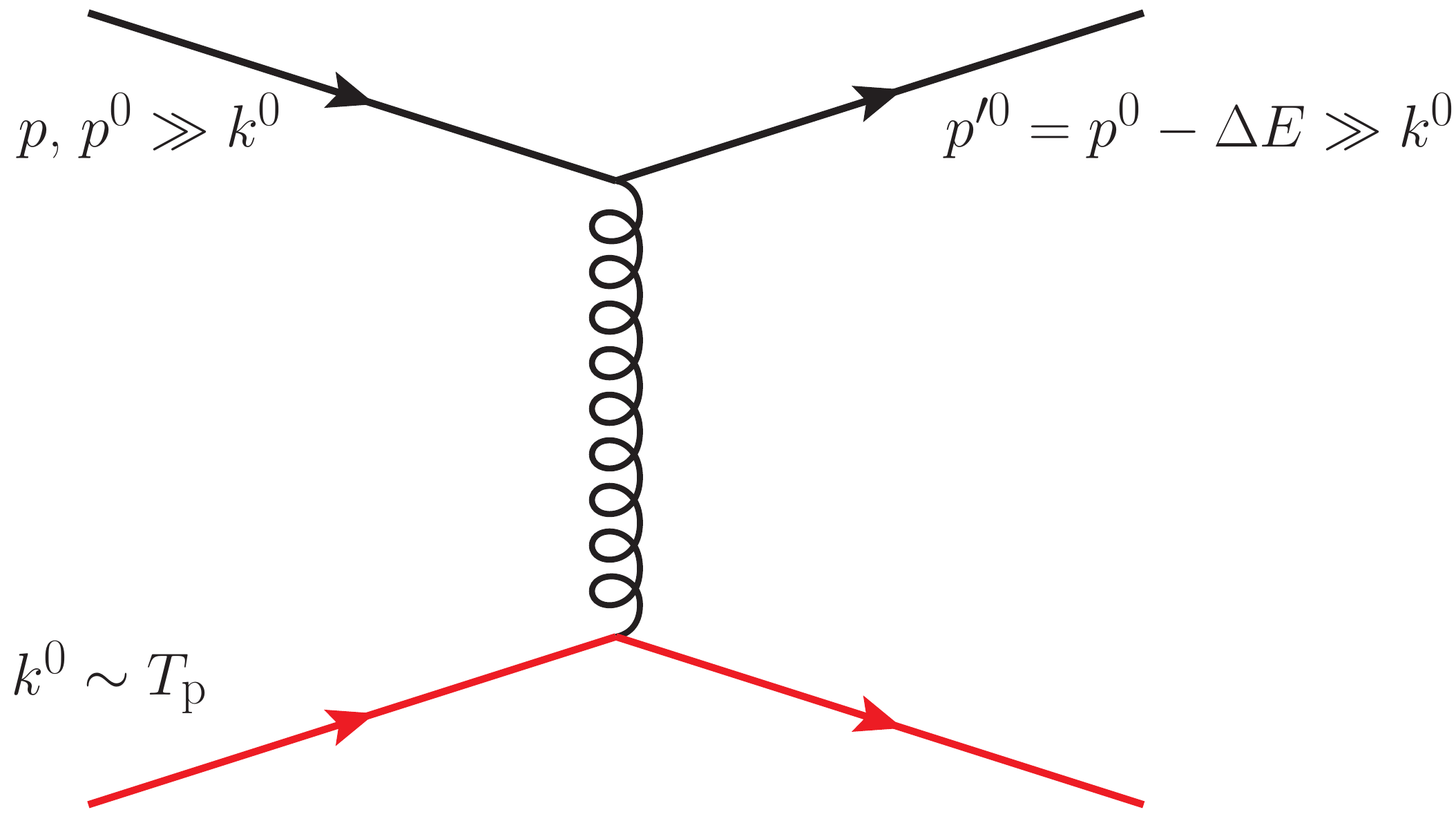}\hfill
 \includegraphics[width=0.48\textwidth]{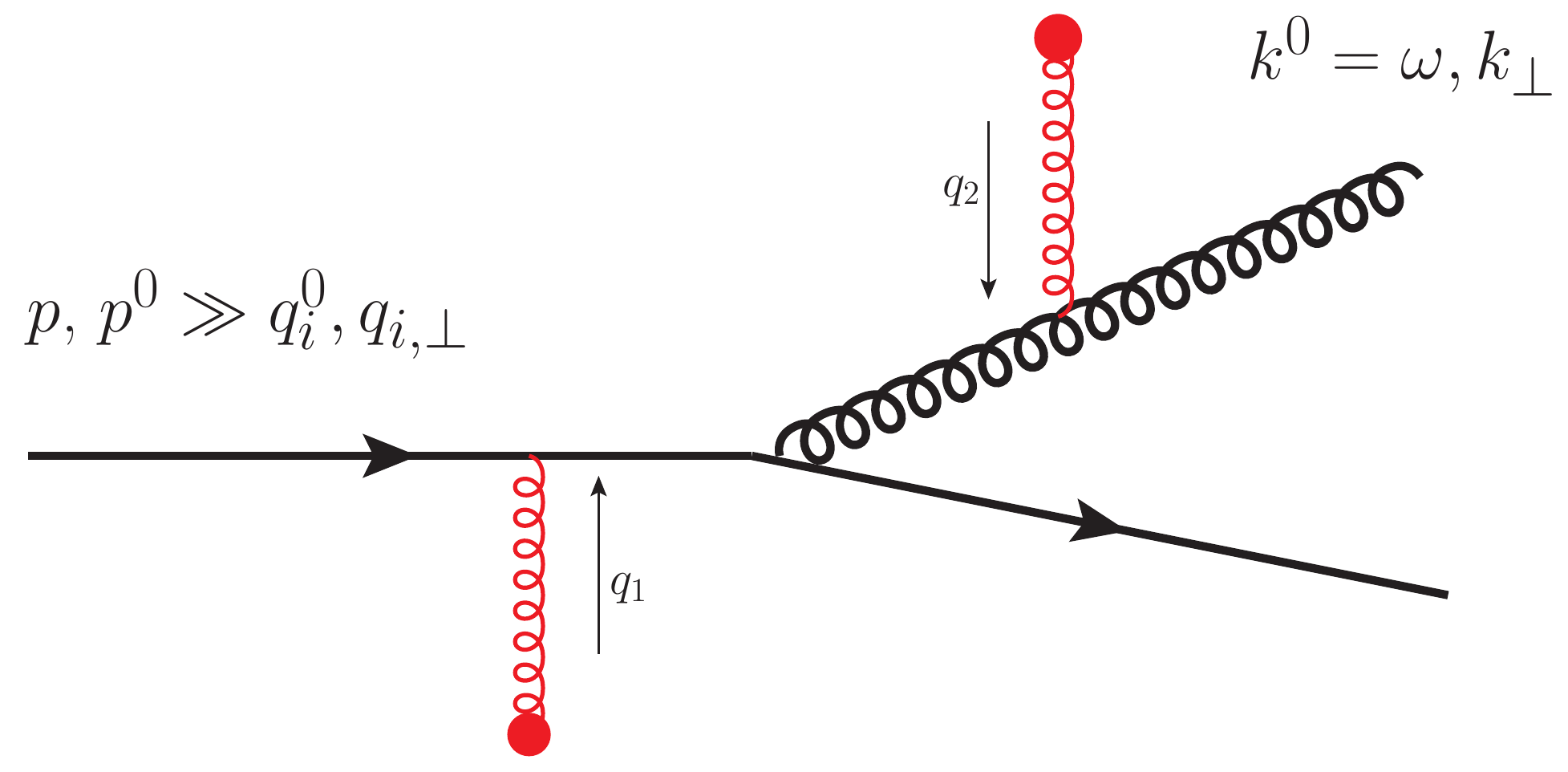}
 \caption{\small(Left) Typical Feynman diagram for the leading order calculation of the elastic energy loss. The red parton is a medium constituent. (Right) Typical Feynman diagram involved in the calculation of the inelastic energy loss. In the opacity expansion scheme, this diagram appears at the $N=2$ order since there are two scattering centers. The BDMPS-Z result resums all orders in opacity, in the regime where all momenta transferred are soft.}
 \label{fig:Fey-process}
\end{figure}

The collisional energy loss is due to the \textit{elastic} scattering of an incoming energetic on-shell parton with a plasma constituent. A typical Feynman diagram for the elastic process is shown Fig.~\ref{fig:Fey-process}-left. Bjorken has given a first estimation of the collisional energy loss for a massless quark, per unit length $x$ in the plasma \cite{Bjorken:1982tu}:
\begin{equation}
\frac{\dif E}{\dif x}=-\frac{8}{3}\pi\alpha_s^2T_{\rm p}^2\big(1+\frac{n_f}{6}\big)\log\Big(\frac{q_{\rm max}}{q_{\rm min}}\Big)
\end{equation}
where $T_{\rm p}$ is the plasma temperature, $n_f$ the number of active quark flavours in the plasma and $q_{\rm min/max}$ the minimal (maximal) momentum transferred by the collision. The minimal momentum transferred is of order of the Debye mass $\mu_D\sim gT_{\rm p}$ which characterizes the plasma (see Chapter \ref{chapter:emissions}, section \ref{sub:Amed-stat}), whereas the maximal momentum transferred is of order $p^0$.

This calculation has then been improved, taking into account quark masses \cite{Braaten:1991we}, plasma screening effects \cite{Thoma:1990fm,Mrowczynski:1991da}, finite length effects \cite{Djordjevic:2006tw} or running coupling corrections \cite{Peshier:2006hi} and corrections beyond $t$-channel scattering \cite{Peigne:2008nd}. Multiple elastic collisions are often included via the linear Boltzmann transport equation with a collision kernel incorporating the elastic $2\rightarrow2$ process, as in the LBT jet quenching model \cite{PhysRevLett.111.062301}.

Collisional energy loss is the dominant process at low particle momentum and can bring non-negligible corrections to radiative processes when quark masses are taken into account \cite{Zakharov:2007pj,Baier:2000mf}. The goal of this thesis is to develop an unified picture for \textit{high}-$p_T$ jet fragmentation. For this reason, direct elastic contributions to the energy loss are neglected (i.e.\ processes like in Fig.~\ref{fig:Fey-process}-left), but the effects of elastic collisions on radiative emissions and transverse distributions are taken into account.

\subsection{Radiative or inelastic energy loss}

In the radiative process, the scattering centers trigger a gluon emission by putting the incoming parton off its mass shell. The process is \textit{inelastic} as an additional parton populates the final state.
We first present the theoretical calculations where only one additional (soft) gluon is emitted. Then, we present the current strategies to go beyond the single gluon emission process.

\subsubsection{Single gluon emission}

 A typical Feynman diagram is shown in Fig.~\ref{fig:Fey-process}-right. Radiative energy loss is the dominant process at high momentum. That is why the second chapter of this thesis is partly dedicated to the computation of the medium-induced gluon spectrum. From the single gluon emission spectrum per unit energy $\om$, noted $\dif N/\dif \om$, it is straightforward to calculate the energy loss:
\begin{equation}
 \Delta E=\int\dif\om\, \om \frac{\dif N}{\dif \om}
\end{equation}
where the integration boundaries depend on the validity regime of the calculation of $\dif N$. To get the energy loss per unit length, one can differentiate the result with respect to the path length of the parton through the medium, noted $L$.

There are several formalisms in the literature related to the calculation of the radiative gluon emission and energy loss. For a detailed comparison between all these formalisms, see \cite{PhysRevC.86.064904}.
\begin{itemize}
 \item The Baier-Dokshitzer-Mueller-Peigne-Schiff and Zakharov (BDMPS-Z) formalism resums multiple scatterings on static colour charges via a path integral that can be computed explicitly in the multiple \textit{soft} scattering regime \cite{Baier:1994bd,Baier:1996vi,Baier:1996kr,Baier:1996sk,Zakharov:1997uu,Zakharov:1998sv,Zakharov:1996fv,Baier:1998kq}. For a static medium with quenching parameter $\qhat$ defined as the mean transverse momentum $k_\perp$ squared acquired via collisions per unit (light cone) time along the eikonal test parton trajectory, $\qhat \Delta t\sim\langle k_\perp^2\rangle$, the gluon emission spectrum per unit energy $\om$ reads:
 \begin{equation}\label{bdmpsz-brick}
 \om\frac{\dif N_{\rm BDPMS-Z}}{\dif \om}\approx\frac{2\alpha_sC_R}{\pi}\left\{
    \begin{array}{ll}
        \sqrt{\frac{\omc}{2\om}} & \mbox{for } \om \ll\omc \\
        \frac{1}{12}\Big(\frac{\omc}{\om}\Big)^2 & \mbox{for }\om \gg\omc
    \end{array}
\right.
\end{equation}
 with $\omc\equiv \qhat L^2/2$. The $\sqrt{\omc/(2\om)}$ tail is characteristic of the Landau-Pomeranchuk-Migdal effect in QCD (see Chapter~\ref{chapter:emissions}, Section~\ref{subsub:integrated-BDMPS}). 
 The angular dependence of the gluon spectrum has been obtained in \cite{Salgado:2003gb,Wiedemann:1999fq}, the generalization to massive quarks in \cite{Armesto:2003jh} and expanding media in \cite{Baier:1998yf,Salgado:2003gb,Salgado:2002cd}.
 \item The opacity expansion scheme organizes the calculation in terms of the number of static scattering centers. This approach has been pioneered by Gyulassy, Levai, and Vitev (GLV) in \cite{Gyulassy:2000er,Gyulassy:2000er,Gyulassy:2001nm} for the first leading terms and Wiedemann in \cite{Wiedemann:2000za} for the all order resummation. This all order resummation reproduces the BDMPS-Z result when the scatterings are soft. However, the gluon spectrum with $N=1$ scattering center gives somehow a significantly different result:
 \begin{equation}
  \om\frac{\dif N_{\rm GLV}}{\dif\om}\approx\frac{\alpha_sC_R}{\pi}\frac{\qhat L}{\mu_D^2}\left\{
    \begin{array}{ll}
        \log\Big(\frac{\mu_D^2L}{2\om}\Big)& \mbox{for } \om \ll\frac{1}{2}\mu_D^2L\\
        \frac{\pi}{4}\Big(\frac{\mu_D^2L}{2\om}\Big) & \mbox{for }\om \gg\frac{1}{2}\mu_D^2L
    \end{array}
\right.
 \end{equation}
 for a medium with quenching parameter $\qhat$.
 \item Whereas the two previous formalisms consider static scattering centers, Arnold, Moore and Yaffe (AMY) calculate gluon production directly in a thermal QCD background \cite{Arnold:2001ms,Arnold:2002ja}.
\end{itemize}

As emphasized in \cite{Mehtar-Tani:2019tvy,Mehtar-Tani:2019ygg}, the modern view to unify the GLV and BDMPS-Z results is the following: the GLV formula gives an accurate description of the hard tail $\om\gg \om_c\sim \qhat L^2/2$ of the medium-induced spectrum coming from hard single scattering and the Bethe-Heitler regime $\om\ll \omBH\sim\qhat\ell^2/2$, with $\ell$ the medium mean free path, whereas the BDMPS-Z result encompasses the multiple soft scattering regime $\omBH\le \om\le \om_c$. 

On the other hand, the AMY formalism gives a more rigorous description of the interaction between the test parton and the scattering centers using results from QCD at finite temperature: instead of relying on the Gyulassy-Wang model \cite{Gyulassy:1993hr} for the elementary scattering cross-section off plasma constituents as in the BDMPS-Z or GLV calculations, the AMY formalism uses an hard-thermal-looped resummed version of this cross-section. In the multiple soft scattering regime, this amounts to a redefinition of the quenching parameter $\qhat$.

In Chapter \ref{chapter:emissions}, we present a detailed calculation of the gluon radiative spectrum within a ``CGC-like'' formalism inspired by the close relation between gluon production in proton-nucleus collisions and parton energy loss in heavy-ion collisions \cite{MehtarTani:2006xq}.

Even if the first QCD calculations of radiative energy loss in plasma date back to the mid nineties, this is still an active topic. The systemic analytic expansion around the harmonic oscillator solution started in \cite{Mehtar-Tani:2019tvy} has been carried out up to the next-to-next-to-leading order in \cite{Barata:2020sav} for the energy spectrum and next-to-leading order for the broadening distribution \cite{Barata:2020rdn}. A novel method for resumming multiple scatterings beyond the standard harmonic approximation can be found in \cite{Andres:2020vxs}.

\subsubsection{Multiple emissions}

Going beyond the single gluon emission in inelastic processes is a complicated task in perturbative QCD, even in the vacuum. Formally, this requires to compute Feynman diagrams with multiple gluons in the final state. For two gluons, such calculations have been performed in the series of papers \cite{Arnold:2015qya,Arnold:2016kek,Arnold:2016mth,Arnold:2016jnq}
with several approximations (static medium and multiple soft scatterings approximations for instance).

That said, there is a physical regime where multiple medium-induced emissions can be seen as a classical branching process, in an approach pioneered by Baier, Mueller, Schiff and Son to describe thermalization in heavy-ion collisions \cite{Baier:2000sb}. In this regime, the resummation of multiple branchings is done via a Boltzmann kinetic equation with a branching rate given by the BDPMS-Z one and a collision kernel taking into account elastic $2\rightarrow2$ processes. This will be studied at length in Chapter \ref{chapter:jet}, section \ref{sec:jets-mie} (see also references therein). Former attempts to resum multiple medium-induced gluons consider that these emissions follow a Poisson process \cite{Baier:2001yt,Salgado:2003gb}. In a certain sense, the effective theory for medium-induced jets described in Chapter~\ref{chapter:jet} Section~\ref{sec:jets-mie} is a generalization of these ideas beyond the Poisson ansatz.

\section{Interplay between virtuality and medium effects}

So far in this introduction, we have implicitly ignored the virtuality of the parton losing part of its energy in the plasma: the incoming test parton was on its mass shell. However, the driven mechanism for jet fragmentation in the vacuum\footnote{In this thesis, by ``vacuum'', we always mean in $e^+e^-$ or $pp$ collisions.} is the initial virtuality of the parton escaping the hard scattering. For a jet fragmenting in a dense environment, this should be also true, at least for high $p_T$ jets. Consequently, a picture combining together virtuality-driven radiations and medium-induced emissions is essential to have a more precise understanding of how jets evolve in the plasma.

\paragraph{Models.} The current treatment of this question is rather incomplete and unsatisfactory. Some models postulate that the fragmentation occurs as in the vacuum up to some arbitrary virtuality scale before quenching effects or that the vacuum-like fragmentation occurs completely outside the medium after the quenching of the virtual leading parton. A more detailed comparison between Monte Carlo event generators is given in Chapter~\ref{chapter:jet}, Section~\ref{sec:MCcomp}. There are some notable counter-examples. For instance, the higher-twist approach \cite{Wang:2001ifa,Majumder:2009zu} in which the vacuum-like fragmentation in the plasma involves modified DGLAP splitting functions including higher-twist contributions due to the medium. The parton shower {\tt Q-Pythia} \cite{Armesto:2009fj} uses also medium modified DGLAP splitting functions taking into account medium induced emissions in the multiple soft scattering regime. Finally, the Monte Carlo event generator {\tt Jewel} \cite{Zapp:2013vla} which uses a selection rule based on formation time arguments in the generation of a vacuum-like branching or a medium scattering.

\paragraph{Colour coherence.} On the theory side, this question has driven recently a lot of activity. The most important results obtained in this direction are related to the concept of coherence. A jet is a quantum object, and as such, it has a typical coherence size, or more precisely a coherence angle in this context. For a jet fragmenting in the vacuum, an emission with a transverse wavelength larger that the opening angle of the leading dipole at the formation time of the emission is emitted coherently by the dipole, irrespectively of its internal structure. For a leading colour singlet dipole, large angle gluon emissions are therefore suppressed. The medium modifies this coherence property: the multiple interactions with the scattering centers break the quantum coherence of the jet after a time called the decoherence time. This phenomenon has been highlighted in \cite{MehtarTani:2010ma,MehtarTani:2011tz,CasalderreySolana:2011rz} from pQCD first principles and some phenomenological applications are considered in \cite{CasalderreySolana:2012ef,Mehtar-Tani:2014yea,Mehtar-Tani:2017web}. Note that a similar idea appears also in the reference \cite{Leonidov:2010he}. Medium-induced decoherence and the coherence property of jets with respect to the plasma show that the (quantum) interplay between virtuality and medium effects is not trivial.

\paragraph{Towards an unified picture of jet fragmentation.} The presence of a finite volume medium addresses the problem of jet fragmentation (or
more generally the decay of virtual particles) in an inhomogeneous space-time. In that respect, jet quenching in heavy-ion collisions is a unique background for understanding QCD jet properties in an open quantum system. There are still open questions but we believe that the theoretical first part of this thesis answers part of them :
\begin{itemize}
 \item the concept of ``off-shellness'' and ``on-shellness'' when a parton is embedded in a dense weakly coupled medium is clarified in Chapter~\ref{chapter:DLApic}, Section~\ref{sec:veto-region}.
 \item In contrast to prior expectations, decoherence has no significant role for the in-medium vacuum-like fragmentation at leading logarithmic accuracy. However, decoherence plays a crucial role for the subsequent fragmentation in the vacuum once the medium is diluted. This is explain in Chapter~\ref{chapter:DLApic}, Section~\ref{sec:DLresum}.
 \item At this accuracy and in the multiple soft scattering regime, the vacuum-like fragmentation factorises from the ``higher-twist'' medium effects. The factorisation scale is not an ad-hoc arbitrary choice: it is given by the virtuality scale that marks the transition between off-shellness and on-shellness. This virtuality scale is set by the medium itself.
\end{itemize}


%% file: chapter2.tex
\chapter{Emissions and decays in a dense QCD medium}
\label{chapter:emissions}

In this chapter, we calculate cross-sections for the decays of off-shell or on-shell partons in the presence of a dense QCD medium. These calculations are performed at leading order in the strong coupling constant $\alpha_s$ when this coupling is not enhanced by the large density of medium scattering centers, and to all orders in $\alpha_s$ in the opposite situation. We provide new analytic formulas for the full dependence of these cross-sections with the emission kinematics. 

Such formulas will be the building blocks for the jet calculations made in the next chapters since these chapters deal with the all order resummation of the simple processes calculated here. The formalism used to derive them is inspired from the Colour Glass Condensate (CGC) formalism \cite{Gelis:2010nm}.
%

The first section provides benchmark results for the decay of off-shell particles in the vacuum. The second section deals with our modelling of the dense QCD medium. In the third section, we discuss medium-induced emissions. In the last section, we examine the medium effects on quantum interferences in emission processes.

\section{Some benchmark results in the vacuum}
\label{sec:vac-bench}

In this section, we recall the main results regarding the decay of a time-like slightly off-shell parton in perturbative QCD. Roughly speaking, a slightly off-shell time-like parton is an internal leg of a Feynman diagram such that the kinematics of the final state particles enforce the 4-momentum squared $q^2$ of the leg to be positive and almost 0 (the propagator ``blows up''). A more precise definition will be given in the context of collinear and soft factorisation.

For simplicity, we consider the $e^+e^-$ annihilation process so that we do not have to deal with incoming space-like partons\footnote{Considering together initial and final state almost on-shell partons rises additional difficulties, see \cite{Catani:2011st}}. As we will only perform leading order calculations in this chapter, we focus on the corresponding leading order calculation in the vacuum. However, many results exist beyond leading order \cite{Bern:1995ix,Kosower:1999xi,Catani:1999ss}.

\subsection{Collinear factorisation at leading order}

In this $e^+e^-$ annihilation process, we assume that the final state is made of $n$ QCD partons, with $4$-momenta labelled by $p_i$, $i\in\{1,..n\}$ --- there is no additional difficulty if the final states particles are non-QCD partons. We are interested in the limit of the $e^+e^-\rightarrow \{n \textrm{ partons}\}$ amplitude squared when two outgoing partons (say $1$ and $2$) become collinear. In this configuration, one defines $p^\mu$ as this collinear direction, with $p^2=0$. The splitting fraction $z$ and the transverse momentum $k_\perp^\mu$ of the splitting are defined thanks to the standard Sudakov decomposition of $p_1$ and $p_2$:
\begin{align}\label{sudakov-dec}
 p_1^\mu&=zp^\mu+k_\perp^\mu-\frac{k_\perp^2}{z}\frac{n^\mu}{2p^\mu n_\mu}\,,\qquad p_2^\mu=(1-z)p^\mu-k_\perp^\mu-\frac{k_\perp^2}{1-z}\frac{n^\mu}{2p^\mu n_\mu}
\end{align}
with $n^\mu$ a light-light vector such that $k^\mu n_\mu=0$. The collinear limit is defined by $k_\perp\rightarrow0$.

The leading order factorisation theorem for the \textit{tree-level} amplitude squared of the process states that \cite{Ellis:1991qj}:
\begin{equation}\label{collinear-factorization}
 |\mathcal{M}_n(p_1,p_2,..,p_n)|^2\underset{k_\perp\rightarrow0}\longrightarrow|\mathcal{M}_{n-1}(p,p_3,..p_n)|^24\pi\alpha_s\frac{z(1-z)}{-k_\perp^2}\Phi_i^{ab}(z)
\end{equation}
in the limit where the partons $1$ and $2$ become collinear.
In this result, the colour sums have been performed as well as the \textit{average} over the polarizations of the decaying parton $i$. $a$ and $b$ are the flavour index of partons $1$ and $2$ whereas $i$ is the flavour index of the parent parton. In the collinear limit $-k_\perp^2= z(1-z)(p_1+p_2)^2\simeq p^0(z(1-z)\th)^2$, with $\th$ the polar angle between the 3-vectors $\vec{p}_1$ and $\vec{p}_2$.

This formula says that in the collinear limit, the amplitude squared of the process factorizes into two subprocesses: the $e^+e^-\rightarrow \{n-1 \textrm{ partons}\}$ process and the decay of the almost on-shell parton $i$ into $a$ and $b$. Therefore, a classical picture for multiple collinear decays emerges from this factorisation theorem.
In the additional soft limit $z\rightarrow0$, this factorisation formula becomes, written now at the level of the full differential cross-section:
\begin{tcolorbox}[ams equation]
 \dif \sigma_n(p_1,p_2,..,p_n)\underset{k_\perp\rightarrow0}\longrightarrow\dif\sigma_{n-1}(p,p_3,..p_n)\frac{\alpha_sC_i}{\pi}\frac{\dif k_\perp^2}{k_\perp^2}\frac{\dif z}{z}
\end{tcolorbox}
\noindent In this thesis, the factor appearing in front of $\dif \sigma_{n-1}$ will often be denoted as the Bremsstrahlung spectrum or vacuum-like emission spectrum. In this chapter and especially in Section~\ref{sec:BDMPSZ}, we aim at calculating this soft-collinear factor in the presence of a dense QCD medium. To calculate this factor, we use the fact that it can be derived considering the $i\rightarrow\{a,b\}$ process alone, where $i$ is the decaying parton created at fixed vertex $x_i^\mu$.

\subsection{Eikonal factor: soft factorisation at leading order}

There is another regime where the same kind of factorisation holds: when one of the final state gluon is soft. The soft limit of, say $p_1$, is defined as $p_1^\mu\rightarrow0$, meaning that all components of the four-vector of the gluon go to zero. In this case, the $n$-parton tree-level amplitude squared discussed above factorizes as \cite{Catani:1999ss}:
\begin{tcolorbox}[ams equation]\label{soft-factorization}
 |\mathcal{M}_n(k,p_2,..,p_n)|^2\underset{k^\mu\rightarrow0}\longrightarrow4\pi\alpha_s\sum\limits_{i,j=2}^n\mathcal{S}_{ij}(k)|\mathcal{M}^{(i,j)}_{n-1}(p_2,..p_n)|^2\,,\qquad \mathcal{S}_{ij}(k)=\frac{p_i^\mu p_{j\mu}}{(p_i^\mu k_\mu)(p_j^\mu k_\mu)}
\end{tcolorbox}
\noindent The sum, which formally breaks the exact factorisation (contrary to Eq.~\eqref{collinear-factorization}), is a consequence of gluons carrying colour and thus producing colour correlations between the outgoing partons. The amplitude squared $|\mathcal{M}^{(i,j)}_{n-1}|^2$ depends on the prong $(i,j)$ which emits the soft gluon:
\begin{equation}
 |\mathcal{M}^{(i,j)}_{n-1}(p_2,..p_n)|^2=[\mathcal{M}^{a_1,..,b_i,..,b_j,..,a_n}_{n-1}(p_2,..p_n)]^\star T^c_{b_ic_i}T^c_{b_jc_j}[\mathcal{M}^{a_1,..,c_i,..,c_j,..,a_n}_{n-1}(p_2,..p_n)]
\end{equation}
with all the colour indices left explicit. The factor $\mathcal{S}_{ij}(q)$ is called the eikonal or soft factor. It encompasses the soft gluon radiation pattern of an off-shell antenna. We calculate it in details in the vacuum and estimate it in the presence of a dense medium in Section \ref{sub:decoherence}. 

This eikonal factor $\mathcal{S}$ has a very interesting property, known as angular ordering. To see this, let us rewrite it in terms of the opening angle $\theta_{ij}$ of the prong and the angles $\th_{gi(j)}$  between the gluon and the parton $i(j)$:
\begin{align}
\mathcal{S}_{ij}(k)&=\frac{1}{{k^{0}}^2}\frac{1-\cos(\th_{ij})}{(1-\cos(\th_{gi}))(1-\cos(\th_{gj}))}\\
 &=\frac{1}{{k^{0}}^2}\Big(\tilde{\mathcal{S}}_{ij}^i+\tilde{\mathcal{S}}_{ij}^{j}\Big)
\end{align}
In the second line, we split the eikonal factor into two symmetric pieces:
\begin{equation}
 \tilde{\mathcal{S}}_{ij}^i=\frac{1}{2}\left(\frac{1-\cos(\th_{ij})}{(1-\cos(\th_{gi}))(1-\cos(\th_{gj}))}+\frac{1}{1-\cos(\th_{gi})}-\frac{1}{1-\cos(\th_{gj})}\right)
\end{equation}
For $\tilde{\mathcal{S}}_{ij}^{j}$, one switches $i\leftrightarrow j$. When the kernel $\tilde{\mathcal{S}}_{ij}^i$ is averaged over the azimuthal angle in the plane transverse to the direction of the 3-momentum of $i$, this vanishes if $\th_{gi}\ge \th_{ij}$ (and similarly if one averages over the azimuthal angle in the plane transverse to $j$). This means that the soft gluon radiation is confined inside a cone of opening angle $\th_{ij}$ around either the $i$ or the $j$ direction. This is the angular ordering property. For a general presentation of colour coherence, we refer the reader to Ref. \cite{Dokshitzer:1991wu}, Chapter 4. A detailed calculation of the azimuthal integral is done in \cite{Ellis:1991qj}, and we will perform a similar calculation in Section~\ref{sub:decoherence}.

\section{Modelling of the dense QCD medium}

In this section, we develop the formalism necessary to set up the calculation of radiative gluon emissions in the presence of a dense QCD medium. To summarize, the medium is a classical background gauge field generated by static colour charges in the rest frame of the plasma. These static colour charges have Gaussian fluctuations to account for the thermal fluctuations of the plasma. A test particle propagating through the plasma undergoes a Coulomb-like interaction with the scattering centers.

\subsection{Infinite momentum frame and conventions}
\label{sub:general-pic}

\paragraph{Boosted frame.} In order to use the technology and computational tools developed in the context of $eA$ and $pA$ collisions \cite{MehtarTani:2006xq}, we always consider the scattering of a projectile moving with the velocity of light in the positive $x^3\equiv z$ direction on a dense medium, the target, boosted very close to the light cone and moving in the negative $z$ direction. The projectiles considered throughout this chapter will be an on-shell or off-shell parton, or a colour dipole. In the off-shell case, we have of course in mind the decay of highly virtual parton coming from the hard scattering process. 

The frame where the dense medium is boosted near the light-cone branch $x^+=0$, named ``boosted frame'' therein, is not the same as the laboratory frame defined in the Introduction~\ref{chapter:intro}. They are related by a boost transformation with parameter $\beta\rightarrow-1$ which leaves the plane $(x,y)=(x^1,x^2)=(X^1,X^2)$ invariant:
\begin{equation}
 \begin{pmatrix}
  x^0\\
  x^3
 \end{pmatrix}
=\begin{pmatrix}
  \cosh(\psi)&\sinh(\psi)\\
  \sinh(\psi)&\cosh(\psi)
 \end{pmatrix}
 \begin{pmatrix}
  X^0\\
  X^3
 \end{pmatrix}\,,\qquad\beta=\tanh(\psi)
\end{equation}
In the end, the differential cross-sections obtained in this chapter are boosted back to the laboratory frame for phenomenological applications. As these cross-sections are boost invariant along the $z$ direction, this transformation is equivalent to use the laboratory frame values of all the physical parameters in the formulas derived in the boosted frame. To say it differently, when plugging numbers into the final cross-section formulas, one can directly use quantities measured in the laboratory frame, such as the quenching parameter (related to the density of scattering centers) or the jet path length, to get the cross-section in the laboratory frame.

\paragraph{Light-cone coordinates.} It is convenient to introduce the light-cone coordinates defined by the following transformation:
\begin{equation}
 x^+=\frac{x^0+x^3}{\sqrt{2}}\,,\qquad
 x^-=\frac{x^0-x^3}{\sqrt{2}}
\end{equation}
where $x^\mu\equiv(x^0,x_\perp,x^3)$ is the coordinate system in the boosted frame. The transverse coordinate $x_\perp=(x^1,x^2)\equiv x^i$ orthogonal to the $z$ direction is unchanged by the light-cone transformation. We adopt the following convention for the metric signature $\eta_{\mu\nu}$:~$(+,-,-,-)$. The product of two four vectors $u\cdot v$ is $\eta_{\mu\nu}u^{\mu}v^{\nu}=u^{+}v^{-}+u^{-}v^{+}-u_\perp v_\perp$ and the Lorentz invariant phase space is:
\begin{equation}
 \dif^4p\,\delta(p^2-m^2)=\frac{\dif p^+\dif^2p_\perp}{2p^+}
\end{equation}
We note $\partial_\mu\equiv \frac{\partial}{\partial x^\mu}$ the derivative operator. In the light cone coordinate system, $\partial_{+/-}=\frac{\partial}{\partial  x^{+/-}}$. Due to the particular form of the metric in light cone coordinates, $\partial^+=\partial_-$ and $\partial^-=\partial_+$ so that $\partial_\mu\partial^\mu=2\partial^+\partial^--\partial_\perp^2$. We point out that the subscript $\perp$ refers to the transverse coordinate with respect to the direction of motion of the projectile which is also the $z$ axis. This notation is adopted in order to make a clear distinction with the transverse coordinate with respect to the \textit{beam} axis, noted with a ``$T$'' subscript. Also, we use contravariant indices for the transverse vectors, meaning that $\partial^i_{x_\perp}\equiv\frac{\partial}{\partial x_i}=-\frac{\partial}{\partial x^i}$. The minus factor is crucial to get the right result for the interference spectrum in Section~\ref{sub:mie-antenna}.

When dealing with an off-shell incoming projectile, we will always call $x_i^\mu$ the coordinate vector of the vertex where the off-shell particle is emitted. The particle enters the plasma at light-cone time $x^+=x_0^+$. We consider massless QCD partons only.

\paragraph{Cross-sections.} In this chapter, we basically compute cross-sections. In order to obtain the right prefactors in front of our formulas, one should be careful with both the normalization of single particle states and the cross-section definition. 

Let us consider that the scattering occurs in a finite 4-dimensional box of volume $V_{\mathfrak{B}}=S_{\mathfrak{B}}\times T_{\mathfrak{B}}^2$ where $S_{\mathfrak{B}}$ is the transverse area of the box and $T_{\mathfrak{B}}$ is the size  of the box in the time and longitudinal direction. In our final formulas, the volume of the box will be sent to infinity. The cross-section is then defined as:
\begin{equation}
 \dif\sigma =\frac{1}{T_{\mathfrak{B}}\Phi_0}\dif P
\end{equation}
where $\Phi_0$ is the flux of incoming particles. In the case at hand, the incoming state is a single particle state moving with the speed of light in the $+z$ direction. The flux is then given by $|\vec{v}|/(S_{\mathfrak{B}}T_{\mathfrak{B}})=1/(S_{\mathfrak{B}}T_{\mathfrak{B}})$ (in high energy physics units, with $c=1$).

The emission probability $\dif P$ reads
\begin{equation}
 \dif P=\frac{|\braket{\textrm{out}|\textrm{in}}|^2}{\braket{\textrm{out}|\textrm{out}}\braket{\textrm{in}|\textrm{in}}}\prod_j\frac{T_{\mathfrak{B}}S_{\mathfrak{B}} }{(2\pi)^3}\dif p_j^3\dif^2p_{j\perp}
\end{equation}
with all the produced particles in the out state labelled by $j$. The factor $T_{\mathfrak{B}}S_{\mathfrak{B}}$ is required to enforce
\begin{equation}
 \int\frac{T_{\mathfrak{B}}S_{\mathfrak{B}} }{(2\pi)^3}\dif p^3\dif^2p_{\perp}=1
\end{equation}
 The single particle states are normalized as follows:
\begin{align}
 \braket{p|p}&=2p^0(2\pi)^3\delta(p^3=0)\delta(p_\perp=0)\\
 &=2p^0\int\dif x^3\int\dif^2 x_\perp\\
 &=2p^0S_{\mathfrak{B}}T_{\mathfrak{B}}\label{norm-state}
\end{align}
Combining all these expressions together when the incoming state is a single particle state one finds:
\begin{align}
 \dif\sigma&=\frac{1}{T_{\mathfrak{B}}}\frac{|\braket{\textrm{out}|\textrm{in}}|^2}{2p^0}\prod_j\frac{\dif p_j^3\dif^2 p_{j\perp}}{(2\pi)^32p_j^0}\\
 &=\frac{1}{T_{\mathfrak{B}}}\frac{|\braket{\textrm{out}|\textrm{in}}|^2}{2p^0}\prod_j\frac{\dif p_j^+\dif^2 p_{j\perp}}{(2\pi)^32p_j^+}
\end{align}
To get the second line, we have recognized the Lorentz invariant phase space for the emitted particles. Note the unusual remaining size factor $T_{\mathfrak{B}}$. Of course, it will cancel in all our final results.

\subsection{The background gauge field $\mathcal{A}_m^\mu$}

The dense medium generates a \textit{classical} background gauge field $\mathcal{A}_m^\mu$. The classical approximation is natural in the weak coupling limit where the QCD coupling constant $g(k_BT_{\rm p})$, evaluated at the typical thermal energy of the in-medium partons, is small. At weak coupling, the tree-level order dominates the projectile-target interactions. As tree-level and classical approximation are equivalent, the weak coupling limit supports the classical nature of the background gauge field. 
%
%
%
This gauge field is produced by a classical current of colour charges. In a non-abelian gauge theory, such a clear separation between sources and gauge field is rather unusual because the colour current is only \textit{covariantly} conserved. The choice of the light-cone gauge leads nevertheless to this simple picture.

We first consider the case of a static medium with a time-independent colour charges density in the laboratory frame. The extension to the case of a longitudinally expanding medium is discussed in Section~\ref{sub:Bjorken}. In the laboratory frame, the 4-current density $\mathcal{J}^\mu_{m,\lab}$ --- in the adjoint representation of $\textrm{SU}(N_c)$ --- reads:
\begin{align}\label{current-restframe}
 \mathcal{J}^\mu_{m,\lab}&=\bar{\rho}_a(x_\perp,X^{3})t^a U^\mu\,,\qquad &U^\mu&=(1,0,0,0)\\
 &=\bar{\rho}_a\big(x_\perp,(X^+-X^-)/\sqrt{2}\big)t^a U^\mu\,,\qquad &U^\mu&=(\underbrace{1/\sqrt{2}}_{+},\underbrace{1/\sqrt{2}}_{-},0,0)
\end{align}
with $\bar{\rho}_a$ the charge density and $U^\mu$ the four velocity of the fluid in the laboratory frame  (which is also the rest frame of the plasma since the fluid is static).

In the boosted frame defined in \ref{sub:general-pic}, with boost parameter $\beta=\tanh(\psi)\rightarrow-1$, the 4-current $\mathcal{J}^\mu_m$ becomes:
\begin{align}
\mathcal{J}^\mu_m&=\bar{\rho}_a\big(x_\perp,(e^{-\psi}x^+-e^{\psi}x^-)/\sqrt{2}\big)t^a u^\mu\,,\quad u^\mu=(\overbrace{e^{\psi}/\sqrt{2}}^{+},\overbrace{e^{-\psi}/\sqrt{2}}^{-},0,0)\\
&\simeq\rho_{a}(x^+,x^-,x_\perp)t^a\delta^{\mu-}\label{current-refframe}
\end{align}
In the second line, we neglect the $+$ component of the 4-velocity of the fluid, since for $\beta\simeq -1$, $\psi\rightarrow-\infty$ and only the minus component remains in this limit. The density $\rho_a$ is now the density in the boosted frame $\rho_a\simeq e^{-\psi}\bar{\rho}_a/\sqrt{2}$ which is larger than $\bar{\rho}_a$ because of the Lorentz contraction in the $z$ direction. Now, we argue that the $x^-$ dependence of the current is irrelevant. We are interested in high energy scatterings where the partons propagating through the medium conserves the $+$ component of their momenta. This boost invariant statement is formalized within the eikonal regime presented Section~\ref{subsub:eikonal-prop}. We will see that neglecting the $x^-$ dependence of the medium gauge field leads to an exact conservation of the $+$ component of the projectiles. Consequently, one can take $\rho_a(x^+,x^-,x_\perp)$ as independent of $x^-$ and evaluated at $x^-=0$ since the right-mover partons propagate near this light cone branch
\begin{equation}\label{rho-refframe}
 \rho_{a}(x^+,x_\perp)\approx \frac{e^{-\psi}}{\sqrt{2}}\bar{\rho}_a\left(x_\perp,\frac{e^{-\psi}}{\sqrt{2}}x^+\right)
\end{equation}
To sum up, the current~\eqref{current-refframe} has only one non-trivial minus component, meaning that all the colour charges of the medium are left movers with the speed of light, and is taken as independent of $x^-\approx0$, simplification valid for eikonal scattering of right-movers. Note however that the standard approximation in pA collision within the CGC formalism $\rho_{a}\propto\delta(x^+)$ is not allowed here for reasons that will become obvious at the end of our calculation: this would neglect emissions with formation time smaller than the projectile in-medium path length and those are precisely the interesting ones in jet quenching physics.

Now, let us choose the light cone gauge $\mathcal{A}^+_m=0$ and the 4-current form given by \eqref{current-refframe}-\eqref{rho-refframe}. In this gauge, the colour current associated with the colour charges of the medium does not couple with the gauge field $\mathcal{A}_m$, leading to a lot of simplifications. Indeed, the 4-current \eqref{current-refframe}-\eqref{rho-refframe} is, in principle, a leading order result which could receive higher order corrections from the feedback of the gauge field on the current. However, it is not the case in this gauge (and in the Lorentz gauge $\partial_\mu\mathcal{A}_m^\mu=0$ as well, see \cite{Iancu:2000hn} or \cite{Blaizot:2004wu}). To see this, let us write the covariant conservation of the colour current $[D_\mu,\mathcal{J}^\mu_m]=0$, which reads in the light cone gauge:
\begin{equation}\label{colour-conservation}
\partial^+\mathcal{J}_m^- -ig[\mathcal{A}_m^+,\mathcal{J}_m^-]=0
\end{equation}
and is automatically satisfied by the form \eqref{current-refframe} with our gauge choice provided that $\partial^+\mathcal{J}_m^-=0$. This means that the colour charge distribution $\rho_a$ in the current~\eqref{current-refframe} is independent of $x^-$ to all orders in $g$. In an other gauge, the current $\mathcal{J}_m^-$ would acquire a higher order $x^-$ dependence via Wilson line conjugation associated with the colour precession of the medium colour charges in the self-generated gluon field. Thus, the light cone gauge $\mathcal{A}^+_m=0$ enables to have a simple picture of the medium in the infinite momentum frame with a clear separation between colour sources and gauge field.

In the light-cone gauge $\mathcal{A}^+_m=0$, it is furthermore possible to solve the classical Yang-Mills equations $[D_\mu,\mathcal{F}_m^{\mu\nu}]=\mathcal{J}^\nu$ for the gauge field \cite{Gelis:2005pt}:
\begin{align}\label{YMeqn1}
\partial^+(\partial_\mu \mathcal{A}_m^\mu)+ig[\mathcal{A}_m^i,\partial^+\mathcal{A}_m^i]&=0\\
[D^-,\partial^+\mathcal{A}_m^-]-[D^i,\mathcal{F}_m^{i-}]&=\mathcal{J}^-\label{YMeqn2}\\
\partial ^+\mathcal{F}_m^{-i}+[D^-,\partial^+\mathcal{A}_{m\perp}]-[D^j,\mathcal{F}_m^{ji}]&=0\label{YMeqn3}
\end{align}
One easily sees that $\mathcal{A}^\mu_m(x^+,x_\perp)$ is a solution of \eqref{YMeqn1}, \eqref{YMeqn2} and \eqref{YMeqn3} if
\begin{equation}\label{YMsol}
 \mathcal{A}_{m,a}^i=0\,,\qquad \partial_{\perp}^2\mathcal{A}^{-}_{m,a}=-g\rho_a(x^+,x_\perp)
\end{equation}
The equation \eqref{YMsol} is very similar to the Poisson equation in electrostatic. It can be solved in Fourier space:
\begin{equation}\label{Amfinal}
\boxed{
 \mathcal{A}^{-}_{m,a}(x^+,x_\perp)=g\int\frac{\dif^2k_\perp}{(2\pi)^2}\frac{e^{-ik_\perp x_\perp}}{k_\perp^2}\rho_a(x^+,k_\perp)}
\end{equation}
This equation is our starting point to calculate the statistical correlations of the gauge field in the next subsection.

\subsection{Statistical properties of $\mathcal{A}_m^\mu$}
\label{sub:Amed-stat}

The collection of colour charges $\rho_a(x^+,x_\perp)$ is a statistical field, as in the McLerran-Venugopalan model \cite{McLerran:1993ni,McLerran:1993ka,McLerran:1994vd}.
In our modelling of the quark-gluon plasma, this accounts for the thermal fluctuations of the distribution. For gauge symmetry reasons, the 1-point correlation function vanishes $ \langle\rho_a\rangle=0$. For an ideal (non-interacting) quark-gluon plasma in thermal equilibrium, the colour charges are uncorrelated so that the 2-points correlation functions read:
\begin{align}\label{2-point-corr-rho}
 \langle \rho_a(x^+,x_\perp)\rho_b(y^+,y_\perp) \rangle&\approx n\delta_{ab}\delta(x^+-y^+)\delta(x_\perp-y_\perp)
\end{align}
 As $\rho_a$, the average squared charge density $n$ is the one measured in the boosted frame, $n=e^{-\psi}\bar{n}/\sqrt{2}$, with $\bar{n}$ the average squared charge density in the plasma rest frame, which scales like the plasma temperature $T^3_{\rm p}$ for an ideal quark-antiquark gas with vanishing baryon chemical potential. For a static, infinite, homogeneous plasma, $\bar{n}$ does not depend on $x^+$ nor $x_\perp$. The form of the correlator \eqref{2-point-corr-rho} in a longitudinally expanding medium is discussed in the next section.

We assume Gaussian statistics meaning that the probability distribution is entirely known from these one-point and two-points correlations. This amounts to neglect the interactions between the scattering centers and the non-linear self interactions. As we shall see, the latter approximation is not valid for very long range correlations, larger than the Debye screening length.

From the linear property of \eqref{Amfinal}, the background gauge field $\mathcal{A}^{-}_m$ is also Gaussian distributed. Combining equations \eqref{2-point-corr-rho} and \eqref{Amfinal} together, one easily finds its 2-point correlator in coordinate space:
\begin{equation}\label{Amed-correlator}
\boxed{
 \langle \mathcal{A}^{-}_{m,a}(x^+,x_\perp)\mathcal{A}^{-}_{m,b}(y^+,y_\perp)\rangle=n\delta_{ab}\delta(x^+-y^+)\Upsilon(x_\perp-y_\perp)}
\end{equation}
with
\begin{equation}\label{Fourier-coulomb}
 \Upsilon(u_\perp)=g^2\int\frac{\dif^2 k_\perp}{(2\pi)^2}\frac{e^{-ik_\perp u_\perp}}{k_\perp^4}
\end{equation}
Note that we ignore the potential transverse dependence of the squared charge density $n$ in this calculation.
The function $\Upsilon$ is the inverse Fourier transform of the Coulomb elastic scattering cross-section $\tilde{\Upsilon}(k_\perp)= g^2/k_\perp^4$. Written like this, the Fourier transform is not well defined due to the infrared singularity as $k_\perp\rightarrow0$. It has to be regulated in some way. Actually, even for an ideal quark-gluon plasma, one cannot ignore screening effects: in the rest frame of the plasma, long range interactions are screened above the characteristic Debye length $\lambda_D=1/\mu_D$ where $\mu_D$ is the Debye mass. This provides a natural infrared cut-off $|k_\perp|\ge\mu_D$ for the integral \eqref{Fourier-coulomb}.

To be more precise and for the sake of completeness, we briefly review the two mains regularizations of\eqref{Fourier-coulomb} encountered in the literature:
\begin{enumerate}
 \item for a thermal plasma in equilibrium, the two-point correlation function of the gauge field \eqref{Amed-correlator} is naturally regularized by the retarded propagator with hard thermal loops resummed. This accounts for Debye screening and Landau damping of the colour fields. The elastic cross-section reads then \cite{Aurenche:2002pd}:
 \begin{equation}\label{HTL}
 \frac{1}{2}\bar{n}\tilde{\Upsilon}_{\rm HTL}(k_\perp)=\frac{\mu^2_DT_{\rm p}}{k_\perp^2(k_\perp^2+\mu^2_D)}
\end{equation}
 \item Another way to regulate the Coulomb divergence is to consider a Yukawa type cross-section by analogy between mass screening and Debye screening \cite{Wang:1991xy}:
  \begin{equation}\label{Yukawa}
 \tilde{\Upsilon}_{\rm Yuk}(k_\perp)=\frac{g^2}{(k_\perp^2+\mu_D^2)^2}
\end{equation}
\end{enumerate}

The emergence of the scale $\mu_D$ in this problem enables to clarify the validity regime of the assumption of instantaneous and independent interactions during the longitudinal motion of a test parton through the medium, namely the $\delta(x^+-y^+)$ in \eqref{2-point-corr-rho}. First of all, from QCD at finite temperature (see \cite{Blaizot:2003tw} for a review), one knows that:
\begin{equation}
 \mu_D \sim gT_{\rm p}
\end{equation}
in the plasma rest frame.
Then, the elastic mean free path $\ell\sim1/(\bar{n}\sigma_{\ell})$ in this frame is parametrically of order $\mu_D^2/(g^4\bar{n})\sim(g^2T_{\rm p})^{-1}$ with $g^4/\mu^2_D$ the order of magnitude of the elastic cross-section $\sigma_{\ell}$. Thus, for a weakly coupled quark-gluon plasma:
\begin{equation}
 \ell\gg \lambda_D
\end{equation}
The elastic mean free path being much larger than the typical duration of plasma interaction, it is allowed to treat collisions as independent and instantaneous, at least in the weak coupling regime.

\subsection{Bjorken expansion}
\label{sub:Bjorken}

In order to include the longitudinal expansion of the quark-gluon plasma in our medium modelling, we use a pragmatic approach relying on the results derived in the static case following \cite{Baier:1998yf,Iancu:2018trm} for example.
By longitudinal expansion, we mean here the expansion of the plasma in the beam direction $X\equiv X^1=x^1$, which is \textit{not} the $X^3=Z$ axis of the coordinate system in the laboratory frame (see Introduction~\ref{chapter:intro}). In the Bjorken model, the plasma is homogeneous in the transverse direction with respect to beam axis and boost invariant along the beam direction. This implies that the fluid state variables depend on the proper time $\tau=\sqrt{{X^0}^2-{X^1}^2}=\sqrt{T^2-X^2}$ of fluid elements only.
Fluid state variables such as temperature or density are well defined if the fluid reaches local thermal equilibrium. 

The temperature field is a function of $\tau$: $T_{\rm p}(X^\mu)=\hat{T}_{\rm p}(\tau)$. For an ideal hydrodynamical expansion, $\hat{T}_{\rm p}(\tau)$ satisfies \cite{Bjorken:1982qr}:
\begin{equation}\label{temp-eq}
 \frac{1}{\hat{T}_{\rm p}}\frac{\dif \hat{T}_{\rm p}}{\dif \tau}+\frac{c_s^2}{\tau}=0
\end{equation}
where $c_s\le 1/3$ is the sound velocity in the plasma, assumed to be constant.
The resolution of \eqref{temp-eq} gives:
\begin{equation}\label{temp-Bjorken}
 \Big(\frac{\hat{T}_{\rm p}}{\hat{T}_0}\Big)^3=\Big(\frac{\tau_0}{\tau}\Big)^{3c_s^2}
\end{equation}
meaning that the average squared charge density $\bar{n}$ (defined in the rest frame of a fluid element) follows also a power law since $\bar{n}$ is proportional to $T^3_{\rm p}$ for an ideal quark-gluon plasma,
\begin{equation}\label{nsquared-rest}
 \bar{n}(\tau)=\bar{n}_0\Big(\frac{\tau_0}{\tau}\Big)^{3c_s^2}.
\end{equation}

Now we argue that the dominant effect of the longitudinal cooling on the results obtained in the static case is to introduce a $x^+$ dependence of the squared charge density $n(x^+)$ in \eqref{2-point-corr-rho} and \eqref{Amed-correlator}. Indeed, if we focus on a space-time region close to the plane $x^1=X=0$ and close to the light-cone branch $x^-=0$, one has the following identities:
\begin{equation}\label{coordinate-eq}
 \tau=T=X^3=\frac{e^{-\psi}}{\sqrt{2}}x^+
\end{equation}
This means that in this space-time domain, the longitudinal cooling can be mimicked by a $X^3$ modulation of a static current. We have already justified why the region $x^-=0$ is relevant in high energy scatterings. By focusing on the region $x^1=0$, we also assume that the highly energetic partons propagate through the medium in the plane transverse to the beam axis. To generalize the present results for dijet production at non-zero rapidity, one can use a boost along the \textit{beam} direction \cite{Iancu:2018trm}.

Therefore, we can refer to the calculations done in the static case to obtain the current and the gauge field in the boosted frame (see Eqs.~\eqref{current-refframe}-\eqref{rho-refframe} and \eqref{Amfinal}). In the end, the only difference relies in the $x^+$ dependence of $n(x^+)$ in the 2-point correlators \eqref{2-point-corr-rho} and \eqref{Amed-correlator}. Using \eqref{nsquared-rest} and \eqref{coordinate-eq}, one finds
%
%
$
 n(x^+)\simeq e^{-\psi}\bar{n}_0(x_0^+/x^+)^{3c_s^2}
$
with $x_0^+=e^{-\psi}\tau_0$ the light cone time at which the projectile enters the plasma in the boosted frame. With $n_0=e^{-\psi}\bar{n}_0$ the initial density in the boosted frame, this is rewritten:
%
%
\begin{tcolorbox}[ams equation]\label{nxplus}
 n(x^+)=n_0\left(\frac{x_0^+}{x^+}\right)^{\gamma}\,,\qquad 0\le\gamma\equiv3c_s^2\le 1
\end{tcolorbox}
\noindent We will use explicitly this formula inside the quenching parameter defined in the next section. The resulting $x^+$ dependence of the quenching parameter is important for the discussions in Chapter \ref{chapter:DLApic}, sections \ref{sub:veto-expansion} and \ref{sub:DLpic-bjorken}.

\section{Medium-induced emissions}
\label{sec:BDMPSZ}

\subsection{Transverse momentum distribution in the eikonal approximation}

Before exploring the medium-induced emission spectrum, let us start with the following more simple exercise as a warm-up. The projectile is a highly energetic on-shell quark with four-momentum $p^\mu=(p^+,p^-,p_\perp)$ and we want to compute the cross-section for producing one quark with four momentum $p'^\mu=(p'^+,p'^-,p'_\perp)$ out of the medium, to all orders in $g\mathcal{A}_m^\mu$ and order 0 in $g$. In the high energy limit, one neglects the quark mass. As usual in quantum field theory, one must first calculate the scattering amplitude $\mathcal{M}(p',p)\equiv\braket{p'|p}$. This amplitude depends also on the spins $s$, $s'$ and colours $j$, $j'$ of the incoming and outgoing quarks. Then, the differential cross-section is given by
\begin{equation}\label{TMDcross}
 \frac{\dif^3\sigma}{\dif p'^+\dif^2p'_\perp}=\frac{1}{T_{\mathfrak{B}}2p^0(2\pi)^3 2p'^+}\left(\frac{1}{2N_c}\sum_{s,s',j,j'}|\mathcal{M}(p',p)|^2\right)
\end{equation}

From the Lehmann-Symanzik-Zimmermann (LSZ) reduction formula, the amplitude $\braket{p'|p}$ is related to the truncated quark propagator in momentum space $D_{tc}(p,p'|\mathcal{A}_m )$ according to:
\begin{equation}\label{TMDamp}
 \braket{p'|p}=\lim_{p^2,p'^2\rightarrow0}\frac{1}{Z_\psi}\bar{u}_s(p')D_{tc}(p,p'| \mathcal{A}_m)u_s(p)
\end{equation}
with $Z_\psi$ the quark wave function renormalization constant. We keep the dependence of the quark propagator on the background field explicit (in the vacuum, this calculation would be pointless since $\braket{p'|p}\propto\delta^{(4)}(p'-p)$ for idealized quark asymptotic states). Hence, to calculate \eqref{TMDcross} to all orders in $g\mathcal{A}_m^\mu$, we need to calculate the truncated quark propagator $D_{tc}(p,p'|\mathcal{A}_m )$.

\subsubsection{In-medium eikonal quark propagation}
\label{subsub:eikonal-prop}

Calculating $D_{tc}(p,p'|\mathcal{A}_m )$ exactly is complicated. However, in the eikonal regime, there exist explicit analytic formulas. In this regime, the incoming quark moves along the light cone $+$ direction with a large $p^+\rightarrow\infty$ component, i.e.\ $p^+\gg |p_\perp|$ (and so $p^-\ll |p_\perp|$) which is conserved during the scattering process $p^+\simeq p'^{+}$. This means that the momentum transferred by the background field, $q=p'-p$ is small: $-q^2\ll p^{+2}$. 

There are several methods to obtain the eikonal truncated quark propagator. The $S$-matrix element \eqref{TMDamp} in the high energy limit has been obtained for instance in \cite{Bjorken:1970ah} in light-cone quantization using the boost invariance of the $S$-matrix (see also \cite{Feldman:1973ps} in conventional quantization). A pedestrian methods consists in calculating and resumming all the Feynman diagrams with external field insertions to all orders. The calculation is sketched in \cite{CasalderreySolana:2007pr} and done precisely in \cite{Hebecker:1999ej,Meggiolaro:2000yf}. One can show that the coupling of the incoming particle with the external field is independent of the representation of the Lorentz group to which the particle belongs. In particular it is the same for scalar particles as for Dirac fermions. We note $\gamma^\mu_{\rm cl}(x^+)=\gamma_{\rm cl}^\mu(0)+u^\mu x^+$ the classical path of the incoming quark through the medium, parametrized by the $x^+$ coordinate along the quark trajectory with $\gamma_{\rm cl}^\mu(0)\equiv(0,x^-,x_\perp)$ and $u^\mu=p^\mu/p^+$ ($u^\mu$ in light cone coordinates). The result of this calculation in the scalar case is then:
\begin{equation}\label{eikonal-prop}
 \boxed{\mathcal{D}_{tc}\left(p,p'|\mathcal{A}_m\right)=2p^+\int\dif^2x_\perp\dif x^-\,e^{ix^-(p'^+-p^+) -ix_\perp(p_\perp-p'_\perp)}[W(\gamma_{\rm cl})]_{j'j}}
\end{equation}
where $W(\gamma_{\rm cl})$ is the Wilson line along the path $\gamma^\mu(x^+)$:
\begin{equation}
 W(x_{\rm cl})=T\left[\exp\left(ig\int_{-\infty}^{\infty}\dif x^+\mathcal{A}_{m,\mu}(\gamma^\mu(x^+))u^\mu\right)\right]
\end{equation}
The operator $T$ is the usual time ordering operator: in expanding the exponential, the colour matrices $\mathcal{A}_{m,\mu}$ are ordered with decreasing time $x^+$ from the left to the right. For fermionic quarks, the truncated propagator reads $\bar{u}_s(p')D_{tc}(p,p'|\mathcal{A}_m) u_s(p)=\delta_{ss'} \mathcal{D}_{tc}(p,p'|\mathcal{A}_m )$, reflecting the fact that quarks conserve their spin in the eikonal limit.

Formula \eqref{eikonal-prop} enables to justify a posteriori why it is allowed to take a $x^-$ independent $4$-current and background gauge field (it is even more transparent in the derivation done in \cite{Bjorken:1970ah}). If we are interested in the regime $p'^{+}\simeq p^+$, neglecting the $x^-$ dependence of $\mathcal{A}_{m,\mu}$ in the integral \eqref{eikonal-prop} gives a $\delta(p'^{+}-p^+)$ factor as requested. This is independent of the gauge choice. It turns out that $\mathcal{A}_{m,\mu}$ does not depend at all on $x^-$ for a current of the form \eqref{current-refframe} in the light cone gauge $\mathcal{A}_{m}^+=0$. Hence, for $u_\perp=0$:
\begin{align}\label{eikonal-prop2}
\bar{u}_s(p')D_{tc}(p,p'|\mathcal{A}_m) u_s(p)&=4\pi p^+\delta_{ss'}\delta(p'^+-p^+)\int\dif^2x_\perp\,e^{-ix_\perp(p'_\perp-p_\perp)}[W(x_\perp)]_{j'j}\\ \label{wilson-line2}
W(x_\perp)&=T\Big[\exp\Big(ig\int_{-\infty}^{\infty}\dif x^+\mathcal{A}^{-}_{m}(x^+,x_\perp)\Big)\Big]
\end{align}

An other approach is to use the worldline formalism to write an exact formal expression for the propagator. In the worldline formalism, the full \textit{scalar} quark propagator in the background field $\mathcal{A}^\mu_m$ reads:
\begin{equation}\label{worldlineprop}
 \mathcal{D}(p,p'| \mathcal{A}_m)=\int_0^{\infty}\dif \tau\int\mathcal{D}x(\tau')\,e^{ip'x(\tau)-ipx(0)}T\left[\exp\left(i\int_0^\tau\dif\tau'\, \frac{1}{2}\dot{x}^2-g\mathcal{A}_{\mu,m}(x)\dot{x}^\mu\right)\right]
\end{equation}
Then, in the eikonal approximation, the path integral \eqref{worldlineprop} is calculated
 with the saddle-point method at the lowest order. The saddle point is the classical trajectory of the particle, and after truncation of the propagator, one gets formula \eqref{eikonal-prop}. This technique explicitly shows that the eikonal approximation is equivalent to neglect quantum fluctuations around the classical path. For non-scalar particles, formula \eqref{worldlineprop} can be modified to take into account the non-trivial Lorentz structure. However, in the eikonal/saddle-point approximation, these terms are irrelevant so Eq.~\eqref{eikonal-prop} is very general.
 
\subsubsection{Transverse momentum broadening}
\label{subsub:TMB}

When taking the square of the amplitude $\braket{p'|p}$, this generates a Dirac delta $\delta(p'^+-p^+)$ squared, which should be understood as:
\begin{align}
 \delta(p'^+-p^+)^2&\equiv\delta(p'^+-p^+)\frac{1}{2\pi}\int \dif x^-\\
&=\delta(p'^+-p^+)\frac{1}{2\pi}\frac{T_{\mathfrak{B}}}{\sqrt{2}}
\end{align}
The second equality comes from the normalization of the one particle states. Indeed, using the result \eqref{wilson-line2} in the vacuum, that is with $W(x_\perp)=1$, and for $p'=p$ one gets:
\begin{align}
 \braket{p|p}&=4\pi p^+ \delta(p^+=0)\int \dif^2x_\perp\\
 &=2p^+S_{\mathfrak{B}}\int\dif x^-=\sqrt{2}p^+S_{\mathfrak{B}}T_{\mathfrak{B}}
\end{align}
The last equality is precisely \eqref{norm-state}. Hence, the volume factor $T_{\mathfrak{B}}$ cancels with the $1/T_{\mathfrak{B}}$ in the cross-section and using $2p^0\simeq \sqrt{2}p^+$ with eikonal accuracy in \eqref{TMDcross}, the two factors $1/\sqrt{2}$ gives a factor $1/2$. Finally, one gets
\begin{equation}\label{dipole-S}
 \frac{\dif^3\sigma}{\dif p'^+\dif^2 p'_\perp}=\frac{\delta(p'^+-p^+)}{(2\pi)^2}\frac{1}{N_c}\int\dif^2x_\perp\dif^2y_\perp e^{-i(x_\perp-y_\perp)(p'_\perp-p_\perp)}\Tr\left(W(x_\perp)W^{\dagger}(y_\perp)\right)
\end{equation}
The wave function renormalization constant is equal to one at this order (see \cite{Meggiolaro:2000yf}).
The last step of the calculation is to average over all statistical configurations of $\mathcal{A}_m$ using the results of Section~\ref{sub:Amed-stat}. Performing the trivial integration over $p'^+$, the previous equation becomes:
\begin{equation}
 \frac{\dif^2\sigma}{\dif^2 p'_\perp}=\frac{1}{(2\pi)^2}\int\dif^2b_\perp\dif^2u_\perp e^{-iu_\perp(p'_\perp-p_\perp)}\frac{1}{N_c}\left\langle\Tr\,W\left(\frac{b_\perp+u_\perp}{2}\right)W^{\dagger}\left(\frac{b_\perp-u_\perp}{2}\right)\right\rangle
\end{equation}
Thus, we are left with the calculation of the average of two Wilson lines with ``impact parameter'' $b_\perp$ and separated by a transverse distance $u_\perp$. The standard way to make this calculation to all orders in $g$ is to expand the product of the two Wilson lines up to the first non trivial order 
in $g$ (leading order in opacity), using the expression for the correlation between the medium gauge field \eqref{Amed-correlator}. One easily finds, for a medium with constant density in the transverse plane,
\begin{equation}
\frac{1}{N_c}\left\langle\Tr\,W\left(\frac{b_\perp+u_\perp}{2}\right)W^{\dagger}\left(\frac{b_\perp-u_\perp}{2}\right)\right\rangle=1-g^2C_R\int_{-\infty}^{\infty}\dif x^+n(x^+)[\Upsilon(0)-\Upsilon(u_\perp)]+\mathcal{O}(g^4)
\end{equation}
$C_R$ depends on the representation of $\mathfrak{su}(N_c)$ in the Wilson line ($C_R=C_F$ here). To all orders in $g\mathcal{A}^\mu_m$, the average of two Wilson lines exponentiates (see \cite{Iancu:2003xm}):
\begin{equation}\label{dipolecross}
\frac{1}{N_c}\left\langle\Tr\,W\left(\frac{b_\perp+u_\perp}{2}\right)W^{\dagger}\left(\frac{b_\perp-u_\perp}{2}\right)\right\rangle=\exp\left(-g^2C_R\int_{-\infty}^{\infty}\dif x^+\,n(x^+)\big(\Upsilon(0)-\Upsilon(u_\perp)\big)\right)
\end{equation}

\paragraph{Dipole cross-section.} In this chapter as well as in Chapter \ref{chapter:jet}, we will often encounter the dipole cross-section appearing inside the exponential in \eqref{dipolecross} and defined by:
\begin{align}\label{dipole-cross-def}
 \sigma_d(u_\perp)&\equiv2\int\frac{\dif^2k_\perp}{(2\pi)^2}\Big(1-e^{ik_\perp u_\perp}\Big)\tilde{\Upsilon}(k_\perp)\\
&= 2\big(\Upsilon(0_\perp)-\Upsilon(u_\perp)\big)
\end{align}
with $\tilde{\Upsilon}$ defined in \eqref{HTL}-\eqref{Yukawa}.
In the case at hand, this dipole is the effective $q\bar{q}$ dipole formed by the quark in the amplitude and the ``antiquark'' in the complex conjugate amplitude.
This dipole cross-section can be estimated in the regime where the transverse size of the dipole $u_\perp$  is smaller than the Debye length. In this case, the integral is controlled by small values of $k_\perp$ between $\mu_D$ and $1/u_\perp$ so that a Taylor expansion of the factor $(1-\exp(ik_\perp u_\perp))$ up to the second order gives the leading behaviour of $\sigma_d(u_\perp)$:
\begin{equation}
 u_\perp\ll \lambda_D\Longrightarrow \sigma_d(u_\perp)=\frac{g^2}{16\pi}u_\perp^2\log\left(\frac{1}{\mu_D^2u_\perp^2}\right)+\mathcal{O}(u_\perp^2)
\end{equation}
With this estimate, one defines the quenching parameter $\qhat_R$ in colour representation $R$:
\begin{equation}\label{def-qhat}
 g^2C_Rn(x^+)\sigma_d(u_\perp)\equiv \frac{1}{2}\qhat_R(x^+,u_\perp)u_\perp^2+\mathcal{O}(u_\perp^2)
\end{equation}
Using \eqref{def-qhat} and \eqref{dipole-cross-def}, one gets:
\begin{equation}
 \qhat_R(x^+,u_\perp)=2\pi\alpha_s^2C_R n(x^+)\log\left(\frac{1}{\mu_D^2u_\perp^2}\right)
\end{equation}
Thus, the quenching parameter depends on the size of the dipole crossing the medium. 

Let us now come back to \eqref{dipolecross}.
In order to avoid complications due to the transverse modelling of the medium, let us consider the differential cross-section per impact parameter. The final result of this subsection is then:
\begin{tcolorbox}[ams align]\label{finalTMDcross}
 \frac{\dif^4\sigma}{\dif^2b_\perp \dif^2 p'_\perp}&=\frac{1}{(2\pi)^2}\int\dif^2u_\perp\exp\left(-iu_\perp(p'_\perp-p_\perp)-\frac{g^2}{2}C_R\int_{-\infty}^{\infty}n(x^+)\sigma_d(u_\perp)\dif x^+\right)
\end{tcolorbox}
We now discuss the physical meaning of this equation. The outgoing quark acquires a transverse momentum after propagation through the medium. One can generally define  the probability distribution $\mathcal{P}(k_\perp,t_1,t_2)$ for an incoming quark to end up with a transverse momentum $k_\perp$ after propagation through the medium between $x^+=t_1$ and $x^+=t_2$:
\begin{equation}\label{Ptransverse}
\boxed{
 \mathcal{P}(k_\perp,t_1,t_2)=\frac{1}{(2\pi)^2}\int\dif^2u_\perp\exp\left(-iu_\perp k_\perp-\frac{g^2}{2}C_R\int_{t_1}^{t_2}n(x^+)\sigma_d(u_\perp)\dif x^+\right)}
\end{equation}
This probability is properly normalized as $\sigma_d(0_\perp)=0$. It is also explicitly boost invariant in the sense that one can use the average density of colour charges $n$ measured in the laboratory frame provided that $t_1$ and $t_2$ refer to the light cone time along the quark trajectory in this frame. This is a generic feature of all the cross-sections calculated in this chapter.

\paragraph{The harmonic approximation.} The harmonic approximation for the quenching parameter is very well suited for analytic approximations and captures the mean features of the multiple soft scattering regime. In this approximation, one neglects the logarithmic dependence of the quenching parameter in \eqref{def-qhat}, by fixing $u_\perp$ at the typical transverse size $R_\perp$ encountered in the calculation at hand,
\begin{equation}
 \qhat_R(x^+)\equiv2\pi\alpha_s^2C_R n(x^+)\log\left(\frac{1}{\mu_D^2R_\perp^2}\right)
\end{equation}
In most of the calculation done in this chapter, the typical transverse size $R_\perp$ is given by the inverse of the maximal transverse momentum acquired via collisions during a propagation time $L$, namely $Q_s=\sqrt{\qhat L}$\footnote{The antenna pattern in Section \ref{sec:decoherence} is an exception, as $R_\perp\sim \thqq L\le 1/Q_s$ if the opening angle $\thqq$ of the dipole is smaller than $\th_c$.}. This enables to define the quenching parameter through the implicit relation:
\begin{equation}\label{qhat-harmonic}
\boxed{
 \qhat_R=2\pi\alpha_s^2C_R n(x^+)\log\left(\frac{\qhat_R L}{\mu_D^2}\right)}
\end{equation}
From this equation, one sees that $\qhat$ is reduced by two powers of $\alpha_s$, however, this reduction is compensated by the large value of $n$ in a \textit{dense} medium. This is characteristic of the calculations performed in this section, to all orders in $\alpha_s^2 n$. With a constant $\qhat_R$, the probability distribution \eqref{Ptransverse} is Gaussian and accounts for the random diffusive walk in the transverse plane of the quark due to the ``kicks'' given by the medium constituents. In this thesis, one should keep in mind that the parameter $\qhat$ is always the \textit{adjoint} $\qhat_A$. Whenever necessary, we restore the index to avoid any ambiguity. Finally, for an expanding medium, $\qhat_R$ depends on $x^+$. Using \eqref{nxplus} in \eqref{qhat-harmonic}, this dependence is written $\qhat(x^+)=\qhat_0(x_0^+/x^+)^\gamma$ in the harmonic approximation. 

\subsection{One-gluon emission spectrum}
\label{sub:one-gluon}

\subsubsection{Effective generating functional for soft gluon emissions}

Now, we turn to the calculation of the one-gluon emission spectrum from an on-shell or off-shell incoming quark. With respect to the previous section, we are now looking for the cross-section for producing one gluon with 4-momentum $k^\mu$ after the propagation of the highly energetic quark through the medium, to leading (first) order in $g$ but all orders in $g\mathcal{A}_m$. We consider a quark, but as the eikonal propagator of a gluon has the same expression, with a Wilson line in the adjoint representation, all the calculations of this section are valid for a gluon emission from an eikonal gluon with the substitution $C_F\rightarrow C_A$.

The basic ingredient of this calculation is the amplitude $\mathcal{M}(k,p',p)\equiv\braket{k,p'|p}$ from which one gets the cross-section:
 \begin{equation}\label{bdmpszcross}
 \frac{\dif^6\sigma}{\dif p'^+\dif^2p'_\perp\dif k^+\dif k_\perp}=\frac{1}{T_{\mathfrak{B}}2p^0 (2\pi)^6  2p'^+ 2k^+}\left(\frac{1}{2N_c}\sum_{s,s',\lambda,j,j',c}|\mathcal{M}(k,p',p)|^2\right)
\end{equation}
where the sum runs over all initial and final colours $j$, $j'$, $c$, spins $s$ and $s'$ and physical polarizations $\lambda$ of the emitted gluon. We use the eikonal approximation for the quark propagation. This means that the emitted gluon cannot be too ``hard''. More explicitly, the constraints on the 4-momentum of the gluon are $k^+\ll p^+\simeq p'^+$ and $k_\perp\ll p_\perp,p'_\perp$.

From the LSZ reduction formula, the amplitude $\mathcal{M}(k,p',p)$ is related to the time ordered three-point Green function of the theory. The generating functional of these Green functions in QCD in the presence of the background field $\mathcal{A}_m$ is 
 \begin{align}
  Z[J^{\mu},\bar{\eta},\eta\mid\mathcal{A}_m]&=\int\mathcal{D}A^{\mu}\mathcal{D}\bar{\psi}\mathcal{D}\psi\,\delta(G[A^{\mu}])\exp\left(i S_{g}[A^{\mu}|\mathcal{A}^{\mu}_{m}]+i\int{\dif^4x\,\bar{\psi}i\slashed{D}\psi}\right.\nonumber\\
  &\hspace{6cm}\left.+i\int{\dif^4x\, (J^{\mu}A_{\mu}+\bar{\eta}\psi+\eta\bar{\psi})}\right)
 \end{align}
  where $S_{g}[A^{\mu}|A^{\mu}_{m}]$ is the part of the Yang-Mills action depending only on the gluon and medium fields and $\delta(G[A^{\mu}])$ is a 
 gauge-fixing prescription. We work in the (physical) light cone gauge $A^+=0$. As we consider an energetic quark which propagates through the medium, we shall need to compute correlation functions with 
 at least one $\bar\psi$ and one $\psi$. It is then convenient to introduce the following effective generating functional:
 \begin{align}
  Z_{\text{eff}}[J^{\mu}|\mathcal{A}_m]&=\Big(-i\frac{\delta}{\delta\bar\eta^{j'}_{s'}(y)}\Big)\Big|_{\bar\eta=0}\Big(-i\frac{\delta}{\delta\eta^j_s(x)}\Big)\Big|_{\eta=0}\,Z[J^{\mu},\bar{\eta},\eta|A^{\mu}_{m}]\\
  &=\int\mathcal{D}\mathcal{A}^{\mu}\mathcal{D}\bar{\psi}\mathcal{D}\psi\,\delta(G[A^{\mu}])\psi^{j'}_{s'}(y)\bar\psi^j_s(x)\exp\Big(i S_{g}[A^{\mu}|\mathcal{A}^{\mu}_{m}]\nonumber\\
  &\hspace{7cm}\left.+i\int{\dif^4x\,\bar{\psi}i\slashed{D}\psi}+i\int{\dif^4x\,J^{\mu}A_{\mu}}\right)
 \end{align}
In the eikonal propagation approximation, it is possible to integrate out exactly the fermionic degrees of freedom using the propagator $D(x,y|A+\mathcal{A}_m)$ discussed in the previous section. It is even more convenient to use the truncated propagator $\delta_{ss'}\mathcal{D}_{tc}(p,p'|A+\mathcal{A}_m)$ in order to reduce the initial and final quarks according to the LSZ procedure. Our effective generating functional is then:
 \begin{equation}
 Z_{\text{eff}}[J^{\mu}|\mathcal{A}_m]=\int{\mathcal{D}A^{\mu}\delta(G[A^{\mu}])\delta_{ss'}\mathcal{D}_{tc}(p,p'|A+\mathcal{A}_m)\exp\left(i S_{g}[A^{\mu}|\mathcal{A}^{\mu}_{m}]+i\int{\dif^4x J^{\mu}A_{\mu}}\right)}
 \end{equation}
 with $\mathcal{D}_{tc}(p,p'|A^\mu+\mathcal{A}^{\mu}_m)$ given by \eqref{eikonal-prop}.
 
From the LSZ reduction formula, the amplitude $\mathcal{M}(k,p',p)$ is now related to the one point function of this effective generating functional:
\begin{align}\label{LSZ-bdmpsz}
\mathcal{M}(k,p',p)&=\lim\limits_{k^2\rightarrow 0}\,\frac{k^2}{iZ_A^{1/2}Z_{\psi}}\bm{\langle} A_{\mu,c}(k)\bm{\rangle}\epsilon^\mu_\lambda(k)\\\label{onepointA}
\bm{\langle} A^{\mu}_c(z)\bm{\rangle}&\equiv\frac{1}{Z_0}\Big(-i\frac{\delta}{\delta J_{\mu,c}(z)}\Big)Z_{\text{eff}}[J^{\mu}|\mathcal{A}_m]\Big|_{J=0}
\end{align}
In the light cone gauge $A^+=0$, only the transverse components of the polarization vector $\epsilon^\mu_\lambda(k)=(0,\epsilon_\perp\cdot k_\perp/k^+,\epsilon_\perp)$ and $A^\mu$ matter. Written explicitly, the above formula is
\begin{equation}\label{<Ai>}
 \bm{\langle} A^{i}_c(z)\bm{\rangle}\propto\frac{1}{Z_0}\int\mathcal{D}A^{\mu}\,\delta(G[A^{\mu}])e^{i S_{g}[A^{\mu}|\mathcal{A^{\mu}}_{m}]}
A^i_c(z)T\Big[e^{ig\int_{-\infty}^{\infty}\dif x^+\,(A^-(x^+,x^-,x_\perp)+\mathcal{A}^-_{m}(x^+,x_\perp))}\Big]_{j'j}
\end{equation}
where the $\propto$ symbol abbreviates all the factors and integrals over $x_\perp$ and $x^-$ in \eqref{eikonal-prop}. As in the transverse momentum distribution calculation, we choose a coordinate system where $x^-$ and $x_\perp$ are constant along the classical path of the incoming quark. We point out that the brackets in \eqref{<Ai>} refer to the \textit{quantum expectation value} of the quantum operator $A$ and should be distinguished from the statistical averages over background field configuration performed later on in this section.

\subsubsection{Gluon propagators beyond the eikonal approximation}
As we shall see after this short digression, the calculation of \eqref{<Ai>} at a given fixed order in $gA^\mu$ requires to know the gluon propagator $G_{ab}^{\mu\nu}(z,x)$ in the background field beyond the eikonal approximation discussed in \eqref{subsub:eikonal-prop}:
\begin{equation}\label{Gprop}
 G_{ab}^{\mu\nu}(z;x|\mathcal{A}_m)\equiv\frac{1}{Z_0}\int{\mathcal{D}A^{\mu}\delta(G[A^{\mu}])e^{i S_{g}[A^{\mu}|\mathcal{A}^{\mu}_{m}]}
A^\mu_a(z)A^\nu_b(x)}
\end{equation}
We focus here only on the $i-$ component of the gluon necessary to pursue this calculation. The other components are given in Appendix~\ref{app:prop} and will be used mainly in Section~\ref{sec:decoherence}.

Since the external field appearing in $S_{g}[A^{\mu}|\mathcal{A}^{\mu}_{m}]$ does not depend on $x^-$, the propagator $G_{ab}^{\mu\nu}(z,x)$ depends on $z^-$ and $y^-$ only through the difference $z^--y^-$. The linearisation of the Yang-Mills equations around the background field leads to the following formula for the $-i$ component\footnote{We do not include the $k^++i\epsilon$ prescription for retarded propagators to keep our expressions compact.} (see \cite{Iancu:2000hn} or \cite{Iancu:2014kga}):
\begin{equation}\label{def-gscalar-prop}
G^{i-}_{ab}(z,x|\mathcal{A}_m)=\partial^i_{x_\perp}\int \frac{\dif k^+}{2\pi}e^{-ik^+(z^--x^-)}\frac{i}{2{k^+}^2}\mathcal{G}_{ab}(z^+,z_\perp;x^+,x_\perp|k^+)
\end{equation}
with the scalar propagator $\mathcal{G}$ satisfying the Green equation
\begin{equation}\label{scalar-green-eqn}
\left(i\delta^{ab}\partial^- +\delta^{ab}\frac{\nabla_{x_\perp}^2}{2k^+}+g\mathcal{A}_{m,d}^-T^d_{ab}\right)\mathcal{G}_{bc}(z^+,z_\perp;x^+,x_\perp|k^+)=i\delta_{ac}\delta(z^+-x^+)\delta^{(2)}(z_\perp-x_\perp)
\end{equation}
This propagator satisfies also the Chapman-Kolmogorov relation:
\begin{equation}\label{ChapKol}
\mathcal{G}_{ab}(z^+,z_\perp;x^+,x_\perp|k^+)=\int{dw_\perp\mathcal{G}_{ad}(z^+,z_\perp;w^+,w_\perp|k^+)\mathcal{G}_{db}(w^+,w_\perp;x^+,x_\perp|k^+)}
\end{equation}
if $x^+<w^+<z^+$.

The solution of the equation satisfied by $\mathcal{G}$ has an elegant representation in term of a path integral as in non-relativistic quantum mechanics. Indeed, Eq.~\eqref{scalar-green-eqn} is formally a two-dimensional Schr\"{o}dinger equation in a non-abelian potential for a particle with effective mass $k^+$. Its Green function reads:
\begin{equation}\label{scalar-prop-subei}
\mathcal{G}(z^+,z_\perp;x^+,x_\perp|k^+)=\int_{r_\perp(x^+)=x_\perp}^{r_\perp(z^+)=z_\perp}\mathcal{D}r_\perp(\xi)\exp\left(\frac{ik^+}{2}\int_{x^+}^{z^+}\dif \xi \,\dot{r}_\perp^2(\xi)\right)\mathcal{W}_{x^+}^{z^+}(r_\perp(\xi))
\end{equation}
In the strict eikonal approximation given by the saddle point evaluation of the path integral, this propagator reduces to a Wilson line over the classical path in the adjoint representation, in agreement with our calculation of the gluon \textit{eikonal} propagator. The propagator \eqref{scalar-prop-subei} includes sub-eikonal corrections by averaging Wilson lines over random quantum motions in the transverse plane. 

One may wonder why the eikonal approximation is not enough for our purposes. The eikonal approximation is given by the saddle point evaluation of \eqref{scalar-prop-subei}, approximation valid when the phase in the exponential is large. For a gluon propagating over a distance $L$ through the medium, this phase is parametrically of order $k^+(\Delta r_\perp/L)^2L$, with $\Delta r_\perp$ the typical variation around the classical trajectory. By the uncertainty principle, $\Delta r_\perp$ is of the same order than $1/k_\perp$ where $k_\perp$ is the measured transverse momentum of the gluon. Consequently, the eikonal approximation is valid when:
\begin{equation}\label{eikonal-validity}
 k^+\Big(\frac{1}{k_\perp L}\Big)^2L\gg1\Leftrightarrow \frac{k^+}{k_\perp^2}\gg L
\end{equation}
In more physical terms, this inequality means that the variation of the gluon transverse coordinate $L\Delta \th \sim Lk_\perp/k^+$ while propagating over a distance $L$ is much smaller than the quantum fluctuation $\sim 1/k_\perp$ in the transverse motion. 
That said, we recognize in \eqref{eikonal-validity} the condition stating that the formation time of the gluon is larger than the gluon in-medium path length. This is a too strong condition, as we do not want to focus only on this kinematic regime: we will see later than many interesting physical phenomena happen when $k^+/k_\perp\ll L$. Therefore, we shall use the propagator \eqref{scalar-prop-subei} for the emitted gluon.

\subsubsection{Details of the amplitude calculation}

Now, we detail the general method for calculating $\bm{\langle} A^i(z)\bm{\rangle}$ at a given order in $g$ and all orders in $g\mathcal{A}_m$.  To proceed, we expand the time ordered Wilson line in $\mathcal{D}_{tc}(p,p'|A+\mathcal{A}_m)$ thanks to the following formula, obtained after discretization of the integral:
\begin{equation}
\label{time_ordered_expansion}
 T\Big[e^{\int_{x_i}^{x_f}{\dif x^+ A(x^+)}}\Big]=\lim\limits_{N\rightarrow\infty}\prod_{n=N..1}{e^{\epsilon A(x_n)}}
\end{equation}
with $\epsilon=(x_f-x_i)/N$, $x_n=x_i+\epsilon n$.
Keeping only the first two lowest order terms in $gA^\mu$, it is easy to see that
\[\prod_{n=N..1}{e^{ig\epsilon(A^-(x^+_n)+\mathcal{A}^-_{m}(x^+_n))}}=\prod_{n}{e^{ig\epsilon \mathcal{A}^-_{m}(x^+_n)}}+ig\sum_n{\prod_{p=N..n+1}\Big[e^{ig\epsilon \mathcal{A}^-_m(x^+_p)}\Big]\epsilon A^-(x^+_n)\prod_{q=n-1..1}\Big[e^{ig\epsilon \mathcal{A}^-_m(x^+_q)}\Big]}\]
Now, if we take the limit $N\rightarrow\infty$, we obtain
\begin{align}
 T\Big[e^{ig\int_{-\infty}^{\infty}{\dif x^+(A^-(x^+,x^-,x_\perp)+\textrm{A}^-_{m,c}(x^+,x_\perp))}}\Big]=&
W_{-\infty}^{\infty}(x_\perp)\nonumber\\
&\hspace{-0.5cm}+ig\int_{-\infty}^{\infty}{\dif y^+W_{y^+}^{\infty}(x_\perp)A^-(y^+,x^-,x_\perp)W_{-\infty}^{y^+}(x_\perp)}
\end{align}
where $W_{t_1}^{t_2}(x_\perp)$ is the Wilson line \eqref{wilson-line2} for a path from $x^+=t_1$ to $x^+=t_2$ with $x_\perp$ fixed.

The first term cancels in the path integral over $A^\mu$ in \eqref{onepointA} because the 1-point quantum correlation function is null by convention (no tadpoles). The second term involves
the gluon propagator $G^{i-}(z;x|\mathcal{A}_m)$ determined in the previous paragraph beyond the eikonal approximation,
which describes the propagation of the gluon inside the medium between the emission point $x=(y^+,x^-,x_\perp)$ and $z$.
Gathering everything together with the colour indices explicit, one gets 
\begin{equation}\label{Aistep1}
\bm{\langle} A^i_c(z)\bm{\rangle}=ig\int_{-\infty}^{\infty}\dif y^+G_{cb}^{i-}(z^+,z^-,z_\perp;y^+,x^-,x_\perp|\mathcal{A}_m)\Big[W_{y^+}^{\infty}(x_\perp)t^bW_{-\infty}^{y^+}(x_\perp)\Big]_{j'j}
\end{equation}
In this formula, the Wilson lines are in the fundamental representation of $\mathfrak{su}(N_c)$. To simplify the colour algebra, it is convenient to express the product of the three matrices inside the bracket as a colour rotation in the adjoint representation of $\mathfrak{su}(N_c)$:
\begin{align}\label{fund-ad-calc}
W_{y^+}^{\infty}(x_\perp)t^bW_{-\infty}^{y^+}(x_\perp)&=W_{-\infty}^{\infty}(x_\perp)W^{\dagger,y^+}_{-\infty}(x_\perp)t^b W_{-\infty}^{y^+}(x_\perp)\\
&=W_{-\infty}^{\infty}(x_\perp)\mathcal{W}_{-\infty,bd}^{y^+}(x_\perp)t^d
\end{align}
In the first line, we have used the unitarity of the Wilson lines and in the second line, the definition of the adjoint representation. Calligraphic letters, as in $\mathcal{W}$, are used for Wilson lines in the adjoint representation. Inserting this relation and the representation \eqref{def-gscalar-prop} of the propagator into the Fourier transform with respect to $z^-$ of the amplitude in coordinate space \eqref{Aistep1}, one finds:
\begin{align}\label{Aistep2}
&\int \dif z^-\,e^{ik^+z^-}\bm{\langle} A^i_c(z)\bm{\rangle}=\frac{-g}{2{k^+}^2}e^{ik^+x^-}W_{-\infty}^{\infty}(x_\perp)\nonumber\\
&\hspace{5cm}\times\int_{-\infty}^{\infty}\dif y^+\,\partial^i_{y_\perp=x_\perp}\mathcal{G}_{cb}(z^+,z_\perp;y^+,y_\perp|k^+)\mathcal{W}_{-\infty,bd}^{y^+}(x_\perp)t^d
\end{align}

The last step consists in taking the Fourier transform with respect to the remaining $z^+$ and $z_\perp$ components of the final position of the gluon and to multiply by $k^2$ in order to follow the LSZ recipe. The multiplication by $k^2$ in Fourier space corresponds to the application of the operator $-2ik^+\partial_z^--\nabla_\perp^2$ in the remaining space coordinates. 

Now let's assume that the background field $\mathcal{A}_m$ goes to $0$ at sufficiently large light cone time $y^+\ge T$. Then we split the integral at $y^+=T$. As $T$ is arbitrary, we can take $T\rightarrow\infty$ at the end of the calculation so that the integral between $T$ and $+\infty$ vanishes. In the $y^+\le T$ piece, we insert a Chapman-Kolmogorov relation for the propagator with intermediate time $T$ and intermediate transverse momentum $w_\perp$:
\begin{equation}
 \mathcal{G}_{cb}(z^+,z_\perp;y^+,y_\perp|k^+)=\int \dif w_\perp \mathcal{G}_{ce}(z^+,z_\perp;T,w_\perp|k^+)\mathcal{G}_{eb}(T,w_\perp;y^+,y_\perp|k^+)
\end{equation}
The propagator $\mathcal{G}(z^+,z_\perp;T,w_\perp|k^+)$ is free so the operator $-2ik^+\partial_z^--\nabla_\perp^2$ gives $-2ik^+\delta_{ce}\delta(z^+-T)\delta(w_\perp-z_\perp)$ leading to a trivial $w_\perp$ integral. After Fourier transform, one gets:
\begin{align}
 &k^2\bm{\langle} A_c^i(k)\bm{\rangle}=\frac{ig}{k^+}e^{ik^+x^-}W_{-\infty}^{\infty}(x_\perp)e^{ik^-T}\nonumber\\
&\hspace{3cm}\times\int_{-\infty}^{T}\dif y^+\,\int \dif^2 z_\perp e^{-ik_\perp z_\perp}\,\partial^i_{y_\perp=x_\perp}\mathcal{G}_{cb}(T,z_\perp;y^+,y_\perp|k^+)\mathcal{W}_{-\infty,bd}^{y^+}(x_\perp)t^d
\end{align}
In order to have a meaningful mathematical expression in the limit $T\rightarrow\infty$, we should have deformed the light cone $z^+$ integration contour in the Fourier transform with respect to $z^+$ to ensure the adiabatic switching of the interactions. This amounts to add a small imaginary part $i \textrm{sign}(z^+)\varepsilon$\footnote{$\varepsilon$ should not be confused with $\epsilon$ defining \textit{retarded} propagators.} to the $k^-$ component of the 4-momentum $k$ in the previous expression. Using the following relation obtained after a trivial integration by part in $y^+$:
\begin{equation}
 \lim_{\varepsilon\rightarrow0}e^{-\varepsilon T}\int_{-\infty}^T\dif y^+\,f(y^+)=\lim_{\varepsilon\rightarrow0}\int_{-\infty}^T\dif y^+\,e^{-\varepsilon y^+}f(y^+)
\end{equation}
one shows that, apart from the pure phase $e^{ik^-T}$ which cancels anyway in the amplitude squared, the limit $T\rightarrow \infty$ is given by:
\begin{tcolorbox}[ams align]
 &k^2\bm{\langle} A_c^i(k)\bm{\rangle}=\frac{ig}{{k^+}}e^{ik^+x^-}W_{-\infty}^{\infty}(x_\perp)\nonumber\\
&\hspace{2.cm}\times\int_{-\infty}^{\infty}\dif y^+\, e^{-\varepsilon|y^+|}\int \dif^2 z_\perp e^{-ik_\perp z_\perp}\,\partial^i_{y_\perp=x_\perp}\mathcal{G}_{cb}(\infty,z_\perp;y^+,y_\perp|k^+)\mathcal{W}_{-\infty, bd}^{y^+}(x_\perp)t^d\label{amplitude-final}
\end{tcolorbox}
\noindent with the implicit limit $\varepsilon\rightarrow0$ taken \textit{after} the integration. This concludes the calculation of the amplitude for the one-gluon emission process at leading order in $g$. However, this result resums to all orders $g\mathcal{A}_m$ by means of Wilson lines. Recall also that the equality here is a small abuse of notation as one should integrate over $x_\perp$ and $x^-$ according to \eqref{eikonal-prop}. 

\subsubsection{The cross-section averaged over $\mathcal{A}_m$}
\label{subsub:1gluon-cross}

The amplitude \eqref{amplitude-final} involves three Wilson lines. This means that we should compute the average of six Wilson lines over background field configurations in order to get the cross-section, which is a very complicated task. However, in the full cross-section \eqref{bdmpszcross}, we are not really interested any more on the distribution over the $4$-momentum $p'$ of the outgoing quark. If we integrate the cross-section \eqref{bdmpszcross} over $p'^+$ and $p'_\perp$, the integral yields a $\delta(x_\perp-\bar{x}_\perp)$ with $x_\perp$, $\bar{x}_\perp$ the transverse coordinate of the quark in the direct amplitude and in the complex conjugate amplitude respectively. This considerably simplifies the calculation of the \textit{inclusive} one-gluon emission cross-section per impact parameter $x_\perp$:
\begin{equation}
 \frac{\dif^5 \sigma_g}{\dif^2 x_\perp\dif k^+\dif^2k_\perp}\equiv\int \dif p'^+\int\dif^2 p_\perp\,\frac{\dif^8\sigma}{\dif^2 x_\perp\dif p'^+\dif^2p'_\perp\dif k^+\dif k_\perp}
\end{equation}
since the transverse coordinate $x_\perp$ of the incoming quark is now frozen in the direct and complex conjugate amplitude. Summing over the polarization states $\lambda$ and spin $s$, $s'$ one gets:
\begin{equation}\label{bdmpsz-inclusive}
 k^+\frac{\dif^5 \sigma_g}{\dif^2 x_\perp\dif k^+\dif^2k_\perp}=\frac{1}{16\pi^3}\frac{1}{N_c}\sum_{j,j',c,i}|k^2\bm{\langle} A_c^i(k)\bm{\rangle}|^2
\end{equation}
The wave function renormalization constants in \eqref{LSZ-bdmpsz} are equal to $1$ at this order and one checks that all the factors in formula \eqref{LSZ-bdmpsz} (and in particular the $1/T_{\mathfrak{B}}$) give a factor $1/(16\pi^3)$ using the same tricks as for the calculation of the transverse momentum distribution.

Before writing explicitly the value of the modulus square, we need to perform the colour algebra:
\begin{align}
 \frac{1}{N_c}[W_{-\infty}^{\infty}(x_\perp)t^d]_{j'j}\times[W_{-\infty}^{\star,\infty}(x_\perp)t^{\star,\bar{d}}]_{j'j}&=\frac{1}{N_c}[W_{-\infty}^{\infty}(x_\perp)t^d]_{j'j}\times[t^{\dagger,\bar{d}}W_{-\infty}^{\dagger\infty}(x_\perp)]_{jj'}\\
 &=\frac{1}{N_c}\Tr(W_{-\infty}^{\infty}(x_\perp)t^dt^{\bar{d}}W_{-\infty}^{\dagger,\infty}(x_\perp))\\
 &=\frac{C_F}{N_c^2-1}\delta^{d\bar{d}}
\end{align}
It is convenient to make appear the $1/(N_c^2-1)$ factor for the future average of Wilson lines in the adjoint representation. Now, the cross-section \eqref{bdmpsz-inclusive} averaged over $\mathcal{A}_m$ reads:
\begin{align}\label{g-cross-step1}
 k^+\frac{\dif^5 \sigma_g}{\dif^2 x_\perp\dif k^+\dif^2k_\perp}&=\frac{\alpha_s C_F}{(2\pi)^2{k^+}^2}2\mathfrak{Re}\int_{-\infty}^{\infty}\dif y^+\int_{-\infty}^{y^+}\dif\bar{y}^+e^{-\varepsilon(|y^+| +|\bar{y}^+|)}\int \dif z_\perp\dif \bar{z}_\perp e^{-ik_\perp(z_\perp-\bar{z}_\perp)}\nonumber\\
 &\hspace{-2cm}\times\frac{1}{N_c^2-1}\Big\langle \Tr\, \partial^i_{y_\perp=x_\perp}\mathcal{G}^{\dagger}(\infty,\bar{z}_\perp;\bar{y}^+,y_\perp|k^+)\partial^i_{\bar{y}_\perp=x_\perp}\mathcal{G}(\infty,z_\perp;y^+,\bar{y}_\perp|k^+)\mathcal{W}_{\bar{y}^+}^{y^+}(x_\perp)\Big\rangle
\end{align}
with $\alpha_s\equiv g^2/(4\pi)$. We used the property $\mathcal{G}(y^+;\bar{y}^+)=\mathcal{G}^{\dagger}(\bar{y}^+;y^+)$ (following the equivalent property for Wilson lines) to write the integral over $\bar{y}^+\ge y^+$ as the complex conjugate of the one over $y^+\ge\bar{y}^+$. Note that the remaining colour precession associated with the Wilson line $\mathcal{W}_{\bar{y}^+}^{y^+}$ occurs between the creation time $\bar{y}^+$ in the complex conjugate amplitude and the absorption time $y^+$ in the direct amplitude.

The latter expression can be further simplified if we manage to use the locality in light cone time $y^+$ of the background field correlations. To do so, we use the Chapman-Kolmogorov relation to write the scalar propagator $\mathcal{G}^\dagger(\infty;\bar{y}^+)$ as the product of two propagators, one between $\bar{y}^+$ and $y^+$, followed by one between $y^+$ up to $+\infty$ with intermediate transverse momentum $u_\perp$. Then,
\begin{align}
 &\frac{1}{N_c^2-1}\Big\langle \Tr\, \mathcal{G}^{\dagger}(\infty,\bar{z}_\perp;\bar{y}^+,\bar{y}_\perp|k^+)\mathcal{G}(\infty,z_\perp;y^+,y_\perp|k^+)\mathcal{W}_{\bar{y}^+}^{y^+}(x_\perp)\Big\rangle=\nonumber\\
 &\frac{1}{N_c^2-1}\int\dif^2 u_\perp \Big\langle \Tr\, \mathcal{G}^{\dagger}(y^+,u_\perp;\bar{y}^+,\bar{y}_\perp|k^+)\mathcal{G}^{\dagger}(\infty,\bar{z}_\perp;y^+,u_\perp|k^+)\mathcal{G}(\infty,z_\perp;y^+,y_\perp|k^+)\mathcal{W}_{\bar{y}^+}^{y^+}(x_\perp)\Big\rangle\nonumber\\
 &=\frac{1}{(N_c^2-1)^2}\int\dif^2 u_\perp\Big\langle \Tr\, \mathcal{G}^{\dagger}(y^+,u_\perp;\bar{y}^+,\bar{y}_\perp|k^+)\mathcal{W}_{\bar{y}^+}^{y^+}(x_\perp)\Big\rangle\nonumber\\
 &\hspace{4cm}\times\Big\langle \Tr\, \mathcal{G}^{\dagger}(\infty,\bar{z}_\perp;y^+,u_\perp|k^+)\mathcal{G}(\infty,z_\perp;y^+,y_\perp|k^+)\Big\rangle\label{chapmantrick}
\end{align}
Thanks to the locality in time of the background field correlations, the average of the four Wilson lines factorizes into the product of averages of only two Wilson lines.
Now, let us define the following general quantities:
\begin{align}\label{def-Sgg}
\tilde{S}_{gg}(k_\perp,a_\perp,b_\perp)&\equiv \frac{1}{N_c^2-1}\int \dif z_\perp\dif \bar{z}_\perp\,e^{-ik_\perp(a_\perp-b_\perp)} e^{-ik_\perp(z_\perp-\bar{z}_\perp)}\nonumber\\
&\hspace{2cm}\times\Big\langle \Tr\, \mathcal{G}^{\dagger}(\infty,\bar{z}_\perp;y^+,a_\perp|k^+)\mathcal{G}(\infty,z_\perp;y^+,b_\perp|k^+)\Big\rangle\\
\mathcal{K}_{qg}(y^+,a_\perp;\bar{y}^+,b_\perp|c_\perp)&\equiv\frac{1}{N_c^2-1}\Big\langle \Tr\,\mathcal{G}^{\dagger}(y^+,a_\perp;\bar{y}^+,b_\perp|k^+)\mathcal{W}_{\bar{y}^+}^{y^+}(c_\perp)\Big\rangle\label{def-Kqg}
\end{align}
where $\tilde{S}_{gg}$ is the equivalent of the Fourier transform of the dipole S-matrix \eqref{dipole-S} with sub-eikonal corrections and $\mathcal{K}_{qg}$ is the effective $qg$ dipole propagator inside the medium between $\bar{y}^+$ and $y^+$. 

One can show (see Appendix~\ref{app:prop}) that for Gaussian statistics, $\tilde{S}_{gg}$ is a function of $a_\perp-b_\perp$ only, that is $\tilde{S}_{gg}(k_\perp,a_\perp,b_\perp)=S_{gg}(a_\perp-b_\perp)$. On top of that, the  Fourier transform of the ``dipole S-matrix'' $S_{gg}$ is not modified by sub-eikonal corrections included in the scalar propagator, namely:
\begin{equation}\label{Sgg}
 S_{gg}(u_\perp)=\exp\left(-\frac{g^2}{2}C_A\int_{y^+}^{\infty}n(x^+)\sigma_d(u_\perp)\dif x^+\right)
\end{equation}
From our calculation of the average of two Wilson lines for Gaussian correlations in \eqref{dipolecross}, the function $\mathcal{K}_{qg}(a_\perp;b_\perp|c_\perp)$ can be written:
\begin{align}\label{kqg-shift}
 \mathcal{K}_{qg}(y^+,a_\perp;\bar{y}^+,b_\perp|c_\perp)&=\mathcal{K}(y^+,a_\perp-c_\perp;\bar{y}^+,b_\perp-c_\perp)
 \end{align}
 with the general function $\mathcal{K}$ defined as a path integral:
 \begin{align}\label{def-K}
 \mathcal{K}(y^+,a_\perp;\bar{y}^+,b_\perp)&=\int_{r_\perp(\bar{y}^+)=b_\perp}^{r_\perp(y^+)=a_\perp}\mathcal{D}r_\perp(\xi)\exp\left(-\int_{\bar{y}^+}^{y^+}\dif\xi\,\frac{i k^+}{2}\dot{r}_\perp^2(\xi)+\frac{g^2C_A}{2}n(\xi)\sigma(r_\perp)\right)
\end{align}

With these notations, and after a shift of the integration variable $u_\perp\rightarrow u_\perp+y_\perp$, the inclusive one-gluon medium-induced cross-section reads:
\begin{tcolorbox}[ams align]\label{final-gluon-cross}
 k^+\frac{\dif^5 \sigma_g}{\dif^2 x_\perp\dif k^+\dif^2k_\perp}&=\frac{\alpha_s C_F}{(2\pi)^2{k^+}^2}2\mathfrak{Re}\int_{-\infty}^{\infty}\dif y^+\int_{-\infty}^{y^+}\dif\bar{y}^+e^{-\varepsilon(|y^+| +|\bar{y}^+|)}\nonumber\\
 &\hspace{-0.5cm}\times\int\dif^2u_\perp e^{ik_\perp u_\perp}S_{gg}(u_\perp)\partial^i_{y_\perp}\partial^i_{\bar{y}_\perp}\mathcal{K}(y^+,y_\perp+u_\perp-x_\perp;\bar{y}^+,\bar{y}_\perp-x_\perp)
\end{tcolorbox}
\noindent with the derivative taken at $y_\perp=\bar{y}_\perp=x_\perp$. This is the main result of this section. It has been obtained in this form by Armesto, Salgado and Wiedemnann in \cite{Wiedemann:2000za,Salgado:2003gb,Armesto:2003jh}, Zakharov \cite{Zakharov:1996fv,Zakharov:1998sv} and BDMPS \cite{Baier:1996kr,Baier:1998kq}.

Even if this cross-section has been obtained from a ``CGC-like'' formalism, it is actually very general and its applicability range goes beyond the traditional underlying assumption of this formalism. For example, it shown in \cite{Wiedemann:2000za,Mehtar-Tani:2019tvy} that using \eqref{final-gluon-cross} with the effective propagator $\mathcal{K}_{qg}$ calculated at the leading order in opacity (i.e truncating the calculation to the first order in $n \sigma_d$), one gets the GLV spectrum \cite{Gyulassy:2000er,Gyulassy:2000er,Gyulassy:2001nm}. 

In the next section, we shall give some analytical results for the cross-section \eqref{final-gluon-cross} in the multiple soft scattering regime and in the harmonic approximation, which are the main underlying assumptions   leading to the BDMPS-Z spectrum. From now on, we shall forget about the dependence of the spectrum on $x_\perp$, the transverse coordinate of the incoming quark. Integrating over it gives the transverse surface of the target, but if we decide to normalize the cross-section $\sigma_g$ by this surface, the factor cancels. We thus define the gluon spectrum as
\begin{equation}\label{crosssec-target}
 k^+\frac{\dif^3N}{\dif k^+\dif^2 k_\perp}\equiv k^+\frac{\dif^5 \sigma_g}{\dif^2 x_\perp\dif k^+\dif^2k_\perp}_{|x_\perp=0_\perp}
\end{equation}
and deal with this quantity for the rest of this section.

\subsection{The multiple soft scattering regime}

In this subsection, we focus on the harmonic approximation which captures the main features of the multiple soft scattering regime. The harmonic approximation enables to perform almost entirely analytically the calculation of the complicated formula \eqref{final-gluon-cross}. 


\subsubsection{Some useful analytic formulas}

Within the harmonic approximation, $S_{gg}$ becomes a simple two-dimensional Gaussian function:
\begin{equation}
  S_{gg}(u_\perp)=\exp\left(-\frac{1}{4}\int_{y^+}^{\infty}\dif x^+\,\qhat(x^+)u_\perp^2\right)
\end{equation}
The in-medium effective propagator $\mathcal{K}$ is delicate to handle written in the form \eqref{def-K}. Fortunately, exact analytical forms are known in the harmonic approximation. In this case, it reads (see \cite{Baier:1998yf} and Appendix~\ref{app:prop}):
\begin{equation}\label{Kqg}
 \mathcal{K}(y^+,u_\perp;\bar{y}^+,x_\perp)=\frac{-k^+}{2\pi iS(y^+,\bar{y}^+)}e^{\frac
 {-ik^+}{2S(y^+,\bar{y}^+)}\big[C(\bar{y}^+,y^+)u_\perp^2+C(y^+,\bar{y}^+)x_\perp^2-2u_\perp x_\perp\big]}
\end{equation}
with $C(t,\xi)\equiv \partial_\xi S(\xi,t)$ and $S(t,\xi)$ is the solution of the second-order linear differential equation:
\begin{align}\label{diff-eq-S}
 \frac{\partial^2S(t,\xi)}{\partial t^2}+\frac{i\hat{q}_A(t)}{2k^+}S(t,\xi)&=0\\
S(\xi,\xi)=0\,,\qquad \frac{\partial S(t,\xi)}{\partial t}_{|t=\xi}&=1
\end{align}
The function $S$ satisfies $S(t,\xi)=-S(\xi,t)$. The function $C$ is also a solution of this differential equation, with different initial conditions though: $C(\xi,\xi)=1$ and $\partial_{t=\xi} C(t,\xi)=0$. Since the Wronskian $\textrm{Wr}(t,\xi)$ of the differential equation is constant:
\begin{equation}\label{Wronskian2}
 \textrm{Wr}(t,\xi)\equiv\partial_tS(t,\xi)C(t,\xi)-S(t,\xi)\partial_tC(t,\xi)=1
\end{equation}
one has \cite{Arnold:2008iy}, 
\begin{equation}\label{Wronskian}\
 -\partial_t\left(\frac{C(t,\xi)}{S(t,\xi)}\right)=\frac{1}{S^2(t,\xi)}
\end{equation}
When $\qhat(t)$ is only piece-wise continuous, the solutions $S$ and $C$ must be chosen in such a way that they are continuous and derivable in their first argument at any points (however the second derivative may not exist everywhere).

In the limit $\hat{q}\rightarrow0$ or $y^+\rightarrow\bar{y}^+$, the effective propagator converges toward the adjoint of the free scalar propagator:
\begin{align}
 \lim_{\qhat\rightarrow0}\mathcal{K}(y^+,u_\perp;\bar{y}^+,x_\perp)&=\mathcal{G}_0^{\dagger}(y^+,u_\perp;\bar{y}^+,x_\perp)\\
 \mathcal{K}(y^+,u_\perp;\bar{y}^+,x_\perp)&\underset{y^+\rightarrow\bar{y}^+}\sim\mathcal{G}_0^{\dagger}(y^+,u_\perp;\bar{y}^+,x_\perp)
\end{align}
with the free scalar propagator given by
\begin{align}
 \mathcal{G}_0(y^+,u_\perp;\bar{y}^+,x_\perp)&\equiv\frac{k^+}{2\pi i (y^+-\bar{y}^+)}\exp\left(\frac{ik^+(u_\perp-x_\perp)^2}{2(y^+-\bar{y}^+)}\right)\label{G0time}\\
 &=\int\frac{\dif^2q_\perp}{(2\pi)^2}e^{iq_\perp(u_\perp-x_\perp)}\exp\left(\frac{-i(y^+-\bar{y}^+)}{2(k^++i\epsilon)}q_\perp^2\right)\label{G0transverse}
\end{align}
Using this property, one easily shows that the spectrum \eqref{final-gluon-cross} vanishes in the limit $\qhat\rightarrow0$. As expected, an on-shell incoming quark cannot radiate in the vacuum.

\subsubsection{Importance of the adiabatic prescription}

Before giving the analytic results for \eqref{final-gluon-cross} using the expression of $S_{gg}$ and $\mathcal{K}$, let us make few comments on the adiabatic regulator $\varepsilon$. In most calculations of the BDMPS-Z spectrum in the case of a finite path length $L$, the calculation of \eqref{final-gluon-cross} is performed by cutting the integral over $y^+$ in pieces: $(-\infty,0)$, $[0,L]$ and $(L,\infty)$ in order to use the vacuum propagator $\mathcal{G}_0$ wherever it is necessary. For the in/in term, corresponding to $y^+\in[0,L]$ and $\bar{y}^+\in[0,y^+]$, the adiabatic regulator is not necessary since the integral is convergent on the compact domain. However, if there is no sharp boundary such as in an expanding infinite medium\footnote{This academic situation is nonetheless considered in Chapter~\ref{chapter:DLApic} Section~\ref{sec:veto} or Chapter~\ref{chapter:jet}, Section~\ref{sub:med-expansion}.}, this procedure is not allowed. As the limit $\varepsilon\rightarrow0$ must be taken \textit{after} the integration, 
this complicates the analytic calculation of the full cross-section in the expanding infinite medium case.

In order to highlight how the adiabatic switching works, we calculate the cross-section \eqref{final-gluon-cross} in the vacuum for an off-shell quark created by a hard process occurring at $y^+=0$. We first take the limit $\hat{q}\rightarrow0$ in the integral. There are two ways to proceed but in any cases, one has to keep track of the $\varepsilon$ dependence. The first way is to calculate 
\begin{align}
k^+\frac{\dif N}{\dif k^+ \dif^2k_\perp}&=\frac{\alpha_s C_F}{4\pi^2{k^+}^2}2\mathfrak{Re}\int_0^\infty \dif y^+\int_0^{y^+}\dif\bar{y}^+ e^{-\varepsilon y^+-\varepsilon\bar{y}^+}\nonumber\\
&\hspace{3cm}\times\int \dif^2u_\perp e^{ik_\perp u_\perp}
 \partial^i_{y_\perp=0_\perp}\partial^i_{\bar{y}_\perp=0_\perp}\mathcal{G}_0^{\dagger}(y^+,y_\perp+u_\perp;\bar{y}^+,\bar{y}_\perp)
\end{align}
integrating first over $u_\perp$. This gives $k_\perp^2\exp(ik^-(y^+-\bar{y}^+))$ so 
 \begin{align}
  k^+\frac{\dif^3N}{\dif k^+ \dif^2k_\perp}&=2\mathfrak{Re}\Big\{\frac{\alpha_s C_F k_\perp^2}{4\pi^2{k^+}^2}\int_0^\infty \dif y^+\int_0^{y^+}\dif\bar{y}^+ e^{-\varepsilon(y^++\bar{y}^+)+ik^-(y^+-\bar{y}^+)}\Big\}\label{vac-eps-example}\\
   &=\lim\limits_{\varepsilon\rightarrow0}\frac{\alpha_s C_F k_\perp^2}{4\pi^2{k^+}^2}\left(\frac{2}{\varepsilon^2+{k^-}^2}-\frac{1}{{k^-}^2+\varepsilon^2}\right)\\
   &=\frac{\alpha_sC_F}{\pi^2}\frac{1}{k_\perp^2}
 \end{align}
which is the standard Bremsstrahlung spectrum in the vacuum.
Changing the time ordering in the double integral, one can also choose to integrate first over $y^+$, and even if the calculation is more difficult, it gives the same result:
 \begin{align}
 k^+\frac{\dif N}{\dif k^+ \dif^2 k_\perp}&=2\mathfrak{Re}\Big\{\frac{\alpha_s C_F k_\perp^2}{4\pi^2{k^+}^2}\int_0^\infty \dif\bar{y}^+e^{-2\varepsilon\bar{y}^+}\int \dif^2u_\perp e^{ik_\perp u_\perp}2\varepsilon \textrm{K}_0\Big[2\Big(\frac{ik^+ u_\perp^2}{2}\Big)^{1/2}\varepsilon^{1/2}\Big]\Big\}\label{vac-eps-example2}\\
 &=\lim\limits_{\varepsilon\rightarrow0}2\mathfrak{Re}\Big\{\frac{\alpha_s C_F k_\perp^2}{4\pi^2\omega^2}\int \dif^2u_\perp e^{ik_\perp u_\perp}\textrm{K}_0\Big[2\Big(\frac{ik^+ u_\perp^2}{2}\Big)^{1/2}\varepsilon^{1/2}\Big]\Big\}\\
 &=\frac{\alpha_sC_F}{\pi^2}\frac{1}{k_\perp^2}
\end{align}
where $\textrm{K}_0$ is the zeroth Bessel function of second kind.

Both ways of doing the calculation give the right answer. However, if we decide to forget about the $\varepsilon$ dependence in the integral over $\bar{y}^+$ in \eqref{vac-eps-example} or in \eqref{vac-eps-example2}, we would have obtained the Bremsstrahlung spectrum with an incorrect factor 2! \cite{Wiedemann:1999fq} We expect this kind of feature to appear also in the medium calculation with $\qhat\neq0$, enlightening the importance of the adiabatic prescription.

With a non zero $\qhat$, the best we can hope is this spurious factor $2$ to be general. Fortunately it is, and we shall use the following lemma in the next section:
\begin{align}
 \lim\limits_{\varepsilon\rightarrow0}\mathfrak{Re}\int_0^\infty\dif y^+\int_0^{y^+}\dif\bar{y}^+\,e^{-\varepsilon(y^+ +\bar{y}^+)}f(y^+,\bar{y}^+)&=\lim\limits_{\varepsilon\rightarrow0}\mathfrak{Re}\int_0^\infty\dif y^+\,e^{-\varepsilon y^+}\nonumber\\
 &\hspace{2cm}\times\int_0^{y^+}\dif\bar{y}^+f(y^+,\bar{y}^+)-\frac{1}{2\phi^2}\label{lemma}
\end{align}
for $f(y^+,\bar{y}^+)\sim\exp(i\phi(y^+-\bar{y^+}))$ as $y^+,\bar{y}^+\rightarrow\infty$, with $\phi$ a real phase.
This lemma is important as it enables to forget about the $\varepsilon$ dependence in the first integral over $\bar{y}^+$. In the medium, one can always find a primitive of the integral over $\bar{y}^+$ without the exponential adiabatic switching, thanks to the Wronskian relation \eqref{Wronskian}. The remaining integral over $y^+$ is even often convergent so the limit $\varepsilon\rightarrow0$ is trivial.

The proof of \eqref{lemma} goes this way: first, we re-order the integrals, integrating first over $\bar{y}^+$. This is allowed as \textit{both} integrals are convergent as long as $\varepsilon>0$. Then, we cut the integral at $\bar{y}^+=T$ with $T$ large enough so that the the difference $|f(y^+,\bar{y}^+)-\exp(i\phi(y^+-\bar{y}^+))|$ is as small as desired. This is possible since by assumption, $f(y^+,\bar{y}^+)\sim\exp(i\phi(y^+-\bar{y}^+))$ for large values of $y^+$ and $\bar{y}^+$. Then, one can remove the $\varepsilon$ prescription in the piece with $\bar{y}^+\le T$, since the integration domain is compact. On the other piece, one notices:
\begin{align}
 \int_T^\infty\dif\bar{y}^+\int_{\bar{y}^+}^{\infty}\dif y^+ e^{i\phi(y^+-\bar{y}^+)-\varepsilon(y^++\bar{y}^+)}&=\frac{i}{2\phi\varepsilon}+\frac{1-2i T\phi}{2\phi^2}+\mathcal{O}(\varepsilon)\\
 \int_T^\infty\dif\bar{y}^+\int_{\bar{y}^+}^{\infty}\dif y^+ e^{i\phi(y^+-\bar{y}^+)-\varepsilon y^+}&=\frac{i}{\phi\varepsilon}+\frac{1-i T\phi}{\phi^2}+\mathcal{O}(\varepsilon)
\end{align}
Taking the real part and then the limit $\varepsilon\rightarrow 0$ concludes the proof. We emphasize the importance of $\mathfrak{Re}$ in the lemma, as well as the condition $\phi\in \mathbb{R}$.

\subsubsection{Transverse momentum dependence of the BDMPS-Z spectrum}
\label{subsub:TMdep-BDMPS}

Without loss of generality, let us assume that the density of scattering centers $n(x^+)$ has support only for $x^+\ge x_0^+$, so that the parton enters into the medium at this time. Then $\qhat(x^+)$ decreases smoothly with $x^+$. We split the integral over $y^+$ and $\bar{y}^+$ in three pieces $I^{b/b}$, $I^{b/i}$ and $I^{i/i}$ with $y^+,\bar{y}^+\in(-\infty,x_0^+]$, $y^+\in[x_0^+,\infty)$, $\bar{y}^+\in(-\infty,x_0^+]$ and $y^+,\bar{y}^+\in[x_0^+,\infty)$ respectively in order to use the expression \eqref{Kqg}:
\begin{equation}
 k^+\frac{\dif^3N}{\dif k^+\dif^2k_\perp}=\frac{\alpha_sC_F}{\pi^2}\big(I^{b/b}+I^{b/i}+I^{i/i}\big)
\end{equation}

If the jet has a finite path length through the medium, meaning that $\qhat(t)$ suddenly vanishes for $t\ge L$, one can obtain even more transparent formulas in terms of $L$. The results for the brick problem or for an eikonal parton with finite path length through an expanding plasma are given in Appendix~\ref{app:B}. Equivalently, if the hard vertex location is distinct from $x_0^+$, one can use the junction method described in Appendix~\ref{app:B} to properly adapt the general formulas presented in this section.

\paragraph{In/in term.} The in/in term is also the spectrum of an off-shell quark created at $y^+=x_0^+$. If we introduce the following quantity:
\begin{align}
 Q_s^2(y^+)&=\int_{y^+}^{\infty}\qhat(\xi)\dif \xi
 \end{align}
 which corresponds to the transverse momentum squared acquired by the gluon during its propagation after $y^+$, the $I^{i/i}$ reads:
\begin{equation}\label{Iiidouble}
 I^{i/i}=2\mathfrak{Re}\int_{x_0^+}^\infty\dif y^+\int_{x_0^+}^{y^+}\dif \bar{y}^+\frac{-e^{-\varepsilon(y^++\bar{y}^+)}}{S^2(y^+,\bar{y}^+)}\frac{Q_s^4+2ik^+\frac{C(\bar{y}^+,y^+)}{S(y^+,\bar{y}^+)}(k_\perp^2+Q_s^2)}{\Big(Q_s^2+2ik^+\frac{C(\bar{y}^+,y^+)}{S(y^+,\bar{y}^+)}\Big)^3}e^{\frac{-k_\perp^2}{Q_s^2+2ik^+\frac{C(\bar{y}^+,y^+)}{S(y^+,\bar{y}^+)}}}
\end{equation}

\begin{figure}[t] 
  \centering
  \begin{subfigure}[t]{0.49\textwidth}
    \includegraphics[page=1,width=\textwidth]{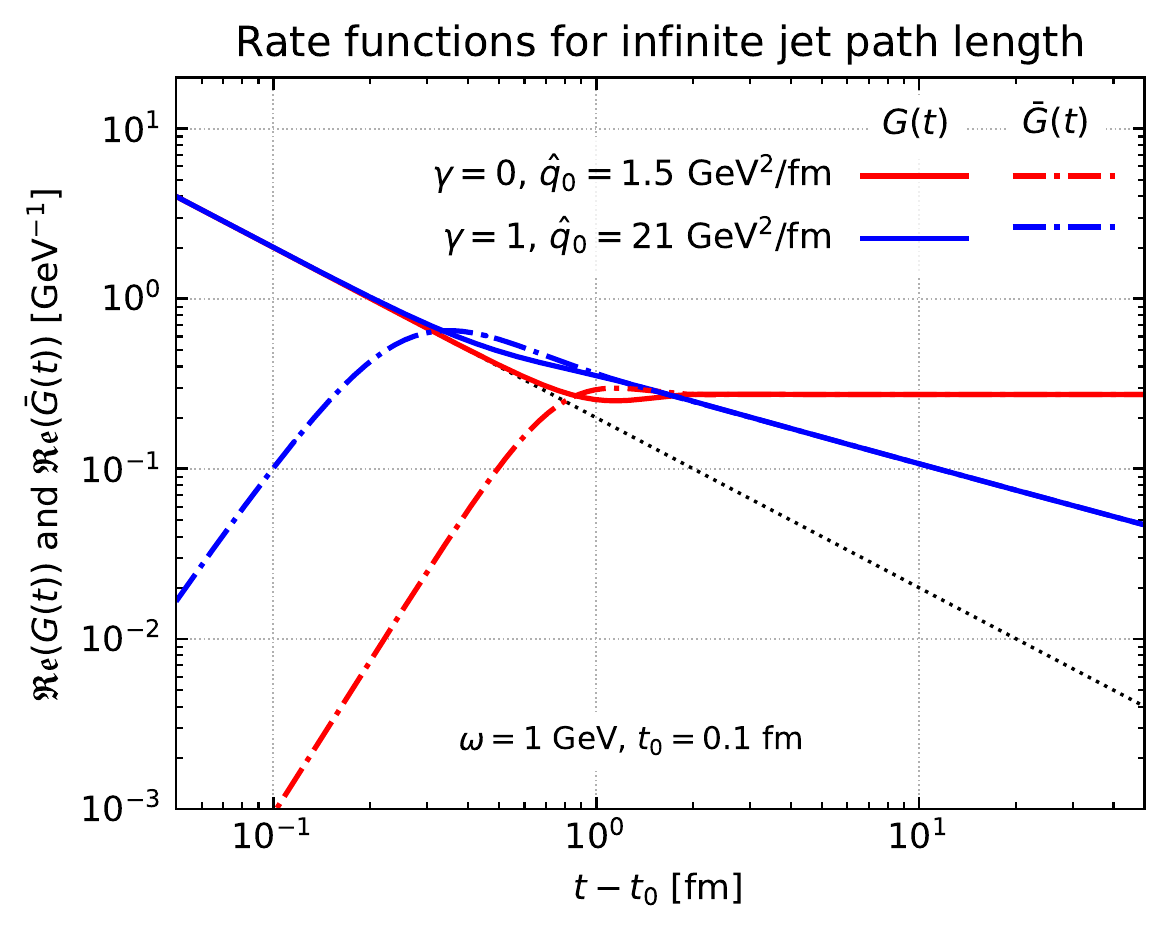}
  \end{subfigure}
  \hfill
  \begin{subfigure}[t]{0.49\textwidth}
    \includegraphics[page=2,width=\textwidth]{./plot-g-gbar.pdf}
  \end{subfigure}
  \caption{\small The functions $G$ (plain curves) and $\bar{G}$ (dashed curves) for a static medium ($\gamma=0$, red curves) and an ideal expanding plasma ($\gamma=1$, blue curves). Only the real parts are represented. On the left hand side, the medium is infinite and $G$, $\bar{G}$ are solutions of \eqref{riccati-g} and \eqref{riccati-gbar} respectively. On the right hand side, $\qhat(t)$ suddenly vanishes for $t\ge L$. The functions $G$ and $\bar{G}$ are the only \textit{continuous} solutions of the equations \eqref{riccati-g} and \eqref{riccati-gbar} (see Appendix \ref{app:B}). The black dotted curve is the $\qhat\rightarrow0$ limit of $G$, i.e.\ $1/(t-t_0)$, which is also the solution of \eqref{riccati-g} when $\qhat=0$.} \label{Fig:G-Gbar}
\end{figure}

Now, one can use our lemma \eqref{lemma} to integrate over $\bar{y}^+$. Indeed, if $\qhat(t)$ goes to 0 at large time, one sees that the integrand becomes a pure vacuum phase of the form $\exp(ik^-(y^+-\bar{y}^+))$ and thus satisfies the condition of \eqref{lemma}. Moreover, one can find a primitive with respect to $\bar{y}^+$ when there is no $\exp(-\varepsilon\bar{y}^+)$ factor using the change of variable $\bar{y}^+\rightarrow C(\bar{y}^+,y^+)/S(y^+,\bar{y}^+)$. 
Naming the function
\begin{align}\label{def-G}
 G(y^+)&=\frac{C(x_0^+,y^+)}{S(y^+,x_0^+)}=\frac{\partial_{y^+}S(y^+,x_0^+)}{S(y^+,x_0^+)},
\end{align}
the final remarkably simple formula for the in/in term, taking into account the correction due to \eqref{lemma} is:
\begin{equation}\label{Iii}
 I^{i/i}=2\mathfrak{Re}\int_{x_0^+}^\infty \dif y^+e^{-\varepsilon y^+}\frac{G(y^+)}{Q_s^2(y^+)+2ik^+G(y^+)}\exp\left(\frac{-k_\perp^2}{Q_s^2(y^+)+2ik^+G(y^+)}\right)-\frac{1}{k_\perp^2}
\end{equation}
To check this result, and in particular the presence of the corrective term in \eqref{Iii}, one verifies easily that in the limit $\qhat\rightarrow0$ where $Q_s^2(y^+)\rightarrow 0$ and $G(y^+)\rightarrow 1/(y^+-x_0^+)$, one recovers the standard Bremsstrahlung spectrum $1/k_\perp^2$.
The function $G^{-1}$ satisfies the following non linear differential equation (of Riccati kind):
\begin{equation}\label{riccati-g}
 \frac{\dif G^{-1}}{\dif y^+}=1+\frac{i\qhat(y^+)}{2 k^+}(G^{-1})^2\,,\qquad G^{-1}(x_0^+)=0
\end{equation}
and is represented Fig.~\ref{Fig:G-Gbar}-left. For a static medium, with constant $\qhat(t)=\qhat_0$, the function $G$ is:
\begin{equation}
 G_{\rm brick}(y^+)=\Omega\,\textrm{cotan}\Big(\Omega(y^+-x_0^+)\Big)\,,\qquad \Omega^2=\frac{i\qhat_0}{2k^+}
\end{equation}

\paragraph{Before/in term.} The calculation of the before/in term is more tedious because of the multiple transverse integrals. However, there is no subtlety due to the limit $\varepsilon\rightarrow0$ as the integrals over $y^+$ and $\bar{y}^+$ decouple. To do the calculation, the trick consists in inserting a Chapman-Kolmogorov relation at intermediate time $y^+=x_0^+$. The effective propagator between $\bar{y}^+$ and $x_0^+$ then reduces to a standard (adjoint) vacuum propagator. Introducing the function:
\begin{equation}\label{Gbar-def}
 \bar{G}(y^+)=\frac{C(x_0^+,y^+)C(y^+,x_0^+)-1}{S(y^+,x_0^+)C(y^+,x_0^+)}=\frac{\partial_{y^+} C(y^+,x_0^+)}{C(y^+,x_0^+)}
\end{equation}
one gets:
\begin{align}
 I^{b/i}&=2\mathfrak{Re}\int_{x_0^+}^\infty \dif y^+e^{-\varepsilon y^+}\frac{\bar{G}(y^+)}{Q_s^2(y^+)+2ik^+\bar{G}(y^+)}\exp\left(\frac{-k_\perp^2}{Q_s^2(y^+)+2ik^+\bar{G}(y^+)}\right)\\
 &-2\mathfrak{Re}\int_{x_0^+}^\infty \dif y^+e^{-\varepsilon y^+}\frac{G(y^+)}{Q_s^2(y^+)+2ik^+G(y^+)}\exp\left(\frac{-k_\perp^2}{Q_s^2(y^+)+2ik^+G(y^+)}\right)
\end{align}
Note the nice cancellation between the in/in term and the second term of the before/in term in the full on-shell spectrum. This term has also the right $\qhat\rightarrow0$ limit $\sim -2/k_\perp^2$ since the function $\bar{G}(y^+)$ goes to zero in this limit.

Also, the function $\bar{G}$ satisfies again a Riccati differential equation:
\begin{equation}\label{riccati-gbar}
 \frac{\dif \bar{G}}{\dif y^+}+\bar{G}^2+\frac{i\qhat(y^+)}{2k^+}=0\,,\qquad \bar{G}(x_0^+)=0
\end{equation}
The behaviour of $\bar{G}$ is very similar to the one of $G$ since both $G$ and $\bar{G}$ satisfy the same differential equation. The main difference lies in the initial condition and the behaviour at $y^+=x_0^+$ since $G(y^+)\sim 1/(y^+-x_0^+)$ whereas $\bar{G}(y^+)\rightarrow0$. This is clear from Fig.~\ref{Fig:G-Gbar}. For a static medium, this function reads:
\begin{equation}
 \bar{G}_{\rm brick}(y^+)=-\Omega\,\tan\Big(\Omega(y^+-x_0^+)\Big)\,,\qquad \Omega^2=\frac{i\qhat_0}{2k^+}
\end{equation}

\paragraph{Before/before term.} Last but not least, the before/before term requires also some care. In this term, the effective propagator $\mathcal{K}(y^+;\bar{y}^+)$ is a free propagator and the Fourier transform of the dipole cross-section $S_{gg}$ is independent of $y^+$ and $\bar{y}^+$ since it is associated with the total transverse momentum acquired during the full propagation. Using the representation \eqref{G0transverse} of the free scalar propagator, this term reads:
\begin{align}
 I^{b/b}&=\mathfrak{Re}\int\frac{d^2q_\perp}{(2\pi)^2}\frac{1}{q_\perp^2}\int\dif^2u_\perp e^{i(k_\perp-q_\perp) u_\perp}S_{gg}(u_\perp)\label{full-befbef}\\
 &=\mathfrak{Re}\int\frac{d^2q_\perp}{(2\pi)^2}\frac{1}{q_\perp^2}\int\dif^2u_\perp\exp\left(iu_\perp (k_\perp-q_\perp)-\frac{g^2}{2}C_A\int_{x_0^+}^{\infty}n(x^+)\sigma_d(u_\perp)\dif x^+\right)
\end{align}
In the harmonic approximation used so far in this section, the integral over $u_\perp$ is the Fourier transform of a Gaussian function giving:
\begin{equation}
  I^{b/b}=\frac{4\pi}{Q_s^2}\,\int\frac{d^2q_\perp}{(2\pi)^2}\frac{1}{q_\perp^2}\exp\left(\frac{-(k_\perp-q_\perp)^2}{Q_s^2}\right)\label{har-befbef-direct}
\end{equation}
with $Q_s^2\equiv Q_s^2(x_0^+)$.

\paragraph{Full results and discussion.} It turns out that the inclusive one gluon emission spectrum for an incoming on-shell quark has a very simple analytical form in the multiple soft scattering limit and harmonic approximation. We first give the result for an on-shell incoming quark. This situation is rather academic since the hard probes of the quark-gluon plasma produced in heavy-ion collisions are off-shell time-like quarks. However, this spectrum will help us to understand the physics contained in the off-shell one (cf. Chapter \ref{chapter:DLApic}). This on-shell spectrum reads:
\begin{tcolorbox}[ams align]\label{onshell-final}
 k^+\frac{\dif^3N^{\textrm{on-shell}}}{\dif k^+\dif^2 k_\perp}&=\frac{\alpha_s C_F}{\pi^2}\left[\frac{4\pi}{Q_s^2}\,\int\frac{d^2q_\perp}{(2\pi)^2}\frac{1}{q_\perp^2}\exp\left(\frac{-(k_\perp-q_\perp)^2}{Q_s^2}\right)\right.\nonumber\\
 &\hspace{-1.5cm}\left.+2\mathfrak{Re}\int_{x_0^+}^\infty \dif y^+\frac{\bar{G}(y^+)}{Q_s^2(y^+)+2ik^+\bar{G}(y^+)}\exp\left(\frac{-k_\perp^2}{Q_s^2(y^+)+2ik^+\bar{G}(y^+)}\right)-\frac{1}{k_\perp^2}\right]
\end{tcolorbox}
\noindent where we have removed the adiabatic prescription for compactness. As expected, this spectrum vanishes in the limit $\qhat\rightarrow0$, since in this limit $\bar{G}(y^+)\rightarrow 0$, so that the last term cancels exactly the before/before term.

On the other hand, the spectrum of an off-shell quark from a hard-vertex located at light cone time $x_i^+$ is:
\begin{tcolorbox}[ams align]\label{offshell-final}
 k^+\frac{\dif^3N^{\textrm{off-shell}}}{\dif k^+\dif^2 k_\perp}&=\frac{2\alpha_s C_F}{\pi^2}\mathfrak{Re}\int_{x_i^+}^\infty \dif y^+\,\frac{G(y^+)}{Q_s^2(y^+)+2ik^+G(y^+)}\exp\left(\frac{-k_\perp^2}{Q_s^2(y^+)+2ik^+G(y^+)}\right)\nonumber\\
 &\hspace{2cm}-\frac{\alpha_s C_F}{\pi^2}\frac{1}{k_\perp^2}
\end{tcolorbox}
\noindent With respect to \eqref{Iii}, note that the lower boundary is now $x_i^+$. If the medium ``starts'' at $x_0^+\ge x_i^+$, the function $G$ is the only continuous function satisfying \eqref{riccati-g} for all $y^+\ge x_i^+$ with $G^{-1}(x_i^+)=0$. This off-shell spectrum is more generally considered in the literature, albeit never written under the simple form \eqref{offshell-final}.

\begin{figure}[t] 
  \centering
  \begin{subfigure}[t]{0.51\textwidth}
    \includegraphics[page=1,width=\textwidth]{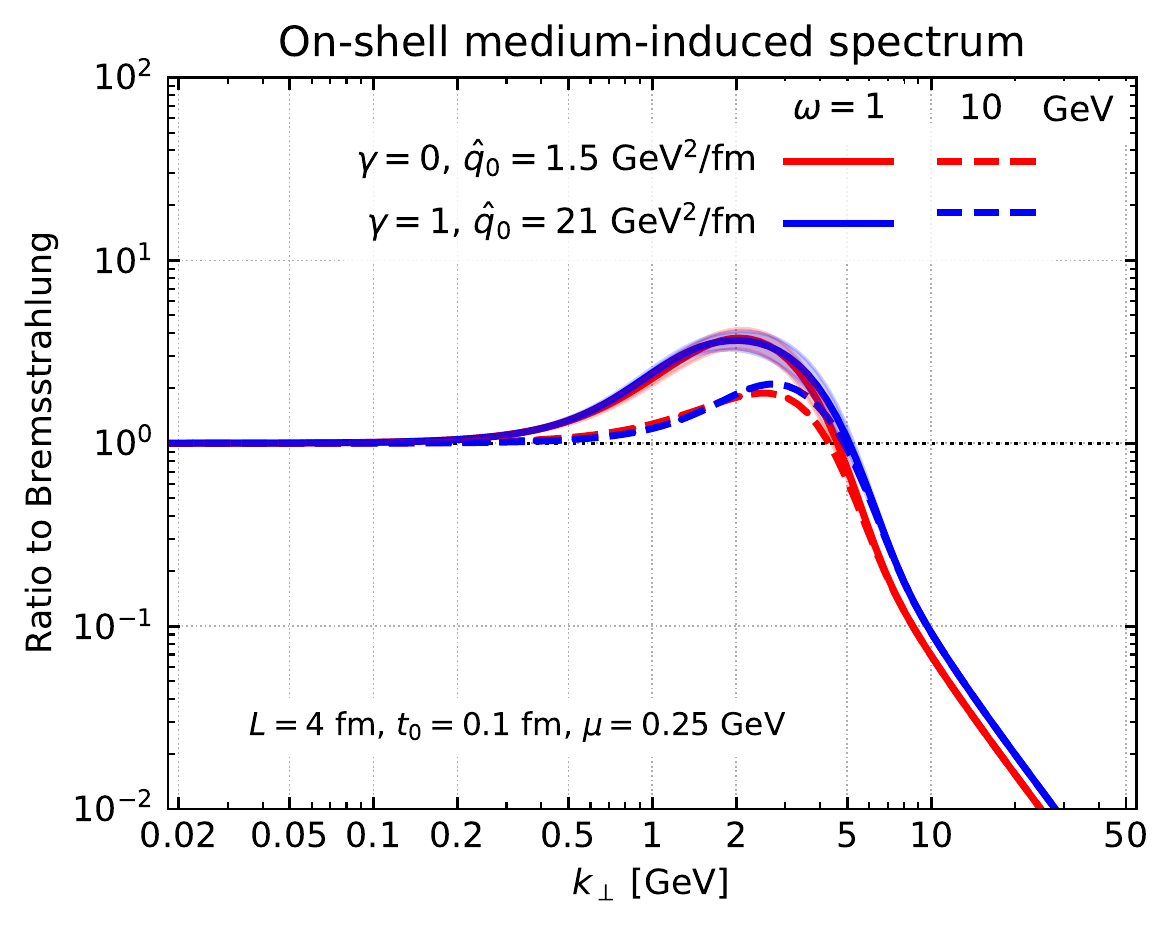}
  \end{subfigure}
  \hfill
  \begin{subfigure}[t]{0.48\textwidth}
    \includegraphics[page=3,width=\textwidth]{./spec.pdf}
  \end{subfigure}
  \caption{\small On-shell (left) and off-shell (right) medium induced spectra as a function of $k_\perp$ for several values of $\om\equiv k^+$. The spectra are normalized by the leading order Bremsstrahlung spectrum. For the off-shell spectrum, the location of the hard vertex is $t_i=0$ fm and the parton enters the medium at $t_0=0.1$ fm. For the on-shell spectrum, the bands correspond to the variation of the regularization scale $\mu$ by a factor $2$. The red curves correspond to the ``brick'' medium with constant $\qhat_0$ density ($\gamma=0$), whereas the blue curves correspond to the Bjorken expansion of an ideal relativistic plasma ($\gamma=1$). For the ideal plasma, the value of $\qhat_0$ is chosen to scale with the brick calculation in the soft limit of the $k_\perp$-integrated BDMPS-Z spectrum (see Fig.~\ref{Fig:LPM-plot} and discussion on this subject in Chapter~\ref{chapter:jet}-Section~\ref{sub:med-expansion}).}\label{Fig:kt-dep-bdmps}
\end{figure}

Whereas the ``off-shell'' spectrum is mathematically well defined, the first term of the on-shell spectrum corresponding to the before/before integral is ill-defined since the integral is divergent. An infrared cut-off $|q_\perp|\ge \mu$ is required for the integral over $q_\perp$ to make sense and then the question of how choosing this cut-off arises. Note that this feature is not related to the harmonic approximation. Indeed, the $1/q_\perp^2$ behavior when $q_\perp\rightarrow 0_\perp$ of \eqref{full-befbef} is independent of the value of $\sigma_d$.

As a matter of fact, as on-shell quarks do not exist in QCD because of confinement, it is natural for the on-shell spectrum to be ill-defined. The divergence of the before/before term corresponds to huge wave-length gluons with very long formation time formed before the interaction with the medium and that subsequently undergo transverse momentum broadening. In a confining theory, there is an upper limit on the formation time, or equivalently an upper limit on the wave-length of any emission that translates into a lower bound of order $\mu \sim\LQCD$ for the $q_\perp$ integral. Note that in the equivalent QED calculation, there is no such singularity as the emitted photon does not further interact with the medium: this divergence is specific to QCD \cite{Wiedemann:2000za}. In Section \ref{sec:decoherence}, where we deal with an initial colour singlet dipole instead of an on-shell incoming quark, we will see more precisely how the divergence of the before/before term is cured in pQCD.

\paragraph{A curiosity.} The $\mu$ regularization of the before/before term is of course not unique. It is interesting to notice that the adiabatic switching prescription provides another natural way to regularize the integral. A calculation of the before/before term leaving unintegrated the integral over $t\equiv y^+-\bar{y}^+$ gives:
\begin{equation}
  I^{b/b}=\frac{-1}{\varepsilon}\mathfrak{Re}\int_0^\infty\dif t\,\frac{Q_s^4t+2ik^+(k_\perp^2+Q_s^2)}{\big(Q_s^2t+2ik^+\big)^3}\exp\left(\frac{-k_\perp^2t}{Q_s^2t+2ik^+}-\varepsilon t\right)
\end{equation}
The integral is well defined as long as $\varepsilon>0$ but of course, does not have a finite limit as $\varepsilon\rightarrow 0$. For any non-zero $\varepsilon$, the integral is cut for $t=|y^+-\bar{y}^+|\ge 1/\varepsilon$. Since $|y^+-\bar{y}^+|$ is also the formation time of the emission, it is clear that the $\mu$ regularization and the $\varepsilon$ regularization are related by:
\begin{equation}\label{curisosity}
 \mu^2 \sim 2k^+\varepsilon
\end{equation}

\subsubsection{Integrated medium-induced spectrum: LPM effect}
\label{subsub:integrated-BDMPS}

Now, we investigate the medium-induced spectrum integrated over $k_\perp$. We first naively attempt to integrate the on-shell spectrum. Indeed, since in this section we are only interested in the purely medium-induced component and not the virtuality driven part, the on-shellness of the quark should prevent the existence of a $k_\perp$-divergent Bremsstrahlung component. However, this is not the case because the medium inevitably puts the quark out of its mass-shell, leading to a Bremsstrahlung tail at small $k_\perp$ (long formation time). We then consider the standard method encountered in the literature, where the $k_\perp$ integrated spectrum is defined as the integral of the off-shell spectrum minus its $\qhat\rightarrow0$ limit. In this case, the collinear singularity of the off-shell spectrum is regularized by the vacuum limit.

\paragraph{Integral of the on-shell spectrum.} Doing a brute force integration of \eqref{onshell-final}, one notices that the two divergent pieces cancel each other exactly in the limit where the infrared cut-off defining the before/before term goes to 0. Indeed, the broadening probability (or the Fourier transform of the dipole cross-section in the more general case) is normalized to $1$ once integrated over $k_\perp$. Thus, we are left with a purely medium induced component, that would vanish in the absence of the external field, given by:
\begin{equation}\label{on-shell-integrated}
 k^+\frac{\dif N^{\textrm{on-shell}}}{\dif k^+}=\frac{2\alpha_s C_F}{\pi}\mathfrak{Re}\int_{x_0^+}^\infty\dif y^+ \bar{G}(y^+)
\end{equation}
Using the Wronskian relation $C\partial_{y^+}S-S\partial_{y^+}C=1$, one can perform the integral over $y^+$:
\begin{equation}
 k^+\frac{\dif N^{\textrm{on-shell}}}{\dif k^+}=\frac{2\alpha_s C_F}{\pi}\lim\limits_{T\rightarrow\infty}\log\Big(\Big|C(T,x_0^+)\Big|\Big)
\end{equation}
As it is, this limit is not defined. We will prove this in Chapter \ref{chapter:jet}, Section \ref{sub:med-expansion}, but this can be understood from Fig.~\ref{Fig:G-Gbar} as well. For a power law decrease of $\qhat(y^+)=\qhat_0(x_0^+/y^+)^\gamma$, $\bar{G}$ behaves like $1/y^+$ at large $y^+$ if $\gamma\ge2$, and is therefore non-integrable. The physical reason for this divergence lies in the collinear singularity at small $k_\perp$. The interactions with the medium put the parton off-shell and thus, the incoming quark can radiate gluons according to the Bremsstrahlung spectrum. We shall also come back to this point in Chapter \ref{chapter:DLApic}. If $\gamma<2$, $G(y^+)\sim\sqrt{\qhat(y^+)/(2i\om)}$ for $y^+\rightarrow\infty$. In this case, $G(y^+)$ is non integrable for a different reason: the amount of radiations induced by the medium is infinite for an infinite jet path length because the medium dilutes too slowly. This suggests that one should integrate the function $\bar{G}$ up to $y^+=x_0^+ +L$ (another argument based on the shape of the $k_\perp$ spectrum is given in Appendix~\ref{app:B}). This integral reads:
\begin{tcolorbox}[ams equation]\label{tildeNmie}
 k^+\frac{\dif \tilde{N}}{\dif k^+}=\frac{2\alpha_s C_F}{\pi}\log\Big(\Big|C(x_0^++L,x_0^+)\Big|\Big)
\end{tcolorbox}
\noindent This spectrum is perfectly well defined, whatever the value of $\gamma$ is. If $\gamma\ge2$, it includes a substantial amount of vacuum-like collinear emissions with $k^2_\perp\gtrsim2k^+/L$.

\paragraph{Integral of the (regularized) off-shell spectrum.} As mentioned above, an alternative way to proceed consists in subtracting the genuine vacuum Bremsstrahlung contribution to the off-shell spectrum of a quark created at $y^+=x_0^+$ (i.e we fix $x_i^+=x_0^+$ in \eqref{offshell-final}):
\begin{equation}
 k^+\frac{\dif N}{\dif k^+}\equiv\int\dif k_\perp^2\,\left(k^+\frac{\dif^3N^{\textrm{off-shell}}}{\dif k^+\dif^2 k_\perp}-\frac{\alpha_sC_F}{\pi^2}\frac{1}{k_\perp^2}\right)
\end{equation}
Both terms in the integral lead to a divergence but the full integral is convergent.
The genuine vacuum spectrum can be written as an integral over $y^+$ using \eqref{Iii} in the limit $\qhat\rightarrow0$. The integral over $k_\perp$ is easy and gives:
\begin{align}\label{integrated-bdmpsz}
k^+\frac{\dif N}{\dif k^+}=\frac{2\alpha_sC_F}{\pi}\int_{x_0^+}^\infty\dif y^+\,\left(G(y^+)-\frac{1}{y^+-x_0^+}\right)e^{-\varepsilon y^+}
\end{align}
The vacuum regulator corresponds to the limit of $G$ when $\qhat\rightarrow0$, and can be directly derived from the defining differential equation \eqref{riccati-g} with $\qhat=0$.

As the integral is convergent, one can now remove the adiabatic switching. In order to control the divergences for $y^+\rightarrow x_0^+,\infty$, let us introduce two cut-offs $\nu$ and $T$:
\begin{align}
 k^+\frac{\dif N}{\dif k^+}&=\frac{2\alpha_sC_F}{\pi}\lim\limits_{\nu\rightarrow0,T\rightarrow\infty}\mathfrak{Re}\int_{x_0^++\nu}^T\dif y^+\,\left(G(y^+)-\frac{1}{y^+-x_0^+}\right)\label{integrant-G}\\
 &=\frac{2\alpha_sC_F}{\pi}\lim\limits_{\nu\rightarrow0,T\rightarrow\infty}\mathfrak{Re}\int_{x_0^++\nu}^T\dif y^+\frac{\partial_{y^+}S(y^+,x_0^+)}{S(y^+,x_0^+)}-\log\big((T-x_0^+)/\nu\big)\\
 &=\frac{2\alpha_sC_F}{\pi}\lim\limits_{T\rightarrow\infty}\log\left(\frac{|S(T,x_0^+)|}{T-x_0^+}\right)=\frac{2\alpha_sC_F}{\pi}\lim\limits_{T\rightarrow\infty}\log\Big(\Big|C(x_0^+,T)\Big|\Big)\label{final-BDMPSZ}
\end{align}
where we used $S(x_0^++\nu,x_0^+)\sim\nu$ when $\nu$ goes to zero. This result has been obtained in \cite{Arnold:2008iy} starting directly from the integral over $k_\perp$ of \eqref{final-gluon-cross} with its vacuum limit subtracted. 

As for \eqref{on-shell-integrated} the limit \eqref{final-BDMPSZ} is not defined if the medium expands too slowly, with a power $\gamma\le2$. In this case, one must also take into account the finite path length $L$ of the parton through the medium. We refer the reader to the calculations presented in Appendix~\ref{app:B} in this case. We show that formula \ref{final-BDMPSZ} is still valid if the argument of the function $C$ is changed:
\begin{tcolorbox}[ams equation]\label{Nmie}
 k^+\frac{\dif N}{\dif k^+}=\frac{2\alpha_s C_F}{\pi}\log\Big(\Big|C(x_0^+,x_0^+ +L)\Big|\Big)
\end{tcolorbox}
\noindent Note the nice symmetry between \eqref{Nmie} and \eqref{tildeNmie}. On the contrary, if $\gamma>2$, the integrand in \eqref{integrant-G} behaves like $\qhat(y^+)y^+/(2i\om)$ at large $y^+$, and is therefore integrable (see again Section~\ref{sub:med-expansion}). The collinear divergence highlighted in the on-shell spectrum is cured by the vacuum subtraction of the same collinear divergence.

\paragraph{The brick case - LPM effect.} To understand the specificity of the BDMPS-Z spectrum, let us consider the brick model where $\hat{q}(t)=\qhat$ for $t<L$ and vanishes for $t\ge L$. As shown in Appendix~\ref{app:B} where we find the solution $C$ of \eqref{diff-eq-S} continuous and derivable for all $t\ge 0$, one gets:
\begin{equation}
 C_{\rm brick}(0,L)=C_{\rm brick}(L,0)=\cos(\Omega L)\,,\qquad \Omega^2=\frac{i\qhat}{2 k^+}
\end{equation}
so that the BDMPS-Z spectrum reads:
\begin{equation}\label{bdmpsz-brick}
 k^+\frac{\dif N^{\textrm{brick}}}{\dif k^+}=\frac{2\alpha_sC_F}{\pi}\log\Big|\cos\Big(\frac{1+i}{2}\sqrt{\frac{\qhat}{k^+}}L\Big)\Big|\approx\frac{2\alpha_sC_F}{\pi}\left\{
    \begin{array}{ll}
        \sqrt{\frac{\omc}{2k^+}} & \mbox{for } k^+ \ll\omc \\
        \frac{1}{12}\Big(\frac{\omc}{k^+}\Big)^2 & \mbox{for } k^+ \gg\omc
    \end{array}
\right.
\end{equation}
with $\omc\equiv \qhat L^2/2$. Note that for the brick model, the spectra \eqref{Nmie} and \eqref{tildeNmie} give the same result because the cosine function is even.

\begin{figure}
   \centering
      \includegraphics[width=0.6\textwidth]{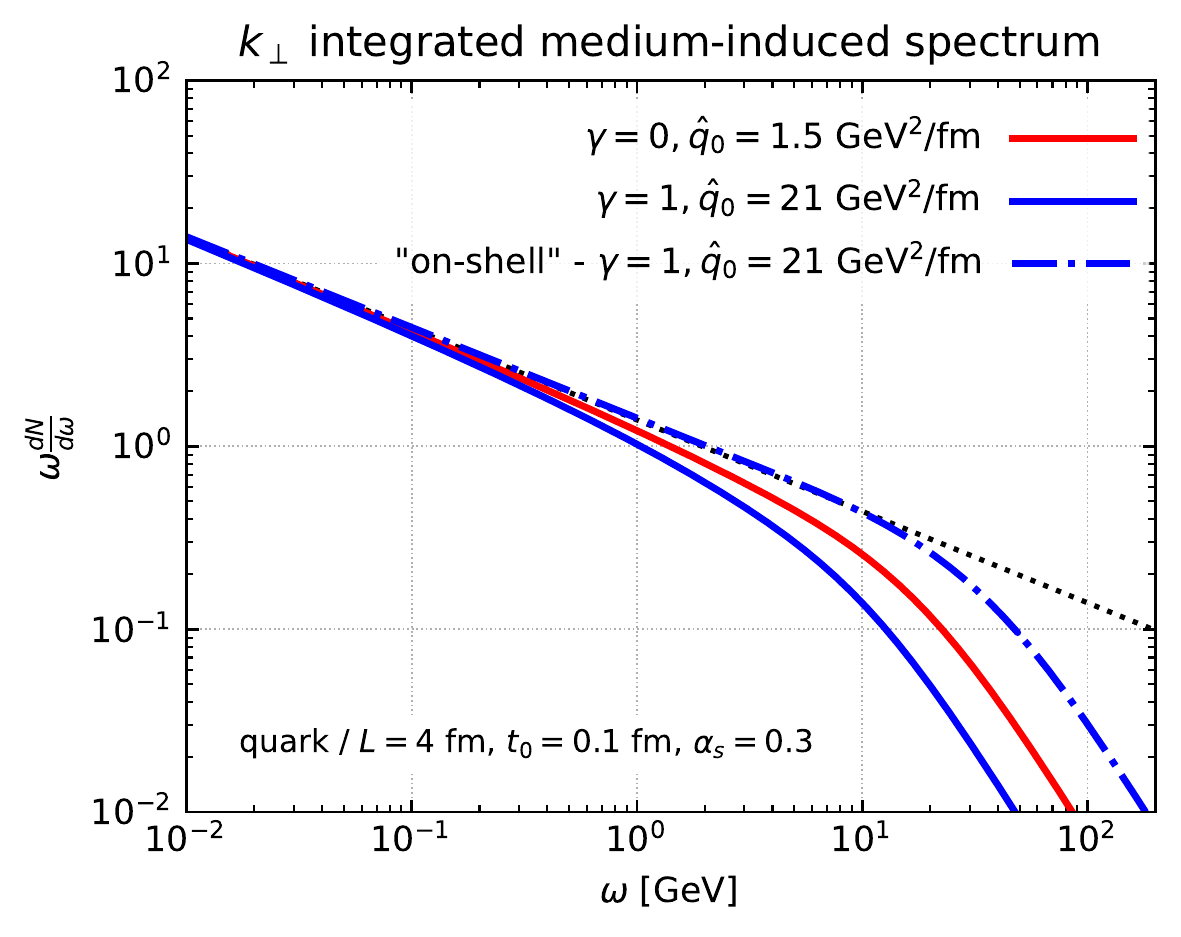}
    \caption{\small Integrated BDMPS-Z spectrum (formula \eqref{Nmie}) as a function of $\om\equiv k^+$, for a static medium ($\gamma=0$) and an expanding ideal plasma ($\gamma=1$). The value of $\qhat_0$ in the latter case is chosen so that both spectra scale exactly in the soft limit. The relation is given by \eqref{qhat-equiv-bjork}. In this soft limit, the asymptote is $\propto\sqrt{\om_c/\om}$ (dotted black line). The dashed blue line is the integrated on-shell spectrum, i.e. formula \eqref{tildeNmie} for comparison.}
    \label{Fig:LPM-plot}
\end{figure}

The $\sqrt{\omc/k^+}$ behaviour is a consequence of the so-called Landau-Pomerantchuk-Migdal (LPM) effect in QCD \cite{Landau:1953gr,Migdal:1956tc}. The formation of the gluon is a quantum process which is not instantaneous. The formation time $t_f$ as given by the uncertainty principle is $t_f\sim 2k^+/k_\perp^2$. During this time, if it is larger than the medium mean free path $\ell$, the quark-gluon pair interacts coherently with many scattering centers so that the gluon acquires a typical transverse momentum of order $k_\perp^2\sim\qhat t_f$ in the multiple soft scattering regime. Inserting the quantum relation for $t_f$, one gets the following relation for the formation time $t_{f,\textrm{med}}$ and transverse momentum $k_{f,\textrm{med}}$ of medium-induced radiations in the multiple soft scattering regime:
\begin{tcolorbox}[ams align]\label{med-scales}
 t_{f,\textrm{med}}=\sqrt{\frac{2k^+}{\qhat}}\,,\qquad k_{f,\textrm{med}}^2=\sqrt{2\qhat k^+}
\end{tcolorbox}
\noindent Hence, for medium induced emissions, the formation time is fixed once the energy $k^+$ of the emission is fixed. The condition $t_f\ge \ell$ translates into a lower limit in $k^+$: $k^+\ge \omBH\sim\qhat\ell^2$ for the validity of our calculation.

The scale $\omc$ should be understood as the maximal energy of a medium induced emission that can develop over a time $L$ from multiple scatterings. Indeed, $t_{f,\textrm{med}}\le L$ implies $k^+<\omc$. Then $\omc$ is also the upper limit of validity of our calculation relying on the multiple soft scattering and harmonic approximation. This upper limit enables also to better constrain the validity of the eikonal approximation for the incoming quark: one should impose $p^+\gg\omc$. One can find an equivalent calculation of the gluon emission spectrum relaxing the eikonal approximation for the incoming parton in \cite{Blaizot:2012fh,Apolinario:2014csa}.

This concludes the study of the BDMPS-Z spectrum in QCD which encompasses the regime $\omBH\le k^+\le\omc$. In the Bethe-Heitler regime, $k^+\le\omBH$, the spectrum $k^+ \dif N/\dif k^+$ scales as $\alpha_s L/\ell$, i.e. an incoherent sum of radiation spectra. In the regime $k^+\ge\omc$, it becomes dominated by the contribution coming from a single hard scattering (GLV spectrum, order 1 in opacity). This regime is captured by the present formalism \cite{Mehtar-Tani:2019tvy} but requires to go beyond the harmonic approximation for the dipole cross-section $\sigma_d$.

\section{Interferences in the multiple soft scattering regime}
\label{sec:decoherence}
The last section was devoted to medium-induced emissions by an incoming eikonal quark or gluon either on its mass shell or infinitely off-shell since created at a vertex with some fixed light cone time $x_i^+$. In this section we shall discuss the scattering of a dipole, i.e. a $q\bar{q}$ pair, off the medium, assuming that the dipole is initially in a colour singlet state. Considering an initial colour singlet state enables to study the effects of interferences on the emission spectrum.

This section is organized as follows: the first part sets up the formalism for the calculation of the emission spectrum from a colour dipole in the medium. The limit $\qhat\rightarrow0$ of the result gives the eikonal factor presented in \ref{sec:vac-bench}. In the second part, the effect of interferences is studied in details for medium-induced emissions, removing by hand the standard Bremsstrahlung emissions. Finally, in the third part, interferences are discussed for the vacuum-like emissions as well.

\subsection{Effective generating functional for soft emissions from a colour-singlet dipole}
\subsectionmark{Generating functional for emissions from a colour-singlet dipole}

\subsubsection{General formalism}

We have already developed the major part of the formalism necessary to set-up the calculation of a gluon emission from a $q\bar{q}$ pair. Hence, we shall give less details in this section and refer the reader to Section \ref{sub:one-gluon}. 
Our starting point will be again an effective generating functional for soft gluon emissions where the fermionic degrees of freedom have already been integrated out using the eikonal propagators for the quark and antiquark in the presence of an external (quantized and/or classical) gauge field.

The main difference with respect to the previous section is the initial colour state of the $q\bar{q}$ pair. This state is the colour singlet state:
\begin{equation}
\ket{\phi}=\frac{1}{\sqrt{N_c}}\sum_{\alpha}\ket{\alpha}\otimes\overline{\ket{\alpha}}
\end{equation} 
with $\{\ket{\alpha}\}$ a basis of the fundamental representation of $\textrm{SU}(N_c)$.
It is the only normed state of a $q\bar{q}$ pair invariant under all colour rotations. The effective generating functional for processes involving a $q\bar{q}$ pair in the initial and in the final state is:
\begin{align}
 Z_{\text{eff}}[J^{\mu}|\mathcal{A}_m]&=\frac{1}{\sqrt{N_c}}\sum_{\alpha}\int\mathcal{D}A^{\mu}\delta(G[A^{\mu}])\mathcal{D}_{tc,\beta\alpha}(p,p'|A+\mathcal{A}_m)\overline{\mathcal{D}}_{tc,\bar{\gamma}\alpha}(\bar{p},\bar{p}'|A+\mathcal{A}_m)\nonumber\\\label{Zeffqqbar}
 &\hspace{6cm}\times\exp\left(i S_{g}[A^{\mu}|\mathcal{A}^{\mu}_{m}]+i\int{\dif^4x J^{\mu}A_{\mu}}\right)
\end{align}
where we have omitted the spin dependence $\delta_{ss'}$, $\delta_{\bar{s}\bar{s}'}$ and written explicitly the colour indices of the truncated propagators in order to highlight the fact that the initial state is fixed. This generating functional fully incorporates the quark degrees of freedom in the eikonal limit, via a colour rotation of the dipole from the state $|\phi\rangle$ to a generic state  $\ket{\beta}\otimes\overline{\ket{\gamma}}$. Note that the Wilson line for the anti-quark truncated propagator is in the conjugate fundamental representation.

We are interested in the gluon emission cross-section with the $q\bar{q}$ pair in the final state, but we do not care about the dependence of this cross-section with respect to the final state momenta $p'$ and $\bar{p}'$. If we integrate over the kinematic variables of the final $q\bar{q}$ pair, the classical paths in \eqref{eikonal-prop} of both partons will be frozen in the amplitude and in the complex conjugate amplitude, as noticed in Section \ref{subsub:1gluon-cross}. For the inclusive gluon emission cross-section, one can directly replace the truncated propagator $\mathcal{D}_{tc}$ in \eqref{Zeffqqbar} by Wilson lines along the frozen classical paths.

Let us then call $\gamma^\mu_{q}(x^+)$ and $\gamma^\mu_{\bar{q}}(x^+)$ the respective classical path of the quark and the antiquark in Minkowski space. It is convenient the choose $x^+$ as the parameter along the path of both particles because $x^+$ plays the role of time in light-cone coordinates for a right mover at the speed of light. In light cone coordinates, these paths are parametrised according to:
\begin{align}
 \gamma^\mu_q(x^+)&=(x^+,\gamma_q^-(x^+),\gamma_\perp^q(x^+))\\
 \gamma^\mu_{\bar{q}}(x^+)&=(x^+,\gamma_{\bar{q}}^-(x^+),\gamma^{\bar{q}}_\perp(x^+))
\end{align}
so that the quark and antiquark Wilson lines read:
\begin{align}
 W_{-\infty,\beta\alpha}^{q,\infty}(\gamma_q|A+\mathcal{A}_m)&=T\Big[\exp\Big(ig\int_{-\infty}^{\infty}\dif x^+\big(A_a^\mu(\gamma_q)+\mathcal{A}_{m,a}^\mu(\gamma_q)\big)t^a u_\mu\Big)\Big]_{\beta\alpha}\label{Wilson-q}\\
\overline{W}_{-\infty,\bar{\gamma}\alpha}^{\bar{q},\infty}(\gamma_{\bar{q}}|A+\mathcal{A}_m)&=T\Big[\exp\Big(-ig\int_{-\infty}^{\infty}\dif x^+\big(A_a^\mu(\gamma_{\bar{q}})+\mathcal{A}_{m,a}^\mu(\gamma_{\bar{q}})\big)t^a\bar{u}_\mu\Big]\Big]_{\alpha\bar{\gamma}}\label{Wilson-qbar}
 \end{align}
with $u^\mu=\dif\gamma^\mu_q/\dif x^+$, $\bar{u}^\mu=\dif\gamma^\mu_{\bar{q}}/\dif x^+$ constant ``4-velocity'' vectors. These 4-vectors are related to the 4-momentum of the quark and the antiquark by $p^\mu\simeq p'^\mu=p^+u^\mu$ and $\bar{p}^\mu\simeq \bar{p}'^\mu=\bar{p}^+\bar{u}^\mu$.

The inclusive one gluon emission probability is related to the one-point vacuum expectation value given by our effective generating functional through the LSZ reduction formula. In the light-cone gauge, only the transverse components $A^i$, $i=1,2$ contribute to the matrix elements for the gluon emission. 
Consequently, we consider
\begin{equation}
\bm\langle A^i_a(z)\bm\rangle = \frac{1}{Z_0}\left(-i\frac{\delta}{\delta J_{i,a}(z)}\right)Z_{\text{eff}}[J^{\mu}|\mathcal{A}^{\mu}_{m}]\Big|_{J=0} 
\end{equation}
in order to obtain
\begin{equation}
 k^+\frac{\dif ^3N}{\dif k^+\dif^2k_\perp}=\lim\limits_{k^2\rightarrow 0}\frac{1}{16\pi^3}\Big\langle\sum_{\beta,\bar{\gamma}}\sum_{a,i}\Big|k^2\bm\langle A^i_a(k)\bm\rangle\Big|^2\Big\rangle
\end{equation}
where the big brackets refer to a statistical average. All the spin sums have already been performed and there is no average over initial colours since the initial state is fixed. As in \eqref{crosssec-target}, the surface of the target has been integrated out.

\subsubsection{Summary of the cross-section calculation}

\paragraph{Amplitude.} Then, we proceed in the same way as in the last section: we develop the generating functional up 
to the first order in $g A^\mu$ thanks to the following expansion of the Wilson line:
\begin{align}
 W_{-\infty,\beta\alpha}^{q,\infty}(\gamma_q|A+\mathcal{A}_m)=&W^{q,\infty}_{-\infty,\beta\alpha}\nonumber\\
 &+ig\int_{-\infty}^{\infty}\dif y^+\Big[W^{q,\infty}_{y^+}A_b^\mu(\gamma_q(y^+))t^b u_\mu W^{q,y^+}_{-\infty}\Big]_{\beta\alpha}+\mathcal{O}(g^2 A^\mu A^\nu u_\mu u_{\nu})
\end{align}
where all the Wilson lines in the right hand side are Wilson lines over the external field only. The generalization for the antiquark Wilson line \eqref{Wilson-qbar} is straightforward.

Consequently, to the first order in $g$ but all orders in $g\mathcal{A}_m$, the vacuum expectation value of $A^i_a$ reads (we have again omitted the tadpole terms):
\begin{align}
\langle A^i_a(z)\rangle =& \frac{ig}{\sqrt{N_c}} \Big[W^{q,\infty}_{-\infty}\int_{-\infty}^{\infty}\dif y^+ G_{ab}^{i\mu}(z,\gamma_q(y^+))u_\mu \mathcal{W}_{-\infty,bc}^{q,y^+}t^c\Big]_{\beta\alpha}\Big[W_{-\infty}^{\bar{q},\infty\dagger}\Big]_{\alpha\bar{\gamma}}\nonumber\\
&-\frac{ig}{\sqrt{N_c}}\Big[W^{q,\infty}_{-\infty}\Big]_{\beta\alpha}\int_{-\infty}^{\infty}\dif x^+G_{a\bar{b}}^{i\nu}(z,\gamma_{\bar{q}}(x^+))\bar{u}_\nu \mathcal{W}_{-\infty,\bar{b}\bar{c}}^{\bar{q},x^+}\Big[t^{\bar{c}}W_{-\infty}^{\bar{q},\infty\dagger}\Big]_{\alpha\bar{\gamma}}
\end{align}
In this expression, we have introduced the gluon propagator $G^{\mu\nu}(z,x)$ defined in \eqref{Gprop}. The sum over $\alpha$ associated with the colour singlet state is implicit. The $y^+$ coordinate corresponds to the emission time of the gluon from the quark, whereas $x^+$ is the emission time from the antiquark. We have also written the conjugation of the matrices $t^b$ and $t^{\bar{b}}$ as colour rotations with Wilson lines in the adjoint representation $\mathcal{W}_{-\infty,bc}^{q,y^+}$ and $W_{-\infty}^{\bar{q},\infty\dagger}$, as in the calculation \eqref{fund-ad-calc}.

Using the representation of $G^{\mu\nu}(z,x)$ in terms of the scalar propagator in the background field and the generalization of formula \eqref{def-gscalar-prop} for the other Lorentz indices given in Appendix~\ref{app:prop}, we perform the Fourier transform with respect to $z^-$:
\begin{align}
&\int \dif z^-\,e^{ik^+z^-}\bm{\langle} A^i_a(z)\bm{\rangle}=\nonumber\\
&\frac{ig}{2k^+\sqrt{N_c}}\int_{-\infty}^{\infty}\dif y^+e^{ik^+\gamma_q^-(y^+)}\Big(\frac{i\partial^i_{y_\perp}}{k^+}+u^i\Big)\mathcal{G}_{ab}(z^+,z_\perp;y^+,y_\perp|k^+)\mathcal{W}^{q,y^+}_{-\infty,bc}\times
 \Big[W^{q,\infty}_{-\infty}t^c\Big]_{\beta\alpha} \Big[W^{\bar{q},\infty\dagger}_{-\infty}\Big]_{\alpha\bar{\gamma}}\nonumber\\
 &-\frac{ig}{2k^+\sqrt{N_c}}\int_{-\infty}^{\infty}\dif x^+e^{ik^+\gamma_{\bar{q}}^-(x^+)}\Big(\frac{i\partial^i_{x_\perp}}{k^+}+\bar{u}^i\Big)\mathcal{G}_{a\bar{b}}(z^+,z_\perp;x^+,x_\perp|k^+)\mathcal{W}_{-\infty,\bar{b}\bar{c}}^{\bar{q},x^+}\Big[W^{q,\infty}_{-\infty}\Big]_{\beta\alpha}\Big[t^{\bar{c}}W^{\bar{q},\infty\dagger}_{-\infty}\Big]_{\alpha\bar{\gamma}}\label{Aizqq-step2}
 \end{align}
 In this formula, it is implicit that the transverse coordinates $y_\perp$ and $x_\perp$ are evaluated at the corresponding transverse position of the quark and the antiquark, i.e. $y_\perp=\gamma^q_\perp(y^+)$ and $x_\perp=\gamma^{\bar{q}}_\perp(x^+)$. 
 With respect to Eq.~\eqref{Aistep2}, the main difference is in the pure phase $e^{ik^+\gamma^-_q(y^+)}$ (and similarly for the antiquark) that appears inside the integral over the emission time and not outside. Indeed, in the present calculation, one cannot generally find a coordinate system where both the quark and the antiquark have a classical path $\gamma(x^+)$ with constant transverse and minus components. These phases have important consequences as we shall see when discussing the result of this calculation.
 
 \paragraph{Colour algebra.} Before completing the Fourier transform of the vacuum expectation value of $A^i_a(z)$, it is convenient to perform the colour algebra when summing over all final colour indices $\beta$ and $\bar{\gamma}$ of the $q\bar{q}$ pair in the amplitude squared. This amplitude squared contains four terms corresponding to a direct emission by the quark, by the antiquark and the two interferences terms. The choice of the initial state of the dipole (the summation over $\alpha$) enables to simplify the colour algebra in the product between the amplitude and its complex conjugate. For instance, the algebra for the interference terms reads:
\begin{align}
 \frac{1}{N_c}&\sum_{\alpha_1,\alpha_2,\beta,\bar{\gamma}}[W^{q,\infty}_{-\infty}t^c]_{\beta\alpha_1} [W^{\bar{q},\infty\dagger}_{-\infty}]_{\alpha_1\bar{\gamma}}
[W^{q,\infty\dagger}_{-\infty}]_{\alpha_2\beta}[W^{\bar{q},\infty}_{-\infty}t^{\bar{c}}]_{\bar{\gamma}\alpha_2}\\
&=\frac{1}{N_c}\sum_{\alpha_1,\alpha_2}t^c_{\alpha_2\alpha_1}t^{\bar{c}}_{\alpha_1\alpha_2}=\frac{1}{N_c}\Tr(t^ct^{\bar{c}})\\
&=\frac{C_F}{N_c^2-1}\delta^{c\bar{c}}
\end{align}
In fact, all the four terms give the same $\delta^{c\bar{c}}C_F/(N_c^2-1)$ factor, so we will not write explicitely the fundamental Wilson lines appearing in Eq.~\eqref{Aizqq-step2} in the following formulas.

Finally, the application of the $k^2$ operator in coordinate space $-2ik^+\partial_z^--\nabla_\perp^2$ and the Fourier transform to momentum space work exactly as in the calculation of a soft gluon emission by a quark. It is again important not to forget the adiabatic switching of interactions, and the analogous of formula \eqref{amplitude-final} is:
\begin{align}\label{amplitudeqq-final}
 &k^2\bm{\langle} A_a^i(k)\bm{\rangle}=g\int_{-\infty}^{\infty}\dif y^+e^{-\varepsilon|y^+|+ik^+\gamma_q^-(y^+)}\int\dif^2z_\perp e^{-ik_\perp z_\perp}\Big(\frac{i\partial^i_{y_\perp}}{k^+}+u^i\Big)\mathcal{G}_{ab}(z^+,z_\perp;y^+,y_\perp|k^+)\mathcal{W}^{q,y^+}_{-\infty,bc}\nonumber\\
 &-g\int_{-\infty}^{\infty}\dif x^+e^{-\varepsilon|x^+|+ik^+\gamma_{\bar{q}}^-(x^+)}\int\dif^2z_\perp e^{-ik_\perp z_\perp}\Big(\frac{i\partial^i_{x_\perp}}{k^+}+\bar{u}^i\Big)\mathcal{G}_{a\bar{b}}(z^+,z_\perp;x^+,x_\perp|k^+)\mathcal{W}_{-\infty,\bar{b}\bar{c}}^{\bar{q},x^+}
\end{align}
The indices $c$ and $\bar{c}$ are free but they will be contracted with $\delta^{c\bar{c}}$ once the amplitude is squared.

\paragraph{Amplitude squared.} The last step consists in squaring this amplitude. We expect that the direct terms by the quark or the antiquark to be very close to \eqref{g-cross-step1}, apart from the phase $e^{ik^+\gamma^-(y^+)}$. The full gluon cross-section is decomposed as:
\begin{equation}\boxed{
 k^+\frac{\dif^3N}{\dif k^+\dif^2 k_\perp}= k^+\frac{\dif^3N^{qq}}{\dif k^+\dif^2 k_\perp}+k^+\frac{\dif^3N^{q\bar{q}}}{\dif k^+\dif^2 k_\perp}+(q\leftrightarrow \bar{q})}
\end{equation}
The two terms $(q\leftrightarrow \bar{q})$ are calculated switching the quark and the antiquark in the formulas. The direct emission spectrum from the quark is:
\begin{align}
 k^+\frac{\dif^3N^{qq}}{\dif k^+\dif^2 k_\perp}&=\frac{\alpha_sC_F}{(2\pi)^2}2\mathfrak{Re}\int_{-\infty}^{\infty}\dif y^+\int_{-\infty}^{y^+}\dif\bar{y}^+e^{-\varepsilon(|y^+| +|\bar{y}^+|)}e^{ik^+(\gamma_q^-(y^+)-\gamma_q^-(\bar{y}^+))}\int \dif z_\perp\dif \bar{z}_\perp e^{-ik_\perp(z_\perp-\bar{z}_\perp)}\nonumber\\
 &\hspace{-2cm}\Big(\frac{i\partial^i_{y_\perp}}{k^+}+u^i\Big)\Big(\frac{-i\partial^i_{\bar{y}_\perp}}{k^+}+u^i\Big)\frac{1}{N_c^2-1}\Big\langle \Tr\, \mathcal{G}^{\dagger}(\infty,\bar{z}_\perp;\bar{y}^+,\bar{y}_\perp|k^+)\mathcal{G}(\infty,z_\perp;y^+,y_\perp|k^+)\mathcal{W}_{\bar{y}^+}^{q,y^+}\Big\rangle\label{direct-step1}
\end{align}
and the derivatives are taken at $y_\perp=\gamma_\perp^q(y^+)$, $\bar{y}_\perp=\gamma_\perp^q(\bar{y}^+)$. 

The $q\bar{q}$ term is a sum of two terms from the amplitude \eqref{amplitudeqq-final} squared:
\begin{equation}
 k^+\frac{\dif^3N^{q\bar{q}}}{\dif k^+\dif^2 k_\perp}\equiv\int_{-\infty}^\infty\dif y^+\int_{-\infty}^{y^+}\dif\bar{x}^+\,\mbox{...}+\int_{-\infty}^\infty\dif \bar{y}^+\int_{-\infty}^{\bar{y}^+}\dif x^+\,\mbox{...}=2\mathfrak{Re}\int_{-\infty}^\infty\dif y^+\int_{-\infty}^{y^+}\dif\bar{x}^+\,\mbox{...}
\end{equation}
with, as usual, the integration variable $x^+$ coming from the complex conjugate amplitude noted $\bar{x}^+$.
Moreover, in the interference term, the product $\mathcal{W}^{q,y^+}_{-\infty}\mathcal{W}^{\bar{q},\bar{x}^+\dagger}_{-\infty}$ cannot be simplified as easily as in the direct terms. This product is responsible for the effect that we will discuss in details in this section: decoherence.
\begin{align}
 k^+\frac{\dif^3N^{q\bar{q}}}{\dif k^+\dif^2 k_\perp}&=\frac{-\alpha_sC_F}{(2\pi)^2}2\mathfrak{Re}\int_{-\infty}^{\infty}\dif y^+\int_{-\infty}^{y^+}\dif\bar{x}^+e^{-\varepsilon(|y^+| +|\bar{x}^+|)}e^{ik^+(\gamma_q^-(y^+)-\gamma_{\bar{q}}^-(\bar{x}^+))}\int \dif z_\perp\dif \bar{z}_\perp e^{-ik_\perp(z_\perp-\bar{z}_\perp)}\nonumber\\
 &\hspace{-2cm}\Big(\frac{i\partial^i_{y_\perp}}{k^+}+u^i\Big)\Big(\frac{-i\partial^i_{\bar{x}_\perp}}{k^+}+\bar{u}^i\Big)\frac{1}{N_c^2-1}\Big\langle \Tr\, \mathcal{G}^{\dagger}(\infty,\bar{z}_\perp;\bar{x}^+,\bar{x}_\perp|k^+)\mathcal{G}(\infty,z_\perp;y^+,y_\perp|k^+)\mathcal{W}_{-\infty}^{q,y^+}\mathcal{W}_{-\infty}^{\bar{q},\bar{x}^+\dagger}\Big\rangle\label{inter-step1}
\end{align}
and the derivatives are taken at $y_\perp=\gamma_\perp^q(y^+)$, $\bar{x}_\perp=\gamma_\perp^{\bar{q}}(\bar{x}^+)$. These results have been obtained independently in \cite{MehtarTani:2012cy} and in \cite{CasalderreySolana:2011rz}.

\subsubsection{Medium average}

In order to go beyond the previous equations, we need to average over the background field describing the medium. The two basic ingredients are the medium average of two Wilson lines calculated in \eqref{dipolecross} with equal initial and final times, and the fact that the background field correlations are local in time. 

In Eq.~\eqref{dipolecross} the two classical paths have a very simple dependence with $x^+$, namely $u^-=0$ and $u_\perp=0_\perp$. The generalization for generic paths is straightforward:
\begin{align}\label{dipolecross-general}
\frac{1}{N_c^2-1}\Big\langle\Tr \mathcal{W}_{y_1^+}^{y_2^+}\big(\gamma\big)\mathcal{W}_{y_1^+}^{y_2^+\dagger}\big(\gamma'\big)\Big\rangle&=\exp\left(-\frac{g^2}{2}C_A\int_{y_1^+}^{y_2^+}\dif x^+\,n(x^+)\sigma_d\big(\gamma_\perp(x^+)-\gamma'_\perp(x^+)\big)\right)\nonumber\\
&\simeq\exp\left(-\frac{1}{4}\int_{y_1^+}^{y_2^+}\dif x^+\,\qhat_A(x^+)\big(\gamma_\perp(x^+)-\gamma'_\perp(x^+)\big)^2\right)
\end{align}
We used the harmonic approximation for the dipole cross section $\sigma_d$ in the second line.

The Chapman-Kolmogorov relation for $\mathcal{G}$ is inserted in \eqref{direct-step1} and \eqref{inter-step1} 
in order to use the locality property. In the direct term, this is done as in \eqref{chapmantrick}. In the interference term, one does the same trick without forgetting to decompose the adjoint Wilson line for the quark colour rotation: $\mathcal{W}_{-\infty}^{q,y^+}=\mathcal{W}_{\bar{x}^+}^{q,y^+}\mathcal{W}_{-\infty}^{q,\bar{x}^+}$. The final result involves then the Fourier transform of the dipole cross-section $S_{gg}$ related to the broadening of the final state gluon defined in \eqref{def-Sgg}, the effective propagators $\mathcal{K}_{qg}$ and $\mathcal{K}_{\bar{q}g}$ which depends on the path of the $q\bar{q}$ pair and a colour dipole in the adjoint representation $S_{q\bar{q}}$:
\begin{equation}
 \boxed{S_{q\bar{q}}(\bar{x}^+,z^+)\equiv\frac{1}{N_c^2-1}\Big\langle\Tr\mathcal{W}_{z^+}^{q,\bar{x}^+}\mathcal{W}_{z^+}^{\bar{q},\bar{x}^+\dagger}\Big\rangle}
\end{equation}
It is interesting to notice that an adjoint dipole appears in relation with a $q\bar {q}$ dipole. This function describes the colour precession of the colour \textit{currents} of the quark and antiquark and not that of their colour states. They are objects belonging to the adjoint representation contrary to the states, which are in the fundamental one. See \cite{CasalderreySolana:2011rz} for more details.

This average can be calculated from the formula \eqref{dipolecross-general}.
With all these pieces, the direct and interference terms for the gluon emission cross-section read\footnote{We have renamed the bound variables $\bar{y}^+=\bar{x}^+$ and $\bar{y}_\perp=\bar{x}_\perp$ in the interference term to highlight the symmetry with the direct term.}:
\begin{tcolorbox}[ams align]
 k^+\frac{\dif^3N^{qq}}{\dif k^+\dif^2 k_\perp}&=\frac{\alpha_s C_F}{(2\pi)^2}2\mathfrak{Re}\int_{-\infty}^{\infty}\dif y^+\int_{-\infty}^{y^+}\dif\bar{y}^+e^{-\varepsilon(|y^+| +|\bar{y}^+|)}e^{ik^+(\gamma^-_q(y^+)-\gamma_q^-(\bar{y}^+))}\nonumber\\
 &\hspace{0cm}\times\Big(\frac{i\partial^i_{y_\perp}}{k^+}+u^i\Big)\Big(\frac{-i\partial^i_{\bar{y}_\perp}}{k^+}+u^i\Big)\int\dif^2v_\perp e^{ik_\perp v_\perp} S_{gg}(v_\perp)\mathcal{K}_{qg}(y^+,y_\perp+v_\perp;\bar{y}^+,\bar{y}_\perp|\gamma_q)\\
  k^+\frac{\dif^3N^{q\bar{q}}}{\dif k^+\dif^2 k_\perp}&=\frac{-\alpha_s C_F}{(2\pi)^2}2\mathfrak{Re}\int_{-\infty}^{\infty}\dif y^+\int_{-\infty}^{y^+}\dif\bar{y}^+e^{-\varepsilon(|y^+| +|\bar{y}^+|)}e^{ik^+(\gamma^-_{q}(y^+)-\gamma_{\bar{q}}^-(\bar{y}^+))}S_{q\bar{q}}(\bar{y}^+,-\infty)\nonumber\\
 &\hspace{0cm}\times\Big(\frac{i\partial^i_{y_\perp}}{k^+}+u^i\Big)\Big(\frac{-i\partial^i_{\bar{y}_\perp}}{k^+}+\bar{u}^i\Big)\int\dif^2v_\perp e^{ik_\perp v_\perp}S_{gg}(v_\perp)\mathcal{K}_{qg}(y^+,y_\perp+v_\perp;\bar{y}^+,\bar{y}_\perp|\gamma_q)\label{final-cross-inter}
\end{tcolorbox}
\noindent with the derivative taken as explained above. 
Because of the non-trivial paths followed by the quark or the antiquark, the effective propagators $\mathcal{K}_{qg}$ (or $\mathcal{K}_{\bar{q}g}$) are a bit more complicated than in \eqref{def-Kqg} as they depend on the classical path $\gamma_{q/\bar{q}}$ followed by the quark or the antiquark.
\begin{align}
 \mathcal{K}_{qg}(y^+,y_\perp;\bar{y}^+,\bar{y}_\perp|\gamma_q)&\equiv\frac{1}{N_c^2-1}\Big\langle\Tr\mathcal{G}^\dagger(y^+,y_\perp;\bar{y}^+,\bar{y}_\perp|k^+)\mathcal{W}_{\bar{y}^+}^{q,y^+}\Big\rangle\\
 &\hspace{-1cm}=e^{-ik^+(u_\perp(y_\perp-\bar{y}_\perp)-u^-(y^+-\bar{y}^+))}\mathcal{K}\big(y^+,y_\perp-\gamma^q_\perp(y^+);\bar{y}^+,\bar{y}_\perp-\gamma^q_\perp(\bar{y}^+)\big)
\end{align}
with $\mathcal{K}$ defined in \eqref{def-K}. The phase appearing in front of the effective propagator used in the calculation of the BDMPS-Z cross-section is a consequence of the geometry of the dipole. Of course, this phase can be one in an adapted coordinate system, but as emphasized before, one cannot generally find a coordinate system where this phase is one for both the quark and the antiquark effective propagators.

\subsubsection{Onium scattering with the medium}
\label{subsub:onium-scatt}

So far, we have deliberately not specified the forms of $\gamma_q^\mu(x^+)$ and $\gamma_{\bar{q}}^\mu(x^+)$ to keep the discussion as general as possible. In this subsection, as a warm up for the antenna calculation, we consider the scattering of an onium with a medium. The onium enters into the medium at light cone time $x_0^+$. This simple model for heavy mesons will provide a solution to the divergence of the on-shell spectrum encountered in Section \ref{subsub:TMdep-BDMPS}.

The onium is defined as an highly energetic (i.e. eikonal) $q\bar{q}$ pair in colour singlet state with fixed transverse separation $X_\perp$. One can then choose a coordinate system such that:
\begin{align}
 \gamma_q^\mu(x^+)&=(x^+,0,0_\perp)\\
 \gamma_{\bar{q}}^\mu(x^+)&=(x^+,0,X_\perp)
\end{align}
Hence, for an onium, $u^-=\bar{u}^-=0$ and $u_\perp=\bar{u}_\perp=0$ and all the new phases enlighten in the previous discussion disappear. Consequently, if the transverse size of the medium is much larger than $|X_\perp|$ and if one assumes that the density of scattering centers is slowly varying on typical scales of order $|X_\perp|$, it means that the calculation we have made in the previous section is correct also for the direct terms. We only need to calculate the interference term which simply reads:
\begin{align}
  k^+\frac{\dif^3N^{q\bar{q}}}{\dif k^+\dif^2 k_\perp}&=\frac{-\alpha_s C_F}{(2\pi)^2 {k^+}^2}2\mathfrak{Re}\int_{-\infty}^{\infty}\dif y^+\int_{-\infty}^{y^+}\dif\bar{y}^+e^{-\varepsilon(|y^+| +|\bar{y}^+|)}S_{q\bar{q}}(\bar{y}^+,-\infty)\nonumber\\
 &\hspace{0cm}\times\int\dif^2u_\perp\,e^{ik_\perp u_\perp}S_{gg}(u_\perp)\partial^i_{y_\perp=0_\perp}\partial^i_{\bar{y}_\perp=X_\perp}\mathcal{K}(y^+,y_\perp+u_\perp;\bar{y}^+,\bar{y}_\perp)
\end{align}

\paragraph{The before/before interference term.} Now, we calculate the before/before component of the interference term, defined by $y^+,\bar{y}^+\in(-\infty,x_0^+]$. In this domain, the effective propagator $\mathcal{K}$ is free and the $q\bar{q}$ dipole factor $S_{q\bar{q}}$ is equal to 1. Using \eqref{G0transverse} to represent the free propagator, one ends up with the following result:
\begin{align}
  k^+\frac{\dif^3N^{q\bar{q},b/b}}{\dif k^+\dif^2 k_\perp}&=\frac{-\alpha_s C_F}{\pi^2}\mathfrak{Re}\int\frac{\dif^2q_\perp}{(2\pi)^2}\frac{e^{iq_\perp X_\perp}}{q_\perp^2}\int\dif^2u_\perp e^{iu_\perp(k_\perp-q_\perp)}S_{gg}(u_\perp)\label{full-befbef-inter}\\
  &\simeq\frac{-\alpha_s C_F}{\pi^2}\left[\frac{4\pi}{Q_s^2}\,\int\frac{\dif^2q_\perp}{(2\pi)^2}\frac{\cos(q_\perp X_\perp)}{q_\perp^2}\exp\left(\frac{-(k_\perp-q_\perp)^2}{Q_s^2}\right)\right]\label{har-befbef-inter}
\end{align}
where we use again the harmonic approximation to get the second line. Combined with the before/before direct terms calculated in \eqref{har-befbef-direct}, the divergence in $1/q_\perp^2$ due to the before/before term of the on-shell spectrum is cured by the $\cos(q_\perp X_\perp)$ factor:
\begin{equation}
k^+\frac{\dif^3N^{b/b}}{\dif k^+\dif^2 k_\perp}=\frac{2\alpha_s C_F}{\pi^2}\left[\frac{4\pi}{Q_s^2}\,\int\frac{\dif^2q_\perp}{(2\pi)^2}\frac{1-\cos(q_\perp X_\perp)}{q_\perp^2}\exp\left(\frac{-(k_\perp-q_\perp)^2}{Q_s^2}\right)\right]
\end{equation}
which is a perfectly well defined integral. The factor $1-\cos(q_\perp X_\perp)$ now cuts the long wavelength emissions with $|q_\perp|\lesssim 1/|X_\perp|$ before scattering. Physically, the very soft gluons do not resolve the onium and are thus suppressed.

\subsection{Medium-induced emissions from a colour singlet antenna}
\label{sub:mie-antenna}

In this section, we consider a colour singlet $q\bar{q}$ antenna. We are interested only in the medium-induced spectrum from this antenna. We leave the study of the vacuum-like pattern to Section \ref{sub:decoherence}. As we are only interested in the medium-induced component, we shall subtract the vacuum limit $\qhat\rightarrow 0$ of our result as in the discussion of the LPM effect in the previous section.

Let us first define an antenna as an eikonal $q\bar{q}$ pair created at a fixed vertex with light cone time $x_i^+=x_0^+$\footnote{The generalization to any value of $x_i^+$ is straightforward.} with fixed opening angle $\theta_{q\bar{q}}$. The corresponding classical trajectories are:
\begin{align}
  \gamma_q^\mu(x^+)&=(x^+,u^-(x^+-x_0^+),u_\perp(x^+-x_0^+))\\
 \gamma_{\bar{q}}^\mu(x^+)&=(x^+,\bar{u}^-(x^+-x_0^+),\bar{u}_\perp(x^+-x_0^+))
\end{align}
Let us choose our $z$ axis aligned with the direction of motion of the center of mass of the $q\bar{q}$ pair.
The opening angle $\theta_{q\bar{q}}$ of the pair is related to the transverse velocities $u_\perp$ and $\bar{u}_\perp$ from the relation $p^2=2 E\bar{E}(1-\cos(\theta_{q\bar{q}}))$ with $p=(p+\bar{p})\simeq (p'+\bar{p}')$ and $E$, $\bar{E}$ the respective energy of the quark and the antiquark. Thus, $2p^\mu\bar{p}_\mu=p^+\bar{p}^+(1-\cos(\theta_{q\bar{q}}))$. Finally, using the relation $p^\mu= p^+u^\mu$, 
\begin{equation}
 2u^\mu\bar{u}_\mu=1-\cos(\theta_{q\bar{q}})\simeq \frac{\theta_{q\bar{q}}^2}{2}
\end{equation}
in the small angle approximation. The product $2u^\mu\bar{u}_\mu=2u^-+2\bar{u}^--2u_\perp\bar{u}_\perp$ is also equal to $(u_\perp-\bar{u}_\perp)^2$ since $u^2=\bar{u}^2=0$ or equivalently  $2u^-=u_\perp^2$ and $2\bar{u}^-=\bar{u}_\perp$ (massless partons).

For an antenna, the $q\bar{q}$ dipole factor $S_{q\bar{q}}$ has a simple expression, in the small angle limit and harmonic approximation. From \eqref{dipolecross-general}, one gets
\begin{tcolorbox}[ams equation]\label{Sqqbar}
 S_{q\bar{q}}(\bar{y}^+)=\exp\left(-\frac{1}{8}\thqq^2\int_{-\infty}^{\bar{y}^+}\dif\xi\,\qhat_A(\xi)(\xi-x_0^+)^2\right)
\end{tcolorbox}

\subsubsection{The energy spectrum}

In a first place, we are not interested in the transverse momentum dependence of the gluon spectrum. In the previous section, we have seen that the off-shell spectrum associated with the emission from a quark created at $x_0^+$ is divergent once integrated over $k_\perp$. To get a meaningful result, a subtraction of the vacuum radiation pattern is necessary. So, we aim at calculating:
\begin{equation}
 k^+\frac{\dif N}{\dif k^+}=\int\dif^2k_\perp\left( k^+\frac{\dif^3N^{i/i}}{\dif k^+\dif^2 k_\perp}-\lim\limits_{\qhat\rightarrow0} k^+\frac{\dif^3N^{i/i}}{\dif k^+\dif^2 k_\perp}\right)
\end{equation}
where the $i/i$ label refers to the in/in term where $y^+,\bar{y}^+\ge x_0^+$, since the antenna is created at this time.

We work within the multiple soft scatterings and harmonic approximations. For symmetry reasons, the $k_\perp$-\textit{integrated} spectrum for direct emissions by the quark or the antiquark is exactly the same as the one calculated in Section \ref{subsub:integrated-BDMPS}. It is given by the expression \eqref{final-BDMPSZ}.

The interference term is more complicated as one cannot avoid the difficulties due to the
phases. As the only dependence on $k_\perp$ is inside the Fourier transform factor $\exp(ik_\perp v_\perp)$, the integration over $k_\perp$ gives a Dirac delta function $(2\pi)^2\delta^{(2)}(v_\perp)$. The final result, for the interference term, is:
\begin{align}
 k^+\frac{\dif N^{q\bar{q}}}{\dif k^+}&\equiv\int\dif^2k_\perp\left( k^+\frac{\dif^3N^{q\bar{q},i/i}}{\dif k^+\dif^2 k_\perp}-\lim\limits_{\qhat\rightarrow0} k^+\frac{\dif^3N^{q\bar{q},i/i}}{\dif k^+\dif^2 k_\perp}\right)\\
&=\frac{-\alpha_s C_F}{\pi}2\mathfrak{Re}\int_{x_0^+}^{\infty}\dif y^+\int_{x_0^+}^{y^+}\dif\bar{y}^+e^{-\varepsilon(y^++\bar{y}^+)}\left(\frac{-S_{q\bar{q}}(\bar{y}^+)}{S^2(y^+,\bar{y}^+)}e^{\phi(y^+,\bar{y}^+)}\big(1+\phi(y^+,\bar{y}^+)\big)\right.\nonumber\\
 &\hspace{6cm}\left.+\frac{1}{(y^+-\bar{y}^+)^2}e^{\phi_0(y^+,\bar{y}^+)}\big(1+\phi_0(y^+,\bar{y}^+)\big)\right)\label{inter-bdmpsz-1}
\end{align}
with the functions $\phi$ and $\phi_0$ defined as (in the small angle approximation):
\begin{equation}
 \phi(y^+,\bar{y}^+)=\frac{-ik^+\theta_{q\bar{q}}^2(\bar{y}^+-x_0^+)}{4}\left(1+\frac{C(y^+,\bar{y}^+)}{S(y^+,\bar{y}^+)}(\bar{y}^+-x_0^+)\right)
\end{equation}
and 
\begin{align}
 \phi_0(y^+,\bar{y}^+)&\equiv\lim\limits_{\qhat\rightarrow0} \phi(y^+,\bar{y}^+)\\
 &=\frac{-ik^+\theta_{q\bar{q}}^2(\bar{y}^+-x_0^+)}{4}\left(1+\frac{\bar{y}^+-x_0^+}{y^+-\bar{y}^+}\right)
\end{align}
From \eqref{inter-bdmpsz-1}, one clearly sees that the unregularised interference term is divergent because of the $1/S^2$ with $S(y^+,\bar{y})\sim y^+-\bar{y}^+$ as $y^+\rightarrow\bar{y}^+$.

Note that this expression is completely symmetric in the exchange $q\leftrightarrow \bar{q}$ so the other $\bar{q}q$ interference term is also equal to \eqref{inter-bdmpsz-1}.
The expression \eqref{inter-bdmpsz-1} can be further simplified using the Wronskian relation \eqref{Wronskian}. The $\varepsilon$ prescription can be removed as the integral is convergent, and one can switch the order of integration, integrating first over $y^+$. If we define the quantity:
\begin{equation}
 G_\infty(\bar{y}^+)\equiv \lim\limits_{T\rightarrow\infty}\frac{C(T,\bar{y}^++x_0^+)}{S(T,\bar{y}^++x_0^+)}
\end{equation}
then, the full interference term can be expressed as a single integral over $\bar{y}^+$:
\begin{tcolorbox}[ams align]
 k^+\frac{\dif N^{q\bar{q}}}{\dif k^+}&=\frac{-2\alpha_sC_F}{\pi}\mathfrak{Re}\int_0^\infty\frac{\dif\bar{y}^+}{\bar{y}^+}\left(S_{q\bar{q}}(\bar{y}^++x_0^+)e^{-\frac{ik^+\thqq^2\bar{y}^+}{4}\big(1+G_\infty(\bar{y}^+)\bar{y}^+\big)}\big(1+G_\infty(\bar{y}^+)\bar{y}^+\big)\right)\nonumber\\
 &\hspace{7cm}+\frac{2\alpha_sC_F}{\pi}\mathfrak{Re}\int_0^\infty\frac{\dif\bar{y}^+}{\bar{y}^+}e^{-\frac{ik^+\thqq^2\bar{y}^+}{4}}\label{final-bdmpsz-inter}
\end{tcolorbox}
\noindent In the limit $\thqq\rightarrow 0$, the interference term is exactly equal to minus the direct term. As expected, colour singlet antenna with vanishing opening angle do not radiate. 

\subsubsection{Time-scales in the interference term}

Formula \eqref{final-bdmpsz-inter} is general and valid for any continuous function $\qhat(t)$. Combined with \eqref{final-BDMPSZ}, it allows for a straightforward numerical calculation of the medium-induced spectrum generated by a colour singlet antenna. However, the different time scales involved in the integral are difficult to extract without assuming a particular time dependence of $\qhat$. In order to have an insight on the physical content of the interference term, let us consider an infinite medium with constant $\qhat(t)=\qhat$ created at $x_0^+=0$. In  this simple model, the function $G_\infty(\bar{y}^+)$ is independent of $\bar{y}^+$, since the medium is invariant by light-cone $+$ translation:
\begin{equation}
 G_\infty(\bar{y}^+)=\frac{1+i}{2}\sqrt{\frac{\qhat}{k^+}}
\end{equation}
Note that an infinite medium with constant $\qhat$ does not satisfy the assumptions made for the calculation of the direct terms in the previous section, as the function $\qhat(t)$ does not vanish at large times. Consequently, the direct terms are ill-defined in this model. Nevertheless, as we shall see, thanks to the suppression mechanism of interference due to decoherence, the following formula with the upper boundary set to the characteristic length of the medium is a good approximation for the interference term in a finite medium. That said, the present discussion for the interference term is meant to illustrate the time scales characteristic of the interference pattern.
The complete formula \eqref{final-bdmpsz-inter} becomes:
\begin{align}
 k^+\frac{\dif N^{q\bar{q}}}{\dif k^+}&=\frac{2\alpha_sC_F}{\pi}\int_0^\infty\frac{\dif t}{t}\left[\exp\left(-\frac{1}{24}\qhat\thqq^2t^3-\frac{1}{8}\sqrt{\hat{q}k^+}t^2\thqq^2\right)\right.\times\nonumber\\
&\left(\sqrt{\frac{\hat{q}}{k^+}}\frac{t}{2}\sin\Big(\frac{k^+t\theta_{q\bar{q}}^2}{4}+\frac{t^2\theta_{q\bar{q}}^2\sqrt{\hat{q}k^+}}{8}\Big)-\Big(1+\sqrt{\frac{\hat{q}}{k^+}}\frac{t}{2}\Big)\cos\Big(\frac{k^+t\theta_{q\bar{q}}^2}{4}+\frac{t^2\theta_{q\bar{q}}^2\sqrt{\hat{q}k^+}}{8}\Big)\right)\nonumber\\
&\hspace{10cm}\left.+\cos\Big(\frac{k^+t\theta_{q\bar{q}}^2}{4}\Big)\right]\label{inter-fixedq}
\end{align}
The last term is the vacuum subtraction required to have a well-defined integral because of the singularity at $t=0$. One can now identify four time scales in the expression \eqref{inter-fixedq} (see also the detailed discussion at the level of the $k_\perp$ unintegrated spectrum in \cite{CasalderreySolana:2011rz}). 

\begin{enumerate}
 \item The vacuum interference time $\tvac$ is the characteristic time scale for interferences of vacuum-like emissions: interference occurs when the transverse size of the dipole $\thqq\tvac$ is shorter than the transverse wavelength $\sim1/k_\perp$ of the emission. For vacuum-like emissions, $k_\perp\simeq k^+\th$. Using $\th=\thqq$ in the relation $1/k_\perp=\thqq \tvac$, this gives the following expression for $\tvac$:
 \begin{equation}
  \tcboxmath{t_{i,\textrm{vac}}=\frac{2}{k^+\thqq^2}}
 \end{equation}
%
This scale is the only one remaining in the limit $\qhat\rightarrow 0$ (that is why it is really a vacuum-like time scale), and corresponds to a purely imaginary phase in formula \eqref{final-bdmpsz-inter} or \eqref{inter-fixedq}.

\item The medium induced interference time $\tint$ is the characteristic time scale for interferences of medium induced emissions. Instead of using $k_\perp\simeq k^+\thqq$ in the relation $1/k_\perp=\thqq \tint$, one uses the typical transverse momentum at formation $k_{f,\textrm{med}}\sim(\qhat k^+)^{1/4}$  of medium-induced emissions, cf. \eqref{med-scales}:
\begin{equation}
\tcboxmath{
 t_{i,\textrm{med}}=\frac{2}{k_f\thqq}=\frac{2^{3/4}}{(\qhat k^+)^{1/4}\thqq}}
\end{equation}
This time scale appears in the exponential suppression factor in \eqref{inter-fixedq} as well as in the imaginary phase seen in the sine and cosine functions.

\item The medium induced formation time $t_{f,\textrm{med}}$ discussed in the previous section:
\begin{equation}
\tcboxmath{
 t_{f,\textrm{med}}=\sqrt{\frac{2k^+}{\qhat}}}
\end{equation}
This time scale is the only one involved in the direct terms (besides the length $L$ of the medium, see below).

\item The decoherence time coming from the dipole $q\bar{q}$ factor $S_{q\bar{q}}$. For a constant $\qhat$, $S_{q\bar{q}}$ can be calculated exactly:
\begin{equation}\label{tcoh}
S_{q\bar{q}}(t)=\exp\left(-\frac{1}{24}\qhat\thqq^2t^{3}\right)=\exp\left(-\frac{1}{6}(t/\tcoh)^3\right)\,,
\end{equation}
\begin{equation}
 \tcboxmath{t_{\textrm{coh}}=\left(\frac{4}{\qhat\thqq^2}\right)^{1/3}}
\end{equation}
It is independent of the energy $k^+$ of the emission.

\end{enumerate}

The numerical prefactors for these time scales have been chosen so that they are equal at a special energy scale $\omega_0(\thqq)$ defined by:
\begin{equation}
 \omega_0(\thqq)\equiv\Big(\frac{2\qhat}{\thqq^4}\Big)^{1/3}
\end{equation}
The ordering of the four time scales is the following:
\begin{equation}\label{time-scale-ordering}
    \begin{array}{ll}
        t_{f,\textrm{med}}\le \tcoh\le \tint\le \tvac & \mbox{ if } k^+ \le\om_0(\thqq) \\
        t_{f,\textrm{med}}\ge \tcoh\ge \tint\ge \tvac& \mbox{ if } k^+ \ge \om_0(\thqq)
    \end{array}
\end{equation}
with equality for $k^+=\om_0(\thqq)$. In terms of these time scales, the interference term for constant $\qhat$ in an infinite medium reads:
\begin{align}
 k^+\frac{\dif N^{q\bar{q}}}{\dif k^+}&=\frac{2\alpha_sC_F}{\pi}\frac{1}{\sqrt{2}t_{f,\textrm{med}}}\int_0^\infty\frac{\dif t}{t}\left\{\exp\Big(-\frac{1}{6}\Big(\frac{t}{\tcoh}\Big)^3-\frac{1}{16\sqrt{2}}\Big(\frac{t}{\tint}\Big)^2\Big)\right.\times\nonumber\\
&\left[t\sin\Big(\frac{t}{2\tvac}+\frac{1}{16\sqrt{2}}\Big(\frac{t}{\tint}\Big)^2\Big)-\Big(\sqrt{2}t_{f,\textrm{med}}+t\Big)\cos\Big(\frac{t}{2\tvac}+\frac{1}{16\sqrt{2}}\Big(\frac{t}{\tint}\Big)^2\Big)\right]\nonumber\\
&\hspace{9cm}\left.+\sqrt{2}t_{f,\textrm{med}}\cos\Big(\frac{t}{2\tvac}\Big)\right\}\label{inter-timescale}
\end{align}

\subsubsection{Discussion}

Now that we have identified the main time scales in the interference term, we would like to discuss the effect of this term on the full medium-induced spectrum as a function of the opening angle $\thqq$ of the dipole and the energy $k^+$ of the medium-induced emission. Thus, we need to re-introduce a characteristic path length $L$ of the antenna through the medium so that $\qhat(t)$ vanishes when $t\ge L$. We shall consider only emissions with $k^+\ll\om_c$, i.e $t_{f,\textrm{med}}\ll L$, so that all the approximations made to derive the BDMPS-Z spectrum (direct terms) are valid.

As a first step, we need to understand how the four time scales involved in the unintegrated medium-induced spectrum are related with the length $L$. We now call this unintegrated medium-induced spectrum the ``medium induced rate'' for the following reason. In the regime $k^+\le\om_c$, the BDMPS-Z spectrum is proportional to $L/t_{f,\textrm{med}}$. Hence one can interpret $t_{f,\textrm{med}}$ as a (constant) rate for medium-induced emissions that can occur everywhere over the characteristic length $L$:
\begin{equation}
 k^+\frac{\dif^2 N^{qq}}{\dif k^+\dif t}\simeq \frac{2\alpha_sC_F}{\pi}\sqrt{\frac{\qhat}{4k^+}}=\frac{2\alpha_sC_F}{\pi}\frac{1}{\sqrt{2}t_{f,\textrm{med}}}
\end{equation}
so that 
\begin{equation}
k^+\frac{\dif N^{qq}}{\dif k^+}=\int_0^L\dif t\,k^+\frac{\dif^2 N^{qq}}{\dif k^+\dif t}=\frac{2\alpha_sC_F}{\pi}\sqrt{\frac{\om_c}{2k^+}}
\end{equation}
as shown in \eqref{bdmpsz-brick}.
Thus, the direct emission rate depends on the medium induced formation time only, whereas the interference emission rate depends on the medium induced formation time and the three other time-scales $\tvac$, $\tcoh$ and $\tint$ appearing in the exponential, either as real or imaginary phases. These three phases depends on $\thqq$. 

Now, from the formula \eqref{inter-fixedq}, the interference rate exactly cancels the direct rate for all $t\le L$ if the exponential is equal to one for $t\le L$, i.e if the three following conditions are satisfied: $\tvac\gg L$, $\tcoh\gg L$ and $\tint\gg L$. 
Fortunately, the conditions $\tcoh\gg L$ implies the other two. Indeed, the condition $\tcoh\ge L$ is equivalent to $\om_0(\thqq)\ge\om_c$ since:
\begin{equation}
 \frac{\om_c}{\om_0(\thqq)}=\Big(\frac{L}{\tcoh}\Big)^2
\end{equation}
As we consider only emissions such that $k^+\le \om_c$, we see from \eqref{time-scale-ordering} that the three time scales $\tcoh$, $\tvac$ and $\tint$ are ordered as follows $\tcoh\le \tint\le \tvac$. Hence, if $\tcoh\gg L$, one has also $\tvac\gg L$ and $\tint\gg L$.
The condition $\tcoh\gg L$ can be expressed more naturally as a condition on the dipole opening angle $\thqq$. Defining the critical angle $\th_c$ such that $\tcoh(\th_c)=L$, the decoherence time is much larger than the characteristic length of the medium if $\thqq\ll\theta_c$, with
\begin{equation}\tcboxmath{
 \th_c\equiv \frac{2}{\sqrt{\qhat L^3}}}
\end{equation}
To sum up, when $\thqq\ll \th_c$, there is no exponential suppression of interferences and the interference term cancels the direct terms: the medium-induced spectrum is strongly suppressed for a colour singlet dipole.
The case $\tcoh\gg L \Leftrightarrow \thqq\ll\th_c$ is simple then: for a colour singlet initial state, the medium-induced spectrum vanishes. For a colour octet initial state, one can show that the medium-induced spectrum is equal to the medium-induced spectrum of a gluon. This leads to the following general statement: when the opening angle of the dipole is smaller than $\th_c$, the dipole emits medium-induced gluons as a single colour charge depending on the initial colour state of the dipole \cite{CasalderreySolana:2011rz,MehtarTani:2012cy,Mehtar-Tani:2017ypq}.

Let us now focus on the other regime $L\gg \tcoh \Leftrightarrow \thqq\gg\th_c$. These inequalities are also equivalent to $\om_0(\thqq)\ll\om_c$. Consequently, the characteristic energy scale $\omega_0(\thqq)$ belongs well to the multiple soft scattering regime. One can trust formula \eqref{inter-fixedq} even when the medium has a finite length since the rate vanishes when $t>L>\tcoh$ due to the exponential decoherence suppression.
Combining all the terms together, the full medium-induced rate from a colour singlet antenna in the regime $\thqq\gg\th_c$ and $k^+\ll\om_c$ reads:
\begin{align}\label{full-rate-antenna}
 k^+\frac{\dif^2 N}{\dif k^+\dif t}&= \frac{4\alpha_sC_F}{\pi}\sqrt{\frac{\qhat}{4k^+}}\left\{1+\exp\Big(-\frac{1}{24}\qhat\thqq^2t^3-\frac{1}{8}\sqrt{\hat{q}k^+}t^2\thqq^2\Big)\right.\times\nonumber\\
&\left[\sin\Big(\frac{k^+t\theta_{q\bar{q}}^2}{4}+\frac{t^2\theta_{q\bar{q}}^2\sqrt{\hat{q}k^+}}{8}\Big)-\Big(\sqrt{\frac{4k^+}{\qhat}}\frac{1}{t}+1\Big)\cos\Big(\frac{k^+t\theta_{q\bar{q}}^2}{4}+\frac{t^2\theta_{q\bar{q}}^2\sqrt{\hat{q}k^+}}{8}\Big)\right]\nonumber\\
&\hspace{9cm}\left.+\sqrt{\frac{4k^+}{\qhat}}\frac{1}{t}\cos\Big(\frac{k^+t\theta_{q\bar{q}}^2}{4}\Big)\right\}
\end{align}

In the regime $L\gg\tcoh$, since $\om_0(\thqq)\ll\om_c$, one has to distinguish two cases: either $k^+\le \om_0(\thqq)$ or $\om_c\ge k^+ \ge \om_0(\thqq)$. In the case $k^+\le\om_0(\thqq)$, the interference rate vanishes for $t>\tcoh$ and the other time $\tint$ and $\tvac$ are irrelevant because of the ordering \eqref{time-scale-ordering}. In the limit $L\gg\tcoh$, after integrating the full rate between $0$ and $L$, 
the net result is an incoherent sum of the emission spectrum by the quark and the antiquark. If $L\gtrsim\tcoh$, one observes numerically a small suppression due to the interference rate integrated for $t\le \tcoh$.

\begin{figure}[t] 
  \centering
  \begin{subfigure}[t]{0.48\textwidth}
    \includegraphics[page=1,width=\textwidth]{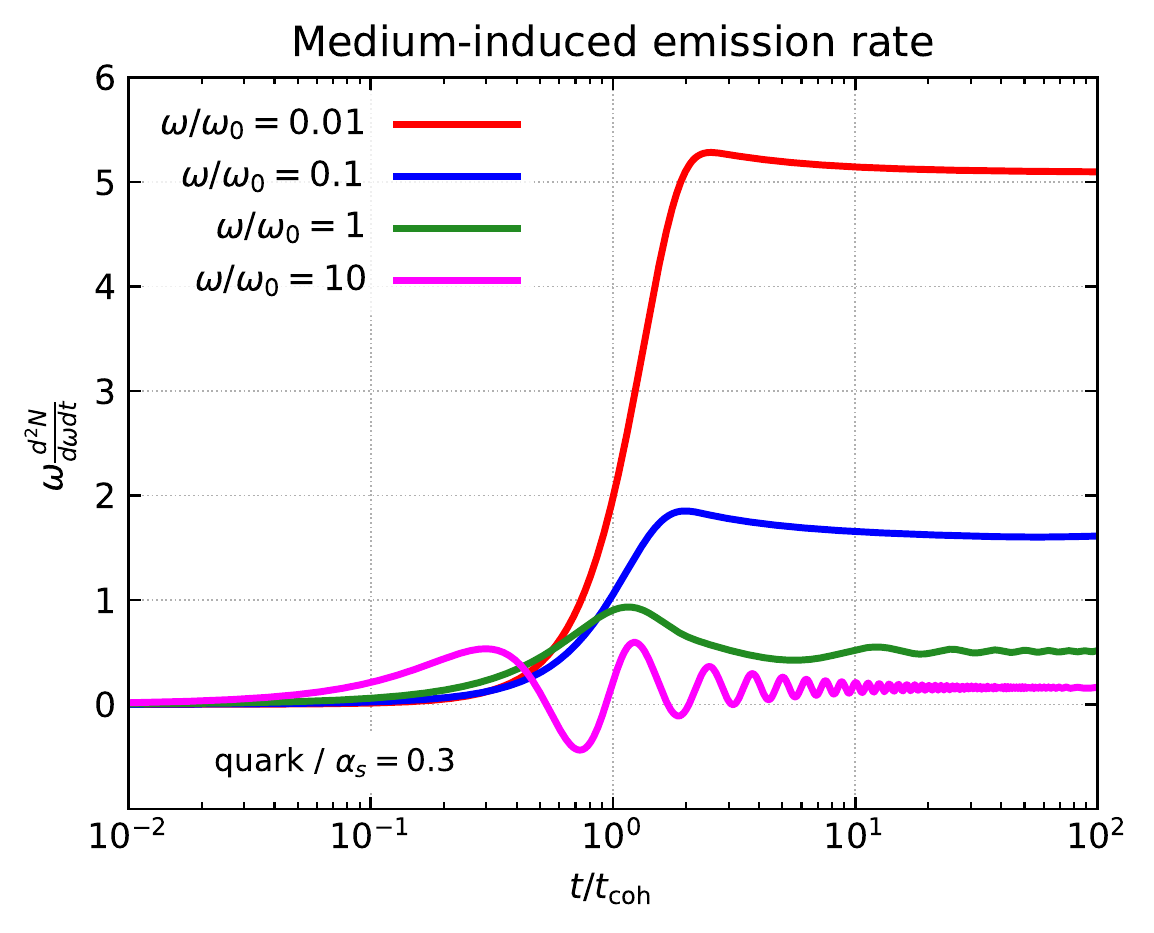}
    \caption{\small Medium induced emission rate from a $q\bar{q}$ antenna.}\label{fig:lpm-antemma-rate}
  \end{subfigure}
  \hfill
  \begin{subfigure}[t]{0.48\textwidth}
    \includegraphics[page=2,width=\textwidth]{./lpm-antenna.pdf}
    \caption{\small BDMPS-Z spectrum from a $q\bar{q}$ antenna.}\label{fig:lpm-antenna-spec}
  \end{subfigure}
  \caption{\small Interference effects in the medium-induced spectrum from a colour singlet antenna for a static medium, in the regime $\om\ll\om_c$. On the left plot, the medium-induced rate \eqref{full-rate-antenna} is represented as a function of the dimensionless time $t/\tcoh$, for several values of $\om\equiv k^+$. The large time limit of the rates is twice the constant BDMPS-Z rate for a single quark. On the right plot, the integrated rate (between $0$ and $L$) is represented as a function of $\om/\om_0(\thqq)$. The dotted lines correspond to the soft limit of the BDMPS-Z spectrum for a static medium multiplied by a $2$.}\label{Fig:lpm-antenna}
\end{figure}

The case $\om_c\ge k^+\ge\om_0(\thqq)$ is more delicate, because the interpretation of \eqref{full-rate-antenna} as a \textit{formation rate} is ambiguous. Indeed, from the ordering \eqref{time-scale-ordering}, the interference rate is controlled by $\tvac$ and thus becomes oscillatory. This is shown Fig.~\ref{fig:lpm-antemma-rate}. The exponential suppression of the interference is controlled by $\tint$. The interference spectrum obtained after integrating the interference rate between $0$ and $L$ gives the average of the oscillatory integral which is typically zero. Consequently, the full spectrum is again twice the spectrum of a single colour charge. Thus, even if the rate interpretation of \eqref{full-rate-antenna} when $\om_c\ge k^+\ge\om_0(\thqq)$ is not allowed, the integrated spectrum is still the incoherent sum of two BDMPS-Z spectra, as shown Fig.\ref{fig:lpm-antenna-spec} because the interference is a quantity proportional to $\tint\ll L$ suppressed with respect to the direct terms.

\subsection{Decoherence effects in the vacuum-like radiation pattern}
\label{sub:decoherence}

In the previous subsection, we have discussed the effect of the interferences on the medium-induced radiations only, as the  standard vacuum radiation pattern was subtracted off in order to have a convergent $k_\perp$ integral. 
The effect of the medium on the vacuum-like radiation pattern from a colour singlet antenna is the subject of this last subsection. We will comment the potential overlap between these two calculations in due course.

In the vacuum, we have seen in Section \ref{sec:vac-bench} that the radiation pattern of a colour singlet dipole presents the property of angular ordering: no radiations are allowed outside the opening angle of the dipole. This is a consequence of quantum interferences associated with the colour coherence of the dipole.
In the medium, the $q\bar{q}$ dipole factor $S_{q\bar{q}}$ washes out the colour coherence of the dipole
so we expect non trivial effects on the vacuum-like radiation pattern.

Calculating the full differential emission spectrum $\dif^3N/\dif k^+\dif^2k_\perp$ from a colour singlet antenna is complicated and the final result is not as simple as the differential emission spectrum from a single quark. However, one can easily understand the effect of the factor $S_{q\bar{q}}$ on the vacuum radiation pattern with a simple calculation. This is the subject of the first subsection, where we explain the effect of decoherence with a simple toy model. In the second subsection, we consider the same antenna as in Section \ref{sub:mie-antenna}, but we constrain the gluon to be vacuum-like by choosing a kinematic regime in which the formation time is either very \textit{short} or very \textit{long}.

\subsubsection{Shockwave limit of the emission cross-section from onium scattering}

Let us first consider again the scattering of the onium briefly discussed in Section \ref{subsub:onium-scatt}. However, the onium is now highly virtual, that is both the quark and the antiquark are created at light cone time $x_0^+=0$. This toy model enables to understand the main features of decoherence in the medium. Since the transverse size of the dipole is fixed to $|X_\perp|$, contrary to the antenna where the transverse size increases with time as $\thqq t$, the interference criterium is simply $|X_\perp|\lesssim |1/k_\perp|$ and does not depend on time.

In the vacuum, the full double differential cross-section, after integration over the azimuthal angle is:
\begin{equation}\label{vac-onium-shock}
 k^+\frac{\dif^2 N^{\vac}}{\dif k^+\dif k_\perp}=\frac{4\alpha_sC_F}{\pi}\frac{1}{k_\perp}\Big(1-\textrm{J}_0\big(X_\perp k_\perp\big)\Big)
\end{equation}
When $k_\perp\gg 1/X_\perp$, the spectrum behaves as an incoherent sum of two Bremsstrahlung spectra. On the contrary, when $k_\perp\ll 1/X_\perp$, the spectrum is suppressed and vanishes in the limit $k_\perp\rightarrow0$.

\begin{figure}[t] 
  \centering
  \begin{subfigure}[t]{0.48\textwidth}
    \includegraphics[page=2,width=\textwidth]{./vle-antenna.pdf}
    \caption{Gluon spectrum from an off-shell onium in the shockwave limit. The red curve is the spectrum in the absence of decoherence effect: one observes a strong suppression for $k_\perp\ll 1/X_\perp$. The blue curves show the numerical evaluation of \eqref{med-onium-shock} for several values of $Q_s$. For large values of $Q_s$ the interference pattern is destroyed, and small $k_\perp$ radiations are copiously emitted.}\label{fig:vle-onium}
  \end{subfigure}
  \hfill
  \begin{subfigure}[t]{0.48\textwidth}
    \includegraphics[page=1,width=\textwidth]{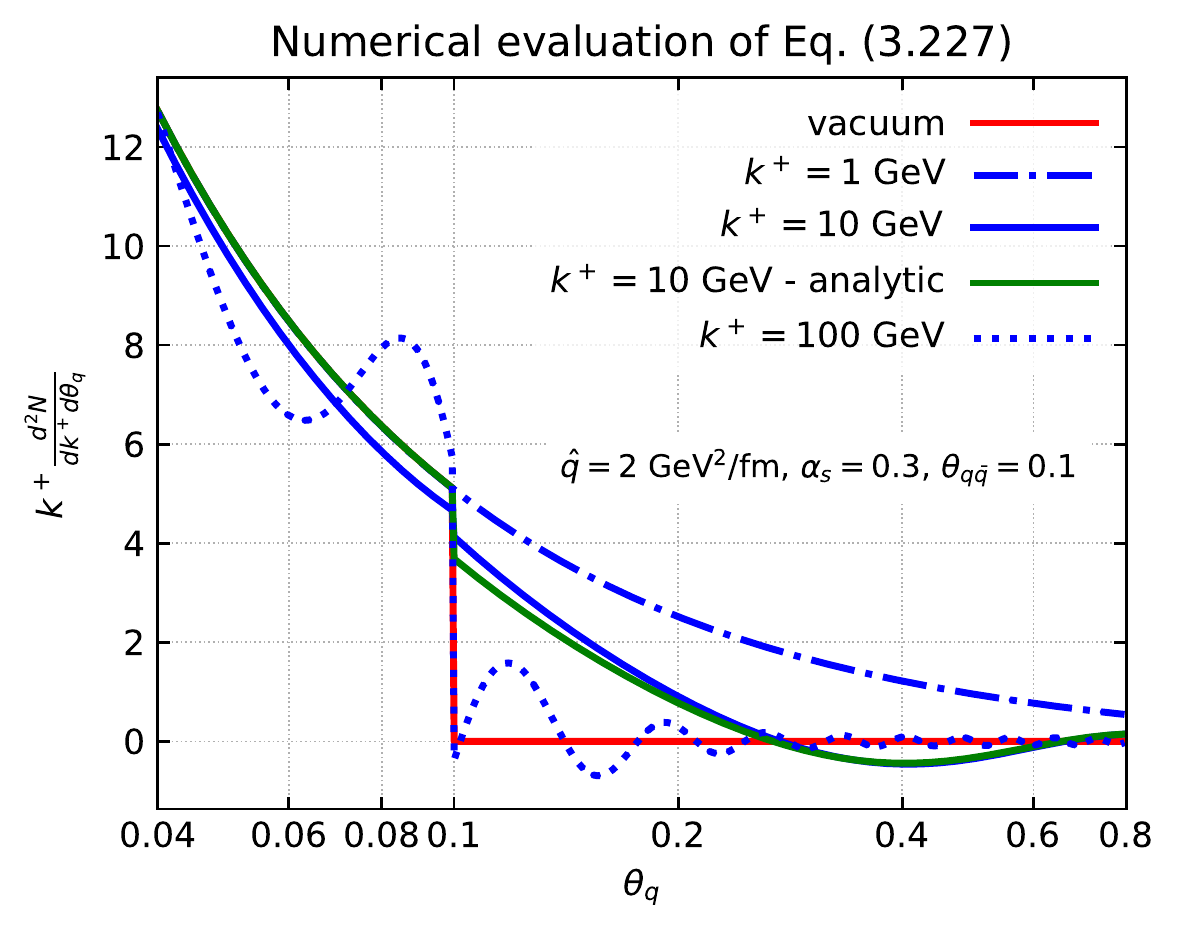}
    \caption{Numerical evaluation of the formula \eqref{vle-dec-numeric}. The green curve corresponds to the approximation \eqref{approx-dec}. The medium parameters are fixed whereas the kinematic of the emission is varied, showing that decoherence induces out-of cone radiations with respect to the vacuum (red curve). We have deliberately chosen $k^+$ not in the regime \eqref{vle-inmed} to highlight the effect. Actually, as shown in Chapter~\ref{chapter:DLApic}, out-of-cone radiations are always suppressed if \eqref{vle-inmed} is enforced (dotted curve).}\label{fig:vle-antenna-spec}
  \end{subfigure}
  \caption{\small Decoherence effects on colour-singlet dipoles with two different geometries: the ``onium'' geometry (left) and the antenna (right).}\label{Fig:vle-antenna}
\end{figure}

In the presence of the medium, the radiation pattern is modified. We are not interested in the medium-induced emission generated by the scattering. Consequently, we can take the shockwave limit of the full spectrum, that is $L\rightarrow 0$ with $Q_s^2=\qhat L$ fixed. The medium is a brick with fixed length $L$ and constant $\qhat$, starting at $x_0^+=0$. In this limit, only the out/out term with $y^+,\bar{y}^+\ge L\rightarrow0$ contributes to the spectrum for a $q\bar{q}$ pair created at $x_0^+=0$. The out/out direct terms are trivial and give one $1/k_\perp^2$ term for the quark and the antiquark. Let us calculate the out/out interference term:
\begin{align}
   k^+\frac{\dif^3N^{q\bar{q},o/o}}{\dif k^+\dif^2 k_\perp}&=\frac{-\alpha_s C_F}{(2\pi)^2 {k^+}^2}2\mathfrak{Re}\int_{L}^{\infty}\dif y^+\int_{L}^{y^+}\dif\bar{y}^+e^{-\varepsilon(y^+ +\bar{y}^+)}S_{q\bar{q}}(\bar{y}^+,-\infty)\\
   &\hspace{3cm}\times\int\dif^2u_\perp e^{ik_\perp u_\perp}\partial^i_{y_\perp=0_\perp}\partial^i_{\bar{y}_\perp=X_\perp}\mathcal{G}_0(y^+,y_\perp+u_\perp;\bar{y}^+,\bar{y}_\perp)\\
   &\underset{L\rightarrow0}=\frac{-\alpha_s C_F}{\pi^2}\exp\Big(-\frac{1}{4}Q_s^2X_\perp^2\Big)\frac{\cos\big(X_\perp k_\perp \big)}{k_\perp^2}
\end{align}
The exponential is entirely due to the factor $S_{q\bar{q}}$ associated with the decoherence of the $q\bar{q}$ pair.
The full off-shell onium spectrum in the shockwave limit reads, again after integration over the azimuthal angle:
\begin{tcolorbox}[ams equation]\label{med-onium-shock}
 k^+\frac{\dif^2 N}{\dif k^+\dif k_\perp}=\frac{4\alpha_sC_F}{\pi}\frac{1}{k_\perp}\Big(\Big(1-\textrm{J}_0\big(X_\perp k_\perp\big)\Big)+\Big(1-e^{-\frac{1}{4}Q_s^2X_\perp^2}\Big)\textrm{J}_0\big(X_\perp k_\perp\big)\Big)
\end{tcolorbox}
\noindent This expression is very simple. For small dipole, with $X_\perp\ll 1/Q_s$, the decoherence factor has no effect as long as $k_\perp>Q_s$ and the spectrum becomes identical to the vacuum spectrum \eqref{vac-onium-shock}. However, if the dipole is large enough, namely $X_\perp\gtrsim 1/Q_s$, the long wavelength radiations, with $k_\perp\ll 1/X_\perp$ are not forbidden anymore by the interferences. Indeed, the spectrum behaves as $(1-\exp(-Q_s^2X_\perp^2/4))/k_\perp$. The factor $\exp(-Q_s^2X_\perp^2/4)$ can be interpreted as the survival probability of the dipole in a colour singlet state, configuration which forbids long wavelength emissions. A numerical evaluation of \eqref{med-onium-shock} is shown Fig.~\ref{fig:vle-onium}.

\subsubsection{Decoherence in the antenna radiation pattern}

Let us now turn to a more physically relevant scenario: the emission of a vacuum-like gluon by a $q\bar{q}$ antenna inside a dense medium. On may wonder why the medium would play any role in the emission of a vacuum-like gluon. The situation is pretty much like for our off-shell onium: the decoherence of the pair may lead to a modification of the vacuum-like radiation pattern, destroying interferences and hence allowing extra radiations.

\paragraph{Vacuum-like emissions with very short formation time.} To understand more precisely the effect of $S_{q\bar{q}}$, we try the following exercise: we replace the effective propagators $\mathcal{K}_{qg}$ and $\mathcal{K}_{\bar{q}g}$ by the free vacuum propagator $\mathcal{G}_0^\dagger$ and we neglect the subsequent transverse momentum broadening. As we are interested in vacuum-like emissions (VLEs), these simplifications do not drastically change the physical picture of the vacuum-like antenna radiation pattern for emissions such that:
\begin{equation}\label{vle-inmed}
 \om_c\ll k^+\ll p^+,\bar{p}^+\,,\qquad Q_s\ll k_\perp \ll p_\perp,\bar{p}_\perp
\end{equation}
Note that that this limit is different from the soft limit considered in \cite{MehtarTani:2010ma,MehtarTani:2011tz}. We focus here on vacuum-like emissions with a very short formation time, hence occurring inside the medium. The formation time is so short that the formation process cannot be triggered by the multiple soft collisions, as it would take at least a time $t_{f,\med}$ for producing it in this way. We will see in Chapter \ref{chapter:DLApic} that the kinematic constraints \eqref{vle-inmed} are actually too strong: in-medium VLEs satisfy $k_\perp\gg(2\qhat k^+)^{1/4}$ instead of \eqref{vle-inmed}. We 
leave for further studies the question of whether  
the calculation below gives a good estimate of the in-medium VLE pattern also in this regime. As a last comment, we emphasize that this kinematic regime was not considered in our discussion of the medium-induced spectrum by an antenna since we assumed $k^+\ll \om_c$.

In this limit, the sum of the direct terms is an incoherent sum of two Bremsstrahlung spectra:
\begin{align}
 k^+\frac{\dif ^3N^{\textrm{direct}}}{\dif k^+\dif^2k_\perp}&=\frac{\alpha_s C_F}{\pi^2}\left(\frac{1}{(k_\perp-k^+u_\perp)^2}+\frac{1}{(k_\perp-k^+\bar{u}_\perp)^2}\right)\nonumber\\
 &=\frac{\alpha_s C_F}{2\pi^2k^{+}}\left(\frac{1}{k^\mu u_\mu}+\frac{1}{k^\mu \bar{u}_\mu}\right)
\end{align}
where the transverse momentum $k_\perp$ has been shifted to take into account the change of coordinate system due to the geometry of the antenna.
The $q\bar{q}$ interference term has the following structure:
\begin{align}
 k^+\frac{\dif^3 N^{q\bar{q}}}{\dif k^+\dif^2k_\perp}&=\frac{-\alpha_s C_F}{4\pi^2k^{+2}}(k_\perp-k^+u_\perp)(k_\perp-k^+\bar{u}_\perp)\nonumber\\
 &\hspace{2cm}\times 2\mathfrak{Re}\int_0^\infty\dif \bar{y}^+\int_{\bar{y}^+}^{\infty}\dif y^+e^{-\varepsilon(y^++\bar{y}^+)}S_{q\bar{q}}(\bar{y}^+)e^{ik^\mu u_\mu y^+-ik^\mu \bar{u}_{\mu}\bar{y}^+}\label{inter-dec}
\end{align}
with $2k^+ k^\mu u_\mu=(k_\perp-k^+u_\perp)^2$.
In the vacuum, i.e without the factor $S_{q\bar{q}}$, the $q\bar{q}$ and the $\bar{q}q$ terms combine to give the famous antenna radiation pattern discussed in Section \ref{sec:vac-bench}:
\begin{align}
 k^+\frac{\dif ^3N^{\textrm{vac}}}{\dif k^+\dif^2k_\perp}&=\frac{\alpha_s C_F}{\pi^2}\left[\frac{1}{(k_\perp-k^+u_\perp)^2}+\frac{1}{(k_\perp-k^+\bar{u}_\perp)^2}-2\frac{(k_\perp-k^+u_\perp)(k_\perp-k^+\bar{u}_\perp)}{(k_\perp-k^+u_\perp)^2(k_\perp-k^+\bar{u}_\perp)^2}\right]\nonumber\\
 &=\frac{\alpha_s C_F}{2\pi^2}\frac{p^\mu\bar{p}_\mu}{(p^\mu k_\mu)(\bar{p}^\mu k_\mu)}\label{antenna-pattern}
\end{align}
If the medium has a size $L$ and constant $\qhat$, the integral \eqref{inter-dec} is cut at $\bar{y}^+=L$. The piece with $\bar{y}^+\ge L$ gives the same contribution as \eqref{antenna-pattern} with an additional factor $S_{q\bar{q}}(L)$ \cite{MehtarTani:2010ma,MehtarTani:2011tz} corresponding to the survival probability of the dipole in a colour singlet state after a time $L$ and a phase $\cos(k^\mu(u_\mu-\bar{u}_\mu)L)$. The piece with $\bar{y}^+\le L$ cannot be calculated analytically:
\begin{align}
 k^+\frac{\dif^3 N^{q\bar{q}+\bar{q}q}}{\dif k^+\dif^2k_\perp}&=\frac{-2\alpha_s C_F}{4\pi^2k^{+2}}(k_\perp-k^+u_\perp)(k_\perp-k^+\bar{u}_\perp)\left[\frac{1}{(k^\mu u_\mu)(k^{\mu}\bar{u}_\mu)}\cos(k^\mu(u_\mu-\bar{u}_\mu)L)S_{q\bar{q}}(L)\right.\nonumber\\
 &\left.+\left(\frac{1}{k^\mu u_\mu}-\frac{1}{k^\mu \bar{u}_\mu}\right)\int_0^L\dif \bar{y}^+\sin\Big(k^\mu(\bar{u}_\mu-u_\mu)\bar{y}^+\Big)\exp\Big(-\frac{1}{6}u^\mu\bar{u}_\mu\qhat\bar{y}^{+3}\Big)\right]\label{inter-dec-final}
\end{align}
with $S_{q\bar{q}}(L)=\exp(-\frac{1}{6}u^\mu\bar{u}_\mu\qhat L^{3})\simeq\exp(-(\thqq/\th_c)^2/6)$.

With this expression, the discussion is pretty much the same as in the onium case. When the opening angle of the dipole $\thqq$ is small, i.e. $\thqq\ll \th_c$, the survival probability $S_{q\bar{q}}(L)\sim 1$ as well as the exponential inside the integral. Consequently, the two terms in \eqref{inter-dec-final} gives the standard vacuum interference spectrum so that the full spectrum is the vacuum one \eqref{antenna-pattern}. Large angle soft emissions are suppressed (angular ordering) because of the interferences.

However, if $\thqq\gg\th_c$ or equivalently $L\gg\tcoh$, the factor $S_{q\bar{q}}(L)$ vanishes. One can safely take the limit $L\rightarrow \infty$ for the integral in the second term of \eqref{inter-dec-final}. After the change of variable $k(\bar{u}-u)\bar{y}^+\rightarrow t$, one gets:
\begin{align}
 k^+\frac{\dif^3 N^{q\bar{q}+\bar{q}q}}{\dif k^+\dif^2k_\perp}&\simeq\frac{-2\alpha_s C_F}{4\pi^2k^{+2}}\frac{(k_\perp-k^+u_\perp)(k_\perp-k^+\bar{u}_\perp)}{(k^\mu u_\mu)(k^\mu \bar{u}_\mu)}\mathcal{I}_{q\bar{q}}(k^\mu)\label{inter-dec-final2}
\end{align}
and the integral $\mathcal{I}_{q\bar{q}}$ defined by:
\begin{equation}
 \mathcal{I}_{q\bar{q}}(k^\mu)=\int_0^\infty\dif t\,\sin(t)\exp\left(-\frac{1}{6}\frac{u^\mu\bar{u}_\mu}{(k^\mu(\bar{u}_\mu-u_\mu))^3}\qhat t^{3}\right)
\end{equation}
With this form where the standard vacuum interference spectrum factorizes, it is clear that the $\mathcal{I}_{q\bar{q}}$ factor plays the same role as the factor $\exp(-Q_s^2X_\perp^2/4)$ for the decoherence of the onium in the shockwave limit. It induces a mismatch between the direct terms and the interference term which allows for large angle radiations outside the opening angle $\thqq$. To see that, we write the full antenna radiation pattern in the limit $L\gg\tcoh$ as:
\begin{tcolorbox}[ams align]\label{antenna-Linf}
 k^+\frac{\dif ^3N}{\dif k^+\dif^2k_\perp}&\simeq\frac{\alpha_s C_F}{2\pi^2}\left(\frac{p^\mu\bar{p}_\mu}{(p^\mu k_\mu)(\bar{p}^\mu k_\mu)}+\big(1-\mathcal{I}_{q\bar{q}}(k^\mu)\big)\frac{k^\mu u_\mu+k^\mu\bar{u}_\mu-k^+u^\mu\bar{u}_\mu}{k^+(k^\mu u_\mu)(k^\mu \bar{u}_\mu)}\right)
\end{tcolorbox}

Now, we want to show that the second term is precisely the large angle contribution due to the rapid decoherence of the $q\bar{q}$ pair while propagating through the medium. There is one difference with respect to the onium case: the factor $1-\mathcal{I}_{q\bar{q}}(k^\mu)$ is \textit{dynamical} since it depends on the kinematic $k^\mu$ of the emission. To understand more precisely how, we estimate this factor neglecting the exponential inside the integral for $t\ge \tcoh$ and replacing it by one for $t\le \tcoh$. One finds:
\begin{align}
 1-\mathcal{I}_{q\bar{q}}(k^\mu)&\simeq 1-\int_0^{k\cdot(\bar{u}-u)\tcoh}\dif t\, \sin(t)\\
 &=\cos\big(k\cdot(\bar{u}-u)\tcoh\big)
 \end{align}
With this approximation, the antenna spectrum reads:
\begin{equation}
 k^+\frac{\dif^3 N}{\dif k^+\dif^2k_\perp}\simeq\frac{\alpha_s C_F}{2\pi^2{k^+}^2}\left[\frac{u^\mu\bar{u}_\mu}{(\tilde{k}^\mu u_\mu )(\tilde{k}^\mu\bar{u}_\mu )}+\cos\big(k\cdot(\bar{u}-u)\tcoh\big)\frac{\tilde{k}^\mu u_\mu+\tilde{k}^\mu\bar{u}_\mu-u^\mu\bar{u}_\mu}{(\tilde{k}^\mu u_\mu)(\tilde{k}^\mu\bar{u}_\mu)}\right]
\end{equation}
with $\tilde{k}=k/k^+$. This spectrum can be decomposed symmetrically into two components, each component having a collinear singularity when $k^\mu$ is aligned with the quark or antiquark direction:
\begin{equation}
 k^+\frac{\dif^3 N}{\dif k^+\dif^2k_\perp}=\left(k^+\frac{\dif^3 N^{k||q}}{\dif k^+\dif^2k_\perp}+k^+\frac{\dif^3 N^{k||\bar{q}}}{\dif k^+\dif^2k_\perp}\right)
\end{equation}
with 
\begin{equation}
 k^+\frac{\dif^3 N^{k||q}}{\dif k^+\dif^2k_\perp}=\frac{\alpha_s C_F}{2\pi^2{k^+}^2}\frac{1}{2}\left(\frac{u^\mu\bar{u}_\mu+\tilde{k}^\mu\bar{u}_\mu-\tilde{k}^\mu u_\mu}{(\tilde{k}^\mu u_\mu )(\tilde{k}^\mu\bar{u}_\mu )}+\cos\big(k^\mu(\bar{u}_\mu-u_\mu)\tcoh\big)\frac{\tilde{k}^\mu u_\mu+\tilde{k}^\mu\bar{u}_\mu-u^\mu\bar{u}_\mu}{(\tilde{k}^\mu u_\mu)(\tilde{k}^\mu\bar{u}_\mu)}\right)
\end{equation}
and similarly for the antiquark term, with $u\leftrightarrow\bar{u}$ exchanged. Now, we average $\dif^3N^{k||q}$ over the azimuthal angle $\phi$ around the quark direction. The following geometrical relations are useful:
\begin{align}
 2u^\mu\bar{u}_\mu&=1-\cos(\th_{q\bar{q}})\\
 2\tilde{k}^\mu u_\mu&=1-\cos(\th_{q})\\
 2\tilde{k}^\mu \bar{u}_\mu&=1-\cos(\th_{\bar{q}})=(1-\cos(\thqq)\cos(\th_q))-\sin(\thqq)\sin(\th_q)\cos(\phi)
 \end{align}
where $\th_{q/\bar{q}}$ is the emission angle with respect to the quark/antiquark direction.
After averaging over $\phi$, the first piece gives the standard angular ordered spectrum, whereas the second is more complicated:
\begin{align}\label{vle-dec-numeric}
 k^+\frac{\dif^2 N^{k||q}}{\dif k^+\dif\th_q}&=\frac{\alpha_s C_F}{\pi}\frac{\sin(\th_q)}{1-\cos(\th_q)}\Big(\Theta\big(\cos(\th_q)-\cos(\thqq)\big)\nonumber\\
 &\hspace{1cm}\left.+\textrm{J}_0(|x_2|)\cos(x_1)+\big(\cos(\thqq)-\cos(\th_q)\big)\int\frac{\dif\phi}{2\pi}\frac{\cos(x_1-x_2\cos(\phi))}{\lambda_1-\lambda_2\cos(\phi)}\right)
\end{align}
with $x_1=\frac{1}{2}k^+\tcoh\cos(\th_q)(1-\cos(\thqq))$, $x_2=\frac{1}{2}k^+\tcoh\sin(\thqq)\sin(\th_q)$, $\lambda_1=1-\cos(\thqq)\cos(\th_q)$ and $\lambda_2=\sin(\thqq)\sin(\th_q)$.
The remaining $\phi$ integral can be estimated neglecting the phase inside the numerator:
\begin{equation}\label{approx-dec}
 \int\frac{\dif\phi}{2\pi}\frac{\cos(x_1-x_2\cos(\phi))}{\lambda_1-\lambda_2\cos(\phi)}\simeq \frac{\textrm{J}_0(|x_2|)\cos(x_1)}{|\cos(\th_q)-\cos(\thqq)|}
\end{equation}
This approximation enables to put the antenna pattern in the following compact form:
\begin{align}\label{decoherence-vle-simple}
 k^+\frac{\dif^2 N^{k||q}}{\dif k^+\dif\th_q}&=\frac{\alpha_s C_F}{\pi}\frac{\sin(\th_q)}{1-\cos(\th_q)}\Big(\Theta\big(\cos(\th_q)-\cos(\thqq)\big)+\textrm{J}_0(|x_2|)\cos(x_1)\Theta\big(\cos(\thqq)-\cos(\th_q)\big)\Big)
\end{align}
With respect to \eqref{antenna-Linf}, the latter formula is even more transparent because the geometry of the emission is explicit: the term proportional to $\Theta(\cos(\thqq)-\cos(\th_q))$ corresponds to the radiation outside the cone, which would be forbidden in the vacuum. The amount of radiation with $\th_q\ge \thqq$ is controlled by the factor $\textrm{J}_0(|x_2|)\cos(x_1)$. In analogy with the factor $1-\exp(-k_\perp^2/Q_s^2)$ for the onium toy model, we call it the decoherence factor $\Delta_{>\th_{q\bar{q}}}$. In the small angle approximation:
\begin{align}\label{decoherence-factor1}
 \Delta_{>\th_{q\bar{q}}}&\simeq\textrm{J}_0\left(\frac{1}{2}k^+\tcoh \th_q\thqq\right)\cos\left(\frac{1}{4}k^+\tcoh\thqq^2\right)\\
                       &=\textrm{J}_0\left(\frac{\tcoh}{t_f}\frac{\thqq}{\th_q}\right)\cos\left(\frac{1}{2}\frac{\tcoh}{\tvac}\right)
\end{align}

Out-of-cone radiations scale with $\tcoh\thqq/(t_f\th_q)$ in the short formation time limit, for $\th_{q\bar{q}}\gg\th_c$. Note the somehow unexpected $\thqq/\th_q$ geometric factor.
In Chapter \ref{chapter:DLApic}-Section \ref{sec:DLresum} we will generalize this calculation to multiple emissions. We will show that the kinematic constraints \eqref{vle-inmed} --- and also the less strong condition $k_\perp\ge(2\qhat k^+)^{1/4}$ --- actually impose that the factor $\Delta_{>\th_{q\bar{q}}}$ vanishes, so that in-medium VLEs with short formation times remain angular ordered!

\paragraph{Vacuum-like emissions in the long formation time limit $t_f\gg L$.} This is the limit considered in \cite{MehtarTani:2010ma,MehtarTani:2011tz}. For these emissions as well, one can approximate the effective propagators $\mathcal{K}_{qg}$ and $\mathcal{K}_{\bar{q}g}$ by the free vacuum propagator $\mathcal{G}_0^\dagger$ and remove the broadening factor $S_{gg}$ because these emissions come from a time integration domain where there is no medium any more. More precisely, the out/out interference term where $y^+\ge L$ and $\bar{y}^+\ge L$ dominates in the infrared limit, and is precisely given by the first term of formula \eqref{inter-dec-final}. Taking the limit $k^\mu(u^\mu-\bar{u}^\mu)L\rightarrow0$ in this out/out term, corresponding to the regime $k^+\th_q^2L$, $k^+\th_{\bar{q}}^2L\ll1$ i.e. $t_f\gg L$, one gets the full spectrum in this limit:
\begin{tcolorbox}[ams align]\label{decoherence-softlimit}
  k^+\frac{\dif ^3N}{\dif k^+\dif^2k_\perp}&\simeq\frac{\alpha_s C_F}{2\pi^2}\left(\frac{p^\mu\bar{p}_\mu}{(p^\mu k_\mu)(\bar{p}^\mu k_\mu)}+\big(1-S_{q\bar{q}}(L)\big)\frac{k^\mu u_\mu+k^\mu\bar{u}_\mu-k^+u^\mu\bar{u}_\mu}{k^+(k^\mu u_\mu)(k^\mu \bar{u}_\mu)}\right)
\end{tcolorbox}
\noindent leading to the following spectrum, once averaged over the azimuthal angle around the quark direction for instance:
\begin{align}
 k^+\frac{\dif^2 N^{k||q}}{\dif k^+\dif\th_q}&=\frac{\alpha_s C_F}{\pi}\frac{\sin(\th_q)}{1-\cos(\th_q)}\Big(\Theta\big(\cos(\th_q)-\cos(\thqq)\big)+\big(1-S_{q\bar{q}}(L)\big)\Theta\big(\cos(\thqq)-\cos(\th_q)\big)\Big)
\end{align}
This time, the decoherence factor $\Delta_{>\th_{q\bar{q}}}=1-S_{q\bar{q}}(L)$ is not \textit{dynamical} (as in the onium toy model), in the sense that it does not depend on the kinematic of the emission in the soft limit. For large antenna $\thqq\gg\th_c$, $1-S_{q\bar{q}}(L)\simeq 1$ whereas for small dipole $1-S_{q\bar{q}}(L)\simeq 0$. The physical interpretation is the following: after propagation through the medium over a distance $L$, the antenna does not remain in a colour singlet state with a probability $1-S_{q\bar{q}}(L)$. Thus, a gluon with formation time $t_f\gg L$ can be emitted at any angle, with a probability $1-S_{q\bar{q}}(L)$ (``anti angular-ordering'' effect \cite{MehtarTani:2010ma,MehtarTani:2011tz}). To conclude, we emphasize that these emissions \textit{are} included in the calculation of Section \ref{sub:mie-antenna} since the subtraction of the limit $\qhat\rightarrow0$ removes only the angular-ordered vacuum-like pattern. The resummation of such emissions is considered in Chapter \ref{chapter:DLApic}-Section \ref{sec:DLresum}.


%% file: chapter3.tex
\chapter{Jets in vacuum and jets from medium-induced emissions}
\chaptermark{Jets}
\label{chapter:jet}

With respect to the previous chapter, where we have computed cross-sections for \textit{one} emission, this chapter deals with multiple emission cross-sections. At first sight, one may wonder why such cross sections are important at all, since multi-particle production comes with sub-leading powers of $\alpha_s$, which is supposed to be small. This reasoning is true to some extent. In certain regions of the phase space for emissions, that we shall specify in this chapter, the smallness of $\alpha_s$ may be compensated by a large dimensionless parameter so that the standard perturbation theory breaks down. Consequently, being able to estimate such cross-sections is crucial to improve the precision of our calculations. 

How is this related to jets? The concept of jets has originally been developed to give a precise meaning to non fully inclusive cross-sections in perturbative chromodynamics, without relying on the concept of \textit{hadrons} which are not under control in this regime. Once a well-suited \textit{jet definition} is provided, a theorist can calculate in pQCD any jet cross-section at a given order in $\alpha_s$. 
Jets are object with a rich ``internal structure'', which can be defined either in terms of the hadronic content of the jets, or even better, in terms of subjets. There is now an increasing interest in such observables related to the internal structure of jets. As we shall see, once one knows how to calculate multiple emission cross-sections, it is straightforward to calculate jet substructure observables.

The first section of this chapter is a textbook recap on jets in the ``vacuum'', that is in electron-positron annihilation and proton-proton collisions. The second section makes the link between multi-particle cross-sections from an off-shell parton in the soft and collinear region of the phase space and jet substructure calculations in the vacuum. The third section is dedicated to the multiple branching regime for medium-induced radiations from an incoming on-shell parton. The emphasis is put on the mathematical similarities between these two distinct regimes. 

\section{Jets in $e^+e^-$ annihilation and $pp$ collisions: generalities}
\sectionmark{Jets in the vacuum: generalities}
\label{sec:jet-general}

Perturbative quantum chromodynamics has quarks and gluons as elementary degrees of freedom. However, due to confinement, these degrees of freedom are never directly measured in experiments as they do not constitute the asymptotic states of the theory. Instead of dealing with quarks and gluons to calculate physical observables such as cross-sections, a way to circumvent this problem is to deal with jets.

\subsection{IRC safety and jet definitions}

A standard calculation and experimental test of pQCD is the \textit{inclusive} $e^+e^-$ annihilation  into hadrons cross-section and its ratio with respect to $\mu^+\mu^-$ production. For this fully inclusive observable, there is a well defined and reliable $\alpha_s$ expansion of the cross-section written in terms of Feynman diagrams with \textit{partonic} final states, even if the experimental final states are hadrons. This is a consequence of the Kinoshita-Lee-Nauenberg theorem \cite{Kinoshita:1962ur,Lee:1964is} which guarantees the cancellation of infrared divergences for inclusive observables over QCD degenerate states and thus, a reduced sensitivity to hadronisation corrections.

If one wants to go one step further and calculate more exclusive processes in pQCD, it is necessary to build observables which are protected by the Kinoshita-Lee-Nauenberg theorem. 
Such observables are generically called infrared and collinear (IRC) safe. The precise definition is the following: let us call $V_{n}(k_1,..,k_n)$ the functional constraints on the $n$-particle phase space ($k_i$ is the 4-momentum of a particle) that define the value of this observable at a given order (related to $n$) in $\alpha_s$. For the cross-section $\sigma_V$ associated with this observable to be well defined, i.e.\ without infrared singularity, to all orders in perturbation theory, $V$ must satisfy:
\begin{enumerate}
 \item $\forall (i,j)$, $V_{n+1}(k_1,..,k_i,..,k_j,..,k_{n+1})\longrightarrow  V_{n}(k_1,..,k_i+k_j,..,k_{n+1})$ as $k_i$ and $k_j$ become collinear, in the sense of \eqref{sudakov-dec}.
 \item $\forall i$, $ V_{n+1}(k_1,..,k_i,..,k_{n+1})\longrightarrow  V_{n}(k_1,..,k_{i-1},k_{i+1},..,k_n)$ as $k_i^\mu\rightarrow0$.
\end{enumerate}

\paragraph{Sterman-Weinberg jets.} Historically, the first IRC safe exclusive observable calculated in pQCD is the Sterman-Weinberg dijet cross-section in $e^+e^-$ annihilation \cite{Sterman:1977wj}. This dijet cross-section is defined as the cross-section for events such that all but a fraction $\xi\sqrt{s}$, $\xi\ll 1$ of the initial energy $\sqrt{s}$ is deposited within two cones of opening angles $R$. This cross-section is fully inclusive with respect to soft and collinear gluon emissions. The famous result for the differential cross-section per solid angle $\dif^2\Omega=\sin(\th)\dif\phi\dif\th$ in the laboratory frame, obtained by Sterman and Weinberg, is
\begin{equation}\label{Sterman-Weinberg-jets}
 \frac{\dif \sigma_{\rm2-jet}}{\dif^2 \Omega}=\frac{\dif \sigma_{0}}{\dif^2 \Omega}\left(1-\frac{\alpha_sC_F}{\pi}\Big(4\log(R)\log(\xi)+3\log(R)+\frac{\pi^2}{3}-\frac{5}{2}+\mathcal{O}(\xi,R)\Big)+\mathcal{O}(\alpha^2_s)\right)
\end{equation}
to first order in $\alpha_s$, with $\alpha_s$ evaluated at $\sqrt{s}$. ($\dif\sigma_0$ is the Born level cross-section.)
Higher order corrections $\mathcal{O}(\alpha_s^2)$ have been calculated \cite{Kramer:1986sg}. This observable is also the first \textit{jet definition}. A jet definition is a systematic way of organizing the final state so that any exclusive \textit{jet} cross-section is always IRC safe.

\paragraph{Modern jet algorithms.} Nowadays, more practical jet definitions are used in the high energy community (see \cite{Salam:2009jx} for a modern review on jets). They have the advantage of being easily implemented in experiments and more convenient for higher order theoretical predictions. The most used are jet definitions from a sequential recombination algorithm, such as the $k_t$, Cambridge/Aachen and anti-$k_t$ algorithms. The basic idea of these recombination algorithms is to cluster final state particles together iteratively, according to a given rule, until no particle remains. The final clusters are called jets.

A sequential recombination algorithm makes use of a distance measure between final state particles and a recombination rule when merging particle together. The distance measures are a little bit different in $e^+e^-$ collisions and in $pp$ or AA collisions, essentially because one does not know a priori the total energy of the collision of the two partons inside the protons. We focus here on the more relevant $pp$ or AA case, and we provide the definition of the ``generalized $k_t$ algorithm'', based on the distance $d_{ij}$ between two particles and a beam distance $d_{iB}$:
\begin{align}
 d_{ij}&=\textrm{min}(p_{T,i}^{2p},p_{T,j}^{2p})\Delta R_{ij}^2\,,\qquad \Delta R_{ij}^2=\Delta \eta_{ij}^2+\Delta \phi_{ij}^2\\
 d_{iB}&=p_{T,i}^{2p} R
\end{align}
where $\Delta R_{ij}$ is the distance between $i$ and $j$ in the rapidity-azimuth plane and $(p,R)$ are free parameters.
Starting from a list of final state particles --- called \textit{pseudo-jets} in this context --- the algorithm then goes as follows, 
\begin{enumerate}
 \item Iteratively find the smallest distance among all the $d_{ij}$ and $d_{iB}$:
 \begin{itemize}
  \item If the smallest distance is of $d_{ij}$ kind then pseudo-jets $i$ and $j$ are removed from the list and recombined using the recombination scheme into a new pseudo-jet $k$, which becomes part of the final state pseudo-jets.
 \item If the smallest distance is a beam distance , pseudo-jet $i$ is called a jet and removed from the list.
 \end{itemize}
  \item Go back to the previous step until all the pseudo-jets in the list have been exhausted.
\end{enumerate}
The standard recombination scheme (``E-scheme'') consists in simply adding the $4$-momenta of particles $i$ and $j$ to get the $4$-momentum of $k$.

For $p=0$, this algorithm is called Cambridge/Aachen \cite{Dokshitzer:1997in,Wobisch:1998wt}. The declustering sequence is ordered in angles. For $p=1$, it is called the $k_t$ algorithm \cite{Catani:1993hr} and the declustering sequence is ordered in $k_t$. For $p=-1$, this algorithm is known as the anti-$k_t$ algorithm \cite{Cacciari:2008gp}. One can check that any $n$-jet cross-section in the sense of this jet definition is an IRC safe observable (provided that a lower cut-off is imposed on the transverse momenta of jets).

\subsection{Resummation of large logarithms and matching}

Even for IRC safe cross-sections with a well defined expansion in powers of $\alpha_s$, each order of the perturbation may go with a large dimensionless parameter of the form $\log(Q/Q_0)$ where $Q$ is a hard scale and $Q_0$ a soft scale (often introduced to make the observable IRC safe). Such large logarithms, remnants of the infrared singularity of the theory, may spoil the convergence of the perturbation series in the region of the phase space where $Q\gg Q_0$. Formula \eqref{Sterman-Weinberg-jets} exhibits this kind of behaviour: for very small jet radii $R$, $|\alpha_s\log(R)|\gg 1$ so that a fixed order calculation is not enough to get an accurate answer. 

Under these circumstances, an all-order \textit{resummation} is required to take into account all the terms in the full perturbation series which are enhanced by large logarithms (for jet cross-sections with small radii, this program has been carried out in \cite{Dasgupta:2014yra}). In the next section, we will obtain resummed results for several jet substructure observables.

It is enlightening to understand more precisely the general structure of a resummed result. Let us first define the cumulative cross-section $\Sigma_V$ for the observable defined by the $V_{n\in\mathbb{N}}$ constraints on the $n$-body phase space:
\begin{equation}
 \Sigma_V(v)=\int_0^v\dif v'\frac{1}{\sigma_{V,0}}\frac{\dif\sigma_V}{\dif v'}
\end{equation}
where $v$ is the value of the observable (without dimension for this discussion) and $\sigma_{V,0}$ is the Born level total cross-section for $V$.
For a certain subclass of IRC safe observables (see \cite{Banfi:2003je,Banfi:2004yd} for more details), and in the region of the phase space where $L\equiv |\log(v)|\gg1$, the resummation \textit{exponentiates} with the following form:
\begin{equation}\label{resum-exponentiation}
 \Sigma_V(x)\simeq\exp\Big(L g_1(\alpha_sL)+g_2(\alpha_sL)+\alpha_sg_3(\alpha_sL)+\mathcal{O}\big(\alpha_s^{n+2}L^n\big)\Big)
\end{equation}
where $\alpha_s$ is evaluated at the hard scale $Q$ of the problem.
The functions $g_i$ are analytic functions that resum to all orders powers of $\alpha_sL\sim1$. From this general structure, one defines the logarithmic accuracy as follows: at leading logarithmic (LL) accuracy, one keeps only the $Lg_1$ term, at next-to-leading logarithmic (NLL) accuracy, one keeps the $Lg_1+g_2$ terms and so on. In this thesis, we define the double logarithmic accuracy as the leading logarithmic approximation where the strong coupling is fixed. This corresponds to the first order expansion of $g_1(x)\simeq g'_1(0)x$, so that:
\begin{equation}
 \Sigma_V(x)=\sum\limits_{n=0}^{\infty} \frac{(g'_1(0))^n}{n!}(\alpha_sL^2)^n
\end{equation}
where the double-logarithmic all order resummation is explicit. The double logarithmic approximation is physically insightful because the resulting formulas are often very simple.

\paragraph{Matching with fixed order calculations.} We emphasize that such a resummation is required only in a region of the phase space where logarithms are large. To obtain an accurate answer all over the phase space, resummed and fixed order results must be combined according to a matching scheme. Such a matching scheme must avoid double counting between terms included in the all order result and terms in the fixed order. 

\section{Jet substructure calculation}
\label{sec:jet-sub}

The previous discussion was rather abstract in order to keep things as general as possible. In this section, we would like first to make concrete this resummation for the special case of jet substructure observables. The ``coherent branching algorithm'' defined in the first subsection provides an intuitive way to resum multiple emissions thanks to an effective classical branching process. The second goal of this section is to give some vacuum benchmark results for the second part of this thesis where we analyse these observables in heavy-ion collisions.

\subsection{The coherent branching algorithm}
\label{sec:mlla}

The coherent branching algorithm enables to calculate multi-partons QCD amplitude \textit{squared} in the phase space where such emissions are enhanced by large collinear/soft logarithms, taking into account the coherence property of soft gluon emissions \cite{Konishi:1979cb,Kalinowski:1980ju,Amati:1980ch,Catani:1990rr}. Its precision depends on the process and the observable under consideration. For example, in $e^+e^-$ annihilation, it is accurate at NLL  in the relevant phase space region where the thrust or the jet mass distributions are enhanced by large logs \cite{Catani:1992ua}.
Yet, it is generally accurate at LL, and some NLL corrections are included as well, but not all of them: single logarithms and non-global logarithms from multiple soft emissions at commensurate angles are missing for instance.

The algorithm relies on the following results, valid at least to LL accuracy: 
\begin{enumerate}
 \item The probability distribution for multi-particle production from a highly virtual parton in the soft and collinear regime can be generated by a classical Markovian branching process.
 \item The evolution variable for this process is the angle of emission with respect to the parent particle, in order to account for the strict angular ordering of soft gluon emissions.
 \item The rate of the branching process is given by the leading order $\mathcal{O}(\alpha_s)$ cross-section (differential in the transverse angle) for producing \textit{one} other parton from a virtual parton, in the collinear regime. This cross-section is averaged over the azimuthal angle of the produced particle:
 \begin{equation}\label{bremrate}
 \rmd^2 \mathcal{P}^{bc}_{a}(z,\theta)= \frac{\alpha_s(k_\perp)}{\pi}\Phi_a^{bc}(z)\dif z \frac{\dif \th}{\th}
 \end{equation}
with $\Phi_a^{bc}(z)$ the DGLAP splitting functions given in Appendix~\ref{app:A}, $z$ the splitting fraction of parton $b$, and $k_\perp=z(1-z)p_T\th$.\footnote{We assume a purely transverse initial parton and work in the small angle approximation.}
\end{enumerate}
To be useful in realistic applications, this algorithm has to be combined with fixed order calculations for producing the initial partons with transverse momentum $p_T$. Hence it is assumed implicitly the factorization between the exclusive hard cross-section and the decay of the produced partons into soft and/or collinear emissions. Note that because of the azimuthal average, informations about the direction of the produced particles in the transverse plane with respect to the initial parton 3-momentum are lost.

Let us define the generating functional for a jet initiated by a parton $i=q,\bar{q},g$, with transverse momentum $p_T$ and time-like virtuality $t=z_0(1-z_0)p^2_T\th^2$. $z_0$ and $\th$ are the energy fraction and opening angle of the first splitting. Yet, the generating functional depends only upon $\th$. 
\begin{equation}\label{Zjet}
 Z_i(p_T,\th\mid u(k))=\sum_{n=0}^{\infty}\int \dif^3k_1...\dif^3k_n\, u(k_1)...u(k_n)P_i^{(n)}(k_1,...,k_n)
\end{equation}
where $P_i^{(n)}(k_1,...,k_n)$ is the probability density for exclusive production of particles with 3-momenta $k_1$,...,$k_n$ and $u(k)$ is an auxiliary profile function. Probability conservation ensures that $Z_i(p_T,\th,\mid u(k)=1)=1$.
The complete knowledge of the generating functional enables to calculate any multi-parton inclusive cross-section. For example, the probability distribution $D_i(k)$ for measuring at least one parton with 3-momentum $k$ in the final state is given by:
\begin{equation}\label{Zjet-ex}
 D_i(k)=\frac{\delta Z_i(p_T,\th\mid u(k))}{\delta u(k)}\Big|_{u=1}
\end{equation}

The Markovian property of the branching process leads to the following master equation in backward form, known as Modified Leading Logarithm Approximation (MLLA) equation \cite{Khoze:1996dn,Dokshitzer:1991wu}:
\begin{tcolorbox}[ams equation]\label{Z-mlla}
 \frac{\partial Z_i(p_T,\th\mid u)}{\partial \log(\th)}=\frac{1}{2!}\sum_{(a,b)}\int_0^1\dif z\, \Phi_i^{ab}(z)\frac{\alpha_s(k_\perp^2)}{\pi}\Big(Z_a(z p_T,\th\mid u)Z_b((1-z)p_T,\th\mid u)-Z_i(p_T,\th\mid u)\Big)
\end{tcolorbox}
\noindent where the sum runs over all \textit{distinct} pairs of partons $(a,b)$.
This master equation has a simple probabilistic interpretation \cite{CVITANOVIC1980429}. The first term is a gain term associated with the splitting of a $i$-jet into two subjets with splitting fraction $z$ and $1-z$. The second term is a loss term associated with all possible splittings of the initial jet $i$. The factor $1/2!$ is a symmetry factor to avoid a double counting of decays.
The logarithmic derivative with respect to $\th$ enforces the angular ordering. In order to regulate the collinear divergence as $\th$ goes to 0, and to avoid the Landau pole in the running coupling, one puts by hand a transverse momentum cut-off $k_T>\ktmin$ in this equation. Then, the initial condition for this differential equation is
\begin{equation}\label{Z-initial}
Z_i(p_T,\th\rightarrow 0\mid u)=u(p_T)
\end{equation}
in agreement with the fact that on-shell partons with $t=0$ cannot radiate.

\subsection{Jet fragmentation function}
\label{sub:frag}

The jet fragmentation function $\mathcal{D}(x)$ is defined by the number $\dif N$ of jets having one constituent with transverse momentum fraction $x=p_{T}/\ptjet$ between $x$ and $x+\dif x$:
\begin{equation}\label{def-frag}
 \mathcal{D}(x)\equiv\frac{1}{\Njet}\frac{\dif N}{\dif x}
\end{equation}
normalized by the total number $\Njet$ of selected jets. This observable is not IRC safe but it remains nonetheless a relatively simple observable. The price to pay is a strong sensitivity of any calculation in pQCD with respect to the infrared cut-offs involved in the calculation. Also, even if $\mathcal{D}(x)$ is itself IRC unsafe, its evolution with the hard scale (i.e. the $p_T$ of the jet) can be obtained from pQCD. 
 
In this section, we investigate the predictions of the coherent branching algorithm for this observable. We will ignore subtleties related to the fixed order calculation and matching with the resummed result. Hence, we focus on the following problem: starting with an off-shell parton with transverse momentum $p_T$ decaying inside a cone of opening angle $\th$, what is the inclusive probability $D_i(x| p_T,\th)$ of measuring a parton with transverse momentum fraction $x$ in the final state? 

This quantity is easily obtained from the generating functional:
\begin{equation}\label{defD}
 D_i(x|p_T,\th)\equiv\int \dif^3k\, \delta(k^0-xp_T)\frac{\delta Z_i(p_T,\th| u(k))}{\delta u(k)}\Big|_{ u(k)=1}
\end{equation}

From \eqref{Z-mlla}, it is straightforward to write the corresponding evolution equation for $D_i(x,Q=p_T\th)$ (the sum over pairs $(a,b)$ is implicit):
\begin{equation}\label{Dvac-mlla}
 \frac{\partial D_i(x,Q)}{\partial \log(Q)}= \int_{0}^1 \dif z\,\frac{\alpha_s(k_\perp^2)}{2\pi}\Phi_i^{ab}(z)\left[\frac{1}{z}D_a\Big(\frac{x}{z},zQ\Big)+\frac{1}{1-z}D_b\Big(\frac{x}{1-z},(1-z)Q\Big)-D_i(x,Q)\right]
\end{equation}
with initial condition $D_i(x,Q=0)=\delta(1-x)$ --- recall that the condition $k_\perp>\ktmin$ is enforced. Once the solution of this equation is found, it should be evaluated for $Q\simeq p_T R$ where $R$ is the opening angle of the jet.

\subsubsection{Solution for $x\sim 1$ with NLL corrections}
\label{subsub:frag-x=1}

In this subsection, we look for the asymptotic form of $D_i(x,Q=p_T R)$ when $x$ is close to 1. In this regime, it is more convenient to consider the {\it cumulative} fragmentation distribution, defined as
\begin{equation}\label{def-sigma}
 \Sigma_i(x,p_TR)\equiv \int_x^1 \dif x' D_i(x',p_TR)
\end{equation} 
When $x\sim 1$, the perturbative expansion of the cumulative fragmentation distribution receives contributions enhanced by two types of logarithms: the collinear logarithm $L_0\equiv\log(p_{T0}R/\ktmin)$ generated by integrating over emission angles within the range $\ktmin/p_{T0} <\theta<R$  and the soft logarithm $L\equiv\log\frac{1}{1-x}$ generated by integrating over soft
gluon emissions with energy fractions $z$ within the range $1-x<z<1$. One has $L_0\ge L$, since all emissions must obey $z\theta p_{T0}>\ktmin$ for any $z\ge 1-x$ and any $\theta \le R$. Taking the value of the running coupling at the largest scale as the small parameter, $\alpha_{0}\equiv\as(p_{T}R)\ll1$, the perturbative series can be organised as follows:
\begin{equation}
\label{log-expansion}
 \log(\Sigma_i(x\mid p_{T}))=Lg_{1,i}(\alpha_{0}L,\alpha_{0}L_0)+g_{2,i}(\alpha_{0}L,\alpha_{0}L_0)+\mathcal{O}(\alpha_{0}^{n+1}\log^n)
\end{equation}
where $L g_{1,i}$ and $g_{2,i}$ resum respectively all the leading-log (LL) terms $\alpha_{0}^n\log^{n+1}$ and the next-to-leading-log (NLL) terms $\alpha_{0}^n\log^{n}$ with $n\ge1$, where $\log$ means either $L$, or $L_0$.

Now, let us sketch the strategy to find the functions $g_1$ and $g_2$. To the accuracy of interest, one can neglect the quark/gluon mixing terms. The MLLA equation for the quark fragmentation function reduces to
\begin{equation}\label{Dlargex-MLLA}
  Q\frac{\partial D_i(x,Q)}{\partial Q}=\int_{0}^{1}\dif z\,\frac{\alpha_s(k_\perp)}{\pi}\Phi_i^{ig}(z)\Theta(k_\perp-\ktmin)\left[\frac{1}{z}D_i\Big(\frac{x}{z},zQ\Big)-D_i(x,Q)\right]
\end{equation}
The standard way to deal with Eq.~\eqref{Dlargex-MLLA} is to go to Mellin space $D_i(x,Q)\rightarrow \tilde{D}_q(j,Q)$ where the integro-differential equation is linear-differential. From the properties of the Mellin transform, $x$ close to 1 corresponds to $j\rightarrow\infty$. 
More precisely, $\log(j)\sim-\log(1-x)$, so we keep all terms of the form $\alpha_0^n\log(j)^n\sim 1$ in the exact solution. The details of the calculation are given in Appendix \ref{app:NLL}.

The results for the resumming functions $g_1$ and $g_2$ are
\begin{tcolorbox}[ams align]\label{g12}
  g_{1,i}(u,v)&=\frac{C_i}{\pi \beta_0}\left[1-\log\Big(\frac{1-2\beta_0u}{1-2\beta_0v}\Big)+\frac{\log(1-2\beta_0u)}{2\beta_0u}\right]\\
  g_{2,i}(u,v)&=\gamma_E\frac{\partial u g_{1,i}}{\partial u}-\log\bigg[\Gamma\Big(1-\frac{\partial u g_{1,i}}{\partial u}\Big)\bigg]+\frac{C_iB_i}{\pi \beta_0}\log(1-2\beta_0v)
\end{tcolorbox}
\noindent where $\Gamma$ is the Euler function, $\beta_0=(11C_A-2n_f)/(12\pi)$, $B_q=-3/4$ and $B_g=-11/12+n_fT_R/(3N_c)$, with $n_f$ the number of active quark flavours.

\begin{figure}
   \centering
      \includegraphics[width=0.6\textwidth]{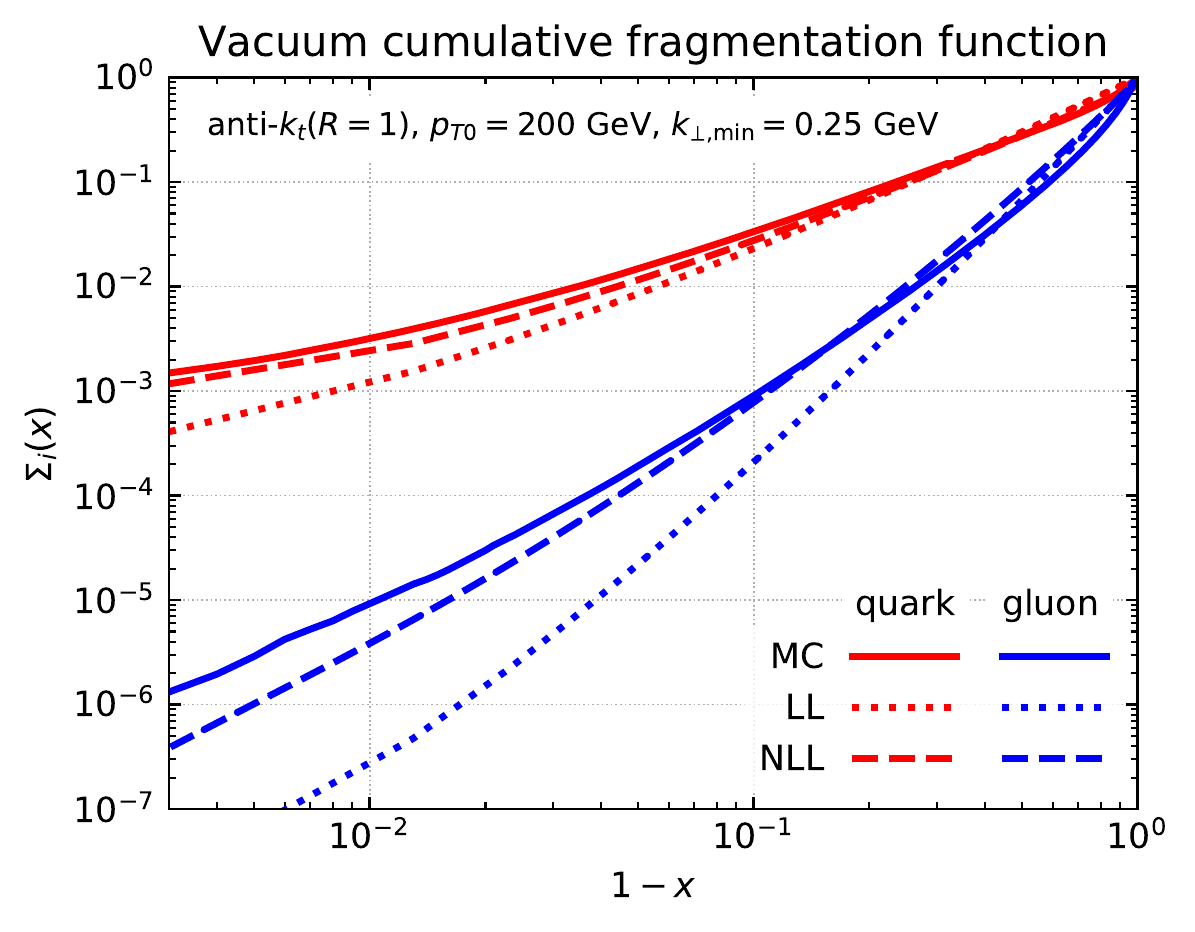}
    \caption{\small   The cumulative fragmentation function $\Sigma_i(x|p_{T0})$
   for quark ($i=q$) and gluon ($i=g$) initiated monochromatic jets in the vacuum.
   The MC calculations from {\tt JetMed} (see Chapter \ref{chapter:MC}) are shown with solid lines, and the two analytic  approximations, LL and NLL, by dotted and dashed lines, respectively. Actually, the gluon NLL curve includes corrections due to quark/gluon mixing which are formally beyond NLL accuracy, as explained in Appendix~\ref{app:NLL}.}
    \label{Fig:frag-largex}
\end{figure}

The  LL piece is rather easy to understand: this is the would-be
standard  double-logarithmic (DL) approximation in which successive emissions are 
strongly ordered both in energy fraction $z$ and in emission angle $\theta$, however amended by the
running of the coupling ($\alpha_s\to \alpha_s(k_\perp)$). To this LL accuracy, one can assume that a single emission --- the first one in the  DL series,
which is soft relative to the leading parton, but much harder than all the other emissions  ---
dominates the jet fragmentation function near $x=1$. Indeed, the even softer emissions do not 
count for the energy balance of the leading parton, hence they are not probed by $D_i(x\mid p_{T}R)$. Accordingly,
the probability \eqref{def-sigma} of finding the leading parton with an energy fraction $x'\ge x$ and $x$ close to 1
 is equal to the probability of having no emissions with an energy fraction larger than $1-x$, which is exactly the Sudakov factor at this accuracy:
 \begin{equation}\label{sigma-vac-LL}
  \Sigma_{i}(x,p_{T}R)=\exp\left(-\frac{2 C_i}{\pi}\int_{1-x}^1 \frac{\dif z}{z}\int_{0}^{R}\frac{\dif \theta}{\theta}\alpha_s(k_\perp)\Theta(k_\perp-\ktmin)\right)
\end{equation}
with $k_\perp\simeq z\theta p_{T}$. Formula \eqref{sigma-vac-LL} gives precisely the function $g_1$ given in \eqref{g12}.

\subsubsection{Solution at small $x$ at DL}
\label{subsub:frag-smallx}

We now turn to the solution of \eqref{Dvac-mlla} at small $x$ to DL accuracy. This solution will be particularly useful when we will discuss the medium modifications of the jet fragmentation pattern. At DLA, one only resums logarithms of the form $\alpha_0\log^2$ where this time, $\log$ means either $\log(1/x)$ or $\log(p_{T}R/\ktmin)$. 

\begin{figure}[t] 
  \centering
  \begin{subfigure}[t]{0.51\textwidth}
    \includegraphics[width=\textwidth]{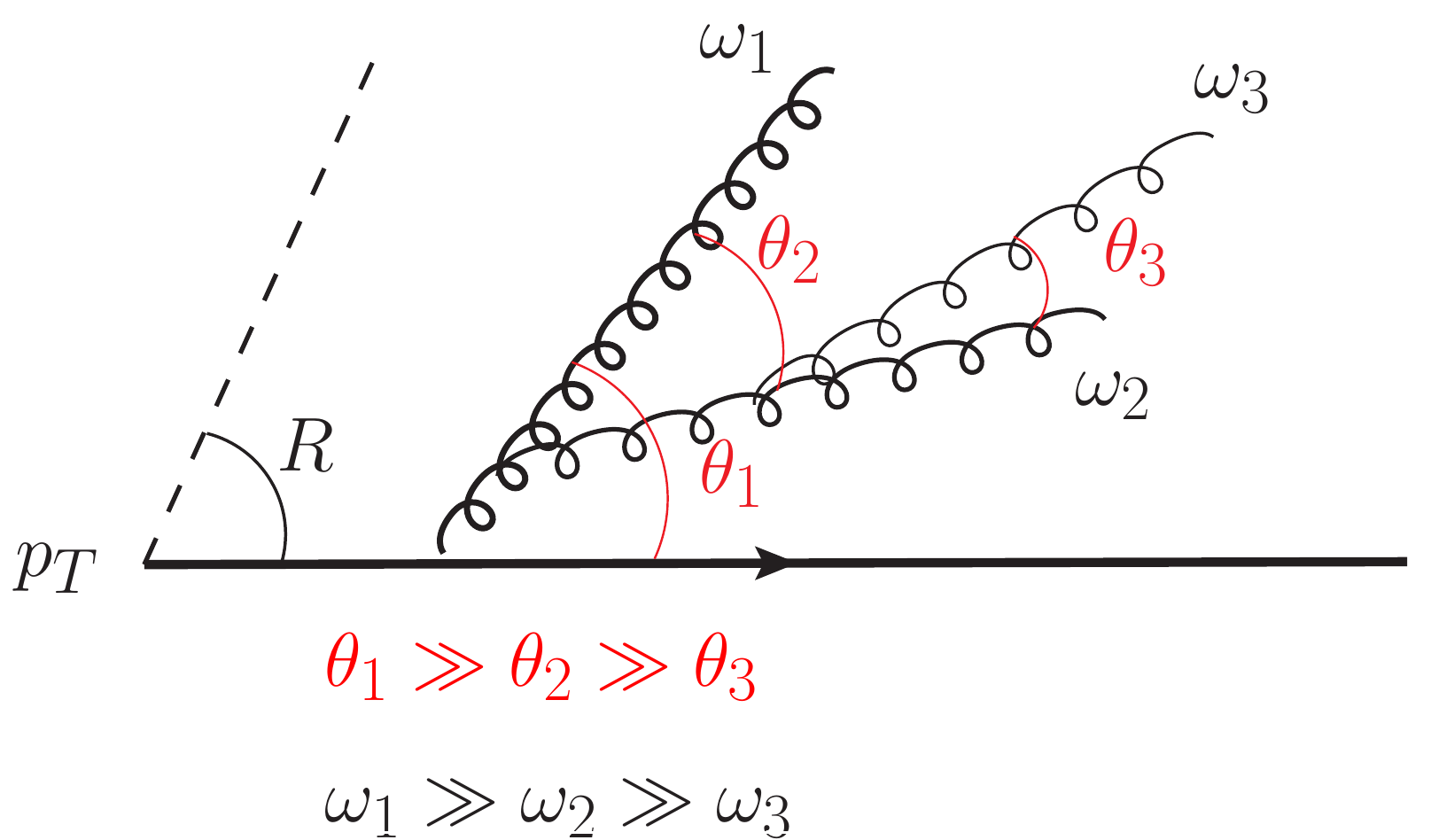}
  \end{subfigure}
  \hfill
  \begin{subfigure}[t]{0.48\textwidth}
    \includegraphics[width=\textwidth]{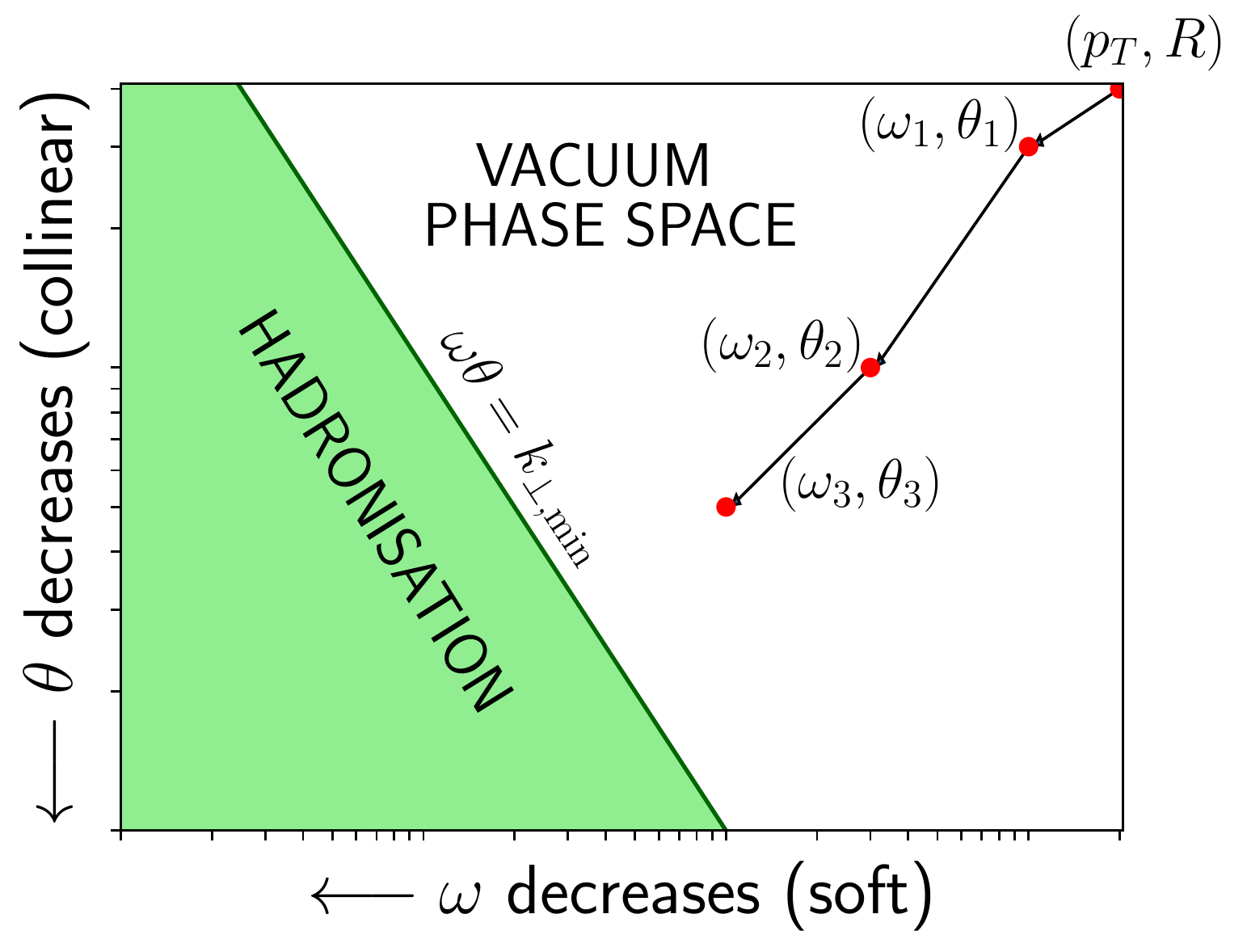}
  \end{subfigure}
  \caption{\small (Left) A given realization of the branching process that matters at DLA for the intrajet multiplicity. The parton sourcing the jet is a quark. (Right) The logarithmic $(\omega,\th)$ phase space for intrajet gluon emissions in the vacuum. Each emission in the cascade shown on the left is represented by a red circle with coordinates $(\om_i,\th_i)$. The green line is the lower cut-off in the evolution, given by $k_\perp\simeq \om\th=\ktmin$. \label{Fig:branching-process}}
\end{figure}

For later convenience, let us introduce the following function $T_i(\om,\th_0\mid p_T,R)$:
\begin{equation}\label{Tdef1}
 T_i(\omega,\theta_0\mid p_T, R) \equiv -\omega\theta_0\frac{\dif \bar{D}_{i,\theta_0}(\omega\mid p_T, R)}{\dif \theta_0}
\end{equation}
where $\bar{D}_{i,\theta_0}(\omega\mid p_T, R)\equiv D_{i,\theta_0}(\omega/p_T,p_T R)$ and $D_{i,\theta_0}$ is the solution of \eqref{Dvac-mlla} with initial condition $D_i(x,p_T\th_0)=\delta(1-x)$. The function $T_i$ has a simple probabilistic interpretation: it is the number density of emissions with energy $\om$ and angle $\th_0$ with respect to the \textit{parent} parton in the branching process. Such a branching 
process is pictured on Fig.~\ref{Fig:branching-process}-left for a quark initiated jet. At DLA, energies and angles with respect to the emitter are strongly decreasing. It is also convenient to represent a given realization of the branching process on the $(\om,\th)$ phase space for emissions, shown Fig.~\ref{Fig:branching-process}-right. In the vacuum, this phase space is only limited by the constraint $k_\perp\ge\ktmin$.

To simplify the discussion, we focus on gluon initiated jets. The corresponding calculation for quark jets is similar. The integrand $I(z)$ in the right-hand side of equation \eqref{Dvac-mlla} has the property $I(z)=I(1-z)$ so we can use the relation
\[\int_{0}^{1}I(z)\dif z=2\int_{0}^{1}(1-z)I(z)\dif z\]
to put the singular behaviour of the splitting function $\Phi_g^{gg}$ at $z=0$ only. At DLA, one neglects the energy recoil of the leading parton by setting $(1-z)p_T\simeq p_T$: the real term with argument $(1-z)p_T\th$ in \eqref{Dvac-mlla} associated with the fragmentation of the hard-branch  cancels against the virtual term. We approximate also $k_\perp\simeq zQ$ so that the evolution equation becomes simply:
\begin{equation}\label{Dsmall-x-NLL}
 \frac{\partial \bar{D}_{g,\theta_0}(\omega\mid p_T,\th)}{\partial \log(\th)}=\int_{\omega/p_T}^1 \dif z\,\frac{\alpha_s(zp_T\th)}{\pi}\Phi^{gg}_g(z)(1-z)\bar{D}_{g,\theta_0}(\omega\mid zp_T,\th)
\end{equation}
The lower bound comes from the fact that $\bar{D}_{g}(\omega\mid p_T,\th)$ obviously vanishes when $\om>p_T$.
Equation \eqref{Dsmall-x-NLL} is convenient to include corrections beyond DLA (see Chapter \ref{chapter:DLApic}-Section \ref{subsub:frag-beyondDLA}). At DLA, it can be further simplified using $\alpha_s(k_\perp)=\alpha_0$ and $\Phi^{gg}_g(z)(1-z)\simeq 2C_A/z$.
By definition, $T_i$ satisfies the same evolution equation, with a different initial condition though: $T_i(\omega,\theta_0\mid p_T,\th_0)=2\alpha_0C_A/\pi \equiv2 \abar$. Integrating over $\th$ between $\th_0$ and  $R$, this equation can be written:
\begin{equation}\label{integral-eq-T}
 T_g(\omega,\theta_0\mid p_T, R) = 2\abar
 +2\abar\int_{\theta_0}^{R}\frac{\dif\theta_1}{\theta_1}\,\int_{\omega/p_T}^{1}\frac{\dif z_1}{z_1}\,T_g(\omega,\theta \mid z_1p_T, \theta_1)
\end{equation}
The series expansion in power of $\bar{\alpha}_s$ of the solution to \eqref{integral-eq-T} can be obtained iteratively using \eqref{integral-eq-T}. One then recognizes the series expansion of the following function:
\begin{tcolorbox}[ams equation]\label{TDLA}
T_g(\omega,\theta_0\mid p_T, R)=2\abar \textrm{I}_0\Big(2\sqrt{2\abar\log(p_T/\omega)\log(R/\th_0)}\Big)
\end{tcolorbox}
\noindent with $\textrm{I}_n$ the modified Bessel function of rank $n$. Equation \eqref{TDLA} is the building block for the calculation of the gluon distribution with a medium at DLA.

From $\eqref{TDLA}$, one can easily obtain the solution of \eqref{Dvac-mlla} with the condition $k_\perp=xp_T\theta_0>\ktmin$ by inverting equation \eqref{Tdef1}, \cite{Dokshitzer:1991wu,Dokshitzer:1992iy,Lupia:1997bs}:
\begin{align}\label{Dg-dla}
 x D_g(x,p_TR)&=\int_{\ktmin/(xp_T)}^{R}\frac{\dif \theta_0}{\th_0}\,T_g(xp_T,\theta_0\mid p_T, R)\\
\label{Dg-dla-b}              &=\sqrt{\abar}\sqrt{\frac{2\abar \log\frac{x p_T R}{\ktmin}\log\frac{1}{x}}
{\abar\log^2\frac{1}{x}}}
\ \textrm{I}_1\left(2\sqrt{2\abar 
\log\frac{1}{x}\,\log\frac{x p_T R}{\ktmin}}\right)
\end{align}
We have written the final result in a way that makes the double-logarithmic resummation clear.
The most striking feature of $D_g(x,p_TR)$ is the maximum at $x_{\textrm{hump}}=\sqrt{\ktmin/(p_TR)}$ and the collapse at small energies. One would naively expect that the multiplicity rises at small $x$ due to the $1/x$ behaviour of the Bremsstrahlung spectrum but colour coherence forbids the multiplication of soft gluon emissions.

\begin{figure}
   \centering
      \includegraphics[width=0.6\textwidth]{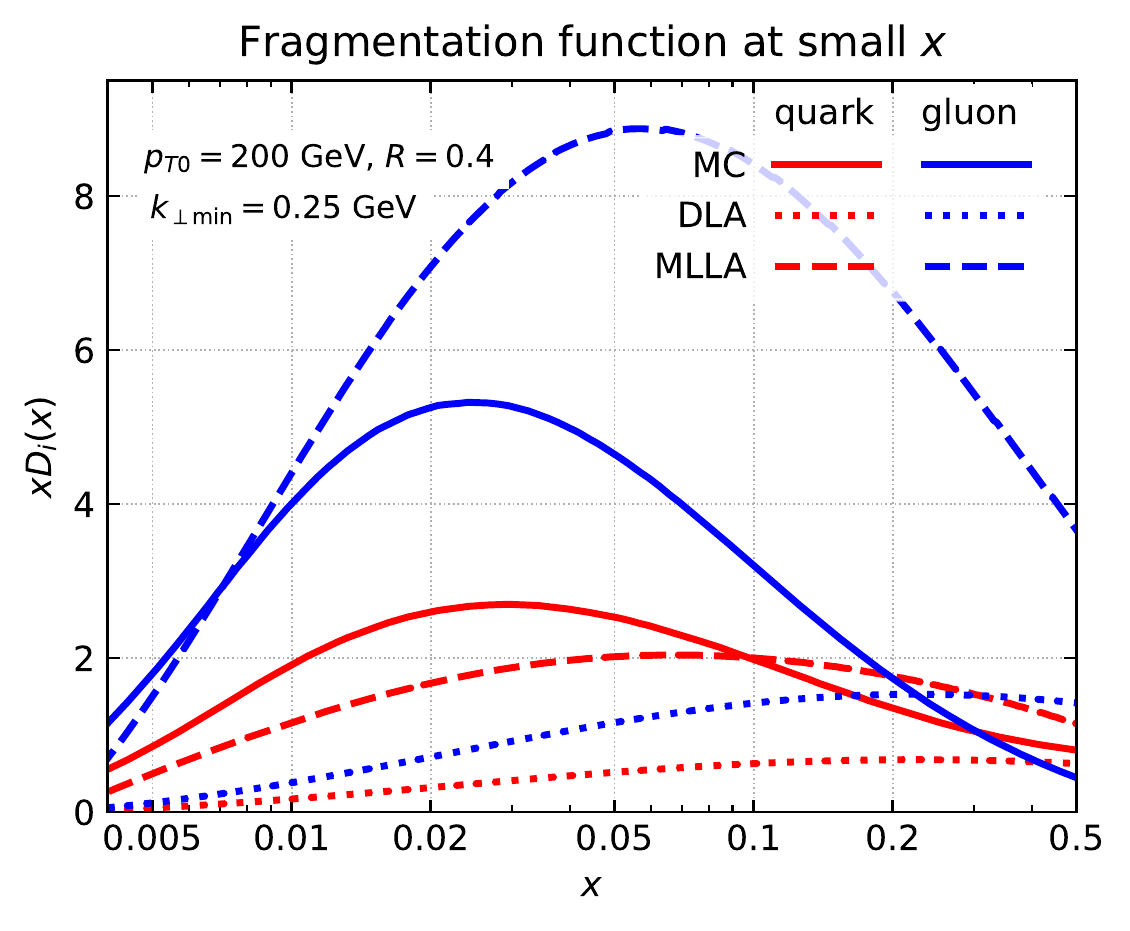}
    \caption{\small Fragmentation function at small $x$. The curves referred as ``DLA'' correspond to the double-logarithmic result \eqref{Dg-dla} whereas the curves referred as ``MLLA'' include running coupling and hard-collinear contributions. The MLLA curve are obtained from a numerical resolution of the partial differential equation \eqref{PDE-mlla} equivalent to the MLLA evolution equation following the method developed in Section~\ref{subsub:frag-beyondDLA}. The solid curves are {\tt JetMed} calculations for comparison (see Chapter \ref{chapter:MC}).}
    \label{Fig:frag-smallx}
\end{figure}

It is possible to include some NLL corrections to formula \eqref{Dg-dla-b} such as those coming from the running coupling or from the hard collinear emissions. As in the $x\sim1$ case, one possible method consists in going to Mellin space and solve the linear-differential equation. However, there is no known analytical solution in $x$ space in terms of usual functions. In Chapter \ref{chapter:DLApic}, Section \ref{subsub:frag-beyondDLA}, we use the partial differential equation method to obtain some NLL corrections shown in Fig.~\ref{Fig:frag-smallx}.

\subsection{Subjet observables}
\label{sub:obs-subjets}

As emphasised in the last section, the fragmentation function defined by Eq.~\eqref{def-frag} is not an IRC safe  observable when computed in perturbative QCD. 
In the experiments, this observable is defined from final state hadrons whose theoretical description transcends  pQCD. In our simplified treatment, 
which ignores hadronisation, the partonic cascades have been forced to terminate at a lowest transverse
momentum $\ktmin$, which mimics the QCD confinement scale $\Lambda_{{\rm QCD}}$. The precise
value of this scale is not well defined and results turn out to be rather strongly sensitive
to its variations (see e.g.\ Eq.~\eqref{Dg-dla}).

In this section, we investigate a set of observables relying on the concept of subjets \cite{Dasgupta:2013ihk} (or \cite{Marzani:2019hun} for a detailed review). Subjets, as jets, are IRC safe objects so there is a possibility to design an observable both IRC safe and sensitive to the intrajet structure. The analytical results obtained to LL accuracy will serve as benchmark formulas when we will discuss the corresponding observables in heavy ion collisions.

\subsubsection{The Soft Drop $z_g$ distribution}
\label{subsub:Soft Drop}

For completeness, we first recall the definition of the Soft Drop (SD)
procedure \cite{Larkoski:2014wba}. For a given jet of radius $R$, SD
first reclusters the constituents of the jet using the
Cambridge/Aachen (C/A)
algorithm \cite{Dokshitzer:1997in}. The ensuing jet is
then iteratively declustered, i.e.\ the last step of the pairwise
clustering is undone, yielding two subjets of transverse momenta
$p_{T1}$ and $p_{T2}$ separated by a distance
$\Delta R_{12}=\sqrt{\Delta y_{12}^2+\Delta \phi_{12}^2}$ in
rapidity-azimuth.
This procedure stops when the SD condition is met, that is when
\begin{equation}\label{zgdef}
 z_{12}\equiv \frac{\mathrm{min}(p_{T1},p_{T2})}{p_{T1}+p_{T2}} > \zc\left( \frac{\Delta R_{12}}{R} \right)^{\beta},
\end{equation}
where $\zc$ and $\beta$ are the SD parameters. If the condition is not satisfied, the subjet with the smaller $p_T$
is discarded and the declustering procedure continues with the harder.
With the above procedure, $\theta_g$ and $z_g$ are defined
respectively as $\Delta R_{12}$ and $z_{12}$ for the declustering
which satisfied the SD condition. When the procedure
exhausts all declusterings without meeting the SD condition, we set $\theta_g$ and $z_g$ to zero. 
Furthermore, one can impose a lower cut-off
$\theta_g>\theta_\text{cut}$. This is commonly used for Pb-Pb
collisions at the LHC.

The distribution is IRC safe when $\beta<0$. The limiting case $\beta=0$, for which the SD procedure coincides with the modified MassDrop Tagger \cite{Dasgupta:2013ihk} is peculiar: the distribution is not IRC safe but ``Sudakov safe'' \cite{Larkoski:2015lea}. 

We then study the differential $z_g$ distribution for a
jet initiated by a parton of flavour $i$ (quark or gluon).
We can consider two possible normalisation for the $z_g$
distribution: the ``self-normalised'' distribution, $p_i(z_g)$, and
the ``$\Njet$-normalised'' distribution, $f_i(z_g)$. The former is
defined such that
\begin{equation}\label{pinorm}
\int_{\zc}^{1/2}\rmd z_g \,p_{i}(z_g) = 1.
\end{equation}
which is equivalent to normalising the $z_g$ distribution to the
number of jets which pass the SD condition and the optional cut on
$\theta_g$.
The latter is instead normalised to the total number of jets,
i.e.\ the normalisation includes jets which fail either the SD
condition or cut on the $\theta_g$.

The coherent branching algorithm enables to calculate the $z_g$ distribution in a intuitive way. Let us calculate this distribution to LL accuracy. At this order, the splitting functions (summed over final states) can be approximated by $2C_i/z$ where $C_i$ is the Casimir factor of the representation of the leading parton. With this simplification, the branching rates \eqref{bremrate} become
\begin{equation}\label{Pvac-sub}
\rmd^2 \mathcal{P}_{i}(z,\theta)= \frac{2C_i\alpha_s(zp_{T}\theta)}{\pi}\,\frac{\dif z}{z}\,
\frac{\rmd\theta}{\theta}
\end{equation}
Then we introduce ``Sudakov factor'' $\Delta_i(R,\theta_g)$, which is the probability to have no emission at any angle between $\theta_g$ and $R$ satisfying the Soft Drop constraint:
\begin{equation}
\label{Ddef}
\Delta_i(R,\theta_g)=\exp\left(-\int_{\theta_g}^{R}{\rmd\theta}\int_{0}^{1/2}\rmd z\,\,\mathcal{P}_{i}(z,\theta)\Theta(z-\zc(\th/R)^\beta)\right)\,.
\end{equation}
The $z_g$-distribution is obtained by considering the probability for
both $z_g$ and $\theta_g$, marginalised over $\theta_g$. The former is
simply expressed as the probability to have no branching between
$\theta_g$ and $R$ times the probability of a branching with
$\theta=\theta_g$ and $z=z_g$, so that
\begin{tcolorbox}[ams align]\label{pzgthetag}
  p_{i, \text{vac}}(z_g)=\frac{1}
  {1-\Delta_i(R,\thetacut)}\int_{\thetacut}^{R}\rmd\theta_g
\,\mathcal{P}_{i}(z_g,\theta_g) \Delta_i(R,\theta_g)\Theta(z_g-\zc(\th_g/R)^\beta),
\end{tcolorbox}
\noindent where we have included an optional cut $\theta_g>\thetacut$. The
overall factor $(1-\Delta_i)^{-1}$ enforces the normalisation condition
\eqref{pinorm} and would be absent for $f_i(z_g)$.

To discuss the underlying physics of the $z_g$ distribution, it is helpful
to consider the fixed-coupling approximation. One can then easily
perform the angular integration in~\eqref{pzgthetag}. We give the results for $\beta\le0$ and $\thetacut=0$ in such a way that the DL resummation is clear \cite{Larkoski:2015lea}:
\begin{align}\label{pzg_fixed_beta}
 z_g p_{i}(z_g)&=\sqrt{\alpha_0}F^{z_g}_{\textrm{\tiny DL}}\big(\alpha_0 \log^2(2 z_g),\alpha_0 \log^2(2 \zc)\big)\Theta(z_g-\zc),\\
F^{z_g}_{\textrm{\tiny DL}}(u,v)&=\sqrt{\frac{C_i}{\beta}}e^{\frac{C_iv}{\pi\beta}}\left[\erf\left(\sqrt{\frac{C_iu}{\pi\beta}}\right)-\erf\left(\sqrt{\frac{C_iv}{\pi\beta}}\right)\right]
\end{align}
with $\erf(x)$ the Gauss error function.
The limiting case $\beta=0$ is singular since there is no apparent $\alpha_0$ dependence. A direct calculation of~\eqref{pzgthetag} when $\thetacut=0$ or the limit $\beta\rightarrow0$ of \eqref{pzg_fixed_beta} give, 
\begin{equation}
\label{pzg_fixed_beta=0}
 p_{i}(z_g)=\frac{1/z_g}{\log(1/(2\zc))}\Theta(z_g-\zc).
\end{equation}
This results makes clear that the $z_g$-distribution provides a direct measurement of the splitting function.

\subsubsection{The $\nSD$ distribution}
\label{subsub:nSD}

The $\nSD$ distribution $p_i(\nSD)$ relies on a generalization of the SD procedure called Iterated Soft Drop \cite{Frye:2017yrw}. Iterated SD proceeds by iterating the Soft Drop procedure, still
following the hardest branch in the jet, until all declusterings have
been exhausted. $n_\text{SD}$ is then defined as the number of
declusterings passing the SD condition. It is a particular case of recursive SD (rSD) which is essentially the same technique: instead of following the hardest branch, rSD proceeds recursively through the full branching tree (see also \cite{Dreyer:2018tjj}).

From the coherent branching algorithm, if we neglect the energy recoil of the leading parton --- assumption which is valid to LL accuracy --- the emission of \textit{primary} partons is a Poisson process where the angle plays the role of time. Thus, the random variable $\nSD$ follows a Poisson distribution whose parameter $\rho_{i,\textrm{LL}}$ is simply given by the integrated rate \eqref{Pvac-sub} over the phase space available for emissions passing the SD cut \cite{Frye:2017yrw}.

\begin{tcolorbox}[ams align]
 p_i(\nSD)&=e^{-\rho_{i,\textrm{LL}}}\frac{\rho_{i,\textrm{LL}}^{\nSD}}{\nSD!}\\
 \rho_{i,\textrm{LL}}&=\frac{2C_i}{\pi}\int_0^R\frac{\dif\th}{\th}\int_0^1\frac{\dif z}{z}\alpha_s(zp_T\th)\Theta_{\textrm{cut}}
\end{tcolorbox}
\noindent with $\Theta_{\textrm{cut}}=\Theta(z-\zc(\th/R)^{\beta})$.
This simple estimate of the $\nSD$ distribution will be useful when we will discuss its nuclear modification in Section~\ref{sub:nsd-pheno}.

\subsubsection{A new fragmentation function: the ISD fragmentation function}
\label{subsub:frag-ISD}

From the ISD procedure, one can build an IRC safe generalisation of the fragmentation function based on subjets and not on final state hadrons: the ISD jet fragmentation function, $\mathcal{D}_{\textrm{ISD}}(z)$. This new observable fixes IRC issues of the fragmentation function while keeping the same fundamental phenomenology. It is defined as the number $\dif N_{\textrm{ISD}}$ of subjets passing the SD cut through the ISD procedure, with a momentum fraction $z$ between $z$ and $z+\dif z$ and normalized by the total number of jets integrated over a suitable range in $p_{T,\textrm{jet}}$:
\begin{equation}
 \mathcal{D}_{\textrm{ISD}}(x)\equiv\frac{1}{N_\textrm{jets}}\frac{\dif N_{\textrm{ISD}}}{\dif z}
\end{equation}
Note that this definition is directly similar to the primary Lund-plane
density~\cite{Dreyer:2018nbf,Lifson:2020gua}, $\rho(\theta,k_\perp)$, integrated over all
angles $\theta$ satisfying the $k_{\perp,\textrm{cut}}$ condition at fixed
$x=k_\perp/(\theta p_{T,\text{jet}})$.

The way the ISD algorithm proceeds requires a slightly different master equation for the generating functional $\mathcal{Z}_{i}(p_T,\th\mid u)$ of \textit{primary} emissions  than the one presented in \eqref{Z-mlla}. Indeed, ISD always follows the hardest branch in the declustering, so the subsequent branchings in the softer branch can be ignored. This leads to a simplification of the master equation, which now reads:
\begin{equation}\label{Z-primary}
\frac{\partial \mathcal{Z}_{i}(p_T,\th\mid u)}{\partial \log(\th)}=\sum_{(a,b)}\int_0^{1/2}\dif z\,\frac{\alpha_s(k_\perp)}{\pi}\Theta_{\rm cut}\Phi_i^{ab}(z)\Big(u(z)\mathcal{Z}_b((1-z)p_T,\th\mid u)-\mathcal{Z}_i(p_T,\th\mid u)\Big)
\end{equation}
With respect to \eqref{Z-mlla}, note the integration range between $0$ and $1/2$ to account for the fact that ISD follows the hardest branch with splitting fraction $1-z$. Moreover, as we will be only interested in the \textit{splitting} fraction $z$ of the emissions, the probing functions $u$ depend only on the splitting fraction $z$.

As for the fragmentation function analysed in details in Section \ref{sub:frag}, let us first consider the ISD fragmentation function $D_{i,\textrm{ISD}}(z\mid p_T,\th)$  of a leading virtual parton of type $i\in\{q,\bar{q}, g\}$ with initial transverse momentum $p_T$ and decaying within an opening angle $\th$, together with its cumulative version $\Sigma_{i,\textrm{ISD}}(z\mid p_T,\th)$:
\begin{equation}\label{sigma-isd}
D_{i,\textrm{ISD}}(z\mid p_T,\th)\equiv\frac{\delta \mathcal{Z}_i(p_T,\th\mid u)}{\delta u(z)}\,,\qquad
 \Sigma_{i,\textrm{ISD}}(z\mid p_T,\th)\equiv\int_z^{1/2}\dif z'\, D_{i,\textrm{ISD}}(z'\mid p_T,\th)
\end{equation}
At small $z$, we expect large logarithms of the form $L=\log(1/z)$ or $\Lc=\log(1/\zc)$ that must be resummed to all orders. This resummation is organised as usual:
\begin{equation}\label{isd-log-expansion}
\Sigma_{i,\textrm{ISD}}(z\mid p_T,R)=Lg_{\textrm{LL}}^{\textrm{\tiny ISD}}\big(\alpha_0L,\alpha_0\Lc\big)+g_{\textrm{NLL}}^{\textrm{\tiny ISD}}\big(\alpha_0L,\alpha_0\Lc\big)+O(\alpha_0^{n+1}\log^n)
\end{equation}
but, interestingly, it does not exponentiate as in \eqref{resum-exponentiation}.

\paragraph{Leading-logarithmic analysis.} At leading logarithmic accuracy, one neglects the energy recoil of the hard branch $\mathcal{Z}_b((1-z)p_T,\th\mid u)\simeq\mathcal{Z}_b(p_T,\th\mid u)$. However, contrary to the case of the fragmentation function at large $x$, the quark-gluon mixing terms matter to LL accuracy. At this level of precision, $k_\perp\simeq zp_T\th$ when appearing in front of the singular part of the splitting functions and $k_\perp\simeq p_T\th$ when appearing in front of the finite part of quark-gluon mixing terms.

From the master equation \eqref{Z-primary}, one easily gets a set of three coupled differential equations for $D_i$ which can be reduced to two coupled differential equations defining $\vec{D}\equiv(D_F,D_A)$ with $D_F=(D_q+D_{\bar{q}})/2$ and $D_A=D_g$. Using the standard variable $Q=p_T\th$,
\begin{equation}\label{diffeq-Disd}
 \frac{\partial \vec{D}_{\textrm{ISD}}(z,Q)}{\partial \log(Q)}=\frac{2\vec{C}_R}{\pi}\frac{\alpha_s(zQ)}{z}\Theta_{\textrm{cut}}+\frac{\alpha_s(Q)}{\pi}\begin{pmatrix}
-f_{F} & f_{F} \\
f_{A} & -f_{A}
\end{pmatrix}\vec{D}_{\textrm{ISD}}(z,Q)
\end{equation}
with $f_{R}$ the finite part of the splitting functions:
\begin{align}\label{finite-mixing}
 f_{F}=\int_0^{1/2}\dif z\, \Phi_q^{qg}(z)\,\qquad f_{A}=2\int_0^{1/2}\dif z\, \Phi_g^{q\bar{q}}(z)
\end{align}
The initial condition of \eqref{diffeq-Disd} is $\vec{D}(z,Q=0)=0$.

One can find a solution of this first order linear differential equation. The complete calculation is given in Appendix \ref{app:ISD}. The full LL result is:
\begin{align}\label{Disd-LL}
 z D_{R,\textrm{ISD}}&(z\mid p_T,R)=\frac{C_F f_A+C_A f_F}{f_F+f_A}f_0(\alpha_0L,\alpha_0\Lc)+\frac{C_R f_R-C_{\tilde{R}} f_R}{f_F+f_A}f_1(\alpha_0L,\alpha_0\Lc)\\
 f_0(u,v)&=\frac{-1}{\pi \beta_0}\log\Big(1+\frac{2\beta_0}{\beta}(v+(\beta-1)u)\Big)\\
 f_1(u,v)&=\frac{(2\beta_0 u)^{\frac{f_R+f_A}{2\pi \beta_0}}}{\pi \beta_0}\left[\mathcal{B}\left(\frac{1+\frac{2\beta_0}{\beta}\big(v-u)}{2\beta_0u},1+\frac{f_R+f_A}{2\pi \beta_0},0\right)-\mathcal{B}\left(\frac{1}{2\beta_0u},1+\frac{f_R+f_A}{2\pi \beta_0},0\right)\right]\nonumber
\end{align}
with $\mathcal{B}(x,a,b)$ the incomplete beta function and $\tilde{R}=A,F$ if $R=F,A$ respectively.
With the full LL result \eqref{Disd-LL}, one can notice that the second term in Eq.~\eqref{diffeq-Disd} associated with quark-gluon mixing is not really numerically important. Consequently, when we will study the nuclear modifications of $D_{i,\textrm{ISD}}(z\mid p_T,R)$, we will approximate the LL formula by:
\begin{tcolorbox}[ams align]\label{Disd-approx}
 z D_{R,\textrm{ISD}}\simeq C_R f_0(\alpha_0L,\alpha_0\Lc)=\frac{2C_R}
{\pi}\int_{0}^R\frac{\dif\theta}{\theta}\,
\alpha_s(zp_T\th)\Theta_{\textrm{cut}}
\end{tcolorbox}
\noindent This result has an interesting property: it coincides with a naive calculation in which one assumes that the jet is made of one soft collinear emission. This is a specificity of the LL approximation that we will extensively use in the rest of this thesis.

\paragraph{NLL corrections.} For completeness, we have also analysed some NLL corrections to the LL result \eqref{Disd-LL}. At NLL, one cannot ignore the energy recoil of the hard branch but one can truncate to the first order the (logarithmic) Taylor expansion of $D_{R,\textrm{ISD}}(z,(1-z)Q)$:
\begin{equation}
D_{R,\textrm{ISD}}(z,\log((1-z)Q))\simeq D_{R,\textrm{ISD}}(z,\log(Q))+\log(1-z)\frac{\partial D_{R,\textrm{ISD}}(z,\log(Q))}{\partial \log(Q)}
\end{equation}
The other NLL corrections included in the master equation \eqref{Z-primary} that can be easily added to the LL piece are the finite part of the splitting functions singular in $z=0$ and the running coupling corrections when $k_\perp$ is small using $k_\perp=zp_T\th$ in the pieces where we ignored the $z$ dependence at LL. With these corrections, formula \eqref{diffeq-Disd} has the following solvable form:
\begin{equation}\label{diffeq-Disd-nll}
 \frac{\partial \vec{D}_{\textrm{ISD}}(z,Q)}{\partial \log(Q)}=\vec{C}_{\textrm{NLL}}(z,Q)+\frac{\alpha_s(Q)}{\pi}\begin{pmatrix}
-f_{F} & f_{A} \\
f_{F} & -f_{A}
\end{pmatrix}\vec{D}_{\textrm{ISD}}(z,Q)
\end{equation}

\begin{figure}
   \centering
      \includegraphics[width=0.6\textwidth]{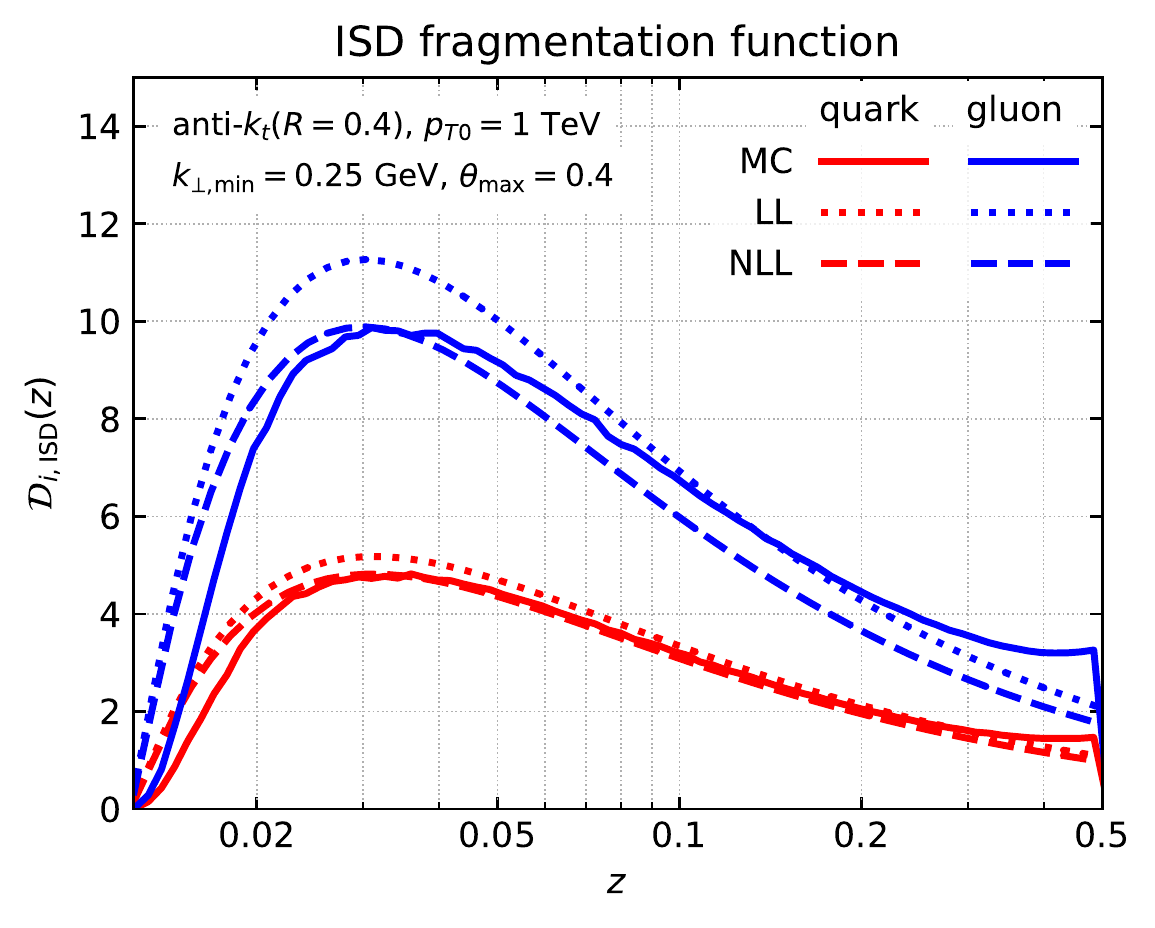}
    \caption{\small The ISD fragmentation function to LL accuracy (dotted curves) and with NLL corrections (dashed curves) compared to {\tt JetMed} Monte Carlo calculations of this observable. The LL results are given by \eqref{Disd-LL} whereas the NLL formula used for this plot is given by the solution \eqref{Disd-NLLsol} of \eqref{diffeq-Disd-nll}.}
    \label{Fig:frag-ISD}
\end{figure}

There are also other NLL corrections which are not taken into account in the master equation \eqref{Z-primary}. First of all, one should use the two loop running coupling and include the singular part at $z=0$ of the two-loops splitting functions wherever it is necessary. This amounts to a simple modification of the function $C_{\textrm{NLL}}(z,Q)$ in formula \eqref{diffeq-Disd-nll}. The full function $C_{\textrm{NLL}}(z,Q)$ and the solution of \eqref{diffeq-Disd-nll} from the variation of parameters method are given in Appendix \ref{app:ISD}.
Moreover, there are also NLL corrections related to clustering effects which occur when a non-primary emission is clustered with the hard branch. These corrections are beyond the scope of this simple analysis. Nevertheless, a complete NLL result would be very interesting for precision jet physics in $pp$ collisions.

\section{Jets from medium-induced emissions}
\label{sec:jets-mie}

The first section of this chapter was dedicated to jets evolving in the vacuum via iteration of Bremsstrahlung emissions triggered by the virtuality of the initial parton coming from the hard event. From a mathematical point of view, iterating vacuum-like soft/collinear emissions is equivalent to an all-orders resummation of the large soft and collinear logarithms. We have seen a couple of examples of such resummations through jet observables that we will study also in heavy-ion collisions.

In this section, we present another kind of ``jet'' that happen specifically in a dense QCD medium and we call it ``jet from medium-induced emissions'' because the emission process which is iterated is not the Bremsstrahlung process associated with an off-shell parton but the medium-induced spectrum studied in Chapter \ref{chapter:emissions} associated with an on-shell parton travelling through a dense medium (see \cite{Blaizot:2015lma} for a review).

\subsection{The multiple branching regime: probabilistic picture}
\label{sub:mie-multiple-branchings}

Before discussing the properties of jets from medium-induced emissions, we first review the main constructing stages of the effective field theory for multiple soft medium-induced emissions in a dense QCD medium \cite{Blaizot:2013vha}.

\subsubsection{The branching energy scale $\ombr$}
\label{subsub:ombr}

In Section \ref{sec:BDMPSZ}, we have calculated the medium-induced gluon spectrum from an on-shell quark or gluon travelling through the medium. Focusing on the behaviour at small energy $\om=k^+$ (in light cone coordinates) in the case of a medium with fixed length $L$ and constant density, this spectrum behaves like:
\begin{equation}\label{BDMPSZ-estimate}
 \omega\frac{\dif \Nmie}{\dif \om}\sim\frac{\alpha_sC_A}{\pi}\sqrt{\frac{2\om_c}{\om}}
\end{equation}
for an initial gluon. We recall that $\om_c=\qhat L^2/2$ is typically the largest medium-induced radiation that can be formed over a time $L$. Without loss of generalities, we focus on gluon splittings in this subsection.

Comparing with the standard Bremsstrahlung spectrum, one notes first that the transverse momentum integral has been carried out: there is no collinear singularity associated with the medium-induced gluon spectrum. However, this spectrum remains singular at small $\om$. Of course, there are several assumptions that lead to formula \eqref{BDMPSZ-estimate}. Among them, one must have $\om>\omBH$, with $\omBH\sim\qhat \ell^2$ the Bethe-Heitler frequency below which the formation time $t_f(\om)=\sqrt{\om/\qhat}$ of the medium-induced parton is smaller than the mean-free path $\ell$ in the medium.

Is there nothing more behind this soft singularity? Actually, the physics related to this singularity happens to be very rich. In the same way that the soft logarithmic singularity of the Bremsstrahlung spectrum must be taken into account beyond the leading order, to all orders in $\alpha_s$, there is a regime in which medium-induced emissions have to be resummed beyond the leading order BDMPS-Z result. This regime occurs when the integrated BDMPS-Z spectrum becomes of order $1$. This trivial integration defines a new energy scale, the branching energy $\om_{\textrm{br}}$:
\begin{equation}\label{ombr}
N(\omega)=\int_{\om}^{\om_c}\dif \om'\, \frac{\dif \Nmie}{\dif \om'}\sim\sqrt{\frac{\abar^2\om_c}{\om}}=\mathcal{O}(1)\Rightarrow \om\sim\ombr\equiv\abar^2\om_c
\end{equation}
Even if $\ombr$ is smaller than $\omc$ by two powers of $\abar\equiv \alpha_sN_c/\pi$, this ``branching'' scale is still hard enough to be larger than $\omBH$ provided that the medium is large enough: $\ell\ll \abar L$. When the latter condition is satisfied, there is a potentially large energy phase space such that $\omBH\ll\om\lesssim\ombr$ in which the emission probability $N(\om)$ is larger than $1$. In this regime, one should not interpret the integrated BDMPS-Z spectrum as a probability, but rather as an average \textit{number} of emissions. Since this number is larger than $1$ below $\ombr$, this means that one must also take into account the single BDMPS-Z process beyond the leading order: this is the so-called multiple branching regime.

To get more intuition on the parameter that will be resummed to all orders through this new branching process, we rewrite the condition \eqref{ombr} as a ratio of time-scales:
\begin{equation}\label{N-time}
 N(\omega)\sim\abar\frac{L}{t_f(\om)}
\end{equation}
Consequently, in the multiple branching regime, the parameter of order 1 which has to be resummed to all orders is $\abar L/t_f$.

\subsubsection{Iteration of medium-induced emissions}
\label{subsub:iteration}

Now that we have recognized the necessity of taking into account multiple medium-induced emissions, the next step is to understand how to calculate cross-sections in the multiple emission regime. The pedestrian method of computing Feynman diagrams with multiple external legs in the presence of the background field as in Chapter \ref{chapter:emissions} would be a gigantic task. Actually, this task is greatly facilitated by doing a couple of parametrically well-controlled approximations that enable to calculate the multiple medium-induced emission probability from a classical Markovian branching process. In this context, there are mainly three approximations necessary to come up with this classical picture.

\paragraph{Colour coherence.} The first one is related to the colour coherence of a medium-induced ``antenna'' that could significantly modify the interference term involved in the calculation of the emission of one gluon by this antenna. By medium-induced antenna, we refers to a two-gluon system associated with a medium-induced branching.  In the vacuum, we have seen that these interferences are responsible for the angular ordering constraint of the subsequent emission, but to LL accuracy at least, this was not an obstacle to see multiple vacuum-like emissions as a classical branching process (ordered in angle though). 

For a medium-induced branching, we now show that the two offspring gluons lose their colour coherence during a time comparable to the duration of the branching $t_f(\om)$. From the antenna calculation done in Section \ref{sec:decoherence}, we know that the (de)coherence time is given by $\tcoh(\th)\sim(\qhat\th^2)^{-1/3}$ where $\th$ is the opening angle of the antenna. For a medium-induced branching, this angle is set by the transverse momentum $k_f$ acquired during the branching process:
\begin{equation}
 k_f^2\simeq \om^2 \th_f^2\sim\qhat t_f(\om) \Rightarrow \th_f\sim\Big(\frac{\qhat}{\om^3}\Big)^{1/4}
\end{equation}
where we have assumed --- without loss of generality --- that one of the offspring gluon with energy $\om$ is softer than the other.
Thus, one has $\tcoh(\th_f)\sim t_f$, equality which precisely means that the medium-induced antenna loses its colour coherence during its formation. 
Now, a subsequent gluon emission can occur with probability of order $(\abar L/t_f)^2=\mathcal{O}(1)$ in the direct terms associated with independent emissions. However the interference term will be reduced by a factor $t_f/L$ since the quantum colour coherence survives only during a time of order $t_f$. Hence, interferences are sub-leading in an calculation aiming at resumming all $\mathcal{O}((\abar L/t_f)^n)$ contributions \cite{Blaizot:2012fh,Apolinario:2014csa}. This is the solution to the first obstacle associated with quantum colour coherence.

\paragraph{Overlapping formation times.} The second obstacle concerns the so-called problem of overlapping formation times \cite{Arnold:2015qya}. If we now only focus on the direct term in the amplitude squared for producing two medium-induced gluons with energy $\om$, the two emissions may overlap in time, changing the naive estimate obtained by squaring the probability for producing one gluon. We would like to bring to light a physical regime enabling to neglect this possibility. 

To do so, we would like to estimate the typical time $\tbr$ between two branchings in a regime where we can assume that they occur independently. Let us rewrite the leading soft behaviour of the BDMPS-Z spectrum \eqref{BDMPSZ-estimate} in a such a way that its interpretation as an emission \textit{rate} is transparent:
\begin{equation}\label{bdmpz-rate}
 \om\frac{\dif \Nmie}{\dif \om\dif t}\sim\frac{\abar}{t_f(\om)}
\end{equation}
One recovers formula \eqref{BDMPSZ-estimate} integrating this constant rate over a time $L$. If the emissions are independent with each other, the no-emission probability $\Delta_{\textrm{mie}}(\Delta t,\om')$ during $\Delta t$ of gluons with energies larger than some energy scale $\om'$ follows the exponential law with parameter given by the rate \eqref{bdmpz-rate} integrated over energies larger than $\om'$:
\begin{equation}
 \Delta_{\textrm{mie}}(\Delta t,\om')=\exp\Big(-\Delta t\int_{\om'}^{}\frac{\dif \om}{\om}\, \frac{\abar}{t_f(\om)}\Big)
\end{equation}
The upper limit in the integration range, of order $\textrm{min}(\om,\om_c)$, is left intentionally blank since the integral is controlled by the lower bound. $\om'$ is a scale which will define the regime of validity of the independent emissions picture. The probability $\Delta_{\textrm{mie}}(\Delta t,\om')$ vanishes for $\Delta t\gg \tbr(\om')$ with 
\begin{equation}\label{tstar-def}
 \tbr(\om')\sim t_f(\om')/\abar
\end{equation} 
As suggested by our notations, $\tbr(\om')$ is the survival time of our parton with energy $\om$ without emitting any gluons with energies larger than $\om'$ (but smaller than $\om$).

With this estimate at our disposal, one can neglect overlapping emissions if the formation time of the parton with energy $\om$ is much smaller than $\tbr(\om')$:
\begin{equation}\label{tstar-cond}
 t_f(\om)\ll \tbr(\om')\Rightarrow\om'\gg\abar^2\om
\end{equation}
This inequality can be summarized as follows: given an emission with energy $\om$, the overlapping formation time problem is harmless if all other subsequent emissions have an energy much larger than $\abar^2\om$. In terms of the splitting fraction $z=\om'/\om$, one must impose $z\ge\abar^2$. This seems to be a strong restriction since $\abar$ is not that small even in a weakly coupled medium. However, we will see that the evolution equation governing jets from medium-induced emissions favours ``democratic'' branchings instead of softer and softer branchings. Consequently, introducing a lower cut off for splitting fractions does not change this picture.

\paragraph{Collinear splittings approximation.} The last main approximation that greatly simplifies the mathematical formulation of the multiple branching regime is the assumption of purely collinear splittings. To be more precise, after the splitting, the two offspring partons share equal fraction, respectively $z$ and $1-z$, of both the initial energy and initial transverse momentum carried by the parent parton. This means that the transverse momentum acquired during the branching process is neglected and that all the transverse momentum of the emitted gluon is acquired during the propagation through the medium --- between the emission and the decay --- via collisions off the medium constituents. 

We now find the regime of validity of this approximation. In the multiple soft scattering regime, the transverse momentum acquired during the formation is $k_f^2=\qhat t_f$ whereas the transverse momentum acquired during propagation is $k_\perp^2=\qhat \tbr$. Hence, the condition $k_\perp\gg k_f$ is equivalent to the condition \eqref{tstar-cond} that we have already discussed. In order to obtain a typical value for the interval between two branchings which depends on $\om$ only, we shall use $\tbr\equiv \tbr(\omega)=t_f/\alpha_s$. Hence, the transverse momentum acquired during formation is suppressed by one power of $\alpha_s$ with respect to the transverse momentum acquired after propagation. We point out that this argument rely on the fact that $\tbr<L\Leftrightarrow\omega<\ombr$ in agreement with the multiple branching regime that we are discussing now. However, even for relatively hard medium-induced emissions in the regime $\ombr<\om<\omc$, this approximation remains valid since in this case, the transverse momentum acquired after emission is typically $Q_s=\sqrt{\qhat L}$ and the condition $\qhat t_f<Q_s^2$ is equivalent to $\om<\om_c$.

Thus, in the collinear splitting approximation, the transverse momentum broadening comes from multiple elastic collisions during propagation. It is worth mentioning that there is another indirect contribution to this transverse momentum coming from the recoil of the parton while radiating gluons \cite{Wu:2011kc,Liou:2013qya,Blaizot:2019muz}. This recoil effect is dominated by hard medium-induced emissions triggered by a single scattering with short formation time. For a parton propagating over a distance $L$, it amounts to an additive correction to the typical momentum transferred $Q_s$ of the form $\alpha_sN_c Q_s\log^2(Lk_BT_{\rm p})/\pi$ where $T_{\rm p}$ is the plasma temperature and $k_BT_{\rm p}$, the thermal energy which is also the maximal energy that a medium constituent can transfer in a single scattering. It is possible to include these corrections to the probabilistic picture that we shall now expose in more details by a renormalization of the quenching parameter: these radiative corrections are local and universal, at least to leading logarithmic accuracy \cite{Blaizot:2014bha,Wu:2014nca,Iancu:2014kga,Iancu:2018trm}. 

\subsubsection{Generating functional and master equation}
\label{subsub:GF-mie-jets}

We turn now to the construction of the master equation for the generating functional of the multiple medium-induced emissions. We choose first to focus on purely collinear processes, so that transverse diffusions due to multiple rescatterings or elastic collisions are integrated out. For observables averaged over transverse coordinates, it generates the right distributions. We will discuss the angular (transverse) structure of jets from medium-induced emissions in the following section.
Within the approximations listed in the previous section, one can calculate the cross-section for multiple medium-induced emissions using a classical and Markovian branching process with the following rules \cite{Blaizot:2013vha}:

\begin{enumerate}
 \item Successive emissions are independent and ordered in time. By time, one refers to the light-cone ``+'' coordinates with the longitudinal axis defined by the direction of motion of the parton triggering the jet. We note $E$ the (large) + component of the 4-vector of the leading parton.
 \item The rate of the branching process is given by the BDMPS-Z spectrum rate \eqref{bdmpz-rate} amended to take into account $q\rightarrow qg$ and $g\rightarrow gg$ splittings and finite $z$ fractions,
 \begin{equation}\label{mie-rate}
  \frac{\dif^2\mathcal{P}^{bc}_{a,\textrm{mie}}}{\dif z\dif t}=\frac{\mathcal{K}_a^{bc}(z)}{2\pi\tbr(z(1-z)E)}\Theta(\omc-z(1-z)E)\,,\qquad \tbr(\om)\equiv\frac{1}{\alpha_s}\sqrt{\frac{\om}{\qhat}}
 \end{equation}
 One can include running coupling corrections using $\alpha_s(k_{\textrm{br}}=(\qhat z E)^{1/4})$. The rate is given for a medium with constant density (i.e. constant $\hat{q}$). The case of a longitudinally expanding medium will be discussed in Section~\ref{sub:med-expansion}. All the ``in-medium'' splitting functions $\mathcal{K}_a^{bc}(z)$ are given in Appendix~\ref{app:A}. Finally, the step function ensures that all medium-induced emissions have an energy smaller than $\omc$ as it should be.
\end{enumerate}

We call $\Zmie_i(t_0,E\mid u(k^+))$ the generating functional associated with this branching process, $t_0$ being the initial time. The definition is the same as the one given in Eq.~\eqref{Zjet}, but the test functions $u(k^+)$ depends only on the + component of the 4-momentum of the emitted partons, since transverse informations have been integrated out. The generating functional $\Zmie$ follows the master equation \cite{Blaizot:2013vha,Escobedo:2016vba}:
\begin{tcolorbox}[ams equation]\label{Z-mie}
 -\frac{\partial \Zmie_i}{\partial t_0}=\frac{1}{2!}\sum_{(a,b)}\int_0^1\dif z\,\sqrt{\frac{\qhat}{E}}\frac{\alpha_s\mathcal{K}_i^{ab}(z)}{2\pi\sqrt{z(1-z)}}\Big[\Zmie_a(t_0,zE)\Zmie_b(t_0,(1-z)E)-\Zmie_i(t_0,E)\Big]
\end{tcolorbox}
\noindent The initial condition for it is $\Zmie_i(t_0=L,E)=u(E)$. As for Eq.~\eqref{Zjet}, the endpoint singularities are regularized with a cut-off $z\ge\epsilon$. From the discussion in \ref{subsub:iteration}, $\epsilon$ should be of order $\alpha_s^2$. Nevertheless, the observable that we will discuss thereafter are finite when $\epsilon\rightarrow 0$, and hence not sensitive to this cut-off.

In order to highlight the similarities and the differences between \eqref{Z-mie} and \eqref{Zjet}, we have written the former in backward form: $t_0\Leftrightarrow\log(1/\th)$. 
Besides the evolution parameter $t_0$ which accounts for the ordering in time of the process, an other important difference with respect to the MLLA evolution equation \eqref{Zjet} is hidden in the endpoint singularities: contrary to the DGLAP splitting functions behaving like $1/z$ at small $z$, the singularity is stronger for the medium-induced evolution $\sim 1/z^{3/2}$. As we shall now see, this will drastically modify the picture of medium-induced cascades with respect to DGLAP cascades.

\subsection{Democratic fragmentation and turbulent energy loss}
\label{sub:frag-mie}

In this section, we simplify the full process described by \eqref{Z-mie} in order to derive analytical results for medium-induced jet fragmentation and get some physical insights on the behaviour of the cascade. We first ignore quarks and assume that only gluons participate to the branching process. We also simplify the branching kernel $\mathcal{K}_g^{gg}(z)$ using:
\begin{equation}\label{K0}
 \frac{1}{2C_A}\frac{\mathcal{K}_g^{gg}(z)}{\sqrt{z(1-z)}}\simeq \mathcal{K}_0(z)\equiv \frac{1}{(z(1-z))^{3/2}}
\end{equation}
Finally, we assume that $E<\omc$ so that one can ignore the step function in \eqref{mie-rate}.
The forward evolution equation for the fragmentation function has a more simple form so we first rewrite \eqref{Z-mie} in forward form using the approximations specific to this section \cite{Blaizot:2013vha}:
\begin{equation}\label{Zmie-forward}
 \frac{\partial \Zmie_g(t,E)}{\partial t}=\int_0^1\dif z\int \dif k^+\,\frac{\abar}{2}\sqrt{\frac{\qhat}{E}}\mathcal{K}_0(z)\Big(u(zk^+)u((1-z)k^+)-u(k^+)\Big)\frac{\delta \Zmie_g(t,E)}{\delta u(k^+)}
\end{equation}
with initial condition $\Zmie_i(t=0,E)=u(E)$.

We want to find the evolution equation for the fragmentation function of medium-induced jets $D_{g,\mie}(x,t,E)$ defined by 
\begin{equation}
 D_{g,\mie}(x,t,E)=x\int \dif k^+\,\delta(k^{+}-xE)\frac{\delta \Zmie_g(t,E)}{\delta u(k^+)}\Big|_{u=1}
\end{equation}
with the initial condition $D_{g,\mie}(x,0,E)=x\delta(1-x)$. From Eq.~\eqref{Z-mie}, it is straightforward to derive it, in terms of the dimensionless variable $\tau\equiv \abar t \sqrt{\qhat/E}$, \cite{Blaizot:2013vha,Blaizot:2013hx}
\begin{equation}\label{frag-mie-evol}
 \frac{\partial D_{g,\mie}(x,\tau)}{\partial \tau}=\int_0^1\dif z\,\mathcal{K}_0(z)\left[\sqrt{\frac{z}{x}}D_{g,\mie}\Big(\frac{x}{z},\tau\Big)-\frac{z}{\sqrt{x}}D_{g,\mie}\Big(x,\tau\Big)\right]
\end{equation}
This is the rate equation for the fragmentation function of medium-induced jets. The corresponding rate equation obtained from \eqref{Z-mie} is of course equivalent. This equation is well-defined without introducing any cut-off $\epsilon$ in $z$ because there is no true endpoint singularity. For $z\rightarrow0$, the first term is harmless because $D_{g,\mie}(x/z,\tau)=0$ for $z<x$ and the second (``virtual'') term is regularized by the $z$ factor. For $z\rightarrow1$, the two terms are singular when separately taken but the singularities cancel in the difference.

\paragraph{Physical discussion.} This equation is particularly remarkable for several reasons \cite{Blaizot:2013hx}.
\begin{itemize}
 \item It admits a set of fixed point solutions $D(x,\tau)=c/\sqrt{x}$ for any constant $c$. In particular, the single-emission BDMPS-Z spectrum belongs to the set of fixed point solutions. They are called Kolmogorov–Zakharov fixed points and they naturally appear in phenomena associated with turbulence.
 \item It is exactly solvable given our initial condition. The solution is 
 \begin{equation}\label{Dmie-sol}
  D_{g,\mie}(x,\tau)=\frac{\tau}{\sqrt{x}(1-x)^{3/2}}e^{-\pi[\tau^2/(1-x)]}
 \end{equation}
Using the final value $\tau_{\textrm{max}}=\abar L/t_f(E)$, one recognises the all order resummation of the parameter $\abar L/t_f$ which defines the multiple branching regime.
\item This solution has the following behaviour: at small times $t\ll\tbr(E)$ it is strongly peaked around $x=1$. As time goes by, after $\tbr$, the leading gluon is progressively degraded into soft gluons that populate the soft modes of the spectrum. This soft part behaves like $\tau\exp(-\pi\tau^2)/\sqrt{x}$. The time dependence apart, the shape of the small $x$ part of the spectrum is close to a fixed point solution. The story is not over though: because of the exponential factor, the amount of energy contained in every $x>0$ bin of the spectrum vanishes in the limit $\tau\rightarrow\infty$. This is paradoxical: recall that energy is conserved along the branching process, even with the simplified kernel $\mathcal{K}_0$. Then, where does the initial energy go? It accumulates in a ``condensate'' located formally in the $x=0$ bin. 
\item One can calculate the energy accumulated in this condensate at the end of the cascade, called $\epsilon_{g,\textrm{flow}}(\tau_{\textrm{max}})$, by energy conservation: it is the initial energy $E$ minus the energy contained in the spectrum:
\begin{tcolorbox}[ams equation]\label{energy-flow}
\epsilon_{g,\textrm{flow}}(\tau_{\textrm{max}})=E-\int_0^1\dif x\,D_{g,\mie}(x,\tau_{\textrm{max}})=E\Big(1-e^{-2\pi\ombr/E}\Big)
\end{tcolorbox}
This relation enlightens the physical role of $\ombr$. If $E\ll \ombr$, almost all the initial energy of the leading gluon will disappear in this condensate after propagation through the medium. On the contrary, for $E\gg\ombr$ (but smaller than $\omc$ by hypothesis), the initial energy is shared between the remnant of the leading gluon and the small $x$ part of the spectrum: in this regime, the typical (event by event) energy loss by the leading gluon is $\ombr$. This is to be contrasted with the average energy loss as given by the BDMPS-Z spectrum which is of order $\abar\omc$. Note that $\epsilon_{\textrm{flow}}$ has also been obtained in the regime $E>\omc$ not discussed in this section \cite{Fister:2014zxa}. In this case, $\epsilon_{g,\textrm{flow}}\simeq \mathnormal{v}\ombr$ with $\mathnormal{v}$ a constant.
\item The endpoint $x=0$ is beyond the regime of validity of Eq.~\eqref{frag-mie-evol}. Recall that we must impose at least $x>\omBH/E$ for instance, since this analysis relies on the BDMPS-Z rate. One can also calculate the amount of energy that goes below the scale $\omBH/E$. As $\omBH$ is of the order of the temperature in a weakly coupled plasma, this is the amount of energy coming from the initial gluon that eventually ``thermalizes'' \cite{Iancu:2015uja}. It happens that the result is very close to $\mathcal{E}_{\textrm{flow}}$ so \eqref{energy-flow} is actually a good estimate of the thermalized energy loss.
\item Equation \eqref{frag-mie-evol} has an other interesting property, related to turbulence. Besides the energy accumulated in the condensate, one can calculate the energy flux $\mathcal{F}(x_0,\tau)$ transferred from modes with $x>x_0$ to modes with $x<x_0$ (counted positively in this direction):
\begin{equation}\label{flux}
 \mathcal{F}(x_0,\tau)\equiv\frac{-\partial}{\partial\tau}\int_{x_0}^{1}\dif x\,D_{g,\mie}(x,\tau)=\underbrace{2\pi\tau e^{-\pi\tau^2}}_{\mathcal{F}_{\textrm{flow}}}+\textrm{``non-flow''}
\end{equation}
For all $x_0$, there is always a non-zero component $\mathcal{F}_{\textrm{flow}}$ of this flux which is independent of $x_0$ and that corresponds to the flow of energy transferred to the condensate. This is characteristic of \textit{turbulent flows} in ideal cascades in which the energy flux at some energy is independent of the energy. This flow does not exist in DGLAP-like cascade such as the cascade describing vacuum-like branchings. A crucial physical ingredient for it is \textit{democratic branching}: whereas DGLAP branchings strongly favour asymmetric splittings, in medium-induced jets, the two daughters partons have comparable energies. Turbulent flow constitutes a very efficient way to transport energy from high modes into arbitrarily soft modes. This scenario is obviously particularly interesting for jet quenching physics. 
 
\end{itemize}

\subsection{Angular structure of medium-induced jets}
\label{sub:angular-structure}

So far, we have only discussed the longitudinal structure of medium-induced jets, ignoring its transverse structure. However, understanding this transverse structure is crucial to answer the following questions: what is the typical angle where the energy flow described in the previous section end up? For instance, if the energy accumulated in the condensate is also transferred to large angles with respect to the leading parton, then we have really an efficient mechanism for large angle energy loss by jets that could eventually explain the behaviour of the experimental data.

\subsubsection{The general picture}
\label{subsub:angular-gen-pict}

In the collinear emission approximation (see Section \ref{subsub:iteration}), the transverse structure is dictated by the transverse momentum acquired during the propagation. Including this transverse momentum in the generating functional formalism as in \eqref{Z-mie} is straightforward. First of all, the generating functional $\Zmie_i(t_0,E,p_\perp)$ and the test functions $u$ depend now on the transverse component $p_\perp$ of the 4-momentum of an emission. Then, the master equation for $\Zmie$ is amended by a term that can change the transverse momentum $p_\perp$ of the initial parton \cite{Blaizot:2013vha}:
\begin{tcolorbox}[ams align]
  \nonumber-\frac{\partial \Zmie_i}{\partial t_0}&=\frac{1}{2!}\sum_{(a,b)}\int_0^1\dif z\,\sqrt{\frac{\qhat}{E}}\frac{\alpha_s\mathcal{K}_i^{ab}(z)}{2\pi\sqrt{z(1-z)}}\Big[\Zmie_a(t_0,zE,zp_\perp)\Zmie_b(t_0,(1-z)E,(1-z)p_\perp)\\
  &\hspace{2.8cm}-\Zmie_i(t_0,E,p_\perp)\Big]+\int \dif^2 q_\perp\,\mathcal{C}_i(q_\perp,t_0)\Zmie_i(t_0,E,p_\perp-q_\perp)\label{Zmie-transverse}
\end{tcolorbox}
\noindent The collision kernel $\mathcal{C}_i(q_\perp,t_0)$ is related to the Fourier transform of the in-medium dipole cross-section $\sigma_d$, cf. Section \ref{sub:Amed-stat}: 
\begin{equation}\label{collision-term}
 \mathcal{C}_i(q_\perp,t_0)= -\frac{g^2}{2}C_i n(t_0)\sigma_d(q_\perp)
\end{equation}
In the multiple soft scattering regime and in the harmonic approximation, this collision kernel simplifies, $\mathcal{C}_i(q_\perp,t_0)\simeq-\frac{1}{4}\qhat_i(t_0)q_\perp^2$ and leads to the diffusion of emissions in the transverse plane. In this section, we will work within this simplified regime.

From \eqref{Zmie-transverse}, one can infer the evolution equation for the double-differential fragmentation function $D_{i,\mie}(x,k_\perp)$. Using the same kind of approximations as those used in the previous section, it is even possible to get analytical results \cite{Blaizot:2014rla}. As we will study this equation numerically via Monte Carlo methods in Chapter~\ref{chapter:MC}, we follow here a different path and give only the physical arguments enabling to get the angular picture of medium-induced jet evolution. 

Actually, all the necessary ingredients for this discussion have already been given in Section \ref{subsub:iteration}. Let us pick a final parton in the distribution $D_{i,\mie}(x,k_\perp)$ with energy $\om=xE$. We want to find its corresponding angle $\th\simeq k_\perp/\om$ with respect to the direction of the parton that has triggered the cascade. This angle is set by the time $\tprop$ of propagation through the medium:
\begin{equation}
 \th(\tprop)=\frac{k_\perp(\tprop)}{\om}=\frac{\sqrt{\qhat \tprop}}{\om}
\end{equation}
The different regimes will depend on the typical time $\tprop$ of propagation. We assume a highly energetic initial parton with $E\gtrsim\omc$.
\begin{enumerate}
 \item If $\om\sim E$ corresponding to the leading parton case, the latter will most likely survive after the medium-induced evolution. Hence, its propagation time through the medium is $\tprop=L$ and its final angle is typically $\th=Q_s/E$.
 \item For $\ombr\lesssim \om \lesssim \omc$, the emission is not in the multiple branching regime. Hence, it corresponds to a single BDMPS-Z like emission emitted directly by the leading parton. We call it a \textit{primary} medium-induced emission. This emission can occur everywhere inside the medium, but the typical propagation time remains the same as for the leading particle $\tprop\sim L$, so that $\th=Q_s/\om$. One sees that if $\om$ is hard-enough, such emissions can remain inside the jet which is defined through a typical opening angle $R$. The criterion is $\om\ge Q_s/R$.
 \item In the multiple branching regime $\om\lesssim \ombr$, the propagation time is precisely the branching time $\tbr$ defined in \ref{subsub:iteration}. Hence, the typical angle of these emissions is 
 \begin{equation}
  \th(\tbr(\omega))=\Big(\frac{\qhat}{\abar^2\omega^3}\Big)^{1/4}
 \end{equation}
 Thus, these soft emissions are most likely deviated at large angles, outside the jet cone. The corresponding criterion is $\om\lesssim 
(\qhat/(\abar^2R^4))^{1/3}\equiv\om_s(R)$. This answers the question which started this section: the energy $\epsilon_{i,\textrm{flow}}$ accumulated in the condensate (formally at $\om=0$) is an energy deviated at very large angles and is consequently always a component of the total energy loss by a medium-induced jet whatever its opening angle is.
\end{enumerate}

From this general picture, one can find a better estimate of the energy loss $\epsilon_i(R)$ by a leading parton with energy $E\gtrsim \omc$ outside a given angle $R$, taking into account both the semi-hard primary large angle emissions and the soft emissions with $x_s(R)\ge\om_s(R)/E$ \cite{Fister:2014zxa}. Actually, one can perfectly neglect the latter contribution as the energy contained in the spectrum between $0$ and $\ombr/E$ is much smaller than the energy accumulated in the condensate. Hence, to the ``turbulent'' component of the jet energy loss given by \eqref{energy-flow}, one must add
the (average) energy taken away by semi-hard gluons whose energies are
larger than $\omega_\text{br}$, yet small enough for the associated propagation angles $\theta\sim
Q_s/\omega$ to be larger than $R$.
The (average) semi-hard contribution to the energy loss is therefore
obtained by integrating the emission spectrum over $\omega$ up to
$\bar\omega\equiv c_*Q_s/R$, with $c_*$ a number smaller than
one. We will see later that an average value for it is $c_*=\sqrt{\pi}/3$.
\begin{equation}\label{Espec}
\epsilon_{i,\rm spec}(E,R)\,\simeq\frac{\alpha_sC_i}{\pi} \int_0^{\bar\omega}\rmd\omega\,
\sqrt{\frac{2\omega_c}{\omega}}\,\textrm{e}^{-\upsilon_0\frac{\ombr}{E}}\,=\,
\frac{2\alpha_sC_i}{\pi}\om_c\sqrt{\frac{2c_*\theta_c}{R}}
\,\textrm{e}^{-\upsilon_0\frac{\ombr}{E}}\,,
\end{equation}
where we have also used $Q_s=\theta_c\om_c$. Note that the above
expression uses the BDMPS-Z spectrum dressed by multiple branchings
computed under the same assumptions as~\eqref{energy-flow} yielding an extra
exponential factor~\cite{Blaizot:2013hx}.
The $R$-dependence here is easy to understand: with increasing $R$,
more and more semi-hard emissions are captured inside the jet
so the energy loss is decreasing.
Note that when $R$ becomes as small as $\theta_c$, all the MIEs are
leaving the jet and the average energy loss by the jet coincides with
that of the leading parton. 

The total energy lost by the jet under the present assumptions is
\begin{equation}\label{eloss-parton2}
\boxed{\epsilon_{i,\text{MIE}}=\epsilon_{i,\rm flow}+\epsilon_{i,\rm
  spec}}
\end{equation}
where the subscript ``MIE'' indicates that for the time
being only MIEs are included. Formula \eqref{eloss-parton2} will be very important to understand the energy loss by a jet including both vacuum-like emissions and medium-induced emissions.


To conclude this discussion on the energy loss by a parton at large angles, let us comment on the Casimir $i$ dependence of $\epsilon_{i,\textrm{flow}}$. The main observation of the previous section obtained for gluon-initiated jets can be generalized to quark-jets as well \cite{Mehtar-Tani:2018zba}: the small-$x$ gluon distribution within a jet initiated by a parton of colour representation $R$ develops a scaling behaviour with $1/\sqrt{x}$, the Kolmogorov-Zakharov (or ``turbulent'') fixed point. For large-enough initial jet energy $E$ this scaling spectrum is identical to the
BDMPS-Z spectrum created by a single emission, hence proportional to the colour representation of the initial parton and to  $\hat q=\qhat_A$ which is the {\it gluonic} jet quenching parameter,
proportional to $C_A$. Consequently, the scale $\ombr$ that appears in Eq.~\eqref{energy-flow} depends on $i$ through the following relation:
\begin{equation}\label{ombr-R}
 \boxed{\ombr^{(i)}=\frac{\alpha_s^2}{\pi^2}C_iC_A\qhat L^2}
\end{equation}
One actually gets a factor $\alpha_s C_i/\pi$ associated with the
emission from the leading parton, whereas the other coupling
$\alpha_s C_A/\pi$ refers to the turbulent energy flux of the emitted
gluons and carried away at large angles.
 
\subsubsection{C/A declustering of medium-induced jets}
\label{subsub:zg-mie}

As an exercise relying on the ideas exposed in the last paragraph, we want to show the following property for medium-induced jets dominated by \textit{primary} medium-induced emissions, that is in the regime $\omega_s(R)\ge\ombr$: the generating functional of a jet constituted by primary medium-induced emissions ordered in time is equal to the generating functional of a fictitious jet with primary emissions ordered in angle and emission rate properly found in order to satisfy this equivalence. 

This fictitious ordering is exactly what a C/A declustering would reconstruct from final states particles. As many substructure observables in the vacuum rely on C/A declusterings, such as $z_g$ or ISD, this statement provides an insight on the values of such observables for medium-induced jets, despite the lack of physical angular ordering. 

To prove this statement, the first step is to rewrite \eqref{Zmie-transverse} using the $z\leftrightarrow1-z$ symmetry to integrate over $z$ values between 0 and $1/2$. As only primary emissions are considered, one replaces the generating functional of the soft subjet by another \textit{independent} generating functional $Z_{a,\textrm{prim}}(t_0,zE,zp_\perp)$ describing primary emissions:
\begin{align}
 \nonumber-\frac{\partial \Zmie_i}{\partial t_0}&=\sum_{(a,b)}\int_0^{1/2}\dif z\,\sqrt{\frac{\qhat}{E}}\frac{\alpha_s\mathcal{K}_i^{ab}(z)}{2\pi\sqrt{z(1-z)}}\Big[Z_{a,\textrm{prim}}(t_0,zE,zp_\perp)\Zmie_b(t_0,(1-z)E,(1-z)p_\perp)\\
 &\hspace{3.1cm}-\Zmie_i(t_0,E,p_\perp)\Big]+\int \dif^2 q_\perp\,\mathcal{C}_i(q_\perp,t_0)\Zmie_i(t_0,E,p_\perp-q_\perp)\label{Zmie-transverse2}
\end{align}
To simplify the discussion, we ignore quark/gluon mixing. We then neglect both the longitudinal and the transverse recoils of its 4-momentum, that is $(1-z)E\simeq E$, $(1-z)p_\perp\simeq p_\perp$ and $q_\perp\ll p_\perp$. This is valid as long as $E\gg\om_c$ since in this case, the splitting fraction $z$ carried away by a primary medium-induced emission is at most $\omc/E\ll1$. In the multiple soft scattering approximation, neglecting $q_\perp\sim Q_s$ amounts to neglect $Q_s$ in front of $p_\perp$. As the integral of $\mathcal{C}_i(q_\perp,t_0)$ over $q_\perp$ is 0 by symmetry, the collision term for the hard branch vanishes. Note that it is relatively easy to take into account the transverse multiple collisions for the hard branch as well, but it is beyond the goal of this calculation.

With these approximations, the master equations \eqref{Zmie-transverse2} for $\Zmie$ is exactly solvable given the initial condition $\Zmie_i(L,E,p_\perp)=u(E,p_\perp)$:
\begin{equation}\label{Zmie-sol}
 \Zmie_i=u(E,p_\perp)\exp\left(\int_{0}^L\dif t_0\int_0^{1/2}\dif z\,\sqrt{\frac{\qhat}{E}}\frac{\alpha_s\mathcal{K}_i^{gi}(z)}{2\pi\sqrt{z(1-z)}}\big(Z_{g,\textrm{prim}}(t_0,zE,zp_\perp)-1\big)\right)
\end{equation}
where we have set $t_0=0$. We must find the generating functional $Z_{g,\textrm{prim}}(t_0,E,p_\perp)$. By definition of primary emissions, we can ignore subsequent splittings. Consequently, $Z_{g,\textrm{prim}}$ follows \eqref{Zmie-transverse} without the first term associated with branchings:
\begin{equation}
 -\frac{\partial Z_{g,\textrm{prim}}(t_0,E,p_\perp)}{\partial t_0}=\int \dif^2 q_\perp\,\mathcal{C}_g(q_\perp,t_0) Z_{g,\textrm{prim}}(t_0,E,p_\perp-q_\perp)
\end{equation}
The solution of this equation with the initial condition $Z_{g,\textrm{prim}}(L,E,p_\perp)=u(E,p_\perp)$ is
\begin{equation}
 Z_{g,\textrm{prim}}(t_0,E,p_\perp)=\int\dif^2k_\perp u(E,k_\perp)\mathcal{P}(k_\perp-p_\perp,t_0,L)
\end{equation}
where $\mathcal{P}(q_\perp,t,L)$ is the probability density for a gluon to acquire a transverse momentum $q_\perp$ during its propagation between $t_0$ and $L$. An explicit calculation of $\mathcal{P}(q_\perp,t,L)$ is given in Section~\ref{sub:Amed-stat}, Eq.~\eqref{Ptransverse}. As $\mathcal{P}$ is normalized to 1, one can write \eqref{Zmie-sol} as
\begin{align}
 \Zmie_i(E,0\mid u)=u(E,0)\exp\left(\int \dif\theta \int_0^{1/2}\dif z \frac{\dif^2K^{\mie}_{i,C/A}}{\dif\theta\dif z}\big(u(zE,k_\perp)-1\big)\right)
\end{align}
This form is precisely the generating functional for a Poisson process ordered in angles, provided that the corresponding rate $K^{i,\mie}_{C/A}$ verifies the relation:
\begin{equation}\label{def-Kca}
\frac{\dif^2K^{\mie}_{i,C/A}}{\dif\theta\dif z}=\int_0^L\dif t_0\int\dif^2k_\perp\,\mathcal{P}(k_\perp,t_0,L)\sqrt{\frac{\qhat}{E}}\frac{\alpha_s\mathcal{K}_i^{gi}(z)}{2\pi\sqrt{z(1-z)}}\delta\left(\th-\frac{|k_\perp|}{z(1-z)E}\right)
\end{equation}
This relation shows the correspondence between the time-average of the probability density for a parton to get a given angle and the angular rate associated with the fictitious angular-ordered branching process.
In the multiple soft scattering regime, and within the harmonic approximation, one can calculate exactly the rate $K^{i,\mie}_{C/A}$ for a medium with constant $\qhat$:
 \begin{tcolorbox}[ams align]\label{KCAmultiple}
 \frac{\dif^2K^{\mie}_{i,C/A}}{\dif\theta\dif z}&=\frac{\alpha_s}{2\pi}\sqrt{\frac{\qhat L^2}{E}}\frac{\mathcal{K}_i^{gi}(z)z^2(1-z)^2E^2}{\sqrt{z(1-z)}}\frac{2\th}{Q_s^2}\Gamma\left(0,\frac{z^2(1-z)^2E^2\th^2}{Q_s^2}\right)\\
 &\simeq\frac{\alpha_sC_i}{\pi}\sqrt{\frac{2\omc}{E}}\frac{1}{z^{3/2}}\frac{2z^2E^2\th}{Q_s^2}\Gamma\left(0,\frac{z^2E^2\th^2}{Q_s^2}\right)\,,\qquad\textrm{for }z\ll1
 \end{tcolorbox}
\noindent The function $\Gamma(0,x)$ is the incomplete gamma function. In Chapter~\ref{chapter:jet-sub}, we will use this expression in the soft $z\ll1$ approximation to understand the phenomenological consequences of relatively hard intrajet medium-induced emissions on substructure observables, as in \cite{Mehtar-Tani:2016aco}. Finally, one can easily find the average angle $\bar{\th}$ of a primary medium-induced emission with respect to the leading parton by calculating the first angular moment of \eqref{KCAmultiple}. The result is $\bar{\th}=c_\star Q_s/\om$ with $c_\star=\sqrt{\pi}/3$.

\subsection{Longitudinally expanding medium}
\label{sub:med-expansion}

To conclude this chapter, we discuss how to include the longitudinal expansion of the medium to the medium induced cascade picture. The naive way of adding the expansion is to modify the emission rate \eqref{mie-rate} using $\qhat(t)$ instead of a fixed $\qhat$:
 \begin{equation}\label{mie-rate-expanding}
  \frac{\dif^2\mathcal{P}^{bc}_{i,\textrm{mie}}}{\dif z\dif t}=\frac{\alpha_s}{2\pi}\frac{\mathcal{K}_i^{bc}(z)}{(z(1-z))^{1/2}}\sqrt{\frac{\qhat(t)}{E}}\underset{z\ll1}\simeq \frac{\alpha_sC_i}{\pi}\frac{1}{z^{3/2}}\sqrt{\frac{\qhat(t)}{E}}
\end{equation}

\subsubsection{The medium-induced formation rate in expanding media}

We would like to show first that this procedure is legitimate under some circumstances. First of all, we recall that the medium-induced rate comes from the following procedure. Once one knows the medium induced spectrum of an incoming parton propagating over a path length $L$ through the medium, the rate is obtained by differentiating this spectrum with respect to $L$ (see \eqref{bdmpz-rate}). As we work within the regime $\om\ll \om_c\sim \qhat L^2/2$, the rate relevant for the master equation \eqref{Z-mie} is obtained in the limit $L\rightarrow \infty$. This large path length limit corresponds to the physical situation $L\gg t_f(\om)\sim\sqrt{2\om/\qhat}$. We have calculated in Chapter~\ref{chapter:emissions} the medium induced gluon spectrum in longitudinally expanding media for a finite jet path length, see~\eqref{Nmie} and Appendix~\ref{app:B}:
\begin{equation}\label{mie-spec-gen}
 \om\frac{\dif N_{\mie}}{\dif \om}=\frac{2\alpha_sC_R}{\pi}\mathfrak{Re}\log\Big(|C(t_0,t_0 +L|\Big)
\end{equation}
where $t_0$ is the light cone time at which the medium is created. 


As mentioned when we did this calculation, the limit $L\rightarrow\infty$ of \eqref{mie-spec-gen} is generally not defined. However, the rate can always be defined after differentiation with respect to $L$:
\begin{equation}\label{mie-rate-exp}
 \om\frac{\dif^2 N_{\mie}}{\dif \om\dif t}=\frac{2\alpha_sC_R}{\pi}\mathfrak{Re}\,\frac{-i\qhat(t)G^{-1}(t)}{2\om}
\end{equation}
From the definition of the function $G$, one shows that it satisfies the following first order (non linear) differential equation, of Riccati kind:
\begin{equation}\label{G-riccati}
 G'+G^2+\frac{i\qhat(t)}{2\om}=0
\end{equation}
The initial condition for $G$ is $G(t)\sim 1/(t-t_0)$ for $t\rightarrow t_0$. We just have to extract from this defining differential equation the asymptotic behaviour of $G(t)$. 

As a comment before finding an equivalent of $G$ as $t\rightarrow\infty$, we recall that one could also perfectly define the medium induced spectrum from the on-shell spectrum as in Section~\ref{subsub:integrated-BDMPS}, leading to (see also Appendix \ref{app:B}):
\begin{equation}\label{mie-spec-gen2}
 \om\frac{\dif \tilde{N}_{\mie}}{\dif \om}=\frac{2\alpha_sC_R}{\pi}\mathfrak{Re}\log\Big(|C(t_0+L,t_0|\Big)
\end{equation}
from which one easily deduces the corresponding rate:
\begin{equation}\label{mie-rate-exp2}
 \om\frac{\dif^2 \tilde{N}_{\mie}}{\dif \om\dif t}=\frac{2\alpha_sC_R}{\pi}\mathfrak{Re}\,\bar{G}(t)
\end{equation}
The function $\bar{G}$ satisfies the same non-linear differential equation as $G$, with a different initial condition $\bar{G}(t_0)=0$. For large values of $t$, the asymptotic behaviour of $\bar{G}$ is the same as the one of $G$. We will comment on the differences between \eqref{mie-rate-exp2} and \eqref{mie-rate-exp} after having found the common asymptotic behaviour of these two functions.

\paragraph{Asymptotic behaviour of $G$ and $\bar{G}$.} We do not claim full mathematical rigour in the demonstration of the following results. Instead, we shall give hand-waving arguments to convince of their validity. In order to make the discussion also more concrete, we use the following form for the time dependence of $\qhat$, inspired by the Bjorken expansion model:
\begin{equation}\label{qhat-assumption}
 \qhat(t)=\qhat_0\Big(\frac{t_0}{t}\Big)^{\gamma}
\end{equation}
In the Bjorken model, $0<\gamma\le 1$. In this discussion, $\gamma$ can take any non-negative values: $\gamma\ge0$. We also invite the reader to look at Fig.~\ref{Fig:G-Gbar} to guide the eye through the following mathematical discussion: on the left figure, the asymptotes of the rates are clear. We first define the following functions:
\begin{equation}
 g(x)=\sqrt{\frac{2\om}{i\qhat_0}}G\left(\sqrt{\frac{2\om}{i\qhat_0}}x\right)\,,\qquad u(x)=g(1/x)
\end{equation}
The function $g$ enables to nondimensionalize the differential equation \eqref{G-riccati}, and the function $u$ is useful to study the large $x$ asymptote of $g(x)$ by looking at the $x\rightarrow0^+$ behaviour of $u(x)$. They satisfy the following differential equations:
\begin{align}
 &g'(x)+g^2(x)+\kappa^\gamma x^{-\gamma}=0\,\qquad \kappa=\sqrt{\frac{i\qhat_0t_0^2}{2\om}}\label{diff-gxka}\\
 &u'(x)=\left(\frac{u(x)}{x}\right)^2+\kappa^\gamma x^{\gamma-2}\label{diff-uxka}
\end{align}
Considering Eq.~\eqref{diff-uxka}, it is natural to distinguish two cases, either $\gamma<2$ or $\gamma>2$:\footnote{For $\gamma=2$, it is possible to solve exactly \eqref{diff-gxka} and one finds that $g(x)\sim(1+\sqrt{1-4\kappa^2})/(2x)$ for $x\rightarrow\infty$.}
\begin{enumerate}
 \item If $\gamma>2$, the second term in \eqref{diff-uxka} proportional to $x^{\gamma-2}$ goes gently to 0 as $x\rightarrow0^+$. We then expect to find the behaviour of $u(x)$ in this limit by solving the vacuum differential equation $u'=(u/x)^2$. The general solution is $u(x)=x/(1-bx)$ with $b$ a constant, so that $u(x)\sim x$ at small $x$. Thus, for $\gamma>2$, $g(x)\sim1/x$ at large $x$ and $G(t)\sim 1/t$ at large times. This is the typical vacuum-like behaviour since $G(t)=1/(t-t_0)$ when $\qhat=0$, meaning that the rate in the presence of a medium with $\gamma>2$ decreases as in the vacuum at large times. We point out that neglecting the $(u/x)^2$ term in \eqref{diff-uxka} instead of the $x^{\gamma-2}$ term would reduce to an absurdity. Indeed, one would get $u'(x)\sim \kappa^\gamma x^{\gamma-2}$. However, in the limit $\kappa\rightarrow0$, this gives $u'(0)=0$ ($u$ is continuous as a function of $\kappa$) in contradiction with the exact solutions found for $\kappa=0$ which predict $u'(0)=1$. Thus, the vacuum solution of \eqref{diff-uxka} strongly constrains the small $x$ asymptote of $u$ when $\gamma>2$.
 \item If $\gamma<2$, the second term on the right hand side of \eqref{diff-uxka} explodes as $x\rightarrow0^+$. Let us \textit{assume} that there exists some power $\alpha\ge0$ such that $u(x)\sim c(\kappa)\, x^\alpha$ at small $x$, with $c(\kappa)$ a number without dimension. To say it differently, we assume that the rate functions $G$ and $\bar{G}$ decrease like a power law at large times. Plugging this power law behaviour inside the differential equation satisfied by $u$, one finds:
\begin{equation}\label{diff-powerlaw}
 c(\kappa) \,x^{\alpha-1}-c^2(\kappa) \,x^{2(\alpha-1)}\simeq\kappa^\gamma x^{\gamma-2}\,,\qquad x\ll1
\end{equation}
One cannot have $\alpha\ge 1$, otherwise the left hand side is convergent in the limit $x\rightarrow0^+$ whereas the left hand side diverges. Moreover, for $\alpha<1$, $x^{\alpha-1}$ is negligible in front of $x^{2(\alpha-1)}$ in the same limit. Consequently, the small $x$ behaviour of $u$ satisfies
\begin{equation}
 -c(\kappa)^2x^{2(\alpha-1)}\sim \kappa^\gamma x^{\gamma-2}\Longrightarrow u(x)\sim (-\kappa x)^{\gamma/2}
\end{equation}
This gives the following large time asymptote for $G$:
 \begin{equation}\label{g-asymptotic}
  G(t)\sim\sqrt{-\frac{i\qhat_0}{2\om}}\Big(\frac{t_0}{t}\Big)^{\gamma/2}
 \end{equation}
\end{enumerate}

\paragraph{Conclusion.} To sum up, if $\gamma>2$, the medium dilutes too fast so $G(t)\sim 1/t$ has a vacuum-like decay law, whereas if $\gamma<2$, $G(t)\sim\sqrt{-i\qhat_0/(2\om)}(t_0/t)^{\gamma/2}$. The function $\bar{G}(t)$ follows the same property since it satisfies the same differential equation and the arguments presented above do not rely on the initial condition for $G$. Therefore, if $\qhat(t)$ decreases at large times slower than $1/t^2$, the large time behaviour of the medium-induced rate using \eqref{g-asymptotic} in \eqref{mie-rate-exp} or \eqref{mie-rate-exp2} is of the form:
\begin{tcolorbox}[ams equation]
 \om\frac{\dif^2 N_{\mie}}{\dif \om\dif t}\underset{t\rightarrow\infty}\simeq\frac{\alpha_sC_R}{\pi}\sqrt{\frac{\qhat(t)}{\om}}
\end{tcolorbox}
\noindent which is precisely the form conjectured from our naive modification of the rate in \eqref{mie-rate-expanding}. 

We point out that if $\gamma>2$, the integral of the rate \eqref{mie-rate-exp2} defined from the on-shell spectrum is divergent due to the vacuum-like $1/t$ tail at large times whereas the rate \eqref{mie-rate-exp} is convergent because of the additional $\qhat(t)$ factor. If $\gamma<2$, both rates agree at large times but they give a divergent time integral. In any case, for finite jet path length $L$, such considerations are meaningless since the time integral is naturally cut at $t=L$.


\paragraph{The exponential case.} In an exponentially decaying medium with $\qhat(t)=\qhat_0e^{-t/\lambda}$ (a rather academic situation), the functions $G(t)$ and $\bar{G}(t)$ behave like $1/t$ at large times since the exponential vanishes faster than any power law. The integral of \eqref{mie-rate-exp} over \textit{all times} is convergent because the additional factor $\qhat(t)$ cuts the integral for $t\ge \lambda$. However, the finite jet path length $L$ is a priori different from the time scale $\lambda$ and the medium induced spectrum should be obtained by integrating the rate up to $t=L$ and not over all times. This can give substantial differences for the medium-induced spectrum if $\lambda \gg L$, a situation where the jet path length is much smaller than the typical decaying time $\lambda$ of the medium. In this case, one can of course use $\qhat(t)\simeq \qhat_0$ in \eqref{G-riccati} as long as $t\ll\lambda$. The emission rate for $\sqrt{2\om/\qhat_0}\ll t\ll \lambda$ behaves like the rate of a static medium with $\qhat$ given by $\qhat_0$.

\subsubsection{Scaling properties of medium-induced jet fragmentation}

Now, we assume that $\qhat(t)$ vanishes slower than $1/t^2$ ($\gamma<2$) as this is the most phenomenologically relevant scenario. Using the dimensionless time $\tau_{\rm exp}$ defined such that:
\begin{equation}\label{reduced-time-exp}
 \frac{\dif\tau_{\rm exp}}{\dif t}=\abar\sqrt{\frac{\qhat(t)}{E}}\Longrightarrow \tau_{\rm exp}=\int_{t_0}^t\dif t'\,\abar\sqrt{\frac{\qhat(t')}{E}}
\end{equation}
in the master equation \eqref{Z-mie} or \eqref{Zmie-forward} with the modified rate \eqref{mie-rate-expanding}, these equations reduce to those of a static medium written in terms of the dimensionless time $\tau=\abar t\sqrt{\qhat_0/E}$. This means, for instance, that the solution of \eqref{frag-mie-evol} given by \eqref{Dmie-sol} is still valid, provided that $\tau$ is replaced by $\tau_{\rm exp}$.

The equation \eqref{reduced-time-exp} enables to answer easily the following question: for a given longitudinal expansion (that is a given $\qhat(t)$ dependence), what is the equivalent static medium with constant $\qhat_{\rm eff}$ such that the fragmentation of medium-induced jet is the same in both cases? 
This equivalent $\qhat_{\rm eff}$ is obtained in order to satisfy the following scaling identity:
\begin{equation}
 \boxed{\tau_{\rm exp}(t_0+L)=\tau(L)}
\end{equation}
so that $\qhat_0$ reads:
\begin{tcolorbox}[ams equation]\label{qhat-equiv}
 \qhat_{\rm eff}=\left(\frac{1}{L}\int_{t_0}^{t_0+L}\dif t'\,\qhat(t')^{1/2}\right)^2
\end{tcolorbox}
\noindent Mathematically, $\qhat_{\rm eff}$ is equal to the $1/2$-norm of the function $\qhat(t)$. This allows for a straightforward generalization of all our analytical results obtained in the static case, provided that one uses $\qhat_{\rm eff}$ defined by \eqref{qhat-equiv} in these results. Written explicitly, the relation between $\qhat_{\rm eff}$ and $\qhat_0$ for a given path length $L$ is:
\begin{align}\label{qhat-equiv-bjork}
 \qhat_{\rm eff}&=\frac{4\qhat_0t_0^\gamma}{L^2(2-\gamma)^2}\Big((t_0+L)^{\frac{2-\gamma}{2}}-t_0^{\frac{2-\gamma}{2}}\Big)^2\\
 &\simeq \frac{4\qhat_0}{(2-\gamma)^2}\Big(\frac{t_0}{L}\Big)^\gamma\qquad\textrm{ if }L\gg t_0
\end{align}
We emphasize that these results are valid as long as one focuses on soft emissions in the medium-induced cascades with $\om\ll\om_{c,\rm eff}=\qhat_{\rm eff}L^2/2$. Beyond this approximation, there exists anoter effective $\qhat$ found numerically in \cite{Salgado:2002cd} defined as 
\begin{equation}
 \qhat'_{\rm eff}=\frac{2}{L^2}\int_{t_0}^{t_0+L}\dif t' (t'-t_0)\qhat(t')
\end{equation}
that provides a better scaling of the BDMPS-Z spectrum in the hard regime $\om\gtrsim\om_c$. Consequently, if the full BDMPS-Z rate \eqref{mie-rate-exp} or \eqref{mie-rate-exp2} is used in the master equation \eqref{Z-mie}, instead of its soft singular limit, the scaling properties of medium-induced cascades are better when considering this effective quenching parameter $\qhat'_{\rm eff}$ \cite{Adhya:2019qse}.


%% file: chapter4.tex
\chapter{A new factorised picture for jet evolution in a dense medium}
\chaptermark{A new factorised picture}
\label{chapter:DLApic}

In Chapter~\ref{chapter:jet}, we have treated separately the vacuum-like emissions triggered by the virtuality, or ``off-shellness'', of the incoming parton and the medium-induced emissions. The leading order calculations as those done in Chapter~\ref{chapter:emissions} combine them on equal footing, but their associated resummation schemes described in Chapter~\ref{chapter:jet} are intrinsically different for the following reasons:
\begin{enumerate}
 \item the large parameter which needs to be resummed to all orders is not the same. It is typically $\alpha_s\log^2(Q_{\textrm{hard}}/Q_{\textrm{min}})$ for a vacuum-like shower between the hard virtuality scale $Q_{\textrm{hard}}$ and a soft virtuality scale $Q_{\textrm{min}}$ --- depending on the observable --- whereas it is typically $\alpha_s (L/t_{f,\med})$ for a medium-induced cascade, with $t_{f,\med}$ the formation time of a medium-induced emission.
 \item the variable playing the role of time in the classical branching process which effectively resums to all orders the large parameter is the angle for vacuum-like emissions and the light-cone time for medium-induced radiations.
\end{enumerate}

In this chapter, we develop a resummation scheme combining both vacuum-like and medium-induced emissions. This scheme relies on physical arguments which are based on the double logarithmic approximation for the vacuum-like series. Beyond the double logarithmic approximation (DLA), one argues that vacuum-like and medium induced emissions factorise in time from each other.

Throughout this chapter, we assume that the dominant medium induced processes are multiple soft collisions so we work within the multiple soft scattering regime and harmonic approximation. This amounts to neglect hard collisions with medium constituents and assume that the mean free path of the medium is small enough compared to the typical formation time of medium-induced emissions.

\section{The veto constraint}
\label{sec:veto-region}

At double-logarithmic accuracy, one aims at resumming to all-orders contributions to a given observable enhanced by both large collinear and soft logarithms. These logarithms are generated by multiple Bremsstrahlung processes with characteristic spectrum: 
\begin{equation}\label{brem}
 \dif^3N^{\brem}=\frac{\alpha_sC_R}{\pi^2}\frac{\dif\om}{\om}\frac{\dif^2 k_\perp}{k_\perp^2}
\end{equation}
for an emission with transverse momentum $k_\perp$ with respect to the emitter, and energy $\om$\footnote{In this chapter, we shall generically call $\om$ the light cone $k^+$ component of a 4-momentum $k^\mu$ and identify it with its energy.}.

Consequently, in the presence of a dense QCD medium, the first task to do before any resummed calculation is to understand where the contributions to the $N$-particle production cross-section enhanced by large soft and collinear logarithms come from. This is the topic of this section. 

The main result of this analysis is the existence of a so-called veto constraint in the phase space for gluon emissions inside the medium. This veto constraint marks the transition between the virtuality driven initial shower and the regime where the partons in the shower can be considered as on their mass shell.  We first give a qualitative but physically insightful argument for the existence of this veto constraint before providing a more mathematically grounded formulation of the proof.

\subsection{Qualitative discussion}
\label{sub:qualitative-veto}
 
To simplify the discussion, we shall consider that the cross-section for multi-particle production 
is controlled by only two medium parameters: the distance $L$ travelled by the jet inside the medium and the jet quenching parameter $\hat{q}$ related to the average 
transverse momentum acquired by multiple collisions during time $\Delta t$ by $\langle k_\perp^2\rangle=\hat{q}\Delta t$. The BDMPS-Z spectrum
$\dif N^{\mie}$ can be approximated by the formula \eqref{bdmpsz-brick},
\begin{equation}
 \dif N^{\mie}\simeq \frac{\alpha_sC_R}{\pi} \sqrt{\frac{2\om_c}{\omega^3}} \dif \omega
\end{equation}
with $\om_c=\qhat L^2/2$.
With respect to the Bremsstrahlung spectrum, the obvious property of the BDMPS-Z spectrum is that collinear and soft radiations are not \textit{logarithmically} enhanced (the angular dependence of the spectrum is roughly a Gaussian distribution with width of order $Q_s$). This will 
have important consequences since working at the double-log accuracy enables to simply ignore such emissions in the multi-particle cross-section.

\paragraph{Vacuum-like emissions inside the medium: the veto constraint.} Medium induced radiations provide a natural constraint on the phase space available for one emission inside the medium.

To see that, we first consider that the medium is very large $L\gg 1/\LQCD$ so that one first neglects all effects related to the finite path length of the jet. 
We recall that $t_f$ is determined by the uncertainty principle, namely the condition that the transverse separation 
$\Delta r \sim \theta t_f$ between the gluon and its parent parton at the time of emission to be as large as the gluon transverse wavelength
$2/k_\perp$, with $k_\perp\simeq \omega\theta$ its transverse momentum with respect to its parent. This argument applies to both vacuum-like
and medium-induced emissions and implies $t_f\simeq 2\omega/k_\perp^2\simeq 2/(\omega\theta^2)$.  Then, gluons emitted inside the medium
have a minimum $k_\perp$ set by the momentum acquired via multiple collisions during its formation, $k_{f,\med}^2=\hat{q}t_f$. Gluons produced inside the medium 
with a transverse momentum smaller than $\hat{q}t_f$ cannot exist. This translates into an upper limit $t_f\le\sqrt{2\omega/\hat{q}}$
on the formation time of any gluon inside the medium. That said, medium induced gluons for which $k_\perp\simeq k_{f,\med}$ are excluded because the emission probability 
is not enhanced by double logarithms. Consequently, at DLA, the only contribution to the intrajet activity inside the medium comes from 
vacuum-like (Bremsstrahlung) emissions with $k_\perp \ge k_{f,\med}$, inequality which becomes strong to our level of accuracy: $k_\perp \gg k_{f,\med}$ or equivalently 
$t_f\ll \sqrt{2\omega/\hat{q}}$. This means that VLEs occur much faster than medium-induced radiations with the same energy. This is the condition for an emission \textit{inside} the medium to be vacuum-like. Thereafter, we shall refer to this condition as the ``in-medium veto constraint''.

\paragraph{Finite path length effects.} Strictly speaking, the veto constraint $k
_\perp\gg k_{f,\med}$ applies for very large path length $L\gg1/\LQCD$. For a realistic path length $L$ in the medium, one should also take into account emissions \textit{triggered by the initial virtuality} but occurring directly outside with a formation time \textit{measured from the hard process} larger than $L$: $t_f\ge L$. Note also that emissions with energies larger than $\om_c$, $\omega\ge\omega_c=\qhat L^2/2$, behave exactly as
in the vacuum: their emission angle can be arbitrarily small in the perturbative domain and their formation time can be larger than $L$.

These two conditions for VLEs, either inside with $t_f\ll \sqrt{2\omega/\hat{q}}$ or outside with $t_f\gg L$ can be written 
in terms of their energies $\omega$ and angles with respect to the emitter $\theta$:
\begin{tcolorbox}[ams align]
\mbox{\textcolor{red}{In-medium veto constraint:}  }\hspace{0.5cm}&t_f\ll \sqrt{2\omega/\hat{q}}&\Longleftrightarrow&\hspace{0.5cm}\omega\gg\omega_0(\theta)=(2\hat{q}/\theta^4)^{1/3}\label{inmed-veto}\\
\mbox{\textcolor{blue}{Medium size constraint:}  }\hspace{0.5cm}&t_f\gg L&\Longleftrightarrow&\hspace{0.5cm}\omega\ll\omega_L(\theta)=2/(\theta^2L)\label{outmed-crit}
\end{tcolorbox}
\begin{figure}
   \centering
      \includegraphics[width=0.6\textwidth]{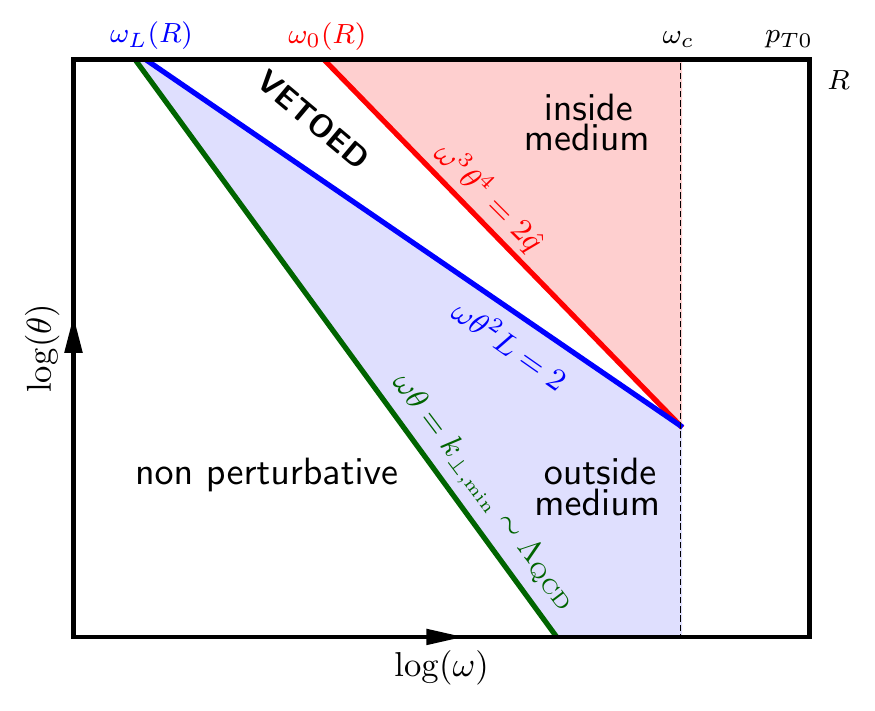}
    \caption{\small Phase space for vacuum-like emissions at DLA considered in this thesis, represented in terms of the energy $\om$ and the angle $\th$ (with respect to the parent) of emission. The red line is the lower boundary of the ``in-medium'' region defined by the veto constraint $k_\perp\ge k_{f,\med}$. The blue line is the line of vacuum formation time equal to the jet path length $L$ through the medium: $t_f=L$. The green line is the hadronisation line $k_\perp=\LQCD$ providing the lower boundary of the perturbative phase space. Between the red and the blue line, the phase space gap appearing is the vetoed region discussed in the main text.}
    \label{Fig:DLA-phase-space1}
\end{figure}
\noindent It is enlightening to represent these conditions on a $(\omega,\theta)$ phase space as in Fig.~\ref{Fig:DLA-phase-space1}. The red curve is the line $k_\perp=k_{f,\med}$ and the blue curve is the line $t_f=L$. Compared to the vacuum case for which the phase space is 
only restricted by the hadronisation line $k_\perp\simeq\omega\theta=\LQCD$, the effect of the medium on the phase space available for VLEs is the presence of a gap or ``vetoed region''. 

\subsection{Derivation of the veto constraint from leading order calculations}
\label{sub:veto-anaytic}
Now, we would like to give a more formal proof of the existence of a veto constraint in the phase space for VLEs. The ultimate goal of this subsection are formulas \eqref{veto-off-form} and \eqref{vle-spec-inf}.

We have seen in Chapter~\ref{chapter:emissions} that the spectrum of an off-shell quark can be written:
\begin{equation}\label{off-shell-recap}
 \om\frac{\dif^3N^{\textrm{off-shell}}}{\dif \om \dif^2k_\perp}=\frac{\alpha_s C_F}{\pi^2}\frac{1}{k_\perp^2}+ \om\frac{\dif^3N^{\textrm{\mie}}}{\dif \om \dif^2k_\perp}
\end{equation}
where the second piece is perfectly integrable over all $k_\perp$ and gives precisely the BDMPS-Z spectrum. Even if such a decomposition is not unique, the integral of $\dif^3N^{\textrm{off}}$ gives always the logarithmic area of the full perturbative phase space plus a piece without collinear logarithms corresponding to the integrated medium-induced spectrum.

We will argue that the $1/k_\perp^2$ term in the expression \eqref{off-shell-recap} comes from different physical mechanisms, not necessarily related to the initial virtuality of the process. Through detailed analytical calculations which have been put in Appendix~\ref{app:off-on} for the sake of brevity, we disentangle these different contributions. We show that {\tt (i)} in an infinite static medium, the transverse momenta of emissions \textit{at formation} is bounded by $k_{f,\med}$, {\tt (ii)} it is physically more insightful to decompose the off-shell spectrum \eqref{off-shell-recap} in another way. In this new decomposition, the first term corresponds to the VLEs with short formation time constrained by the veto $k_\perp>k_{f,\med}$ whereas the second term is ``medium-induced''. We finally argue that Bremsstrahlung emissions with $t_f\gg L$ contained in the latter component are the only one that must be resummed to all orders in a leading twist approximation.

In all this subsection, we assume a static medium and $\om\le \om_c$ since all the interesting features of the medium appear in this regime.

\subsubsection{The off-shell spectrum at formation in the infinite path length limit}

We first emphasize that the above discussion of the phase space for vacuum-like emissions refers only to their kinematic at formation. Final state broadening and energy loss subsequently move the emissions on this phase space. This leads to the concept of off-shell spectrum \textit{at formation}: in this spectrum, one removes by hand the final state broadening of the emitted gluon by setting:
\begin{equation}\label{broadening-delta}
 \mathcal{P}(k_\perp-q_\perp,t_1,t_2)=\delta(k_\perp-q_\perp)
\end{equation}
In the next subsection, we will relax this approximation and take also into account the final state broadening.

Let us then rewrite the off-shell spectrum obtained in \eqref{offshell-final} (more precisely \eqref{Iiidouble}) where the final state broadening appears as a convolution:
\begin{align}\label{off-double-v2}
 \om\frac{\dif^3 N^{\textrm{off-shell}}}{\dif\om\dif^2 k_\perp}&=\frac{\alpha_s C_F}{\pi^2}2\mathfrak{Re}\int_0^\infty\dif y^+\,e^{-\varepsilon y^+}\int\dif^2q_\perp \mathcal{P}(k_\perp-q_\perp,y^+,\infty)\nonumber\\ &\hspace{6cm}\times \frac{q_\perp^2}{(2i\om)^2}\int_0^{y^+} \dif\bar{y}^+\,\frac{-e^{-\varepsilon\bar{y}^+}}{C^2(\bar{y}^+,y^+)}e^{\frac{-q_\perp^2S(y^+,\bar{y}^+)}{2i\om  C(\bar{y}^+,y^+)}}
\end{align}
where the position of the hard vertex is here $0$.
Using \eqref{broadening-delta} in \eqref{off-double-v2}, one finds:
\begin{align}\label{off-formation}
 \om\frac{\dif^3 N^{\textrm{off-shell,f}}}{\dif\om\dif^2 k_\perp}&=\frac{\alpha_s C_F}{\pi^2}\frac{k_\perp^2}{(2\om)^2}2\mathfrak{Re}\int_0^\infty\dif y^+\int_0^{y^+}\dif\bar{y}^+\,e^{-\varepsilon(y^+ +\bar{y}^+)} \frac{1}{C^2(\bar{y}^+,y^+)}e^{\frac{-k_\perp^2S(y^+,\bar{y}^+)}{2i\om  C(\bar{y}^+,y^+)}}
\end{align}
where the superscript ``$\rm f$'' is put to highlight the difference between this off-shell spectrum at formation and the complete off-shell spectrum.

Instead of elucidating the functions $S$ and $C$ for a static medium (they are given in Appendix~\ref{app:B} though), we show that for $y^+-\bar{y}^+\ll t_{f,\med}\equiv\sqrt{2\om/\qhat}$, these two functions have their vacuum-like form, that is:
\begin{align}\label{S-C-vac}
 S_{\rm vac}(y^+,\bar{y}^+)&=y^+-\bar{y}^+\\
 C_{\rm vac}(\bar{y}^+,y^+)&=1
\end{align}
To proceed, we find, thanks to the defining differential equation \eqref{diff-eq-S} the Taylor expansion of these functions when $y^+$ is close to $\bar{y}^+$ (short formation times). We will use the same method to derive the veto constraint in an expanding plasma.
From the differential equation \eqref{diff-eq-S}, the Taylor expansion of $S$ and $C$ around $y^+=\bar{y}^+$ gives:
\begin{align}
 S(y^+,\bar{y}^+)&=(y^+-\bar{y}^+)-\frac{1}{6}\frac{i\qhat}{2\om}(y^+-\bar{y}^+)^3+\mathcal{O}\big((y^+-\bar{y}^+)^4\big)\\
 C(\bar{y}^+,y^+)&=1-\frac{1}{2}\frac{i\qhat}{2\om}(y^+-\bar{y}^+)^2+\mathcal{O}\big((y^+-\bar{y}^+)^3\big)
\end{align}
Thus, when $t_f=|y^+-\bar{y}^+|\ll t_{f,\med}$ one can use $C(\bar{y}^+,y^+)\simeq 1$ and $S(y^+,\bar{y}^+)\simeq y^+-\bar{y}^+$, and neglect the sub-leading terms in the Taylor expansion. In the off-shell spectrum at formation, one then split the integral into two pieces: one with $y^+\le t_{f,\med}$ (where $y^+-\bar{y}^+\le t_{f,\med}$ since $\bar{y}^+\le y^+$), and the other with $y^+\ge t_{f,\med}$ but $y^+-t_{f,\med}\le \bar{y}^+\le y^+$. Using the vacuum approximations for the functions $S$ and $C$ and keeping only the terms with a leading $1/k_\perp^2$ singularity, this calculation gives:
\begin{align}
 \om\frac{\dif^3 N^{\textrm{off-shell,f }t_f\le t_{f,\med}}}{\dif\om\dif^2 k_\perp}&\simeq
 \frac{\alpha_s C_F}{\pi^2}2\mathfrak{Re}\int_0^{t_{f,\med}}\dif y^+\frac{1}{2i\om}\Big(-1+e^{\frac{-k_\perp^2y^+}{2i\om}}\Big)\nonumber\\
 &+\frac{\alpha_s C_F}{\pi^2}2\mathfrak{Re}\int_{t_{f,\med}}^\infty\dif y^+\,e^{-2\varepsilon y^+}\frac{\varepsilon}{k_\perp^2}\Big(-1+e^{\frac{-k_\perp^2t_{f,\med}}{2i\om}}\Big)
\end{align}
In this calculation, it is again very important to keep track of the $\varepsilon$ dependence up to the first order in the piece with $y^+$ unbounded. The second integral over $y^+$ then gives a factor $1/2$ in the limit $\varepsilon\rightarrow0$. The first integral is easy and gives the same value as the second piece with an extra factor $-2$, leading to the final result:
\begin{equation}
 \om\frac{\dif^3 N^{\textrm{off-shell,f }t_f\le t_{f,\med}}}{\dif\om\dif^2 k_\perp}\simeq
 \frac{\alpha_s C_F}{\pi^2}\frac{1}{k_\perp^2}\mathfrak{Re}\Big(1-e^{\frac{-k_\perp^2t_{f,\med}}{2i\om}}\Big)
\end{equation}
in the collinear limit. This spectrum is suppressed for $|k_\perp^2t_{f,\med}/(2i\om)|\ll 1$ and averages to 1 for $|k_\perp^2t_f/(2i\om)|\gg 1$, giving precisely the in-medium veto constraint:
\begin{equation}\label{veto-off-form}
 \om\frac{\dif^3 N^{\textrm{off-shell,f }t_f\le t_{f,\med}}}{\dif\om\dif^2 k_\perp}\simeq
 \frac{\alpha_s C_F}{\pi^2}\frac{1}{k_\perp^2}\Theta(k_\perp-k_{f,\med})
\end{equation}

To sum up, the off-shell spectrum at formation has a Bremsstrahlung component for $k_\perp\gg k_{f,\med}$ coming from the short formation time emissions with $t_f=y^+-\bar{y}^+\ll t_{f,\med}$. To prove the result asserted in our qualitative discussion of the in-medium veto constraint, one should also show that in the infinite path length limit, it is the \textit{only} Bremsstrahlung-like component. This is easy. In the infinite path length limit, the function $S$ and $C$ for a static medium are given by
\begin{align}
 S(y^+,\bar{y}^+)&=\Omega^{-1}\sin\Big(\Omega(y^+-\bar{y}^+)\Big)\\
 C(\bar{y}^+,y^+)&=\cos\Big(\Omega(y^+-\bar{y}^+)\Big)
\end{align}
with $\Omega^2=i\qhat/(2\om)$. Thus, the phase inside the exponential of formula \eqref{off-formation} reads:
\begin{equation}
 \frac{S(y^+,\bar{y}^+)}{C(\bar{y}^+,y^+)}=\bar{\Omega}^{-1}\tanh\Big(\bar{\Omega}(y^+-\bar{y}^+)\Big)\underset{y^+-\bar{y}^+\ge t_{f,\med}}\simeq \bar{\Omega}^{-1}
\end{equation}
which does not lead to a collinear singularity at small $k_\perp$:
\begin{align}\label{mie-spec-atformation}
 \om\frac{\dif^3 N^{\textrm{off-shell,f } t_{f,\med}\le t_f}}{\dif\om\dif^2 k_\perp}&\simeq\frac{\alpha_s C_F}{\pi^2}\frac{k_\perp^2}{(2\om)^2}2\mathfrak{Re}\,\left[\frac{\mathcal{N}}{\Omega^2}\exp\left(-\frac{1-i}{\sqrt{2}}\frac{k_\perp^2}{k_{f,\med}}\right)\right]\\
 \mathcal{N}&=\Omega^2\int_{t_{f,\med}}^{\infty}\dif y^+\int_0^{y^+-t_{f,\med}}\dif\bar{y}^+\frac{e^{-\varepsilon(y^++\bar{y}^+)}}{C^2(\bar{y}^+,y^+)}
\end{align}
where $\mathcal{N}$ is a $k_\perp$ independent dimensionless factor which diverges in the limit $\varepsilon\rightarrow0$ since for an infinite medium path length, the amount of induced radiations blows up in a static medium\footnote{Rigorously speaking, we have made the assumption that $\qhat$ vanishes at sufficiently large times to derive the off-shell spectrum derived in Chapter \ref{chapter:emissions}, so one may wonder if our reasoning and especially formula \eqref{off-double-v2} remain valid for an infinite path length through a static medium. Actually, one can assume that $\qhat(t)$ vanishes very slowly (the criterion is slower than $1/t^2$ at large times, see Section \ref{sub:med-expansion}), so that formula \eqref{off-double-v2} applies as well as the results \eqref{veto-off-form} and \eqref{mie-spec-atformation} with $t_{f,\med}$ and $k_{f,\med}$ calculated in Section \ref{sub:veto-expansion}. Of course, even in this case, the factor $\mathcal{N}$ is still divergent when $\varepsilon\rightarrow0$, but the $k_\perp$ dependence of the spectrum we find is right.}.
Thus, for $t_f>t_{f,\med}$, the off-shell spectrum at formation is strongly peaked around $k_\perp\simeq k_{f,\med}$ due to the exponential suppression of the form $\exp(-k_\perp^2/k_{f,\med})$ in \eqref{off-formation}. In physical terms, in the infinite medium length limit: either the emission spectrum is vacuum-like with a double logarithmic singularity at small $k_\perp$ and $\om$ but constrained by $k_\perp\gg k_{f,\med}$, or it saturates the condition $k_\perp=k_{f,\med}$. However, in the latter case, the spectrum has no collinear singularity. In this formal limit, no emission occurs with $k_\perp\le k_{f,\med}$.

\subsubsection{Broadening and finite path length effects}

The previous discussion aimed at proving the existence of the red line in the phase space (at formation) for vacuum-like emissions shown in Fig.~\ref{Fig:DLA-phase-space1}. To discuss the rest of this phase space, beyond the infinite path length limit, we find extremely convenient to define the following quantity, that we call ``vacuum-like spectrum in a dense medium'':
\begin{equation}\label{vle-spec-def}
 \om\frac{\dif^3 N^{\textrm{VL}}}{\dif \om\dif^2k_\perp}\equiv \om\frac{\dif^3N^{\textrm{off-shell}}}{\dif\om\dif^2 k_\perp}-\om\frac{\dif^3N^{\textrm{on-shell}}}{\dif\om\dif^2 k_\perp}
\end{equation}
\begin{figure}
   \centering
      \includegraphics[width=0.6\textwidth]{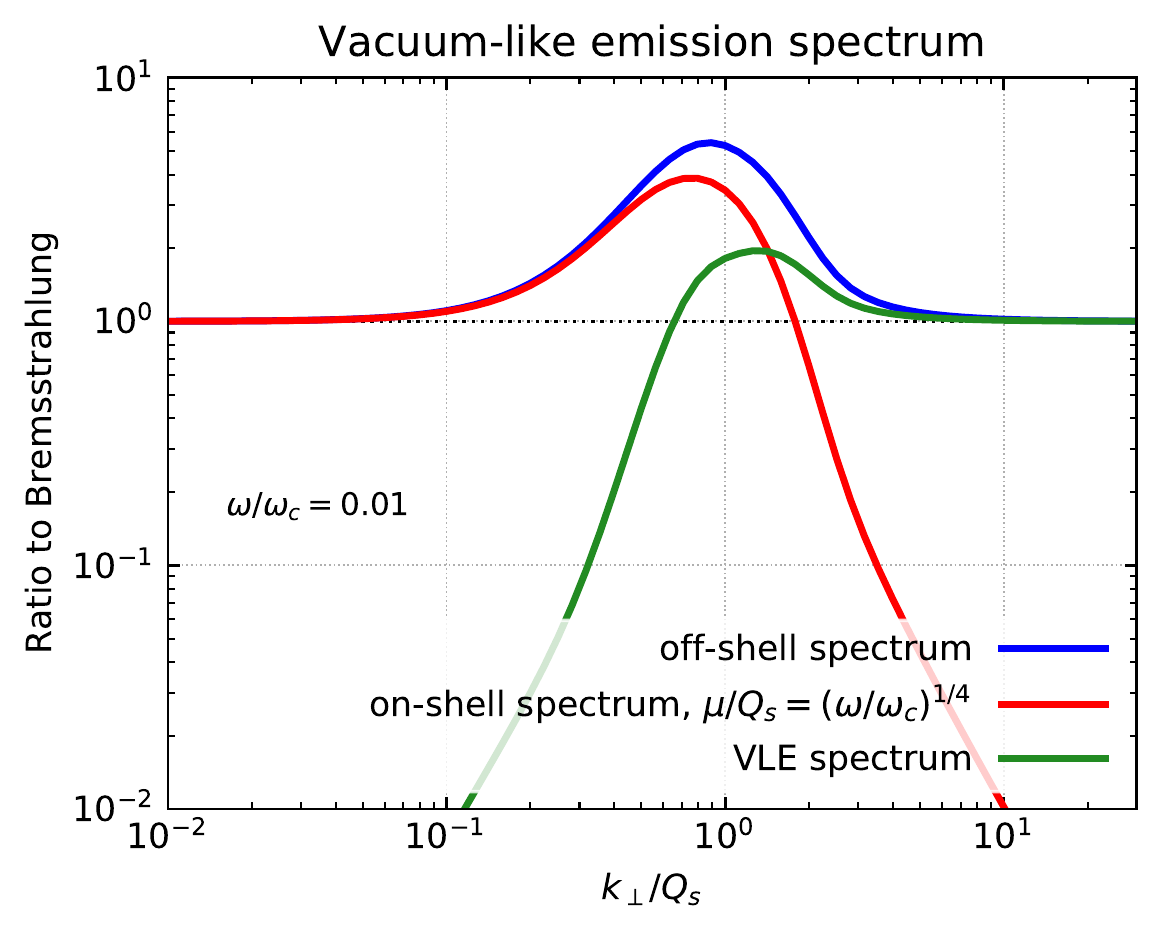}
    \caption{\small Vacuum-like emission spectrum \eqref{vle-spec-def} (green curve) as a function of $k_\perp/Q_s$ for a static medium, compared to the off-shell (blue curve) and on-shell (red curve) spectra. The spectra are normalized by the vacuum Bremsstrahlung spectrum. The choice $\mu/Q_s=(\om/\om_c)^{1/4}$ for the regularisation of the on-shell cross-section corresponds to $\mu=k_{f,\med}$. At small $k_\perp$, the off-shell and the on-shell spectra coincide.}
    \label{Fig:VLEspec}
\end{figure}
This definition naturally subtracts all the emissions which are not related to the initial virtuality of the parton. Then, the off-shell spectrum is rewritten:
\begin{tcolorbox}[ams align]\label{off-new}
 \om\frac{\dif^3 N^{\textrm{off-shell}}}{\dif \om\dif^2k_\perp}=\om\frac{\dif^3 N^{\textrm{VL}}}{\dif \om\dif^2k_\perp}+\om\frac{\dif^3N^{\textrm{on-shell}}}{\dif\om\dif^2 k_\perp}
\end{tcolorbox}
\noindent This expression has a clear physical interpretation: the first term is to be associated with emissions triggered by the virtuality of the leading parton, whereas the second term corresponds to emissions which are all ``medium-induced'' since an on-shell quark would not radiate in the vacuum. Nevertheless, in spite of being medium-induced, they could have a Bremsstrahlung-like form. This is what we want to understand.

The expression \eqref{vle-spec-def} has an explicit $\mu$ regularization dependence, remnant of the $\mu$ dependence of the on-shell cross section discussed in Chapter \ref{chapter:emissions}, Section \ref{subsub:TMdep-BDMPS}. Of course, by construction this $\mu$ dependence cancels in the off-shell spectrum \eqref{off-new} so we could choose $\mu$ as we want. The crucial point here is that there is a choice of $\mu$ which makes the physical content of \eqref{off-new} transparent. We now justify this choice of $\mu$.

In QCD, as strictly on-shell partons do not exist, the meaning of ``on-shellness'' is \textit{relative to a given virtuality scale}. For instance, if the virtuality of an incoming parton is much smaller than the hard scattering scale, it makes sense to consider it as an idealized on-shell asymptotic state. In the scattering of an on-shell quark with a dense medium that we are discussing now, the minimal momentum transferred in the process is of order $k_{f,\med}$ so the ``on-shellness'' of the quark is relative to this scale.

With this perspective, we point out that $\mu$ is of order of the small virtuality of the incoming almost on-shell parton. This is clear from the ``onium regularization'' studied in Section \ref{subsub:onium-scatt}. The divergence is regulated by the frozen transverse size $X_\perp$ of the onium: $\mu \sim 1/X_\perp$, itself related to the virtuality $Q$ of the photon sourcing the onium by $X_\perp\sim 1/Q$. 
Consequently, for the calculation \eqref{vle-spec-def} to make sense, the small virtuality $\mu$ of the on-shell incoming quark should not be larger than $k_{f,\med}$. For reasons that will become clear afterwards, we choose $\mu$ that saturates this constraint, i.e.\ $\mu=k_{f,\med}$: 
\begin{equation}
  \om\frac{\dif^3 N^{\textrm{VL}}}{\dif \om\dif^2k_\perp}\equiv \om\frac{\dif^3N^{\textrm{off-shell}}}{\dif\om\dif^2 k_\perp}-\om\frac{\dif^3N^{\textrm{on-shell}}}{\dif\om\dif^2 k_\perp}\Big|_{\mu=k_{f,\med}}
\end{equation}
This is somehow our working hypothesis: \textit{the scale $k_{f,\med}$ must be considered as the transition scale between ``offshellness'' and ``onshellness'' in the presence of a dense medium}.

With this definition, we show in Appendix~\ref{app:off-on} that the leading behaviour of \eqref{vle-spec-def} in the soft and collinear limit is:
\begin{tcolorbox}[ams align]\label{vle-spec-inf}
 \om\frac{\dif^3 N^{\textrm{VL}}}{\dif \om\dif^2k_\perp}&\simeq\frac{\alpha_sC_F}{\pi^2}\int\frac{\dif^2q_\perp}{q_\perp^2}\mathcal{P}(k_\perp-q_\perp,0,L)\Theta(|q_\perp|-k_{f,\med})
\end{tcolorbox}
\noindent This expression has a clear probabilistic picture for vacuum-like emissions in a dense medium in agreement with our initial qualitative discussion. These emissions have a very short formation time $t_f\ll \sqrt{2\om/\qhat}$ and are created \textit{inside} so that they subsequently undergo momentum broadening over the full path length $L$. The medium only restricts the phase space so that only emissions with $k_\perp>k_{f,\med}$ are allowed. This is precisely the veto constraint on in-medium vacuum-like emissions discussed in the previous subsection, with the final state broadening of the emitted gluon included.

The spectrum \ref{vle-spec-inf} does not account for the finite path length $L$ of the off-shell parton in the medium and especially the vacuum-like radiations outside with $t_f\gg L$ discussed in section \eqref{sub:qualitative-veto}. Actually, these emissions are all included in the on-shell term in \eqref{off-new}! In mathematical terms, the on-shell spectrum with $\mu=k_{f,\med}$ reads (see Appendix~\ref{app:off-on} for the detailed calculations): 
\begin{align}\label{on-shell-physic}
 \om\frac{\dif^3N^{\textrm{on-shell}}}{\dif\om\dif^2 k_\perp}\simeq\textrm{``BDMPS-Z''}+\frac{\alpha_sC_F}{\pi^2} \frac{1}{k_\perp^2}\Theta(k_{f,\med}-k_\perp)
\end{align}
The BDMPS-Z term is a $k_\perp$ integrable term that gives the BDMPS-Z spectrum once integrated over $k_\perp$. The second term has the Bremsstrahlung form, but is bounded from above by $k_\perp\le k_{f,\med}$. Such a contribution is expected, as the interactions with the medium put the initially on-shell quark off its mass-shell, so that it radiates afterwards according to the Bremsstrahlung law. What is somehow unexpected is that the upper boundary is given by $k_{f,\med}$ and not $Q_s$, the maximal transverse momentum acquired via multiple soft collisions during the propagation of the quark through the medium over a time $L$.

Now the question of the \textit{iteration} or resummation of these Bremsstrahlung-like emissions appearing in \eqref{on-shell-physic} arises. We remind the reader that we are looking for \textit{leading twist} modifications of the intrajet radiation pattern, that is emissions triggered by the virtuality of the parent parton and not by the scattering with the medium. In this sense, the spectrum \eqref{on-shell-physic} \textit{alone} is not leading-twist since the initial quark is on its mass shell, and therefore should not be iterated. However, in \eqref{off-new}, this on-shell spectrum is a component of the off-shell one.  This suggests that there are also leading twist emissions in the second term of \eqref{off-new}.

We now justify that only the contributions with $t_f\gg L$, that is $k_\perp^2\ll2\om/L$ should be considered as leading twist in the second term of \eqref{on-shell-physic}. There is clearly a competition between two kinds of Bremsstrahlung in the single gluon emission spectrum from an off-shell quark:
\begin{itemize}
 \item the Bremsstrahlung associated with the initial large virtuality of the quark,
 \item the Bremsstrahlung associated with the scattering of the quark with the medium, giving a smaller virtuality of order $k_{f,\med}$ to the quark. This explains the presence of the scale $k_{f,\med}$ in \eqref{on-shell-physic}.
\end{itemize}
In our language, the former is said ``leading twist'', the latter is ``higher twist''. To simplify the discussion, let us consider that the full medium acts as a \textit{single scattering center} transferring a typical momentum $k_{f,\med}$ and located at light cone time $L$. One knows that there exist two regimes according to the ratio between $L$ and $t_f=2\om/k_\perp^2$:
\begin{enumerate}
 \item If $t_f\ll L$, the Bremsstrahlung radiations associated with the initial virtuality and the final state radiations associated with the scattering add up \textit{incoherently}. With the constraint $t_f\ll L$, the support of the Bremsstrahlung spectrum associated with the initial virtuality is $k_\perp\gg k_{f,\med}$ since we have just shown that in-medium leading twist Bremsstrahlung are bounded from below by $k_{f,\med}$. The support of the Bremsstrahlung associated with the scattering is constrained by $(2\om/L)^{1/2}\ll k_\perp\ll k_{f,\med}$. Hence, there is no overlap between the two mechanisms in the phase space $k_\perp^2\gg 2\om/L$.
 \item If $t_f\gg L\Leftrightarrow k_\perp^2\ll 2\om/L$, the two Bremsstrahlung \textit{interferes}, as in the LPM effect, and cannot be distinguished from each other. Physically, the wavelength of the gluon is so large that it does not resolve the source of the virtuality of its parent. Nevertheless, this is a leading twist effect which must be iterated.
\end{enumerate}
The interested reader should go to Appendix \ref{app:off-on} to check these results in the simple case of the shockwave limit of the off-shell spectrum. In this calculation, the medium-induced spectrum defined as in \eqref{off-shell-recap} reads:
\begin{equation}
  \om\frac{\dif N^{\textrm{mie, SW}}}{\dif\om}\simeq\frac{2\alpha_sC_F}{\pi}\log\left(\frac{LQ_s^2}{2\om}\right)
\end{equation}
with $L$ the light-cone time location of the shockwave. This is precisely the logarithmic area of the vetoed region in the shockwave limit where the transverse scale $k_{f,\med}$ is replaced by $Q_s$. This shows that a higher twist effect can produce logarithms in the multiple soft scattering approximation, at the level of the single gluon spectrum.

The final decomposition of the off-shell spectrum relevant for the resummation performed in the next section is then:
\begin{align}
 \om\frac{\dif^3 N^{\textrm{off-shell}}}{\dif \om\dif^2k_\perp}&=\underbrace{\frac{\alpha_sC_F}{\pi^2}\int\frac{\dif^2q_\perp}{q_\perp^2}\mathcal{P}(k_\perp-q_\perp,0,L)\Theta(|q_\perp|-k_{f,\med})+\frac{\alpha_sC_F}{\pi^2} \frac{1}{k_\perp^2}\Theta(2\om/L-k_\perp^2)}_{\textrm{leading-twist}}\nonumber\\
 &\hspace{3cm}+\underbrace{\textrm{``BDMPS-Z''}+\frac{\alpha_sC_F}{\pi^2} \frac{1}{k_\perp^2}\Theta(k_{f,\med}-k_\perp)\Theta(k_\perp^2-2\om/L)}_{\textrm{higher twist}}
\end{align}
We believe that this decomposition is much more physical (as far as one is concerned with the resummation of vacuum-like emissions in the presence of a dense medium) than the naive one \eqref{off-shell-recap} from which we have started this section.

The conclusion to drawn from this discussion is that an apparent ``Bremsstrahlung'' form for an emission spectrum is not a sufficient condition to iterate such emissions in the leading twist approximation. The latter condition imposed, i.e.\ if one aims at resumming only emissions driven by the high virtuality of the process, the phase space for vacuum-like emissions is the one described in Section \ref{sub:qualitative-veto}.
That said, we leave for further studies the incorporation of this medium-induced Bremsstrahlung component in the phase space $2\om/L\ll k_\perp^2\ll k^2_{f,\med}$ to our global pQCD picture (see discussion in~\ref{sub:beyondDLA}).

\subsection{Veto constraint in longitudinally expanding media}
\label{sub:veto-expansion}
\subsubsection{The general case}

The calculations made in the previous subsection in order to obtain the veto constraint for VLEs in a dense static medium can be easily generalised for longitudinally expanding media. The method is straightforward: in the leading order off-shell spectrum \textit{at formation}, one finds the leading soft and collinear singularity coming from the integration domain $|y^+-\bar{y}^+|\le t_{f,\med}$, for some $t_{f,\med}$, where the functions $S$ and $C$ have their vacuum form. The leading $1/k_\perp^2$ behaviour extracted in this way will have the signature of the constraints on VLEs inside the medium. 
%

We first consider the case where the off-shell parton is created at light cone time $t_0$ inside the medium. Then the medium expands longitudinally. The easiest way to do the calculation is to use the expression \eqref{off-formation} and the Taylor expansion method. In an expanding medium, $\qhat$ depends on $y^+$:
\begin{align}\label{taylor-exp}
 S(y^+,\bar{y}^+)&=(y^+-\bar{y}^+)-\frac{1}{6}\frac{i\qhat(\bar{y}^+)}{2\om}(y^+-\bar{y}^+)^3+\mathcal{O}\big((y^+-\bar{y}^+)^4\big)\\
 C(\bar{y}^+,y^+)&=1-\frac{1}{2}\frac{i\qhat(y^+)}{2\om}(\bar{y}^+-y^+)^2+\mathcal{O}\big((\bar{y}^+-y^+)^3\big)
\end{align}
Note that one can also use $\qhat(y^+)$ in the Taylor expansion of $S$ since the difference $\qhat(y^+)-\qhat(\bar{y}^+)$ is of order $\mathcal{O}((y^+-\bar{y}^+))$ if $\qhat(t)$ is derivable.

The condition to use $C(y^+,\bar{y}^+)\simeq 1$ and $S(y^+,\bar{y}^+)\simeq y^+-\bar{y}^+$ is:
\begin{equation}
 (y^+-\bar{y}^+)^2\le (2\om)/\qhat(y^+)\equiv t^2_{f,\med}(y^+)
\end{equation}
In the expanding case, the only extra difficulty is that the medium induced formation time depends on time.
Let us call $t^\star_{f,\med}(\om)$ the smallest solution of the equation $y^+=t_0+t_{f,\med}(y^+)$ larger than $t_0$ (if such solution exists):
\begin{equation}\label{def-tfmed-star}
t_{f,\med}^\star-t_0=\sqrt{\frac{2\om}{\qhat(t_{f,\med}^\star)}}
\end{equation}
By construction, one has $y^+-t_{f,\med}(y^+)\le t_0$ for $y^+\le t^\star_{f,\med}$ and $y^+-t_{f,\med}(y^+)\ge t_0$ for $y^+\ge t^\star_{f,\med}$ (at least in a neighbourhood of $t^\star_{f,\med}$).

Consequently, coming back to \eqref{off-double-v2}, the integral over $y^+$ can be split into two pieces: $y^+\le t^\star_{f,\med }$ and $y^+\ge t^\star_{f,\med}$. The two pieces give the following result in the soft and collinear limit:
\begin{align}\label{expanding-veto-calculation}
 \om\frac{\dif^3 N^{\textrm{off-shell,f }t_f\le t_{f,\med}}}{\dif\om\dif^2 k_\perp}&\simeq
 \frac{\alpha_s C_F}{\pi^2}2\mathfrak{Re}\int_{t_0}^{t^\star_{f,\med}}\dif y^+\frac{1}{2i\om}\Big(-1+e^{\frac{-k_\perp^2(y^+-t_0)}{2i\om}}\Big)\nonumber\\
 &\hspace{-1.5cm}+\frac{\alpha_s C_F}{\pi^2}2\mathfrak{Re}\int_{t^\star_{f,\med}}^\infty\dif y^+\,e^{-2\varepsilon y^+}\frac{\varepsilon}{k_\perp^2}\Big(-1+e^{\frac{-k_\perp^2t_f(y^+)}{2i\om}}\Big)
\end{align}
With respect to the static case, the second term is weird because $t_{f,\med}(y^+)\rightarrow\infty$ at large times in a diluting medium. This is because the condition $y^+-\bar{y}^+\le t_{f,\med}(y^+)$ does not make sense when $y^+$ is large. To avoid this difficulty, we take our cue from the static calculation and fix $t_{f,\med}(y^+)$ to $t_{f,\med}(t_{f,\med}^\star)=t_{f,\med}^\star-t_0$ in the second term of \eqref{expanding-veto-calculation}\footnote{For a cleaner proof, we show in Appendix~\ref{app:off-on} that the second term in \eqref{expanding-veto-calculation} is actually irrelevant to get the veto constraint. The latter can always be read in the first term alone. The price to pay is that a factor $2$ appears in front of the spectrum, but this factor $2$ cancels when using the spectrum \eqref{vle-spec-def}.}.
Then, as in the fixed $\qhat$ case, the piece with $y^+\ge t^\star_{f,\med}(\om)$ gives a factor $1/2$ and the first term can be integrated explicitly:
\begin{equation}\label{veto-expansion}
  \om\frac{\dif^3 N^{\textrm{off-shell,f }t_f\le t_{f,\med}}}{\dif\om\dif^2 k_\perp}\simeq \frac{\alpha_s C_F}{\pi^2}\frac{1}{k_\perp^2}\Theta\left(k_\perp-\Big(\frac{2\om}{t^{\star}_{f,\med}-t_0}\Big)^{1/2}\right)
\end{equation}
The step function in \eqref{veto-expansion} gives the veto constraint on vacuum-like emissions in an expanding medium. This can be rewritten in a form similar to the static case:
\begin{tcolorbox}[ams align]\label{veto-exp}
k_\perp\ge (2\qhat^\star\om)^{1/4}
\end{tcolorbox}
\noindent with $\qhat^\star=\qhat(t^\star_{f,\med})$.

\paragraph{Illustrative example: exponential decay} The most simple time dependent $\qhat$ is the exponential decay, with $t_0=0$:
\begin{equation}
 \qhat(t)=\qhat_0\exp(-t/\lambda)
\end{equation}
In this case, the scale $t^\star_{f,\med}(\omega)$ is solution of the implicit equation:
\begin{equation}
 \exp(-t^\star_{f,\med}/\lambda)t^{\star2}_{f,\med}=2\om/\qhat_0
\end{equation}
When $\om\ll \qhat_0 \lambda^2$, i.e $t_{f,\med}^\star\ll \lambda$, one can approximate the exponential by one, so that
\begin{equation}
 t_{f,\med}^\star(\om)\sim \sqrt{\frac{2\om}{\qhat_0}}
\end{equation}
and then, the transverse momentum of vacuum-like emissions inside the medium is such that $k_\perp\ge (2\qhat_0\om)^{1/4}$.
Note than one would obtain the same result using the argument given in Section~\ref{sub:qualitative-veto} amended to take into account the time dependence of $\qhat$. During its formation $t_f=2\om/k_\perp^2$, an emission acquires a typical transverse momentum $k_{f,\med}$ given by:
\begin{equation}
 k^2_{f,\med}=\int_0^{t_f}\dif t\, \qhat(t)=\qhat_0 \lambda(1-\exp(-t_f/\lambda))\underset{t_f\ll \lambda}\simeq \qhat_0t_f
\end{equation}
so that the condition $k_\perp^2\ge k^2_{f,\med}$ gives also $k_\perp\ge (2\qhat_0\om)^{1/4}$.

\subsubsection{Bjorken expansion} 
\label{subsub:bjork}

For a Bjorken expansion, the time dependence of $\qhat$ is:
\begin{equation}
 \qhat(t) =\qhat_0\Big(\frac{t_0}{t}\Big)^\gamma\Theta(t-t_0)
\end{equation}
with $0\le\gamma\le 1$ for realistic expansion, and $t_0>0$ by convention (one cannot take $t_0=0$ in the general case). One must consider two different scenarii according to the location of the hard vertex: either $t_i=t_0$ or $t_i<t_0$.

\paragraph{Hard vertex at $t_i=t_0$.}

The first scenario is more simple as one can directly rely on the calculation made above. The characteristic medium formation time satisfies:
\begin{equation}\label{tstar-bjork}
 t^\star_{f,\med}-t_0=\sqrt{\frac{2\om}{\qhat_0t_0^\gamma}}t^{\star\gamma/2}_{f,\med}
\end{equation}
The extreme case $\gamma=0$, where one can safely use $t_0=0$, corresponds to a static medium. One finds again the familiar relation $t_{f,\med}^\star=\sqrt{2\om/q_0}$. The other extreme $\gamma=1$ is the ideal relativistic plasma. The solution of \eqref{tstar-bjork} is, for $\gamma=1$,
\begin{equation}\label{eq-tstar-bjork}
 t^\star_{f,\med}-t_0=\frac{1}{\qhat_0t_0}\Big(\om+\sqrt{\om^2+2\qhat_0t_0^2\om}\Big)
\end{equation}
One can then distinguish two different regimes according to the ratio between $\om$ and the energy scale $\qhat_0t_0^2/2$:
\begin{enumerate}
 \item If $\om\gg\qhat_0 t_0^2/2$, then one can neglect the second term under the square root and one gets $t^\star_{f,\med}-t_0\simeq 2\om/(\qhat_0t_0)$. The veto constraint is then:
 \begin{equation}
  k_\perp\ge\sqrt{\qhat_0t_0}
 \end{equation}
 This non trivial result comes entirely from the study of the off-shell spectrum for short formation times.
%
\item On the other hand, if $\om\ll\qhat t_0^2/2$, one gets $t_{f,\med}-t_0\simeq \sqrt{2\om/\qhat_0}$ which has the same form as in the static case with constant $\qhat_0$. Physically, these emissions have such a short formation time that they do not ``see'' the expansion of the medium. Hence, the veto constraint is the same as for a static medium with constant $\qhat_0$.
\end{enumerate}
The condition $k_\perp^2\ge \qhat_0t_0$ for $\om\ge \qhat_0t_0^2$ is interesting as the transverse  momentum scale $\sqrt{\qhat_0t_0}$ is naturally related to the saturation scale $Q_{s,A}$ of the incoming nuclei in $A-A$ collisions.

In the general $\gamma>0$ case, it also possible to get the boundary of the in-medium region for vacuum-like emissions in the two regimes elucidated for $\gamma=1$. For this, it is more convenient to write \eqref{eq-tstar-bjork} as an equation for $\qhat^\star$:
\begin{equation}
 \left(\frac{\qhat_0}{\qhat^\star}\right)^{1/\gamma}-\sqrt{\frac{2\om}{\qhat_0t_0^2}}\left(\frac{\qhat_0}{\qhat^\star}\right)^{1/2}-1=0
\end{equation}
When $\om\ll \qhat_0t_0^2/2$, one can neglect the second term in this equation, so that the solution is $\qhat^\star=\qhat_0$. The constraint on $k_\perp$ is the same as in a static medium with constant $\qhat_0$: $k_\perp\ge(2\qhat_0\om)^{1/4}$.
When $\om\gg\qhat_0t_0^2/2$, one can neglect the minus 1 term and the solution is:
\begin{tcolorbox}[ams align]\label{qstar}
 \qhat^\star=\qhat_0\left(\frac{\qhat_0t_0^2}{2\om}\right)^{\frac{\gamma}{2-\gamma}}
\end{tcolorbox}
\noindent Plugging this relation into \eqref{veto-exp} gives the boundary of the veto region for $\om\gg\qhat_0t_0^2/2$.

As a last comment, we point out that the generalisation of the condition $k_\perp\ge\qhat t_f$ for expanding media:
\begin{equation}\label{bjork-qual2}
 k_\perp^2\ge\int_{t_0}^{t_f}\dif t\, \qhat(t)
\end{equation}
gives the same power dependence of the in-medium boundary as a function of $\om$. Indeed, for $\gamma<1$,
\begin{equation}\label{bjork-qual}
 \int_{t_0}^{t_f}\dif t\, \qhat(t)=\frac{\qhat_0t_0^{\gamma}}{1-\gamma}\Big(t_f^{1-\gamma}-t_0^{1-\gamma}\Big)
\end{equation}
Then, when $t_f\simeq t_0$, one can use $t_f^{1-\gamma}-t_0^{1-\gamma}\simeq (1-\gamma)t_0^{-\gamma}(t_f-t_0)$. One can then take the limit $t_0\rightarrow0$, and one recovers the condition $k_\perp\ge(2\qhat_0\om)^{1/4}$ corresponding to the regime $\om\ll\qhat_0t_0^2/2$. This also means that this regime is the regime of formation times of the same order as $t_0$. If $t_f\gg t_0$, one can neglect the minus term in \eqref{bjork-qual} leading to 
\begin{equation}
 k_\perp^2\ge \frac{1}{1-\gamma}\qhat_0t_0^\gamma t_f^{1-\gamma}
\end{equation}
As emphasized, once $t_f$ is replaced by $\sim2\om/k_\perp^2$, this leads to the same dependence in $\om$ of the $k_\perp$ boundary (except the uninteresting prefactor). Nevertheless, the case $\gamma=1$ is ambiguous from the condition \eqref{bjork-qual2} because of a logarithmic prefactor of the form $\log(t_f/t_0)$.

\paragraph{Hard vertex at $t_i<t_0$.} We would like to understand how these results are modified once the position of the hard vertex creating the off-shell parton is located at $0<t_i<t_0$. In the limit $t_i\rightarrow t_0$, one should recover the results obtained above. 

More precisely, when $\Delta t\equiv t_0-t_i\ll\sqrt{2\om/\qhat_0}$, we expect the boundary found in the previous paragraph not to be modified since $\sqrt{2\om/\qhat_0}$ is the medium-induced time scale when $t_f\sim t_0$. Consequently, we focus on the regime $\Delta t\gg \sqrt{2\om/\qhat_0}\Leftrightarrow \om\ll\qhat_0\Delta t^2/2$.

There are two possible ways to do the calculation. The first method consists in cutting the off-shell cross-section in three pieces with $t_i\le y^+,\bar{y}^+\le t_0$ or $y^+,\bar{y}^+\ge t_0$ or $y^+\ge t_0$, $t_i\le\bar{y}^+\le t_0$. The other method consists in finding a continuous and derivable solution over the full range $[t_i,\infty] $ of the differential equation satisfied by $S$ and $C$. If this is possible, the calculation leading to \eqref{off-double-v2} and \eqref{off-formation} remains valid. It happens that such solutions exist and can be built explicitly (cf.\ Appendix \ref{app:B} for a discussion of the junction method). In this paragraph, $\gamma$ is arbitrary and all the conclusions drawn are independent of the value of $\gamma$. Performing the integral over $\bar{y}^+$ in \eqref{off-formation} using lemma \eqref{lemma}, one finds:
\begin{equation}\label{off-formation2}
  \om\frac{\dif^3 N^{\textrm{off-shell,f}}}{\dif\om\dif^2 k_\perp}=\frac{\alpha_s C_F}{\pi^2}\left[2\mathfrak{Re}\frac{1}{2i\om}\int_{t_i}^\infty \dif y^+\exp\left(\frac{-k_\perp^2 G^{-1}(y^+)}{2 i\om}\right)-\frac{1}{k_\perp^2}\right]
\end{equation}
The function $G(y^+)=C(t_i,y^+)/S(y^+,t_i)$ is continuous and derivable even in $y^+=t_0$ and verifies:
\begin{align}
 G^{-1}(y^+)&=y^+-t_i\,\qquad\mbox{ if }t_i\le y^+\le t_0
\end{align}
For $y^+\ge t_0$, one can approximate the functions $S(y^+,t_i)$ and $C(y^+,t_i)$ by their Taylor expansion around $y^+=t_0$:
\begin{align}
 S(y^+,t_i)&=y^+-t_i-\frac{i\qhat_0\Delta t}{4\om}(y^+-t_0)^2+\mathcal{O}\big((y^+-t_0)^3)\\
 C(t_i,y^+)&=1-\frac{i\qhat_0\Delta t}{2\om}(y^+-t_0)+\mathcal{O}\big((y^+-t_0)^2)
\end{align}
The continuity and derivability of the solutions $C$ and $S$ in $y^+=t_0$ are essential to get this result. Let us comment these relations. First of all, when $\Delta t\rightarrow 0$, the sub-leading term vanishes, meaning that one should go beyond the first order calculation to obtain the range of validity of the vacuum-like expressions for $S$ and $C$. This is precisely the situation studied in the case $t_i=t_0$ in the previous paragraph. Our assumption $\Delta t\gg \sqrt{2\om/\qhat_0}$ is in agreement with the fact that $\Delta t$ cannot be too small for this calculation to make sense.

That said, when 
\begin{equation}
 \frac{\qhat_0\Delta t}{2\om}(y^+-t_0)\ll 1 \Leftrightarrow y^+\ll t^\star_{f,\med}\equiv t_0+\frac{2\om}{\qhat_0\Delta t}
\end{equation}
one can neglect the $\mathcal{O}(y^+-t_0)$ terms and the calculation of \eqref{off-formation2} in the integration domain $y^+\le t^\star_{f,\med}$ gives\footnote{As we have integrated over all $\bar{y}^+$, the final result is asymmetric (see also Appendix~\ref{app:off-on} for a  discussion of this peculiarity). Anyway, the $-1/k_\perp^2$ should be decomposed into two pieces, one cancelling the factor 2 in \eqref{off-formation2} and another with $k_\perp\le (2\om)/(t^\star_{f,\med}-t_i)$ which is not related to short formation times vacuum-like emissions.}:
\begin{equation}
  \om\frac{\dif^3 N^{\textrm{off-shell,f } t_f<t_{f,\med}}}{\dif\om\dif^2 k_\perp}=\frac{\alpha_s C_F}{\pi^2}\frac{1}{k_\perp^2}\Theta\left(k_\perp-\Big(\frac{2\om}{t^\star_{f,\med}-t_i}\Big)^{1/2}\right)
\end{equation}
The step function is the signature of the constraint on in-medium Bremsstrahlung emissions. As $\Delta t\gg \sqrt{2\om/\qhat_0}$ by assumption, $t_{f,\med}^\star-t_i=\Delta t+ 2\om/(\qhat_0\Delta t)\simeq \Delta t$.
%
The veto constraint reads then:
 \begin{equation}
  k_\perp\ge \sqrt{\frac{2\om}{\Delta t}}
 \end{equation}
In this case, the vacuum-like emissions have a formation time short enough to happen entirely in the vacuum before entering into the medium, so that the veto does not depend on $\qhat_0$.
When $\om\ll \qhat \Delta t^2/2$, the scale $\sqrt{2\om/\qhat_0}$ has disappeared in this scenario, contrary to the situation where the hard scattering occurs directly inside the medium.

To sum up, in a Bjorken expanding plasma, the transverse momentum of vacuum-like emissions is constrained by:
\begin{tcolorbox}[ams equation]
 k_\perp\ge \textrm{min}\Big((2\qhat^\star(\om)\om)^{1/4},(2\om/\Delta t)^{1/2}\Big)
\end{tcolorbox}
\noindent for $\Delta t\le t_0$ and $\qhat^\star(\om)$ given by $\qhat_0$ for $\om\le \qhat_0t_0^2/2$ and \eqref{qstar} for $\om\ge \qhat_0t_0^2/2$.

\sectionmark{Resummation at DLA}
\section{Resummation at double logarithmic accuracy and in the large $N_c$ limit}
\sectionmark{Resummation at DLA}
\label{sec:DLresum}

In the previous section, we have identified the phase space for one vacuum-like emission in the presence of a dense QCD medium. In this section, we explain how such emissions are iterated. The main obstacle with respect to a straightforward generalisation of the coherent branching algorithm is the decoherence induced by the medium, studied in Section \ref{sub:decoherence}. As we have seen, decoherence does not forbid soft emissions outside the opening angle of a colour singlet antenna. This means that the coherent branching algorithm fails to correctly resum to all orders soft and collinear emissions for a jet evolving in the medium.

We shall see, however, that one can build an effective branching algorithm for VLEs amended to take into account the effect of medium decoherence. The main results are:
\begin{enumerate}
 \item the branching process is angular ordered inside the medium,
 \item angular ordering can be violated for the first emission outside the medium.
\end{enumerate}
These statements are demonstrated within the double logarithmic approximation (DLA) of pQCD and in the large $N_c$ limit. In the last section of this chapter, we provide arguments in order to extend the validity of the branching process beyond DLA and the large $N_c$ limit.

\subsection{Angular ordering inside the medium}

For simplicity, we consider a jet which starts as a colour-singlet
quark-antiquark antenna with a small opening angle $\thqq \ll 1$,
e.g. produced by the decay of a boosted $W/Z$ boson or a virtual photon.
It also applies to the evolution of a quark or gluon jet,
provided one replaces $\thqq$ by either the jet radius or the angle of
the first emission.
The quark and the antiquark are assumed to have equal energies:
$E_q=E_{\bar q}\equiv E$.
As emphasized, we focus on the double-logarithmic approximation (DLA) where
subsequent emissions are strongly ordered in both energies and emission
angles. Furthermore, in the large $N_c$ limit, the antenna is the only degree of 
freedom in this context as the emission of gluon from the initial $q\bar{q}$ antenna is equivalent 
to the splitting of this antenna into two other antennae.

In this subsection, we assume that the medium is infinite, with constant quenching parameter $\qhat_0$. We relax this approximation in the next subsection. In this limit, the phase space available for vacuum-like splittings from the initial $q\bar{q}$ pair is (see \eqref{inmed-veto} or \eqref{vle-spec-inf}):
\begin{equation}\label{tfvac}
 k_\perp^2\gg\sqrt{\qhat_0\om}\quad\Longleftrightarrow \quad\om\gg \Big(\frac{2\qhat_0}{\th^4}\Big)^{1/3}\equiv\om_0(\th)
\end{equation}
where the inequalities are stong at DLA. $\om$ is the energy of the soft daughter antenna and $\th$ its opening angle.

\paragraph{Colour decoherence.} For emissions by a colour-singlet antenna, even a vacuum-like emission
obeying \eqref{tfvac} could be still affected by the medium, via {\em
  colour decoherence} \cite{MehtarTani:2010ma,MehtarTani:2011tz,CasalderreySolana:2011rz,CasalderreySolana:2012ef}.
In the vacuum, gluon emissions at large angles $\theta\gg \thqq$ are
suppressed by the destructive interferences between the quark and the
antiquark. This argument can be iterated to conclude that successive
emissions in the vacuum are {\em ordered in angles},
$\theta_{i+1} \lesssim\theta_i$, an ordering which becomes {\em
  strong} ($\theta_{i+1} \ll \theta_i$) at DLA. But, as seen in Chapter~\ref{chapter:emissions}-Section \ref{sub:decoherence}, an antenna propagating through a
dense quark-gluon plasma can lose its coherence via rescattering off
the medium: the quark and the antiquark suffer independent colour
rotations, hence the probability that the antenna remains in a colour
singlet state decreases with time.
The two legs of the antenna start behaving like independent
colour sources after the time $t\sim t_{\rm coh}$ defined in \eqref{tcoh}
\begin{equation}\label{tdec}
  t_{\rm coh}(\thqq)\,\equiv\,\left(\frac{4}{\hat
      q\thqq^2}\right)^{1/3}
\end{equation}
In principle, then, emissions with formation time $t_f=2/(\om\th^2)$ larger than $t_{\rm coh}$ could violate the angular ordering property. 

Now comes the crucial point. We show that to the order of interest, 
the first emission by the leading $q\bar{q}$ still has an angle $\th$ smaller than $\thqq$, despite the 
potential decoherence of the parent antenna. To see this, note that
\begin{equation}\label{exp}
\frac{t_f}{t_{\rm coh}}
\,=\,
\frac{(2\hat q\thqq^2)^{1/3}}{\omega\theta^2}
\,=\,\frac{\omega_{0}(\theta)}{\omega}\,
     \left(\frac{\thqq}{\theta}\right)^{2/3}
\end{equation}
\noindent where $t_f$ is the formation time of the daughter antenna. 
The loss of colour coherence may only affect the emissions at
sufficiently large angles, $\theta\gtrsim \thqq$, which overlap with
both sources.
For VLEs satisfying (\ref{tfvac}), this implies
${t_f}\ll{t_{\rm coh}}$ meaning that the antenna is still coherent at
the time of the emission and the would-be large-angle emissions
are killed by the interference.
In fine, only emissions with $\theta\lesssim\thqq$ are allowed whether
or not they occur at times larger than the decoherence time~(\ref{tdec}).

The attentive reader will notice that this time scale argument is based on the ratio $t_f/\tcoh$ and not $t_f\th/(\tcoh\thqq)$ as one would expect from our discussion of the decoherence factor calculated in Section~\ref{sub:decoherence} (see \eqref{decoherence-vle-simple} and \eqref{decoherence-factor1}). Nevertheless, this more refined order parameter was derived in the regime $\om\gg\om_c$ and $k_\perp\gg Q_s$. In this regime, using the equality
\begin{equation}
\frac{\tcoh}{t_f}\frac{\thqq}{\th}=\frac{\om}{\om_c}\left(\frac{L}{\tcoh}\right)^{2}\left(\frac{\th}{\thqq}\right),
\end{equation}
one shows that the argument inside the Bessel function in \eqref{decoherence-factor1} is large for in-medium VLEs with $\th\gtrsim\thqq$, and thus large angle emissions are suppressed ($L/\tcoh\gg1$ in the formal infinite medium length limit we are discussing). The question of whether the right order parameter is $t_f/\tcoh$ or $t_f\th/(\tcoh\thqq)$ in the regime $\om_0(\th)\ll\om\ll\om_c$ deserves further studies from first principle calculations as those performed in Section~\ref{sub:decoherence}.

These arguments, given for the first emission from the leading $q\bar{q}$ pair, can be easily generalised by induction for any vacuum-like cascade in the phase space \eqref{tfvac}. The leading logarithmic behaviour of in-medium parton showers comes
from cascades which are strongly ordered in energies and angles,
i.e. from cascades with $n$ VLE's satisfying
$\thqq\gg\theta_1\gg\cdots \gg\theta_n$ and
$E\gg\omega_1\gg\cdots\gg\omega_n\gg\omega_{0}(\theta_n)$.
First, note that the formation times $t_i=2/(\omega_i\theta_i^2)$ are
strongly increasing from one emission to the next. 
%
It means that the condition~\eqref{tfvac} is satisfied by all the
gluons in the cascade if it is satisfied by the last one.
To validate the above picture, we now show by induction that colour coherence
guarantees the angular ordering.

For any antenna in the cascade, say with opening angle $\theta_i$, $i\in \mathbb N^\star$, one can apply
the same argument about angular ordering as for the original antenna
with angle $\thqq$: VLEs at larger angles $\theta > \theta_i$ are
strongly suppressed because their formation times are smaller that the
decoherence time $t_{\rm coh}(\theta_i)$ of that antenna.

\paragraph{Energy loss during formation.}

Besides colour decoherence, another difficulty due to the medium is the potential energy loss during the formation of emissions. This effect could change the energy $\om$ in such a way that they could get out of the phase space \eqref{tfvac}. As our conclusions strongly relies on the condition \eqref{tfvac}, one must show that the energy lost via medium-induced
radiation remains negligible during the development of a vacuum-like
cascade.

The hardest medium-induced emission that can occur over the time $t$
has an energy $\omega_c(t)\simeq\hat q t^2/2$ and a probability of
order $\alpha_s$. For
$t=t_n\equiv 2/(\omega_n\theta^2_n)$
\eqref{tfvac} implies $\omega_c(t_n)\ll\omega_n$, i.e. the maximal
energy loss is small compared to the energy of the {\it softest} gluon
in the cascade. The {\em average} energy loss, of order
$\alpha_s\omega_c(t_n)$, is even smaller.
This argument also shows that, over their formation time, the gluons
from the vacuum-like parton showers do not contribute to the energy loss of
the jet. However, at the end of the in-medium
  partonic cascade, each of the final emissions will act as an
  additional source for medium-induced radiations and loses energy. We shall return to this point in the next subsection as it intimately related to the finite path length of the jet in the medium.

\subsection{Angular ordering violation and finite length effects}
\label{sub:finite-L-DL-picture}

\begin{figure}
   \centering
      \includegraphics[width=0.6\textwidth]{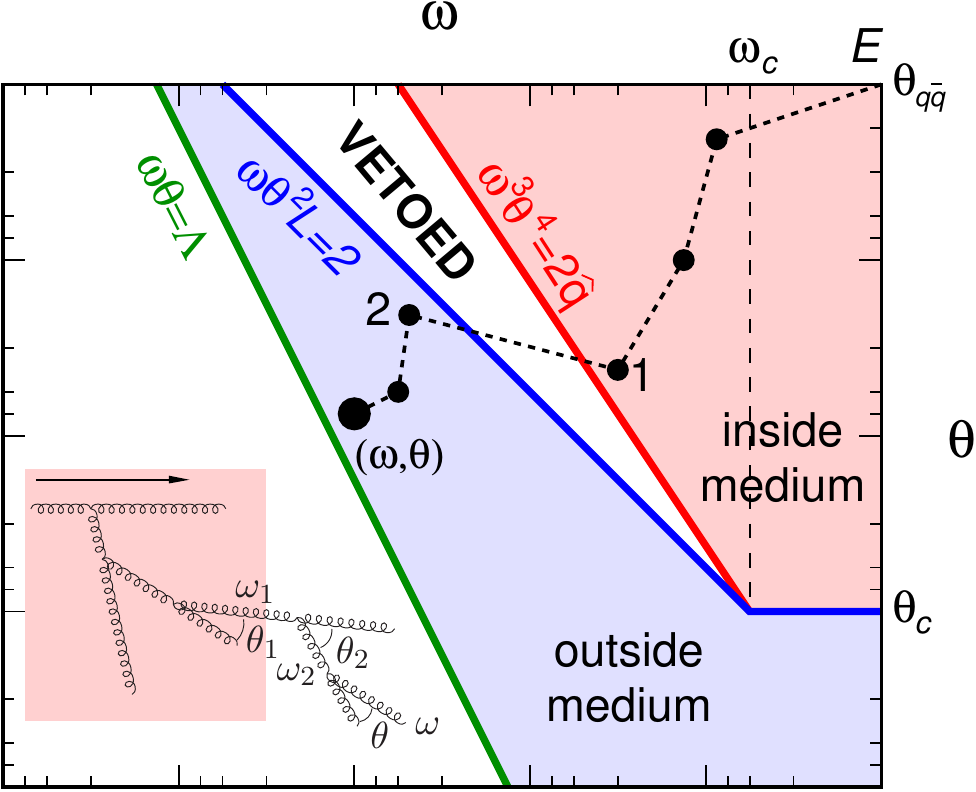}
    \caption{\small Complete double-logarithmic phase space taking into account the decoherence of large angle antennae during their propagation through the medium. With respect to Fig.~\ref{Fig:DLA-phase-space1}, the only difference is the introduction of the angular scale $\th_c$ that modifies the shape of the in-medium region for energies larger than $\om_c$. The figure includes an example of a cascade with “1” the last emission inside the medium and “2” the first emission outside. This cascade is pictured on the left. This figure should also be compared to the corresponding vacuum phase space for emissions shown Fig.~\ref{Fig:branching-process}. \label{Fig:DLA-phase-space}}
    
\end{figure}

Before exploring the consequences of the finite path length $L$ of the jet through the medium, let us summarize the resummation scheme of VLEs in an infinite medium: the coherent branching algorithm still correctly resums to all orders the soft and collinear double logarithms provided that all the emissions belong to phase space \eqref{tfvac}. Introducing the path length scale $L$ changes this picture in two respects:
\begin{enumerate}
 \item the phase space for VLEs is extended to emissions with $t_f\gg L$,
 \item two other scales come into play: the energy $\om_c=\qhat L^2/2$ of the hardest medium-induced emission that develops over a time $L$ and its corresponding angle $\th_c=2/\sqrt{\qhat L^3}$.
\end{enumerate}
When discussing the shape of the phase-space for VLEs in Section \ref{sub:qualitative-veto}, we have said that emissions with $\om\ge\om_c$ are necessarily vacuum-like since the maximal energy of a medium-induced emission in the multiple soft scattering regime is $\om_c$. Similarly, the angular scale $\th_c$ should be thought as the minimal angle of an antenna resolvable by the medium. Indeed, the coherence time scale $\tcoh$ becomes comparable to $L$ when
$\thqq\sim\theta_c$. Antennas with smaller opening angles $\thqq\lesssim \theta_c$ cannot
lose their coherence, hence their radiation pattern within the medium
is exactly the same as in the vacuum. For this reason, the phase space for vacuum-like emissions shown in Fig~\ref{Fig:DLA-phase-space1} is amended to take into account this property: any antenna with opening angle smaller than $\th_c$ is considered as ``outside'' (i.e.\ as in the vacuum), even if the emission occurs physically inside. This modification of the phase space is shown in Fig.~\ref{Fig:DLA-phase-space}.

\paragraph{Violation of angular ordering for the first emission outside the medium} We remind that the gluons produced inside the medium are not yet on-shell: their virtualities are as large as their transverse momenta, themselves
bound by the multiple scattering inside the medium:
$k_\perp^2\gg\sqrt{\omega\hat q}\gg\LQCD^2$, with $\LQCD$ the QCD
confinement scale. These partons will thus continue radiating. Their next VLE must occur {\em outside} the medium, with a large
formation time ${2}/({\omega\theta^2})\gg L$, i.e. with an energy
$\omega\ll \omega_L(\theta)\equiv {2/(L\theta^2)}$, as discussed in Section \ref{sec:veto-region}. As seen in the first section, this implies the
existence of a gap in the energy of the VLEs, between the lower limit
$\omega_{0}(\theta)$ on the last gluon emitted {\em inside} the
medium, and the upper limit $\omega_L(\theta)$ on the first gluon
emitted {\em outside} the medium. 
Since $\omega_{0}(\theta)=\omega_L(\theta)=\omega_c$ for
$\theta=\theta_c$ the gap exists only for $\omega < \omega_c$, as shown Fig.~\ref{Fig:DLA-phase-space}.

As explained in Chapter \ref{chapter:emissions}, Section \ref{sub:decoherence}, the medium has an important
effect on emissions with $t_f\gg L$: this first emission outside the medium can violate angular
ordering \cite{Mehtar-Tani:2014yea}. Approximating the decoherence factor $1-S_{q\bar{q}}(L)$ that appears in \eqref{decoherence-softlimit} by $\Theta(\th-\th_c)$, one sees that if the parent antenna of the emission with $t_f\gg L$ has an angle $\th$ larger than $\th_c$, the angle of the emission is not bounded any more by $\th$.
In physical terms, all the in-medium sources with $\theta\gg\theta_c$ satisfy
$t_{\rm coh}(\theta)\ll L$ and thus lose colour coherence after
propagating over a distance $L$ in the medium.
These sources can then radiate at {\em any} angle.
%
%
%


\paragraph{Energy loss and broadening after formation.} After being created inside
the medium via VLEs, the partons cross the plasma over a distance of
order $L$ and hence lose energy via medium-induced radiations and undergo
$p_\perp$-broadening. The pattern of fragmentation produced by the in-medium VLEs, serving as multiple
  sources for medium-induced emissions and energy loss, is a cornerstone of many phenomenological observations made in Part \ref{part:phenomenology}.
  
%
%

\paragraph{Emissions from sources created outside the
  medium.}  After a first emission outside the medium, the subsequent
emissions follow, of course, the usual pattern of vacuum-like
cascades, with angular ordering (and energy ordering in our DLA
approximation). The evolution stops when the transverse momentum
$k_\perp\simeq \omega\theta$ becomes comparable to the hadronisation
scale $\LQCD$. This implies a lower boundary,
$\omega\gtrsim\LQCD/\theta$, on the
energy of the produced gluons, shown in Fig.~\ref{Fig:DLA-phase-space} (green curve)
together with the other boundaries introduced by the medium.
The most interesting region for gluon production --- the most
sensitive to medium effects highlighted above --- is the ``outside
medium'' region at energies $\omega<\omega_c$.

\subsection{Summary of the fundamental QCD picture}
\label{sub:fac-summary}

A very simple picture for the development of a partonic cascade in the
medium emerges from the above observations and within our approximations. The full cascade can be
factorised in three major steps.
\begin{enumerate}
\item {\it in-medium vacuum-like cascade}: an angular-ordered
  vacuum-like cascade governed by the standard DGLAP splitting
  functions occurs inside the medium up to
  $t_f(\omega, \theta) = t_{f, \rm med}(\omega)$. During this process,
  the only effect of the medium is to set the constraint \eqref{tfvac} on the formation time;
\item {\it medium-induced emissions and broadening}: every parton
  resulting from the in-medium cascade travels through the medium,
  possible emitting (a cascade of) MIEs and acquiring momentum
  broadening;
\item {\it outside-medium vacuum-like cascade}: each parton exiting
  the medium at the end of the previous step initiates a new
  vacuum-like cascade outside the medium, down to a non-perturbative
  cut-off scale. The first emission in this cascade can happen at an
  arbitrary angle.
\end{enumerate}

\subsection{The double-logarithmic picture in Bjorken expanding media}
\label{sub:DLpic-bjorken}

In this subsection, we generalise the double logarithmic picture for a Bjorken expanding medium. More precisely, we show that the all-order factorised picture summarized above holds for a medium in expansion. We check that the boundaries of the veto region found in Section \ref{subsub:bjork} lead to the same properties for the in-medium parton shower: angular ordering is preserved and energy loss during formation is negligible. Transverse momentum broadening during formation is negligible \textit{by construction} of the veto region, see \eqref{bjork-qual2}.

\paragraph{Angular ordering.} As in the brick case, we first consider a leading $q\bar{q}$ pair with opening angle $\thqq$. We also simplify the discussion of subsection \ref{subsub:bjork} assuming that this antenna is created at $t_i=0$, meaning that $\Delta t=t_0-t_i=t_0$. In this situation, the constraint on short formation times VLEs is 
\begin{tcolorbox}[ams align]\label{veto-bjork}
\om&\ge \frac{2}{t_0\th^2}\qquad&\textrm{ if }\om\le \frac{1}{2}\qhat_0 t_0^2\\
\om&\ge\om_0(\th)\equiv\left(\frac{2\qhat_0(t_0/2)^\gamma}{\th^{4-2\gamma}}\right)^{1/(3-\gamma)}\qquad&\textrm{ if }\om\ge \frac{1}{2}\qhat_0 t_0^2
\end{tcolorbox}
\noindent The first condition is not really relevant in practice since the energy scale $\frac{1}{2}\qhat_0t_0^2$ is 
very small for typical values of $t_0\sim 0.1$ fm.

We want to extract the characteristic decoherence time as a function of $t_i$. It is given by the calculation of the $q\bar{q}$ dipole S-matrix $S_{q\bar{q}}(t)$ defined by \eqref{Sqqbar}:
\begin{align}
\label{Sqq-bjork-def} S_{q\bar{q}}(t)&=\exp\left(-\frac{1}{8}\thqq^2\int_{t_0}^{t}\dif \xi\, \qhat(\xi)(\xi-t_i)^2\right)\\
 &\label{Sqq-bjork}=\exp\left(-\frac{1}{8(3-\gamma)}\thqq^2\qhat_0t_0^\gamma\big(t^{3-\gamma}-t_0^{3-\gamma}\big)\right)
\end{align}
for $t_i=0$. Clearly, this calculation is interesting in the regime where $t_f\ge t_0$. Otherwise, the emission happens before $t_0$, i.e in the vacuum where angular ordering holds.
At DLA, one can further consider $t_f\gg t_0$. Neglecting the $-t_0^{3-\gamma}$ term in the result \eqref{Sqq-bjork}, one finds the characteristic (de)coherence time:
\begin{equation}\label{tcoh-bjork}
 \tcboxmath{\tcoh=\left(\frac{4}{\thqq^2\qhat_0t_0^\beta}\right)^{1/(3-\gamma)}}
\end{equation}
For $t_f\gg t_0$, one has then:
\begin{equation}\label{exp-bjork}
 \frac{t_f}{\tcoh}=\frac{\om_0(\th)}{\om}\left(\frac{\thqq}{\th}\right)^{2/(3-\gamma)}
\end{equation}
which looks pretty much like \eqref{exp} except for the exponent which is positive for $\gamma\le 1<3$. This means that for emissions satisfying $\om\ge \om_0(\th)$, the antenna is still coherent at $t_f$ so that large angle emissions are forbidden.

In this reasoning, which is a simple transposition of the argument given for the brick medium, we have used $t_f\gg t_0=\Delta t$. This means that the emission should not be sensitive to position of the hard vertex $t_i$. For consistency, we thus need to check that the angular ordering property remains valid also when $t_i=t_0$, that is when the antenna is created at the same time as it enters into the medium. The calculation of \eqref{Sqq-bjork-def} gives in this case, for $\gamma<1$:
\begin{equation}\label{Sqq-ti=t0}
 S_{q\bar{q}}(t)=\exp\left(-\frac{1}{8}\thqq^2\qhat_0 t_0^\gamma\left[\xi^{1-\gamma}\left(\frac{t_0^2}{1-\gamma}-\frac{2t_0\xi}{2-\gamma}+\frac{\xi^2}{3-\gamma}\right)\right]_{\xi=t_0}^{\xi=t}\right)
\end{equation}
According to our discussion in Section \ref{subsub:bjork}, there are two regimes that we should discuss: either $t\simeq t_0$ or $t\gg t_0$. When $t_f\simeq t_0$, the dipole S-matrix reads:
\begin{equation}
 S_{q\bar{q}}(t)\underset{t\simeq t_0}\simeq\exp\Big(-\frac{1}{24}\thqq^2\qhat_0(t-t_0)^3\Big)
\end{equation}
The limit $t_0\rightarrow0$ is not singular any more, and the characteristic coherence time is the same as in the fixed $\qhat$ case: $\tcoh=(\qhat_0\thqq^2/4)^{-1/3}$. We have seen that for $t_f\simeq t_0$, the constraint on $k_\perp$ is not given by \eqref{veto-bjork} but rather by $k_\perp\ge(2\qhat_0\om)^{1/4}$ when $\om\le \qhat_0t_0^2/2$. Consequently, for emissions in this phase space, the argument for angular ordering developed in the previous section when $\qhat$ is constant applies in the same way.  

In the second regime $t\gg t_0$, the exponent in \eqref{Sqq-ti=t0} is dominated the leading power in $t$:
\begin{equation}
 S_{q\bar{q}}(t)\underset{t\gg t_0}\simeq\exp\Big(-\frac{1}{8(3-\gamma)}\thqq^2\qhat_0t_0^\gamma t^{3-\gamma}\Big)
\end{equation}
One recovers the characteristic coherence time \eqref{tcoh-bjork}, so that the calculation in \ref{exp-bjork} remains valid. 

To sum up, whatever the position of the hard-vertex with respect to the parameter $t_0$ is, the angular ordering property for the first emission by the leading $q\bar{q}$ pair holds, as long as this emission belongs to the in-medium phase space. The generalisation for a sequence of $n$ emissions in this phase space proceeds by mathematical induction.

\begin{figure}
   \centering
      \includegraphics[width=0.6\textwidth]{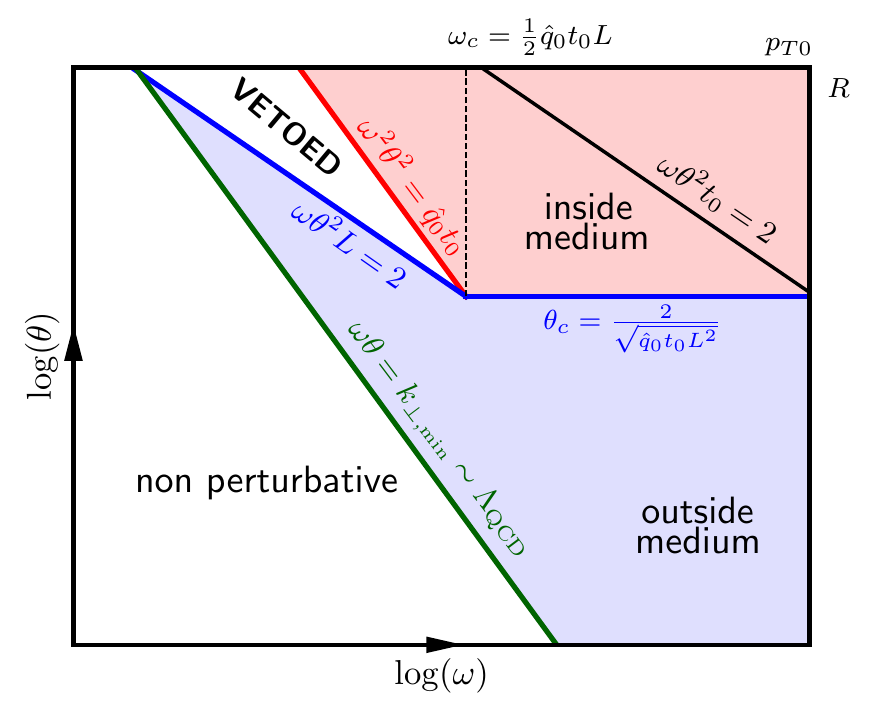}
    \caption{\small Phase space for vacuum-like emissions at DLA in a Bjorken expanding medium with $\gamma=1$}
    \label{Fig:DLA-phase-space-exp}
\end{figure}

\paragraph{Energy loss during $t_f$.} In a Bjorken expanding medium, the hardest medium-induced emission that can develop during the formation time $t_f$ of an in-medium vacuum-like emission has an energy
\begin{equation}
\om_c(t_f)\simeq\frac{1}{2}\qhat(t_f)t_f^2=\frac{1}{2}\qhat_0t_0^\gamma t_f^{2-\gamma} 
\end{equation}
As in static media, the condition $\om\gg\om_0(\th)$ for $t_f\gg t_0$ implies $\om_c(t_f)\ll \om$ meaning that one can neglect the energy loss via medium-induced radiations during the formation of an in-medium vacuum-like emission.

\paragraph{Angular ordering violation and critical angle $\th_c$.} From the definition \eqref{tcoh-bjork} of the coherence time in a Bjorken expanding medium, one can easily deduces the parametric dependence of the critical angle $\th_c$ for a jet getting out of the medium at time $t_L=t_0+L\gg t_0$. This critical angle is given by the condition $\tcoh(\th_c)=t_L$:
\begin{equation}
 \tcboxmath{\th_c=\frac{2}{\sqrt{\qhat_0t_0^\gamma t_L^{3-\gamma}}}=\frac{2}{\sqrt{\qhat(t_L)t_L^3}}}
\end{equation}
An antenna with an angle smaller than $\th_c$ does not
lose its colour coherence when travelling through the diluting medium. As in the static case, a source inside
the medium with $\th\ge\th_c$ can then radiate at any angle in the out-medium phase space.

\subsection{Beyond the double-logarithmic picture}
\label{sub:beyondDLA}

Now that we have exposed the basic picture for the development of
parton showers in the presence of a medium, we show that
several sub-leading corrections, beyond DLA, can easily be taken into account. First of all, the large $N_c$ limit can be easily relaxed (in the collinear limit) taking into account the 
color representation of the emissions thanks to the usual DGLAP splitting functions.

The validity of our factorisation between VLEs and MIEs relies on
strong inequalities between the formation times. Clearly, these
inequalities do still hold if the strong ordering refers only to the
emission angles ($\theta_n\ll\theta_{n-1}$) but not to the
parton energies (as is the case beyond the soft limit $z\ll
1$). There is nevertheless some loss of accuracy with respect to a strict
single logarithmic approximation associated with the uncertainties in
the boundaries of the vetoed region in phase-space. Notably the
condition  $\tf(\omega, \theta)=t_{f,\rm med}(\omega)$ defining the
upper boundary is unambiguous only at DLA. For a generic splitting
fraction $z$, the formation times also depend upon the energy $xE$ of
the parent parton and not just upon the energy $\omega=xzE$ of the
soft daughter gluon. For a generic $1\to 2$ splitting where
the ``vacuum-like'' formation time $\tf\equiv \tf(x, z, \theta)$ is
given by
\begin{equation}\label{tvac}
 t_{f}(x,z,\theta)\,\simeq\,\frac{2xE}{Q^2}\,\simeq\,\frac{2}{z(1-z)xE\theta^2}\,\simeq\,\frac{2z(1-z)xE}{k_\perp^2}\,,
\end{equation}
where $z$ and $\theta$ (assuming $\theta\ll 1$) are the
energy fraction and opening angle of the partonic decay and
$k_\perp\simeq z(1-z)\theta xE$ is the (relative) transverse momentum
of any of the two daughter partons
with respect to the direction of the leading parton.

The corresponding ``medium''
formation time $t_{f,\rm med}(x,z)$ is different for different partonic
channels. For example, for a $g\to gg$ splitting, it reads
\begin{equation}\label{tmed}
t_{f,\rm med}(x,z)\big|_{g\to gg}\,=\,\sqrt{\frac{2z(1-z)xE}{ \hat
    q_{\text{eff}}(z)}}\
   \overset{z\ll 1}{\approx}\ \sqrt{\frac{2\omega}{\hat q}}\,,\qquad \hat q_{\text{eff}}(z)\equiv\hat q \big[1-z(1-z)\big],
\end{equation}
with $\omega=zxE$. One could in principle use these more accurate
estimates for $\tf$ and $t_{f,\rm med}$ in  \eqref{tfvac}. One would then
need to deal with the difficulty that the evolution phase-space
depends explicitly on $xE$, $z$ and $\theta$ and not just on $\omega$
and $\theta$. The corresponding generalisation of \eqref{tfvac} would
also be different for different partonic channels. Last but not least,
the distinction between VLEs and MIEs according to their formation
times only holds so long as the {\it strong} inequality $\tf\ll t_{\rm
  med}$ is satisfied, meaning that the precise form of the boundary
could also be sensitive to subleading corrections. In practice, our
strategy to deal with this ambiguity (notably, in the Monte Carlo
simulations presented in Part \ref{part:phenomenology}) is to stick to the simpler form of the boundary in
\eqref{tfvac}. In phenomenological applications, one should keep in mind that the
  dependence on $\qhat$ and $L$ of the VLEs is only leading-logarithmic.

\paragraph{Limitations.} The resummation scheme developed in this chapter has of course its own limitation and domain of applicability. We will not discuss here the limitations due to our simple modelling of the medium and geometry of the collision. This will be detailed in the next chapter about the Monte Carlo parton shower built from the present picture. However, there are intrinsic theoretical limitations in our approach. We provide an non-extensive list here.
\begin{enumerate}
 \item The multiple soft scattering approximation is maybe the strongest assumption in the perturbative treatment of the parton shower. Hard scattering off medium scattering centers are neglected. Combining consistently such scatterings with both the initial virtuality and the multiple soft collisions is beyond the scope of the present thesis, but should not be too difficult to include in a Monte Carlo framework. This is also a widely developed topic in the literature.
 
 \item Even in the multiple soft scattering regime, there are other emissions which are not taken into account in the double-logarithmic cascade: the BDMPS-Z emissions and the ``Bremsstrahlung-like'' emissions triggered by the medium discussed in Section \ref{sec:veto-region}. 
 A complete picture should include the medium-induced emissions \`{a} la BDMPS-Z and the medium-induced emissions \`{a} la Bremsstrahlung appearing in \eqref{on-shell-physic}. The former are currently included in the Monte Carlo {\tt JetMed} presented in Chapter~\ref{chapter:MC}, but not the latter. 
 
 \item The double logarithmic resummation performed so far does not account for running coupling single logarithmic corrections. Including such corrections is straightforward. This is done semi-analytically in the next section for the jet fragmentation from vacuum-like emissions and numerically in the Monte Carlo parton shower {\tt JetMed} presented in Chapter~\ref{chapter:MC}. 
 
 \item While crossing the medium, the hard parton sourcing the jet can transfer energy and momentum to the medium constituents that may finally belongs to the reconstructed jet. This is the so-called medium response effect \cite{CasalderreySolana:2004qm,Ruppert:2005uz,Li:2010ts,Wang:2013cia,He:2015pra} (see \cite{Cao:2020wlm} for a review). This requires a detailed modelling of the medium itself which clearly goes beyond our current simple approach.
\end{enumerate}

\section{Analytic fragmentation function and jet energy loss}

In this section, as a first step towards the detailed phenomenological study made in Part \ref{part:phenomenology}, we provide analytic results for the fragmentation function and the jet energy loss obtained directly from our new factorised picture. This results are important since they enable to understand the important physical mechanisms at play in the modification of the jet fragmentation pattern.

\subsection{Fragmentation function from vacuum-like emissions}

Our first calculation is the intrajet multiplicity for a jet evolving in the medium with constant $\qhat$. As usual, we note $L$ the path length of the jet through the medium. This calculation is a double logarithmic resummation in the strict sense, meaning that one completely neglects medium-induced emissions even for the last vacuum-like emission at the end of the in-medium shower. We recall that medium-induced emissions do not \textit{formally} matter at DLA as the process is not enhanced by collinear logarithms. 

Also, $p_\perp$-broadening effects, while important in general, can safely
be neglected when computing the gluon multiplicity since the latter is
only sensitive to the angle of emission and not to a change in the
direction of the emitter.

\subsubsection{Strict double logarithmic calculation}
\label{subsub:DLA-FF}

Within the present approximation, and in the large $N_c$ limit, it is straightforward to compute the
gluon distribution generated by VLEs. 
To that aim we compute the double differential distribution,
\begin{equation}\label{Tdef}
  T(\omega,\theta|E, R^2)\,\equiv\,\omega
  \theta^2\frac{\rmd^2 N}{\rmd\omega\rmd\theta^2}\,,
\end{equation}
which describes the gluon distribution in both energies and emission
angles. We note $R$ the opening angle of the jet that we identify with the opening angle $\thqq$ of the leading $q\bar{q}$ pair. 
In the vacuum, this distribution has been calculated in Section \ref{subsub:frag-smallx}, Eq.~\eqref{TDLA} thanks to the coherent branching algorithm. At DLA, successive gluon emissions are strongly ordered in both energy and emission
angle and one finds
\begin{equation}
\label{Tvac}
 T_{i}^{\textrm{vac}}(\omega,\theta^2|E,R^2)=\frac{\as C_i}{\pi}\,
 \textrm{I}_0\left(2\sqrt{\abar 
\log\frac{E}{\omega}\,\log\frac{R^2}{\theta^2}}\right)
+\om\theta^2\delta(E-\om)\delta(R^2-\theta^2)
\end{equation}
where $\abar = \alpha_sC_A/\pi$ and $ \textrm{I}_0(x)$ is the modified Bessel function of rank 0 which increases
exponentially for $x\gg 1$. The second term in the r.h.s.\ which is not present in \ref{subsub:frag-smallx} represents
the leading parton and the first term is associated with subsequent
gluon emissions. The fragmentation function can be obtained from $T_i$ by integrating
over all the angles in the jet, as explained in Section \ref{subsub:frag-smallx} (with $\theta_{\textrm{min}}=
\ktmin/\omega$)
\begin{equation}\label{DfromT}
\om D_{i}(\omega,\theta^2|E,R^2)  =
 \int_{\theta^2_{\textrm{min}}}^{R^2}\frac{\dif \theta^2_1}{\theta^2_1}
\, T_{i}(\omega,\theta^2_1|E,R^2)
\end{equation}
with $\theta_{\textrm{min}}=
\ktmin/\omega$.

We want to do the same calculation taking into account the modifications of the branching process due to the medium.
In the presence of the medium, the DLA
calculation is modified by two effects: the
presence of a vetoed phase-space for VLEs inside the medium, and the colour decoherence allowing for the
violation of angular ordering by the first emission outside the
medium.
It is helpful to split the medium fragmentation function
$T^{\textrm{med}}_{i}$ in two contributions:
\begin{equation}
\label{fullT}
 T^{\textrm{med}}_{i}(\omega,\theta^2|E,R^2)=\Theta_{\textrm{in}}(\omega,\theta^2)T_{i}^{\textrm{vac}}+\Theta_{\textrm{out}}(\omega,\theta^2)T_{i,\textrm{out}}
\end{equation}
where the step functions $\Theta_{\textrm{in/out}}$ enforces that an
emission $(\omega,\theta^2)$ belongs to the ``inside'' or ``outside''
region.
The first term,
$\Theta_{\textrm{in}}(\omega,\theta^2)T_{i}^{\textrm{vac}}$,
corresponding to the in-medium contribution, is unmodified compared to
the vacuum.
The outside-medium, $T_{i,\textrm{out}}$, contribution can be
expressed as the product of a vacuum-like cascade inside the medium,
up to an intermediate point $(\omega_1,\theta^2_1)$, followed by a
first emission outside the medium at $(\omega_2,\theta^2_2)$ (possibly
violating angular ordering), and by a standard vacuum cascade from
$(\omega_2,\theta^2_2)$ to the final point $(\omega,\theta^2)$:
\begin{align}
\label{master-eq}
 T_{i,\textrm{out}}(\om,\theta^2|p_{T0},R^2)=  \abar\int_\omega^{p_{T0}}\frac{\dif \omega_1}{\omega_1}
 \int_{\theta^2_c}^{R^2}\frac{\dif \theta^2_1}{\theta^2_1}\,\Theta_{\textrm{in}}(\omega_1,\theta^2_1) \int_\omega^{\omega_1}\frac{\dif \omega_2}{\omega_2}\int_{\theta^2}^{R^2}\frac{\dif \theta^2_2}{\theta^2_2}\,\Theta_{\textrm{out}}(\omega_2,\theta^2_2)&\nonumber\\
 T_{i}^{\textrm{vac}}(\om_1,\theta_1^2|p_{T0},R^2)T_{g}^{\textrm{vac}}(\om,\theta^2|\omega_2,\theta^2_2)
 &
\end{align}
The integral over $\theta_2^2$ is not constrained by the angle
$\theta_1^2$  of the previous emission due to absence of angular
ordering for the first emission outside the medium.

The two angular integrations in~\eqref{master-eq} can be performed
analytically. The resulting formula is well suited for numerical evaluation of
the double differential gluon distribution $T$ as well as the fragmentation function $D$.

\begin{figure}
   \centering
      \includegraphics[width=0.6\textwidth]{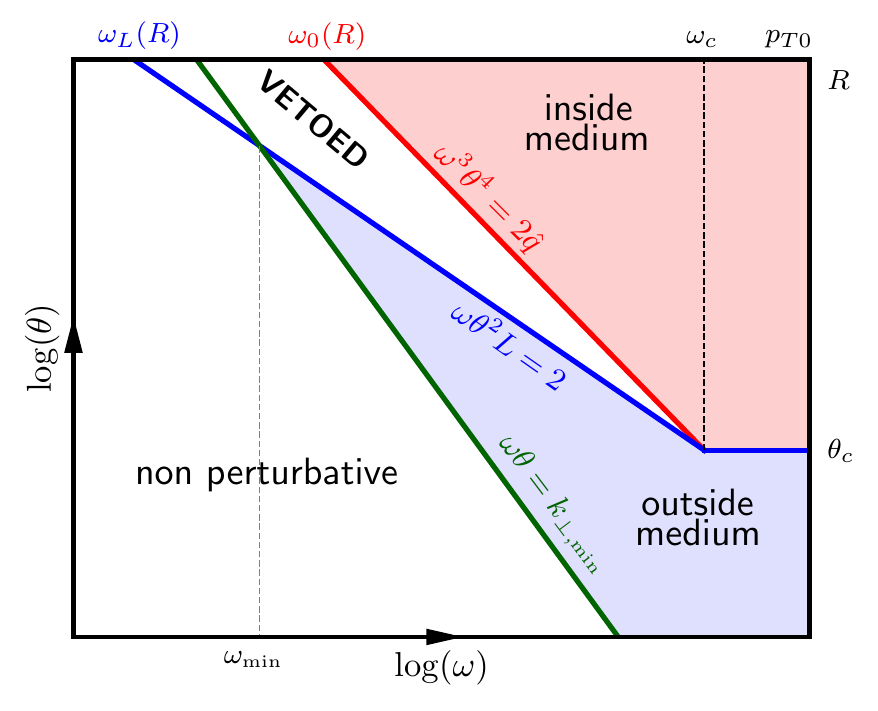}
    \caption{\small The phase space considered in our double-logarithmic resummation of vacuum-like emissions in the presence of a dense QCD medium. The relevant energy scales for the calculation of the vacuum-like fragmentation function are represented.}
    \label{Fig:DLA-phase-space3}
\end{figure}

To gain more physical intuition, we now develop an analytic
approximation, which is valid when both the energy and angular
logarithms are larger than $1/\sqrt{\as}$.
We give here the main ingredients of the calculation and defer details
to Appendix~\ref{app:DLA}.

In the limit of interest, the $\delta$ contribution to
$T^{\textrm{vac}}$ (the second term in~(\ref{Tvac})) can be neglected
in both $T^{\textrm{vac}}$ factors in~\eqref{master-eq}, the Bessel
functions can be approximated by their (exponential) asymptotic
behaviour and the integrations can be evaluated in the saddle-point
approximation.

For definiteness, let us consider parameters such that
$\om_L(R) < \ktmin/R$, meaning that the hadronisation line
$\om\theta =\ktmin$ and the medium boundary
$\om_L(\theta)=2/(L\theta^2)$ intersect at
$\om_{\textrm{min}}=L\ktmin^2/2$.
In practice we are interested in the fragmentation function
at energies $\om$ within the range $\om_{\textrm{min}}\ll\om\ll
\om_c$.
The saddle points for $\omega_1$ and $\omega_2$ integrals
are respectively found to be
\begin{tcolorbox}[ams align]\label{omSP}
 \omega_{1}^{\star}=\sqrt{\frac{p_{T0}(2\qhat)^{1/3}}{R^{4/3}}} = \sqrt{p_{T0}\omega_0(R)}
 \,,\qquad
 \omega_{2}^{\star}=\sqrt{\frac{2\omega}{L\theta^2}}=\sqrt{\om\om_L(\theta)}\,,
\end{tcolorbox}
\noindent with $\omega_0(\theta)\equiv(2\qhat/\theta^4)^{1/3}$ such that
$\omega_0(R)$ is the lowest possible energy for a VLE inside
the medium.
 
Several conditions are needed for these saddle points to control the energy integrations.
First, the integration ranges must be wide enough, $p_{T0}\gg\omega_0(R)$ and
$\om_L(\theta)\gg\om$, to allow for large enough logarithmic
contributions. This translates into the following conditions:
\begin{equation}\label{lnPS}
\sqrt{\abar}\,\log\frac{p_{T0}}{\omega_0(R)}\,\gtrsim\,1
\qquad\mbox{and}\qquad
\sqrt{\abar}\,\log\frac{\om_L(\theta)}{\om}\,\gtrsim\,1\,
\end{equation}
Second, for $\omega_{1}^{\star}$ to be a genuine saddle point, it must remain smaller than $\omega_c$, 
meaning
\begin{equation}\label{eq:cdt-pt0-omc}
p_{T0} < \om_c \left(\frac{R}{\theta_c}\right)^{4/3}\,=\,\frac{\qhat^{5/3}L^4R^{4/3}}{2^{7/3}}
\end{equation}
When this condition is satisfied (which is always the case in practice), the integral over $\om_1$ is dominated by relatively
low-energy emissions with $\om_0(\theta) <\om_1<\om_c$, i.e. by the
triangular region of the ``inside medium'' phase-space with energies
below $\om_c$, see Fig.~\ref{Fig:DLA-phase-space3}. In the opposite situation,
  which would occur for sufficiently large $p_{T0}$, the dominating
  region in phase-space is the rectangular region at
  $\omega_c\le\omega_1\le p_{T0}$ and $\theta_c<\theta_1<R$; see
  Appendix~\ref{app:DLA} for details.

Third, energy conservation in Eq.~\eqref{master-eq} requires 
$\omega_{2}^{\star}\le \omega_{1}^{\star}$ which implies a
$\theta$-dependent upper limit on $\om$.
When computing the fragmentation function using Eq.~(\ref{DfromT}), 
this condition must be satisfied for all the angles $\theta$ that are
integrated over, including lower bound
$\theta_{\textrm{min}}= \ktmin/\omega$.
This defines a critical energy $\omega_{cr}$, obtained for
$\theta=\theta_{\textrm{min}}$, below which the saddle point method works:
\begin{equation}\label{ocr}
  \omega < \omega_{cr}=\big(p_{T0}\om_0(R)\om_{\textrm{min}}\big)^{1/3}=
  \left(\frac{p_{T0}L\ktmin^2(2\hat{q})^{1/3}}{2R^{4/3}}\right)^{1/3}=
  \left(\frac{p_{T0}\ktmin^2}{R^2}\right)^{1/3}\left(\frac{R}{\theta_c}\right)^{2/9}.
\end{equation}

When the conditions in Eqs.~\eqref{lnPS}--\eqref{ocr} are satisfied, the 
saddle point method gives a meaningful approximation for the double differential gluon distribution
in  \eqref{master-eq}, which reads
\begin{tcolorbox}[ams equation]
\label{Tsaddle}
 T_{i,\textrm{out}}(\om,\theta^2|p_{T0},R^2)\simeq
 \frac{\alpha_s C_i}{4 \pi}\exp\left\{\sqrt{\frac{3\abar}{2}}\,\log
  \frac{p_{T0}}{\omega_0(R)}\right\}
  \exp\left\{\sqrt{\abar}\,\log\frac{\omega_{L}(\theta)}{\omega}\right\}
\end{tcolorbox}
\noindent The first exponential comes from the integrations over $\theta^2_1$ and
$\omega_1$, i.e. over the ``inside'' region, and can be interpreted as the number of partonic
sources generated via VLEs.
The second exponential represents the number of gluons generated by
each of these sources via gluon cascades developing outside the
medium.  This simple factorisation between the ``inside'' and the
``outside'' jet dynamics holds strictly speaking only in the saddle
point approximation (and for energies $\om\le\omega_{cr}$) and is
ultimately a consequence of the colour decoherence which washes out
any correlation between the emission angles outside and inside the
medium.

Integrating~\eqref{fullT} over $\theta$ using Eq.~(\ref{DfromT}) we find
the fragmentation function for $\om\le\omega_{cr}$:
\begin{tcolorbox}[ams align]
\label{frag-DLA}
\om D^{\textrm{med}}_i(\omega)\simeq\frac{\sqrt{\abar}C_i}{4C_A}\exp\left\{\sqrt{\abar}\left(\sqrt{\frac{3}{2}}\log\,\frac{p_{T0}R^{4/3}}{(2\qhat)^{1/3}}+\log\,\frac{2\omega}{\ktmin^2L}\right)\right\}.
\end{tcolorbox}
\noindent The integration is dominated by the lower limit, $\theta=\ktmin/\omega$. The
  respective contribution of the first term
  $\propto T_{i}^{\textrm{vac}}$ in \eqref{fullT}, that would be
  non-zero only for $\om>\om_0(R)$, is
  comparatively small, since it lacks the evolution outside the
  medium.
Since  ${2\omega}/{\ktmin^2L}=\om/\om_{\textrm{min}}\gg 1$, the second
logarithm in~\eqref{frag-DLA} is positive and $\om D^{\textrm{med}}_i(\omega)$  decreases
when decreasing $\om$.

\begin{figure}[t] 
  \centering
  \begin{subfigure}[t]{0.48\textwidth}
    \includegraphics[page=2,width=\textwidth]{./ff.pdf}
    \caption{Fragmentation function at DLA}\label{fig:DLA-dec-D}
  \end{subfigure}
  \hfill
  \begin{subfigure}[t]{0.48\textwidth}
    \includegraphics[page=1,width=\textwidth]{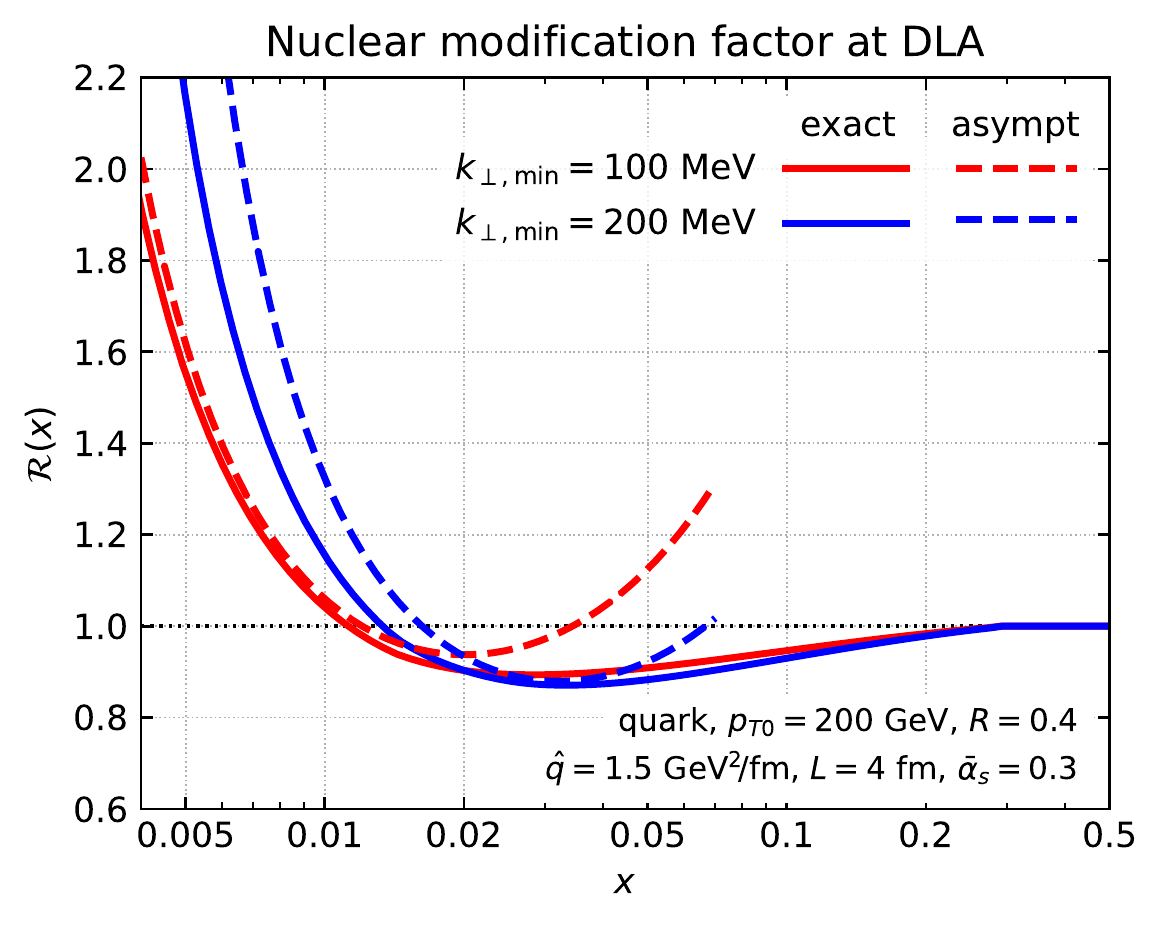}
    \caption{Nuclear modification $\mathcal{R}(x)$ at DLA}\label{fig:DLA-dec-R}
  \end{subfigure}
  \caption{\small Comparison of the exact calculation of fragmentation
    functions (solid lines) and the asymptotic approximations (dashed
    lines). }\label{Fig:DLA-dec}
\end{figure}

Our predictions are shown in Fig.~\ref{Fig:DLA-dec} for the
fragmentation function in Fig.\ref{fig:DLA-dec-D} and the nuclear
modification factor $R_i(x|p_{T0})$ in
Fig.~\ref{fig:DLA-dec-R}. This nuclear modification factor is defined by:
\begin{equation}
 R_i(x|p_{T0})=\frac{D^{\textrm{med}}_i(\omega)}{D^{\textrm{vac}}_i(\omega)}
\end{equation}
with $D^{\textrm{vac}}_i(\omega)$ given by \eqref{Dg-dla} (or \eqref{DfromT}).
These plots compare the exact results at DLA based on Eq.~\eqref{Tvac} and
the numerical integration of Eq.~(\ref{master-eq}) for the vacuum
and medium results respectively, to their asymptotic counterparts. The
latter are obtained by taking the asymptotic behaviour of~\eqref{Tvac}
in the vacuum case and by using the saddle-point approximation
Eq.~\eqref{frag-DLA} for the medium results.
In Fig.~\ref{fig:DLA-dec-R} we consider two different values for the
IR cut-off $\ktmin$ (blue: $\ktmin=200$~MeV, red: $\ktmin=100$~MeV).
 Overall we see a good agreement, which is
moreover improving when $\ktmin$
decreases, i.e.\ when the phase-space increases and the saddle point
method becomes more reliable.

The fact that the ratio $R_i(x|p_{T0})$ increases at small
$\om$ can be traced back to angular ordering and the associated
humpback plateau.
Unlike the double-differential gluon distribution \eqref{Tvac} which
keeps increasing when decreasing $\omega$ at fixed $\theta$, the
vacuum fragmentation function $\om D^{\textrm{vac}}_i(\omega)$ in
\eqref{Dg-dla} develops a maximum at
$\omega\simeq \om_{\rm hump}=(E\ktmin/R)^{1/2}$ and decreases very
fast for $\om$ below $\om_{\rm hump}$.
This is due to the fact that the angular phase-space at
$\ktmin/\om<\theta <R$ permitted by angular ordering shrinks to zero
when decreasing $\om$. For sufficiently small $\om$, namely such
that\footnote{The upper limit $p_{T0} \ktmin^2/R^2$ is smaller than
  $\omega_{cr}^3$ guaranteeing the validity of the saddle-point method.}
$\om^3\lesssim p_{T0} \ktmin^2/R^2$, the denominator
$\om D^{\textrm{vac}}_i(\omega)$ in the medium/vacuum ratio
$R_i(x|p_{T0})$ decreases faster with $1/\om$ than the respective
numerator $\om D^{\textrm{med}}_i(\omega)$ (see also
Fig.~\ref{fig:DLA-dec-D}), so the ratio is increasing.

\subsubsection{Including running coupling and hard collinear corrections}
\label{subsub:frag-beyondDLA}

The calculation made in the previous subsection does not take into account single logarithmic corrections coming from the running of the coupling constant and the hard collinear splittings. We briefly discuss two ways of including semi-analytically such corrections. Each one has its own advantages and drawbacks.

The first way is to calculate the convolution \eqref{master-eq} in Mellin space. In Mellin space, the fixed coupling fragmentation function \eqref{DfromT} or \eqref{Dg-dla} in vacuum reads:
 \begin{equation}\label{Dvac-mellin}
  \om D_g(\om|E,R^2)=\int \frac{\dif j}{2\pi i}C_j(\abar)\exp\Big(\int_{\ktmin^2}^{(ER)^2}\frac{\dif Q^2}{Q^2}\gamma_j(\abar)+(j-1)\log(E/\om)\Big)
 \end{equation}
with $\gamma_j$ the anomalous dimension, and $C_j$ the coefficient function such that:
\begin{equation}
    \begin{array}{l}
        \gamma_j(\abar)=\frac{1}{4}\Big(-(j-1)+\sqrt{(j-1)^2+8\abar}\Big)\\
        C_j(\abar)     =\frac{1}{2}\Big(1+\frac{j-1}{\sqrt{(j-1)^2+8\abar}}\Big)
    \end{array}
\end{equation}
From this expression, running coupling corrections can be included using $\abar(Q^2)$ instead of $\abar$ in the $Q^2$ integral corresponding to the evolution, while $\abar$ remains fixed (evaluated at $Q=ER$) inside the coefficient function. Corrections from hard collinear splittings are added using:
\begin{equation}
 \gamma^{\rm MLLA}_j(\abar)=\gamma_j(\abar(Q^2))+2B_{g}\frac{\abar}{j^2+8\abar}
\end{equation}
instead of $\gamma_j$ in the evolution integral. $B_{g}$ is the finite part of the gluon splitting function. 

Then the convolution \eqref{master-eq} between the inside evolution and the outside evolution is made using \eqref{Dvac-mellin} amended as explained. In the evolution inside the medium, the lower limit of the $Q^2$ integration is the in-medium boundary of the veto region, while for the outside evolution, the upper limit is the out-medium boundary. Such a calculation is complicated, especially because of the convolution and the final inverse Mellin transform. However, as in the fixed coupling case, one may obtain a good approximation using the saddle point method. Instead of taking this path, we explain now a different approach more suitable for a final numerical evaluation.

The idea is to start with \eqref{Dsmall-x-NLL} where running coupling effects and hard-collinear contributions are already included \cite{Caucal:2018epd}. From this equation, one obtains the following equation for the function $T$ \cite{Dokshitzer:1992iy,Lupia1998}:
\begin{equation}
\label{master}
T(\omega,\theta^2\mid E, R^2) = \abar(\omega^2\theta^2) \frac{\omega}{E} P_{gg}(\frac{\omega}{E})
 +\int_{\theta^2}^{R^2}\frac{d\theta_1^2}{\theta_1^2}\int_{\omega/E}^{1}dz_1\abar(z_1^2E^2\theta_1^2)P_{gg}(z_1)T(\omega,\theta^2\mid z_1E, \theta_1^2)
\end{equation}
with $P_{gg}(z)=(1-z)\Phi_g^{gg}(z)/(2C_A)$. Note that we still work within the large $N_c$ limit in order to avoid complications due to quark/gluon mixing.

Let us first take into account the running of the coupling at one loop, while $P_{gg}(z)$ is approximated by its soft behaviour $P_{gg}(z)\simeq1/z$.
The Landau pole in the running coupling is regularised by adding a constant $C_{\rm reg}$ such that
\begin{equation}
 \abar(k_\perp^2)=\abar(\omega^2\theta^2)=\frac{1}{\bar{\beta}_0}\frac{1}{\log(\omega^2\theta^2/\Lambda^2+C_{\rm reg})}
\end{equation}
where $\bar{\beta}_0=\beta_0\pi/N_c$. In intermediate steps, we shall use $C_{\rm reg}=0$. Taking the logarithmic derivative with respect to $E$ and $R^2$, one finds the following partial differential equation
\begin{equation}
 ER^2\frac{\partial^2 T}{\partial E \partial R^2}=\frac{1}{\bar{\beta}_0}\frac{T}{\log(ER^2/\Lambda^2)}
\end{equation}
Then, changing to the variables $\kappa = \log(E/\omega)$ and $\lambda = \log(R^2/\theta^2)$ so that 
\begin{equation}
 \mathcal{T}(\kappa,\lambda,\mu)\equiv T(\omega,\theta^2\mid E, R^2)
\end{equation}
one gets
\begin{tcolorbox}[ams equation]\label{PDE-rc}
 \frac{\partial^2 \mathcal{T}}{\partial \kappa \partial \lambda}=\frac{1}{\bar{\beta}_0}\frac{\mathcal{T}}{2\kappa + \lambda + \mu}
\end{tcolorbox}
\noindent with $\mu=\log(\omega^2\theta^2/\Lambda^2)$ and the initial conditions $\mathcal{T}(0,\lambda)=\mathcal{T}(\kappa,0)=1/(\bar{\beta}_0\mu)$.

Including hard collinear splittings consists in 
adding the integral between 0 and 1 of the finite part of the splitting function, namely using
\begin{equation}
P_{gg}(z)\simeq\frac{1}{z}+\int_{0}^{1}\dif z\text{ }\Big(P_{gg}(z)-\frac{1}{z}\Big)=\frac{1}{z}-\frac{11}{12}\equiv\frac{1}{z}+B_g
\end{equation}
With this approximated splitting function and the running of the coupling, the evolution equation for $T(\omega,\theta^2\mid E, R^2)$ or equivalently $\mathcal{T}(\kappa,\lambda,\mu)$
can also be put into a partial differential equation form,
\begin{tcolorbox}[ams equation]\label{PDE-mlla}
 \frac{\partial^3\mathcal{T}}{\partial \kappa^2\partial\lambda}+\frac{\partial^2\mathcal{T}}{\partial \kappa \partial \lambda}=(1+B_g)\frac{\partial \abar \mathcal{T}}{\partial\kappa}+\abar\mathcal{T}
\end{tcolorbox}
\noindent with the initial conditions 
\begin{align}
\mathcal{T}(\kappa,0) &= \frac{1}{\bar{\beta}_0\mu}(1+B_ge^{-\kappa})\\
\mathcal{T}(0,\lambda) &= \frac{1}{\bar{\beta}_0\mu}(1+B_g)\\
\frac{\partial \mathcal{T}}{\partial\kappa}(0,\lambda)&=
\frac{1}{\bar{\beta}_0\mu}\Big(-B_g+\frac{(1+B_g)^2}{\bar{\beta}_0}\log\big(1+\lambda/\mu\big)\Big)\\
\frac{\partial \mathcal{T}}{\partial\lambda}(\kappa,0)&=\frac{1}{\bar{\beta}_0^2\mu}\int_0^\kappa d\kappa_1\frac{(1+B_ge^{-\kappa_1})(1+B_ge^{\kappa_1-\kappa})}{2(\kappa-\kappa_1)+\mu}
\end{align}
since equation \eqref{master} gives not only the differential equation but also the initial conditions. The partial differential equation (PDE) is then solved numerically on a two-dimensional grid, to get the function $\mathcal{T}(\kappa,\lambda,\mu)$ for several values of $\mu$.

This PDE is similar to the one given in Gerwick et al \cite{Gerwick:2012fw} in term of logarithmic accuracy but it has the advantage of being mathematically equivalent to the master equation \eqref{master} so that the solution $T$ satisfies the Chapman-Kolmogorov relation. This enables to use \eqref{fullT} and \eqref{master-eq} in order to calculate the medium-modified version of $T$.

\begin{figure}
   \centering
      \includegraphics[width=0.6\textwidth]{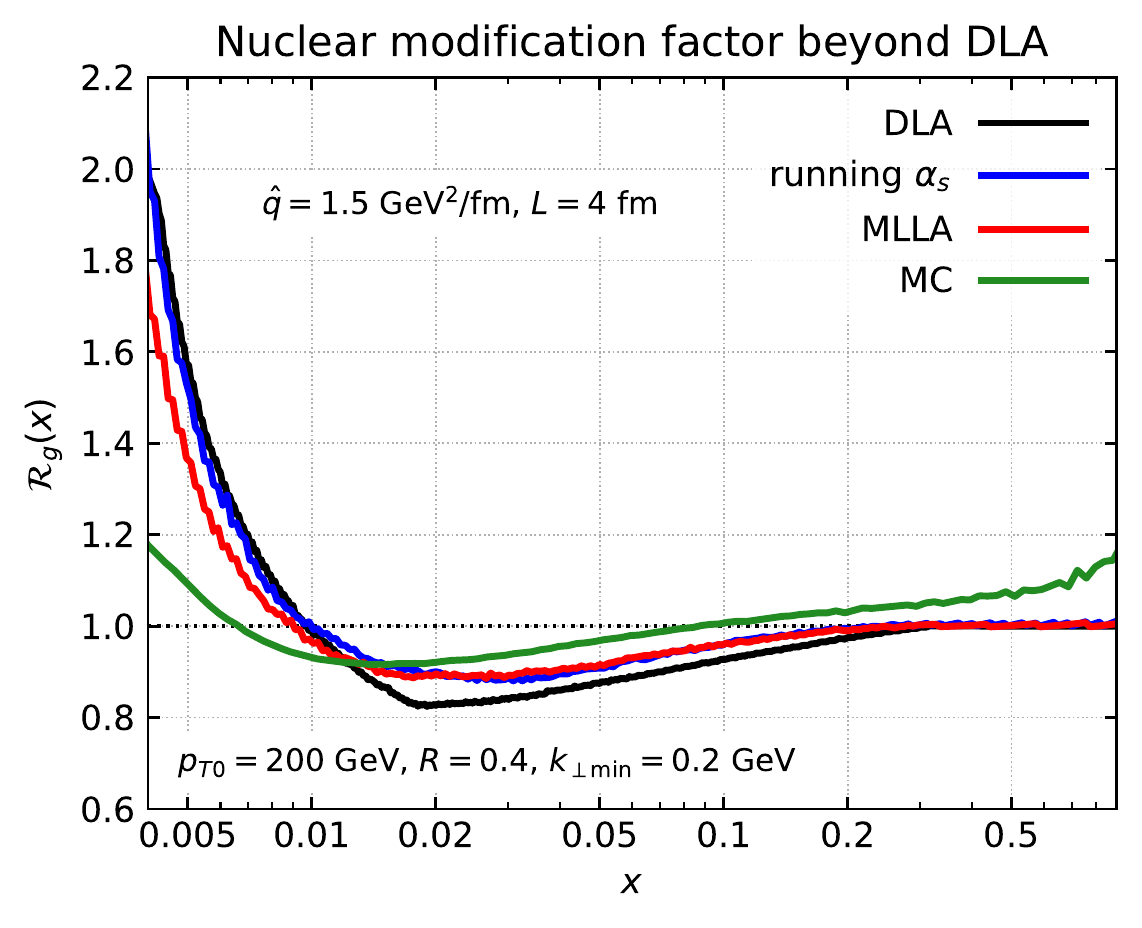}
    \caption{\small Nuclear modification factor for the fragmentation function beyond DLA}
    \label{Fig:R-beyondDLA}
\end{figure}

Numerical results are shown Fig.~\ref{Fig:R-beyondDLA} for the nuclear modification factor of the jet fragmentation function. Four calculations are compared: {\tt (i)} the DLA result already shown above, {\tt (ii)} the effect of the running coupling alone using the solution of \eqref{PDE-rc}, {\tt (iii)} the MLLA approximation including both running coupling and hard collinear radiations thanks to the PDE \eqref{PDE-mlla}, {\tt (iv)} a MC calculation with full DGLAP splitting function and energy conservation (see Chapter \ref{chapter:MC}).
As expected, DLA overestimates the number of soft gluons inside the jet but the enhancement
at small energies is still significant if we include the effects of the running coupling and the hard-collinear emissions. This shows the robustness of decoherence to generate soft gluons inside jets in
the presence of a medium.

\subsection{Jet average energy loss}
\label{sub:jet-eloss-th}

As the last exercise of this chapter, we estimate the jet energy loss within our factorised picture. So far, we have focused on the vacuum-like emissions since our arguments were grounded at DLA where one neglects the medium induced radiations. As explained in Section \ref{sub:finite-L-DL-picture}, this is correct for the vacuum-like series inside the medium, except for the set of partons at the end of the VLE cascade inside the medium which \textit{by definition} cross all the medium before radiating outside. During this travel, such emission triggers medium-induced jets as those studies in Chapter~\ref{chapter:jet}, Section \ref{sec:jets-mie} so that the jet loses energy from large angles radiations.

For the intrajet multiplicity, at first sight, it is legitimate to neglect this energy loss since it does not change the number of partons produced but merely shifts their final energy in the spectrum (recall also that a double-logarithmic cascade does not satisfy energy conservation anyway). It is certainly not correct for jet observables which are strongly sensitive to the energy of the final state particles. For such quantities, one cannot ignore medium-induced processes. However, one can often rely on the factorisation between vacuum-like emissions and medium-induced jets to estimate them.

\paragraph{Jet cross-section in A-A collisions.} In this subsection, we deal with the average jet energy loss $\mathcal{E}$. This quantity is ambiguous as it is not generally a well defined quantity in pQCD. It is not even measurable in experiments. The reason for this is that measuring an energy loss requires the knowledge of the energy of the leading parton sourcing the jet or some ``vacuum-equivalent'' jet. From a theoretical point of view, this is not a well defined problem whereas it is clearly not experimentally realisable. 

Nevertheless, one can use this quantity as an intermediate tool in order to obtain an estimation of a well defined observable: the jet cross-section. We start then by explaining how the average jet energy loss enters into the calculation of the jet cross-section in nucleus-nucleus collisions. Jets are defined in this subsection with a single parameter $R\lesssim1$, which should be thought as the jet radius in the anti-$k_t$ algorithm or the opening angle of the effective cone in cone-based jet definitions.

We call $\dif \sigma_{q/g}/\dif p_{T0}$ the Born level cross-section for producing respectively a quark or a gluon with transverse momentum $p_{T0}$ at central rapidity in $pp$ collision. At the lowest order in $\alpha_s$, the inclusive jet cross-section in $pp$ collisions can be approximated by:
\begin{equation}
 \frac{\dif \sigma^{pp}_{\rm jet}}{\dif p_{T}}=\frac{\dif \sigma_q}{\dif p_{T}}+\frac{\dif \sigma_g}{\dif p_{T}}+\mathcal{O}(\alpha_s^3)
\end{equation}
At this order, it is independent of $R$.

Now, let us assume that energy loss is the dominant effect of the medium in the calculation of the inclusive jet cross section in A-A collisions (this is of course a very strong assumption). A jet initially made of a single leading parton $i\in\{q,g\}$ with energy $p_{T0}$ evolves via vacuum-like and medium-induced emissions in the medium. We introduce the probability $P_i(\mathscr{E}|p_{T0},R)$ for the  reconstructed jet to have an energy $p_{T0}-\mathscr{E}$. In principle, this probability is non trivial even in the vacuum as the leading parton can radiate many soft gluons at angles larger than $R$. However, if $R$ is large enough, such contributions can be neglected. Therefore, $p_{T0}$ can be thought as the energy of an equivalent jet which would have evolved in the vacuum, so that $P_i(\mathscr{E}|p_{T0},R)$ is the probability for a jet to loose an energy $\mathscr{E}$.

From the probability $P$, one can calculate the jet yield in A-A collisions as follows:
\begin{equation}\label{AA-jet-cross}
 \frac{1}{\textrm{T}_{AA}}\frac{\dif \sigma_{\rm jet}^{AA}}{\dif p_T}\simeq\sum\limits_{i\in\{q,g\}}\int_0^{\infty}\dif p_{T0}\int_0^\infty\dif\mathscr{E}\,\delta(p_T-p_{T0}+\mathscr{E})P_i(\mathscr{E}|p_{T0},R)\frac{\dif \sigma_i}{\dif p_{T0}}
\end{equation}
The factor $\textrm{T}_{AA}$ accounts for the geometric enhancement of the jet yield due to the larger number of nucleons in a nucleus $A$. It depends only on the centrality of the collision and on the nature of the nucleus. Equation \eqref{AA-jet-cross} states that the initial \textit{parton} with transverse momentum $p_{T0}$ ends up as final state jet with energy $p_{T0}-\mathscr{E}$ with probability $P$. An other underlying assumption of \eqref{AA-jet-cross} is the absence of nuclear parton distribution function effects since we use the same Born-level quark/gluon cross-section as in the $pp$ baseline. This formula is hence an estimation for phenomenological applications rather than a rigorous pQCD calculation with a controlled $\alpha_s$ expansion.

Performing the $p_{T0}$ integration in \eqref{AA-jet-cross} and a Taylor expansion of $P_i\sigma_i$ for $\mathscr{E}/p_{T}\ll 1$, one gets:
\begin{align}\label{quenching-meth}
 \frac{1}{\textrm{T}_{AA}}\frac{\dif \sigma_{\rm jet}^{AA}}{\dif p_T}&=\sum\limits_{i\in\{q,g\}}\,\frac{\dif \sigma_i}{\dif p_T}+\frac{\partial }{\partial p_T}\Big(\mathcal{E}_i(p_{T},R)\frac{\dif\sigma_i}{\dif p_T}\Big)+\mathcal{O}\big((\mathscr{E}/p_T)^2\big)\\
 &\simeq \sum\limits_{i\in\{q,g\}}\frac{\dif \sigma_i(p_T+\mathcal{E}_i(p_{T},R))}{\dif p_T}
\end{align}
where the $\mathcal{O}\big((\mathscr{E}/p_T)^2\big)$ term makes sense only if the probability $P$ as a finite support $\mathcal{E}<\mathcal{E}_{\rm max}\le p_{T0}$ so that $\mathcal{O}\big((\mathscr{E}/p_T)^2\big)=\mathcal{O}\big((\mathscr{E}_{\textrm{max}}/p_T)^2\big)$. This means that if the typical maximal energy loss is negligible in front the initial $p_{T0}$ of the parton, the jet cross-section in A-A can be understood as a simple shift of the partonic cross-section. Yet, this shift is different according to the ``flavour'' of the leading parton and is given by the average jet energy loss $\mathcal{E}_i(p_{T0},R)$:
\begin{equation}
 \mathcal{E}_i(p_{T0},R)\equiv\int_0^\infty \dif\mathscr{E}\,\mathscr{E}P_i(\mathscr{E}|p_{T},R)
\end{equation}
To get the second line, we have neglected the logarithmic derivative of $\mathcal{E}$ with respect to $p_T$ in front of the logarithmic derivative of the differential cross-section. This is allowed because the differential cross-section $\dif \sigma_i$ is a steeply falling function of $p_T$ whereas we will see that $\mathcal{E}_i(p_T)$ increases with a $\alpha_s$ suppressed power of $p_T$. This steeply falling property of $\dif \sigma_i$ as a function of $p_T$ implies that $R_{AA}\le1$.

\paragraph{Calculation of $\mathcal{E}_i$ at DLA for the vacuum-like series.}

Let us now turn to an analytic estimation of the quantity $\mathcal{E}_i(p_{T0},R)$. In Sections~\ref{sub:frag-mie} and \ref{subsub:angular-gen-pict}, we have investigated the energy loss by a leading parton via medium-induced jets. We have argued that the energy lost by the jet via
medium-induced radiations at large angles ($\theta > R$) is controlled
by multiple branchings (resummed to all orders) and the
associated characteristic scale is $\ombr$. The results of \eqref{eloss-parton2} are summarized as follows:
\begin{itemize}
\item For jets with energies $p_{T0}\gg \upsilon\ombr$, the jet energy loss
  via MIEs becomes independent of $p_{T0}$:
  \begin{equation}\label{ElossHigh}
  \epsilon_{i,\rm MIE}(R)\,\simeq\,\upsilon\ombr+2\abar\om_c\sqrt{\frac{2c_*\theta_c}{R}}\qquad\mbox{when}\quad
  p_{T0}\gg \upsilon\ombr\,.
  \end{equation}
  %
\item Jets with $p_{T0}\lesssim \upsilon\ombr$ lose their whole energy via democratic branchings:  $\epsilon_{i,\rm MIE}(R)\simeq \epsilon_{i,\rm flow}\simeq p_{T0}$.
\item For a large jet radius
  $R\gtrsim \theta_c/\abar^2$, the flow component dominates over the
  spectrum component,
  $\epsilon_{i,\rm flow}\gg\epsilon_{i,\rm spec}$, for any energy
  $E$, and the energy loss
  $\epsilon_{i,\text{MIE}}\simeq \epsilon_{i,\rm flow}$ becomes
  independent of $R$.
\end{itemize}

We can now consider the generalisation of the above results to the
full parton showers, including both VLEs and MIEs. In the sequential
picture described in Section~\ref{sec:DLresum}, in which the two kinds of emissions are factorised in time,
each VLE occurring inside the medium acts as an independent source of MIEs
and hence the energy loss by the full jet can be computed by
convoluting the distribution of partonic sources created by the VLEs
in the medium with the energy loss via MIEs by any of these sources.
Assuming that all the in-medium VLEs are collinear with the jet axis
(which is the case in the collinear picture described in Section
\ref{sec:DLresum}), the energy lost by the {\it
  full} jet is computed as
\begin{tcolorbox}[ams align] \label{logeloss}
 \mathcal{E}_{i}(p_{T0},R)&\simeq  \int_{\omega_0(R)}^{p_{T0}} \rmd\omega\, \frac{\rmd N_{\text{VLE}}}
 {\rmd \omega}\,\epsilon_{\text{MIE}}(\omega,R)\,,\nonumber\\
 &\simeq\epsilon_{i,\text{MIE}}(p_{T0},R) + \frac{2\alpha_s C_i}{\pi}\int_{\theta_c}^R\frac{\rmd\theta}{\theta}
  \int_{\omega_0(\theta)}^{p_{T0}}\frac{\rmd\omega}{\omega}\,\textrm{I}_0\left(2\sqrt{\abar 
\log\frac{p_{T0}}{\omega}\,\log\frac{R^2}{\theta^2}}\right)\epsilon_{g,\text{MIE}}(\omega,R)\,
\end{tcolorbox}
 \noindent where ${\rmd N_{\text{VLE}}}/ {\rmd \omega}$ is the energy
 distribution of the partons created via VLEs inside the medium. The second line follows after using the DLA
 result for the gluon multiplicity, \eqref{TDLA}. This is of course a
 rough approximation which overestimates the number of sources, but it
 remains useful to get a physical insight. Similarly, for qualitative purposes, one can use the simple estimate for $\epsilon_{i,\text{MIE}}(\omega,R)$ given by the sum of Eqs.~\eqref{energy-flow}--\eqref{Espec}. 
 
\begin{figure}[t] 
  \centering
  \includegraphics[width=0.48\textwidth,page=1]{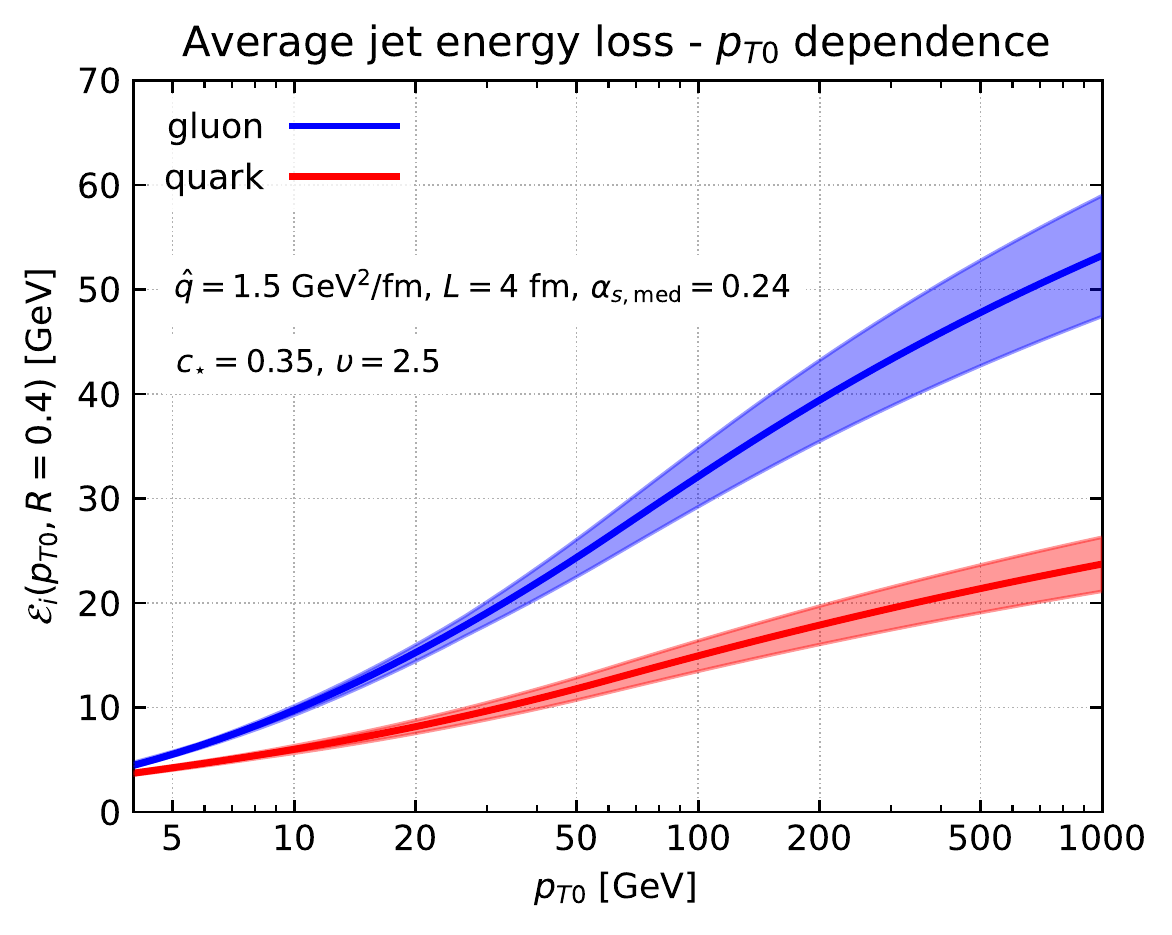}\hfill%
  \includegraphics[width=0.48\textwidth,page=2]{./jet-eloss.pdf}\hfill
 \caption{\small Double-logarithmic estimate of the jet energy loss as a function of $p_{T0}$ and $R$, obtained from a numerical integration of \eqref{logeloss}. The bands correspond to 25\% variations of the parameter $c_\star$ and $\upsilon$ around $c_\star=0.35$ and $\upsilon=2.5$. Note that \eqref{logeloss} predicts an exact Casimir scaling of the quark and gluon jet energy loss.}
 \label{Fig:jeteloss-DLA}
\end{figure}

 In Fig.~\ref{Fig:jeteloss-DLA}, we show the numerical results of Eq.~\eqref{logeloss} as a function of the initial $p_{T0}$ of the parton and the jet opening angle $R$\footnote{The precise description of the $R$ dependence of the jet energy loss for $R\sim1$ would require a proper treatment of initial state radiations, which is beyond our current
  framework.}. As expected, $\mathcal{E}_i(p_{T0},R)$ rises significantly as $p_{T0}$ increases, as a consequence of the in-medium fragmentation. More surprisingly, we observe a very mild increase of the jet energy loss as $R$ increases. One would naively expect that the more the jet opening angle is large, the more the energy loss is recovered at large angles (or equivalently, the spectrum component \eqref{Espec} decreases). However, this effect is compensated by the increase of the in-medium phase space for radiations. This compensation phenomenon has also been pointed out in \cite{Pablos:2019ngg} within the hybrid strong/weak coupling model with medium response included. The figure \ref{Fig:jeteloss-DLA} should also be compared with the Monte Carlo calculations of this function made in Chapter \ref{chapter:RAA}. There is a very nice qualitative agreement between the simple analytic estimate \eqref{logeloss} and these Monte Carlo results.

It is enlightening to understand the asymptotic behaviour of \eqref{logeloss} at large $p_{T0}$. This is straightforward from the analysis made for the double-logarithmic fragmentation function. Recall that when $p_{T0}\le \om_c(R/\th_c)^{4/3}$, the in-medium multiplicity is controlled by the position of saddle point $\om_1^\star$. If this saddle point is larger than $\ombr$, one can approximate $\epsilon_{i,\rm MIE}$ by $\ombr$ in \eqref{logeloss} meaning that the typical energy loss of an in-medium source is $\ombr$ since its initial energy satisfies $\om\gg\ombr$. The condition $\om_1^\star\ge\ombr$ is equivalent to $p_{T0}\ge \abar^4\om_c(\th_c/R)^{4/3}$.  Thus, \eqref{logeloss} has the following large $p_{T0}$ behaviour:
\begin{equation}\label{eloss-asymptot}
 \mathcal{E}_{i}(p_{T0},R)\sim \ombr^{(A)}\frac{C_i}{2C_A}\left(\frac{p_{T0}}{\om_0(R)}\right)^{\sqrt{\frac{3\abar}{2}}}\,\qquad\textrm{ if }\abar^4\om_c\left(\frac{R}{\th_c}\right)^{4/3}\ll p_{T0} \le \om_c\left(\frac{R}{\th_c}\right)^{4/3}
\end{equation}
This power-like increase of the energy loss with $p_T$ is a simple consequence of the larger phase space available for radiations inside the medium as $p_{T0}$ increases. It is a crucial ingredient to explain the flatness of the $R_{\rm AA}$ ratio for jet cross-section even at large $p_T$. Last but not least, \eqref{logeloss} and \eqref{eloss-asymptot} predicts that the jet energy loss is proportional to the
Casimir of the initial parton.

The condition $p_{T0}\le\om_c(R/\th_c)^{4/3}$ is not very restrictive for phenomenological applications as $\om_c(R/\th_c)^{4/3}=\mathcal{O}(1 \textrm{ TeV})$ for typical values of $\qhat$ and $L$. For completeness we give also the asymptotic behaviour of \eqref{logeloss} when $p_{T0}$ is larger than $\om_c(R/\th_c)^{4/3}$. In this case, the multiplicity is controlled by the phase space above $\om_c$ and $\th_c$ so that the energy loss increases like:
\begin{equation}
 \mathcal{E}_{i}(p_{T0},R)\sim \ombr^{(A)}\frac{C_i}{\sqrt{4\pi}C_A}\frac{\exp\Big(2\sqrt{\abar \log(p_{T0}/\om_c)\log(R/\th_c)}\Big)}{(\abar\log(p_{T0}/\om_c)\log(R/\th_c))^{1/4}}\,\qquad\textrm{ if } p_{T0} \ge \om_c\left(\frac{R}{\th_c}\right)^{4/3}
\end{equation}

 \begin{figure}
   \centering
      \includegraphics[width=0.48\textwidth,page=3]{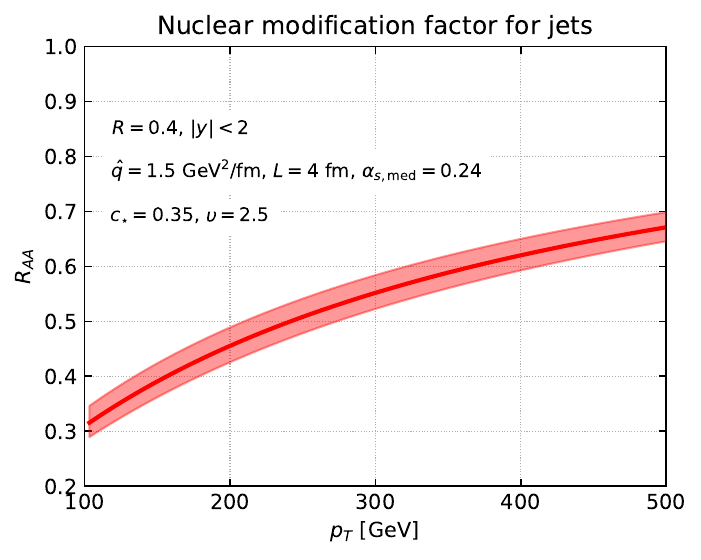}
    \caption{\small Analytic estimation of the nuclear modification factor for jets $R_{AA}$. The bands correspond to 25\% variations of the parameter $c_\star$ and $\upsilon$ around $c_\star=0.35$ and $\upsilon=2.5$. For the quark and gluon Born level cross-sections, we used a power law decrease $A_i (\hat{p}_T/p_T)^{n_i+m_i\log(p_T/\hat{p}_T)}$ from a fit of the numerical cross-sections over $p_T\in[100,500]$ GeV and $|y|<2$ with LHAPDF6 parton distribution functions \cite{Buckley:2014ana}.}
    \label{Fig:RAA-DLA}
\end{figure}

\paragraph{Jet quenching vs. hadron quenching.} To conclude, let us comment about the difference between the jet yield and the \textit{hadron} yield in nucleus-nucleus collisions. For a steeply falling cross-section, the method \eqref{quenching-meth} would give a wrong answer if the probability $P$ is such that the mean $\mathcal{E}$ is significantly larger than the typical energy loss corresponding to the energy where $P$ is maximal \cite{Baier:2001yt}. This is a consequence of the steeply falling cross-section which favours typical events rather than the mean scenario. In this situation, one cannot truncate the Taylor expansion up to the first order derivative. The alternative strategy is to expand \eqref{AA-jet-cross} as a function of $1/n$ where $n$ is the (large) power of the power suppressed cross-section $\dif \sigma_i$ \cite{Baier:2001yt}. For the hadron cross-section, such scenario occurs since we have seen in Chapter~\ref{chapter:jet}, Section~\ref{sec:jets-mie} that the mean \textit{parton} energy loss is of order $\alpha_s\om_c$ whereas the typical event by event energy loss is $\ombr=\alpha_s^2\om_c\ll \alpha_s\om_c$. 

For jets, this is different though. Introducing the parameter $R$ enforces the energy loss by the jet to be of the order of the event by event energy loss $\ombr$ so that the gluons carrying away this energy end up at angles larger than $R$. 
For jets, the typical and the mean energy loss are of the same order so that one could expect Eq.~\eqref{quenching-meth} to give a good estimate of the \textit{jet} cross-section in AA.

In Fig.~\eqref{Fig:RAA-DLA}, we show an analytic estimation of the $R_{AA}$ ratio based on Eqs.~\eqref{quenching-meth} and the average energy loss function \eqref{logeloss}. The general trend is good, namely a strong suppression, slightly increasing with $p_T$. However, the suppression is largely overestimated with respect to the data and the Monte Carlo results obtained in Chapter~\ref{chapter:RAA}. One explanation is that the assumption $\mathcal{E}_i\ll p_{T0}$ is not really realised as one can see from Fig.~\ref{Fig:jeteloss-DLA}. Another explanation is that large angle energy loss fluctuations are as large as the mean $\sim\ombr$. Moreover, for jets, this energy loss is enhanced by the multiplicity of vacuum-like sources, and this multiplicity has also large $p_T$-dependent event-by-event fluctuations. Thus, the first order truncation in \eqref{quenching-meth} is not a good approximation and the quenching weight method would be a better option if we knew the full probability distribution $P_i(\mathscr{E}|p_{T},R)$.


%% file: chapter5.tex
\chapter{JetMed: a Monte Carlo parton shower for jets in the medium}
\chaptermark{A new MC for in-medium parton showers}
\label{chapter:MC}

Chapter \ref{chapter:DLApic} was dedicated to the theoretical foundation of our factorised picture for the parton showers in the presence of a dense QCD medium. In the last section of this chapter, we have tried to include analytically the sub-leading single logarithmic corrections for the vacuum-like shower as well as the medium-induced emissions in order to quantify the energy lost by jets in the plasma. Another approach consists in numerically implementing the classical branching process in a so-called Monte Carlo event generator that mimics the evolution of the jet. This numerical approach is equivalent to the resolution of the evolution equation for the generating functional of the exclusive multi-particle production cross-sections so that the large terms in the $\alpha_s$ expansion of the cross-section --- due to either large soft/collinear logarithms for the vacuum-like emissions or large medium path length over formation time ratio for the medium-induced emissions ---are correctly resummed to all orders.

In this chapter, we describe the Monte Carlo parton shower for jet fragmentation in a dense QCD medium as it emerges from the ideas developed in the two previous chapters. The first Section~\ref{sec:veto} recalls the basics of the Sudakov veto algorithm used to generate the ``times'' in the branching processes. The two following Sections~\ref{sec:vac-show} and \ref{sec:med-show} describe in details the two main modules of the full parton shower: the vacuum shower which solves the coherent branching evolution equation \eqref{Z-mlla} and the medium induced shower which solves the evolution equation for medium-induced jets \eqref{Zmie-transverse}. These two modules are combined in the full medium shower according to the principles exposed in Chapter~\ref{chapter:DLApic}: this is the topic of Section~\ref{sec:full-show}. In this section, we also briefly present the code architecture. The last section \ref{sec:MCcomp} compares our parton shower with other Monte Carlo generators.

\section{The Sudakov veto algorithm}
\label{sec:veto}

A Monte Carlo parton shower simulates a classical branching process. In this section, we call $t$ or ``time'' the ordering variable of the process, while keeping in mind that the physical sense of $t$ may change according to the branching process at hand. The building block of a branching process is a time branching rate that we call $\Gamma(t)$. The probability $\dif P$ for a branching to occur between $t$ and $t+\dif t$ is, by definition:
\begin{equation}\label{branching-rate}
 \dif P\equiv \Gamma(t)\dif t
\end{equation}
From $\Gamma$, one can calculate the probability density $\mathcal{B}(t)$ for a branching to occur at time $t$, \textit{knowing that} no branching occurred between $t_0$ and $t$. This is given by the exponential decay law:
\begin{equation}\label{branching-prob}
 \mathcal{B}(t)=\Gamma(t)\exp\left(-\int_{t_0}^t\dif t'\,\Gamma(t')\right)\Theta(t-t_0)
\end{equation}
This probability density is properly normalised over the time interval $[t_0,\infty]$. If the branching process has an intrinsic maximal time $t_{\rm max}$, the normalised probability density $\mathcal{B}$ is given by \eqref{branching-prob} divided by its integral between $t_0$ and $t_{\rm max}$.

In a Monte Carlo parton shower, one must be able to select a time according to the probability density $\mathcal{B}$. This gives the time of the subsequent splitting, if the previous one occurred at time $t_0$. The purpose of this section is to recall the standard methods used to sample $\mathcal{B}$.

\subsection{Basics of random variable sampling}

We assume that we are able to generate numerically random numbers following the uniform distribution $\mathcal{U}(0,1)$ over $[0,1]$. We give here two general methods to sample $x$ according to the density $\mathcal{B}(x)$ for $x\in[x_0,x_{\rm max}]$.

\paragraph{Inverse transform method.} If the repartition function $B(x)$
\begin{equation}
 B(x)=\int_{x_0}^x\dif x'\,\mathcal{B}(x')
\end{equation}
 is strictly increasing and if one can find its inverse $B^{-1}$, then it is straightforward to generate $x$ according to the probability density $\mathcal{B}$:
\begin{equation}
 x=B^{-1}(u)
\end{equation}
with $u$ following $\mathcal{U}(0,1)$. For $\mathcal{B}$ of the form \eqref{branching-prob}, if one knows a primitive $\gamma$ of $\Gamma$ and its inverse $\gamma^{-1}$, then the inversion methods reads:
\begin{equation}
 t=\gamma^{-1}\Big(\gamma(t_0)-\log(u)\Big)
\end{equation}

\paragraph{Hit-or-miss (rejection) method.} When one cannot find analytically either the repartition function or its inverse, one can rely on the hit-or-miss method. If one finds a function $\mathcal{S}(x)$ and a constant $M\ge1$ such that:
\begin{equation}
 \mathcal{B}(x)\le M \mathcal{S}(x)
\end{equation}
for all $x$ in $[x_0,x_{\rm max}]$, and such that one knows how to sample $x$ following the density $\mathcal{S}(x)$, the variable $x$ is selected according to the following procedure:
\begin{itemize}
 \item Step 1: select $x$ according to $\mathcal{S}$, and $u$ according to $\mathcal{U}(0,1)$ independently,
 \item Step 2: while 
 \begin{equation}
  u> \frac{\mathcal{B}(x)}{M\mathcal{S}(x)}
 \end{equation}
 go back to step 1, otherwise stop and select $x$.
\end{itemize}
The final $x$ that comes out of this algorithm follows the density $\mathcal{B}$.

Note that if the function $\mathcal{B}$ is bounded by $M$ on $[x_0,x_{\rm max}]$, one can use $\mathcal{S}=\mathcal{U}(x_0,x_{\rm max})$. Of course, the more $M$ is small, the more the hit-or-miss algorithm is efficient (one can show that the expected value of the number of iteration is precisely $M$), so that it is advantageous to choose the envelop function $\mathcal{S}$ as close of $\mathcal{B}$ as possible. 

\subsection{Sudakov veto method}

The hit-or-miss method is very general. Actually, we did not use the fact that the density $\mathcal{B}$ is of the form \eqref{branching-prob}. The Sudakov veto method is a generalisation of the hit-or-miss method that makes use of the specific form \eqref{branching-prob} of the branching density probability. 

We first consider the case where there is no maximal time $t_{\rm max}=+\infty$.
The veto algorithm relies also on an envelop function $R(t)$, yet directly for the branching \textit{rate} and not the full probability density \eqref{branching-prob}. That is, we assume that there exists a function $R(t)$ such that:
\begin{equation}
 \Gamma(t)\le R(t)\qquad\textrm{ for }t_0\le t
\end{equation}
and such that one knows how to generate $t$ following the probability density $\mathcal{S}_i(t)$:
\begin{equation}\label{S-def}
 \mathcal{S}_i(t)=R(t)\exp\left(-\int_{t_i}^t\dif t'\,R(t')\right)\Theta(t-t_i)
\end{equation}
where the index $i$ refers to the initial time in the lower boundary of the integral and have been introduced for future convenience. 
From our discussion of the inversion method, one sees that it is interesting to look for a function $R$ such that its primitive $r$ and its inverse $r^{-1}$ are analytically known.

Now, let us give the veto algorithm that generates $t$ following the density $\mathcal{B}$:
\begin{itemize}
 \item Step 1: Initialize the index $i=0$, and $t_{i=0}=t_0$,
 \item Step 2: While $i\ge 0$, select $t_{i+1}$ according to $\mathcal{S}_{i}$ (hence $t_{i+1}>t_i$) and $u_{i+1}$ in $\mathcal{U}(0,1)$. If 
 \begin{equation}
  u_{i+1}>\frac{\Gamma(t_{i+1})}{R(t_{i+1})}
 \end{equation}
 continue, otherwise stop and select $t=t_{i+1}$.
 \end{itemize}
 
With respect to the hit-or-miss algorithm, one notices two main differences: firstly, the veto condition is on the ratio of the rates and not on the probability densities, and secondly the candidates $t_i$ are picked according to a probability density that changes from one step to another. A proof of the validity of this method can be found in \cite{Sjostrand:2006za}.

Finite $t_{\rm max}$ values are handled by stopping the veto algorithm if $t_{i+1}>t_{\rm max}$. Also, in all the applications of the veto algorithm discussed in this chapter, the rate $\Gamma(t)$ has the following general form:
\begin{equation}
 \Gamma(t)=\int \dif x\,\tilde{\Gamma}(t,x)
\end{equation}
where the boundaries of the $x$ integral may depend on $t$. Here, $x$ is an other random variable over which the rate $\tilde{\Gamma}(t,x)$ is marginalised. In principle, if one knows how to select a time $t$ distributed according to $\Gamma$, the generation of the other random variable $x$ proceeds afterwards. One first selects $t$ and then $x$ following the distribution $\tilde{\Gamma}(t,x)$ properly normalised over the allowed $x$-range. If one does not know how to generate $t$ directly, one can use an extension of the veto algorithm relying on an envelop rate $\tilde{R}(t,x)$ such that
\begin{equation}
 \tilde{\Gamma}(t,x)\le \tilde{R}(t,x)\qquad\textrm{ for all } x \textrm{ and } t_0\le t
\end{equation}
and such that the associated probability density $\mathcal{S}_i(t)$ given by \eqref{S-def} --- with $R$ given by $\tilde{R}(t,x)$ marginalised over $x$ --- is easy to generate.
The properly adapted veto algorithm becomes:
\begin{itemize}
 \item Step 1: Initialize the index $i=0$, and $t_{i=0}=t_0$,
 \item Step 2: While $i\ge 0$, select $t_{i+1}$ according to $\mathcal{S}_{i}$, $x_{i+1}$ following the (normalised) density $\tilde{R}(t_{i+1},x)$, and $u_{i+1}$ in $\mathcal{U}(0,1)$. If 
 \begin{equation}
  u_{i+1}>\frac{\tilde{\Gamma}(t_{i+1},x_{i+1})}{\tilde{R}(t_{i+1},x_{i+1})}
 \end{equation}
 continue,
 \item Step 3: else, select  $t=t_{i+1}$ and $x=x_{i+1}$ and stop.
 \end{itemize}
 
This algorithm is the standard tool in Monte Carlo parton showers. In the following sections, we shall describe more precisely how this algorithm is used to generate the splitting angles in the vacuum shower and the splitting times in the medium-induced shower.

\section{Vacuum shower}
\label{sec:vac-show}

We start by describing the implementation of the vacuum-like shower. This shower serves first as a vacuum baseline for quantifying jet quenching when medium effects are introduced. Most importantly, this module enters into the full medium shower to iterate vacuum-like emissions both in the inside and outside phase space.

\paragraph{Generic kinematic.}
We will represent the massless 4-vectors corresponding to emissions
using their transverse momentum $p_{Ti}$, their rapidity $y_i$ and
their azimuth $\phi_i$.
Since our physical picture is valid in the collinear limit, we will
often neglect differences between physical emission angles $\theta$
and distances $\Delta R=\sqrt{\Delta y^2+\Delta\phi^2}$ in the
rapidity-azimuth plane.
All showers are considered to be initiated by a single parton of given
transverse momentum $p_{T0}$, rapidity $y_0$ and azimuth $\phi_0$, and
of a given flavour (quark or gluon).

\paragraph{Implementation of the shower.}
Still working in the collinear limit, we will generate our partonic
cascades using an angular-ordered approach, starting from an initial
opening angle $\theta_\text{max}$. Hence the ``time'' variable of the branching process in the sense
of the previous section is the angle $\th$ always decreasing, or equivalently $\log(1/\th)$. The initial parton can thus be seen
as having $\theta=\theta_\text{max}$ and a relative transverse
momentum $k_{\perp}=p_{T0}\theta_\text{max}$.
To regulate the soft divergence of the splitting functions, we
introduce a minimal relative transverse momentum cut-off
$k_{\perp,\text{min}}$. This corresponds to the transition towards the
non-perturbative physics of hadronisation.
Note that for a particle of transverse momentum $p_{T0}$, the condition
$\kt >k_{\perp,\text{min}}$ imposes a minimal angle $\theta>\theta_\text{min}=k_{\perp,\text{min}}/p_{T0}$, and a minimal transverse momentum $p_T>p_{T,\rm min}=\ktmin/\theta_{\rm max}$ for the
next emission.

The shower is then generated using the Sudakov veto algorithm described in Section~\ref{sec:veto}. For the purposes of the
  subsequent discussion, $p_{T0}$ and $\th_0$ denote respectively the transverse
  momentum and the angle of a {\it generic} parent parton, which is not necessarily
  the {\it leading} parton. Its relative transverse momentum is defined by $k_{\perp0}=p_{T0}\th_0$.
  
With the notation used in Section~\ref{sec:veto}, the rate of the branching process is (see also \eqref{bremrate} for the description of the coherent branching algorithm) reads
\begin{tcolorbox}[ams equation]
 \tilde{\Gamma}^{\rm vac}_i(\th,z)=\frac{\alpha_s(k_\perp)}{\pi}P_i(z)\frac{1}{\th}
\end{tcolorbox}
\noindent where $i=(\text{g,\,q})$ is a flavour index, $z\equiv p_T/p_{T0}$ is the $p_T$ fraction of the softest emission with respect to the transverse momentum $p_{T0}$ of the parent and $k_\perp=zp_{T0}\th=p_T\th$ the relative transverse momentum of the emission. $P_i(z)$ is the splitting function of the parton $i$ defined in Appendix \ref{app:A}. $z$ and $\th$ are random variables, and we aim at first generating $\th\le \th_0$ according to the probability density:
\begin{equation}
 \mathcal{B}_i^{\vac}(\th)=\frac{1}{\th}\int_{p_{T,\rm min}/p_{T0}}^1\dif z\,\frac{\alpha_s(zp_{T0}\th)}{\pi}P_i(z)\exp\left(-\int_\th^{\th_0}\frac{\dif \th'}{\th'}\int_{p_{T,\rm min}/p_{T0}}^1\dif z\,\frac{\alpha_s(zp_{T0}\th)}{\pi}P_i(z)\right)
\end{equation}
and then, $z$ is generated with probability density $\tilde{\Gamma}^{\rm vac}_i(\th,z)$ normalised over $z\in[p_{T,\rm min}/p_{T0},1]$.

The envelop rate used for the veto algorithm is:
\begin{tcolorbox}[ams equation]\label{simple-rate}
 \tilde{R}_i^{\vac}(\th,z)=\frac{2C_i}{\pi}\frac{\alpha_s(zp_{T0}\th )}{\th}\frac{1}{z}
\end{tcolorbox}
\noindent where $C_\text{g}\equiv C_A$, $C_\text{q}\equiv C_F$. After the change of variable $k_\perp=zp_{T0}\th$, the marginalisation of $\tilde{R}(\th,z)$ over $z$ gives:
\begin{align}\label{marg-simple-rate}
 R_i^{\vac}(\th)&=\frac{1}{\theta}\int_{k_{\perp,\text{min}}}^{k_{\perp 0}} \frac{\rmd k_\perp'}{k_\perp'}\frac{2\alpha_s(k_\perp')C_i}{\pi}\\
  &=\frac{1}{\th}\frac{C_i}{\pi\beta_0}\log\Big(\frac{1-2\alpha_s\beta_0\log(k_{\perp 0}/M_Z)}{1-2\alpha_s\beta_0\log(k_{\perp,\text{min}}/M_Z)}\Big)
\end{align}
To obtain the second line, we used a 1-loop running coupling
$\alpha_s(k_\perp)=\tfrac{\alpha_s}{1-2\alpha_s\beta_0\log(k_\perp/M_Z)}$ with
$\alpha_s\equiv\alpha_s(M_Z)$ the running coupling at the $Z$ mass,
fixed to 0.1265, and $\beta_0=(11C_A-2n_f)/(12\pi)$ the 1-loop QCD
$\beta$ function. We consider a fixed $n_f=5$ flavours of massless quarks. The probability density $\mathcal{S}^{\vac}(\th)$ given by this rate and \eqref{S-def} is sampled using the inversion method since one can easily find the primitive of $R_i^{\vac}$ with respect to $\th$ and its inverse. In equations, given $u$ uniformly distributed over $[0,1]$, $\th$ is picked using:
\begin{equation}
 \th = \th_0\exp(\log(u)/R_i^{\vac}(1))
\end{equation}

The $k_\perp$ of the emission is then generated between $k_{\perp,\text{min}}$
and $k_{\perp 0}$ following the distribution
$\alpha_s(k_\perp)\dif k_\perp/k_\perp$. The allowed range for $\kt$ in \eqref{marg-simple-rate} has been chosen in order to simplify the analytic calculation of the $\th$ integration and does not correspond to the $z$ physical boundaries $p_{T,\rm min}/p_{T0}\le z\le 1$.
This procedure thus neglects finite effects in the splitting function and
momentum conservation as the splitting fraction
$z\equiv\tfrac{p_T}{p_{T0}}
=\tfrac{k_\perp}{k_{\perp 0}}\tfrac{\theta_0}{\theta}$
 associated with the emission of the gluon $\theta$ and $k_\perp$ in
\eqref{marg-simple-rate} can take values larger than one.
This is simply taken into account by vetoing emissions with $z>1$ and
accepting those with $z\le 1$ with a probability
$\tfrac{z}{2C_i}P_i(z)$ with $P_i(z)$ the targeted splitting
function. A similar trick allows us to select between the
  $g\to gg$ and $g\to q\bar q$ channels for gluon splittings. This trick is detailed in the next section as it is also used in the medium-induced shower.
If any of these vetoes fails, we set $\theta\to \theta_0$ and $k_\perp\th/\th_0\to
k_{\perp 0}$ --- so that $p_{T0}=k_{\perp 0}/\th_0$ is preserved --- and reiterate the procedure, following the Sudakov veto algorithm.

After a successful parton branching, both daughter partons are further
showered.
The procedure stops when the generated angle $\theta$ is smaller than
the minimal angle $\theta_\text{min}$.

\paragraph{4-vectors reconstruction.} To fully specify the procedure we still need to specify how to
reconstruct the daughter partons from the parent.
For this, we use $z\equiv \tfrac{p_T}{p_{T0}}=\tfrac{k_\perp}{k_{\perp 0}}\tfrac{\theta_0}{\theta}
$ and also generate an azimuthal angle
$\varphi$ around the parent parton, randomly chosen between 0 and $2\pi$.
We then write
\begin{equation}\label{eq:splitting-kinematics}
  \text{parent: }(p_{T0},y_0,\phi_0)
  \longrightarrow \begin{cases}
    \text{daughter 1: }\big(z\,p_{T0},y_0+(1-z)\theta\cos(\varphi),\phi_0+(1-z)\theta\sin(\varphi)\big)\\
    \text{daughter 2: }\big((1-z)p_{T0},y_0-z\,\theta\cos(\varphi),\phi_0-z\,\theta\sin(\varphi)\big)
  \end{cases}.
\end{equation}
This is strictly valid in the small-angle limit and power corrections in $\theta$ would appear beyond our small-angle approximation.

For examples of use of the vacuum shower, we refer the reader to Chapter~\ref{chapter:jet} where the fragmentation function and other jet substructure observables are calculated using the vacuum module of {\tt JetMed}. These calculations are compared to analytical predictions.

\section{Medium-induced shower}
\label{sec:med-show}

The medium-induced module develops a medium-induced parton shower from a leading parton with transverse momentum $p_{T0}$, as those described in Section \ref{sec:jets-mie}. The evolution parameter in the case of medium induced cascades is the ``time'' $t=x^+$ in light-cone coordinates with the longitudinal axis defined by the direction of motion of the leading particle. The generation of the shower is made in two steps. In the first step, all the partons produced are exactly collinear to the leading parton. In the second step, an angle is assigned to each parton produced by the collinear cascade assuming that all their relative transverse momentum comes from momentum broadening.

\subsection{Implementation of the collinear shower}

The collinear shower is mathematically equivalent to the resolution of the equation \eqref{Z-mie}.

\paragraph{The splitting time probability density.}
The rate $\tilde{\Gamma}^{\med}_i(t,z)$ of the Markovian branching process is given
\begin{tcolorbox}[ams equation]\label{mie-MC-rate}
 \tilde{\Gamma}^{\med}_i(t,z)=\frac{\alpha_{s,\med}}{2\pi}\sqrt{\frac{\qhat}{p_{T}}}\frac{\mathcal{K}_i(z)}{(z(1-z))^{1/2}}
\end{tcolorbox}
\noindent as in \eqref{mie-rate}. The functions $\mathcal{K}_i$ are given in Appendix \ref{app:A}. Here $i\in\{q,g\}$ and $p_T$ are respectively the flavour and the transverse momentum of the parent that splits, which is \textit{a priori} different from the leading parton $p_{T0}$ which defines the direction of the light cone coordinate $t$. In {\tt JetMed}, the coupling constant $\alpha_{s,\med}$ is frozen during the medium induced evolution and is a parameter of the Monte Carlo. A straightforward extension of the rate \eqref{mie-MC-rate} consists in evaluating $\alpha_s$ at the typical branching transverse momentum $k_{\rm br}=(\qhat z(1-z)p_T)^{1/4}$. The rate \eqref{mie-MC-rate} is constant with time since there is no explicit time dependence. The soft divergence in $\mathcal{K}_i$ is regulated with a minimal \textit{energy fraction} $zp_T/p_{T0}>z_{\rm min}$. 
With respect to \eqref{mie-rate}, we have dropped the step function $\Theta(\om_c-z(1-z)p_T)$. This constraint is handled directly by the veto algorithm.

It is convenient to use a dimensionless evolution variable, so we promote the ``reduced time'' $\tau$ defined by
\begin{equation}
 \boxed{\tau\equiv \frac{\alpha_{s,\med}}{\pi}\sqrt{\frac{\qhat/C_A}{p_{T0}}}t}
\end{equation}
as our new evolution variable. (We recall that, by convention, we use the adjoint quenching parameter to define $\qhat\equiv\qhat_A$.) The evolution is made from some $\tau_0$, the reduced splitting time of the parent parton to $\tau_L$ when $t=L$, that is when the leading parton escapes the medium. For the leading parton triggering the cascade, we set $\tau_0=0$.

If $\tau_0$ is the splitting of the parent parton, the targeted probability distribution for the subsequent splitting time $\tau>\tau_0$ is:
\begin{equation}\label{Bmed2}
 \mathcal{B}_i^{\med}(\tau)=\int_{x_{\rm min}}^{1-x_{\rm min}}\dif z\, \frac{\sqrt{C_A}}{2\sqrt{x}}\frac{\mathcal{K}_i(z)}{(z(1-z))^{1/2}}
 \exp\left(-\int_{\tau_0}^{\tau}\dif \tau'\int_{x_{\rm min}}^{1-x_{\rm min}}\dif z\, \frac{\sqrt{C_A}}{2}\frac{\mathcal{K}_i(z)}{(z(1-z))^{1/2}}\right)
\end{equation}
with  $x=p_T/p_{T0}$ and $x_{\rm min}=z_{\rm min}/x$.
The $z$-fraction of the splitting is chosen with probability $\tilde{\Gamma}_i(\tau,z)$ with $z$ between $x_{\rm min}$ and $1-x_{\rm min}$. Since it is not possible to use the inversion method for the probability density \eqref{Bmed2}, we use the Sudakov veto method. 

\paragraph{Sudakov veto method for gluon splittings.} The envelop rate $\tilde{R}_g^{\med}(\tau,z)$ for gluons makes use of the approximated kernel $\mathcal{K}_0(z)$ \eqref{K0} for which analytical solutions from the generating functional method are known. For convenience, the rate \eqref{Bmed2} is symmetrised so that $z$ can be chosen between $x_{\rm min}$ and $1/2$. This cancels the factor $1/2$ in \eqref{mie-MC-rate}. Then, one checks that
\begin{tcolorbox}[ams align]
\frac{\sqrt{C_A}}{\sqrt{x}}\frac{\mathcal{K}_g(z)}{(z(1-z))^{1/2}} \le \tilde{R}^{\med}_g(\tau,z)\equiv\frac{C_A^{3/2}}{\sqrt{x}}\mathcal{K}_0(z),\qquad \mathcal{K}_0(z)=(z(1-z))^{-3/2}
\end{tcolorbox}
\noindent Marginalising over $z\in[x_{\rm min},1/2]$ this envelop rate, one finds the following expression:
\begin{equation}
 R_g^{\med}(\tau)=\frac{C_A^{3/2}}{\sqrt{x}}\frac{2-4x_{\rm min}}{\sqrt{x_{\rm min}(1-x_{\rm min})}}
\end{equation}
The probability density $S_i^{\med}(\tau)$ built from this rate is sampled by the inversion method using:
\begin{equation}
 \tau=\tau_i-\log(u)/R_g^{\med}(0)
\end{equation}
with $u$ distributed as $\mathcal{U}(0,1)$. Then, $z$ is randomly chosen with probability density $\mathcal{K}_0(z)$ normalised between $x_{\rm min}$ and $1/2$, again using the inversion method
\begin{equation}
 z=\frac{v}{2\sqrt{16+v^2}},\qquad v=u'\left(\frac{2-4x_{\rm min}}{\sqrt{x_{\rm min}(1-x_{\rm min})}}\right)
\end{equation}
with $u'$ following $\mathcal{U}(0,1)$. With these expressions, one knows how to generate the branching process with the simplified kernel $\mathcal{K}_0(z)$. We are left with the implementation of the veto algorithm, that we detail to highlight how the condition $\om\le \om_c$ for medium-induced emissions is imposed and how the gluon/gluon and quark/antiquark splittings are handled:
\begin{itemize}
 \item Initialize $\tau_{i=0}=\tau_0$.
 \item While $i\ge 0$:
 \begin{itemize}
 \item select $\tau_{i+1}$ distributed as $S_i^{\med}(\tau)$, $z_{i+1}$ between $x_{\rm min}$ and $1/2$ distributed as $\mathcal{K}_0(z)$ and $u_{i+1}$ in $\mathcal{U}(0,1)$.
 \item If $\tau_{i+1}\ge \tau_L$, stop.
 \item Else if
 \begin{equation}
  z_{i+1}xp_{T_0}\ge \om_c
 \end{equation}
 continue,
 \item Else if
  \begin{equation}
  u_{i+1}\le\frac{\mathcal{K}_g^{q\bar{q}}(z_{i+1})/\sqrt{z_{i+1}(1-z_{i+1})}}{C_A\mathcal{K}_0(z_{i+1})}
 \end{equation}
 stop, and select $\tau_{i+1}$ as the next branching time. The branching is a $q\bar{q}$ splitting with momentum fraction $z_{i+1}$.
 \item Else if
 \begin{equation}
  u_{i+1}\le\frac{\mathcal{K}_g^{gg}(z_{i+1})/\sqrt{z_{i+1}(1-z_{i+1})}}{2C_A\mathcal{K}_0(z_{i+1})}
 \end{equation}
stop, and select $\tau_{i+1}$ as the next branching time. The branching is a gluon-gluon splitting with momentum fraction $z_{i+1}$.
 \item Else continue.
 \end{itemize}
\end{itemize}
The first condition guarantees that the subsequent branching time is well smaller than $L$ whereas the second veto condition ensures that the energy of the soft emission is smaller than $\om_c=\qhat_A L^2/2$. 

\paragraph{Sudakov veto method for quark splittings.} For quark, one puts the singular behaviour of the splitting function in $z=0$ only using:
\begin{equation}
 \mathcal{K}_q(z)=\mathcal{K}_{q}^{gq}(z)
\end{equation}
thanks to the $z\leftrightarrow1-z$ symmetry. $z$ is then generated between $x_{\rm \min}$ and $1$ with the envelop rate $\tilde{R}^{\med}_q(\tau,z)$ such that:
\begin{tcolorbox}[ams align]
 \frac{\sqrt{C_A}}{2\sqrt{x}}\frac{\mathcal{K}_q(z)}{(z(1-z))^{1/2}} \le \tilde{R}^{\med}_q(\tau,z)\equiv\frac{C_FC_A^{1/2}}{\sqrt{x}}\frac{1}{z(z(1-z))^{1/2}}
\end{tcolorbox}
\noindent The marginalisation of $\tilde{R}^{\med}_q(\tau,z)$ over $z\in[x_{\rm min},1]$ gives:
\begin{equation}
 R_q^{\med}(\tau)=\frac{C_FC_A^{1/2}}{\sqrt{x}}2\sqrt{\frac{1-x_{\rm min}}{x_{\rm min}}}
\end{equation}
Then, as for gluons, the branching time $\tau_i$ following $\mathcal{S}_i^{\med}(\tau)$ is picked using, with $u$ distributed as $\mathcal{U}(0,1)$:
\begin{equation}
 \tau = \tau_i-\log(u)/R_q^{\med}(0)
\end{equation}
Finally $z$ is randomly chosen between $x_{\rm min}$ and $1$ using (with again $u'$ given by $\mathcal{U}(0,1)$):
\begin{equation}
 z=\frac{1}{1+v^2/4}\,,\qquad v=u'\left(2\sqrt{\frac{1-x_{\rm min}}{x_{\rm min}}}\right)
\end{equation}
These formulas enable to generate the branching process given by the envelop rate $\tilde{R}_q^{\med}(\tau,z)$. The Sudakov veto algorithm is run to generate the branching given by the true rate $\tilde{\Gamma}_q^{\med}(\tau,z)$. This goes as for gluon splittings with the further simplification that there is only one decay channel for quarks.

\subsection{Implementation of the angular structure}

In the cascade described above, all the splittings are considered to
be exactly collinear. The angular pattern is generated afterwards via
transverse momentum broadening, cf.\ Sect.~\ref{sub:angular-structure}. 
For this, we go over the whole cascade and, for each parton, we generate
an opening angle $\theta$ and azimuthal angle $\varphi$ according to
the two-dimensional Gaussian distribution:
\begin{equation}\label{GaussianDis}
 \frac{\dif^2 \mathcal{P}_{\rm broad}}{\dif^2 k_\perp}=\frac{1}{\pi\qhat_R\Delta t}\exp\left(-\frac{k_\perp^2}{\qhat_R\Delta t}\right)
\end{equation}
where $\Delta t$ is the lifetime of the parton with colour representation $R$ in the cascade, and $k_\perp\simeq \om \th$.
Once we have the transverse momenta and angles of each parton in the
cascade, we use~(\ref{eq:splitting-kinematics}) to reconstruct the
kinematics.
Partons which acquire an angle larger than $\theta_\text{max}$ via
broadening are discarded together with their descendants.
We could study more involved broadening distributions in the future, taking into account hard scattering contributions beyond the Gaussian approximation (see e.g. \cite{Kutak:2018dim,Blanco:2020uzy}).

\begin{figure}
 \centering
\includegraphics[width=0.6\textwidth]{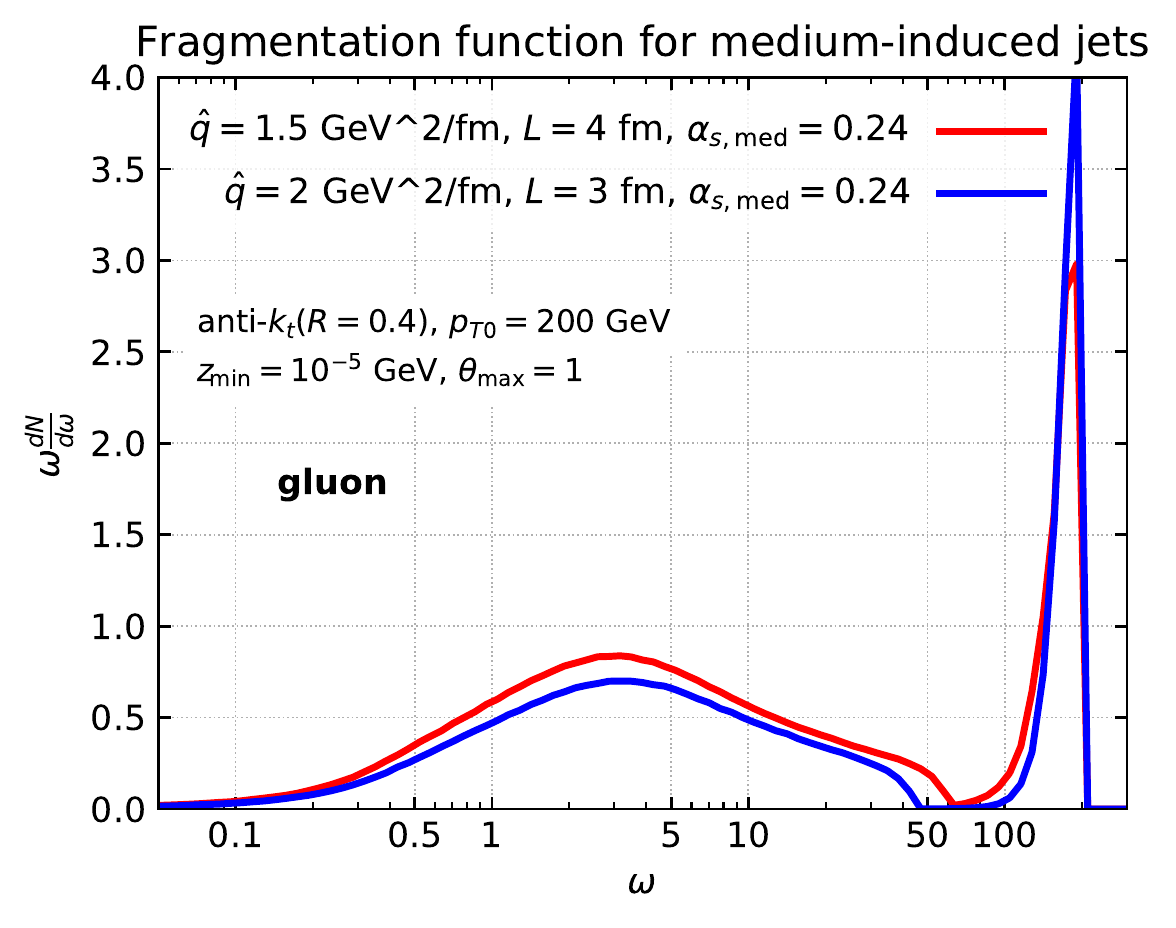}
\caption{\label{Fig:plotmodules}\small The 
energy spectrum of particles produced by medium-induced splittings from a leading gluon with $p_{T0}=200$ GeV resulting from the module {\tt MediumInducedShower}, for two 
different values of $\hat{q}$ and $L$. The parameters $z_c$ and $\theta_{\rm max}$ are fixed respectively to $10^{-5}$ and 1.}
\end{figure}

As an example, we show on figure \ref{Fig:plotmodules}, the resulting energy spectrum (or fragmentation function) of a medium-induced jet triggered by an initial gluon with energy $p_{T0}=200$ GeV. After the evolution of the leading gluon via the medium-induced shower, all the produced particles are clustered with the anti-$k_\perp$ algorithm with $R=0.4$ and only the hardest resulting jet is taken into account into the calculation of the fragmentation pattern. The large narrow peak near $\omega=200$ GeV is obviously the remnant of the leading particle. The second broad peak corresponds to the accumulation around the characteristic energy scale $Q_s/R=(\hat{q} L)^{1/2}/R$ of soft gluons produced via medium induced 
radiations, due to momentum broadening. This spectrum agrees with our study of the fragmentation of medium-induced jets in Chapter~\ref{chapter:jet}, Section~\ref{sec:jets-mie}.

\section{Full medium shower}
\label{sec:full-show}

\subsection{The global picture}
The in-medium shower is generated in three stages, according to the
factorisation discussed in Section~\ref{sub:fac-summary}.
\begin{enumerate}
 \item The first step is to generate in-medium VLEs. This is done exactly as
for the full vacuum shower except that each emission is further tested
for the in-medium conditions $k_\perp^3\theta>2\hat{q}$ and
$\theta>\theta_c$. If any of these two conditions fails, the emission
is vetoed.
\item The second step is to generate MIEs for each of the partons obtained
at the end of the first step, following the procedure described above. We neglect the formation time of the vacuum-like sources since for them, $t_f\ll L$. Consequently, all the medium-induced cascades triggered by the in-medium vacuum-like emissions develop over a time $L$.
\item The third step is to generate the VLEs outside the medium.
For this, each parton at the end of the MIE cascade is taken and
showered outside the medium. This uses again the vacuum shower,
starting from an angle $\theta_\text{max}$ since decoherence washes
out angular ordering for the first emission outside the medium. Each
emission which satisfies either $k_\perp\theta<2/L$ or $\theta<\theta_c$
is kept, the others are vetoed.

\end{enumerate}

The full parton shower can be converted to 4-vectors suited for any
analysis. Final-state (undecayed) partons are taken massless with a
kinematics taken straightforwardly from
Eq.~(\ref{eq:splitting-kinematics}). If needed, the 4-vectors of the
other partons in the shower are obtained by adding the 4-momenta of
their daughter partons. This requires traversing the full shower
backwards.

Whenever an observable requires to cluster the particles into jets and
manipulate them, we use the {\tt FastJet} program
(v3.3.2)~\cite{Cacciari:2011ma} and the tools in {\tt fjcontrib}~\cite{fastjet-contrib}.
In particular, the initial jet clustering is always done using the
anti-$k_\perp$ algorithm~\cite{Cacciari:2008gp} with $R=0.4$ unless
explicitly mentioned otherwise.

\subsection{Code architecture}

The code is written in {\tt C++} (v3 or higher). It requires the installation of the {\tt FastJet} program
(v3.3.2)~\cite{Cacciari:2011ma}.

\paragraph{Event classes.} Besides generating the branching of a given parton described in the previous sections, the Monte Carlo records the full history of the branching process. One can access to any branching information thanks to the three following classes:
\begin{enumerate}
 \item the class {\tt Parton} which handles the leaves in the shower. This class gathers all the necessary informations about the partons generated during the cascade. These informations can be physical: one can access to the kinematic or the ``flavour'' of any given {\tt Parton}. There are also unphysical informations related to the location of the {\tt Parton} in the full {\tt Event}.
 
 \item the class {\tt Vertex} which handles the nodes of the branching process. This class records the physical informations associated with the splitting at the node, as well as the location of the parent and daughters {\tt Parton} in the {\tt Event}.

 \item the class {\tt Event} contains a vector of {\tt Vertex}, {\tt Parton} and {\tt PseudoJet} (class from {\tt FastJet}) corresponding to the full tree of the branching process. A set of methods enables to handle the branching of a parton coming from a given shower to the full event or to access to the final states partons. 
\end{enumerate}

\paragraph{Shower classes.} The two kind of showers described in the previous sections, vacuum-like or medium-induced constitute two independent classes {\tt VacuumShower} and {\tt MediumInducedShower}. A final class {\tt MediumShower} combines together the vacuum-like shower and the medium-induced shower according to the principles of our factorisation. In order to highlight the specificity of this {\tt MediumShower} with respect to other possible implementation, it is interesting to develop fictitious unphysical showers where one of the important ingredients of our picture is missing. Thus, a parton can be showered also via:
\begin{enumerate}
 \item the class {\tt HardMediumInducedShower}: in this shower, the angular pattern of the medium induced shower is not generated. All the medium-induced partons produced via the purely collinear shower are discarded in the event (and not further showered), because their angles are set to a value larger than $\theta_{\rm max}$ by hand. Only the resulting energy loss of the leading parton is recorded. This class enables to isolate the effects of the energy loss via medium-induced radiations, and by contrast, to understand the effect of the medium-induced emissions with relatively small angles.
 
 \item the class {\tt MediumAOShower}: this shower is the same as {\tt MediumShower} except that there is no reopening of the phase space for the first decay given by the {\tt VacuumShower} in the outside phase space. The angle of this emission is constrained by the opening angle of the previous one, so that all \textit{vacuum-like} branchings are angular ordered.
 \end{enumerate}

\subsection{{\tt JetMed} free parameters}
\label{sub:JetMed-params}

We give here the list of parameters that enter into a {\tt JetMed} calculation. These parameters can be divided into two sets: the physical parameters that are, in principle, measurable and the unphysical parameters which enter into the calculation in order to regulate the soft or collinear singularities.

\begin{table}[!h]
  \centering
  \begin{tabular}{|l|c|c|}
  \hline
  Shower                     &     Unphysical parameters           & Physical parameters \\
  \hline
  {\tt VacuumShower}         &   $\theta_{\rm max}$, $\ktmin$      & $\alpha_s(M_Z)$     \\
  \hline
  {\tt MediumInducedShower}  &   $\theta_{\rm max}$, $z_{\rm min}$ & $\qhat$, $L$, $\alpha_{s,\med}$ \\
  \hline
  \end{tabular}
    \caption{\small Parameters of the Monte Carlo {\tt JetMed}. The unphysical parameters $\ktmin$ and $z_{\rm min}$ are used to regulate the infrared divergences in the probability distribution for generating a given emission. They are not measurable quantities contrary to the the physical parameters listed in the second column. }
\end{table}

Among these parameters, one could easily avoid the use of the frozen coupling $\alpha_{s,\med}$ using the value of $\alpha_s$ at the transverse momentum scale $k_{\rm br}$ (see Section~\ref{sec:med-show}).

\section[Limitations and comparison with other Monte Carlo event generators]{Limitations and comparison with other Monte Carlo event generators%
\sectionmark{Limitations and comparison with other MC}}
\label{sec:MCcomp}
\sectionmark{Limitations and comparison with other MC}

\subsection{The missing ingredients}

The Monte Carlo generator that is described above is of course very
simplistic and has a series of limitations. We list them here for the
sake of completeness. 
\paragraph{Hadronisation and initial state radiations.} First of all, we only generate a partonic
cascade, neglecting non-perturbative effects like hadronisation. Even
if one can hope that these effects are limited --- especially at large
$p_T$ --- our description remains incomplete and, for example,
track-based observables are not easily described in our current framework.
Additionally, our partonic cascade only includes final-state
radiation. Including initial-state radiation is left for future work. This would be needed, for
example, to describe the transverse momentum pattern of jets recoiling
against a high-energy photon.

\paragraph{Medium and nucleus-nucleus collision modelling.} Our description of the medium is also simplified: several effects like
medium expansion, density non-uniformities and fluctuations, and the
medium geometry are neglected. The geometry of the collision is a crucial missing ingredient if we aim at understanding the dependence of Monte Carlo calculations with the centrality and the energy of the collision.
This can to a large extent be hidden into an adjustment of the few parameters we have left, but we
would have to include all these effects to claim a full in-medium generator.

\paragraph{Hard scatterings off medium constituents and medium response.} As discussed in Section~\ref{sub:beyondDLA}, there are other medium effects which are not included in the current implementation of {\tt JetMed}. First of all, the medium response of the jet propagation is known to produce sizeable effects on medium modified jet shapes. Including the medium response of the jet at the level of our Monte Carlo would require a complete  knowledge of the hydrodynamic evolution of the medium. 

Hard scatterings between the leading parton that triggers the jet (or any of its descendent) and a medium scattering center are also neglected in the multiple soft scattering approximation. In terms of Monte Carlo modelling, i.e. without relying on a theoretical proof of the validity of the Monte Carlo calculation in a given approximation scheme, hard scattering contributions could be added using the full transverse momentum probability distribution $\mathcal{P}(k_\perp,\Delta t)$ defined in \eqref{Ptransverse} instead of its Gaussian approximation \eqref{GaussianDis} used to generate the angular pattern of the medium induced shower (or equivalently, using the analytic approach of \cite{Barata:2020rdn}). This should at least correct the angular distributions, so that the Monte Carlo would really solve \eqref{Zmie-transverse} with the right collision kernel. Note that the Monte Carlo implementation of the branching process associated with \eqref{Zmie-transverse} for gluons only has also been pursued in the parton shower {\tt MINCAS} \cite{Kutak:2018dim}. 

Furthermore, hard scattering also modifies the emission rate. Hence, a complementary improvement would be to use the recent results of \cite{Mehtar-Tani:2019ygg}. In this paper, the authors provide a rate for medium induced splittings that encompasses at the same time the Bethe-Heitler regime, the multiple soft scattering regime and the single hard scattering regime.

\paragraph{Shortcoming improvements.} Among all of these approximations, some of them could easily be circumvented. As argued, the longitudinal expansion of the medium requires a modification of the veto region for vacuum-like emissions. This veto region has been calculated in Chapter~\ref{chapter:DLApic}, so  the implementation should be straightforward. Moreover, the effect of the longitudinal expansion on the medium-induced cascade discussed in \ref{sub:med-expansion} amounts to a change in the relation between the dimensionless parameter $\tau$ and the proper time $t$ along the direction of motion of the leading parton.

\subsection{{\tt JetMed} in the landscape of in-medium event generators}

Over the past few years, a large variety of Monte Carlo parton showers and event generators has appeared in the heavy-ion community, stimulated by the large amount of data associated with jet quenching effects. In this subsection, we would like to precise the location of {\tt JetMed} among all these parton showers. {\tt JetMed} is based on the factorised picture exposed in Chapter \ref{chapter:DLApic}, which allows for a fast implementation of an in-medium parton shower. With this perspective, {\tt JetMed} is unique as it is the only parton shower that includes decoherence effects in the vacuum-like cascade and the multiple branching regime for medium-induced radiations as predicted by perturbative QCD.

The factorisation between the vacuum-like shower and the medium effects is not unique. Actually, many in-medium parton showers start with a shower triggered by the virtuality of the hard-process up to some cut-off virtuality scale $Q_{\med}$ where medium effects take over. This is the case for {\tt MARTINI} \cite{Schenke:2009vr} for instance, for which the vacuum-like shower is generated by {\tt Pythia} 8.1. Nevertheless, in our approach, this virtuality scale $Q_{\med}$ is dictated by the physical parameters of the medium. Besides the treatment of the vacuum-like cascade, it is not clear whether the {\tt MARTINI} approach encompasses the multiple branching regime discussed in Chapter \ref{chapter:jet}, Section \ref{sec:jets-mie}, even if the transition rates used in {\tt MARTINI} account for both the Bethe-Heitler and LPM regimes.

In the framework {\tt MATTER}+{\tt LBT} \cite{Majumder:2013re,Cao:2017qpx,He:2018xjv,Luo:2018pto}, there is also a virtuality transition scale between the virtuality-driven shower and the outgoing partons-medium interactions. This transition scale is chosen of order $\mathcal{O}(1\textrm{ GeV})$. The {\tt MATTER} shower includes the so-called higher twist (HT) contributions to the one-gluon emission cross-section from a single scattering, generalised to include multiple emissions in a DGLAP-like formalism with medium-modified splitting functions. When the shower reaches the typical virtuality scale generated by medium scatterings, the {\tt LBT} shower takes over. This shower implements a kinetic rate equation, whose rates are again given by a higher-twist approach. In {\tt JetMed}, higher-twist contributions to emission cross-sections and rates are neglected.

In the way the dominant medium effects are implemented, {\tt MATTER} is close to the {\tt Q-PYTHIA} and {\tt Q-HERWIG} implementations of the in-medium parton shower \cite{Polosa:2006hb,Armesto:2007dt,Armesto:2009fj,Armesto:2009ab} (or \cite{Dainese:2004te} which uses a similar idea) since medium-modified splitting functions are used. These medium-modified splitting functions incorporate the effect of medium-induced emissions via an additive term to the usual DGLAP splitting functions. This additive term is given by the off-shell spectrum calculated in Chapter \ref{chapter:emissions} (with the vacuum limit subtracted). This implementation is different from the one in {\tt JetMed} since our factorisation enables to ignore either the vacuum or the medium-induced component according to the stage of the evolution. That said, at the level of a single emission, both calculations agree in the soft limit.

Our factorised approach is in principle closer to the implementation of the Monte Carlo {\tt JEWEL} \cite{Zapp:2013vla} in which ``the interplay between competing radiative processes is governed by the formation times of the emissions'' meaning  that, for instance, in the generation of a splitting, the choice between a vacuum-like and a medium-induced one is made according to the criterion of shortest formation time. This is precisely the argument made to derive the veto region in Section \ref{sub:qualitative-veto}. Hence, {\tt JetMed} focuses somehow on the drastic scenario where the in-medium vacuum-like cascade happens very shortly before triggering a bunch of medium-induced emissions. Since this scenario is the dominant one to single logarithmic accuracy for vacuum-like emissions, the {\tt JetMed} approach is considerably simplified.

Last but not least, {\tt JetMed} is completely orthogonal to the hybrid strong/weak coupling model \cite{Casalderrey-Solana:2014bpa}, as the underlying medium in {\tt JetMed} is assumed to be weakly coupled. Nevertheless, there are some physical similarities since the hybrid model takes into account a characteristic resolution length \cite{Hulcher:2017cpt} of the plasma, a concept which is closed to the coherence angle derived in weak coupling calculation in Chapter~\ref{chapter:emissions}.

To summarise, we provide a \textit{non-exhaustive} Table \ref{Tab:MCs} recapitulating the main ingredients of all the parton showers discussed in this subsection. 

\begin{table}[t]
\small
  \centering
  \begin{tabular}{|l|c|c|c|c|c|c|}
  \hline
  \textbf{Monte Carlo}                 & {\tt JetMed} & {\tt MARTINI} & {\tt MATTER}+{\tt LBT} & {\tt Q-PYTHIA} & {\tt JEWEL} &  Hybrid\\
  \hline
  Factorisation scale         & $\checkmark$ &$\checkmark$ &$\checkmark$&$\xmark$ &$\xmark$ &$\xmark$ \\ \hline
  Decoherence                 & $\checkmark$ &$\xmark$     &$\xmark$  &$\xmark$   &$\xmark$ &$\xmark$ \\ \hline
  LPM effect                  & $\checkmark$ & $\checkmark$&$\xmark^{(1)}$& $\checkmark$& $\checkmark$& $\xmark$\\ \hline
  Multiple branching regime   & $\checkmark$ & ?           &$\xmark$    &$\xmark$ &? &$\xmark$  \\ \hline
  Hadronisation               &$\xmark$      &$\checkmark$ &$\checkmark$&$\checkmark$ &$\checkmark$ &$\checkmark$ \\ \hline
  Medium geometry/expansion        &$\xmark$      &$\checkmark$ &$\checkmark$&$\xmark^{(2)}$&$\checkmark$ &$\checkmark$ \\ \hline
  Hard single scatterings     &$\xmark$      &$\checkmark$ & $\checkmark$&$\xmark$ &$\checkmark$ & $\xmark$\\ \hline
  Medium response             &$\xmark$      &$\xmark$     &$\checkmark$&$\xmark$   &$\checkmark$ &$\checkmark$ \\ \hline
  HT splitting functions  & $\xmark$     &$\xmark$     & $\checkmark$&$\xmark$ &$\xmark$ &$\xmark$ \\ \hline
  Strongly coupled energy loss&$\xmark$      &$\xmark$     &$\xmark$&$\xmark$   &$\xmark$ & $\checkmark$\\ \hline
  \end{tabular}
  \caption{\small A personal comparison of several Monte Carlo in-medium parton showers. We apologize if they are not faithfully represented as it can be sometimes difficult to disentangle the physical ingredients behind the numerical implementations. Notes: (1) A modified-Boltzmann approach is proposed in \cite{Ke:2018jem} to take into account the LPM regime. (2) {\tt Q-PYTHIA} can be interfaced to an optical Glauber model \cite{vanLeeuwen:2015upu}.}
  \label{Tab:MCs}
\end{table}


%% file: intro-part2.tex
\chapter{Introduction: jet quenching observables}
\label{chapter:intro-part2}

In this introductory chapter, we first aim at giving a large (but of course non exhaustive) survey of the measurements related to jet quenching phenomena which will be studied in more details in the following chapters. Furthermore, we explain qualitatively how the physical picture for jet evolution in a dense QCD medium discussed in Part~\ref{part:theory} enables to understand the salient features of these experimental data. 

So far, there are still important missing ingredients in this picture. These missing ingredients do not allow for a complete description of all the current data. Each time we discuss a measurement, we explain why we expect this measurement to be or not to be properly understood (at least qualitatively) within our theory. 

As usual in jet physics, we also expect better agreement between the theory and the data when there is a large separation of scales between the hard process and the soft sector: hadronisation, underlying event, flow in heavy-ion collisions, etc. That is why we focus mainly on LHC data, as the center of mass energy is large enough to produce high $p_T\gtrsim 100$ GeV jets in the final state. A non-exhaustive list of references regarding jet measurements at RHIC is given in the first section.

One can divide the measurements done so far regarding jets in heavy-ion collisions into two main categories: global and substructure observables. For global jet observables, all the informations about the inner properties of jets are discarded, and only the global properties (transverse momentum, mass, etc) are compared to the proton-proton collision case. On the contrary, for substructure observables, one looks for modifications of the inner properties of jets after propagation through the quark-gluon plasma. Therefore, this chapter is divided into three sections. The first two discuss these global and substructure observables independently. In the last section, we argue that one can get substantial physical informations by correlating these two classes.

\section{Global jet observables}

\subsection{Inclusive jet cross-sections}

The first natural jet observable to look at in nuclei-nuclei collisions is the total jet cross-section. Since the seminal suggestion of Bjorken \cite{Bjorken:1982tu}, such quantities have been extensively measured both at RHIC \cite{Adcox:2001jp,Adler:2003qi,Adler:2003cb,Adler:2003au,Adler:2005ee,Adare:2008qa,Adare:2008ae,Adler:2002xw,Adler:2002tq,Adams:2003kv,Adams:2003am,Adams:2003im,Adams:2005ph,Adams:2006yt,Adamczyk:2016fqm,Adamczyk:2017yhe,Adam:2020wen} and at the LHC \cite{Aad:2014wha,Aad:2015wga,ATLAS:2017rmz,Chatrchyan:2011sx,Chatrchyan:2011pb,CMS:2012aa,Chatrchyan:2012gw,Chatrchyan:2012nia,Khachatryan:2016odn,Abelev:2012hxa,Abelev:2013kqa,Abelev:2014laa,Adam:2015kca,Adam:2015ewa,Acharya:2017goa,Acharya:2018qsh} (and other references below). We do not discuss the inclusive hadron cross section which has been the first historical evidence for jet quenching in heavy-ion collisions at RHIC. Instead we focus on the inclusive \textit{jet} cross section and its nuclear modification factor as a function of the transverse momentum $p_T$ and the rapidity $y$, defined by:
\begin{equation}
 R_{AA}(p_T,y)\equiv \frac{1}{\langle T_{AA}\rangle}\frac{\frac{\dif^2\sigma_{\rm jet}}{\dif p_T\dif y}\Big|_{AA}}{\frac{\dif^2\sigma_{\rm jet}}{\dif p_T\dif y}\Big|_{ pp}}
\end{equation}
The normalisation factor $\langle T_{AA}\rangle$ is the mean nuclear thickness function which accounts for the geometric enhancement of the hard scattering rate \cite{Miller:2007ri}. In the absence of nuclear effect, this normalisation factor guarantees $R_{AA}(p_T,y)=1$. Jets are usually reconstructed using the anti-$k_t$ algorithm with parameter $R$ \cite{Cacciari:2008gp}.

\begin{figure}[t] 
  \centering
  \begin{subfigure}[t]{0.50\textwidth}
    \includegraphics[width=\textwidth]{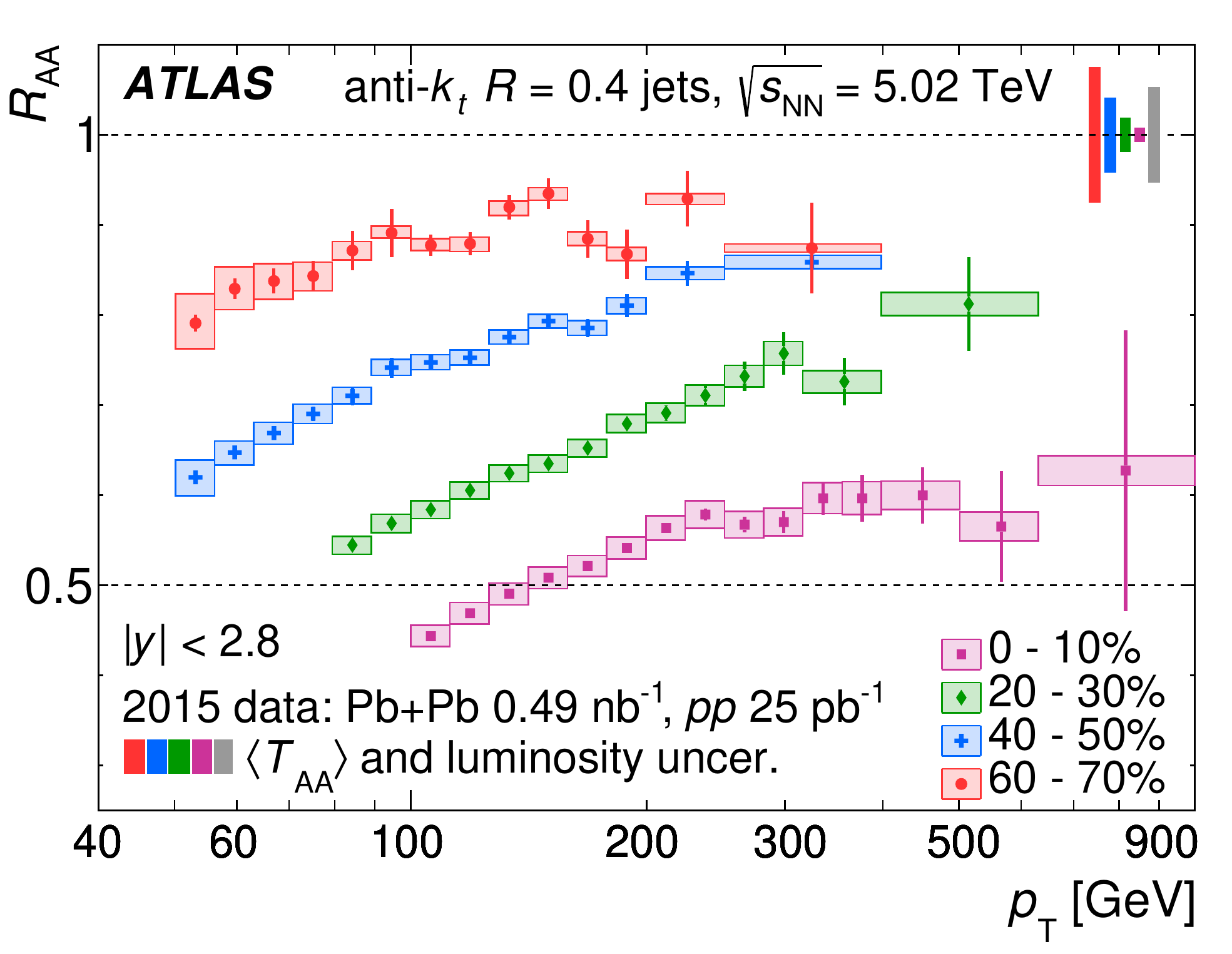}
    \caption{\small ATLAS - Centrality dependence. Figure from \cite{Aaboud:2018twu}.}\label{fig:ATLAS-RAA}
  \end{subfigure}
  \hfill
      \begin{subfigure}[t]{0.46\textwidth}
    \includegraphics[width=\textwidth]{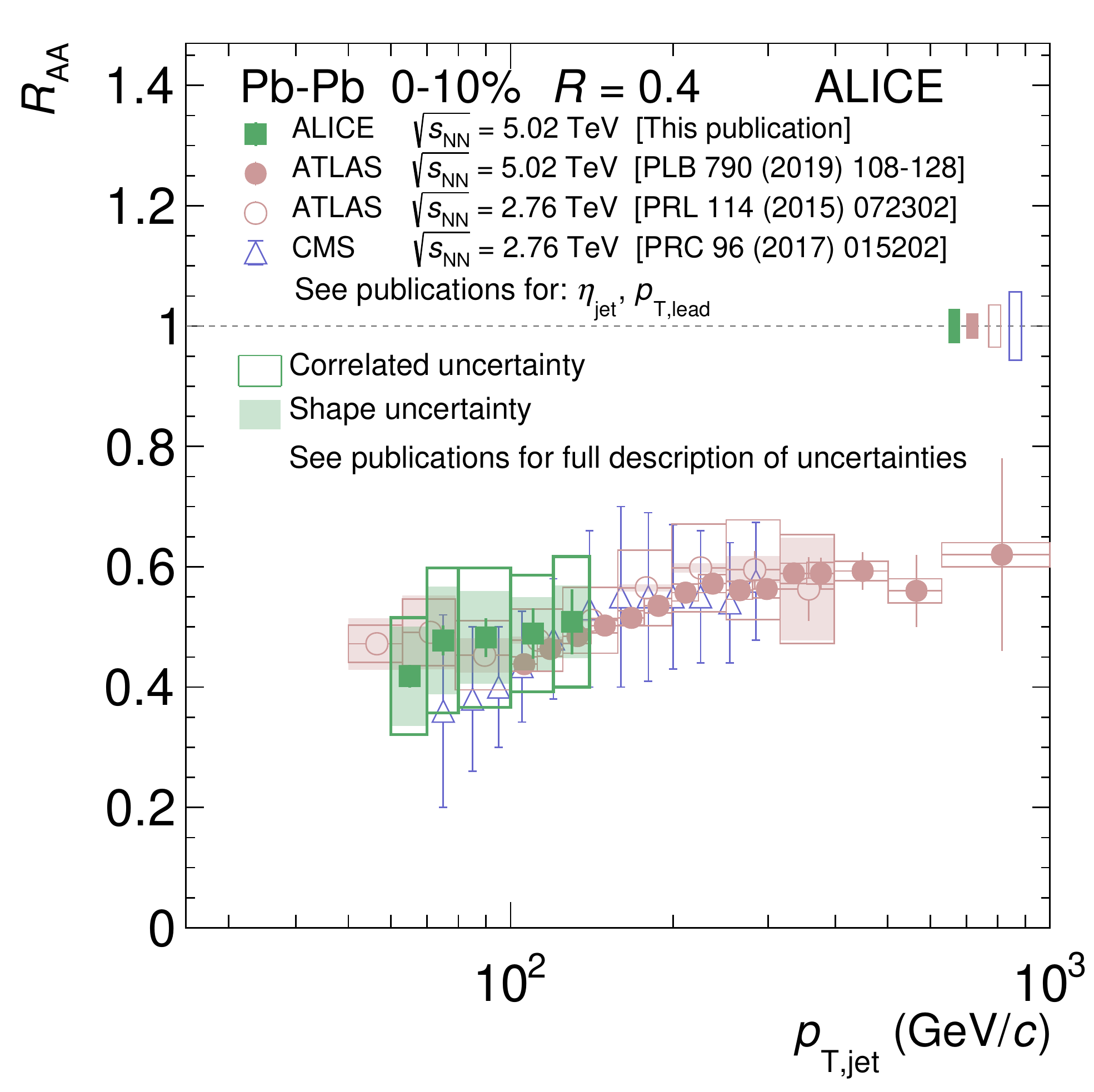}
    \caption{\small ALICE data \cite{Acharya:2019jyg} compared to ATLAS \cite{Aad:2014bxa,Aaboud:2018twu} and CMS \cite{Khachatryan:2016jfl}. Figure from \cite{Acharya:2019jyg}.}\label{fig:ALICE-RAA}
  \end{subfigure}
  \caption{\small Measurements of the nuclear modification factor for jets. The left plot shows the dependence of the observable with the centrality of collisions. The right plot compares several jet $R_{AA}$ measurements at the LHC from different detectors and collision energy $\sqrt{s_{NN}}$.}
\label{Fig:RAA-ATLAS-CMS} 
\end{figure}

In Fig.~\ref{Fig:RAA-ATLAS-CMS}-left, we show the experimental results obtained by the ATLAS collaboration \cite{Aaboud:2018twu}. The $R_{AA}$ ratio is plotted as a function of $p_T$ for several centrality classes. For the most central collisions, one observes a strong suppression of the jet yield over all the $p_T$ range, i.e even for very large $p_T$. This suppression is understood as a consequence of jet energy loss: jets produced with a final $p_T$ come from hard partons with a larger transverse momentum in AA collisions than in pp collisions. As the hard scattering cross-section is steeply falling with $p_T$, producing such hard partons is less likely, leading to a suppression of the jet yield in AA with respect to pp. 

This simple explanation hides two subtleties, which are easily explained within our theory for in-medium jet fragmentation:
\begin{itemize}
 \item contrary to hadrons, jets are reconstructed with a parameter $R$, the jet radius. Thus, the energy flow must be deviated at large angles, larger than this jet radius, to effectively yields a jet suppression. The multiple branching regime, which occurs in the second step of the jet evolution inside the medium is a natural mechanism for such a large angle energy loss. \textit{Event by event}, the leading parton loses an energy of order $\ombr=\alpha^2\qhat L^2$ deviated and thermalized outside the jet cone.
 
 \item this is not the end of the story, as a constant energy loss independent of $p_T$ would not account for the flatness of $R_{AA}$ over such a large range of $p_T$. This is where the \textit{in-medium} vacuum-like evolution comes into play: this evolution increases the number of sources that subsequently lose a typical energy $\obr$. The higher $p_T$ is, the higher the number of these sources is. 
\end{itemize}

Combining together these two mechanisms explains the behaviour of the ATLAS data, as we shall see with more quantitative arguments in the next chapter. We expect our approach to give reasonable results for the jet $R_{AA}$ because the inclusive jet cross-section is an IRC safe observable, which is qualitatively well described by leading order pQCD calculations in $e^+e^-$ or $pp$ collisions. Moreover, one expects that for the most central collisions, all the distinct geometries of collision can be averaged and thus absorbed into our single parameter $L$. Therefore, it is a natural observable to compute in a first place. 

\subsection{Dijet asymmetry and $\gamma$-jet correlations}
\label{sub:dijet-asymm}

Among global jet quenching observables, another important class corresponds to the correlations between the global properties of a jet and its recoil partner in two-jets or $\gamma$-jet events. Typical quantities related to this class are the dijet asymmetry distribution $\sigma_{A_J}$ and the $x_{\gamma J}$ distribution $\sigma_{x_{\gamma J}}$ defined respectively by:
\begin{align}\label{sigma_AJ}
 \sigma_{A_J}&\underset{\textrm{2-jets ev.}}\equiv\frac{1}{\Njets}\frac{\dif N}{\dif A_J}\,,\qquad A_J=\frac{p_{T1}-p_{T2}}{p_{T2}+p_{T1}}\\
  \sigma_{x_{\gamma J}}&\underset{\gamma\textrm{-jets ev.}}\equiv\frac{1}{\Njets}\frac{\dif N}{\dif x_{\gamma J}}\,,\qquad x_{\gamma J}=p_{T,\rm jet}/p_{T,\gamma}\label{sigma_xJ}
\end{align}
where $p_{T1}>p_{T2}$ and $p_{T2}$ are the transverse momenta of the two jets in a dijet event, and $p_{T,\rm jet}$ and $p_{T,\gamma}$ are respectively the transverse momenta of the jet and photon in a $\gamma$-jet event.

\begin{figure}[t] 
  \centering
  \begin{subfigure}[t]{0.70\textwidth}
    \includegraphics[width=\textwidth]{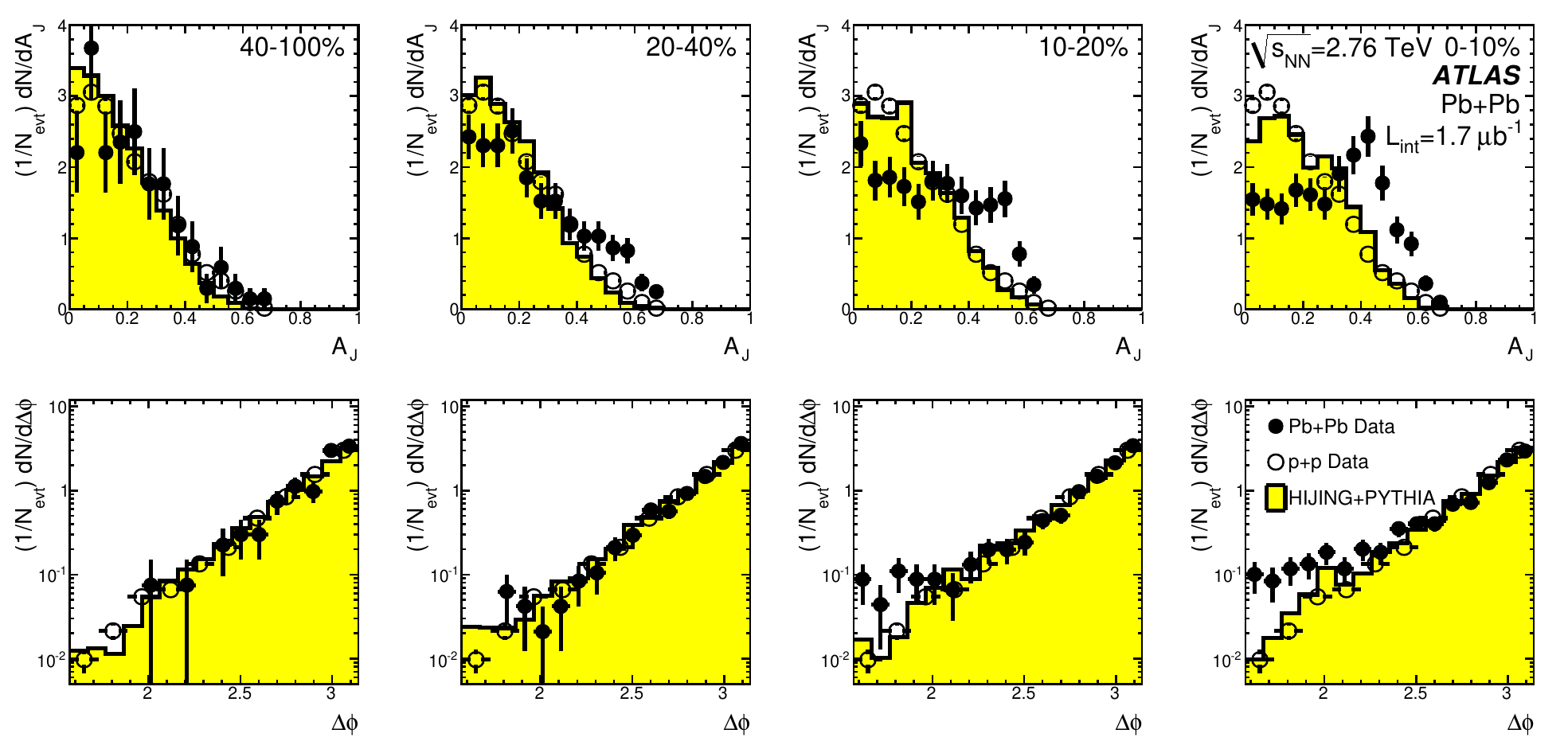}
    \caption{\small ATLAS - dijet asymmetry. Figure from \cite{Aad:2010bu}}\label{fig:ATLAS-dijet}
  \end{subfigure}
      \begin{subfigure}[t]{0.7\textwidth}
    \includegraphics[width=\textwidth]{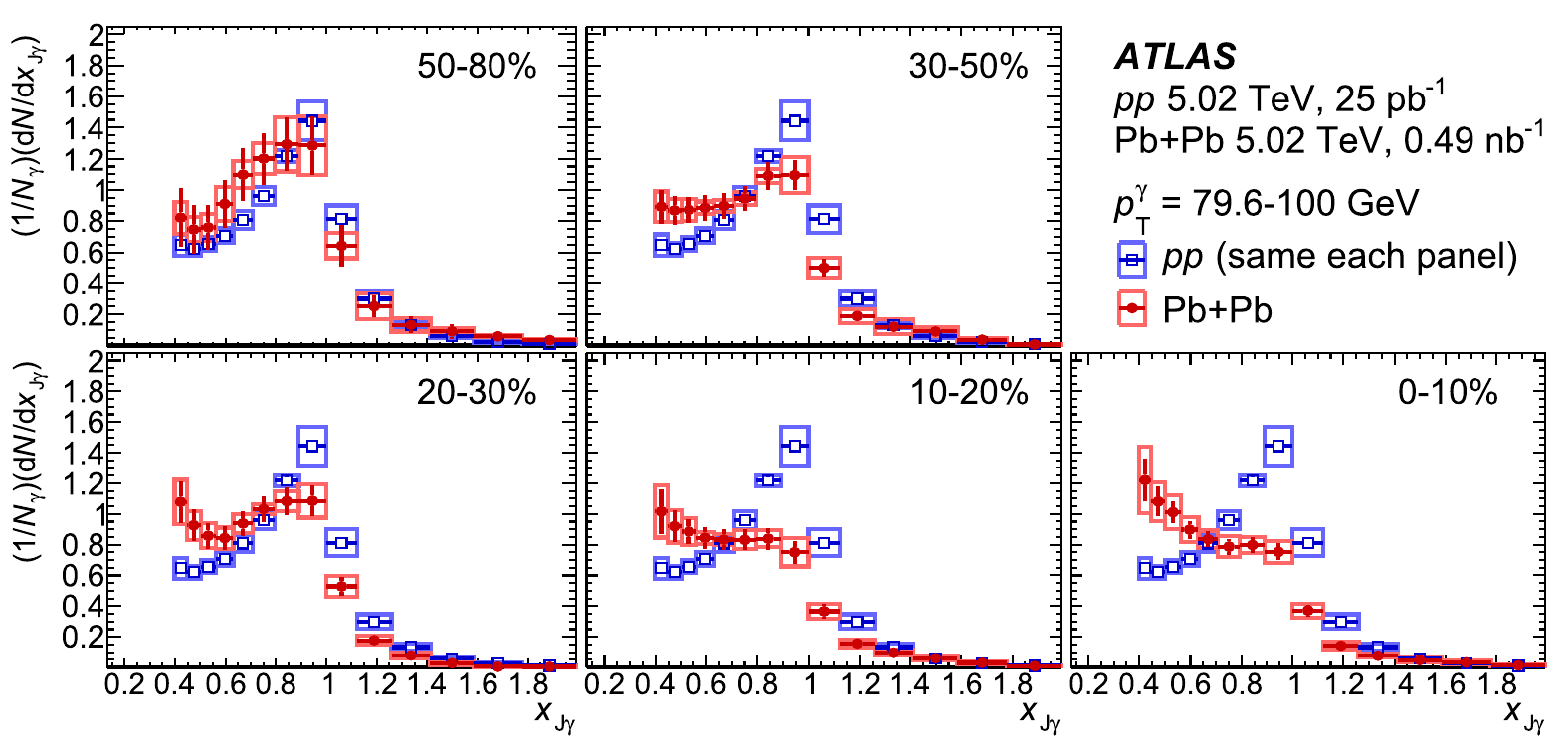}
    \caption{\small ATLAS - $\gamma$-jet correlation. Figure from \cite{Aaboud:2018anc}}\label{fig:ATLAS-gamma}
  \end{subfigure}
  \caption{\small Dijet and $\gamma$-jet correlations measured by the ATLAS experiment. The upper panel of Fig.~\ref{fig:ATLAS-dijet} corresponds to the definition of $\sigma_{A_J}$ given in \eqref{sigma_AJ}. The results are provided for four centrality selections in $pp$ and PbPb. The lower panel of Fig.~\ref{fig:ATLAS-dijet} is a measurement of azimuthal correlations ($\Delta\phi$ is defined as the azimuthal separation between the two jets). The bottom figure shows the distribution defined in \eqref{sigma_xJ} for five centrality classes in $pp$ and PbPb. In most central collisions ($0-10$\% centrality), all these distributions are strongly modified.}
\label{Fig:dijet-ATLAS} 
\end{figure}

By energy conservation, we expect these distributions in $pp$ collisions to be peaked around $0$ for $\sigma_{A_J}$ and $1$ for $\sigma_{x_{\gamma J}}$. The results from the ATLAS experiment \cite{Aad:2010bu,Aaboud:2017eww,Aaboud:2018anc} are shown in Fig.~\ref{Fig:dijet-ATLAS}. Both distributions are modified, and the expected peak is no longer visible. One observes a strong suppression of symmetric dijet and $\gamma$-jet events in PbPb collisions. The CMS collaboration has made a similar measurement for the photon-jet correlations \cite{Chatrchyan:2012gt,Sirunyan:2017qhf} and the dijet asymmetry \cite{Khachatryan:2015lha} which shows the same trend.

For dijet events in AA collisions, there are two possible mechanisms (not necessarily incompatible with each other) explaining the observed dijet asymmetry.
\begin{itemize}
 \item The geometry of the collision may lead to the following situation: one of the two jet has a shorter path length through the medium and thus loses less energy, whereas the other has a much longer path length and loses a substantial amount of its initial energy.
\item Even if the geometry is perfectly symmetrical, the event by event fluctuations in the jet substructure and its energy loss induce also a dijet asymmetry. 
 \end{itemize}
 Making predictions for the dijet asymmetry taking into account both possible mechanisms requires a good modelling of the geometry of the nucleus-nucleus collision. This is currently not included in {\tt JetMed}. Note however that event by event fluctuations in jet substructure and energy loss are included in our picture, so we expect to see a dijet asymmetry in our calculation. As other models predict a minor effect of the geometry fluctuation \cite{Milhano:2015mng,Brewer:2018mpk}, this might suggest that dijet asymmetry would be also qualitatively well described by a {\tt JetMed} calculation. 

 For $\gamma$-jet events in AA, the underlying physics is even more simple to understand as the highly energetic photon does not interact with the plasma so the quantity $x_{\gamma J}$ can be thought as the fraction of energy lost by the leading parton at large angles (either via vacuum-like or medium-induced emissions). Unfortunately, the $x_{\gamma J}$ distribution is very sensitive to initial state radiations. That is why this observable lies (so far) beyond the regime of validity of our calculation. 
 
\section{Substructure jet observables}

In this section, we discuss jet measurements related to their substructure properties. Again, it is enlightening to divide the actual set of data into two classes: the IRC unsafe substructure observable and the IRC safe substructure observables. While IRC unsafe observable are interesting to have an insight on the main qualitative differences between the fragmentation pattern of jets in AA with respect to $pp$, their very definition does not allow for controlled comparisons between pQCD and data, or to say it differently, any prediction becomes highly model-dependent. On the contrary, IRC safe observable are often more delicate to interpret physically (we shall spend a lot of time in Chapter \ref{chapter:jet-sub} doing it) but are much more robust for quantitative comparisons with pQCD predictions.

\subsection{IRC unsafe}

\paragraph{Definitions.} Given a reconstructed jet, one can always define the following two-dimensional distribution of its hadron content:
\begin{equation}
 f_h(x,\Delta R)=\frac{1}{\Njets}\frac{\dif^2 N_h}{\dif x \dif \Delta R}
\end{equation}
where $\dif N_h$ is the number of hadrons inside the jet with ``energy fraction'' between $x$ and $x+\dif x$ and ``angle'' with respect to the jet axis between $\Delta R$ and $\Delta R+\dif \Delta R$:
\begin{equation}
 x\equiv \frac{p_T\cos(\Delta R)}{p_{T,\rm jet}}\,,\qquad \Delta R\equiv\sqrt{(\Delta\eta)^2+(\Delta\phi)^2}
\end{equation}
This two-dimensional distribution carries all the informations about the longitudinal and transverse structure of the hadron content within jets. Marginalising over the angle $\Delta R$, one gets the so-called jet fragmentation function, while marginalising over $x$ gives the jet shape:
\begin{equation}
 \mathcal{D}(x)=\frac{1}{\Njets}\frac{\dif N_h}{\dif x},\,\qquad P(\Delta R)=\frac{1}{\Njets}\frac{\dif N_h}{\dif \Delta R}
\end{equation}
Such substructure observables are not IRC safe. They are built from final state \textit{hadrons} which are not under control in pQCD. In Chapter \ref{chapter:jet}, Section \ref{sub:frag}, we have seen that a transverse momentum cut-off is required in order to regulate the collinear divergence in the calculation of the fragmentation function. The need for such a cut-off is a consequence of non IRC safety.

\begin{figure}[t] 
  \centering
      \includegraphics[width=0.5\textwidth]{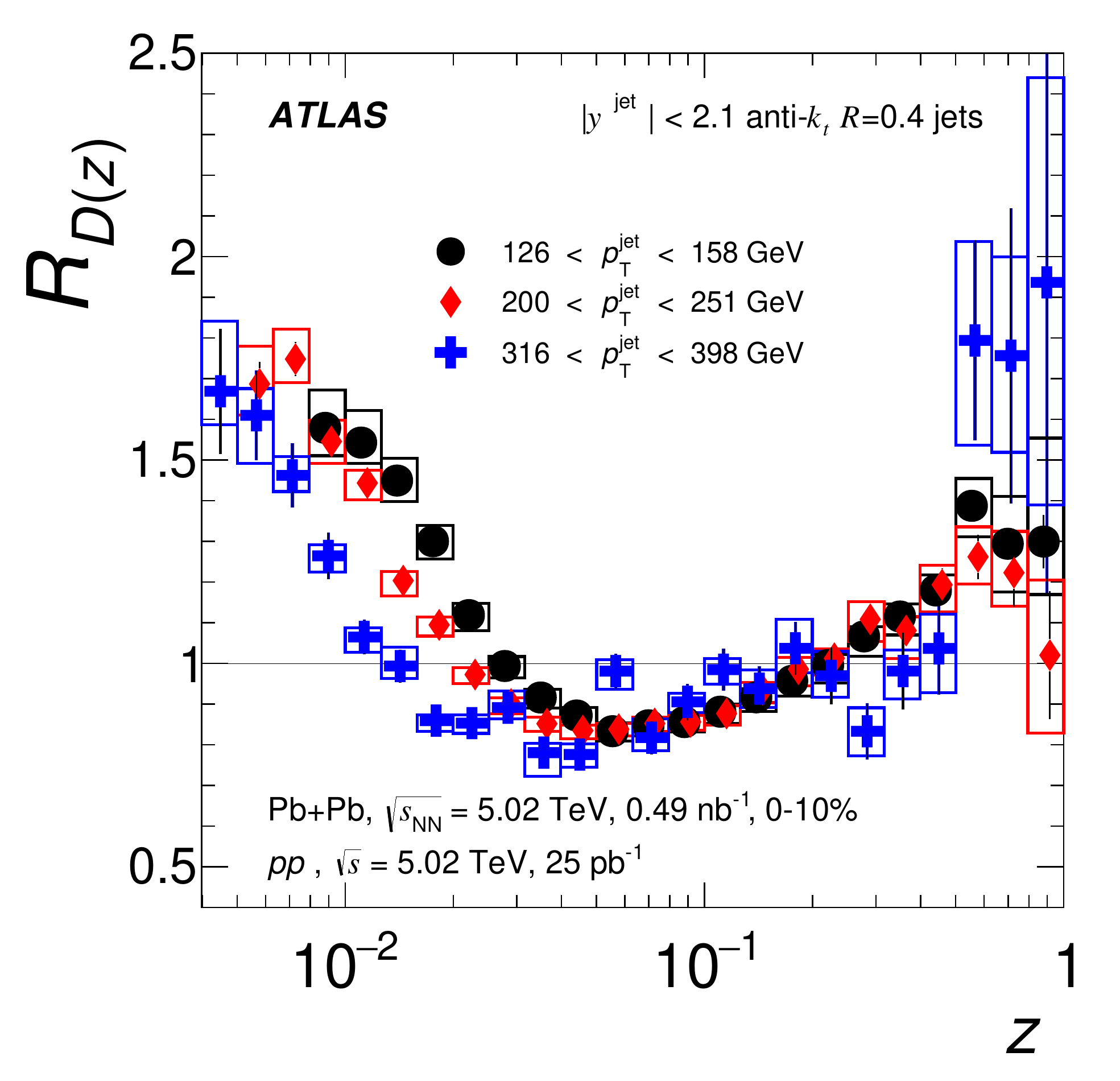}

    \caption{\small ATLAS - fragmentation function. Note that the variable $z$ is called $x$ in this thesis and especially in Chapter \ref{chapter:FF}. Three jet $p_T$ ranges are shown, suggesting an approximate scaling of the large $x$ part of the fragmentation function with the jet transverse momentum. Figure from \cite{Aaboud:2018hpb}.}
    \label{Fig:ATLAS-FF}
\end{figure}
\begin{figure}[t]
\centering
    \includegraphics[width=0.75\textwidth]{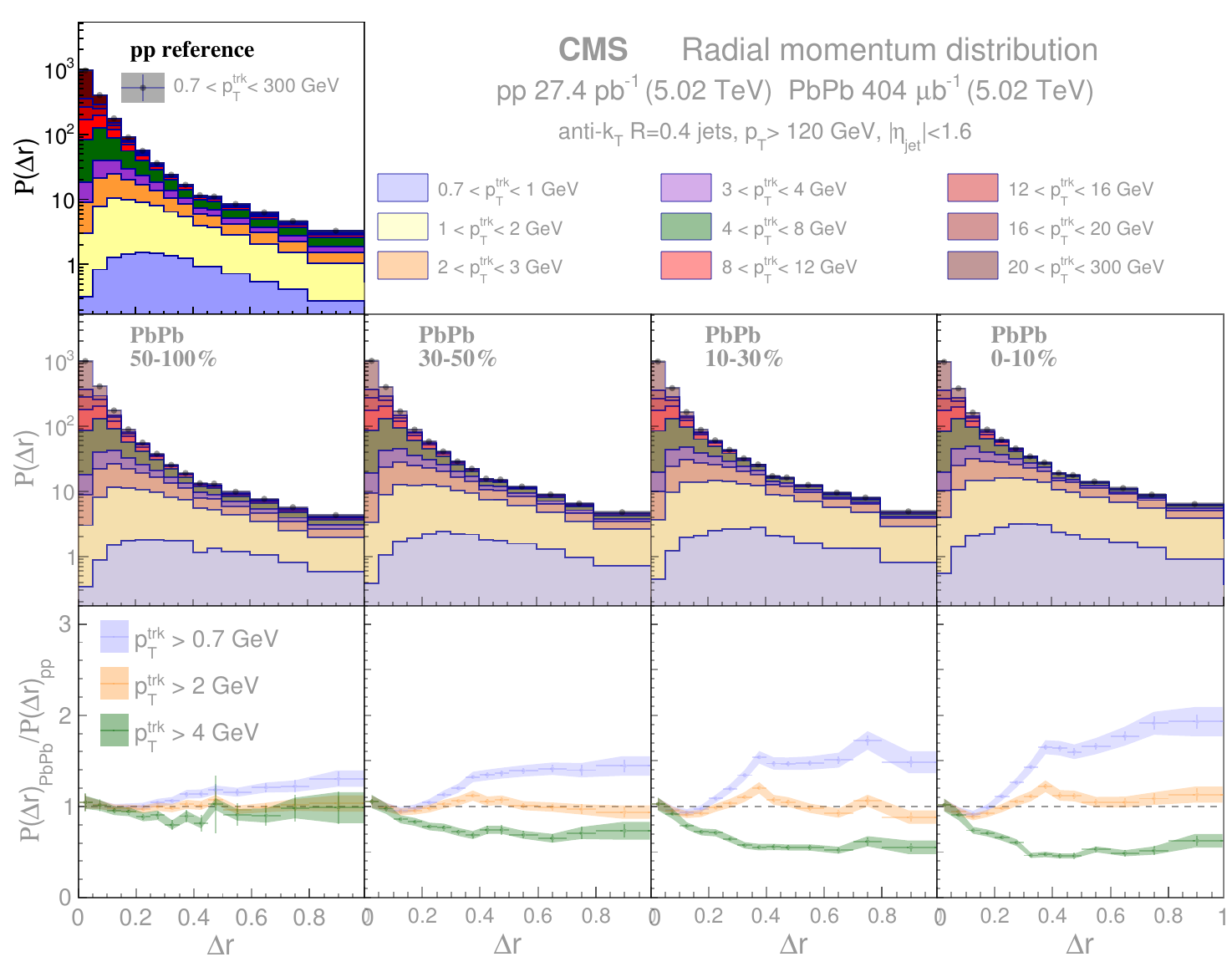}
    \caption{\small CMS - jet radial profile. The upper panel is the $pp$ reference. The middle panels show the radial distribution of jet constituents binned in terms of the transverse momenta of these constituents. The lower panels present the nuclear modification factor for the jet angular shape. Each distribution is given for three centrality classes. Figure from \cite{Sirunyan:2018jqr}.}
    \label{Fig:CMS-jet-shape}
\end{figure}

\paragraph{The experimental data.} Before discussing what we can say about these observables with our Monte Carlo and its limitations, let us present some experimental data available. 
For the fragmentation function, the nuclear modification factor measured by ATLAS \cite{Aaboud:2018hpb} and shown Fig.~\ref{Fig:ATLAS-FF} has an interesting pattern. There is a significant excess of soft and hard hadrons within jets produced in Pb-Pb collisions with respect to jets produced in $pp$ collisions. A similar pattern is observed (at least for the enhancement of soft constituents) in $\gamma$-jet events \cite{Aaboud:2019oac,Sirunyan:2018qec} and at lower $\sqrt{s}$ \cite{Chatrchyan:2014ava,Aaboud:2017bzv}. Regarding the jet shape, the CMS data \cite{Chatrchyan:2013kwa,Sirunyan:2018ncy,Sirunyan:2018jqr}, as those shown in \ref{Fig:CMS-jet-shape}, show an excess of intrajet \textit{very soft} particles at large angles in PbPb for the most central collisions. This excess is stronger around the edges of the jets. Note that the ALICE collaboration also measured jet radial profiles in qualitative agreement with the CMS results \cite{Acharya:2018uvf,Acharya:2019ssy}.

The main obstacle for quantitative comparisons between our MC calculations and these experimental data is the lack of hadronisation model in {\tt JetMed}. Hadronisation has a large impact on the jet shape \cite{Cal:2019hjc}, especially for large $\Delta R$ where the medium effects are known to be important from the CMS data. This large $\Delta R$ enhancement comes mainly from hadrons with $0.5\lesssim p_T\lesssim 2$ GeV, a kinematic region which is dangerously close to the hadronisation scale. Note that such effects also limit our predictive power for the large $R$ dependence of the $R_{AA}$ ratio.

The fragmentation function is also sensitive to hadronisation effects even if the local parton-hadron duality hypothesis has been proven to give reasonably good descriptions of multiplicity distributions within jets produced in $e^+e^-$ annihilation. Furthermore, one sees already substantial nuclear effects for hadron constituents with $p_T\gtrsim 2$ GeV well beyond the $\LQCD$ scale. That is why we believe that our calculation for the partonic fragmentation function in PbPb collisions captures at least qualitatively the trend of the hadronic fragmentation function. This observable is explored in Chapter \ref{chapter:FF}.  Nevertheless, this study would not be complete without a detailed investigation of hadronisation corrections on the nuclear modification factor.

\paragraph{Physics of in-medium fragmentation.} Regarding the physics of the fragmentation function, we argue in Chapter \ref{chapter:FF} that the enhancement seen at large $x$ in the fragmentation function is essentially a filter effect caused by the medium. Indeed, looking at the large $x$ part of this observable means looking at jets with a hard constituent inside. Such jets have a reduced evolution (otherwise, the leading particle would not carry a large $x\sim 1$ value), and thus lose less energy. Then, there is this generic property of the energy loss effect combined with the steeply falling cross-section: jets losing less energy are favoured. Consequently, hard fragmenting jets are preferably produced in AA relatively to other classes, simply because they lose less energy. This mainly explains the enhancement seen at large $x$.

The low $x$ part is less trivial. As this enhancement is already present for tracks with $p_T\sim 3$-$4$ GeV $\gg \LQCD$ at the LHC, we expect a pQCD mechanism to explain this pattern. In Chapter \ref{chapter:DLApic}, we have seen that the violation of angular ordering leading to a reopening of the angular phase space for radiations is a powerful mechanism to produce soft intrajet particles. Indeed, we confirm in Chapter \ref{chapter:FF} that this is a crucial ingredient for the nuclear enhancement of the fragmentation function at low $x$. On top of that, relatively hard medium-induced emissions with energies between $\obr$ and $\oc$ also play an important role. Whereas medium induced emissions with $\om\lesssim \obr$ account for the event by event jet energy loss (at large angles), hard medium induced emissions are rare events, but they remain inside the jet cone and contribute to the fragmentation function at low $x$.

\subsection{IRC safe}

\begin{figure}[t] 
  \centering
  \begin{subfigure}[t]{0.40\textwidth}
    \includegraphics[width=\textwidth]{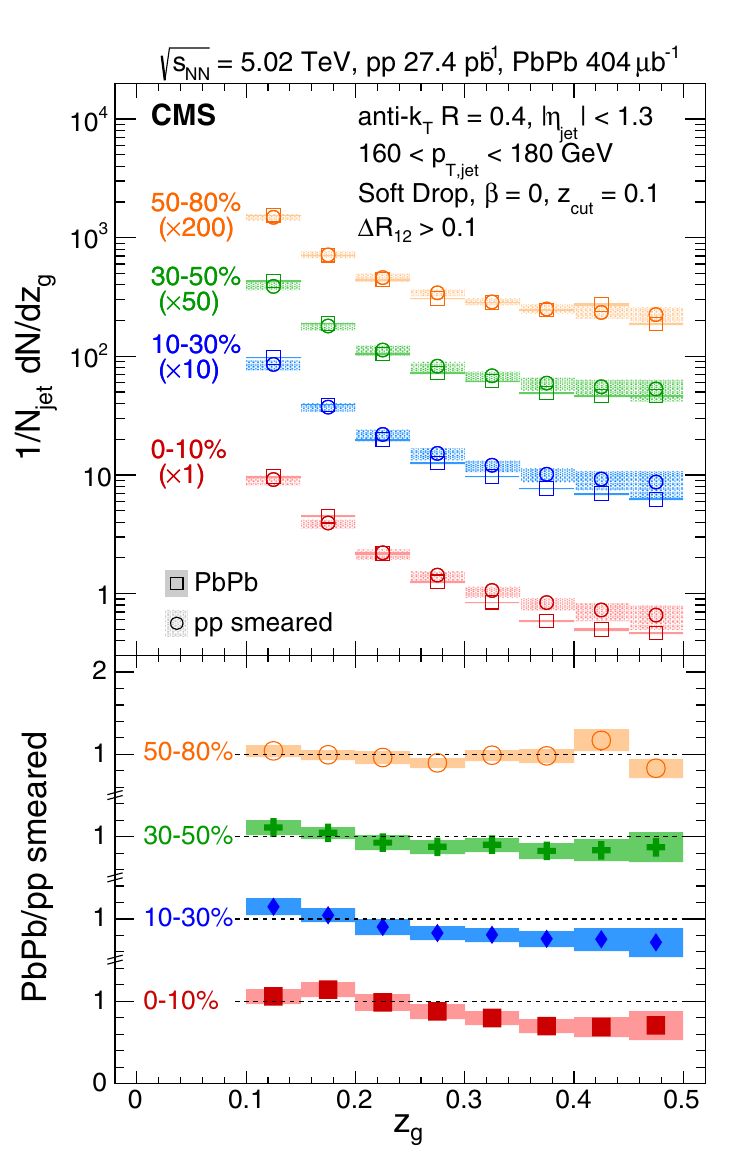}
    \caption{\small CMS - self normalised $z_g$ distribution. Figure from \cite{Sirunyan:2017bsd}.}\label{fig:CMS-zg}
  \end{subfigure}
  \hfill
      \begin{subfigure}[t]{0.46\textwidth}
    \includegraphics[width=\textwidth]{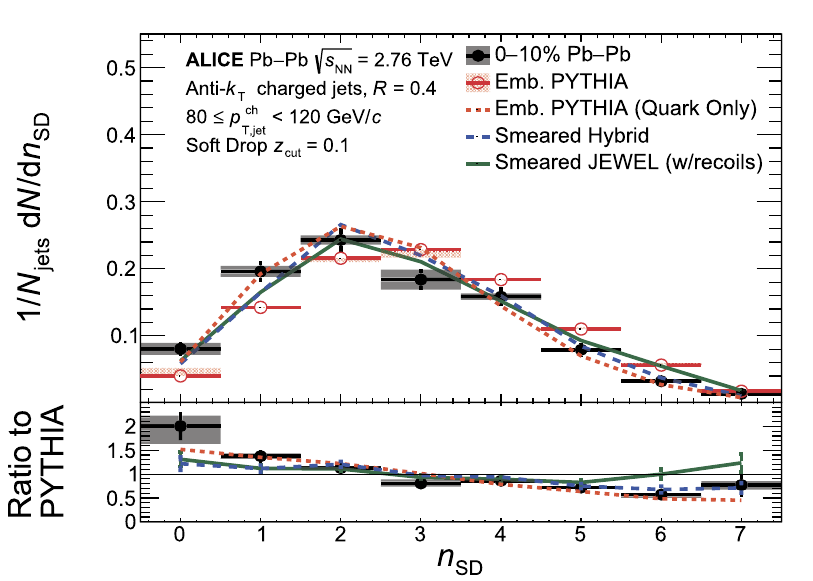}
    \caption{\small ALICE - $\nSD$ distribution. Figure from \cite{Acharya:2019djg}.}\label{fig:ALICE-nsd}
  \end{subfigure}
  \caption{\small Substructure jet observables measured in heavy-ion collisions at the LHC. The figure on the left shows the self normalised $z_g$ distributions defined in section \eqref{subsub:SoftDrop} measured by the CMS collaboration for four centrality classes, and their nuclear modification factors PbPb/($pp$ \textit{smeared}). The smearing of $pp$ data takes into account detector effects. The figure on the right is a measurement of the Iterative Soft Drop multiplicity distribution (see section \ref{subsub:nSD}) in PbPb central collisions by the ALICE collaboration and its ratio to Pythia embedded in heavy-ion collisions. The measurement is compared to theoretical predictions from the Hybrid model and {\tt Jewel}.}
\label{Fig:zg-nsd} 
\end{figure}

We turn now to IRC safe substructure observables. A common feature of these observables is to rely on subjets (in a given declustering procedure) and not on hadrons. There are a lot of possibilities for defining such observables. Among them, an interesting class relies on the primary Lund plane density \cite{Dreyer:2018nbf}. The Lund plane density is defined as the number density of primary subjet emissions obtained after a C/A declustering of the jets:
\begin{equation}
 \rho(\th,k_\perp)=\frac{1}{\Njets}\frac{\dif^2N_{\sub}}{\dif \log(1/\th) \dif\log( k_\perp)}
\end{equation}
with $\th$ the angle of the subjet with respect to the hard branch and $k_\perp$ its transverse momentum. Choosing a kinematic cut on this Lund plane which avoids the non-perturbative region enables to build IRC safe observables. We refer the reader to Chapter \ref{chapter:jet}, Section \ref{sub:obs-subjets} for the definitions of the $z_g$ and $\nSD$ distributions and the subjet fragmentation function. One can easily check that such distributions are built from the Lund plane density with an IRC safe kinematic cut.

\paragraph{A glimpse at the experimental data.} There is a lot of experimental activity around the measurement of substructure observables from grooming techniques \cite{Sirunyan:2018gct,Adam:2020kug}. There is a measurement by the CMS collaboration of the $z_g$ distribution in PbPb collisions \cite{Sirunyan:2017bsd}. Unfortunately, this measurement is not unfolded and fully corrected from detector effects. This prevents a quantitative comparison between theory and data. Nevertheless, the results are shown in Fig.~\ref{Fig:zg-nsd}. The ratio is decreasing with $z_g$ for most central collisions. Note that in this CMS measurement, $z_g$ distributions are normalised to unity, so that if the ratio is larger than $1$ in some $z_g$ range, it has to be smaller than $1$ in another range. Consequently, one can not conclude that there is an enhancement of soft splitting in the medium, or a reduction of symmetric splittings since it could just be a normalisation artefact. 

\begin{figure}[t]
\centering
    \includegraphics[width=0.95\textwidth]{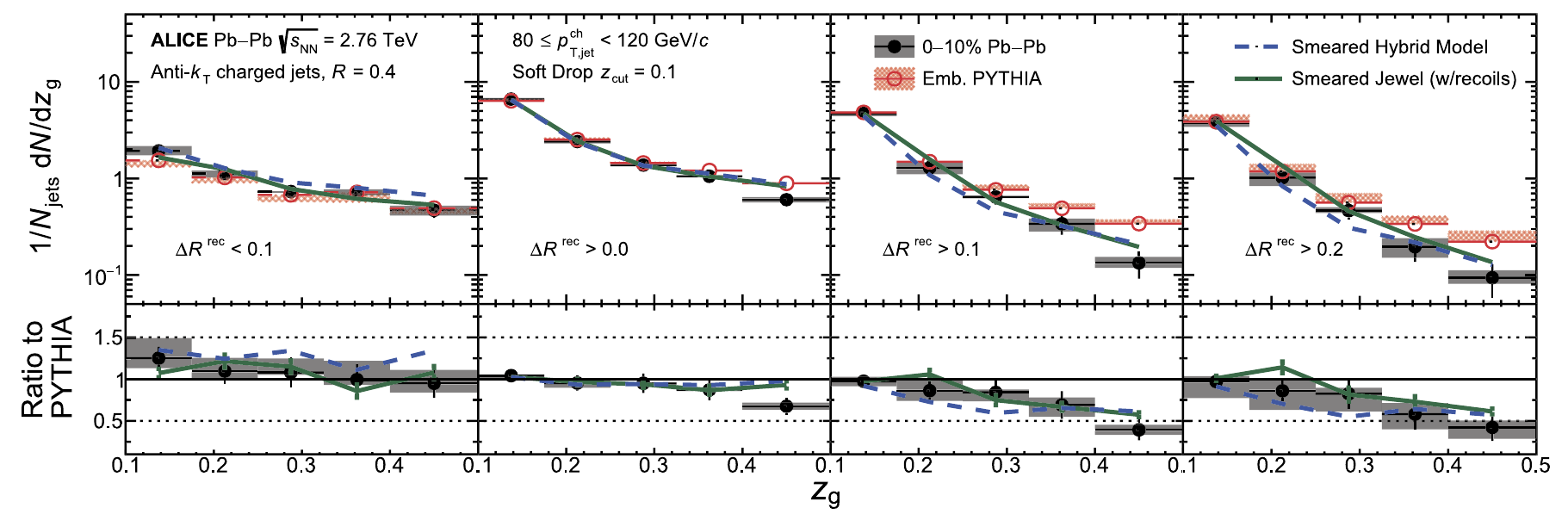}
    \caption{\small ALICE - \njets-normalised $z_g$ distribution in PbPb central collisions. As in Fig.~\ref{fig:ALICE-nsd}, the ratio shown in the bottom panel is taken with Pythia predictions embedded in heavy-ions collisions. The difference between the four panels lies in the angular cut (called $\thetacut$ in Section \ref{subsub:Soft Drop}) imposed for the Soft Drop procedure. Figure from \cite{Acharya:2019djg}.}
    \label{Fig:ALICE-zg}
\end{figure}

The ALICE measurement for this observable shown Fig.~\ref{Fig:ALICE-zg} uses the \njets-normalisation prescription. Also, additional cuts on the splitting angles are applied. For $\Delta R_{12}=\th_g\ge 0.1$ or $0.2$, the decreasing behaviour is still visible but the overall normalisation leads to a reduction in the number of wide angles SD splitting in PbPb with respect to pp. Again these measurements must be interpreted with care as they are not unfolded from detector effects. More recent measurements of the $z_g$ distribution by the ALICE collaboration without cut on $\Delta R_{12}$ and fully corrected from detector effects show a very mild medium modification, even if the uncertainties are quite large. 

ALICE also measured the nuclear modification factor of the $\nSD$ distribution \cite{Acharya:2019djg}. The result is shown \ref{fig:ALICE-nsd}. They measure a relative increase in the number of jets with few SD splittings in PbPb collision and a relative reduction of jets with a large number of SD splittings.

\paragraph{Brief physics discussion.} Thanks to their IRC safety, these observables are less sensitive to hadronisation corrections. Moreover, the geometry of the collision is not the driven mechanism that would explain a nuclear modification. The underlying event is groomed away in the Soft Drop or Iterative Soft Drop procedure. Consequently, this improves the reliability of our calculations. The physics a play in these Lund plane based observables consists in the following four ingredients:
\begin{itemize}
 \item jets with a hard substructure are more suppressed because they have a larger in-medium evolution, and hence lose more energy.
 \item As for the fragmentation function, the small angle (hard) medium-induced emissions contribute to the Lund plane density at small $z$ and $k_\perp\sim Q_s$. As the single BDMPS-Z emission spectrum is more singular at small $z$ than the standard Bremsstrahlung spectrum, this often produces a peak at small $z$ in the PbPb/pp ratio of integrated distributions.
 \item On the contrary, the veto region induces a suppression of the density of vacuum-like splittings in this region of the Lund plane.
 \item Finally, for in-medium splittings, subsequent energy loss changes the value of the kinematic associated with that spitting. Thus the difference between the physical kinematic of this emission and the \textit{measured} one must be taken into account when calculating the probability of emission.
\end{itemize}
Such effects will be discussed in details in Chapters \ref{chapter:jet-sub} and \ref{chapter:FF}.

\subsection{Correlating global and substructure jet measurements}

Whereas we were dealing so far with global and substructure jet observables rather independently, we discuss here the correlation between these two kinds of measurement. 
The idea that the jet energy loss is strongly correlated with jet substructure can be tested experimentally by measuring the nuclear modification factor for different classes of jets according to their substructure. A proposal for such a measurement is presented in Chapter \ref{chapter:jet-sub}, Section \ref{sub:correl-raa-zg}. On the experimental side, a measurement at the LHC correlating $R_{AA}$ and substructure has been performed by the ATLAS collaboration \cite{ATLAS:2019rmd}, but is still preliminary. 

Beyond the correlation of $R_{AA}$ and substructure, one can correlate dijet or $\gamma$-jet asymmetry with substructure, in the spirit of the correlation between jet shape and leading versus sub-leading jet in dijet events as done in \cite{Khachatryan:2016tfj}. Such measurements could disentangle the two mechanisms discussed in Section \ref{sub:dijet-asymm} for the dijet asymmetry and understand the relative importance of the path length dependence mechanism versus the jet substructure/energy loss event by event fluctuation.


%% file: chapter6.tex
\chapter{Jet energy loss and the jet nuclear modification factor $R_{AA}$}
\chaptermark{The jet nuclear modification factor $R_{AA}$}
\label{chapter:RAA}

In this chapter we present our Monte Carlo results for the jet nuclear modification
factor $R_{AA}$. We first discuss the case of a monochromatic
leading parton, for which we compute the jet average energy loss defined in Chapter \ref{chapter:DLApic}, Section \ref{sub:jet-eloss-th}, and then turn to $R_{AA}$ itself, using a Born-level jet spectrum for the
hard process producing the leading parton.

\section{Choice of parameters}

The implementation of in-medium  partonic cascades described in Chapter~\ref{chapter:MC} has
5 free parameters: two unphysical ones, $\theta_\text{max}$ and
$k_{\perp,\text{min}}$, essentially regulating the soft and collinear
divergences, and three physical parameters, $\hat{q}$, $L$ and
$\amed$, describing the medium.
In our phenomenological studies, we will make sure that our results
are not affected by variations of the unphysical parameters and we
will study their sensitivity to variations of the medium parameters.

The different sets of parameters we have used are listed in
Table~\ref{tab:parameters}. The first line is our default setup. It
has been chosen to give a reasonable description of the $R_{AA}$ ratio
measured by the ATLAS collaboration (see Sect.~\ref{sec:RAA} below).
The next 3 sets are variants which give a similarly good description of
$R_{AA}$ and can thus be used to test if other observables bring an
additional sensitivity to the medium parameters compared to $R_{AA}$.
The last 6 lines are variations that will be used to probe which
physical scales, amongst $\theta_c$, $\omega_c$ and
$\omega_\text{br}$, influence a given observable.

\begin{table}[t]
  \centering
  \begin{tabular}{|l|ccc|ccc|}
    \hline
    & \multicolumn{3}{|c|}{parameters} 
    & \multicolumn{3}{|c|}{physics constants} \\
    \cline{2-7}
    Description
    & $\hat{q}$ [GeV$^2$/fm] & $L$ [fm] & $\amed$
    & $\theta_c$ & $\omega_c$ [GeV] & $\omega_\text{br}$ [GeV] \\
    \hline
    default
    & 1.5   & 4     & 0.24  & 0.0408 &  60 & 3.456 \\
    \hline
    & 1.5   & 3     & 0.35  & 0.0629 & 33.75 & 4.134 \\
    similar $R_{AA}$
    & 2     & 3     & 0.29  & 0.0544 & 45    & 3.784 \\
    & 2     & 4     & 0.2   & 0.0354 & 80    & 3.200 \\
    \hline
    \multirow{2}{*}{vary $\theta_c$}
    & 0.667 & 6     & 0.24  & 0.0333 & 60    & 3.456 \\
    & 3.375 & 2.667 & 0.24  & 0.05   & 60    & 3.456 \\
    \hline
    \multirow{2}{*}{vary $\omega_c$}
    & 0.444 & 6     & 0.294 & 0.0408 & 40    & 3.456 \\
    & 5.063 & 2.667 & 0.196 & 0.0408 & 90    & 3.456 \\
    \hline
    \multirow{2}{*}{vary $\omega_\text{br}$}
    & 1.5   & 4     & 0.196 & 0.0408 & 60    & 2.304 \\
    & 1.5   & 4     & 0.294 & 0.0408 & 60    & 5.184 \\
    \hline
  \end{tabular}
  \caption{\small Table of medium parameters used in this paper. The default
    set of parameters is given in the first line. The next 3 lines
    are parameters which give a similar prediction for
    $R_\text{AA}$. The last 6 lines are up and down variations of
    $\theta_c^2$, $\omega_c$ and $\omega_\text{br}$ by 50\%, keeping
    the other two physics constants fixed.}\label{tab:parameters}
\end{table}

\section{The jet average energy loss}
\label{sec:MC-jet-eloss}

To study the jet energy loss we start with a single hard parton of
transverse momentum $p_{T0}$ and shower it with the Monte Carlo
including either MIEs only, or both VLEs and MIEs.
The jet energy loss is defined as the difference between the energy of
the initial parton and the energy of the reconstructed jet. To avoid
artificial effects related to emissions with an angle $\theta$ between the jet
radius $R$ and the maximal opening angle $\theta_\text{max}$ of the
Monte Carlo, we have set $\theta_\text{max}=R$. Furthermore, for the
case where both VLEs and MIEs are included, we have subtracted the
pure-vacuum contribution (which comes from clustering and other edge
effects and is anyway small for $\theta_\text{max}=R$).

\begin{figure}[t] 
  \centering
  \includegraphics[width=0.48\textwidth,page=1]{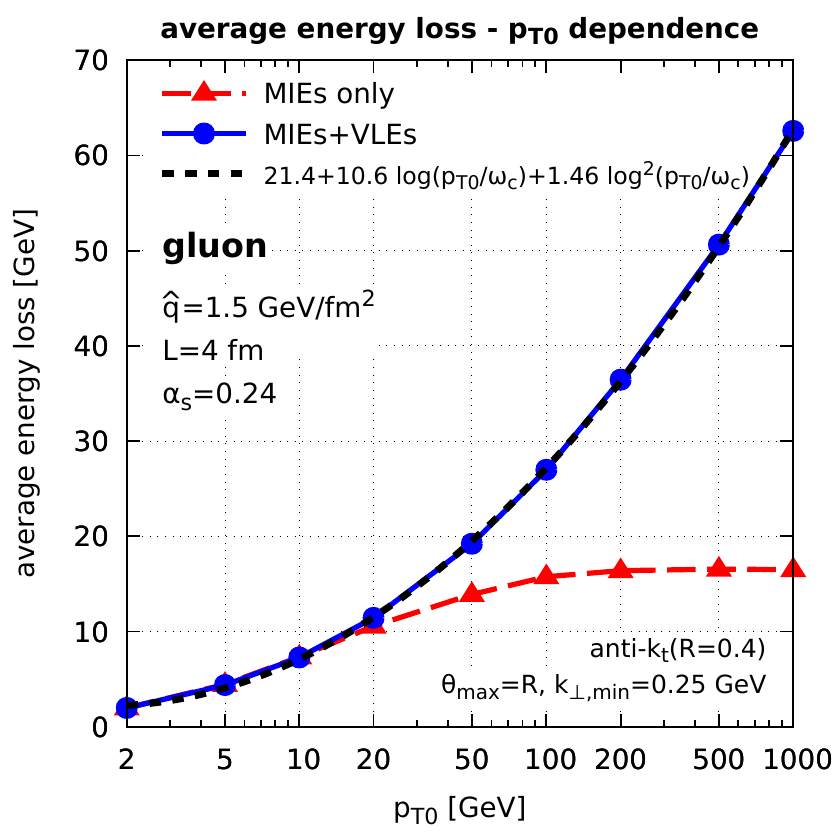}\hfill%
  \includegraphics[width=0.48\textwidth,page=1]{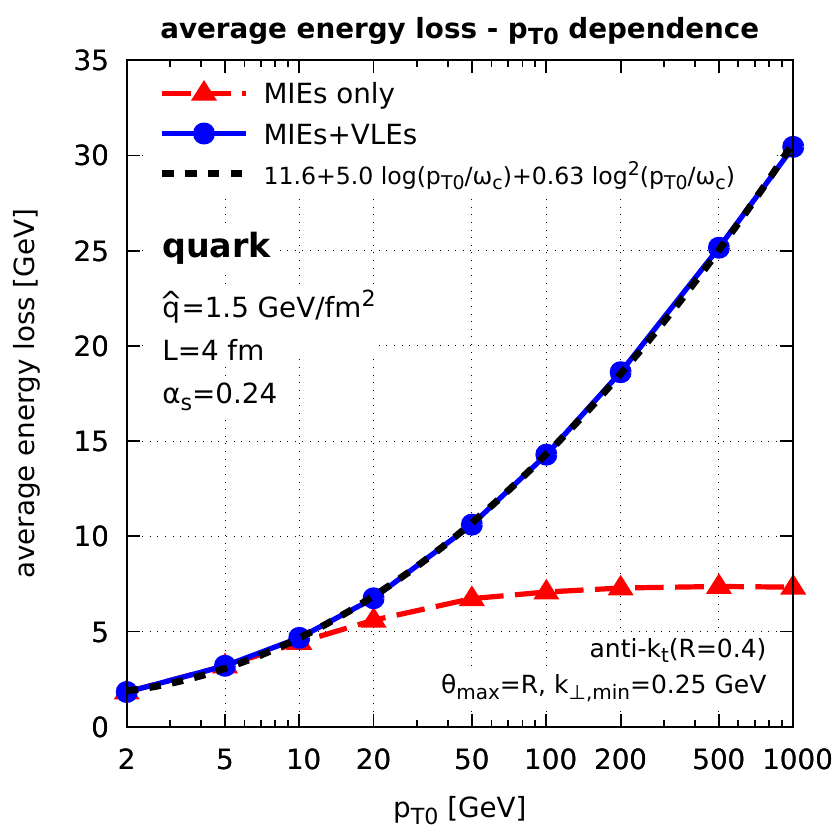}
 \caption{\small The MC results for the average energy loss by a gluon-initiate jet (left), respectively,
 a quark-initiated one (right), is displayed as a function of the initial energy $p_{T0}$ of the leading parton,
 for two scenarios for the jet evolution: jets with MIEs only (triangles) and full showers with both MIEs and VLEs (circles). The dashed line shows the quadratic fit to the energy loss by the full parton shower.}
 \label{Fig:jeteloss-v-pt}
\end{figure}

\begin{figure}[t] 
  \centering
  \includegraphics[width=0.48\textwidth,page=2]{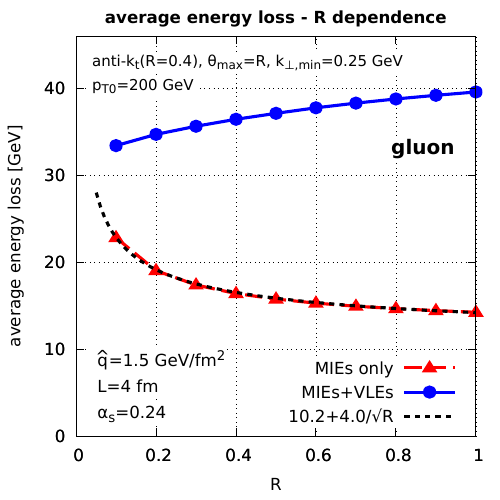}\hfill%
  \includegraphics[width=0.48\textwidth,page=2]{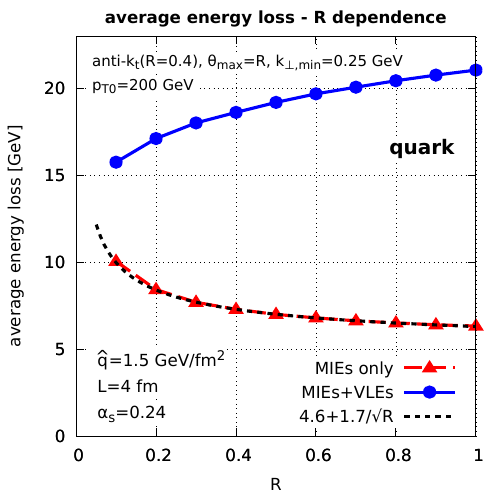}
 \caption{\small The MC results for the average energy loss by the jet are displayed
 as a function of the jet angular opening $R$, for the same choices as in Fig.~\ref{Fig:jeteloss-v-pt}.  We also show in dashed line a fit (inspired by the theoretical estimate in \eqref{Espec}) for the jet built with MIEs alone.}
 \label{Fig:jeteloss-v-R}
\end{figure}

Our results for the energy loss are shown in
Fig~\ref{Fig:jeteloss-v-pt} as a function of $E\equiv p_{T0}$ and in
Fig.~\ref{Fig:jeteloss-v-R} as a function of $R$, for both gluon- and
quark-initiated jets. For these plots we have used the default values
for the medium parameters (see the first line of
Table~\ref{tab:parameters}).
Overall, we see a good qualitative agreement with the features expected
from the theoretical discussion in Sections~\ref{sub:angular-structure}
and~\ref{sub:jet-eloss-th}.

For the jets involving MIEs only, we see that the energy loss first
increases with $p_{T0}$ and then saturates, as predicted by
Eqs.~\eqref{energy-flow}--\eqref{Espec}. Also, as a function of $R$ for
fixed $p_{T0}=200$~GeV, it decreases according to the expected
$1/\sqrt{R}$ behaviour, cf.\ \eqref{Espec}.
The agreement is even quantitatively decent. Indeed, with the physical
parameters given in Table~\ref{tab:parameters}, the fitted
$R$-dependence for gluon-initiated jets,
$\epsilon_{\text{MIE}}(R)\simeq \epsilon_0+\epsilon_1/\sqrt{R}$,
with $ \epsilon_0=10.2$~GeV and $ \epsilon_1=4.0$~GeV, corresponds to
the prediction in~\eqref{ElossHigh} provided one chooses
$ \upsilon\simeq 2.95$ and $c_*\simeq 0.38$, which are both
reasonable.

Consider now the full parton showers, with both VLEs and
MIEs. Although we do not have accurate-enough analytic results to
compare with (only the DLA estimate~\eqref{logeloss}), the curves
``MI+VLEs'' in Figs.~\ref{Fig:jeteloss-v-pt}
and~\ref{Fig:jeteloss-v-R} show the expected trend: the energy loss
increases with both $p_{T0}$ and $R$, due to the rise in the
phase-space for VLEs. For later use, we have fitted the dependence on
$p_{T0}$ with a quadratic polynomial in $\ln(p_{T0}/\omega_c)$ and the
resulting coefficients are shown on Fig.~\ref{Fig:jeteloss-v-pt}.

It is also striking from Figs.~\ref{Fig:jeteloss-v-pt}
and~\ref{Fig:jeteloss-v-R}, that the average energy loss obeys a
surprisingly good scaling with the Casimir colour factor of the leading
parton: the energy loss by the quark jet is to a good approximation
equal to $C_F/C_A=4/9$ times the energy loss by the gluon jet. Such a
scaling, natural in the case of a single-gluon emission, is very
non-trivial in the presence of multiple branchings.
Let us first recall the explanation of this scaling for the case of
MIEs alone. We have seen that the branching scale that appears in \eqref{energy-flow} depends on the Casimir 
of the leading parton through the relation \eqref{ombr-R}:
\begin{equation}\label{gluonR}
\obr^{(R)}\equiv \frac{\alpha_s^2}{\pi^2} \,C_A C_R\,  \frac{\hat q L^2}{2}\,,
\end{equation}
where $\hat q$ is the gluonic jet quenching parameter,
proportional to $C_A$.

From~\eqref{gluonR} it is easy to show the energy loss~\eqref{ElossHigh}
of a jet initiated by a parton in an arbitrary colour representation
$R$ scales linearly with $C_R$: the first term in~\eqref{ElossHigh} is
proportional to $C_R$ as it is proportional to $\obr^{(R)}$ and the
second term is also proportional to $C_R$ in the general case. All the other factors only refer to
gluons and are independent of $R$.
This justifies the Casimir scaling visible in
Figs.~\ref{Fig:jeteloss-v-pt} and \ref{Fig:jeteloss-v-R} for the
cascades with MIEs only.
For the full cascades including VLEs, the linear dependence on $C_R$
can be argued based on~\eqref{logeloss}, assuming $p_{T0}\gg \obr$. The
first term in the r.h.s.\ of~\eqref{logeloss} is the energy lost by the
leading parton and is by itself proportional to $C_R$, as just
argued. The second term in \eqref{logeloss}, which refers to the
additional ``sources'' created via VLEs, one can assume that most of
these ``sources'' are gluons, so they all lose energy (via MIEs) in
the same way; the only reference to the colour Casimir of the leading
parton is thus in the overall number of sources, which is indeed
proportional to $\alpha_s C_R/\pi$ as it is clear from Eq.~\eqref{eloss-asymptot}.

\section{The nuclear modification factor $R_{AA}$}
\label{sec:RAA}

We now consider the physically more interesting jet nuclear
modification factor $R_{AA}$, which is directly measured in the
experiments.  In order to compute this quantity, we have considered a
sample of Born-level $2\to 2$ partonic hard scatterings. We
  have used the same hard-scattering spectrum for both the pp baseline
  and the PbPb sample. This means that we neglect the effects of
  nuclear PDF, which can sometimes be as large as 15-20\% and can be
  added in a more phenomenologically-oriented study. For each event,
both final partons are showered using our Monte Carlo. Jets are
reconstructed using the anti-$k_\perp$
algorithm~\cite{Cacciari:2008gp} as implemented in {\tt FastJet}
v3.3.2~\cite{Cacciari:2011ma}.
All the cuts are applied following the ATLAS measurement from
Ref.~\cite{Aaboud:2018twu}.

\begin{figure}[t] 
  \centering
  \includegraphics[width=0.48\textwidth]{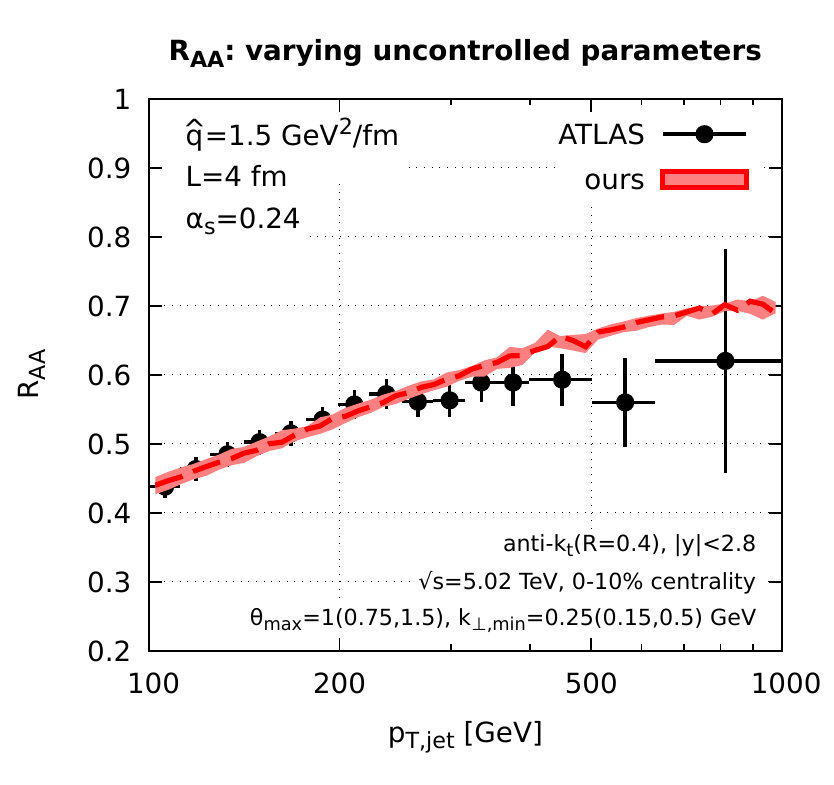}\hfill
  \includegraphics[width=0.48\textwidth]{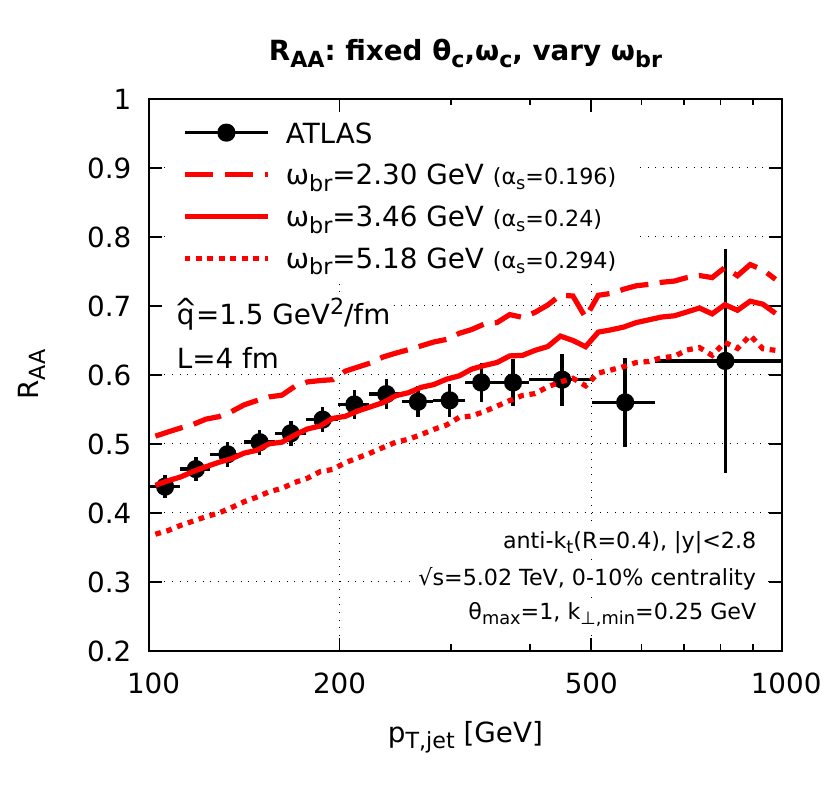}
  \caption{\small Our MC predictions for $R_\text{AA}$ are compared with the results
  of an experimental analysis by ATLAS \cite{Aaboud:2018twu} (shown as dots with error bars).
  Left: the sensitivity of our results to changes in the kinematic cuts $\theta_{\rm max}$ and $k_{\perp,\text{min}}$. Right: the effect of varying $\obr$ (by $\pm 50\%$) at fixed values for $\oc$ and $\theta_c$.}
  \label{Fig:RAA}
\end{figure}

\begin{figure}[t] 
  \centering
  \includegraphics[width=0.48\textwidth]{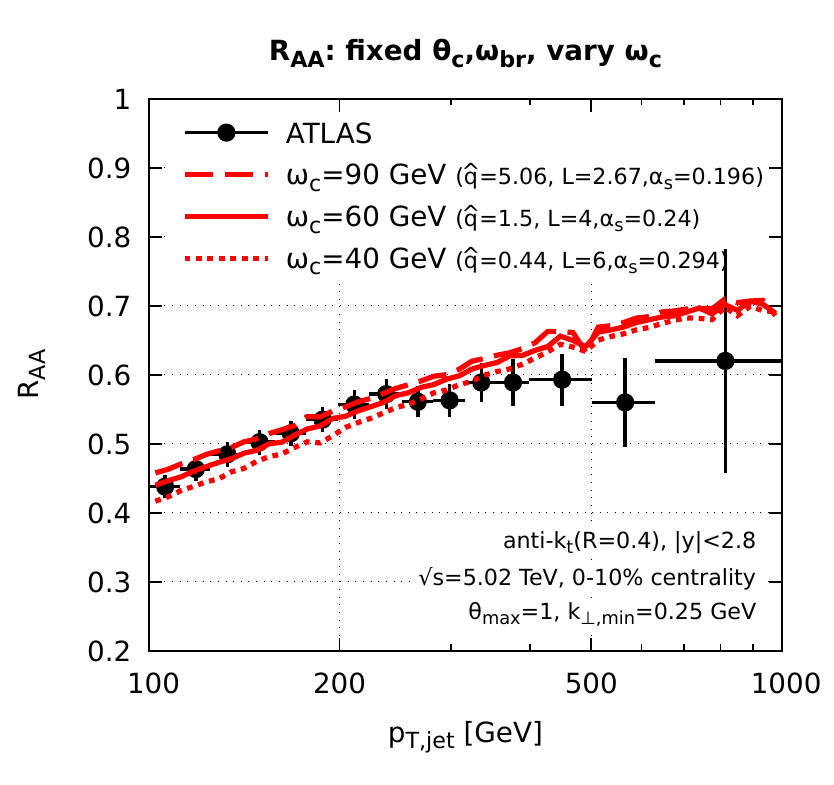}\hfill
  \includegraphics[width=0.48\textwidth]{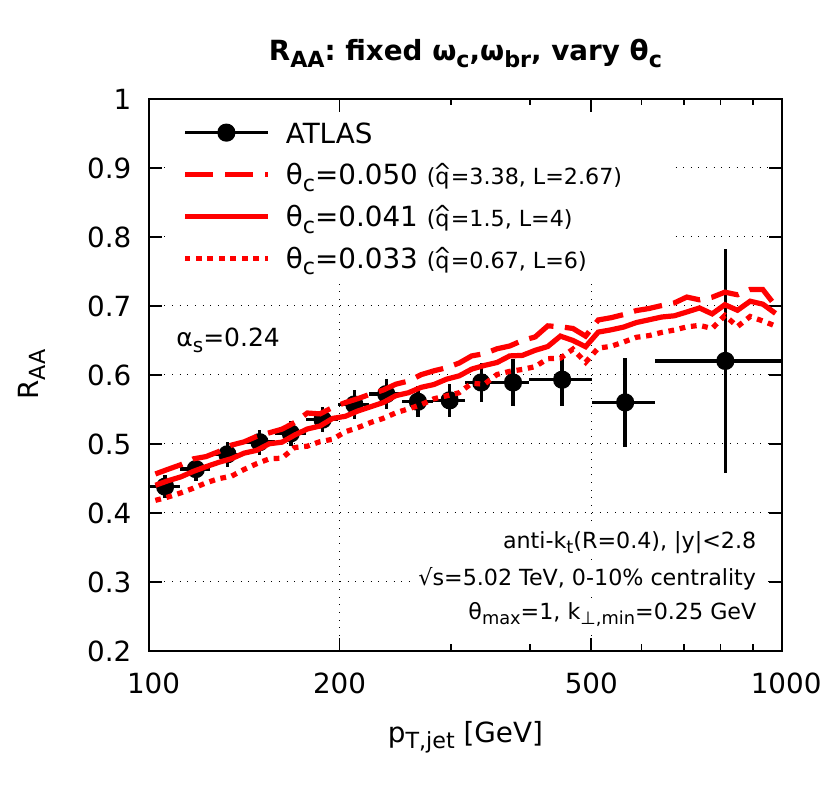}
  \caption{\small  The effects of varying the medium parameters $\hat q$, $L$ and $\amed $ in such a way to keep a constant value for $\ombr$ (the same as the central value in  the right plot in Fig.~\ref{Fig:RAA}, i.e. $\obr=3.46$~GeV). Left: we vary $\om_c$ by $\pm 50\%$ at fixed $\theta_c$. Right: we vary $\theta_c^2$ by $\pm 50\%$ at fixed $\omc$.}
  \label{Fig:variab}
\end{figure}

Figs.~\ref{Fig:RAA}--\ref{Fig:variab} show our predictions together
with the LHC (ATLAS) data \cite{Aaboud:2018twu} as a function of the transverse momentum
$p_{T,{\rm jet}}$ of the jet (that we shall simply denote as $p_T$ in this section).
As discussed in Sect.~\ref{sub:JetMed-params}, our calculation involves 5 free
parameters: the 3 ``physical'' parameters $\hat q$, $L$ and $\amed$
which characterise the medium properties and 2 ``unphysical''
parameters $\theta_{\rm max}$ and $k_{\perp,\text{min}}$ which
specify the boundaries of the phase-space for the perturbative parton
shower. Our aim is to study the dependence of our results under
changes of these parameters.

The first observation, visible in Fig.~\ref{Fig:RAA}~(left) is that the
$R_{AA}$ ratio appears to be very little sensitive to variations of
the ``unphysical'' parameters $\theta_{\rm max}$ and $k_{\perp,\text{min}}$.
Although our results for the inclusive jet spectrum do depend on these
parameters,\footnote{In particular on $\theta_\text{max}$ as the parton
  shower would generate collinear logarithms of $\theta_\text{max}/R$
  to all orders.} the impact on $R_\text{AA}$ remains well within the
experimental error bars when changing $\theta_{\rm max}$ by a factor
2 and  $k_{\perp,\text{min}}$ by a factor larger than 3.
 
In Fig.~\ref{Fig:RAA} right and in the two plots of
Fig.~\ref{Fig:variab} we fix the ``unphysical'' parameters and vary
the medium ones.
The variations are done, following Table~\ref{tab:parameters}, so as
to keep two of the three physical scales $\ombr$, $\omc$ and $\theta_c$
fixed while varying the third. It is obvious from the figures that
$R_\text{AA}$ is most sensitive to variations of $\obr$
(Fig.~\ref{Fig:RAA}, right) and shows only a small dependence on $\oc$
and $\theta_c$.
This is in perfect agreement with the expectations from
section~\ref{sub:jet-eloss-th} that the jet energy loss is mostly
driven by the scale $\ombr$.
Furthermore, the small variations of $R_\text{AA}$ with changes in $\oc$
and $\theta_c$ can be attributed to the slight change in the
phase-space for VLEs leading to a corresponding change in the number
of sources for energy loss (see section~\ref{sub:jet-eloss-th} and,
in particular, \eqref{logeloss}).

One remarkable fact about the LHC measurements is the
fact that  $R_\text{AA}$ increases very slowly with the jet
$p_T$. This implies that the jet energy loss $\mathcal{E}_{\rm{jet}}$
must itself increase with $p_T$ to avoid a fast approach of
$R_\text{AA}$ towards unity. In our picture, such an increase is indeed
present, as manifest in Fig.~\ref{Fig:jeteloss-v-pt}, and is associated with the steady 
rise of the phase-space for VLEs leading to an increase in the 
number of sources for MIEs (cf.~\eqref{logeloss}).

\begin{figure}[H] 
  \centering
  \begin{subfigure}[t]{0.48\textwidth}
    \includegraphics[width=\textwidth]{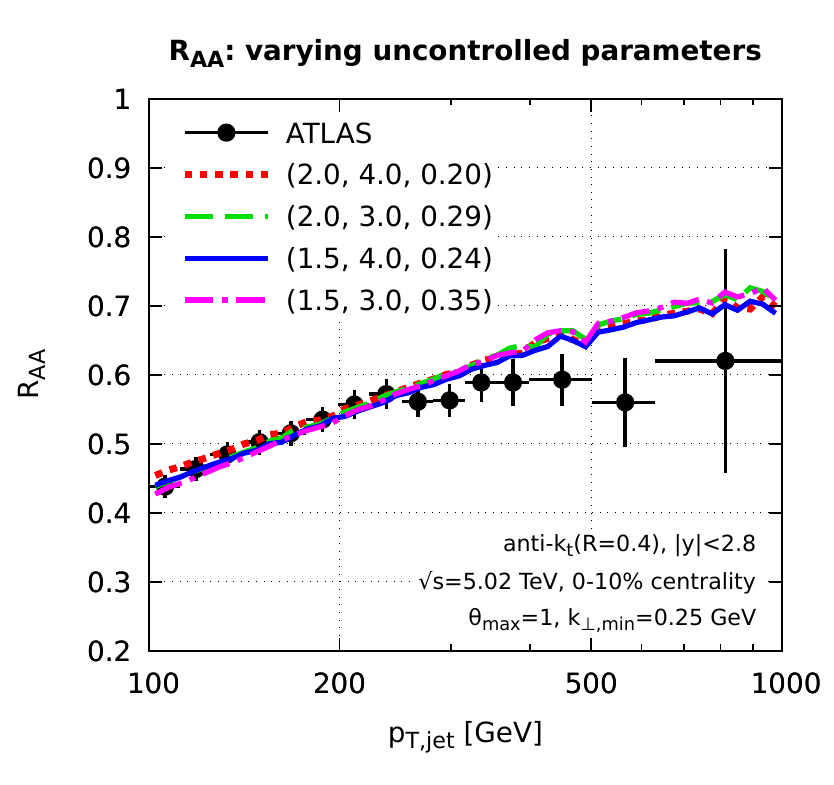}
    \caption{\small Our MC results for jet $R_{AA}$ and for 4 sets of
 medium parameters which give quasi-identical predictions are compared to
 the ATLAS data \cite{Aaboud:2018twu}.}
 \label{Fig:RAA-degeneracy}
 \end{subfigure}\hfill
 \begin{subfigure}[t]{0.48\textwidth}
        \includegraphics[width=\textwidth]{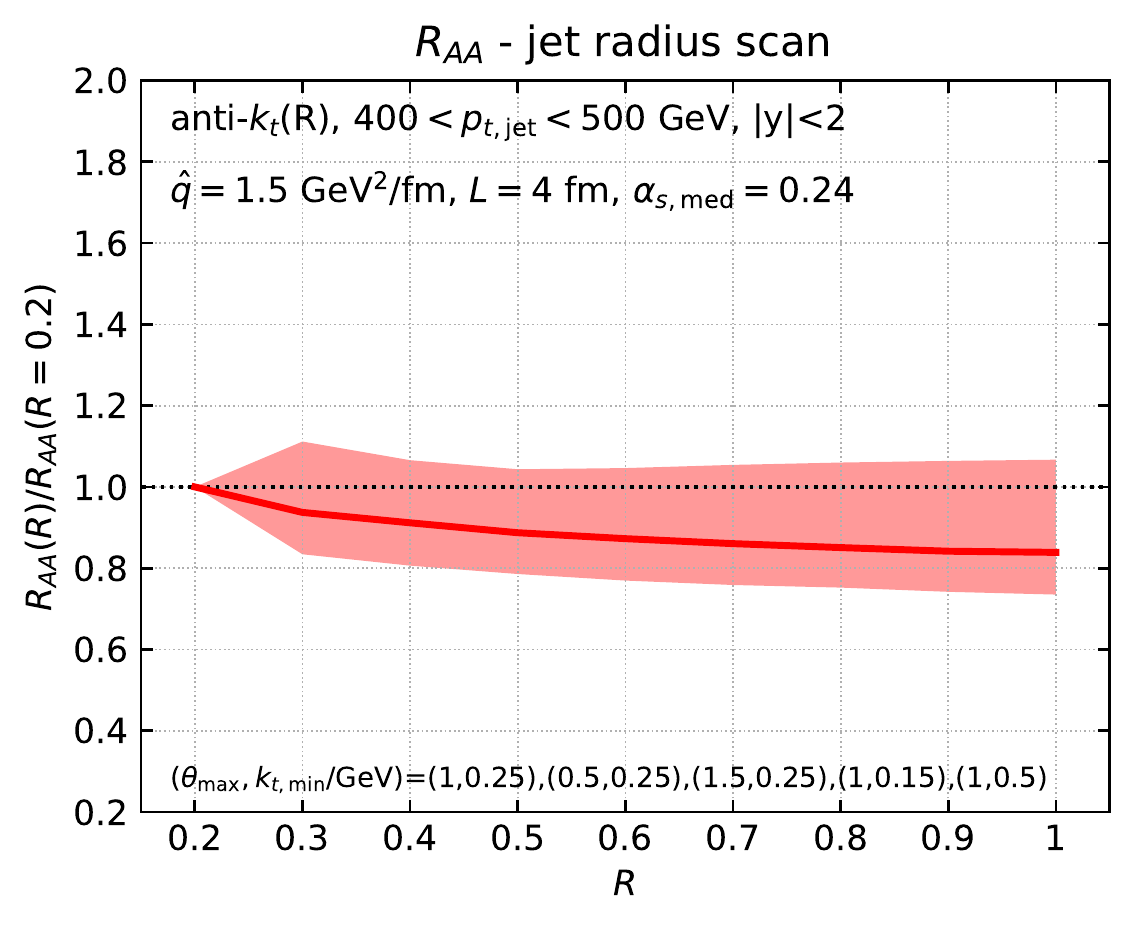}
       
 \caption{\small Dependence of $R_{AA}$ with 
 the jet reconstruction radius $R$ in a ``CMS-like'' set-up.}
 \label{Fig:RAA-Rdep}
 \end{subfigure}
 \caption{\small Phenomenology of the nuclear modification factor for jets with {\tt JetMed}.}
\end{figure}

\subsection{Medium parameters degeneracy}
\label{subsub:param-degeneracy}

In Fig.~\ref{Fig:RAA-degeneracy}, we show a selection of 4 sets of medium parameters,
$\hat q$, $L$ and $\amed$ which provide a good description of the LHC data
\cite{Aaboud:2018twu} for the jet $R_{AA}$ ratio.
We see that our 4
choices of medium parameters correspond to
somewhat different values for the physical medium scales $\obr$, $\oc$
and $\theta_c$. They therefore lead to different
predictions both for the average energy lost {\it by a single parton}
at large angles, dominated by $\obr$, and for the number and
distribution of sources, which is controlled via the phase-space
boundaries for VLEs by $\oc$ and $\theta_c$.
While the $R_{AA}$ ratio is most sensitive to variations in $\obr$,
small variations in $\obr$ ($\sim 30$\% between our extreme values)
can be compensated by larger variations of $\oc$ and $\theta_c$ (a
factor $\sim 2$ between our extreme values).

\subsection{Dependence on the jet reconstruction parameter $R$}
\label{sub:RAA-R}

In the last subsection of this chapter, we would like to comment on the $R$ dependence of the $R_{AA}$ ratio predicted by {\tt JetMed}. The dependence of $R_{AA}$ with the jet reconstruction radius has driven a lot of experimental and theoretical activities over the past few months. Indeed, it has been realised that this measurement has an interesting discriminatory power between jet quenching models. 
At the time we are  writing these lines, the preliminary results of the CMS collaboration show a very mild increase of the jet cross-section ratio $\mathscr{R}(R)$ defined by
\begin{equation}
 \frac{R_{AA}(R)}{R_{AA}(0.2)}=\frac{\sigma^R_{\rm jet, PbPb}}{\sigma^{R=0.2}_{\rm jet PbPb}}
\end{equation}
where $\sigma^R_{\rm jet}$ is the integrated jet cross-section over $p_T=[400,500]$ GeV and pseudo-rapidity $|\eta|<2$ for jets reconstructed with anti-$k_t(R)$.

From the analytic estimations of the $R$ dependence of the average jet energy loss made in Chapter \ref{chapter:DLApic}, Section \ref{sub:jet-eloss-th} and from the corresponding Monte Carlo calculations in Section \ref{sec:MC-jet-eloss}, we expect also to observe a very mild modification of the function $R_{AA}(R)/R_{AA}(0.2)$. Indeed, the average energy loss is only slightly increasing with $R$. As explained in Section \ref{sub:jet-eloss-th}, this is a consequence of the competition between the recovered energy loss and the increase of in-medium vacuum-like sources. This slight increase explains the slight \textit{decrease} of $R_{AA}(R)/R_{AA}(0.2)$ seen in the central curve of Fig.~\ref{Fig:RAA-Rdep}.

That said, the way the large angle energy loss is recovered is strongly model dependent and not really under control in the Monte Carlo. By varying the parameter $\theta_{\rm max}$ of {\tt JetMed} which should be thought as the angle where the turbulent medium-induced flow ends up, one can access to the uncertainty related to the lack of modelling of thermalization and plasma-jet interactions (e.g. medium response.) As shown in Fig.~\ref{Fig:RAA-Rdep} with the red band, this uncertainty is quite large. We have checked that the more $\theta_{\rm max}$ is small, the more the energy loss recovery effect is important, so that the upper boundary of the bands in Fig.~\ref{Fig:RAA-Rdep} corresponds to the minimal value of $\theta_{\rm max}$. Thus, one must be cautious when interpreting these {\tt JetMed} results, as the physics at play in the $R$ dependence of $R_{AA}$ certainly involves a lot of non perturbative modelling which are not currently included in the Monte Carlo.


%% file: chapter7.tex
\chapter{The Soft Drop $z_g$ distribution in heavy-ion collisions}
\label{chapter:jet-sub}

 Nowadays, there is an increasing interest in jet observables dealing with the inner structure of jets \cite{Andrews:2018jcm}. By requiring infrared and collinear safety for these substructure observables, one can hope for controlled calculations in pQCD even in the complex environment of a nucleus-nucleus collision and therefore quantitative comparisons with experiments. In this chapter, we explore jet substructure, and in particular the Soft Drop $z_g$ distribution within our picture of jet fragmentation in a dense QCD medium. The emphasis is put on the main physical ingredients that explain the behaviour of our results. Thus, all the numerical results obtained from the Monte Carlo {\tt JetMed} are supported by analytical phenomenological formulas that capture these ingredients. 
 
 To understand the nuclear modifications of a jet observable, it is often enlightening to adopt the following methodology:
\begin{itemize}
 \item starting with a leading parton with \textit{fixed} initial transverse momentum $p_{T0}$, one first tries to understand the effects of the medium on this observable. In this configuration, one looks for effects coming from the modification of the jet fragmentation pattern itself. This first approach also enables to use simple analytical estimates.
 \item Then, one adds a realistic initial hard scattering spectrum. This causes additional effects, mostly because large angle energy loss changes the statistics in the transverse momentum of the initial partons.
\end{itemize}

Therefore, our discussion is divided into two main sections. Section \ref{sec:zgmed} only investigates the monochromatic hard spectrum case, whereas Section \ref{sec:data} deals with the {\tt JetMed} calculations with a realistic initial spectrum. In the last section, we discuss other Soft Drop related observables, such as the Iterated Soft Drop multiplicity.

\section{The $z_g$ distribution for monochromatic hard spectra}
\label{sec:zgmed}

For the benchmark vacuum results, definitions of the $z_g$ distribution and notations, we refer the reader 
to Chapter \ref{chapter:jet}, section \ref{subsub:Soft Drop}.
  
\subsection{Two regimes: high-$p_T$ and low-$p_T$ jets} 

Let us start with a presentation our Monte Carlo results for the $z_g$-distribution
created by ``monochromatic'' jets which propagates through the
quark-gluon plasma. From these results, we will have to distinguish two regimes --- high $p_T$ and low $p_T$ --- according to the value of the transverse momentum $p_{T0}$ of the leading parton with respect to the medium-induced scale $\oc=\qhat L^2/2$. Detailed analytic calculations are presented in Sections~\ref{sec:high} for high-energy jets and~\ref{sec:lowpt} for
lower-energy jets.

We focus on the \njets-normalised distribution
$f_{i, \text{med}}(z_g)$ (see \ref{subsub:Soft Drop}), which carries more information. We define
the corresponding nuclear modification factor,
$\mathcal{R}_i(z_g)\equiv f_{i, \text{med}}(z_g)/f_{i,
  \text{vac}}(z_g)$. Similarly we define
$\mathcal{R}_i^\text{(norm)}(z_g)\equiv p_{i,
  \text{med}}(z_g)/p_{i,\text{vac}}(z_g)$ as the nuclear modification
factor of the self-normalised $z_g$ distributions.
We study four different values for the initial $p_{T0}$ spanning a
wide range in $p_{T0}$, from $100$~GeV to $1$~TeV.  We use the same SD
parameters as in the CMS analysis~\cite{Sirunyan:2017bsd}, namely
$\beta=0$ and $\zc=0.1$, together with a cut $\theta_g>\thetacut=0.1$.
%

\begin{figure}[t] 
  \centering
  \includegraphics[page=2,width=0.48\textwidth]{./zg-parton-ptdep.pdf}\hfill
  \includegraphics[page=1,width=0.48\textwidth]{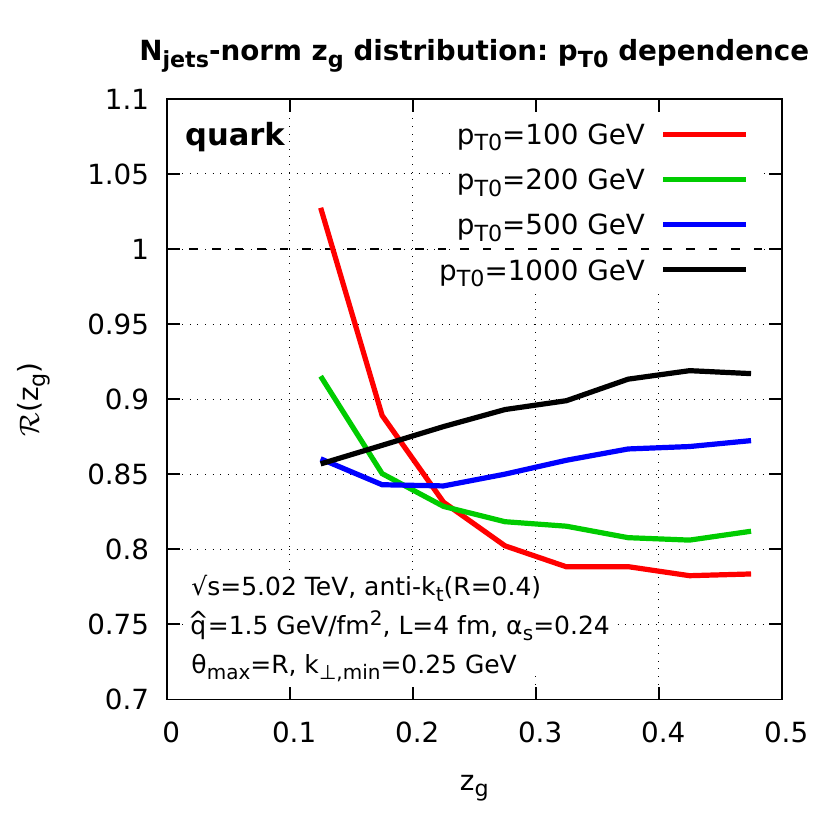}
  \caption{\small A summary of our MC results  for the medium/vacuum ratio $\mathcal{R}_i(z_g)$ of the 
$N_\text{jets}$-normalised distributions, for monochromatic jets initiated either by a gluon (left figure), or by a quark (right figure), and for 4  different values for the initial transverse momentum $p_{T0}$. }
  \label{Fig:zgvar}
\end{figure}

Our results are shown in Fig.~\ref{Fig:zgvar}, separately for jets
initiated by a gluon (left plot) and by a quark (right plot), using
our default MC parameters (cf.\ Table~\ref{tab:parameters}).
As for our study of energy loss for monochromatic jets in
Section~\ref{sec:eloss}, we set the angular cutoff scale of our Monte Carlo to $\theta_\text{max}=R$ with $R=0.4$ the jet radius.
Each of the plots in Fig.~\ref{Fig:zgvar} show qualitatively different
behaviours between our lowest $p_{T0}$ (100~GeV) value and the largest
one (1~TeV).
More precisely, for the highest energy jets, $p_{T0}=1$~TeV, the ratio
$\mathcal{R}(z_g)$ is always smaller than one, indicating a nuclear
suppression, and it increases monotonously with $z_g$, meaning that
the nuclear suppression is larger at small $z_g$.
Conversely, for lower $p_{T0}$, while the nuclear suppression becomes
stronger at large $z_g$, a peak develops at small $z_g$ where
$\mathcal{R}(z_g)$ can even become larger than one, indicating a
nuclear {\em enhancement}.

\begin{figure}[t] 
  \centering
  \includegraphics[width=0.45\textwidth]{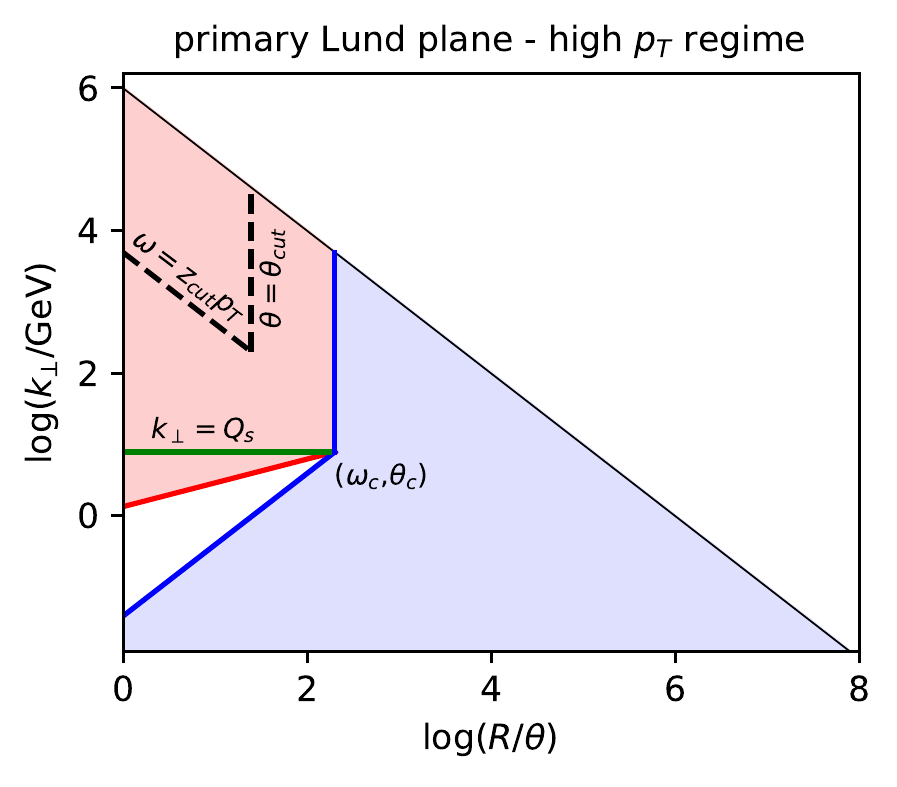}\hfill
  \includegraphics[width=0.45\textwidth]{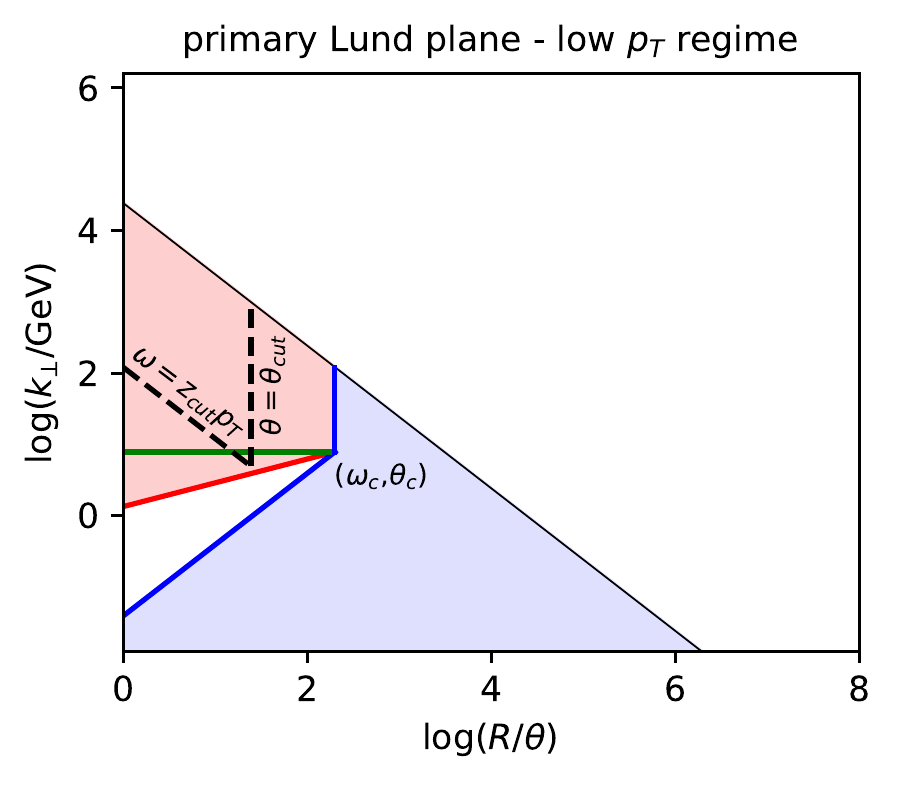}
  \caption{\small  The kinematic regions in the Lund plane that are covered by the SD algorithm
  in the case of a high-energy jet ($\zc p_T >\oc$) in the left figure and of a low-energy jet ($\zc p_T <\oc$) in the
  right figure. As suggested by the pictorial representation in the right figure, the most interesting situation for ``low-energy jets'' is such that there is only little overlap
  between the kinematic region for SD and the phase-space for MIEs.}
  \label{Fig:zgLund}
\end{figure}

Let us first discuss the behaviour at large $p_{T0}$, focusing on
$p_{T0}=1$~TeV.
In this case, the softest radiation that can be captured by Soft Drop
has an energy $\zc p_{T0}=100$~GeV which is still larger than the
hardest medium-induced emissions which have energies
$\omega\sim\oc=60$~GeV. Hence, for jets with high-enough $p_{T0}$, SD
can only select vacuum-like emissions. To illustrate this, we show in
the left plot of Fig.~\ref{Fig:zgLund} the phase-space selected by
SD. Under these circumstances, the only nuclear effect on the
$z_g$-distribution is the energy lost by the two (hard, $z_g > \zc$)
subjets passing the SD condition.
Due to this energy loss, the {\it
  effective} splitting fraction $z_g$ measured by SD turns out to be 
slightly smaller than the {\it physical} splitting fraction $z$ at the
branching vertex (see Sect.~\ref{sec:high} for details).
If we call for now this shift $\Delta z=z-z_g > 0$, we have
(cf.~\eqref{pzg_fixed_beta=0})
\beq
\mathcal{R}(z_g) \approx \frac{\bar{P}_g(z=z_g+\Delta z)}{\bar{P}_g(z_g)}\,\simeq\,\frac{z_g}{z_g+\Delta z}\,\simeq\,1-\frac{\Delta z}{z_g}\quad
\mbox{for \ $\Delta z\ll z_g\ll 1$}\,,
\eeq
which explains the pattern (smaller than one and increasing with
$z_g$) observed at large $p_{T0}$ in Fig.~\ref{Fig:zgvar}.

The above discussion also suggests what changes when moving to the
opposite regime of (relatively) low energy jets, say
$p_{T0}=100$~GeV. In this case, the energy interval covered by SD,
that is $\omega$ between $\zc p_{T0}=10\,{\rm GeV}$ and
$p_{T0}/2=50$~GeV, fully overlaps with the BDMPS-Z spectrum for
medium-induced radiation which has $\omega\lesssim \oc=60$~GeV (cf.\
the right plot of Fig.~\ref{Fig:zgLund}). Consequently, the SD
condition can now be triggered either by a vacuum-like splitting, or
by a medium-induced one. Since the BDMPS-Z spectrum increases rapidly
at small $z$ (faster than the vacuum splitting function), this
naturally explains the peak in the ratio  $\mathcal{R}(z_g)$ at small
$z_g$, visible in Fig.~\ref{Fig:zgvar}.
The nuclear suppression observe at large $z_g$ suggests that in this
region, the energy loss effects dominate over the BDMPS-Z emissions.

The above arguments show that the $z_g$ distribution is best discussed
separately for high-energy and low-energy jets, where the separation
between the two regimes is set by the ratio $\zc p_{T0}/\oc$. The
high-energy jets, for which $p_{T0}>\oc/\zc$, are conceptually simpler
as the in-medium $z_g$ distribution is only affected by the energy
loss via MIEs. For low-energy jets, i.e.\ jets with $p_{T0}< \oc/\zc$,
the $z_g$ distribution is affected by the medium both {\it directly}
when the SD condition is triggered by a MIE, and {\it indirectly} via the
energy loss of the two subjets emerging from the hard splitting.
This second case is more complex for a series of reasons and notably
because MIEs do not obey angular ordering.

\begin{figure}[t] 
  \centering
  \includegraphics[width=0.48\textwidth]{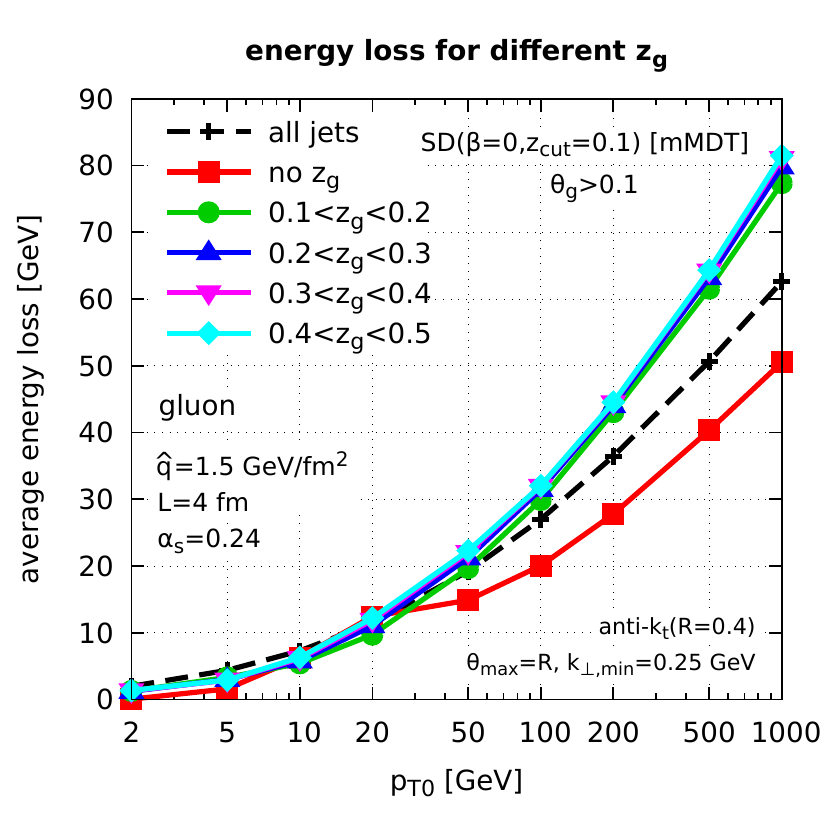}
 \caption{\small Our MC results for the average energy loss by a gluon-initiated jet are displayed
 as a function of the initial gluon energy $p_{T0}$ in bins of $z_g$. The inclusive (all jets) result  is also shown, by the dashed line.}
 \label{Fig:elosszg}
\end{figure}

Since $z_g$ is intrinsically tied to energy loss effects, it is
interesting to study how the average jet energy loss correlates with
$z_g$.
Our numerical results are presented in Fig.~\ref{Fig:elosszg}.
The dashed curve shows the MC results for the inclusive jets (all
values of $z_g$), the one denoted ``no $z_g$'' refers to jets which
did not pass the SD criterion or failed the cut $\theta_g>\thetacut$,
and the other curves correspond to different bins of $z_g$.
One clearly sees a distinction between the ``no $z_g$'' jets, which
lose much less energy than the average jet, and those which passed SD,
whose energy loss is larger than the average and quasi-independent of
$z_g$.
The main reason for this behaviour is that jets passing the SD
condition are effectively built of two relatively hard subjets. Since
the angular separation between these two hard subjets is larger than
$\thetacut=0.1 > \theta_c\simeq 0.04$, they lose energy (via MIEs) as
two independent jets, giving a larger-than-average energy loss. This
is mostly controlled by the geometry of the system, with only a
limited sensitivity to the precise sharing of the energy between the
subjets. On the other hand, the jets which did {\it not} pass SD are
typically narrow one-prong jets with either no hard substructure
or with some substructure at an angle smaller that $\thetacut=0.1$
(i.e.\ at an angle $\simeq \theta_c$). These jet therefore lose less
energy that an average jet.
The fact that the angular cut-off $\thetacut=0.1$ is close to the
critical value $\theta_c\simeq 0.04$ is clearly essential for the
above arguments.

A last comment concerns the difference between the $z_g$-distributions
for gluon- and quark-initiated jets, as shown in the left and right
plots of Fig.~\ref{Fig:zgvar}, respectively. The deviation of the
medium/vacuum ratio from unity appears to be larger for quark jets
than for gluon jets. This might look surprising at first sight given
that the average energy loss is known to be larger for the gluon jet
than for the quark one (cf.\ Figs.~\ref{Fig:jeteloss-v-pt}
and~\ref{Fig:jeteloss-v-R}). However we will show in
Section~\ref{sec:high} that the $z_g$ distribution is mostly
controlled by the energy loss of the softest subjet, which is
typically a {\it gluon} jet even when the leading parton is a
quark. The difference between quark and gluon jets in
Fig.~\ref{Fig:zgvar} is in fact controlled by ``non-medium'' effects,
like the difference in their respective splitting functions.

\subsection{High $p_T$ jets: VLEs and energy loss}
 \label{sec:high}
 
We begin our investigations of the nuclear effects on the
$z_g$ distribution with the case of a high-energy jet,
$p_{T0}>\oc/\zc$
In this case, the SD condition is triggered by an in-medium VLE that
we call the ``hard splitting'' in what follows. This splitting occurs
early (since $\tf\ll L$, cf. \eqn{tfvac}) and the daughter partons
propagate through the medium over a distance of order $L$. During
their propagation, they evolve into two subjets, both via VLEs (which
obey angular ordering, except possibly for the first emission outside
the medium) and via MIEs (which can be emitted at any angle).

Since the C/A algorithm is used by the SD procedure, both
subjets have an opening angle of order\footnote{On average, the two
  subjets have an (active) area $\approx 0.69\pi\theta_g^2$, while a
  single jet of radius $\theta_g$ has an area
  $\approx 0.81\pi\theta_g^2$, cf.~Fig~8 of
  Ref.~\cite{Cacciari:2008gn}. Each subjet thus have an effective
  radius of order $\sqrt{0.69/0.81}\theta_g\approx 0.92\theta_g$ which
  is very close to $\theta_g$.} $\theta_g$, with $\theta_g$ the angle
of the hard splitting.
Consequently, the emissions from the two subjets with angles larger
than $\theta_g$ --- either MIEs, or VLEs produced outside the medium
--- are {\it not} clustered within the two subjets. Accordingly, their
reconstructed transverse momenta $p_{T1}$ and $p_{T2}$ are generally
lower than the initial momenta, $\omega_1$ and $\omega_2$, of the
daughter partons produced by the hard splitting. This implies a
difference between the {\it reconstructed} splitting fraction
$z_g=p_{T,1}/(p_{T,1}+p_{T,2})$ and the {\it physical} one,
$z\equiv \omega_1/(\omega_1+\omega_2)$. This difference is controlled
by the energy lost by the two subjets.

Let us first mention that the energy loss via VLEs outside the medium
at angles $\theta >\theta_g$ can be neglected. Indeed, since these emissions
have $\tf \sim 1/(\omega\theta^2) > L$, they are soft and only
give very small contributions to the energy loss.
We have checked this explicitly with MC studies of the VLEs {\em
  alone}: we find that the effect on the $z_g$ distribution of the
vetoed region and of the violation of angular ordering for the first
emission outside the medium are much smaller than those associated
with the energy loss via MIEs.

We then discuss the role of {\it colour coherence} for the energy loss
via MIEs. If the splitting angle $\theta_g$ is smaller than $\theta_c$
the daughter partons are not discriminated by the medium. This is a
case of {\em coherent} energy loss where the MIEs at angles
$\theta>\theta_g$ are effectively sourced by their parent
parton~\cite{MehtarTani:2010ma,MehtarTani:2011tz,CasalderreySolana:2011rz,CasalderreySolana:2012ef,Mehtar-Tani:2016aco}, so that $z=z_g$.
On the other hand, for larger splitting angles
$\theta_g \gg \theta_c$, the colour coherence is rapidly washed out, so
the two daughter partons act as independent sources of MIEs. 
In this case, one can write
$\omega_i=p_{Ti}+ \mathcal{E}_i(\omega_i, \theta_g)$, where $i=1,2$
and $\mathcal{E}_i(\omega_i, \theta_g)$ is the average energy loss for a
jet of flavour $i$, initial energy $\omega_i$ and opening angle
$\theta_g$ (cf.\ e.g.~\eqn{logeloss}).
It would be relatively straightforward to deal with generic values of
$\theta_g$, both coherent and incoherent. In practice, all existing
measurements at the LHC imposes a minimal angle
$\theta_g \ge \thetacut$, with $\thetacut=0.1$. Since
$\theta_\text{cut}$ is larger than $\theta_c$ for all our choices of
parameters, we only consider the incoherent case
$\theta_g>\theta_\text{cut}$ in what follows.

That said, the relation between the measured $z_g$ and the physical
splitting fraction $z$ can be written as (assuming $p_{T1}<p_{T2}$)
\begin{tcolorbox}[ams align]\label{zinzg}
  z_g
  =\frac{p_{T1}}{p_{T1}+p_{T2}}
  = \frac{zp_T-\mathcal{E}_1(zp_T, \theta_g)}{p_T-\mathcal{E}_1(zp_T, \theta_g)
    -\mathcal{E}_2((1-z)p_T, \theta_g)}
  \,\equiv\, \mathcal{Z}_g(z, \theta_g)\,,
\end{tcolorbox}
\noindent where $p_{T}\equiv \omega_1+\omega_2$ is the energy (or transverse
momentum) of the parent parton at the time of the ``hard''
branching. In what follows we approximate $p_{T}\simeq p_{T,0}$. This
is valid as long as one can neglect two effects: \texttt{(i)} the
transverse momentum of the partons which have been groomed away during
previous iterations of the SD procedure, and \texttt{(ii)} MIEs prior
to the hard branching. The former is indeed negligible as long as we
work in the standard limit $\zc\ll 1$, and the latter is also
negligible based on our short formation time arguments in
Chapter~\ref{chapter:DLApic}.

For a given average energy loss $\mathcal{E}(p_{T}, \theta_g)$ one
can, at least numerically, invert \eqn{zinzg} to obtain the physical
splitting fraction $z=\mathcal{Z}(z_g, \theta_g)$ corresponding to the
measured $z_g$. The kinematic constraint $z_g>\zc$ thus implies a
constraint on $z$,
$z\ge \mathcal{Z}(\zc, \theta_g)$,\footnote{Here we assume
  that $\mathcal{Z}_g(z, \theta_g)$ is a monotonously increasing
  function of $z$.} and the in-medium $z_g$ distribution in this
high-energy regime becomes a straightforward generalisation of
\eqn{pzgthetag}
\begin{tcolorbox}[ams align]
 \label{zghighpt}
  p_{i, \text{med}}(z_g)= \mathcal{N}
 \int_{\thetacut}^{R}\rmd\theta_g\,
\Delta_i^{\text{VLE}}  (R,\theta_g)
  \int_0^{1/2}\rmd z\, \mathcal{P}_{i,\text{vac}}(z,\theta_g)\,\delta\big(z_g\minus \mathcal{Z}_g(z, \theta_g)\big)
   \Theta(z_g\minus\zc),
\end{tcolorbox}
\noindent where the Sudakov factor is formally the same as in the vacuum,
\eqn{Ddef}, but with the new, medium-dependent, lower limit
$\mathcal{Z}(\zc, \theta)$ on $z$:
\begin{equation}\label{DeltaVLE}
\Delta_i^{\text{VLE}}
(R,\theta_g)=\exp\left(-\int_{\theta_g}^{R}{\rmd\theta}\int_{0}^{1/2}\rmd
  z\,\,\mathcal{P}_{i,\text{vac}}(z,\theta)\,
  \Theta\big(z-\mathcal{Z}(\zc, \theta)\big)\right)\,
\end{equation}
The normalisation factor $\mathcal{N}$ in~(\ref{zghighpt}) is given by
$(1-\Delta_i^{\text{VLE}})^{-1}$. The \njets-normalised distribution
$f_{i, \text{med}}(z_g)$ is obtained by simply removing this factor
$\mathcal{N}$.

In practice we will replace $\theta_g$ by $R$ in the argument of the
energy loss. This is motivated by two facts.
Firstly, due to the SD procedure, we know that the jet is free of
emissions with $\omega>\zc p_{T0}$ at angles between $\theta_g$ and
$R$, simply because such an emission would have triggered the
SD condition. The remaining emissions between $\theta_g$ and $R$ are
therefore soft and we neglect them.
Secondly, the angular phase-space $\thetacut < \theta < R$ is
relatively small and $\mathcal{E}$ is slowly varying over this domain.
With this approximation, both $\mathcal{Z}_g$ and $\mathcal{Z}$
becomes independent of $\theta_g$ and the Sudakov factor
\eqref{DeltaVLE} simplifies to the vacuum one, \eqn{Ddef}, evaluated
at $\zc\to\mathcal{Z}(\zc)$.

The above picture can be further simplified by noticing that the
energy losses are typically much smaller than $zp_T$ and
$(1-z)p_T$. This means that the difference between $z$ and $z_g$ is
parametrically small and we can replace $z$ by $z_g$ in the arguments
of the energy loss in~(\ref{zinzg}) so that 
  \begin{equation}\label{zvszg}
z \equiv \mathcal{Z}(z_g, \theta_g) \simeq\,z_g+\,\frac{\mathcal{E}_1-z_g(\mathcal{E}_1+\mathcal{E}_2)}{p_T}\,,
\end{equation}
with $\mathcal{E}_1\equiv \mathcal{E}_1(z_gp_T)$ and $\mathcal{E}_2\equiv \mathcal{E}_2((1-z_g)p_T)$. 
Since $z_g<1/2$, this shows that the physical $z$ is
typically\footnote{Small deviations from this behaviour can happen close to
  $z_g=1/2$ when $\mathcal{E}_2\neq \mathcal{E}_1$. In this limit, our
  assumption that the softer physical parton ($z<1/2$) matches with
  the softer measured subjet ($z_g<1/2$) has to be reconsidered
  anyway.} larger than $z_g$.

As before, it is useful to consider the fixed-coupling scenario where  
the $z_g$-dependence of \eqn{zghighpt} factorises from the integral
over $\theta_g$. After dividing out by the vacuum distribution
$\propto \bar{P}_{i}(z_g)$, we find
\begin{equation}\label{Rdef}
  \mathcal{R}(z_g)
  \equiv \frac{f_\text{med}(z_g)}{f_\text{vac}(z_g)}
  \simeq \mathcal{J}(z_g)\,
  \frac{\bar{P}_g(\mathcal{Z}(z_g))}{\bar{P}_g(z_g)}\,,\qquad\mbox{with}\quad
\mathcal{J}(z_g)\equiv \bigg |\frac{\rmd  \mathcal{Z}(z_g)}{\rmd z_g}
\bigg |
\simeq 1-\frac{\mathcal{E}_1+\mathcal{E}_2}{p_T}\,,
\end{equation}
where $\mathcal{J}$ is a Jacobian and the last equality
in~\eqref{Rdef} is obtained using the simplified expression~(\ref{zvszg}).

At this level, it becomes necessary to specify the energy lost by a
subjet.
At high energy, both $zp_T$ and $(1-z)p_T$ are large and the energy
lost by the subjets is sensitive to the increase in the number of
partonic sources for MIEs (cf.\ Sections~\ref{sub:jet-eloss-th}-\ref{sec:MC-jet-eloss} and
Fig.~\ref{Fig:jeteloss-v-pt}).
To test this picture, we consider two energy loss scenarios.
First, the case of an energy loss which captures the increase in the
number of sources for MIEs and increases with the jet $p_T$, as in
Eq.~(\ref{logeloss}).
Since Eq.~(\ref{logeloss}) is not very accurate we will instead use
$\mathcal{E}_j=\mathcal{E}_{j,\text{fit}}$ corresponding to the fit of
the Monte Carlo result shown in Fig.~\ref{Fig:jeteloss-v-pt}.
The second scenario corresponds to what would happen in the absence of
VLEs, i.e.\ when only MIEs from the leading parton in each subjet are
included. This gives an energy loss which saturates to a constant
$\mathcal{E}_j=\varepsilon_j$ at large $p_T$ (see again
Fig.~\ref{Fig:jeteloss-v-pt}).
Clearly, the first scenario is the most physically realistic.

\begin{figure}[t] 
  \centering
  \includegraphics[page=1,width=0.48\textwidth]{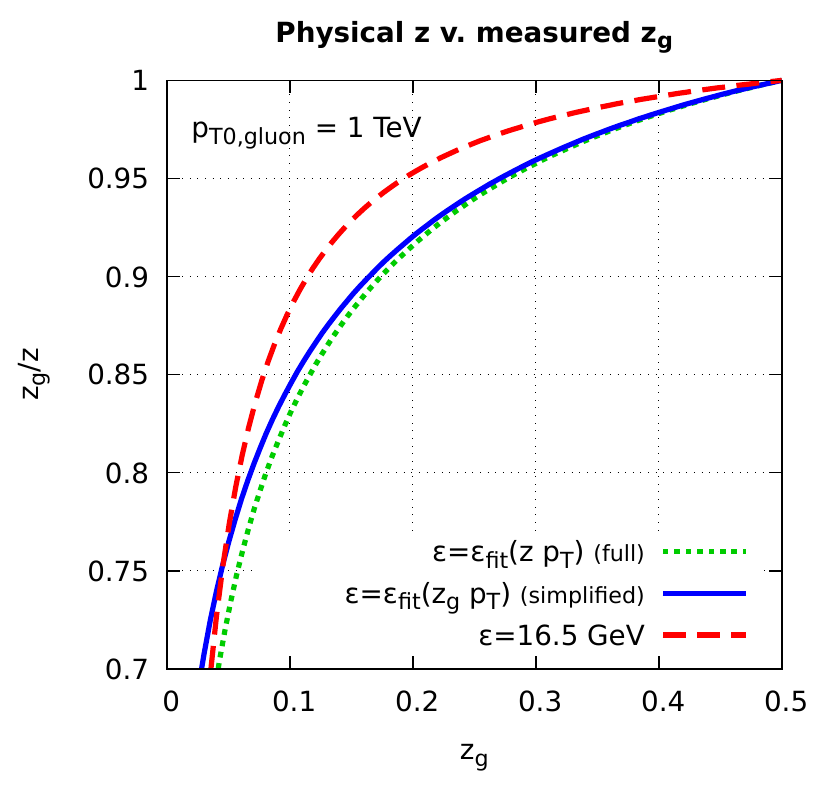}
  \hfill
  \includegraphics[page=2,width=0.48\textwidth]{./delta_zg.pdf}
  \caption{\small The ratio $z_g/z$ (left figure) and the difference $z-z_g$ (right figure) between the
  splitting fraction $z_g$ measured by SD and the physical splitting fraction $z$ for the hard 
  splitting.  This difference is given by Eq.~(\ref{zinzg}) that we evaluate with
   two scenarios for the energy loss by the subjets: a constant energy loss and a $p_T$-dependent
  one; for the second scenario, we also show the predictions of the simplified relation~(\ref{zvszg}).
   }
\label{Fig:Deltazg} 
\end{figure}

For definiteness, let us first consider the case of a 1-TeV
gluon-initiated jet.\footnote{We only included the dominant partonic
  channel $g\to gg$ in our analytic calculation.}
Fig.~\ref{Fig:Deltazg} shows the relation between the physical
splitting fraction $z$ and the measured $z_g$, with the ratio $z_g/z$
plotted on the left panel and the difference $z-z_g$ on the right
panel. We see that $z$ is larger than $z_g$ in both energy-loss
scenarios. The difference $z-z_g$ decreases when increasing $z$ (at
least for $z>\zc$), while the ratio $z_g/z$ gets close to 1. The
effects are roughly twice as large for the full energy-loss scenario
than for a constant energy loss.
The dotted (green) curve shows the result obtained using the ``full''
relation~(\ref{zinzg}) while the solid (blue) line uses the simplified
version, Eq~(\ref{zvszg}). As expected, they both lead to very similar
results and we therefore make the simplified version our default from
now on.

\begin{figure}[t] 
  \centering
  \includegraphics[page=1,width=0.48\textwidth]{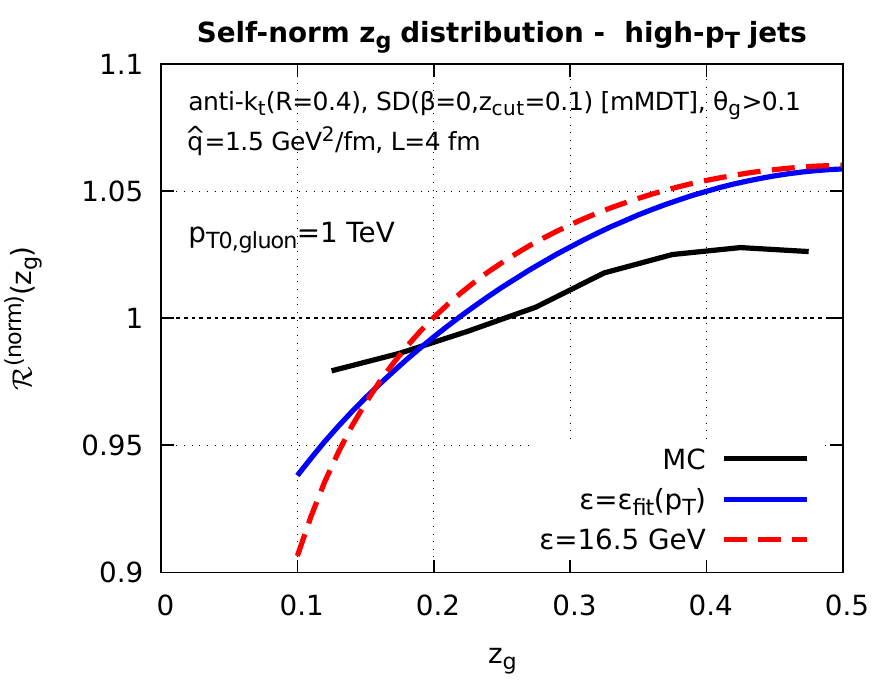}\hfill
  \includegraphics[page=2,width=0.48\textwidth]{./plot-highpt-ratio.pdf}
 \caption{\small Our MC results for the medium/vacuum ratio of the $z_g$ distributions associated with
 gluon-initiated jets with initial energy $p_{T0}=1$~GeV are compared to our analytic
 predictions for the two scenarios of energy loss described in the text. The left plot refers to the
 self-normalised distributions, cf. \eqn{zghighpt}, and the right plot to the $N_\text{jets}$-normalised ones.} \label{Fig:zghighpt}
\end{figure}

The nuclear modification factor for the $z_g$ distribution obtained
from our analytic calculation~(\ref{zghighpt}), including
running-coupling corrections, is shown in Fig.~\ref{Fig:zghighpt} for
both the self-normalised distribution $p_\text{med}(z_g)$ (left) and the
$N_\text{jets}$-normalised one $f_\text{med}(z_g)$ (right).
We see that $\mathcal{R}(z_g)$ increases with $z_g$. This is expected
since, at small $z_g$, $\bar{P}_g(z)/\bar{P}_g(z_g)\simeq z_g/z$ which
increases with $z_g$ (see e.g.\ Fig.~\ref{Fig:Deltazg}, left).
Furthermore, the medium/vacuum ratio of the $N_\text{jets}$-normalised distributions
(Fig.~\ref{Fig:zghighpt}, right) is always smaller than one.
With reference to the fixed-coupling estimate in Eq.~(\ref{Rdef}), 
this is a combined effect of $z$ being larger than $z_g$, hence
$\bar{P}_g(z)<\bar{P}_g(z_g)$, and of the extra Jacobian in front of
Eq.~(\ref{Rdef}). 

Regarding the comparison between the two scenarios for the energy loss, we see that
although they produce similar results for the self-normalised $z_g$
distribution, the (physical) ``full jet'' scenario predicts a larger
suppression than the ``constant'' one for the medium/vacuum ratio of the 
$N_\text{jets}$-normalised distributions.
In particular, the former predicts a value for $\mathcal{R}(z_g)$ 
which remains significantly smaller
than one even at $z_g$ close to 1/2.
This behaviour is in also in better agreement with our Monte Carlo simulations.
Generally speaking, it is worth keeping in mind that the
\njets-normalised ratio is better suited to disentangle between different
energy-loss models than the self-normalised ratio which is bound to cross one
by construction.

Since the nuclear modification of the $z_g$ distribution appears to be
so sensitive to the energy loss, it is interesting to check whether
this observable follows the Casimir scaling of the jet energy loss.
We show that this is not the case and that the nuclear modification
is even slightly larger for quark than for gluon jets.
In practice, the $z_g$ distribution is controlled by the energy loss
of the softest among the two subjets created by the hard splitting,
which is typically a gluon independently of the flavour of the initial
parton. Let us then consider \eqn{zvszg} in which
we take $\mathcal{E}_1= \mathcal{E}_{\text g}$ and
$\mathcal{E}_2=\mathcal{E}_R$, with $R={\text q}$ or ${\text g}$ depending on the
colour representation of the leading parton. Simple algebra yields
\beq\label{zGszQ}
z^{(\text{g-jet})}(z_g)\,\simeq\,z^{(\text{q-jet})}(z_g)+z_g\,\frac{\mathcal{E}_{\text g}-\mathcal{E}_{\text q}}{p_T}
\eeq
where $z^{(R\text{-jet})}(z_g)$ is the physical splitting fraction $z$
corresponding to a measured fraction $z_g$ for the case of a leading
parton of flavour $R$, and the energy loss functions $\mathcal{E}_R$
are evaluated at $(1-z_g)p_T$.
Since $\mathcal{E}_{\text g}\simeq  2\mathcal{E}_{\text q}$ the second term
in~\eqref{zGszQ} is positive and thus $z^{(\text{g-jet})}(z_g) >z^{(\text{q-jet})}(z_g)$
as expected on physical grounds. Yet, the difference between
$z^{(\text{g-jet})}(z_g)$ and $z^{(\text{q-jet})}(z_g)$ is weighted by $z_g$,
hence it suppressed at
    small $z_g$, where the energy loss effects should be more important.
Furthermore, the effects of the difference $z^{(\text{g-jet})}(z_g)-
z^{(\text{q-jet})}(z_g)$ are difficult to distinguish in practice
since there are other sources of differences between the  $z_g$
distributions of quark and  gluon jets like the non-singular terms in
the splitting functions and the different Sudakov factors. In practice
these effects appear to dominate over difference between
$z^{(\text{g-jet})}(z_g)$ and $z^{(\text{q-jet})}(z_g)$

\begin{figure}[t] 
  \centering
  \includegraphics[width=0.48\textwidth,page=2]{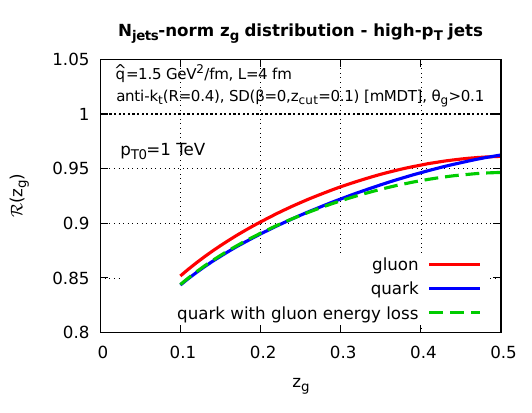}
 \caption{\small The results of the analytic calculations for the medium/vacuum ratio 
 $\mathcal{R}(z_g)$ for both gluon- and quark-initiated jets, together with the fictitious
 case where a quark-initiated  subjet loses the same amount of energy as a gluon-initiated one.}
 \label{Fig:QvsG}
\end{figure}

This is confirmed by our analytic calculations in
Fig.~\ref{Fig:QvsG}. Together with our previous results for a
gluon jet, we show two scenarios for quark jets: \texttt{(a)} a realistic
scenario which takes into account the different quark and gluon energy
losses (cf.\ Fig.~\ref{Fig:jeteloss-v-pt}), and  \texttt{(b)} a
fictitious case, which assumes that a quark subjet loses the same
energy as a gluon one, i.e.\ $\mathcal{E}_{\text q}=\mathcal{E}_{\text g}$.
In both cases, the nuclear suppression of the $z_g$ distribution
appears larger for quark-initiated jets than for gluon-initiated
jets, in qualitative agreement with our Monte Carlo findings (recall 
 Fig.~\ref{Fig:zgvar}).
This is clearly driven by effects beyond the energy loss difference
between quarks and gluons (cf our case \texttt{(b)}), even though this
difference has indeed the effect of slightly increasing
$\mathcal{R}(z_g)$, especially close to $z_g=1/2$, as visible by comparing
the curves corresponding to the cases \texttt{(a)} and \texttt{(b)}.

\paragraph{Summary.} To sum up, the main lessons one draws from our study of high-energy
jets are as follows:
\texttt{(i)} the incoherent energy loss by the
two subjets created by the hard splitting leads to a suppression in
the nuclear $z_g$ distribution which is larger at small $z_g$;
\texttt{(ii)} the MC results are sensitive to the evolution of
the subjets multiplicity via VLEs which leads to an energy loss
increasing with the subjet $p_T$;
\texttt{(iii)} this last effect may be hidden when studying the
self-normalised $z_g$ distribution; in that respect, the \njets-normalised
distribution is better suited to disentangle between different
energy-loss models.

\subsection{Low-$p_T$ jets: MIEs and energy loss}
\label{sec:lowpt}

We now turn to more phenomenologically-relevant case of ``low energy''
jets, $p_{T0} <  \omega_c/\zc$ (with $\omega_c/\zc=600$~GeV for 
our default choice of medium parameters), for which the ``hard'' 
emission that triggers the SD condition can be either vacuum-like or medium-induced.
In both cases, the  two ensuing subjets lose energy via MIEs, which, as 
explained in the previous section, implies 
that the measured value $z_g$ is different (typically slightly
smaller) than the physical value $z$.

Our main goal in this section is to develop analytic approximations
which qualitatively and even semi-quantitatively capture this complex
dynamics.
For definiteness, we focus on gluon-initiated jets with
$p_{T0}=200$~GeV. This value is at the same time low enough to be
representative for the low-energy regime and large enough to justify
some convenient approximations, like the single emission approximation
for the MIEs captured by SD.
For pedagogical reasons it is convenient to first consider two
simplified situations --- a jet built with MIEs alone and a jet in
which the SD condition can only be triggered by a VLE (as in the
``high-energy'' case) ---, before addressing the full picture in
Sect.~\ref{sec:full}.

\subsubsection{Low-$p_T$ jets: medium-induced emissions only}
\label{sec:low}

To study the case where the SD condition is triggered by a MIE, we
consider jets generated via MIEs only, disabling VLEs.
Since the emission angles of MIEs are controlled by their transverse
momentum broadening (cf. Sect.~\ref{sub:angular-structure}), they are not ordered in
angle.
Hence, the reclustering of the jet constituents with the C/A algorithm
does not necessarily respects the physical ordering of the MIEs in
time.
In particular, the branching selected by the SD procedure may not be a
{\it primary} emission, i.e.\ a direct emission by the leading
parton.
However, as long as $\zc p_{T0}$ is sufficiently large compared to the
characteristic scale $\obr=3.46$~GeV for multiple branching --- which
is definitely the case for our 200~GeV jets, ---  the
probability to select a non-primary branching is suppressed by
$\amed$. From now on we can therefore assume that SD selects a {\it primary} MIE.

Next, we can argue that the MIEs captured by the SD algorithm are soft and
located in a small corner near $z=\zc$ and $\theta=\thetacut$.
Indeed, the bulk of the MIEs lies below the line $\kt=Q_s=2.4$~GeV and
the smallest value of $\kt$ that can be selected by SD, namely
$\zc p_{T0}\thetacut= 2$~GeV, is only slightly smaller than
$Q_s$. This is visible in the phase-space diagram of
Fig.~\ref{Fig:zgLund} (right).
Together with the fact that the BDMPS-Z rate~\eqref{mie-MC-rate} grows
quickly as $z\to 0$, this means that MIEs contribute to the $z_g$
distribution only at small $z_g$.
After SD, one therefore has a soft subjet of transverse momentum
$p_{T1}$ corresponding to the MIE and a harder subjet of
momentum $p_{T2}$ corresponding to the leading parton.

\begin{figure}[t] 
  \centering
    \includegraphics[width=0.48\textwidth]{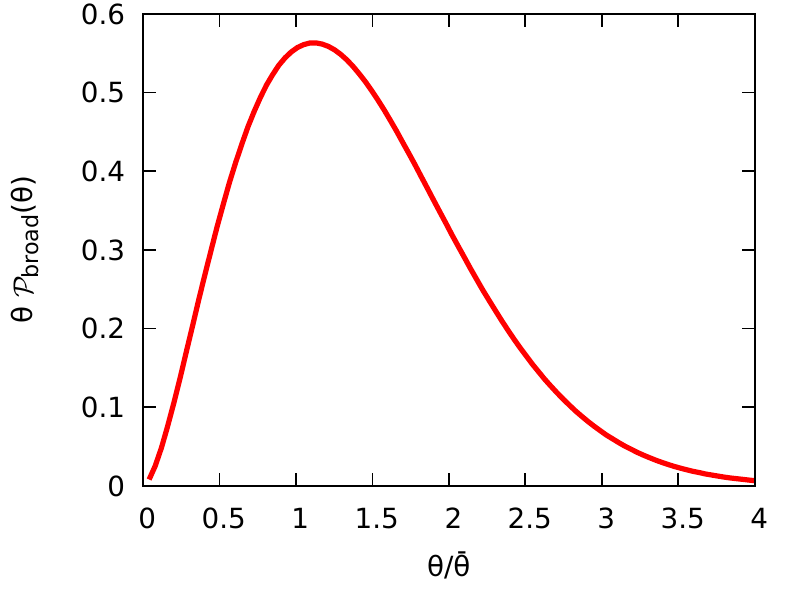}\hfill%
    \includegraphics[width=0.48\textwidth]{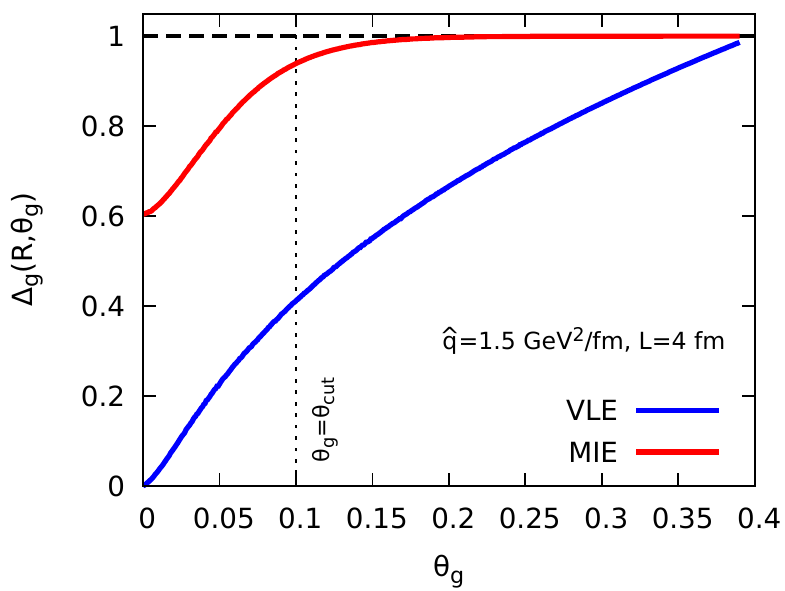}
 \caption{\small  Left: after multiplication by $\theta$, the angular distribution $\mathcal{P}_{\text{broad}}$ in \eqn{pbroad} scales as a function of the ratio $\theta/\bar\theta$, with  $\bar\theta=(\sqrt{\pi}/3)Q_s/\omega$. Right: the Sudakov factors $\Delta_{\text g}^{ \text{MIE}} (R,\theta_g)$ for MIEs, cf.~\eqn{Deltamed}, and 
 $\Delta_{\text g}^{\text{VLE}}(R,\theta_g)$ for VLEs, cf.~\eqn{DeltaVLE}, for a gluon-initiated jet with
 $ p_{T0}=200$~GeV and $R=0.4$.  }
 \label{Fig:broad}
\end{figure}

The differential probability for the emission of a primary MIE with
$\omega_1\equiv\omega \ll p_{T0}$ is given by the BDMPS-Z
spectrum~\eqref{BDMPSZ-estimate} multiplied by the angular distribution
produced via transverse momentum broadening after emission,
\eqn{Ptransverse}. The latter depends on the distance $\Delta t=L-t$,
with $t$ the emission time, travelled by the two
subjets through the medium.
In principle one should therefore work differentially in $t$. Since
this would be a serious complication, we rather use a picture in which
we average over all the emission times $t$, distributed with uniform
probability over the interval $0<t<L$. This picture is further supported by the angular structure
of jets with primary medium-induced emissions after a C/A declustering. In Section \ref{subsub:zg-mie}, we have seen that in 
the regime $\omega_s(R)=(\qhat/(\abar^2R^4))^{1/3}\ge \obr$, a jet with only primary medium-induced emissions can be seen as an effective angular ordered branching process, with a differential rate
for the ``hard'' splitting taking the form:
\begin{equation}
\label{Pmed}
\rmd^2 \mathcal{P}_{i, \text{med}}(\omega,\theta)= \frac{C_i\amed}{\pi}\Theta\left({\oc}-\omega\right)
\sqrt{\frac{2\oc}{\omega^{3}}}\,\mathcal{P}_{\text{broad}}(\omega,\theta)\,{\rmd \omega}
\rmd\theta
 \,\equiv \,\mathcal{P}_{i,\text{med}}(\omega,\theta) \rmd \omega \rmd\theta,
\end{equation}
where~\cite{Mehtar-Tani:2016aco}
  \begin{equation}
\label{pbroad}
 \mathcal{P}_{\text{broad}}(\omega,\theta)\equiv \frac{1}{L}\int_0^L\rmd t\,\frac{2\omega^2\theta}{\hat{q}(L-t)}\exp\left\{-\frac{\omega^2\theta^2}{\hat{q}(L-t)}\right\}=
 2\theta\,\frac{\omega^2}{Q_s^2}\,\Gamma\left(0,\frac{\omega^2\theta^2}{Q_s^2}\right),
\end{equation}
This distribution predicts an average value
$\bar k_\perp=\tfrac{\sqrt{\pi}}{3}Q_s$ for $k_\perp\equiv \omega\theta$. It
shows a peak near $\bar k_\perp$, and a rather wide tail at
larger values $\kt > \bar k_\perp$ (see Fig.~\ref{Fig:broad}, left).
Since we have just argued that SD selects emissions in a narrow range
in $k_\perp$, close to $Q_s$, the tail of this distribution plays an
important role in our discussion. This is amplified by the fact that
$\bar k_\perp$ is slightly smaller than $Q_s$.
Note that in terms of the emission angle, this argument means
that the emissions selected by SD will need to acquire a $\theta$
larger than the peak value $\bar\theta(\omega)=\bar k_\perp/\omega$ 
from broadening in order to pass the
$\theta_\text{cut}$ condition.
In future work, it will be interesting to study how a description of
broadening beyond the Gaussian approximation affects quantitatively
our results.

Additionally, we need to account for the fact that both the MIE that
triggers the SD condition and the leading parton lose energy. 
The situation is mostly the same as for our earlier high-energy case
except that now the medium-induced gluon emission can occur anywhere
inside the medium, i.e.\ at any time $t$ with $0< t<L$.
For an emitted gluon of energy $\omega$, one can write
$p_{T1}=\omega-\varepsilon_{\text g}(\omega,\theta_g, \Delta t)$, where the
energy loss depends explicitly on the distance $\Delta t=L-t$
travelled by the subjet through the medium.
In our kinematic range, this energy loss is relatively small,
$\varepsilon_{\rm g}\ll \omega$, and therefore varies slowly with
$\omega$ (cf.\ \eqn{ElossHigh} so we can neglect the $\omega$
dependence.
It depends however quadratically on $\Delta t$.
For simplicity, we use a time-averaged picture in which
$\langle t\rangle \simeq \langle\Delta t\rangle \simeq L/2$ and
therefore
$\varepsilon_{\text g}(\Delta t)\approx\varepsilon_{\text g}(L/2)\simeq
\tfrac{1}{4}\varepsilon_{\text g}(L)\equiv \bar\varepsilon_{\text g}$, with
$\varepsilon_{\text g}(L)\propto L^2$ the energy loss corresponding to a
distance $L$.

The situation for the harder subjet matching with the leading parton, is more complex, owing to
differences between the time ordering of MIEs and the
(angular-ordered) C/A clustering used by SD.
Indeed, MIEs from the leading parton at times smaller than $t$ and
angles smaller than $\theta_g$ will be clustered by the C/A algorithm
in the harder subjet.
These emissions can carry a substantial amount of energy so that,
without a full picture of the time evolution of the jet, it is
delicate to even define a physical transverse momentum, $\omega_2$,
for the harder subjet at the vertex where the MIE triggering the SD
condition is emitted.
We can however take a different approach and realise that, by
definition, the difference $p_{T0}-(p_{T1}+p_{T2})$ corresponds to the
energy $\varepsilon_i(p_{T0},\theta_g)$ lost by the initial parton at
angles larger than $\theta_g$. As for the high-energy case, we can
neglect the energy lost between $\theta_g$ and $R$ and hence write
$p_{T1}+p_{T2}\simeq p_{T0}-\varepsilon_i(p_{T0},R)$.

In fine, the measured value of $z_g$ is related to the initial energy
$\omega$ of the gluonic MIE subjet via
\beq\label{zgMIE}
z_g\,\simeq\,\frac{\omega-\bar\varepsilon_{\text g}(\omega,\theta_g)}{p_{T0}-\varepsilon_{i}(p_{T0},R)}\,
\qquad\text{ with }\,\bar\varepsilon_{\text g}(\omega,\theta_g)=\frac{1}{4}\varepsilon_{\text g}(\omega,\theta_g)
\eeq
Since both energy losses in~\eqref{zgMIE} are small, one can ignore
their $p_T$ dependence and use the fits to the MC results shown in Figs.~\ref{Fig:jeteloss-v-R}.
The $z_g$ distribution created via MIEs can then be computed using a formula similar to that used for VLEs in the previous subsections, cf. \eqn{zghighpt}, namely 
\begin{tcolorbox}[ams align]
 \label{zglowpt}
  f_{i,\text{med}}(z_g)= \int_{\thetacut}^{R}\rmd\theta_g\,
\Delta_i^{ \text{MIE}} (R,\theta_g)
  \int\rmd \omega\, \mathcal{P}_{i,\text{med}}(\omega,\theta_g)\delta\left(z_g-\frac{\omega-\bar\varepsilon_{\rm g}(\theta_g)}{p_{T0}-\varepsilon_{i}(R)}\right)
   \Theta(z_g-\zc)
\end{tcolorbox}
\noindent with the Sudakov factor $\Delta_i^{ \text{MIE}} (R,\theta_g)
$ accounting for the probability to have no primary MIEs with
$\omega>\omega_{\text{cut}}\equiv \bar\varepsilon_{\rm
  g}+\zc(p_{T0}-\varepsilon_{i}(R))$ and $\theta >\theta_g$ at any
point  inside the medium:
\begin{equation}\label{Deltamed}
\Delta_i^{ \text{MIE}} (R,\theta_g)=\exp\left(-\int_{\theta_g}^{R}{\rmd\theta}\int\rmd \omega\,\,\mathcal{P}_{i,\text{med}}(\omega,\theta)\, \Theta\left(\omega-\omega_{\text{cut}}\right)\right)\,
\end{equation}
The self-normalised distribution can be computed as
$p_{i, \text{med}}(z_g)= \mathcal{N} f_{i, \text{med}}(z_g)$ with
$\mathcal{N}=\tfrac{1}{1-\Delta_i^{ \text{MIE}} (R,\thetacut)}$.
The Sudakov factor is plotted for both MIEs, \eqn{Deltamed}, and VLEs,
\eqn{DeltaVLE}, as a function of $\theta_g$ in the right plot of
Fig.~\ref{Fig:broad}. In both plots, we use $p_{T0}=200$~GeV and a gluon-initiated
jet.
While this factor is clearly important for VLEs at all $\theta_g$, it
remains very close to one for MIEs. It is mostly irrelevant for the
shape of the $z_g$ distribution and only has a small impact on its
overall normalisation.

\begin{figure}[t] 
  \centering
  \includegraphics[width=0.48\textwidth]{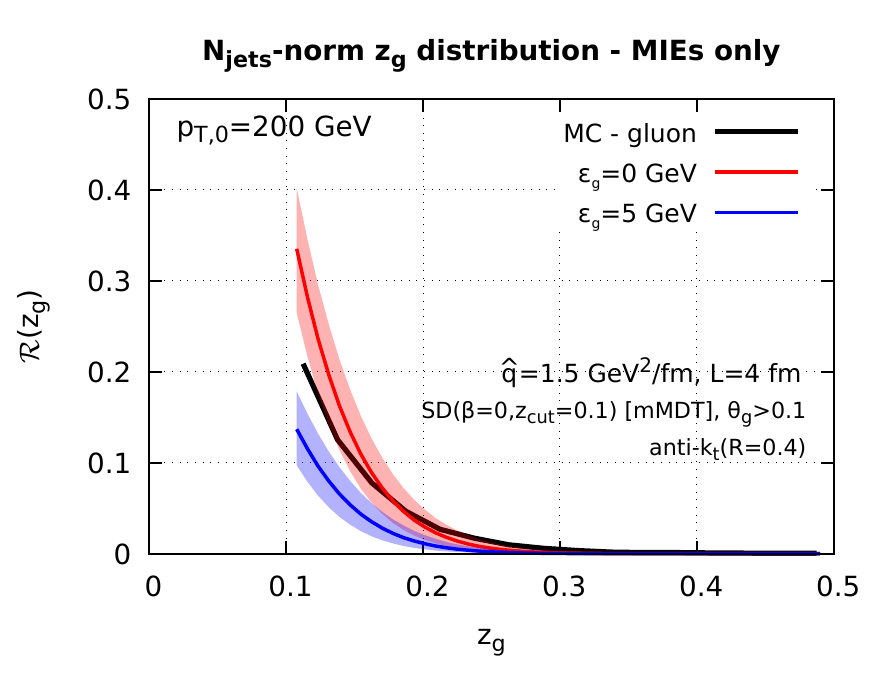}\hfill%
  \includegraphics[width=0.48\textwidth]{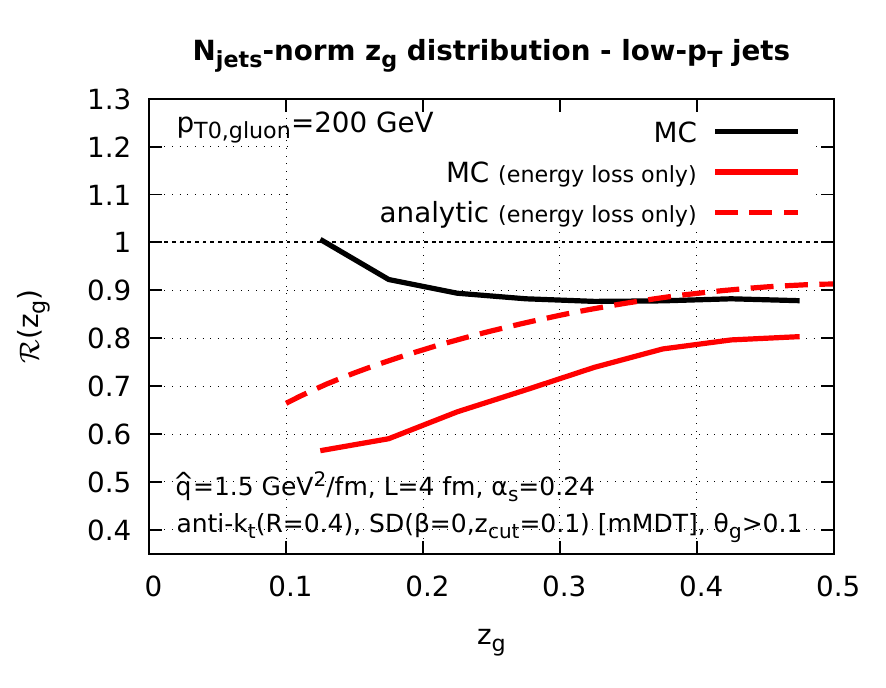}
  \caption{\small Separation of the nuclear effects on the $z_g$
    distribution of low-energy jets into a contribution due to the
    MIEs (left) and a contribution due to the VLEs with energy loss
    (right). Left: MIEs only (MC vs.\ analytic calculations).
    Right: Monte Carlo results for the full parton shower (black) vs.\
    the case where only VLEs with energy loss are included (red). We
    also show an analytic result for the second case (dashed line).}
  \label{Fig:lowptMIEVLE}
\end{figure}

In Fig.~\ref{Fig:lowptMIEVLE}(left) we compare our analytic approximation for
the ratio\footnote{Here, the medium/vacuum ratio is not
  a genuine nuclear modification factor. For example, in the absence of medium effects, it would be equal to zero, not to one.} $f_{\text{med}}(z_g)/f_{\text{vac}}(z_g)$ of the
\njets-normalised $z_g$ distributions corresponding the a gluon-initiated
jet to the MC results obtained by ``switching off'' the VLEs from the
general numerical code.
The energy losses are estimated from the 
 fit to the MC results shown in the left plot of Fig.~\ref{Fig:jeteloss-v-R}, which yields
$\varepsilon_{\rm g}\simeq 16.5$~GeV for the whole jet with $R=0.4$ and $\bar\varepsilon_{\rm g}\simeq 5$~GeV for the subjet with $\theta_g\simeq\thetacut=0.1$. Given the large uncertainty in the calculation of $\bar\varepsilon_{\rm g}$, we present  two sets of results, one corresponding to $\varepsilon_{\rm g}=0$~GeV and the other one to $\varepsilon_{\rm g}=5$~GeV.
For each of these 2 choices we indicate by a band the uncertainty associated with $\pm 10\%$ variations in the saturation scale $Q_s$ around its central value $Q_s=\sqrt{\hat q L}=2.4$~GeV. This variation corresponds to the fact that the relation \eqref{zgMIE} between $z_g$ and $\omega$ is only approximate and the associated uncertainty in the value of $\omega$ has consequences, via \eqn{pbroad}, on the angular distribution; this uncertainty was mimicked by varying $Q_s$.
 
Fig.~\ref{Fig:lowptMIEVLE}(left) shows a qualitative agreement between the MC
and the analytic calculations: all the curves have a visible rise
at small $z_g$, reflecting the fact that the BDMPS-Z
spectrum behaves like $z^{-3/2}$ which is more singular than the
vacuum spectrum $\propto z^{-1}$.
This being said, our analytic study is still too poor to
quantitatively reproduce the MC results, or to discriminate between
various scenarios for the energy loss. In particular, the ``zero
energy loss'' scenario is not in clear disagreement with the MC
results. This may be related to the fact that the angular distribution
in \eqn{pbroad} favours small values $z\sim\zc$ which biases the
distribution towards events with a smaller-than-average energy loss.

\subsubsection{Low-$p_T$ jets: energy loss only} 
\label{sec:lowpteloss}
 
In this section, we consider the situation (opposite to the previous
section) where the SD condition is triggered by a VLE.
In this case, we turn off the direct contribution of the MIEs to SD,
but only keep their (indirect) effect associated with incoherent
energy loss of the subjets found by the SD procedure.
The physical situation is similar to the high-energy case studied in
Sect.~\ref{sec:high} where the ``direct'' contribution of the MIEs to
SD was negligible by definition.

To artificially remove the direct contribution of the MIEs from the MC
simulations, we have enforced that all the partons generated via MIEs
propagate at angles $\theta \gg R$.
This obviously overestimates the jet energy loss, simply because some
of the partons which would have remained within the jet are
artificially moved outside. This is fine as long as we only focus on
illustrating the qualitative effects of the energy loss on the $z_g$
distribution.

In Fig.~\ref{Fig:lowptMIEVLE}(right), we compare Monte Carlo
results obtained with this artificial removal of the direct
contribution of MIEs (``energy loss only'', red,  curve) to the
full simulation (``full'', black, curve), where both VLEs and MIEs
contribute directly.
Fig.~\ref{Fig:lowptMIEVLE}(right) also shows the prediction of an analytic
calculation which ignores the direct contribution of the MIEs to SD.
This calculation is the same as the one presented in
Sect.~\ref{sec:high} for the case of a high-energy jet, i.e.\ it is
based on Eqs.~\eqref{zinzg}, \eqref{Pvac} and~\eqref{DeltaVLE}, now
applied to $p_{T}\simeq p_{T,0}=200$~GeV.  The respective Sudakov
factor is plotted in the right plot of Fig.~\ref{Fig:broad}.
By inspection of these curves, we first notice that the effect of energy loss alone is the same for
low-energy jets as it was for high-energy jets: it leads to a strong
nuclear suppression\footnote{This suppression appears to be larger for the respective MC calculation than for the analytical one 
because, for the former, the energy loss is artificially amplified.} of the $z_g$ distribution with larger effects at small $z_g$. 
Second, adding the direct contribution of the MIEs to SD changes
the picture significantly: the medium/vacuum ratio is now increasing at small $z_g$
and can even becomes close to one.
The difference between the two curves is, at least qualitatively,
consistent with an additional peak at small $z_g$ from MIEs (see e.g.\
Fig.~\ref{Fig:lowptMIEVLE}, left).

\subsubsection{Low-$p_T$ jets: full parton showers} 
\label{sec:full}

Now that we have studied both effects separately, we can provide an
analytic calculation for the complete $z_g$ distribution for a
low-energy jet, including both VLEs and MIEs.

Due to angular ordering, VLEs selected by the SD procedure are
necessarily primary gluon emissions from the leading parton.
However, whenever SD selects a MIE with energy fraction $z$ and
emission angle $\theta_g$, this emission can be emitted by any of the
partonic sources created via VLEs with energy $\omega>zp_{T0}$. Since,
by definition, SD selects the largest-angle emission with $z_g$ above $\zc$,
these sources of MIE can have any angle in the range
$\theta_c<\theta<\theta_g$. Such emissions are formally clustered by
the C/A algorithm together with the subjet corresponding to the leading
parton.

The $z_g$ distribution for a full parton shower generated by a parton
of type $i=(\text{q,\,g})$  is obtained by incoherently summing up the
probabilities for SD to select either a VLE or a MIE:
\begin{tcolorbox}[ams align]  \label{full}
 f_i(z_g)& = \int_{\thetacut}^{R}\rmd\theta_g\,
\Delta_i^{\text{VLE}}  (R,\theta_g) \,\Delta_i^{\text{MIE}} (R,\theta_g) \\
 &\hspace{-1cm}\times
 \int_{0}^{1/2}\rmd
   z\Big[\mathcal{P}_{i,\text{vac}}(z,\theta_g)\delta\big(
   \mathcal{Z}_{g,\text{vac}}(z,\theta_g)-z_g\big)+\mathcal{P}_{i,\text{med}}(z,\theta_g)\delta\big(
   \mathcal{Z}_{g,\text{med}}(z,\theta_g)-z_g\big)\Big]\Theta(z_g-z_{\text{cut}})
   \nonumber
 \end{tcolorbox}
Here, we have used different functions, $ \mathcal{Z}_{g,\text{vac}}(z,\theta_g)$ and $ \mathcal{Z}_{g,\text{med}}(z,\theta_g)$, for the relation between the measured splitting fraction $z_g$ and the physical one $z$, to take into account the fact that the energy loss is generally different for the subjets produced by a VLE or a MIE.  
The function $ \mathcal{Z}_{g,\text{vac}}(z,\theta_g)$ is given by
\eqn{zinzg}, with the $p_T$ of the parent gluon identified with the
$p_{T0}$ of the leading parton. As before, we replace $\theta_g$ by $R$ in
the energy loss and take $\mathcal{E}_{1}(zp_T,R)$ and
$\mathcal{E}_{2}((1-z)p_T,R)$ from the fits in Fig.~\ref{Fig:jeteloss-v-pt}.
In the case of a medium-induced splitting, we use a generalisation of \eqn{zgMIE}, that is
\beq\label{zgfull}
z_g\,\simeq\,\frac{zp_{T0}-\bar\varepsilon_{\rm g}(\theta_g)}{p_{T0}-\mathcal{E}_{i}(p_{T0},R,z_g)}
\,\equiv  \mathcal{Z}_{g,\text{med}}(z,\theta_g)\,
\eeq
The main difference w.r.t.\ \eqn{zgMIE} refers to the energy loss by
the whole jet, i.e.\ the function $\mathcal{E}_{i}(p_{T0},R,z_g)$ in
the denominator: not only this has now a strong dependence upon
$p_{T0}$, due to the rise in the number of partonic sources via VLEs,
but this must be evaluated for the special jets which include a hard
splitting with a given value $z_g>0.1$ and with any
$\theta_g>\thetacut=0.1$. From our MC calculation illustrated in
Fig.~\ref{Fig:elosszg}, we know that it is larger than the
average energy loss and largely independent of $z_g$. We use
$\mathcal{E}_{\text g}=43$~GeV for  $p_{T0}=200$~GeV and $R=0.4$. As for
$\bar\varepsilon_{\text g}(\theta_g)$, we use 5~GeV  as in Sect.~\ref{sec:low}
and study the sensitivity of our results to variations around this value.

The vacuum splitting probability density $\mathcal{P}_{i,\text{vac}}$
and the associated Sudakov factor $\Delta_i^{\text{VLE}}$ take the
same form as in Sect.~\ref{sec:high}, Eqs.~\eqref{Pvac}
and~\eqref{DeltaVLE}.
The corresponding medium-induced probability density
$\mathcal{P}_{i,\text{med}}$ takes a form similar to Eq.~\eqref{Pmed}
modified to account for the fact that each VLE produced in the medium
can act as a source of MIE. We therefore write
\begin{align}
\label{Pfullmed}
  \mathcal{P}_{i,\text{med}}(z,\theta_g)&=\nu(z,\theta_g)\,\frac{\alpha_{s,\text{med}} C_i}{\pi}
  \sqrt{\frac{2\oc}{p_{T0}}}\,z^{-3/2}\,\mathcal{P}_{\text{broad}}(z,\theta_g)
\end{align}
The number of MIE sources $\nu(z,\theta_g)$ is obtained from the
density $\tfrac{\rmd^2N_{\text{VLE}}^{\text{(in)}}}{\rmd\omega
  \rmd\theta}$ of VLEs produced inside the medium:
\beq
\label{nudef}
\boxed{\nu(z,\theta_g)\equiv 1+ \int_{z p_{T0}}^{p_{T0}}\rmd\omega\int_{\theta_c}^{\theta_g}\rmd\theta \frac{\rmd^2N_{\text{VLE}}^{\text{(in)}}}{\rmd\omega \rmd\theta}\,}
\eeq
In this last expression, the first term corresponds to the leading
parton and the integration boundaries in the second term impose that
an MIE which triggers the SD condition has to come from a source of
larger energy at an angle between $\theta_c$ and $\theta_g$. 
The associated Sudakov factor $\Delta_i^{\text{MIE}}$ is then constructed in terms of $\mathcal{P}_{i,\text{med}}$ as in \eqn{Deltamed}.
We note however that, although for the case with only MIEs, the
exponentiation of the emission probability in the Sudakov factor
$\Delta_i^{\text{MIE}}$ is relatively straightforward, this
does not obviously hold in the presence of multiple sources of
MIEs.
However, since the Sudakov factor $\Delta_i^{\text{MIE}}$ only introduces a
small correction (see Fig.~\ref{Fig:broad}, right), 
we have kept the exponential form,
Eq.~(\ref{Deltamed}), for simplicity and as an easy way to maintain
the conservation of probability for MIEs.

\begin{figure}[t] 
  \centering
  \begin{minipage}[t]{0.45\textwidth}
    \mbox{ }\\
    \includegraphics[page=1,width=\textwidth]{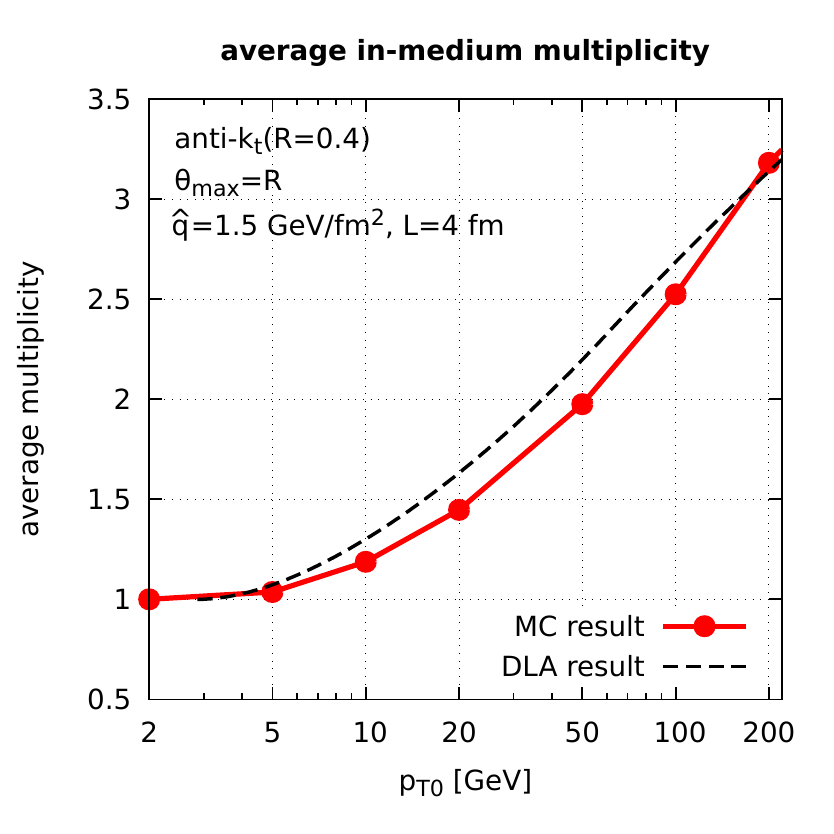}
  \end{minipage}
  \hfill%
  \begin{minipage}[t]{0.43\textwidth}
    \mbox{ }\\
    \vspace*{1.0cm}

    \caption{\small The average multiplicity of partons created inside
      the medium via VLEs by a leading gluon with $p_{T0}=200$~GeV: the
      full Monte Carlo results shown by (red) points are compared with a
      running-coupling extension of the DLA estimate in \eqn{TDLA},
      integrated over the ``inside'' region of the phase-space of
      Fig.~\ref{Fig:DLA-phase-space}.} \label{Fig:multi}
  \end{minipage}
\end{figure}

\begin{figure}[t] 
  \centering
  \includegraphics[page=1,width=0.48\textwidth]{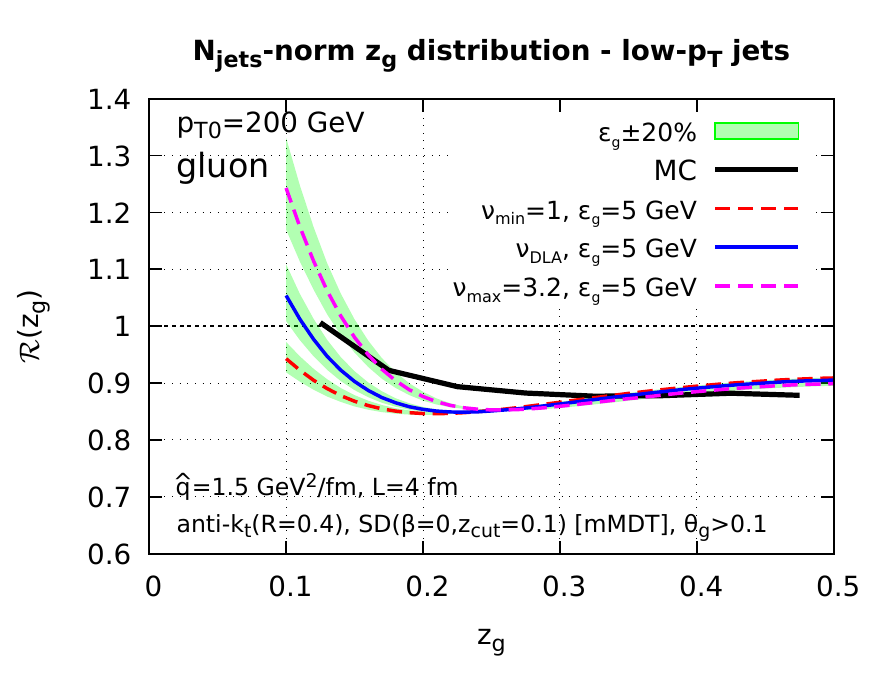}\hfill%
  \includegraphics[page=2,width=0.48\textwidth]{./plot_zg-low_pT.pdf}\hfill%
  \caption{\small
    $\mathcal{R}(z_g)$ for a leading parton (left: gluon, right:
    quark) with initial transverse momentum $ p_{T0}=200$~GeV. The
    Monte Carlo results are compared to analytic calculations
    corresponding to 3 different approximations for the number $\nu$
    of sources emitting MIEs (see the text for details). }
  \label{Fig:zglowpt}
\end{figure}

In practice, we test three different approximations\footnote{Note also that the formal
limit $\nu\to 0$, in which one keeps only the ``direct'' contribution of the VLEs to \eqn{Pfullmed},
corresponds to the analytic result shown with dashed line in Fig.~\ref{Fig:lowptMIEVLE}(right).}
for $\nu$: \texttt{(i)} $\nu\equiv\nu_{\rm min}=1$ includes only the leading
parton, \texttt{(ii)} $\nu\equiv\nu_{\rm max}=3.2$ is our MC estimate
for the average multiplicity of partons created inside the medium via
VLEs by a leading gluon with $p_{T0}=200$~GeV,
cf. Fig~\ref{Fig:multi}. This has to be seen as a maximal value since
it ignores the kinematic limits in \eqn{nudef}). \texttt{(iii)}
$\nu\equiv\nu_{\rm DLA}$ obtained by evaluating \eqn{nudef} with a DLA
estimate for the gluon multiplicity, corresponding to \eqn{TDLA} with
$R\to\theta_g$ and taking the coupling $\abar$ at the
scale $k_\perp=E\theta_g$. We see in Fig.~\ref{Fig:multi} that this DLA
approximation gives a reasonable description of the VLE multiplicity in a
jet (setting $\theta_g=R$).
For the case $\nu=\nu_{\rm DLA}$ we show no
variation band in Fig.~\ref{Fig:zglowpt} to avoid overlapping bands 
in an already complicated plot, but it is quite clear
what should be the effects of varying $\bar\varepsilon_{\rm g}$ and
$Q_s^2$.

In Fig.~\ref{Fig:zglowpt}, we show our MC results for a full shower
generated by a leading parton, gluon (left) or quark (right), with
$ p_{T0}=200$~GeV, together with our analytic approximation based on
\eqn{full} using the three different approximations for $\nu$.
In each case, the central curve corresponds to the average values
$\bar\varepsilon_{\text g}=5$~GeV for the subjet energy loss in \eqn{zgfull}  and $Q_s^2=\hat q
L=6$~GeV$^2$ for the saturation momentum squared
in~\eqref{pbroad}. The bands around these central curves correspond to
variations by 20\% of $\varepsilon_{\rm g}$. Note that varying $Q_s^2$
by 20\% has a smaller effect.
The unphysical case $\nu_{\rm min}=1$ is disfavoured by the comparison
with the MC results as it underestimates the peak associated with the
direct MIE contribution. The other two cases are at least
qualitatively consistent with the MC results.

It is also interesting to notice the dependence of the results upon the flavour
of the leading parton. The rise of the nuclear distribution at small $z_g$ appears
to be stronger  for the gluon-initiated jet than for the quark-initiated one and this
difference is rather well captured by our analytic approximations, where it is due
to a change in the number  $\nu$ of partonic sources for  a ``hard'' MIE: one has
indeed $\nu-1\propto C_R$. 
 At large $z_g$ on the other hand, the analytic approximation
appears to be less satisfactory for the quark-initiated jet --- most likely, because
it underestimates the energy loss by the subjets resulting from a hard VLE. As a matter
of facts, a similar difficulty occurs in the high-energy case, as can be seen by comparing
the quark-jet results for $p_{T0}=1$~TeV  in Figs.~\ref{Fig:zgvar} and \ref{Fig:QvsG} respectively.

The main conclusions that we can draw from in Fig.~\ref{Fig:zglowpt}
and from the overall discussion in this section is that $z_g$
distribution for ``low energy jets'' is a superposition of two main
effects: \texttt{(i)} incoherent energy loss for the subjets created
by a vacuum-like splitting; this controls the $z_g$ distribution at
moderate and large values of $z_g$, where it yields a nuclear
modification factor which slowly increase with $z_g$, and
\texttt{(ii)} sufficiently hard medium-induced emissions, with
$z\gtrsim \zc$, which leads to a significant growth of the $z_g$
distribution at small values $z_g\sim\zc$.
This behaviour is qualitatively reproduced by our simple analytic
calculations which shows, for example, that including multiple (VLE)
sources of MIEs is important. It is however more delicate to draw more
quantitative conclusions as several effects entering the calculation
would require a more involve treatment.

\section{$z_g$ distribution with realistic initial jet spectra}
\label{sec:data}

Even if studying monochromatic jets is helpful to understand
the dominant physical effects at play, any realistic measurement would
instead impose cuts on the $p_{T,{\rm jet}}\equiv p_T$ of the final jet.
For this we need the full $p_{T0}$ spectrum of the hard scattering.
Here, we follow our prescription from Section~\ref{sec:RAA} and use a
LO dijet spectrum where both final partons are showered using our Monte
Carlo. One can then cluster and analyse the resulting event.

In the case of the $z_g$ distribution, it is interesting to note that
we expect a competition between two effects.
On one side, due to the steeply-falling underlying $p_{T0}$ spectrum,
cutting on the jet $p_T$ tends to select jets which lose less energy
than average.
On the other side, we have seen from Fig.~\ref{Fig:elosszg} that jets
with $z_g>\zc$ and $\theta_g>\thetacut$ lose more energy than average.

Below, we first study the case of the $N_\text{jets}$-normalised $z_g$
distribution which is best suited to discuss the underlying physical
details highlighted in section~\ref{sec:zgmed}.
Our distributions are also qualitatively compared to a recent experimental 
analysis by ALICE~\cite{Acharya:2019djg}.
We then consider the self-normalised $z_g$ distribution which is more
easily compared to the CMS measurements~\cite{Sirunyan:2017bsd}.

\subsection{Phenomenology with the \njets-normalised $z_g$ }

\begin{figure}[t] 
  \centering
  \includegraphics[width=0.48\textwidth]{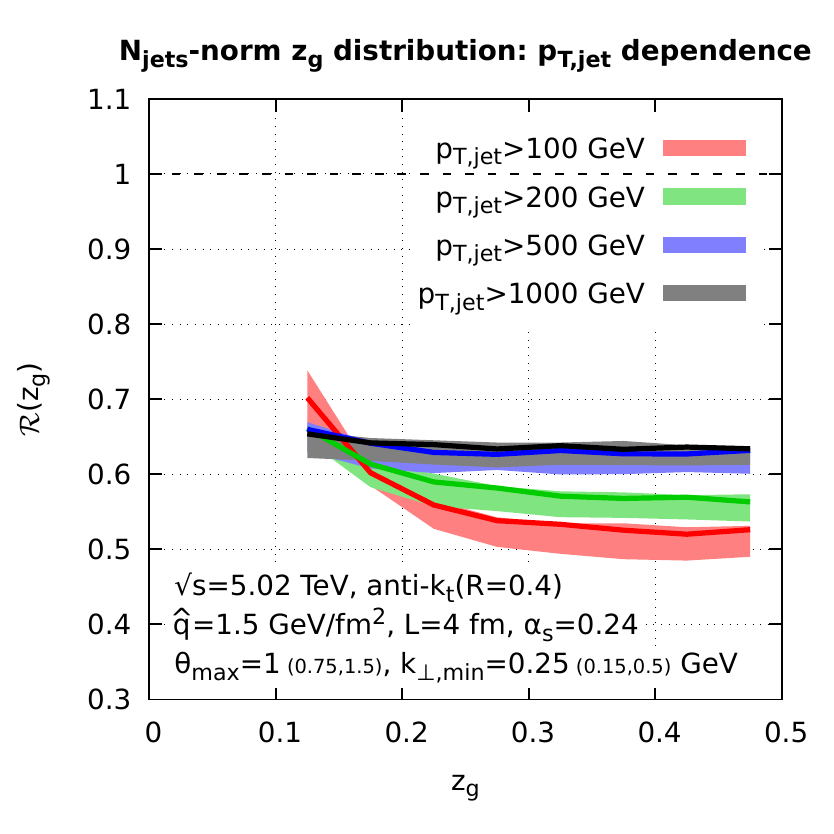}\hfill%
  \includegraphics[width=0.48\textwidth]{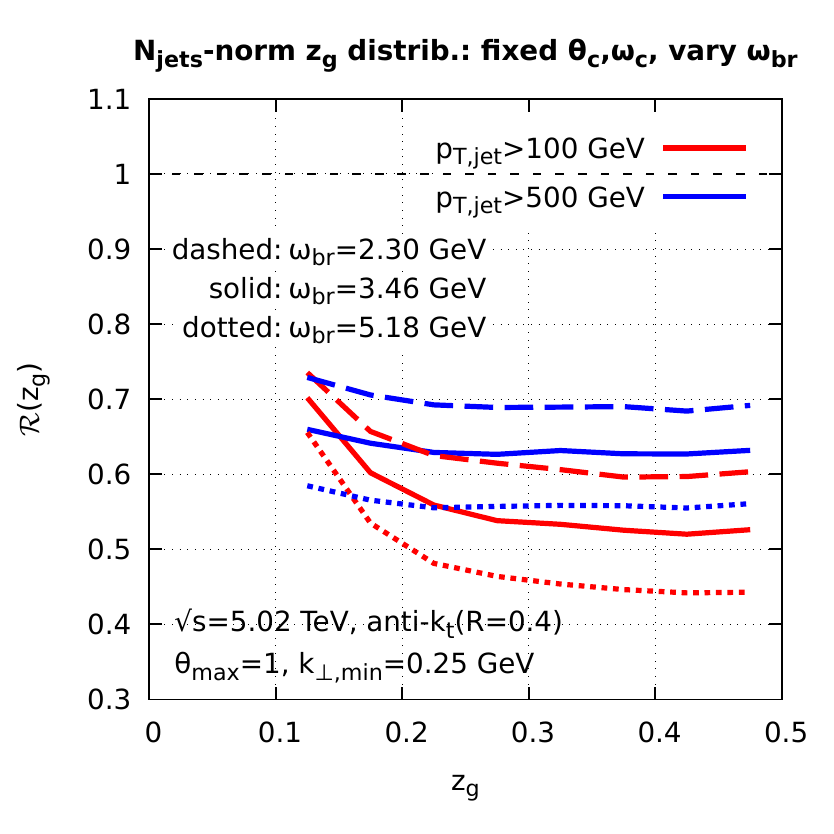}
\caption{\small Our full MC predictions for the medium/vacuum ratio $\mathcal{R}(z_g)$ of the 
$N_\text{jets}$-normalised $z_g$ distributions, including the convolution with the initial jet spectrum.   Left: the sensitivity of our results to changes in the kinematic cuts $\theta_{\rm max}$ and $k_{\perp,\text{min}}$. Right: the effect of varying $\obr$ (by $\pm 50\%$) at fixed values for $\oc$ and $\theta_c$.}
\label{Fig:zgvar1}
\end{figure}

\begin{figure}[t] 
  \centering
  \includegraphics[width=0.48\textwidth]{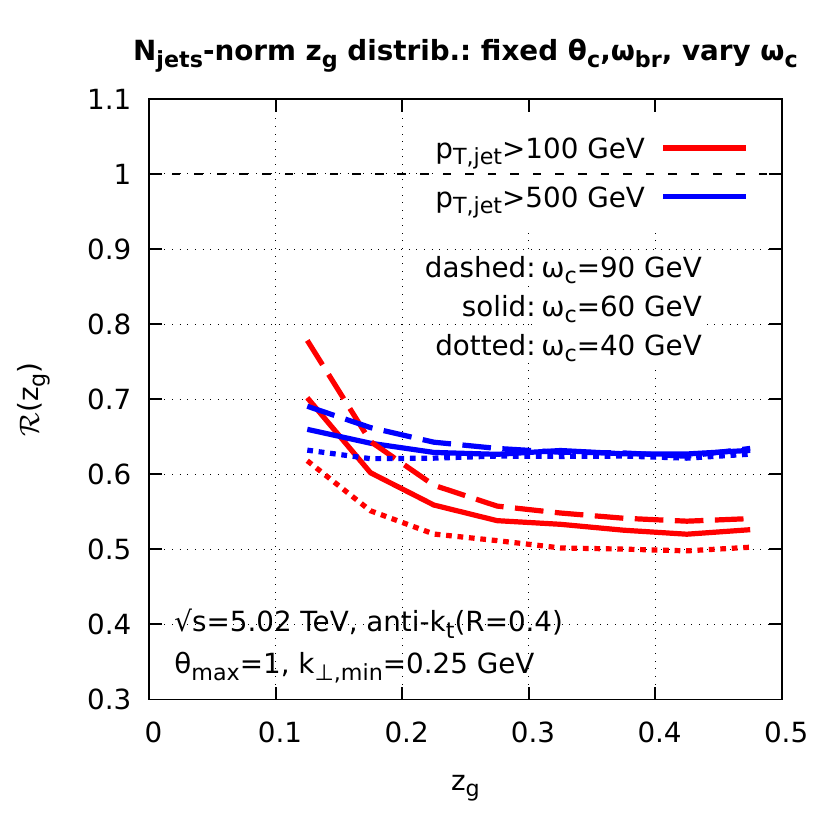}\hfill%
  \includegraphics[width=0.48\textwidth]{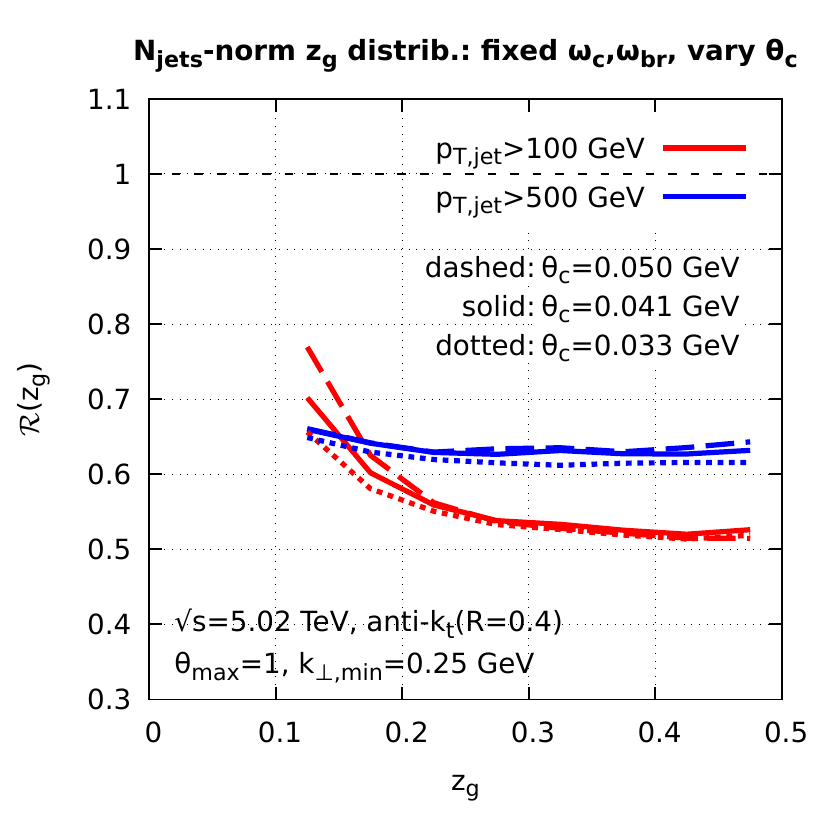}
  \caption{\small The effects of varying $\hat q$, $L$ and $\amed $
    keeping $\obr=3.46$~GeV fixed at its central value (cf.\ the right
    plot of Fig.~\ref{Fig:RAA}). Left: we vary $\oc$ by $\pm 50\%$ at
    fixed $\theta_c$. Right: we vary $\theta_c^2$ by $\pm 50\%$ at
    fixed $\oc$.}
\label{Fig:zgvar2}
\end{figure}

Our main results for the $N_\text{jets}$-normalised $z_g$ distribution
are plotted in Figs.~\ref{Fig:zgvar1}--\ref{Fig:zgvar2}. They are the
analogue of the results for $R_{AA}$ shown in
Figs.~\ref{Fig:RAA}--\ref{Fig:variab}: they highlight the
$p_T$-dependence of our predictions together as well as their
sensitivity to changes in the physical ($\hat q$, $L$ and $\amed$) and
unphysical ($\theta_{\rm max}$ and $k_{\perp,\text{min}}$)
parameters.
The various curves shown in these figures have been obtained by integrating
the $z_g$ distribution over all the values of $p_{T}$ above a lower cut-off 
$p_{T,{\rm min}}$ (explicitly shown for each curve). 
In practice, our choices for this cut-off are the same
as the values taken for $p_{T0}$ in Fig.~\ref{Fig:zgvar}.
Since the jet $p_T$ spectrum falls rapidly with $p_T$ and the jet
energy loss is relatively small compared to the jet $p_T$, it makes
sense to compare the respective results.

First of all, we see from Fig.~\ref{Fig:zgvar1}, left, that our
predictions are robust w.r.t.\ variations of the unphysical parameters in
our Monte Carlo.
Then, based on the analyses from Section~\ref{sec:zgmed}, we expect
the $z_g$ distribution to be mostly sensitive to changes in the
multiple-branching energy scale $\obr$ which controls both the energy
loss and the rate for SD to be triggered by a MIE.
When varying $\obr$ by 50\% around its central value, keeping $\oc$
and $\theta_c$ fixed, the $z_g$ distribution is indeed strongly
affected, see Fig.~\ref{Fig:zgvar1}, right.
The effects of changing either $\oc$ or $\theta_c$, at fixed $\obr$,
are much less pronounced as seen in Fig.~\ref{Fig:zgvar2}.
The residual variations observed when varying $\oc$ or $\theta_c$ can be
mainly attributed to variations in the phase-space for VLEs, which
affect the multiplicity of sources for MIEs and the energy loss (cf.\
\eqn{Pfullmed}). Besides, for the low-$p_T$ jets, a change in $\oc$
can have a sizeable effect on the phase-space for MIEs that are
accessible to SD (cf.\ Fig.~\ref{Fig:zgLund}). This is indeed seen in
Fig.~\ref{Fig:zgvar2} which shows a stronger dependence on $\oc$ than
on $\theta_c$, especially for the peak at low $p_T$ and small $z_g$.

Comparing now to the results for monochromatic jets shown in
Fig.~\ref{Fig:zgvar}, we observe important differences that can be
understood as follows. When studying the \njets-normalised ratio, the
deviation of
$\mathcal{R}(z_g)= f_{\text{med}}(z_g)/f_{\text{vac}}(z_g)$ from unity
is proportional to the ratio, $N_{\rm SDjets}/N_{\rm jets}$, between
the number of jets which passed the SD condition and the total number
of jets. This ratio is considerably smaller when using a realistic jet
spectrum, Fig.~\ref{Fig:zgvar1} (left), than for monochromatic jets,
Fig.~\ref{Fig:zgvar}.
This is explained by the fact that jets passing the SD condition lose
more energy than average jets and therefore have a more suppressed
production rate.
Moreover, among the jets which have passed SD with a given $z_g>\zc$,
the initial cross-section favours those where the subjets have lost
less energy,
leading to a flattening in the shape of the ratio $\mathcal{R}(z_g)$
at large $z_g$, in agreement with Fig.~\ref{Fig:zgvar1} left.
Finally, imposing a lower $p_T$ cut on jets introduces a bias towards quark
jets, which lose less energy than gluon jets. Since the former have a
smaller $\mathcal{R}(z_g)$ than the latter (cf.  Fig.~\ref{Fig:zgvar}), this further reduces
$\mathcal{R}(z_g)$ for jets.

Another interesting feature of Fig.~\ref{Fig:zgvar1} left is the fact
that the ratio $\mathcal{R}(z_g)$ is almost identical for
$p_T=500$~GeV and $p_T=1$~TeV. We believe that this purely
fortuitous. First, the normalisation factor
$N_{\rm SDjets}/N_{\rm jets}$ penalises the jets with
$p_T=1$~TeV more than those with $p_T=500$~GeV, thus reducing an
initially-small difference between the respective results in
Fig.~\ref{Fig:zgvar}.
Second, as $p_T$ increases so does the fraction of quark-initiated
jets, thus contributing to a reduction of $\mathcal{R}(z_g)$.

\begin{figure}
  \centering
  \begin{minipage}[t]{0.50\textwidth}
    \mbox{ }\\
    \includegraphics[width=\textwidth]{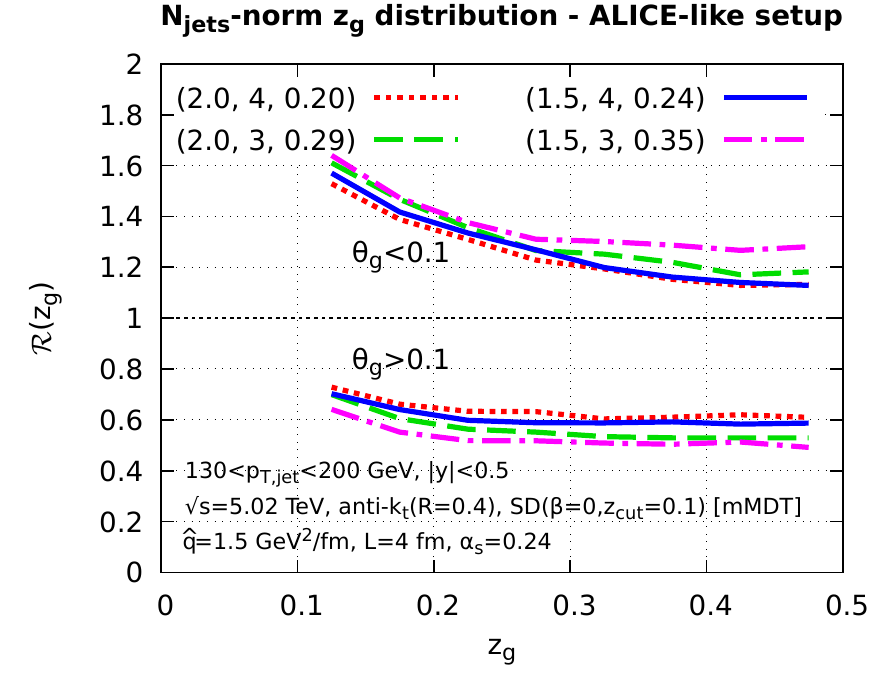}%
  \end{minipage}
  \hfill%
  \begin{minipage}[t]{0.43\textwidth}
    \caption{\small Predictions of our Monte Carlo generator for the $z_g$
      distribution obtained with a setup similar to the one used in the
      ALICE measurement of Ref.\cite{Acharya:2019djg}. We included the
      distribution obtained with either $\theta_g<0.1$ (bottom set of
      curves), or $\theta_g>0.1$ (top set of curves). In each case we
      show the result for different sets of medium parameters,
      $\hat{q}$, $L$ and $\alpha_s$ as indicated in the
      legend.}\label{Fig:ALICE}
  \end{minipage}
\end{figure}

At this point, it is interesting to compare our predictions with the
measurements by the ALICE collaboration~\cite{Acharya:2019djg} at the
LHC.
This is not immediately straightforward as the ALICE measurement is at
a different collider energy than what we have considered so far, uses
only charged tracks which are not accessible in our parton-level
shower, and is not unfolded for the detector effects  and residual
background fluctuations.
For simplicity, we keep the collider energy at 5.02~TeV. Since the
charged and full transverse momenta of jets are roughly proportional
to one another, we scale
the acceptance region for the jet $p_T$ from $[80,120]$~GeV to
$[130,200]$~GeV and work with all the particles.
The discussion below should therefore, at best, be considered as
qualitative.

Our findings are presented in Fig.~\ref{Fig:ALICE} where, following
Ref.~\cite{Acharya:2019djg} (see the first and third plots in Fig.~3),
we have considered both the case $\theta_g > 0.1$ and the case
$\theta_g < 0.1$.
Our predictions are shown for a range of medium parameters (see also
Table~\ref{tab:parameters} of Fig.~\ref{Fig:zgnorm}).
In all cases, our results are qualitatively similar to those of the
experimental analysis: the ratio $\mathcal{R}(z_g)$ is decreasing with
$z_g$, it shows nuclear suppression ($\mathcal{R}(z_g)<1$) for the
large-angle case $\theta_g > 0.1$ and nuclear enhancement
($\mathcal{R}(z_g) > 1$) for the small-angle case $\theta_g < 0.1$.
Within our framework, the enhancement observed for $\theta_g < 0.1$
and the rise at small $z_g$ are both associated with medium-induced
emissions\footnote{Since with our parameters the minimal angle
  for MIEs is $\theta_c\approx 0.04$, MIEs can pass the SD condition
  even for $\theta_g < 0.1$.} being captured by SD.
The suppression visible for $\theta_g > 0.1$ is a
consequence of incoherent energy loss as seen in Sect.~\ref{sec:zgmed}.

\subsection{Self-normalised $z_g$ distribution and CMS data}

We want to compare the predictions of our Monte Carlo generator to the
measurement of the self-normalised $z_g$ distribution by the CMS
collaboration in Ref.~\cite{Sirunyan:2017bsd}.
This comparison should however be taken with care since the CMS
results are not unfolded for detector (and residual Underlying Event)
effects and are instead presented under the form of ``PbPb/smeared
pp'' ratios. 
Without a proper dedicated study, it is delicate to assess the precise
effects of this smearing on $\mathcal{R}^\text{(norm)}(z_g)$.

\begin{figure}[t] 
  \centering
  \includegraphics[width=0.48\textwidth]{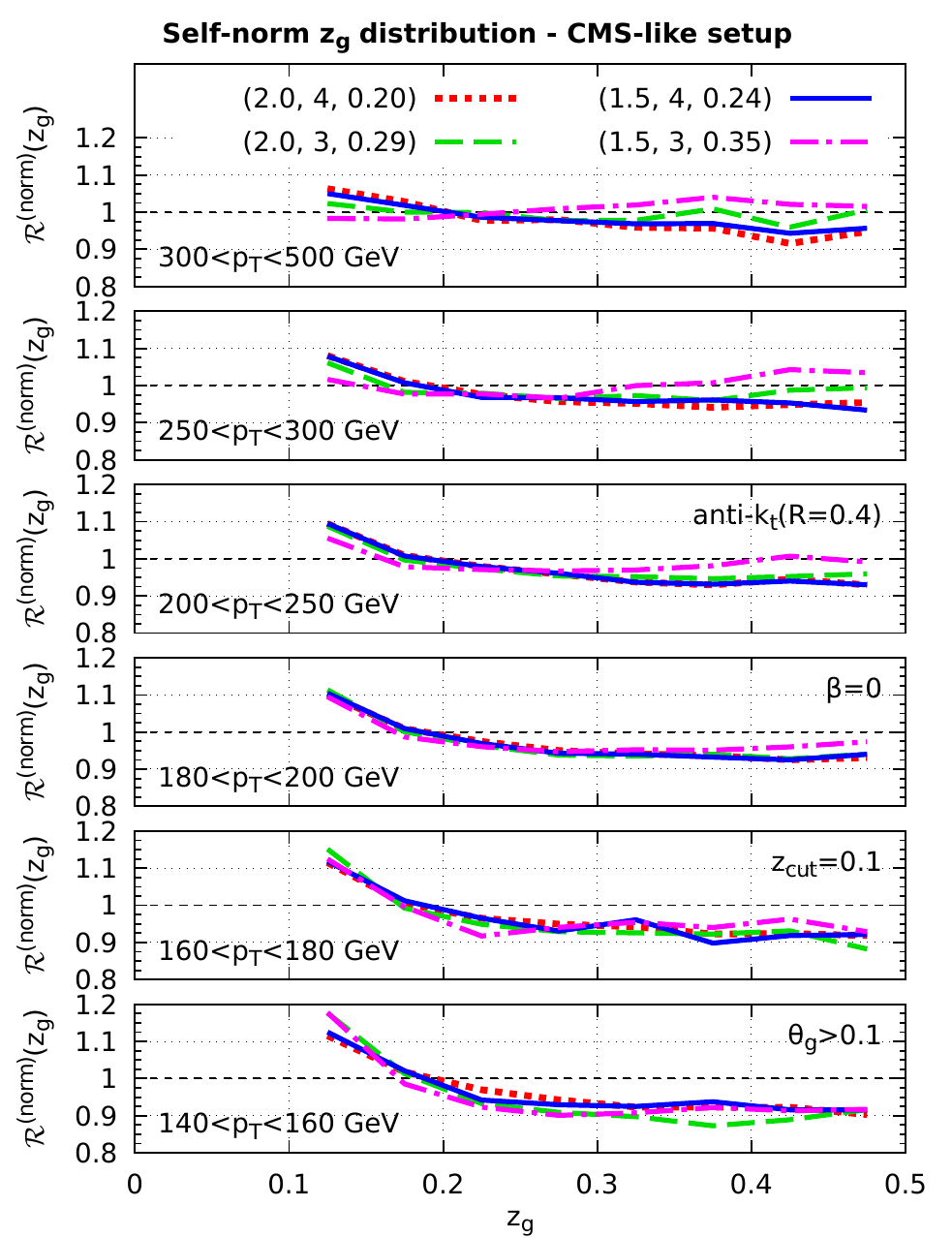}
 \caption{\small  Left: our MC results for jet $R_{AA}$ and for 4 sets of
 medium parameters which give quasi-identical predictions are compared to
 the ATLAS data \cite{Aaboud:2018twu} (black dots with error bars). Right:
 the MC predictions for the medium/vacuum ratio $\mathcal{R}^\text{(norm)}(z_g)$ of the 
self-normalised $z_g$ distributions are presented in bins of $p_T$ for
the same 4 sets of medium parameters as in the left figure.}
 \label{Fig:zgnorm}
\end{figure}

Our findings are shown in Fig.~\ref{Fig:zgnorm}(right).
%
%
In this plot, we show the predictions for the $z_g$
nuclear modification factor for the selection of 4 sets of medium parameters discussed in Section \ref{subsub:param-degeneracy}, $\mathcal{R}^\text{(norm)}(z_g)$, using
the same bins and cuts as in the CMS analysis.

%
%
Since the interplay between the 3 scales $\obr$,  $\oc$ and $\theta_c$
is different for $R_{AA}$ and $\mathcal{R}^\text{(norm)}(z_g)$, our 4
sets of parameters predict different behaviours for the latter.
However, both observables are predominantly controlled by the energy loss
of the jet, so the spread in $\mathcal{R}^\text{(norm)}(z_g)$ remains
limited.
Some differences are nonetheless observable, in particular for the two
bins with the largest $p_T$. The predictions obtained with a larger
$\obr$ --- i.e.\ larger single-parton energy loss but smaller
phase-space for VLEs --- show a pattern dominated by energy loss,
similar to what was seen in Sect.~\ref{sec:zgmed} for high-$p_T$
jets. Conversely, the predictions obtained with a smaller $\obr$ ---
i.e.\ smaller single-parton energy loss but larger phase-space for
VLEs --- show an enhancement of the small-$z_g$ peak associated to
MIEs.

If we compare these results with the CMS measurements (see e.g.\
Fig.~4 of Ref.~\cite{Sirunyan:2017bsd}), we see that the two agree
within the error bars for both the pattern and the magnitude of the
deviation from one.
In particular, the CMS data too indicate that
$\mathcal{R}^\text{(norm)}$ decreases quasi-monotonously with $z_g$ at
low $p_T$ and become flatter and flatter, approaching unity, when
increasing $p_T$.
This supports our main picture where the nuclear effects on the $z_g$
distribution are a combination of incoherent energy loss affecting a
vacuum-like splitting and a small-$z_g$ peak associated with the SD
condition being triggered by a MIE.
With increasing $p_T$ the first mechanism dominates over the over,
yielding a flatter distribution, in agreement with the CMS data. 
That being said, the current experimental uncertainty does not allow one
to distinguish between different sets of medium parameters.

\section{Other substructure observables}

\begin{figure}[t] 
  \centering
  \includegraphics[page=1,width=0.48\textwidth]{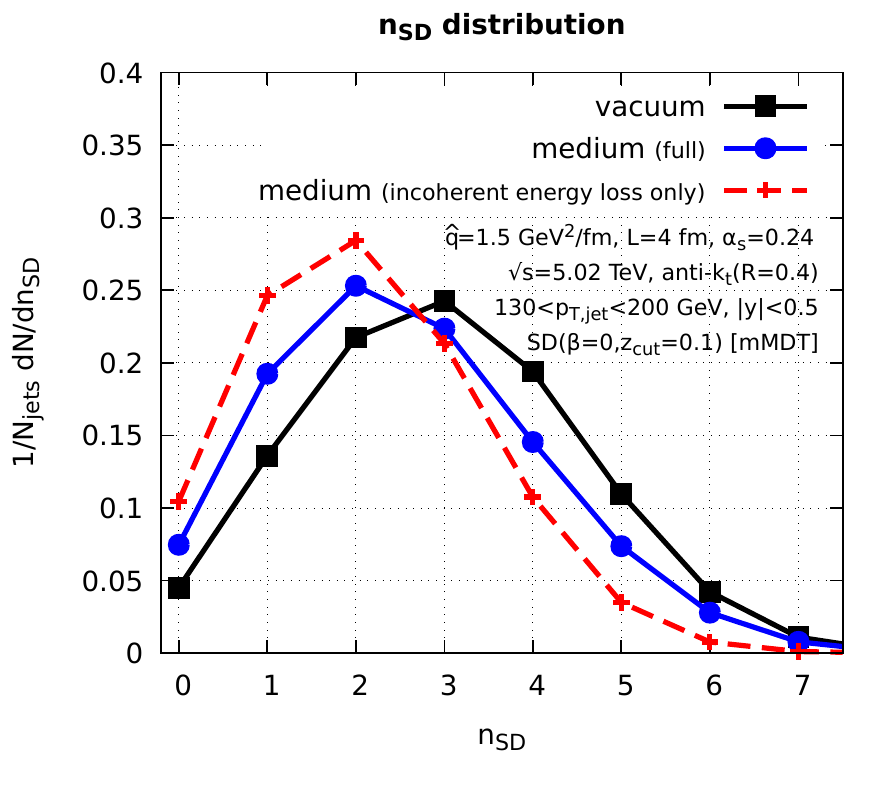}\hfill%
  \includegraphics[page=2,width=0.48\textwidth]{./nsd-ALICE.pdf}
 \caption{\small $n_\text{SD}$ distributions emerging from our Monte Carlo simulations. Left: the 
 distributions themselves, for the vacuum shower, for the full in-medium parton shower, and also
 for the case where the MIEs contribute only to the energy loss (but not directly to SD). Right: the
 medium/vacuum ratios.}
 \label{Fig:nsd}
\end{figure}
\begin{figure}[t] 
  \centering
  \includegraphics[page=1,width=0.48\textwidth]{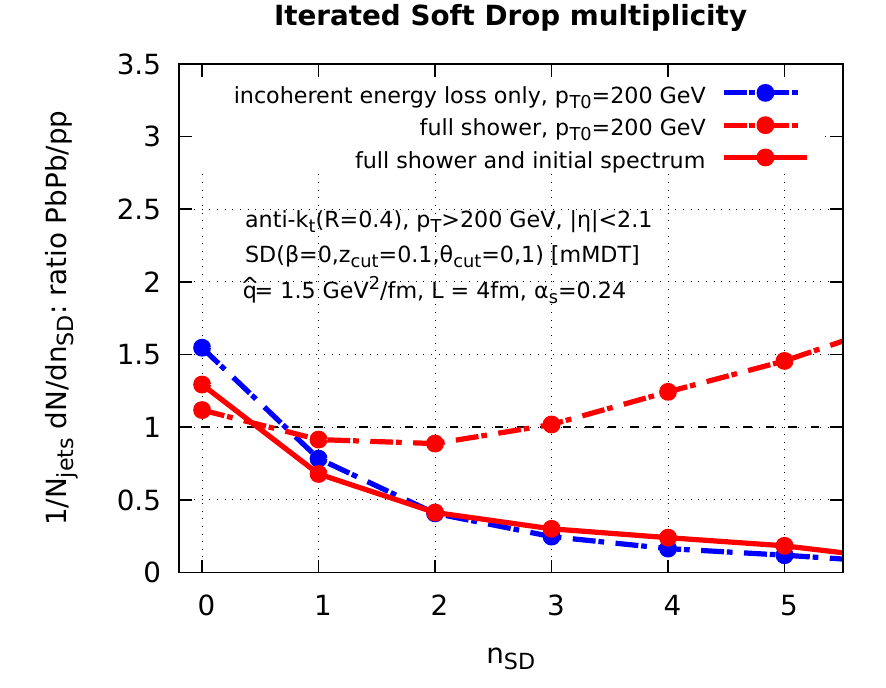}
  \caption{\label{zgnsd} \small{Nuclear modification factor $n_{\textrm{\tiny SD}}$ distribution calculated with our MC. For the dashed curves, the calculation is done for a gluon with $p_{T0}=200$ GeV. For the plain curves, we use the Born-level spectrum in $p_{T0}$. The red curves correspond to the full parton shower. For the blue curves, all the MIEs are artificially sent at very large angles to remove the effect coming from intrajet MIEs.}}

\end{figure}

Our final section discusses two substructure observables related to the $z_g$ distribution.

\subsection{Iterated SD multiplicity.}
\label{sub:nsd-pheno}
The first observable we consider is the Iterated SD
multiplicity~\cite{Frye:2017yrw}, $n_\text{SD}$, which has also been
measured on track-jets by the ALICE
collaboration~\cite{Acharya:2019djg}\footnote{Our comparison to this
  measurement is subject to the same caveats that for the $z_g$
  distribution in the same paper.}.
%
This observable has been defined in Chapter \ref{chapter:jet}, Section \ref{subsub:nSD}.

Our results for the $n_\text{SD}$ distribution are presented in
Fig.~\ref{Fig:nsd} and show the same trend as the ALICE measurements
(Fig.~4 of Ref.~\cite{Acharya:2019djg}).
In particular, the $n_{\rm SD}$ distribution is shifted to smaller
values for jets created in PbPb collisions compared to pp collisions.
This might seem puzzling at first sight since in the
low-$p_{T,\text{jet}}$ range probed by the measurement, one could
naively expect an enhancement of $n_{\rm SD}$ due to the additional MIEs
passing the SD condition.
However, we believe that the dominant mechanism at play is the
incoherent energy loss which, as discussed in Sect.~\ref{sec:data} and Sect.~\ref{sec:high},
results {\tt (i)} in a statistical bias towards jets with small $n_{\rm SD}$ values as they lose less energy, scenario favored by the steeply-falling initial spectrum, {\tt (ii)} in an effective $z_g$ fraction smaller than the actual
momentum fraction $z$ at the splitting. These effects lower the number
of measured hard splittings.
To support this argument, we have run a variant of our MC simulations
where all the partons created via MIEs are moved outside the jet and
hence only contribute to the energy loss.
The corresponding results, shown as crosses in Fig.~\ref{Fig:nsd}
demonstrate as expected an even stronger reduction in the average
value of $n_{\rm SD}$, which is only partially compensated by MIEs
captured by the Iterated SD procedure.

\paragraph{Analytic estimate.} It is always enlightening to see how these results compare with simple analytic formulas 
based on the same ideas as those developed in the investigation of the $z_g$ distribution.  For the vacuum baseline, the leading-logarithmic estimate is given in Chapter \ref{chapter:jet}, Section \ref{subsub:nSD}. When a monochromatic hard spectrum is considered, we identify essentially three physical mechanisms: the veto of vacuum-like emissions, the shift of the splitting fraction caused by large angle energy loss and the existence of relatively hard medium-induced emissions captured by ISD. We believe that any observable relying on the primary Lund plane can be analysed in the same way as the following brief analysis of the $n_{\rm SD}$ distribution\footnote{In the next Chapter, we will disentangle the nuclear modification factors for the large $x$ part of the fragmentation function in this way.}.
\begin{enumerate}
 \item \textbf{Vetoed region.} The vetoed region included in {\tt JetMed} has an impact on the $n_{\rm SD}$ distribution at DLA e.g. by reducing the area $\rho_{i,\LL}$ in the parameter of the Poisson law. However, as shown Fig.~\ref{Fig:zgLund}, the phase space area probed by SD does not overlap with this vetoed region for our choice of SD parameters. Consequently, the vetoed region can be neglected at leading logarithmic accuracy.
 
 \item \textbf{Incoherent large angle energy loss.} Because of the energy loss shift as given by \eqref{zvszg}, the logarithmic area probed by SD in phase space is reduced, hence the medium/vacuum ratio for the $\nSD$ distribution decreases with $\nSD$. This is shown in the blue curve, Fig.~\ref{zgnsd}. 
 
 \item \textbf{Intrajet semi-hard MIEs.} The semi-hard MIEs with emission angles $\theta<R$ remain inside the jet and can trigger the ISD condition as for the SD $z_g$ distribution. To estimate the order of magnitude of this effect, we rely again on the C/A declustering of jets from primary medium-induced emissions established in Section \ref{subsub:zg-mie}. For a jet evolving via primary MIEs only, $\nSD$ is also Poisson distributed with average value $\rho_{i,\med}$ and 
\begin{align}
\label{poissonmie}
 \rho_{i,\med}(\theta_{\rm cut})&=\int_0^{\omega_c/p_{T0}}\frac{\dif z}{z^{3/2}}\int_{\theta_{\textrm{cut}}}^R\dif\th\frac{\dif^2 \mathcal{P}_{i,\med}}{\dif z\dif \th}\Theta(z-\zc(\theta/R)^{\beta})\\
 &\sim\frac{2\alpha_sC_i}{\pi}\sqrt{\frac{\omega_c}{2p_{T0}}}\int_0^{\omega_c/p_{T0}}\frac{\dif z}{z^{3/2}}\int_{\theta_{\textrm{cut}}}^R\dif\theta\delta\Big(\theta-\frac{Q_s}{zp_{T0}}\Big)\Theta(z-\zc(\theta/R)^{\beta})
\end{align}
where we approximate the angular distribution \eqref{pbroad} with a delta centered around $k_\perp=Q_s\equiv\sqrt{\qhat L}$. 
If $\zc p_{T0}\theta_{\rm cut}\lesssim Q_s$, one finds $\rho_{i,\med}\propto(\zc p_{T0}/\omega_c)^{-1/2}$ which is non-negligible compared to $\rho_{i,\LL}$ when $\zc p_{T0}\ll \omega_c$. In this case, $\nSD$ follows again a Poisson distribution with average value $\rho_{i,\LL}+\rho_{i,\med}$. Hence the medium/vacuum ratio increases with $\nSD$ as in the red dashed curve in Fig.~\ref{zgnsd}-right. 
\end{enumerate}
\begin{figure}[t] 
  \centering
  \includegraphics[page=1,width=0.48\textwidth]{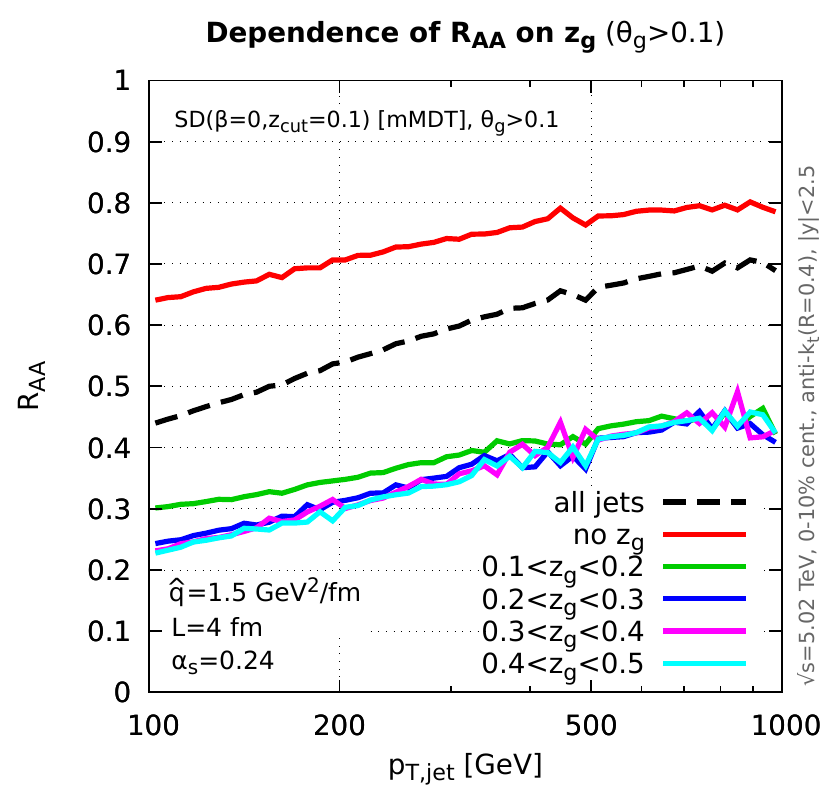}\hfill%
  \includegraphics[page=2,width=0.48\textwidth]{./RAA-v-zg}
 \caption{\small  Our MC results for the jet  $R_{AA}$ as a function of $p_{T,{\rm jet}}$ are shown
 in bins of $z_g$ (left figure) and in bins of $\theta_g$ (right figure). The inclusive (all-$z_g$, respectively
 all-$\theta_g$) results are shown with dashed lines.}
 \label{Fig:zgRAA} 
\end{figure}
As expected, there is a strong effect of the initial cross-section, as shown with the red plain curve of Fig.~\ref{zgnsd}. Jets with large $\nSD$ are highly suppressed and the enhancement seen at large $\nSD$ in the monochromatic case, due to additional MIEs, is no longer visible. Such a compensation implies that one must be cautious when interpreting a measurement of $\nSD$ with $\beta=0$ \cite{Acharya:2019djg}.

\subsection{Correlation between $R_{AA}$ and $z_g$}
\label{sub:correl-raa-zg}
Given that both $R_{AA}$  and the $z_g$ distribution are primarily
controlled by the jet energy loss, it is interesting to study the
correlation between these 2 variables~\cite{Andrews:2018jcm} (similarly to
Fig.~\ref{Fig:elosszg} for the energy loss of monochromatic jets). 
To
that aim,  we show in Fig.~\ref{Fig:zgRAA} the ratio $R_{AA}$ as a
function of $p_T\equiv p_{T,{\rm jet}}$ for different bins in $z_g$
(imposing $\theta_g>0.1$) (left plot) and for different bins in
$\theta_g$ (right plot). For reference, the inclusive $R_{AA}$ ratio is shown by
the ``all jets'' curve.
The curve labelled as ``no $z_g$'' in the left plot includes both the
events which did not pass the SD criteria and the events which failed
the $\theta_g>0.1$ constraint.
Correspondingly, the curve labelled ``$\theta_g<0.03$'' in the right
plot includes both the events with a genuine splitting passing the SD
condition with $\theta_g<0.03$ and the events which did not pass the
SD condition.

The remarkable feature in both plots is the striking difference
between the events which passed SD and those which did not.
As explained when we discussed Fig.~\ref{Fig:elosszg}, this difference
reflects the fact that, on average, two-prong jets lose more energy
than single-prong ones.
These results also reveal the role played by colour (de)coherence and
the emergence of a critical angle $\theta_c$. With our choice of
parameters, $\theta_c\simeq 0.04$ corresponds to the region in
$\theta_g$ where $R_{AA}$ changes significantly. For example, the
curve corresponding to $\theta_g<0.03 < \theta_c$ receives almost
exclusively contributions from single-prong jets --- even more so
than the ``no $z_g$ curve in the left plot --- and thus shows a nuclear
factor $R_{AA}$ close to unity.
This suggests that measuring $R_{AA}$ in bins of $\theta_g$ can be
interesting to better characterise the propagation of jets in the
quark-gluon plasma.



%% file: chapter8.tex
\chapter{Jet fragmentation functions in nucleus-nucleus collisions}
\chaptermark{Jet fragmentation in AA collision}
\label{chapter:FF}

The last chapter of this thesis deals with the jet fragmentation function in nucleus-nucleus collision. At first sight, the fragmentation function looks like an ideal observable to study the jet structure
in terms of parton showers and its modifications by the interactions with the medium. One should however be cautious as the jet fragmentation function is
not a well-defined IRC safe quantity in pQCD as explained at length in Chapter \ref{chapter:jet}. This means that its theoretical
predictions are strongly sensitive to non-perturbative (confinement) physics like the modelling of the
hadronisation mechanism. Another potential drawback of the fragmentation function, already recognised in the literature \cite{Spousta:2015fca,Casalderrey-Solana:2018wrw}, is that the nuclear enhancement seen in the LHC data at $x\gtrsim0.5$ is not necessarily an evidence for new physics in the jet fragmentation at large $x$, but merely a consequence of the overall energy loss by the jet together with the bias introduced by the initial spectrum for jet production via hard
(nucleon-nucleon) scatterings.

After a brief presentation of the {\tt JetMed} results for the IRC unsafe fragmentation in Section~\ref{sec:MCmain}, we disentangle the several medium effects in the large $x$ domain of the fragmentation function in Section~\ref{sec:x=1nuc} and in the small $x$ region in Section~\ref{sec:smallx-MC-results}. The last Section~\ref{sec:frag-subjets} is dedicated to the study of the nuclear modification of the fragmentation function into subjets, which is IRC safe and defined in a similar way as in Section~\ref{subsub:frag-ISD}. We show that it is sensitive to medium effects while under better perturbative control.

\section{Monte Carlo results}
\label{sec:MCmain}
We start by presenting our Monte Carlo results for the fragmentation function
and the associated nuclear modification factor.
We want to pay a special attention to their dependence on the two
``unphysical'' parameters of the Monte Carlo, $\theta_{\rm max}$ and
$k_{\perp,\text{min}}$, and to the 3 ``physical'' parameters,
$\hat q$, $L$ and $\amed$.
We recall that the dependence on the former can be viewed as an uncertainty in our
underlying parton-level theoretical description and a large
uncertainty would signal a strong dependence of the
observable on  non-perturbative effects such as hadronisation.
Conversely, the dependence on the ``physical'' medium parameters sheds
light on the role and importance of the medium effects at play.

\subsection{General set-up}
\label{sec:defin}

In order to describe $pp$ and PbPb collisions at the LHC, we consider
jets with an initial spectrum given by a $pp$ collision  with center-of-mass energy $\sqrt{s}=5.02$ TeV computed at
leading-order, i.e.\ with Born-level $2\to 2$ partonic hard
scatterings. This is similar to our study of the $R_{AA}$ ratio in Chapter \ref{chapter:RAA}, Section \ref{sec:RAA}.
A key property of this initial parton (or dijet) spectrum is that it
is steeply falling with the partons'
transverse momentum $p_{T0}$:  $\dif N^{\textrm{hard}}/{\dif p_{T0}}\propto 1/p_{T0}^n$
with $n\gtrsim 5$. For each event, both final partons are showered using our Monte Carlo. 
Jets are reconstructed using the anti-$k_\perp$ algorithm~\cite{Cacciari:2008gp} as implemented in {\tt FastJet}
v3.3.2~\cite{Cacciari:2011ma}.  The final jets are characterised by their transverse momentum 
$p_{T,\textrm{jet}}$, which is generally different from the initial momentum $p_{T0}$,
in particular for jets in PbPb collisions which suffer energy loss.
The $pp$ baseline is obtained by using the vacuum limit of our Monte Carlo.

We denote the final jet spectrum by
$\dif N_{\textrm{jets}}/{\dif p_{T,\textrm{jet}}}$ and use the upper
scripts ``med'' and ``vac'' to distinguish between jets in the medium
(PbPb collisions) and jets in the vacuum ($pp$ collisions),
respectively.  The jets can be initiated by either a quark or a
gluon. In practice, one often considers the jet yield integrated over
an interval in $p_{T,\textrm{jet}}$, that is,
\beq\label{pTrange}
N_{\textrm{jets}} (p_{T,{\rm min}}, p_{T,{\rm max}})\,=
\int_{p_{T,{\rm min}}}^{p_{T,{\rm max}}}\dif p_{T,\textrm{jet}}\,
\frac{\dif N_{\textrm{jets}}}{\dif p_{T,\textrm{jet}}}\,.
\eeq
For a given jet with transverse momentum $p_{T,\textrm{jet}}$, 
we characterise its fragmentation in terms of the longitudinal
momentum fraction
\begin{equation}\label{eq:def-x}
  x\equiv \frac{p_T\cos(\Delta R)}{p_{T,\textrm{jet}}},
\end{equation}
where $p_T$ is the transverse momentum of a constituent of the jet and
$\Delta R =\sqrt{(\Delta y)^2+(\Delta\phi)^2}$, with $\Delta y$ and
$\Delta\phi$ the differences between the jet axis and the particle
direction in rapidity and azimuth. Note that since our Monte Carlo
does not include hadronisation, the jet constituents are partons.

The jet fragmentation function $\mathcal{D}(x)$ 
and its nuclear modification factor $\mathcal{R}(x)$ are defined as in Chapter \ref{chapter:jet}:
\begin{equation}
\label{frag-def}
 \mathcal{D}(x)=\frac{1}{N_{\textrm{jets}}}\frac{\dif N}{\dif x}\textrm{ , }\qquad 
 \mathcal{R}(x)=\frac{\mathcal{D}^{\textrm{med}}(x)}{\mathcal{D}^{\textrm{vac}}(x)}\,,
\end{equation}
with $N_{\textrm{jets}}$ the number of jets (in the considered
$p_{T,\textrm{jet}}$ range) and ${\dif N}/{\dif x}$ the number of
jet constituents with a given momentum fraction $x$. 

For later conceptual studies, we shall also consider ``monochromatic
jets'' produced by a well identified parton, quark or gluon, with a fixed initial transverse
momentum $p_{T0}$. In such a case, we denote the fragmentation function by
$D_i(x|p_{T0})$, where $i\in\{q,g\}$ refers to the flavour of the leading parton.
The corresponding medium/vacuum ratio is defined as 
${\cal {R}}_i(x|p_{T0})\equiv D^{\textrm{med}}_i(x|p_{T0})/D^{\textrm{vac}}_i(x|p_{T0})$. 

\subsection{Variability with respect to the unphysical cut-offs}

\begin{figure}[t] 
  \centering
  \begin{subfigure}[t]{0.48\textwidth}
    \includegraphics[page=1,width=\textwidth]{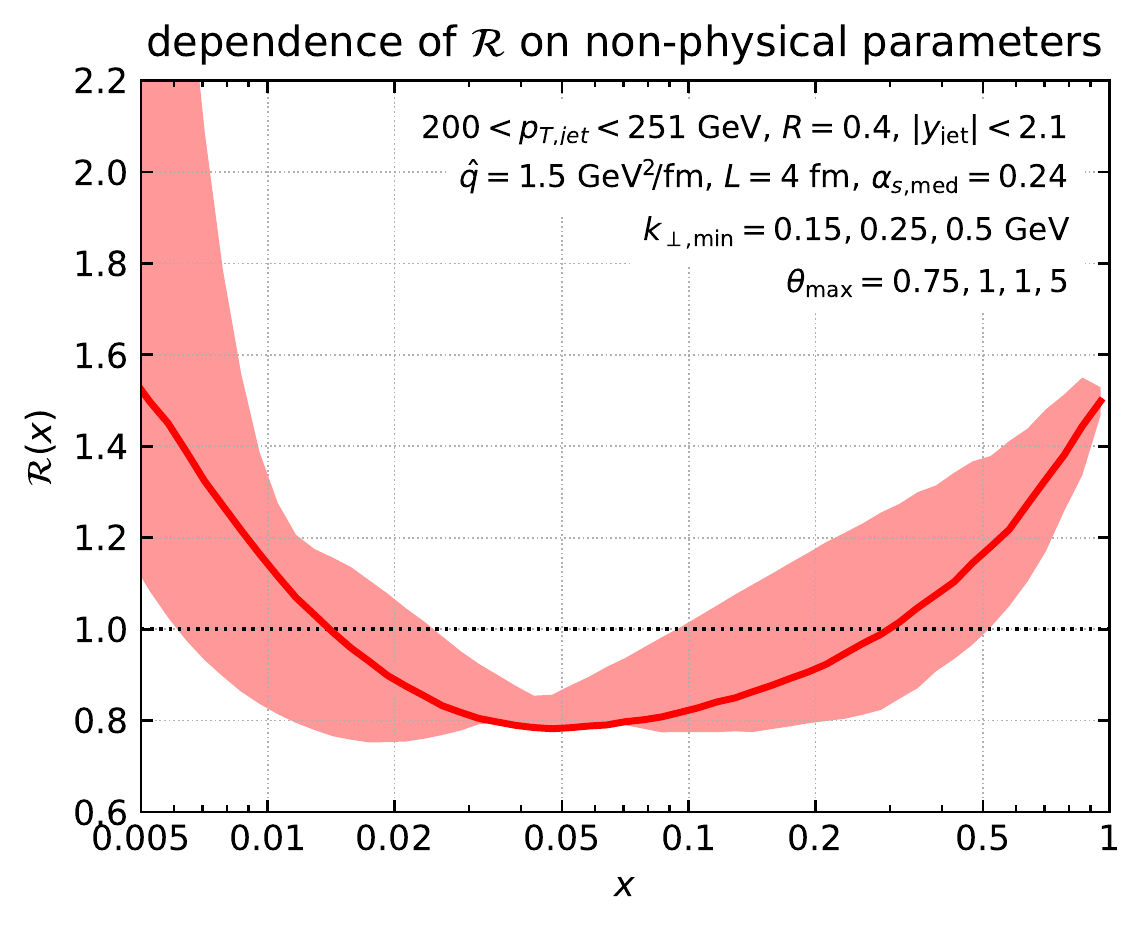}
    \caption{\small Variations in  $\theta_{\rm max}$ and $k_{\perp,\text{min}}$.}\label{Fig:MCunphys} 
  \end{subfigure}
  \hfill
  \begin{subfigure}[t]{0.48\textwidth}
    \includegraphics[page=1,width=\textwidth]{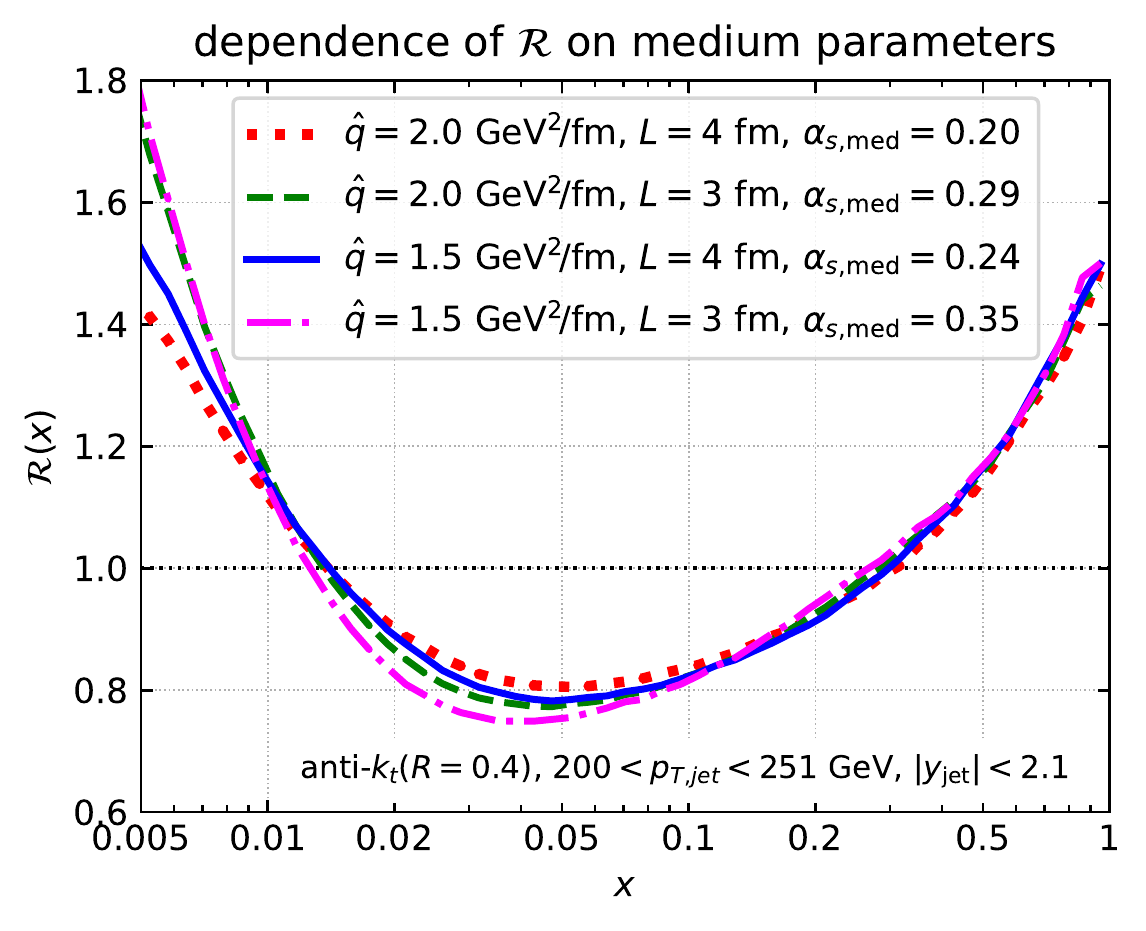}
    \caption{\small Variations in  $\hat q$, $L$ and $\amed$.}
    \label{Fig:MC-pheno}
  \end{subfigure}
  \caption{\small The variability of our MC results for the ratio $\mathcal{R}(x)$ w.r.t. changes
  in the ``unphysical'' (left) and ``physical'' (right) parameters. The 4 sets of values for
  the ``physical'' parameters are correlated in that they provide similarly good descriptions
  of the LHC data~\cite{Aaboud:2018twu} for the ``standard'' nuclear  modification factor for jets $R_{AA}$ (see the discussion in Chapter \ref{chapter:RAA}).    }
  \label{fig-R-param-dependence}
\end{figure}

Fig.~\ref{Fig:MCunphys} displays the sensitivity of our MC  results for  $\mathcal{R}(x)$  
to variations of the ``unphysical'' parameters 
around their central values $\theta_{\rm max}=1$ and $k_{\perp,\text{min}}=0.25$~GeV,
for fixed values of $\hat q$, $L$ and $\amed$.

The first observation from Fig.~\ref{Fig:MCunphys} is reassuring: the
distribution shows a strong enhancement both at small $x$ and at large
$x$, with a nuclear suppression at intermediate values of $x$. This is
in qualitative agreement with experimental measurements (see e.g.~\cite{Aaboud:2018hpb}).

However, the variations w.r.t. the unphysical parameters appear to be
very large. We have checked that they were strongly dominated by
variations in $k_{\perp,\text{min}}$.
This should not come as a surprise since the fragmentation function,
measured directly on individual constituents, is not an
infrared-and-collinear (IRC) safe observable.
The sizeable variations in the small-$x$ region directly come from the
variations of the available phase-space for radiating soft gluons when
varying $k_{\perp,\text{min}}$.
The large variations in the radiation of soft particles directly
affect the spectrum of hard particles in the jet, hence the large
uncertainty in the large-$x$ region.
Only a proper description of hadronisation (including varying
hadronisation parameters) would (hopefully) reduce this uncertainty.
This should be kept in mind when studying the dependence of our
results on the medium parameters and when comparing our MC results in
this work with actual experimental data.

\subsection{Variability with respect to the medium parameters}

We now fix the unphysical parameters to their central value and study
how $\mathcal{R}(x)$ depends on the medium parameters $\hat{q}$, $L$, and
$\alpha_{s,\textrm{med}}$.
We first consider 4 different sets of values, given in
Table~\ref{tab:parameters} together with the angular and energy scales $\theta_c$,
$\om_c$ and $\ombr$ characterising the medium-induced radiation.

The plot in Fig.~\ref{Fig:MC-pheno} shows our new results
for $\mathcal{R}(x)$ for the 4 sets of values for the
physical parameters. 
For large values of $x$, $x\gtrsim 0.1$, the small variations in
$\ombr$ (see Table~\ref{tab:parameters}) are compensated by relatively
large variations of $\om_c$ and $\theta_c$.
This is similar to what happens for $R_{AA}$, as discussed at length
in Chapter \ref{chapter:RAA}.
This suggests that for largish $x\gtrsim 0.1$, the nuclear effects on
jet fragmentation and on the inclusive jet production are strongly
correlated and in particular that they are both controlled by the jet
energy loss.  Such a correlation has been already pointed out in the
literature \cite{Spousta:2015fca,Casalderrey-Solana:2018wrw} and used to
provide a simple and largely model-independent argument for explaining
the enhancement  in the ratio $\mathcal{R}(x)$ at $x\gtrsim 0.5$,
as observed both in the LHC data~\cite{Aaboud:2018hpb} and in our MC results in
Fig.~\ref{Fig:MC-pheno}.  This argument will be completed in Sect.~\ref{sec:bias}.

Turning to smaller $x$ values, $x\le 0.01$, the situation becomes
different. There is a clear lift of degeneracy between the 4 sets of
values, with two of them --- corresponding to the smallest medium size
$L=3$~fm, but larger values for $\amed$ --- yielding results that are
significantly larger than those predicted by the two other sets (with
$L=4$).  In Section \ref{sec:smallx-MC-results}, we provide physical explanations for these
trends.

\subsection{Dependence on the jet $p_T$}

\begin{figure}[t] 
  \centering
  \includegraphics[page=1,width=0.48\textwidth]{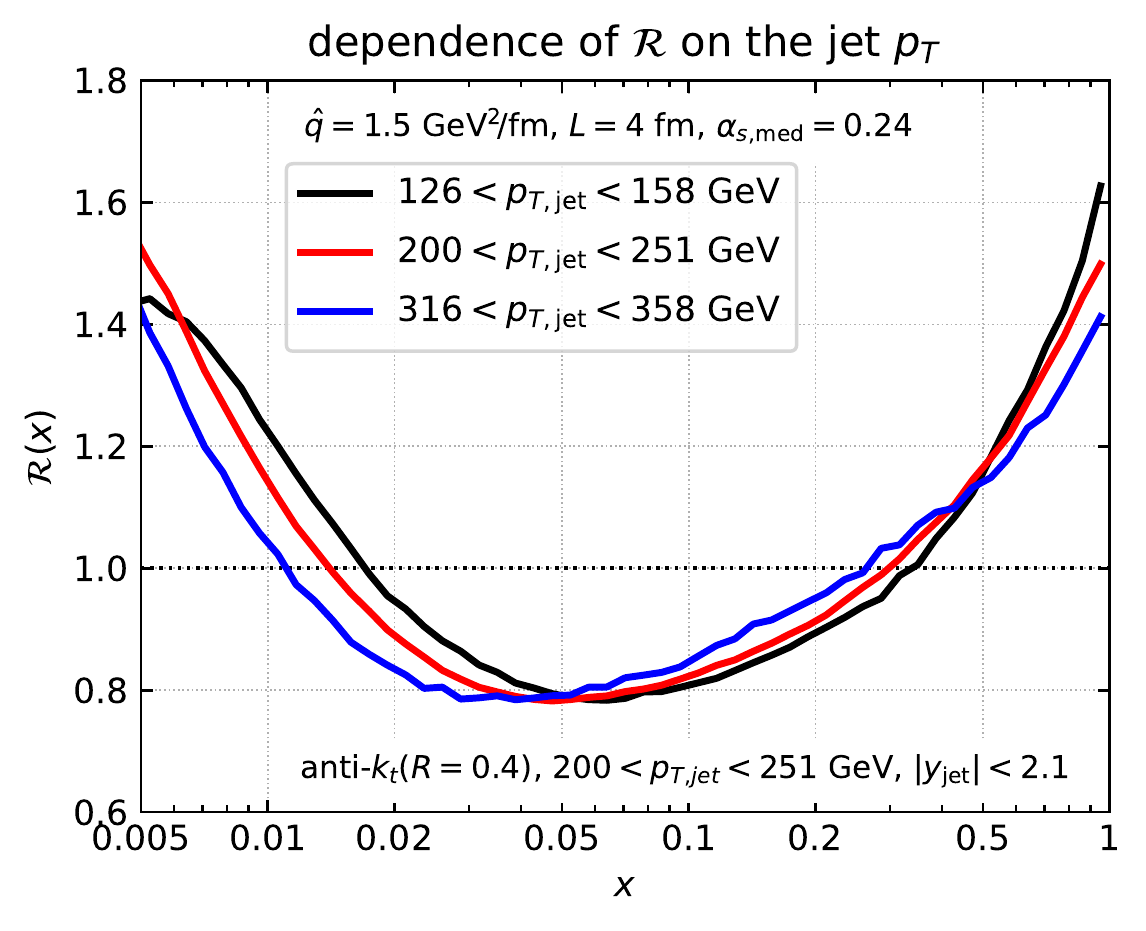}
  \hfill
  \includegraphics[page=2,width=0.48\textwidth]{./ff-ptdep.pdf}
  \caption{\small Our MC results for the nuclear modification factor $\mathcal{R}(x)$ 
 shown as a function of the energy fraction $x$ of a jet constituent (left) and of its
  transverse momentum $p_T$  (right), for 3 bins of the jet $p_{T,\textrm{jet}}$.
  }
\label{Fig:MC-ptdep} 
\end{figure}

Our Monte Carlo predictions for the nuclear modification $\mathcal{R}(x)$
are shown in Fig.~\ref{Fig:MC-ptdep} for three bins of
$p_{T,\textrm{jet}}$ and for the default set of (medium and
unphysical) parameters, cf.\ the first line in Table~\ref{tab:parameters}.
Following the experimental analysis by ATLAS~\cite{Aaboud:2018hpb}, we
have separately plotted our results as a function of $x$ (left plot)
and of the parton $p_T$ (right plot).
The left-hand plot shows only a mild dependence of $\mathcal{R}(x)$ on
$p_{T,\textrm{jet}}$ for $x\gtrsim 0.1$ when increasing. 
This suggests a weak
$p_{T,\textrm{jet}}$-dependence 
which is likely correlated to the similarly weak dependence observed
for $R_{AA}$.
At small $x$, the $x$ scale below which the ratio is larger than
$1$ decreases with $p_{T,\textrm{jet}}$, but the corresponding $p_T$
scale increases with $p_{T,\textrm{jet}}$. These trends are in
qualitative agreement with the respective ATLAS
results~\cite{Aaboud:2018hpb}.

\section{Nuclear effects on the fragmentation function near $x=1$}

\label{sec:x=1nuc}

With this section, we start our analytic investigations of the nuclear effects on the jet
fragmentation function when $1-x\ll1$. Since the main goal of the two following subsections \ref{sec:vetoed} and \ref{sec:MIE} is to discuss the nuclear effects neglecting the steeply falling initial jet spectrum, we mostly work with monochromatic jets with a given
initial transverse momentum $p_{T0}$. In the last subsection \ref{sec:bias}, the effect of the hard scattering cross-section combined with jet energy loss is taken into account. To guide the reader, we emphasize that the latter effect turns out to be the dominant one in the hard fragmentation region.

We therefore first focus on the jet fragmentation function $D_i(x|p_{T0})$
with $i\in\{q,g\}$, which can be conveniently computed as a derivative
of the {\it cumulative} fragmentation distribution
\begin{equation}
\label{def-sigma}
\Sigma_i(x|p_{T0})\equiv \int_x^1 \dif x' D_i(x'|p_{T0})\,.
\end{equation} 
For the vacuum calculation of this distribution, we refer the reader to Section \ref{subsub:frag-x=1}.

To discuss medium-induced effects, it is sufficient to work in the LL
approximation where jet fragmentation function near $x=1$ is dominated by a single, relatively soft,
gluon emitted by the leading parton.
From this two-parton system we then have to take three effects into account:
\texttt{(1)} emissions in the vetoed region of Fig.~\ref{Fig:DLA-phase-space} are forbidden,
\texttt{(2)} the leading parton and the emitted gluon can both lose energy via MIEs at large angles,
\texttt{(3)} the gluon emission can be a MIE remaining inside the jet.
We consider the effect of the vetoed region before the other two.

\subsection{Effect of the vetoed region}
\label{sec:vetoed}

The effect of the vetoed region in Fig.~\ref{Fig:DLA-phase-space} can be implemented
as a $\Theta$-function excluding this particular region from the phase-space for VLEs.
At LL accuracy, this amounts to having an extra factor
\begin{equation}\label{step-veto}
 \Theta_{\textrm{veto}}=1-\Theta(\sqrt{2\hat{q} zp_{T0}}-k_\perp^2)\Theta(k_\perp-2zp_{T0}L^{-1}),
\end{equation}
in the integrand of~\eqref{sigma-vac-LL}.
The first (second) $\Theta$-function in the r.h.s.\
of~\eqref{step-veto} corresponds to the upper (lower) boundary of the
vetoed region.
For a fixed-coupling approximation, we find assuming for simplicity $1-x\le2/(Lp_{T0}R^2)$ (see
Appendix~\ref{sec:rc-effects} for the result including running coupling)
\begin{equation}\label{eq:g1-veto}
  L g_{1,i}^{\textrm{veto}}(\alpha_s L, \alpha_sL_0)
  = L g_{1,i}^{\text{vac}}(\alpha_s L, \alpha_sL_0)
  + \frac{2\alpha_sC_i}{3\pi}\ln^2\frac{R}{\theta_c}\,.
\end{equation}
NLL corrections, $g_{2,i}^{\textrm{veto}}$, can be
obtained using~\eqref{g12}.
In particular, the hard-collinear term proportional to $B_i$ is not
modified by the veto region and therefore cancels in the medium/vacuum
ratio.

Our analytic estimate for the ratio $R_i(x|p_{T0})$ is shown in
Fig.~\ref{fig:large-x-analytic}~left in green
  for $p_{T0}=200$ GeV.
For comparison, we also show the corresponding MC result, which only
includes VLEs (the green curve in Fig.~\ref{fig:large-x-MC}). These
results agree well with each other and they both predict a nuclear
enhancement near $x=1$.
This enhancement can be easily understood on the basis of \eqref{eq:g1-veto},
which implies
\begin{tcolorbox}[ams align]
\ln\frac{\Sigma^{\textrm{veto,LL}}_{i}(x)}{\Sigma^{\textrm{vac,LL}}_{i}(x)}
=\frac{2\alpha_{0}C_i}{3\pi}\ln^2
\frac{R}{\theta_c}\,> 0\,,
\end{tcolorbox}
\noindent meaning $\Sigma^{\textrm{med}}_{i}(x) \simeq \Sigma^{\textrm{veto}}_{i}(x) > \Sigma^{\textrm{vac}}_{i}(x)$
and hence $R_i(x)>1$ when $x\to 1$. Indeed, the presence of the vetoed region reduces the phase-space
allowed for the decay of the leading parton.

\subsection{Effect of medium-induced emissions} 
\label{sec:MIE}

The medium-induced emissions (MIEs), as triggered by the interactions with
the plasma constituents, affect differently the total jet momentum
$p_{T,\textrm{jet}}$ and the energy $\om_\textrm{LP}$ carried by its
leading parton. This implies a nuclear modification
$\mathcal{R}(x)$ at large $x\equiv
\om_\textrm{LP}/p_{T,\textrm{jet}}$.

For convenience, we focus on the case where $x$ is not
{\it too} close to one, such that $\ombr/p_{T0}\ll 1-x\ll 1$, with $\ombr\sim \alpha_s^2\hat q L^2$
the characteristic scale  for multiple branchings.
For jets with $p_T\ge 200$~GeV, a phenomenological region $0.80
\lesssim x \lesssim 0.95$ translates into
$(1-x)p_{T0}\gtrsim 10$~GeV
which is indeed larger than $\ombr\sim 4$~GeV (cf.\ Table~\ref{tab:parameters}).

Within this regime, the medium-induced emissions which control the energy
loss by the leading parton are relatively hard,  with energies $\om\gg \ombr$.
Thus, they remain inside the jet and can be accurately computed
in the single emission approximation. This situation 
is similar to the one discussed for
jets in the vacuum at double-logarithmic accuracy: 
the parton distribution near $x=1$ is controlled by a single intra-jet
emission, with an energy of the order of $(1-x)p_{T0}$. This emission
can be either vacuum-like, or medium-induced. This
``semi-hard'' emission is accompanied by an arbitrary number of soft
MIEs, with energies $\om\lesssim \ombr$, which propagate outside the
jet and take energy away from the jet constituents.
The in-medium fragmentation function near $x=1$ can therefore be
evaluated as:
\begin{tcolorbox}[ams align]  \label{Elossred}
 D^{\textrm{med}}_i(x|p_{T0})\,\simeq\,
  \int\dif\omega \,\Delta_i^{\text{VLE}}  (\omega) \,\Delta_i^{\text{MIE}}  (\omega)\,
  \left[\frac{\del\mathcal{P}_{i, \text{vac}}}{\del \omega} 
  +\frac{\del\mathcal{P}_{i, \text{med}}}{\del \omega}\right]
 \delta\left(x-\frac{p_{T0}-\omega-\varepsilon_i}
{p_{T0}-\mathcal{E}_i}\right)
\end{tcolorbox}
\noindent In this expression,
${\del\mathcal{P}_{i, \text{vac}}}/{\del \omega}$ is the differential
probability for emitting a soft gluon with energy $\omega$ at any
emission angle $\theta$ (with $\ktmin/\om< \theta < R$) and
$\Delta_i^{\text{VLE}} (\omega)$ is the Sudakov factor forbidding VLEs
with energies larger than $\om$ (including the condition
\eqref{step-veto} for the phase space gap), i.e.\
\begin{equation}\label{Pvac}
 \Delta_i^{\text{VLE}}  (\omega) =
 \Sigma^{\text{veto}}\left(1-\frac{\omega}{p_{T0}}\right)
 \qquad \text{ and } \qquad
   \frac{\del\mathcal{P}_{i, \text{vac}}}{\del \omega} 
=\frac{\dif \ln\Delta_i^{\text{VLE}}}{\dif \omega} 
\simeq \frac{2 \alpha_sC_i}{\pi}\frac{1}{\om}\, \ln\left(\frac{\om R}{\ktmin}\right),
\end{equation}
where the second expression for ${\del\mathcal{P}_{i, \text{vac}}}/{\del\omega}$, shown only
for illustration, 
holds for the case of a fixed coupling $\alpha_s$ and ignores the constraints 
introduced by the vetoed region.

Furthermore, ${\del\mathcal{P}_{i, \text{med}}}/{\del \omega}$ and
$\Delta_i^{\text{MIE}} (\omega)$ are the corresponding quantities for
the semi-hard MIE inside the jet ($\theta_c <\theta < R$).
Its energy is restricted to $\bar\om< \om<\om_c$, where
$\om_c=\hat q L^2/2$ and $\bar\om$ is a cutoff of order $\ombr$,
separating between ``semi-hard'' and ``soft'' MIEs.\footnote{The
  precise value of this cut-off is not important: as we will show below
  the energy integration is controlled by the $\delta$-function, and
  since the energy losses are relatively small one roughly has
  $\omega \simeq (1-x)p_{T0}\gg \bar\om\sim \ombr$.}
In this regime, one can safely use the single emission approximation,
i.e. (compare to \eqn{BDMPSZ-estimate})
\begin{align}\label{Pmed}
 \frac{\del\mathcal{P}_{i, \text{med}}}{\del
 \omega_m} \simeq \frac{\amed C_i}{\pi}\sqrt{\frac{2\omega_c}{\omega^3_m}}\,,
 \qquad\quad \Delta_i^{ \text{MIE}} (\om_m) 
 =\exp\left(-\int_{\omega_m}^{\omega_c}\dif\omega\,  \frac{\del\mathcal{P}_{i, \text{med}}}{\del
 \omega} \right)\,.
\end{align}
Next, $\varepsilon_i$ and $\mathcal{E}_i$ refer to the energy loss via soft
MIEs outside the jet ($\theta > R$), for the leading parton and for
the jet as a whole, respectively.
Finally, the $\delta$-function in \eqn{Elossred} encodes the fact that, in our
present approximation, the energy of the leading parton is the
energy $p_{T0}$ of the parton initiating the jet minus the energy of
the semi-hard emission and the partonic energy loss $\varepsilon_i$,
while the energy of the jet is $p_{T,\textrm{jet}}= p_{T0}-\mathcal{E}_i$.

For more clarity, we study separately the two types of medium effects included in
\eqn{Elossred}, namely energy loss at large angles and energy redistribution via intra-jet MIEs.

\subsubsection{Energy loss at large angles}
\label{sec:eloss}

To study the  energy loss effects alone, we temporarily neglect the contribution of
the intra-jet MIEs to \eqn{Elossred}, which then  simplifies to (with
$\omega_s$ the energy of the soft VLE)
\begin{align}  \label{VL+Eloss}
 D^{\textrm{med}}_i(x|p_{T0})\Big |_{\text{e-loss}}= 
  \int\dif\omega_s \frac{\del\mathcal{P}_{i, \text{vac}}}{\del
 \omega_s}\,\Delta_i^{\text{VLE}}  (\omega_s) \,
 \delta\left(x-\frac{p_{T0}-\omega_s-\varepsilon_i}
{p_{T0}-\mathcal{E}_i}\right).
\end{align}

In the absence of VLEs, a single parton with initial energy $\omega_0$ 
loses energy by radiating MIEs at large angles ($\theta\gtrsim
\theta_c/\abar^2$).
This is associated with the ``turbulent'' component of the medium-induced cascades,
associated with very soft partons of
energies $\omega\lesssim \ombr$, which are deflected at large angles
via collisions with the plasma.
The average energy loss is estimated by \eqref{energy-flow}
\begin{equation}\label{eloss-flow}
  \varepsilon_i(\omega_0) =\omega_0 \big[1-\rme^{-v_0\ombr/\omega_0}\big],
  \qquad\text{ with} \quad
  \ombr
  = \left(\frac{\amed}{\pi}\right)^2C_AC_i\,\frac{\hat q L^2}{2}\,.
\end{equation}
%
%
For a full jet, the energy loss receives contributions of the form of
\eqn{eloss-flow} from both the leading parton (LP) and each of the
(vacuum-like or medium-induced) {\it intra-jet} emissions ($\theta<R$)
 which are radiated within the medium. An estimate is given in \eqref{logeloss}.

For a hard-fragmenting jet made of only two partons (the LP and a relatively soft
VLE, as in \eqn{VL+Eloss}), we have to consider two options:
\begin{itemize}
 \item If the VLE is emitted outside the medium, i.e.\
either with $\theta<\theta_c$ or with $t_{\rm f}=2/(\om \theta^2)>L$,
only the LP loses energy and we have
$\mathcal{E}_i= \varepsilon_i$. For $\theta<\theta_c$, the two
  partons lose energy coherently, so one can see the energy loss as coming only from the 
  LP.
 \item If the VLE occurs inside the medium, both partons lose energy and we
have $\mathcal{E}_i=\varepsilon_i+\varepsilon_g$, with $\epsilon_g$ the
energy lost by the VLE. In this case, $t_{\rm f}\ll L$ so
  the VLE travels a length or order $L$ through the medium.
\end{itemize}

\begin{figure}[t] 
  \centering
  \begin{subfigure}[t]{0.48\textwidth}
    \includegraphics[page=1,width=\textwidth]{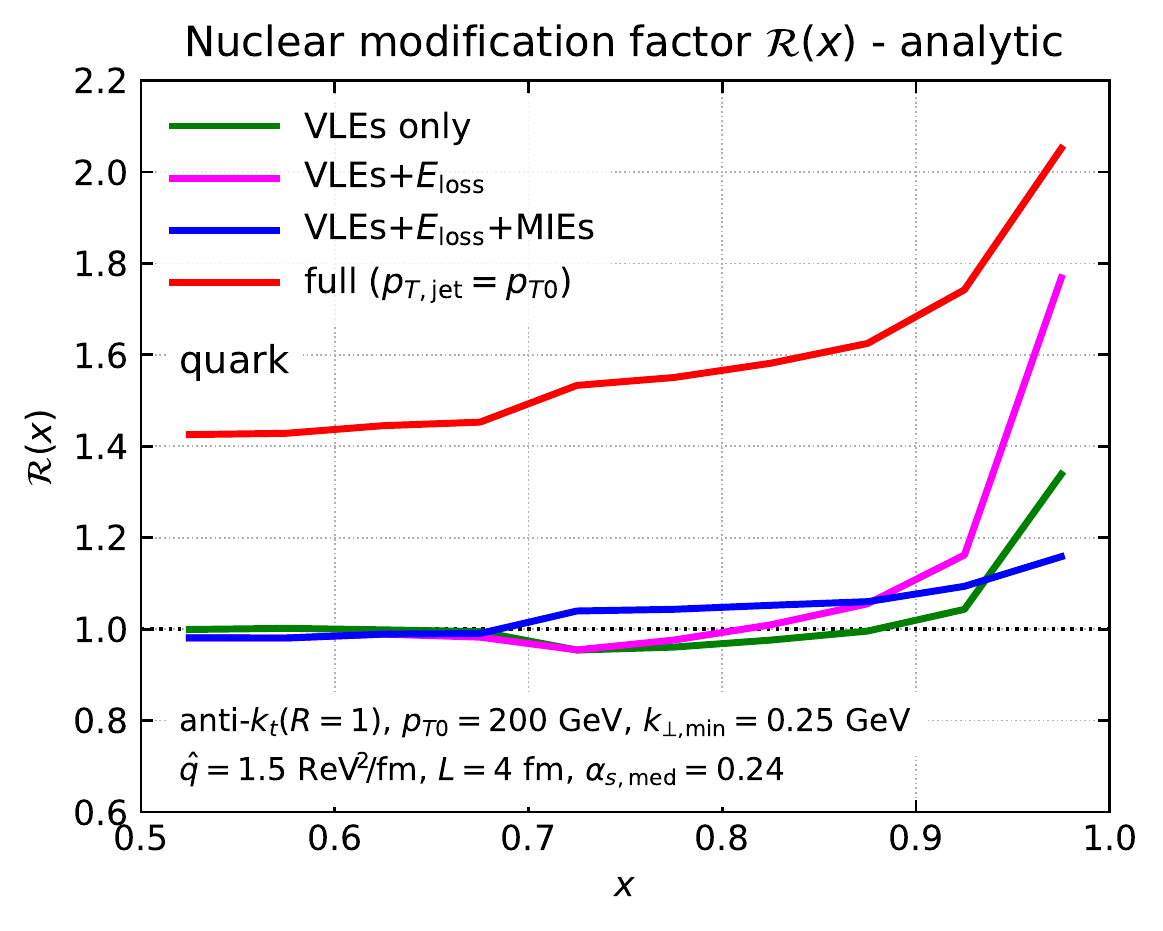}
    \caption{\small Semi-analytic estimates.}\label{fig:large-x-analytic}
  \end{subfigure}
  \hfill
  \begin{subfigure}[t]{0.48\textwidth}
    \includegraphics[page=2,width=\textwidth]{./plot-frag-ratio-largex.pdf}
    \caption{\small Monte Carlo simulations.}\label{fig:large-x-MC}
  \end{subfigure}
  \caption{\small Nuclear effects on the fragmentation function at
    large $x$ for monochromatic jets.
    Three increasingly more physical scenarios are considered: \texttt{(i)}
    VLEs only (only the nuclear effects from the vetoed region are
    included), \texttt{(ii)} adding energy loss via soft MIEs at large
    angles (not shown on the right plot), and \texttt{(iii)} further
    adding semi-hard MIEs inside the jet.
    Additionally, we show the ``full'' curve in red which includes the bias
    introduced by the initial hard spectrum and is manifestly the
    dominant effect.}
\label{Fig:analytic} 
\end{figure}

For a VLE inside the medium, the $\delta$-function in \eqn{VL+Eloss} can be equivalently rewritten as
\beq
\delta\left(1- x-\frac{\omega_s - (\mathcal{E}_i-\varepsilon_i)}{p_{T0}-\mathcal{E}_i}\right)
\simeq
\delta\left(1- x- z+ \frac{ (1-z)\mathcal{E}_i-\varepsilon_i}{p_{T0}}\right)\,,
\eeq
with $z\equiv\om_s/p_{T0}$ the splitting fraction of the VLE. We have used the fact that the energy loss is 
relatively small, $\mathcal{E}_i\ll p_{T0}$.
The effect of the in-medium energy loss is a small increase of the splitting fraction, from its initial value
in the vacuum, $z_{\textrm{vac}}=1-x$, to
\beq\label{z-Eloss}
z = 1-x+ \frac{ (1-z)\mathcal{E}_i-\varepsilon_i}{p_{T0}}
\simeq 1-x+\frac{x\varepsilon_g-(1-x)\varepsilon_i}{p_{T0}}
\simeq 1-x+\frac{\varepsilon_g}{p_{T0}} > z_\text{vac}.
\eeq
In the second equality we have used $\mathcal{E}_i=\varepsilon_i+\varepsilon_g
$ and $z\simeq 1-x$.
For the third equality we have used $x\simeq 1$
and $\varepsilon_g\ge \varepsilon_i$, making clear that the dominant effect is
the energy loss by the soft gluon. \footnote{Interestingly,
  for a VLE outside the medium, we can
  set $\varepsilon_g\to 0$ to get $z = 1-x-(1-x)\varepsilon_i/{p_{T0}}$
with $1-x\ll 1$. The energy loss effect is therefore much smaller than
for an in-medium VLE and with an opposite sign.}

The fact that $z>z_\text{vac}\equiv 1-x$ means
that the probability $P(z)\propto 1/z$ of its emission is smaller, so
there is an {\em enhancement} in the
probability for the leading parton to survive at large $x$. This effect is reinforced by the 
associated Sudakov factor: when $\om_s=zp_{T0} > (1-x)p_{T0}$, there is a reduction in  
the phase-space for emissions by the leading parton and therefore
$\Delta_i^{\text{VLE}}  (\omega_s) >\Delta_i^{\text{VLE}}  ((1-x)p_{T0}) $.

The purple curve in Fig.~\ref{Fig:analytic}-left shows
a calculation of $\mathcal{R}_q(x|p_{T0})$
based on \eqn{VL+Eloss} together with  $\mathcal{E}_q=\varepsilon_q+\varepsilon_g$ 
and with \eqn{eloss-flow} for the partonic energy loss. 
Compared to the green curve in the same figure, which
includes solely the effect of the vetoed region, the purple curves indeed shows a larger
enhancement near $x=1$.

\subsubsection{Energy redistribution via a hard MIE}
\label{sec:redis}

A semi-hard MIE with energy $\omega \gg \ombr$ and which remains inside
the jet can modify the fragmentation function
$D^{\textrm{med}}_i(x|p_{T0})$
near $x=1$ in two ways. On one hand, it brings a positive contribution
via the term proportional to $\partial\mathcal{P}_{i,\text{med}}/\partial\omega$
in \eqn{Elossred}.
On the other hand, the additional Sudakov factor
$\Delta_i^{\text{MIE}}(\om)$ induces an extra suppression.
These two effects are competing with each other. It turns out that
the second effect is stronger, resulting in a {\it decrease} of
$D_i^\text{med}(x|p_{T0})$ near $x=1$ as compared to the 
vacuum, and hence a decrease of the medium/vacuum ratio
$\mathcal{R}_i(x|p_{T0})$.

We can actually estimate these two contributions to \eqn{Elossred}.
To that aim, we can neglect the effects of the energy loss
at large angles.\footnote{Indeed, in this case, the intra-jet MIE is the dominant medium
effect, whereas the energy loss at large angles is a subdominant effect since
  $\mathcal{E}_i\sim\varepsilon_i\sim\ombr$ are much smaller than $\om\simeq
  (1-x)p_{T0}$.}
Using the $\delta$-function to 
perform the integral over $\omega$ we find
\begin{align}  \label{hardMIE}
 D^{\textrm{med}}_i(x|p_{T0})\Big |_{\text{MIE}}& = p_{T0}
   \left[\frac{\del\mathcal{P}_{i, \text{vac}}}{\del \omega} 
  +\frac{\del\mathcal{P}_{i, \text{med}}}{\del \omega}\right]
\Delta_i^{\text{VLE}}  (\omega) \,\Delta_i^{\text{MIE}}  (\omega)\Big |_{\om=
(1-x)p_{T0}}.
\end{align}
We need to show that the ``medium'' Sudakov effect on the VLE (first term in the
square bracket) is larger in absolute value than the direct
contribution from MIEs (second term in the square
bracket):
\beq\label{ineq}
\frac{\del\mathcal{P}_{i, \text{vac}}}{\del \omega} \left[1-\Delta_i^{\text{MIE}} 
  (\omega)\right] > \frac{\del\mathcal{P}_{i, \text{med}}}{\del \omega}
\, \Delta_i^{\text{MIE}}  (\omega)\,.
\eeq
At leading-order accuracy for the MIE, one can set
$\Delta_i^{\text{MIE}} \simeq 1$ in the r.h.s.\ of the above inequality, whereas in
the l.h.s. one must also keep the linear term in its Taylor expansion:
\begin{align} 1-\Delta_i^{\text{MIE}} 
  (\omega)\,\simeq\,
\frac{2\alpha_s C_i}{\pi}\sqrt{\frac{2\omega_c}{\om}}\,.
\end{align}
Using a fixed-order approximation for the vacuum emission probability
(cf.\ \eqn{Pvac}), together with \eqn{Pmed} for the medium-induced,
one finds after simple algebra that \eqn{ineq} is equivalent to
\beq\label{log}
\frac{4 \alpha_sC_i}{\pi} \,\ln\left(\frac{(1-x)p_{T0}R}{\ktmin}\right) \,>\,1\,.
\eeq 
This is satisfied both parametrically and numerically under our
working assumptions that collinear logarithms are large.
For the parameters used in Fig.~\ref{Fig:analytic}, namely $p_{T0}=200$~GeV, $R=1$, and
$\ktmin=0.25$~GeV, and with $x=0.9$ and $\alpha_s=0.3$, one finds that
the l.h.s.\ of \eqn{log} is about $5.3$.

These considerations are confirmed by the explicit 
numerical integration of~\eqn{Elossred}.
The blue curve in Fig.~\ref{fig:large-x-analytic} includes all the medium effects
discussed in this section (the vetoed region, the energy loss at large
angles and the effects of semi-hard MIEs).
Comparing it to the purple curve which does not include the effects of
semi-hard MIEs, we see that the latter reduce the ratio
$\mathcal{R}_i(x|p_{T0})$ near $x=1$, as expected.
This plot also shows that the three medium effects appear to be of
similar magnitude and to almost compensate each other, leaving only
a modest enhancement at $x\gtrsim 0.9$.
This pattern is in very good agreement with what we see from our MC
simulations, Fig.~\ref{fig:large-x-MC}.
Whereas the details of this compensation depend on the specific
parameters used in our calculation, we have checked using our MC that
such a competition between comparable but opposite effects is
a relatively robust prediction from our pQCD scenario.

One can view this conclusion as a little bit deceptive since it shows
that the fragmentation function has a reduced sensitivity to nuclear
effects associated with the internal dynamics of the jets.

\subsection{Bias introduced by the steeply falling jet spectrum} 
\label{sec:bias}

The behaviour at large $x$ is in fact largely controlled by the physics of energy loss and its interplay
with the initial production spectrum, as we now explain.

A jet which, after crossing the medium, is measured with a
transverse momentum $p_{T,\textrm{jet}}$ has originally been produced
from a hard quark or gluon emerging from a hard process with a larger momentum $p_{T0}=
p_{T,\textrm{jet}}+\mathcal{E}(p_{T0})$, where $\mathcal{E}(p_{T0})$ is the energy
lost by the jet via MIEs at large angles  
(see Ref.~\cite{Caucal:2019uvr} for an extensive discussion of this quantity). 
While the energy lost by a {\it parton} with momentum
$p_T\gg \ombr$ saturates at a value $\epsilon\sim\ombr$, which is
independent of $p_T$ \cite{Blaizot:2013hx}, 
the average energy lost by a jet keeps increasing with $p_{T0}$, 
because of the rise in the phase space for VLEs and hence in the number of partonic sources for medium-induced radiation.
%

Due to the steeply-falling underlying $p_{T0}$ spectrum, cutting on the jet $p_T$ tends to select 
jets which lose less energy than average. In particular, this bias favours the 
``hard-fragmenting'' jets which contain a  parton with large $x$ (say, $x>0.5$). Such jets
correspond to rare configurations,  in which the radiation from leading parton is strongly limited 
in order to have a final $x$ fraction close to one.  
Since they contain only few partons, the hard-fragmenting jets  suffer very little energy loss,
of the order of the partonic energy loss $\epsilon\sim\ombr$. They are therefore less
suppressed than the average jets by the steeply-falling initial spectrum.   In other terms,
the medium acts as a filter which enhances the proportion of hard-fragmenting  jets
compared to the vacuum.

This bias has already consequences for the {\it inclusive} jet production, 
as measured by $R_{AA}$: 
the fraction of hard-fragmenting jets among the total number of jets 
(say, in a given bin in $p_{T,\textrm{jet}}$)
is larger in $AA$ collisions than in $pp$ collisions. The effects of this bias are
however expected to become even stronger for the jet distribution ${\dif N}/{\dif x}$ at large $x$, 
which by definition selects {\it only} hard-fragmenting jets. 
This stronger bias towards hard-fragmenting jets has been proposed as an explanation for the nuclear
enhancement in the fragmentation function 
observed in the LHC data~\cite{Aaboud:2018hpb} at large $x\gtrsim 0.5$.
This argument is very general: it applies to
a large variety of microscopic pictures for the jet-medium interactions, assuming either
weak coupling \cite{Milhano:2015mng,KunnawalkamElayavalli:2017hxo}, 
or strong coupling \cite{Chesler:2015nqz,Rajagopal:2016uip}, or a hybrid scenario 
\cite{Casalderrey-Solana:2016jvj,Casalderrey-Solana:2018wrw,Casalderrey-Solana:2019ubu}.
All these scenarios naturally predict that hard-fragmenting jets lose less energy
towards the medium than average jets, for the physical reason that we already mentioned:
hard-fragmenting jets contain less partonic sources for in-medium energy loss. 
This physical argument is manifest in both the pQCD~\cite{Milhano:2015mng,KunnawalkamElayavalli:2017hxo}
and the hybrid approaches~\cite{Casalderrey-Solana:2016jvj,Casalderrey-Solana:2018wrw,Casalderrey-Solana:2019ubu}, 
which explicitly include
a vacuum-like parton shower. It is also implicit in the strong coupling scenario  in~\cite{Chesler:2015nqz,Rajagopal:2016uip} which
is tuned such as to reproduce the angular distribution of jets in p+p collisions at the LHC
(itself well described by PYTHIA).



In this section, we argue that this is also the main explanation for the rise seen in our
results in Fig.~\ref{Fig:MC-pheno} at $x\gtrsim 0.5$. 
Within our pQCD approach
this is not entirely obvious since our scenario also
allows for nuclear modifications of the fragmentation process itself, via medium-induced emissions and energy loss effects. Similar ingredients
are {\it a priori} present in other scenarios, like JEWEL, but their relative importance
has not been explicitly studied to our knowledge. In the previous subsections, we have performed
an extensive study of these effects, via both analytical and
numerical (MC) methods. 


%
%

\begin{figure}
  \centering
  \begin{subfigure}[t]{0.48\textwidth}
    \includegraphics[width=\textwidth]{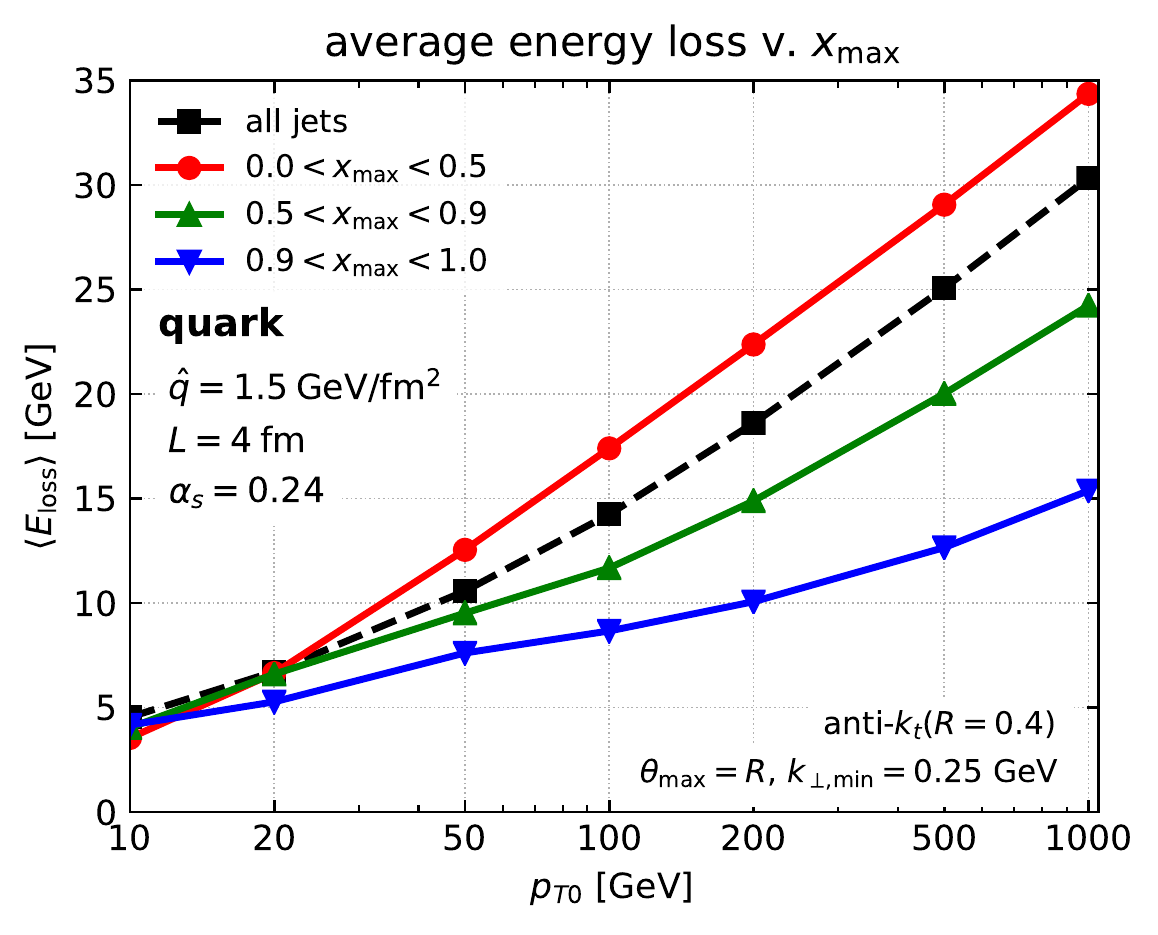}
    \caption{Energy loss as a function of the initial parton
      transverse momentum $p_{T0}$.}\label{fig:eloss-v-xmax}
  \end{subfigure}
  \hfill
  \begin{subfigure}[t]{0.48\textwidth}
    \includegraphics[width=\textwidth]{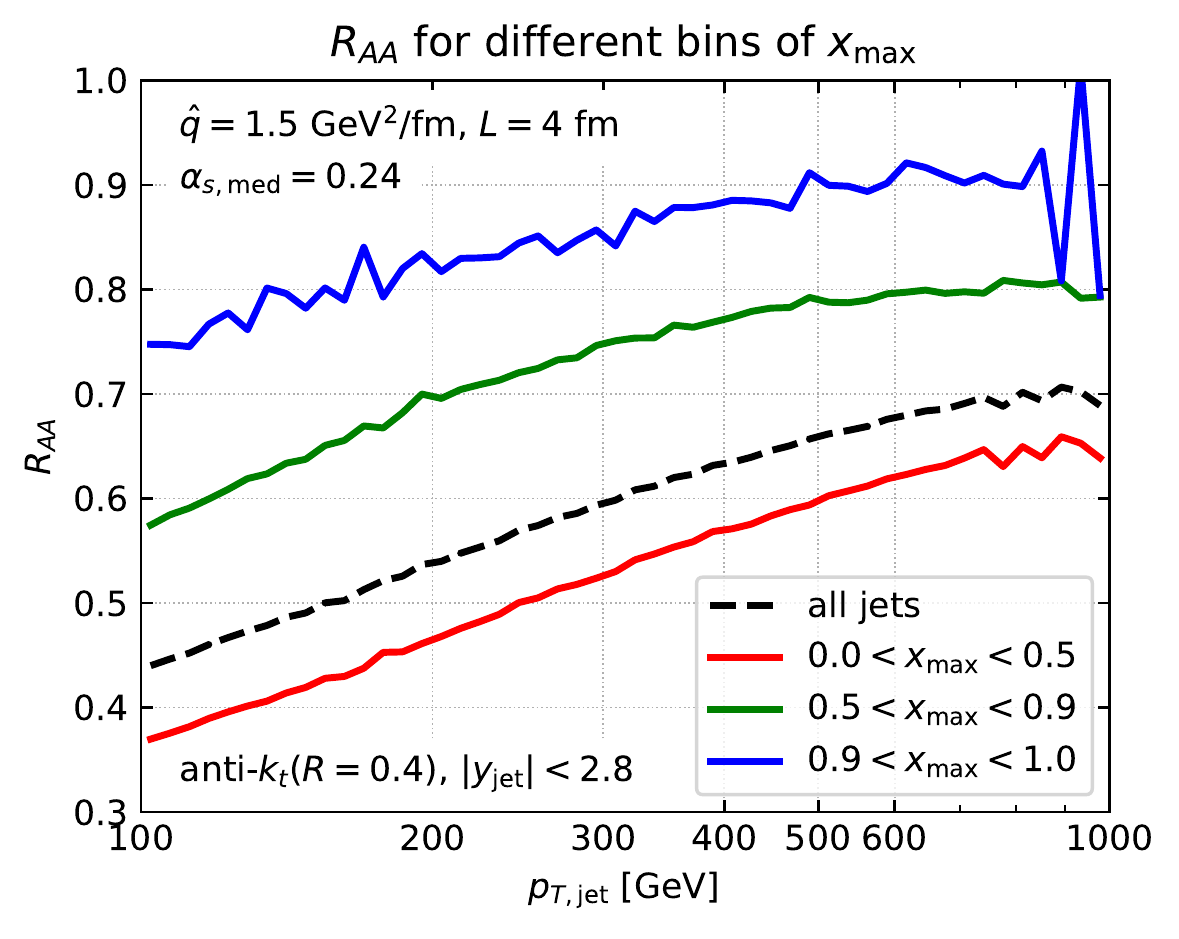}
    \caption{Jet nuclear modification factor $R_{AA}$ as a function of
      the jet $p_T$.}\label{fig:RAA-v-xmax}
  \end{subfigure}
  \caption{Energy loss and $R_{AA}$ for different bins of
    $x_\text{max}$, the momentum fraction of the jet harder
    constituent.}\label{fig:quenching-v-xmax}
\end{figure}
%

Before we discuss the fragmentation function {\it per se}, let us first
demonstrate that,  in our picture too,  a hard-branching jet loses less energy than
the average one. We have numerically verified this, by selecting (in our MC events)
jets for which the harder parton carries a momentum fraction
$x_\text{max}$ in a restricted window.
These results are presented in Fig.~\ref{fig:eloss-v-xmax} for the
energy loss of monochromatic jets and in Fig.~\ref{fig:RAA-v-xmax}
for the jet $R_{AA}$,  for the 3 bins in $x_\text{max}$ and (for comparison)
also for the inclusive jets.
Focusing first on the left figure, we find indeed that the energy lost by
jets with $x_\text{max}>0.9$, i.e.\ hard-fragmenting jets, is both
considerably smaller and also less rapidly growing with $p_{T0}$ then
for the average jets.
As $x_\text{max}$ decreases, both the energy loss and its $p_{T0}$
growth increase.
This tendency is confirmed by the study of $R_{AA}$,
Fig.~\ref{fig:RAA-v-xmax}, where jets with a large $x_\text{max}$
show a smaller-than-average nuclear suppression.
 It would be interesting to experimentally measure the correlation between the jet  $R_{AA}$
and the momentum fraction $x_\text{max}$ and compare to our above predictions 
(see also \cite{Casalderrey-Solana:2018wrw} for a related observable, which compares
the nuclear suppression for high-$p_T$ hadrons and inclusive jets).
%

To have a more quantitative argument, let us focus on a
single bin in $p_T\equiv p_{T,\textrm{jet}}$ with a (vacuum)
Born-level $p_T$ spectrum.
The vacuum fragmentation function can then be easily estimated as
\beq\label{Dvac}
{\mathcal{D}^{\textrm{vac}}(x|p_T)}\simeq \frac{N_q(p_T)D^{\textrm{vac}}_q(x|p_{T})
+N_g(p_T)D^{\textrm{vac}}_g(x|p_{T})}{N_q(p_T)+N_g(p_T)}\,,
\eeq
where $N_i(p_T)\equiv \dif N^{\textrm{hard}}_i/{\dif p_{T}}\propto 1/p_{T}^n$
are the initial spectra for quarks ($i=q$) and gluons ($i=g$) and the fragmentation
functions for monochromatic jets have been introduced at the end of
Sect.~\ref{sec:defin}.
To write down the corresponding formula for jets in the medium, let us assume
that the only medium effect on the jet production is the energy
loss. One can thus write
\beq\label{Dmed}
{\mathcal{D}^{\textrm{med}}(x|p_T)}\simeq 
\frac{\sum\limits_{i\in\{q,g\}} N_i (p_{T}+\varepsilon_i (x))
D^{\textrm{med}}_i(x|p_{T}+\varepsilon_i(x))}
{\sum\limits_{i\in\{q,g\}} N_i (p_{T}+\mathcal{E}_i(p_T))}\qquad\mbox{for $x\simeq 1$}.
\eeq
The  quantity $\varepsilon_i(x)$  in the numerator is the energy loss
of a hard-fragmenting jet.
It depends on $x$ because the focus on large values 
$x>0.5$ selects special configurations in which jets are made with only few partons.
Its precise $x$--dependence is not important for what follows. Rather, it suffices
to know that $\varepsilon_i(x)$  is a partonic energy loss, of order $\ombr$, 
and to a good approximation is independent of the jet $p_T$.
The corresponding quantity in the denominator, $\mathcal{E}_i(p_T)$,
is the average energy loss by a jet with transverse momentum $p_T$.
It is much larger than $\varepsilon_i(x)$ and increases with $p_T$.
This difference between the {\it partonic} energy loss  $\varepsilon_i(x)$ in the numerator of 
 \eqn{Dmed} and the {\it average} energy loss $\mathcal{E}_i(p_T)$ in its denominator,
 together with the rapid decrease of $N_i (p_{T})$ when increasing $p_T$, are the origin
 of the nuclear bias towards hard-fragmenting jets at large $x$, discussed
 at the beginning of this section.\footnote{Strictly speaking, the ``average'' energy loss 
 $\mathcal{E}_i(p_T)$ in the denominator is influenced too by this bias, since it should be
 computed as an average over an inclusive sample of jets produced in $AA$ collisions.
 However, this bias is less important for the inclusive sample
 than for the large-$x$ distribution in the numerator of \eqn{Dmed}.}

On top of their bias towards less energy loss, hard-fragmenting jets also favour
quark-initiated jets. There are two reasons for
this~\cite{Spousta:2015fca,Caucal:2019uvr}:
\texttt{(i)} a quark radiates less than a gluon due to its reduced
colour charge ($C_F < C_A$), resulting in a larger probability to
contribute at large $x$, and
\texttt{(ii)} quark-initiated jets typically contain less partons than
gluon-initiated jets and hence lose less energy
($\varepsilon_q < \varepsilon_g$); this feature together with
 the steeply-falling $p_T$ spectrum favours their production
 in $AA$ collisions.
We can therefore only keep the quark contribution to the numerators of
Eqs.~\eqref{Dvac} and \eqref{Dmed} and write
\beq\label{RQ}
\mathcal{R}(x|p_T)\,\simeq\, \frac{f^\textrm{med}_q(x|p_T)}{f^\textrm{vac}_q(p_T)}
\,
\mathcal{R}_q(x|p_T)\,,
\eeq
with the following definitions:
\beq\label{fq}
 f^\textrm{vac}_q(p_T)\equiv \frac{N_q(p_T)}
{N_q(p_T)+N_g(p_T)}\,,\qquad
f^\textrm{med}_q(x|p_T)\equiv \frac{N_q (p_{T}+\varepsilon_q(x))}
{{\sum\limits_{i\in\{q,g\}} N_i (p_{T}+\mathcal{E}_i(p_T))}}\,.
\eeq
For jets in the vacuum, $ f^\textrm{vac}_q(p_T)$ is simply the 
fraction of quark-initiated jets. However, the corresponding quantity for jets in the medium 
is generally {\it not} a fraction, because of the different energy losses appearing in the
numerator and in the denominator of $f^\textrm{med}_q(x|p_T)$.

The condition of hard fragmentation ($x\sim 1$) only plays a role in the case of the
medium, where it distinguishes between the ``partonic'' energy loss  $\varepsilon_q(x)$ 
in the numerator and the jet energy loss $\mathcal{E}_q(p_T)$ in the
denominator.
As already discussed, the physical observation that $\varepsilon_q(x)\ll
\mathcal{E}_i(p_T)$ implies that the  fraction of hard-fragmenting jets in
the medium is larger than that in the vacuum, i.e., 
${f^\textrm{med}_q(x|p_T)}/{f^\textrm{vac}_q(p_T)}>1$, which in turn
causes $\mathcal{R}(x|p_T)$ to go above one
for $x\lesssim 1$.
As $x$ decreases, the energy loss of jets contributing at this value
of $x$ increases, becoming closer to $\mathcal{E}(p_T)$ and the
nuclear enhancement is less pronounced.

It is enlightening to go one step further: let us present a more detailed (numerical) argument, 
based on simple 2-parton jets, which supports
 \eqn{RQ}.
\eqn{RQ} relies on the ``fraction'' $f_i(x|p_T)$ of hard-fragmenting jets 
with one constituent having an energy of at least $xp_T$.
In practice, we define (cf.\ \eqn{fq})
\begin{equation}
\label{quenching-factor}
f^\textrm{vac}_q(p_T)=\frac{\frac{\dif\sigma_q}{\dif p_{T}}}
{\sum\limits_{i\in\{q,g\}}\frac{\dif\sigma_i}{\dif p_T}}\,,\qquad
f^\textrm{med}_q(x|p_T)=\frac{\frac{\dif\sigma_q}{\dif p_{T0}}\big |_{p_T+\mathcal{E}^{\textrm{n=2}}_q}}{\sum\limits_{i\in\{q,g\}}\frac{\dif\sigma_i}{\dif p_{T0}}\big|_{p_T+\mathcal{E}_i(p_{T0})}}\,,
\end{equation}
where  $\dif\sigma_i/\dif p_{T0}\propto p_{T0}^{-n_i}$ is the initial
jet spectrum.
$n_q=5$ and $n_g = 5.6$ give a decent description over the kinematic
range covered in this paper.
$\mathcal{E}_i(p_{T0})$ is the average energy loss by a jet with
initial transverse momentum $p_{T0}$ and is numerically extracted from
MC simulations in Chapter~\ref{chapter:RAA}, Section~\ref{sec:MC-jet-eloss}. $\mathcal{E}^{\textrm{n=2}}_q$
is the energy lost by a simple two-parton jet (a leading quark of
energy fraction $x\sim 1$ and a relatively soft gluon of energy
fraction $1-x$). The dominant
contribution (cf.\ Sect.~\ref{sec:eloss}) comes from events where the
quark and gluon lose energy
independently of each other\footnote{Strictly speaking, the energy
  argument of $\varepsilon_g$ and $\varepsilon_q$ should be $zp_{T0}$
  and $(1-z)p_{T0}$, respectively, with $z$ the gluon splitting
  fraction, cf.  \eqn{z-Eloss}, but to the accuracy of interest one
  can replace $z\simeq 1-x$ and $p_{T0}\simeq p_T$.}:
$\mathcal{E}^{\textrm{n=2}}_q= \varepsilon_q(xp_{T0}) +
\varepsilon_g((1-x)p_{T0})$, with $\varepsilon_g$ and $\varepsilon_q$
given by \eqn{eloss-flow}.

By combining \eqn{quenching-factor} for the fractions of
hard-fragmenting jets with our previous calculations of the ratio
$\mathcal{R}_q(x|p_T)$ for monochromatic jets, we can provide a
semi-analytic estimate for the physical observable
$\mathcal{R}(x|p_T)$ using \eqn{RQ}.
This is shown by the red curve in Fig.~\ref{fig:large-x-analytic}, that
should be compared to the corresponding MC result in
Fig.~\ref{fig:large-x-MC}.
The two red curves are both in good agreement with each other and with
the general trend seen in the LHC data~\cite{Aaboud:2018hpb}.
For $x$ very close to 1 (mainly the last
bin in our plots), the pattern observed in our MC calculations is
a combination of the bias induced by the jet spectrum and of the
medium effects on the internal jet dynamics $\mathcal{R}_q(x|p_T)$,
with a strong domination of the former.
The current experimental uncertainties in this region of $x$ are too
large to draw a stronger conclusion, notably concerning  the relative importance of 
the nuclear effects associated with $\mathcal{R}_q(x|p_T)$, i.e.\ with the
medium modifications of jet fragmentation itself.

\section[Small-$x$ enhancement: colour decoherence and medium-induced radiations]{Small-$x$ enhancement: colour decoherence and \\
medium-induced radiations}

\sectionmark{Small-$x$ enhancement}
\label{sec:smallx-MC-results}
\subsection{Qualitative discussion}
Let us now consider the situation at small $x\lesssim 0.01$, where our numerical results in
Fig.~\ref{Fig:MC-pheno} show a pronounced medium enhancement of the fragmentation
function, in qualitative agreement with the experimental
observations~\cite{Aaboud:2018hpb}.
These results also exhibit a (partial) lift of the degeneracy between the various sets of values 
for the medium parameters, suggesting a weaker correlation between
$\mathcal{R}(x)$ and  the jet nuclear modification factor $R_{AA}$.
This section provides explanations for these observations within our
framework.

We first note that,
for the considered range in $p_{T,\textrm{jet}}$, $x\lesssim 0.01$ corresponds to momenta
$p_T\lesssim 2$~GeV for the emitted partons, which are smaller than the characteristic medium
scale $\ombr$ for multiple branching.
In our framework, such soft emissions are dominated by 
VLEs outside the medium since MIEs with energies $\om\lesssim \ombr$
would fragment into very soft gluons propagating at angles larger than the jet
radius (i.e.\ outside the jet).
The medium enhancement of VLEs outside the medium has two main
origins: \texttt{(i)} the
violation of angular ordering by the first emission outside the
medium, which opens the angular phase-space beyond what is allowed in
the vacuum \cite{Mehtar-Tani:2014yea,Caucal:2018dla}, and \texttt{(ii)} the presence of MIEs with $\om> \ombr$
which remain inside the jet and can radiate VLEs outside the medium~\cite{Caucal:2019uvr}.
Our analytic study in Section~\ref{subsub:DLA-FF} and our numerical investigations in the following section
show that
both effects contribute to explaining the enhancement visible in the
MC results.

The above interpretation of the nuclear enhancement at small $x$ as
additional VLEs outside the medium does explain the differences
between the various choices of medium parameters seen in Fig.~\ref{Fig:MC-pheno}.
A smaller value for $L$ increases the energy phase-space for the
parton cascades developing outside the medium because the energy of the first emission
outside the medium, $\omega\sim 2/(L\theta^2)$, with an emission angle
$\theta\le R$, increases with $1/L$.
Furthermore, a larger value of $\amed$ enhances the rate for MIEs and
hence the number of sources for VLEs outside the medium.

Even though our MC results at small $x$ show the same qualitative trend as  
the relevant LHC data \cite{Aaboud:2018hpb}, one must remain cautious when
interpreting this agreement. Indeed, our current formalism lacks some
important physical ingredients, which are known to influence the soft region of
the fragmentation function: the hadronisation and the medium response
to the energy and momentum deposited by the jet. Whereas one may expect the 
effects of hadronisation to at least partially compensate when forming the medium-to-vacuum
ratio $\mathcal{R}(x)$, the medium-response effect --- i.e.\ the fact
that the experimentally reconstructed jets also include soft particles 
originating from the wake of moving plasma trailing behind the jet 
(and not only from the jet itself) --- is clearly missing in our approach and its inclusion
should further enhance the ratio $\mathcal{R}(x)$ at small $x$. Indeed, we know from other
approaches~\cite{Casalderrey-Solana:2016jvj,Tachibana:2017syd,KunnawalkamElayavalli:2017hxo,Chen:2017zte}, where the medium response is the only (or at least the main) mechanism  
for producing such an enhancement, that this effect by itself is comparable with the
enhancement seen in the data (see also \cite{Chien:2015hda} for a different
picture). 

Of course, it is of utmost importance to complete our formalism with a more
realistic description of the medium, including its feedback on the jet.
(We shall return to this point in the concluding section.)
Before such a more complete calculation is actually performed, it 
is difficult to anticipate what should be the combined effect of both mechanisms
on the behaviour of $\mathcal{R}(x)$ at small $x$.

\subsection{Monte Carlo tests}

We would like to extend the DLA arguments exposed in Chapter \ref{chapter:DLApic}, Section \ref{subsub:DLA-FF} to include all the ingredients in our physical picture of jet
quenching.
Our ultimate goal is to provide a deeper understanding of the MC results
presented in Section.~\ref{sec:MCmain} and discussed qualitatively in the previous section. 

For this purpose, it is convenient to think in terms of the factorised
picture emerging from our DLA calculation which allows us to write (for $\om\le\om_{cr}$, cf. \eqn{ocr})
\begin{equation}
\label{factorization}
 \om D^{\textrm{med}}(\om)\simeq \mathcal{N}_{\textrm{in}}\times\Big(\om \frac{\dif N^{\textrm{out}}}{\dif \om}\Big)
\end{equation}
where $\mathcal{N}_{\textrm{in}}$ is the multiplicity of partonic sources produced by the
jet evolution inside the medium and $\om\dif N^{\textrm{out}}/\dif\om$ is the fragmentation function
generated outside the medium by any of these sources.
This picture is a consequence of colour decoherence which allows the
first out-of-medium emission to be emitted at any angle.
This factorisation is not expected to hold beyond DLA, but can still
be used for qualitative considerations.

Beyond DLA, several competing expects should be considered.
\texttt{(i)} 
VLEs are emitted with the full (DGLAP) splitting functions
(including energy conservation) and with a running coupling.
These effects are expected to reduce both factors in
\eqn{factorization}.
\texttt{(ii)} Adding the intra-jet MIEs enhances the
multiplicity $\mathcal{N}_{\textrm{in}}$ of the partonic sources.
\texttt{(iii)} Direct contributions of the MIEs to the fragmentation
function $D^{\textrm{med}}(\om)$ are also possible, but are expected
to be a small effect for the jet kinematics ($p_{T0}\sim 200$~GeV,
$x\le 0.02$) and medium parameters (see Table.~\ref{tab:parameters})
considered in this paper. Indeed, the relevant energies
$\omega\lesssim 2$~GeV are softer than the medium scale
$\ombr\sim 4$~GeV for multiple branching meaning that these MIEs would
be deviated outside the jet.
  
To test these expectations under realistic conditions, we
perform MC simulations for inclusive jets (using the full Born-level
hard spectrum) with $200\le p_T\le 251$ GeV and $|y|\le2.1$, and with
three different scenarios:
\texttt{(a)} the partons from the hard scattering are showered via VLEs only;
\texttt{(b)} the partons from the hard scattering are showered via both VLEs
and MIEs, but angular ordering is enforced all along the shower,
including for the first emission outside the medium (labelled ``no
decoherence'');
\texttt{(c)} the physical case where the partons from the hard scattering are showered via both VLEs and MIEs and the angle of the first emission outside the medium is unconstrained.

\begin{figure}[t] 
  \centering
  \begin{subfigure}[t]{0.48\textwidth}
    \includegraphics[page=1,width=\textwidth]{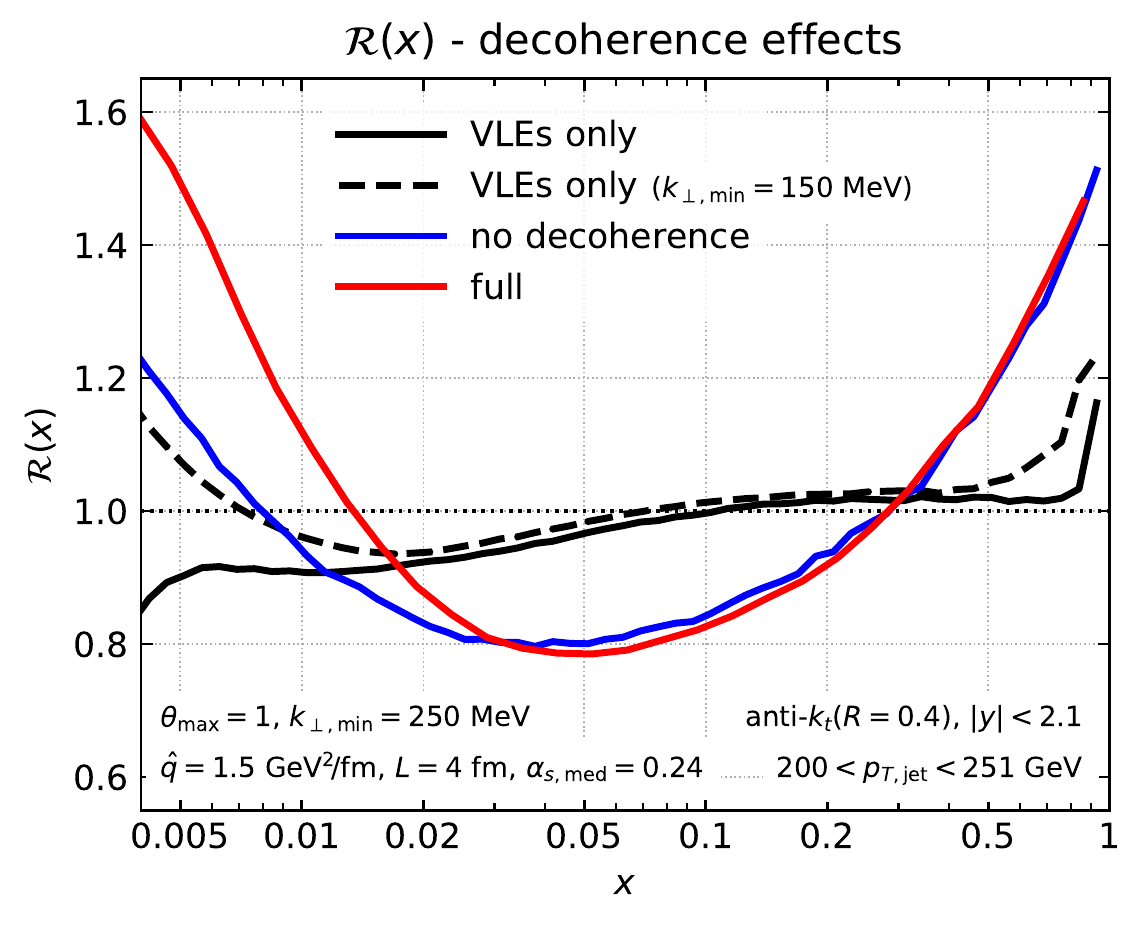}
    \caption{Different physical scenarios: with and without MIEs, and with and without
  violation of angular ordering.}\label{Fig:MC-dec-decoherence}
  \end{subfigure}
  \hfill
  \begin{subfigure}[t]{0.48\textwidth}
    \includegraphics[page=2,width=\textwidth]{./plot-MC-decoherence.pdf}
    \caption{Comparison between quark- and gluon-initiated
      monochromatic jets and ``full'' jets including a  convolution with the initial, hard spectrum.}\label{Fig:MC-dec-spectrum}
  \end{subfigure}
  \caption{\small Nuclear effects on the fragmentation function at
    small $x$. Left figure: 3 different physical scenarios,
  }\label{Fig:MC-dec} 
\end{figure}

The MC results for $\mathcal{R}(x)$ are shown in
Fig.~\ref{Fig:MC-dec-decoherence} for each of these three setups.
The black curves correspond to setup \texttt{(a)} for two 2 different IR
cutoffs (solid: $\ktmin=200$~MeV, dashed: $\ktmin=150$~MeV).
compared to the DLA results in Fig.~\ref{fig:DLA-dec-R} the medium
enhancement is strongly reduced and can even be replaced by a
suppression for larger values of $\ktmin$.

Switching on MIEs leads to a robust nuclear
enhancement as visible from the blue curve which corresponds to setup
\texttt{(b)} with $\ktmin=200$~MeV.
This enhancement is even more pronounced for setup \texttt{(c)}
corresponding to the red curves in Fig.~\ref{Fig:MC-dec-decoherence}.
This new enhancement is easily associated with the fact that the first
``outside'' emission can be sourced by any ``inside'' emissions while
in setup \texttt{(b)} it can only be sourced by ``inside'' emissions
at larger angles.\footnote{For setup \texttt{(b)} the
  factorisation~(\ref{factorization}) is obviously violated as
  ``inside sources'' and ``outside emissions'' are correlated by
  angular ordering.}
Incidentally, the comparison between the blue and the red curves also
shows that the decoherence has no sizeable effects at $x\sim 1$.

For a more detailed understanding, we compare in
Fig.~\ref{Fig:MC-dec-spectrum} the results for $\mathcal{R}(x)$ with
the ratio $\mathcal{R}_i(x|p_{T0})$ corresponding to monochromatic
jets with $p_{T0}= 200$~GeV, for both quark-initiated ($i=q$, magenta,
dashed-dotted curve) and gluon-initiated ($i=g$, green, dashed, curve)
jets.
The small-$x$ enhancement appears to be stronger in the case where the
LP is a quark, rather than a gluon. Although this might look
surprising at first sight, one should recall that the
dominant $C_i$-dependence for monochromatic jets cancels out in the
medium/vacuum ratio $\mathcal{R}_i(x|p_{T0})$.
The differences between the quark and gluon curves visible in
Fig.~\ref{Fig:MC-dec-spectrum} is attributed to more subtle
sub-leading effects.
For example,  a gluon jet loses more energy than a quark jet via MIEs at large
angles and hence has a (slightly) smaller energy phase-space for
radiating outside the medium (and inside the jet).

\section{Jet fragmentation into subjets}\label{sec:frag-subjets}

The fragmentation function defined by Eq.~\eqref{frag-def} is not an 
infrared-and-collinear (IRC) safe observable.
It is sensitive to the details of hadronisation which is not included
in our present approach.
This translates in the strong dependence, observed in
Fig.~\ref{Fig:MCunphys}, on the cut-off scale $\ktmin$
which regulates the infrared behaviour of our partonic cascade.
This strong dependence on $\ktmin$ is also present in the analytic
calculations of sections~\ref{sec:x=1nuc} and~\ref{subsub:DLA-FF}.

To circumvent this theoretical problem, we propose in this section a
different observable which uses subjets instead of individual hadrons
to characterise the jet fragmentation, as explained in Section~\ref{subsub:frag-ISD}.
This observable is IRC-safe by construction and is therefore expected
to be less sensitive to non-perturbative effects in general and to
our $\ktmin$ cut-off in particular.
There are several ways to define a jet fragmentation function in terms
of subjets, e.g.\ using different jet algorithms or keeping different
branches of the clustering tree.
The definition we propose below, as in Section~\ref{subsub:frag-ISD}, relies on the Cambridge/Aachen
algorithm~\cite{Dokshitzer:1997in,Wobisch:1998wt}.
While other approaches, like those based on the $k_t$
algorithm~\cite{Catani:1993hr}, show a similar behaviour,
using the Cambridge/Aachen algorithm appears to be slightly more
sensitive to medium effects and easier to study analytically.

\subsection{Definition and leading-order estimate in the vacuum}

The fragmentation function $\mathcal{D}_{\textrm{sub}}(z)$ for jet
fragmentation into subjets discussed in this section is defined as follows. For a given jet with
transverse momentum $p_{T,\textrm{jet}}$, we iteratively decluster the
jet using the Cambridge/Aachen algorithm following the hardest branch
(in $p_T$).
At each step, this produces two subjets $p_1$ and $p_2$, with
$p_{T1}>p_{T2}$.
When the relative transverse momentum of the splitting, $k_\perp = p_{T2}\sqrt{\Delta
  y_{12}^2+\Delta\phi_{12}^2}$, is larger than a (semi-hard) cut-off
$k_{\perp,\textrm{cut}}$, we compute and record the splitting
fraction $z=\frac{p_{T2}}{p_{T1}+p_{T2}}$ of the splitting ($0 < z < 1/2$).
The procedure is iterated with the harder branch $p_1$ until it can no
longer be de-clustered.
The fragmentation function into subjets is then defined as the density
of subjets passing the $k_\perp>k_{\perp,\textrm{cut}}$ criterion
normalised by the total number of jets:\footnote{We use the notation
  $z$  for the splitting fraction to emphasise that it is defined with respect to
  the parent subjet, in contrast with the longitudinal momentum
  fraction $x$ used in the previous sections which is defined as a
  fraction of the total jet momentum $p_{T,\textrm{jet}}$.}
\begin{equation}\label{def-dsub}
 \mathcal{D}_{\textrm{sub}}(z) \equiv \frac{1}{N_\textrm{jets}}\frac{\dif N_{\textrm{sub}}}{\dif z}
\end{equation}

The cut-off scale $k_{\perp,\textrm{cut}}$ regulates the infrared
behaviour, guaranteeing that $\mathcal{D}_{\textrm{sub}}(z)$ be an
IRC-safe observable.
As long as $k_{\perp,\textrm{cut}}\gg \ktmin\sim \Lambda_{_{\rm QCD}}$
we therefore expect small non-perturbative effects and a small
dependence on the (non-physical) $\ktmin$ parameter.

This definition is very close to the ISD fragmentation function discussed in Chapter~\ref{chapter:jet}, Section~\ref{subsub:frag-ISD}: the only difference is the stopping condition that involves the $k_\perp$ of the splitting. This difference produces negligible effects at leading logarithmic accuracy, so that 
%
in the soft-and-collinear approximation, 
we can use the result obtained in Section \ref{subsub:frag-ISD} for $\mathcal{D}_{\textrm{sub}}(z)$:
\begin{align}\label{fragISDvac2}
  \mathcal{D}^{\textrm{vac}}_{\textrm{sub}}(z)
  & \simeq\left[\int_{0}^R\frac{\dif\theta}{\theta}\,
  \frac{2\alpha_s(z\theta p_{T,\textrm{jet}})}
  {\pi z}\Theta(z\theta p_{T,\textrm{jet}} -
  k_{\perp,\textrm{cut}})\right]
  \times\sum_{i=q,g} C_i\, f_i^{\textrm{vac}}(p_{T,{\textrm{jet}}}),\nn
  & \overset{\text{f.c.}}\simeq 
  \frac{2\alpha_s}{\pi z} \log\left(\frac{z Rp_{T,\textrm{jet}}}{k_{\perp,\textrm{cut}}} \right)
  \times\sum_{i=q,g} C_i\, f_i^{\textrm{vac}}(p_{T,{\textrm{jet}}}),
\end{align}
where
$f_{q(g)}^{\textrm{vac}}(p_{T,{\textrm{jet}}})$ is the Born-level
cross-section for quark (gluon) production with transverse momentum
$p_{T,{\textrm{jet}}}$ normalised to the total number of jets, as
defined in Eq.~\eqref{quenching-factor}.
The second line in the above equation gives the result for a
fixed-coupling approximation.

\subsection{Nuclear modification for $\mathcal{D}_{\textrm{sub}}(z)$: Monte Carlo results}

In this section, we provide Monte Carlo results for the nuclear
modification factor for the fragmentation function into subjets,
defined as
$\mathcal{R}_{\textrm{sub}}(z)\equiv
\mathcal{D}^{\textrm{med}}_{\textrm{sub}}/\mathcal{D}^{\textrm{vac}}_{\textrm{sub}}$.

\begin{figure}[t] 
  \centering
  \includegraphics[page=1,width=0.48\textwidth]{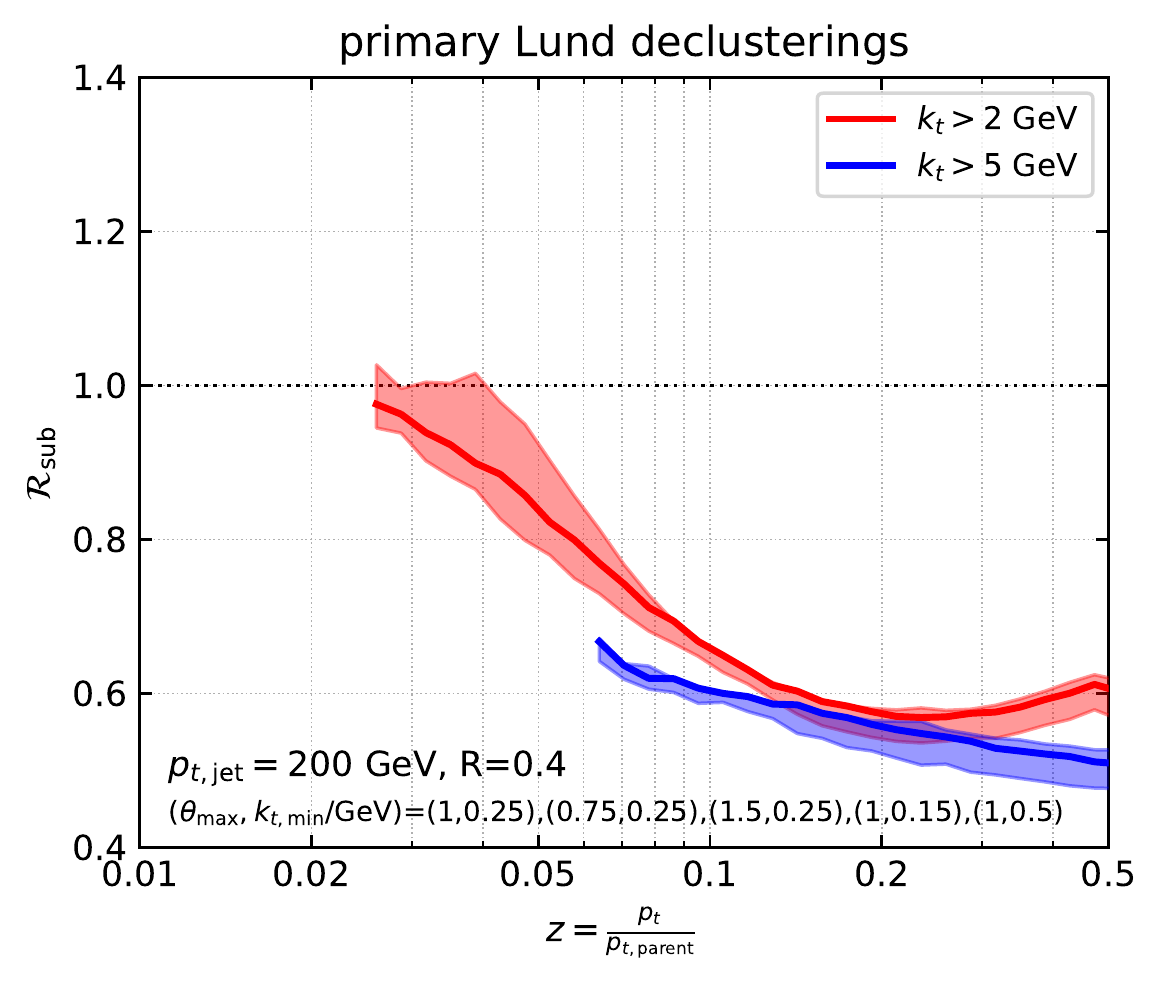}
  \hfill
   \includegraphics[page=2,width=0.48\textwidth]{./plot-subjets.pdf}
  \caption{\small Monte Carlo
  results for the nuclear modification factor  $\mathcal{R}_{\textrm{sub}}(z)$
  for the fragmentation function into subjets, 
  for jets with $p_{T,{\textrm{jet}}}>200$~GeV (left)
  and $p_{T,{\textrm{jet}}}>500$~GeV (right) and for 2 values of the lower momentum
  cut-off $k_{\perp,\textrm{cut}}$ (2 and 5~GeV). The bands show the 
  variability of our results w.r.t. changes  in the ``unphysical'' parameters  around their
central values  $\theta_{\textrm{max}}=1$ and $\ktmin=250$~MeV.}
\label{Fig:fragsubjet} 
\end{figure}

\begin{figure}[t] 
  \centering
  \includegraphics[page=3,width=0.48\textwidth]{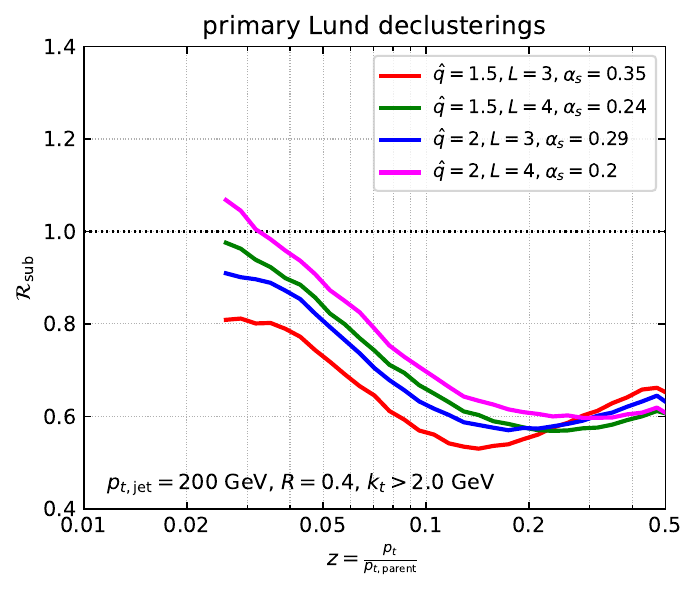}
  \hfill
   \includegraphics[page=5,width=0.48\textwidth]{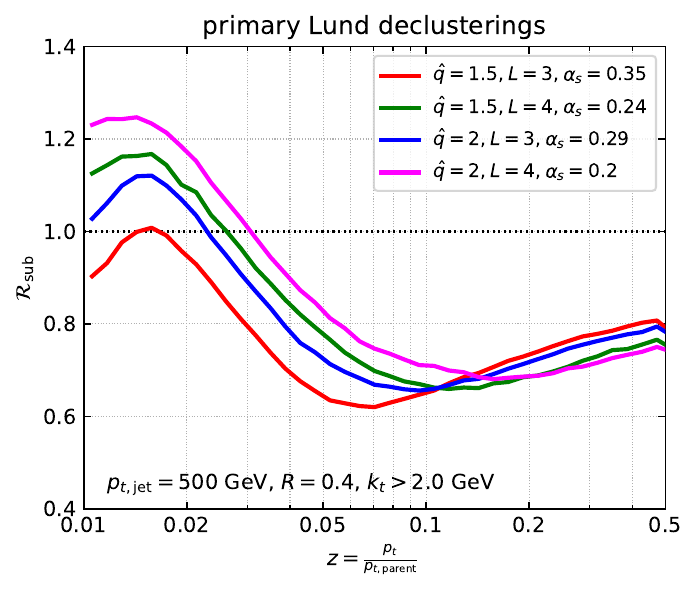}
  \caption{\small  Monte Carlo results for the nuclear modification factor 
  $\mathcal{R}_{\textrm{sub}}(z)$ for the values of the medium parameters that reproduce the
ATLAS $R_{AA}$ ratio (cf. Fig.~\ref{Fig:MC-pheno}), for the same two
ranges in $p_{T,\textrm{jet}}$ as in~Fig.~\ref{Fig:fragsubjet} and for
$k_{\perp,\textrm{cut}}=2$~GeV. The unphysical parameters are fixed to 
$\theta_{\textrm{max}}=1$ and $\ktmin=250$~MeV.
   }
\label{Fig:subjet-med} 
\end{figure}

As for the study of the jet fragmentation function $\mathcal{D}(x)$,
we first study the dependence of the the fragmentation function into
subjets, $\mathcal{D}_{\textrm{sub}}(z)$, on the non-physical
parameters $\theta_{\textrm{max}}$ and $\ktmin$ of our Monte Carlo.
This is shown in Fig.~\ref{Fig:fragsubjet} for two different jet $p_T$
cuts (200 and 500~GeV) and two different lower cut-offs
$k_{\perp,\textrm{cut}}$ (2 and 5~GeV).
The medium parameters are taken as their default values (cf.\
Table~\ref{tab:parameters}) and the non-physical parameters are varied
as for Fig.~\ref{Fig:MCunphys}.
As expected, the uncertainty bands in Fig.~\ref{Fig:fragsubjet} are
much smaller than what was observed in Fig.~\ref{Fig:MCunphys},
confirming that the (IRC-safe) fragmentation function into subjets
$\mathcal{D}_{\textrm{sub}}(z)$ is under much better perturbative
control than (the IRC-unsafe) $\mathcal{D}(x)$.

That said, we must keep in mind that taking $k_{\perp,\textrm{cut}}$
large-enough to guarantee $k_{\perp,\textrm{cut}}\gg \ktmin \sim
\Lambda_{\text{QCD}}$  also cuts some of the medium effects occurring
below this cut.
E.g., it removes the direct contributions to
$\mathcal{D}_{\textrm{sub}}(z)$ coming from MIEs with transverse
momenta $k_\perp\lesssim k_{\perp,\textrm{cut}}$.
One should therefore choose the free parameter
$k_{\perp,\textrm{cut}}$ such as to simultaneously minimise the effects
of hadronisation and highlight the interesting medium effects.

In Fig.~\ref{Fig:subjet-med}, we show the subjet fragmentation
function for the values of the medium parameters that reproduce the
ATLAS $R_{AA}$ ratio (cf. Fig.~\ref{Fig:MC-pheno}), for the same two
values of $p_{T,\textrm{jet}}$ as in~Fig.~\ref{Fig:fragsubjet} and for
$k_{\perp,\textrm{cut}}=2$~GeV.
Compared to Fig.~\ref{Fig:MC-pheno}, we notice that the curves are
less degenerate at small and intermediate values of $z$. Most
importantly, the dependence on the medium parameters is larger than
the uncertainty bands related to non-physical parameters shown
in~Fig.~\ref{Fig:fragsubjet}.

\subsection{Analytic studies of the nuclear effects}
\label{sec:DsubAnalytic}

In this section, we would like to disentangle, based on physics considerations
and simple analytic calculations, the various nuclear effects contributing to the behaviour
observed in the MC results in Fig.~\ref{Fig:subjet-med}.
To understand how  Eq.~\eqref{fragISDvac2} is affected by the medium, it is sufficient
  to consider jets made of a single splitting (i.e.\ two subjets) with
  $k_\perp\ge k_{\perp,\textrm{cut}}$.
For definiteness, all the numerical results shown in this subsection
correspond to $k_{\perp,\textrm{cut}}=2$~GeV.

\begin{figure}[!h] 
\centering
\includegraphics[width=0.45\textwidth]{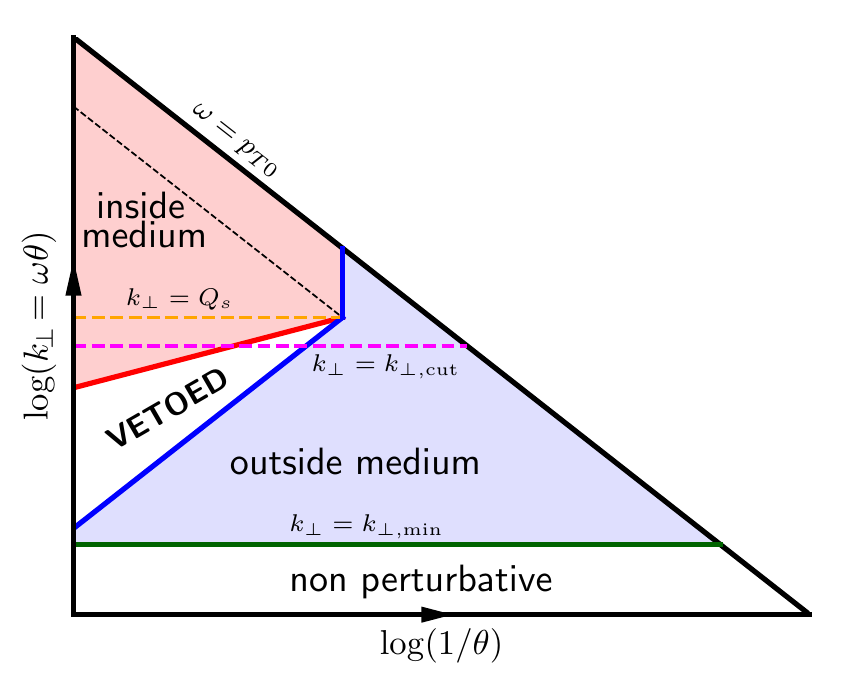}
\caption{\small The phase-space for vacuum-like gluon emissions by a jet propagating
through a dense QCD medium, in logarithmic units. In this plot, the variables are the relative transverse momentum $k_\perp\simeq \omega\theta$ and the inverse of the angle $1/\theta$.}
 \label{Fig:LundPS2}
\end{figure}

\paragraph{Vetoed region.} When only VLEs are taken into account, the
leading medium effect is the vetoed region.
Its effect is straightforwardly
included in \eqn{fragISDvac2} by inserting the 
step-function $\Theta_{\notin\textrm{veto}}$ defined in
Eq.~\eqref{step-veto} within the integrand.
The largest $k_\perp$ in the vetoed region is $Q_s\equiv (2\hat{q} \om_c)^{1/4} =
(\hat q L)^{1/2}$ which is about 2.4~GeV for our default choice of
medium parameters.
The vetoed region has thus no effect for
$k_{\perp,\textrm{cut}}=5$~GeV and only a small effect for
$k_{\perp,\textrm{cut}}=2$~GeV (see
Fig.~\ref{Fig:LundPS2} for an illustration).

\begin{figure}[!h] 
  \centering
  \begin{subfigure}[t]{0.48\textwidth}
    \includegraphics[page=2,width=\textwidth]{./frag-ISD-analytic.pdf}
    \caption{Anaytic results}\label{Fig:fragISD-LL-analytic}
  \end{subfigure}
  \hfill
  \begin{subfigure}[t]{0.48\textwidth}
    \includegraphics[page=1,width=\textwidth]{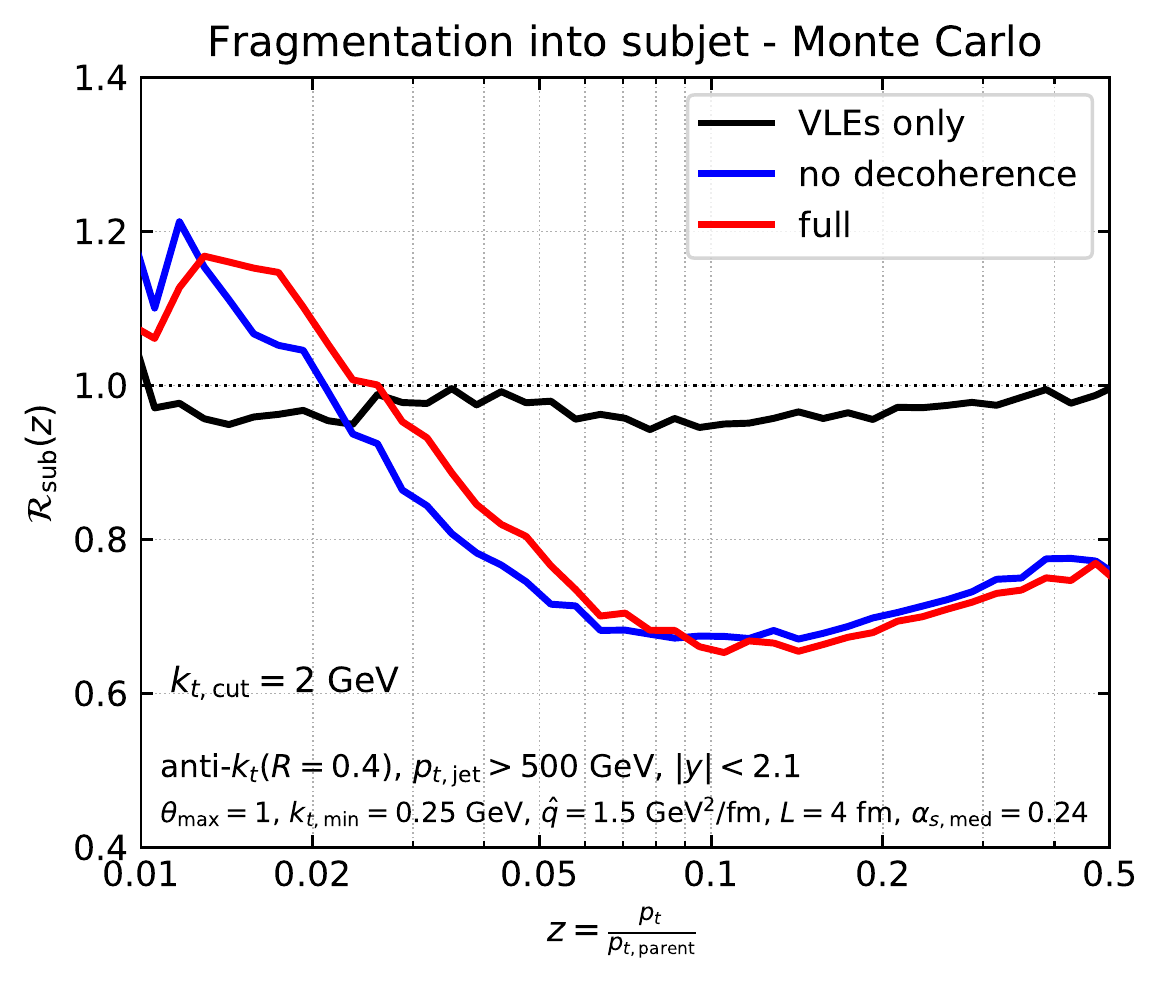}
    \caption{Results from Monte Carlo simulations}\label{Fig:fragISD-LL-mc}
  \end{subfigure}
  \caption{\small  Disentangling nuclear effects on the subjet fragmentation function.
  Left: analytic approximations illustrating the effects of the vetoed region, the energy
  loss at large angles, and the intra-jet MIEs. Right: MC calculations which illustrate
  the importance of  MIEs and the lack of sensitivity to violations of angular ordering.}
\label{Fig:fragISD-LL} 
\end{figure}

This is confirmed both by our analytic calculations, based on \eqn{fragISDvac2} 
with the additional constraint $\Theta_{\notin\textrm{veto}}$, 
and by MC simulations with only VLEs shown as the black curves in
Fig.~\ref{Fig:fragISD-LL}.
Of course,
one could enhance the effect of the vetoed region by decreasing the value of $k_{\perp,\textrm{cut}}$,
but this would also amplify the sensitivity of $\mathcal{D}_{\textrm{sub}}(z)$  
to the non-perturbative, soft, emissions.

Incidentally, the previous discussion also shows that, for the ranges
of $k_{\perp,\textrm{cut}}$ considered here, the VLEs which control
$\mathcal{D}_{\textrm{sub}}(z)$ do either occur in the ``inside''
region of the phase-space in Fig.~\ref{Fig:LundPS2}, or at very small
angles $\theta\lesssim \theta_c$ in the ``outside'' region. They are
therefore not significantly affected by colour decoherence. To check
that, we have performed MC calculations with and without the effects
of decoherence (i.e.\ by enforcing or not angular ordering for the first
outside emission). The results, shown by the red and blue
curves in Fig.~\ref{Fig:fragISD-LL-mc}, respectively, are indeed
very close to each other.

\paragraph{Energy loss at large angles.}   
From the discussion in Sect.~\ref{sec:x=1nuc}, we already know
that the energy loss by a (sub)jet via MIEs at large angles $\theta\gtrsim R$ may have two main effects
on a substructure observable such as  $\mathcal{D}_{\textrm{sub}}(z)$: \texttt{(i)} a shift between the measured
value $z$ of the splitting fraction and the respective value at the time of splitting, and  \texttt{(ii)} a bias introduced by the steeply falling initial spectrum which favours jets losing less energy 
than average jets, with the second effect being larger than the first one.
The same two effects are still at play for $\mathcal{D}_{\textrm{sub}}(z)$. As in the case of the
standard fragmentation function discussed  in Sect.~\ref{sec:x=1nuc}, we expect the effects of the energy
loss to be more important for relatively large values $z\gtrsim 0.1$ of the splitting fraction. However,
their effects on $\mathcal{R}_{\textrm{sub}}(z)$ is opposite to those on $\mathcal{R}(x)$: unlike the hard-fragmenting jets, which lose {\it less} energy than
the average jets (leading to an enhancement in $\mathcal{R}(x)$ at $x\gtrsim 0.5$), the 
jets selected by $\mathcal{D}_{\textrm{sub}}(z)$ lose {\it more} energy than the average jets, so we expect
a {\it nuclear suppression}, $\mathcal{R}_{\textrm{sub}}(z)< 1$, at sufficiently large $z$.
The main reason for this larger energy loss is the following: the jets included in $\mathcal{D}_{\textrm{sub}}(z)$
involve at least two (relatively hard) subjets with  $z\gtrsim 0.1$ and $k_\perp > k_{\perp,\text{cut}}$.
For the typical values of $z$ and $k_\perp$, the angle $\theta\simeq k_\perp/p_{T2}$ between
these two subjets is larger than the critical angle $\theta_c$ characterising the angular resolution
of the plasma ($\theta_c\lesssim 0.06$, see Table~\ref{tab:parameters}).
Accordingly the two subjets lose energy independently from each other and the whole jet
loses more energy than a typical jet from the inclusive sample $N_\textrm{jets}$
\cite{Mehtar-Tani:2017ypq,Caucal:2019uvr} which also includes single-prong jets, as well as two-prong configurations with $\theta <
\theta_c$.

This discussion is in qualitative agreement with the MC results in
Fig.~\ref{Fig:fragISD-LL-mc}, except at very small $z$ where new
effects discussed below contribute. 
For a more quantitative argument, 
we notice that, if one neglects the
shift in the value of $z$, then the energy loss at large angles
affects only the quark- and gluon-jet ``fractions'' $f_i^{\textrm{med}}$
in~\eqn{fragISDvac2}. These should be computed following
Eq.~\eqref{quenching-factor}, with different energy losses in the
numerator and respectively the denominator.
In the numerator, $\mathcal{E}_i^{n=2}$ is the
energy loss of jets having two subjets with transverse
momentum balance $z$ and angle $\theta$ ($p_T\equiv p_{T,\textrm{jet}}$)
\begin{equation}
\label{incoh-subjet-eloss}
 \mathcal{E}_i^{n=2}(z,\theta)=\mathcal{E}_i((1-z)p_{T},R)+\mathcal{E}_g(z p_{T},R) \qquad
 \textrm{ if }(z,\theta)\in\textrm{ inside region,}
\end{equation}
whereas in the denominator, $ \mathcal{E}_i= \mathcal{E}_i(p_{T},R)$.
Using the energy loss as a function of $p_T$ and $R$ extracted from the MC
simulations in Ref.~\cite{Caucal:2019uvr} in
Eqs.~\eqref{quenching-factor} and~\eqn{fragISDvac2}, one obtains the
dashed, green, curve in Fig.~\ref{Fig:fragISD-LL-analytic}.
This indeed shows a nuclear suppression,
$\mathcal{R}_{\textrm{sub}}(z)<1$. The suppression is more pronounced
at large $z$, as anticipated,
 since the discrepancy (in terms of energy loss) between the special jets selected by 
$\mathcal{D}^{\textrm{med}}_{\textrm{sub}}(z)$ and the average jets increases with $z$.

\paragraph{Intrajet MIEs.} A relatively hard subjet with $k_\perp > k_{\perp,\textrm{cut}}$ may also be 
created by a semi-hard MIE, with energy $\om\gtrsim\ombr$, which remains inside the jet.
To leading order, the respective contributions from VLEs and MIEs can be simply added together,
as in \eqn{Elossred}.
Compared to the latter, the calculation of
$\mathcal{D}^{\textrm{med}}_{\textrm{sub}}(z)$ must also keep the
information about the emission angle, in order to ensure the condition
$k_{\perp}>k_{\perp,\textrm{cut}}$.
We therefore write
\begin{equation}
\label{frag-isd-tot}
 \mathcal{D}^{\textrm{med}}_{\textrm{sub}}(z)=\left[
\int_{0}^R\dif\theta\left(\frac{2\alpha_s(k_\perp)}{\pi z \theta}\Theta_{\notin\textrm{veto}}+\sqrt{\frac{2\omega_c}{p_{T,\textrm{jet}}}}\frac{\alpha_{s,\textrm{med}}}{\pi z^{3/2}}\mathcal{P}_{\rm broad}(z,\theta)\right)\Theta(k_\perp-k_{\perp,\textrm{cut}})\right]\sum_{i=q,g} C_i f_i^{\textrm{med}}
\end{equation}
where $k_\perp=z\theta p_{T,\textrm{jet}}$ and
$\mathcal{P}_{\rm broad}(z,\theta)$ is the
angular distribution due to transverse momentum broadening after
emission calculated in \ref{subsub:zg-mie} and given explicitly by \eqref{pbroad}.
In writing \eqn{frag-isd-tot}, we have assumed for simplicity that the
energy loss at large angles is given by \eqn{incoh-subjet-eloss} for
both the vacuum-like and medium-induced emissions that generates the
subjets.
This rough approximation could be relaxed in practice,
but is sufficient for our illustrative purposes.
The distribution $\mathcal{P}_{\rm broad}(z,\theta)$ for MIEs is
rather strongly peaked near $k_\perp\sim Q_s$ (recall the discussion for the $z_g$ distribution in \ref{sec:low}) so
its corresponding contribution to \eqn{frag-isd-tot} is expected to be
important only when $k_{\perp,\textrm{cut}}\lesssim Q_s$, in which
case it should be rapidly increasing at small $z$.
This is in agreement with the MC results in
Figs.~\ref{Fig:fragsubjet} and~\ref{Fig:subjet-med}, which show an
enhancement at small $z$ for $k_{\perp,\textrm{cut}}=2$~GeV and no
visible enhancement for $k_{\perp,\textrm{cut}}=5$~GeV.
  
\eqn{frag-isd-tot} includes all the medium effects discussed in this
section.
The red curve in Fig.~\ref{Fig:fragISD-LL-analytic} shows the result
of numerically evaluating the integral in~\eqn{frag-isd-tot}. The new
enhancement at small $z$ compared to the dashed, green, curve is due
to the intrajet MIEs.
The overall behaviour agrees well with the full MC results shown in
Fig.~\ref{Fig:fragISD-LL-mc} as well as with
Figs.~\ref{Fig:fragsubjet} and~\ref{Fig:subjet-med}.


%% file: conclusion.tex
\chapter{Conclusion}
\label{chapter:conclusion}

In this thesis, we have presented a new picture, emerging from
perturbative QCD, for the parton showers created by an energetic parton
propagating through a dense quark-gluon plasma. This picture is
factorised in time with vacuum-like emissions occurring first and
creating sources for subsequent medium-induced radiations.
Both types of emission are Markovian processes, yielding a modular
Monte Carlo implementation of our picture in the parton shower {\tt JetMed}. This allows us to study
separately various aspects of the dynamics of jet quenching and assert
their relative importance in jet observable measured at the LHC in heavy-ion collisions.  

We have focused on three observables for which we believe
our approximations to be robust: the jet nuclear modification factor
for $R_{AA}$, the nuclear effects on the $z_g$ distribution given
by the Soft Drop procedure and the jet fragmentation function (and notably its new version infrared and collinear safe).
For all these observables, we obtained good qualitative and
semi-quantitative descriptions of the respective LHC data and
we discussed the physical interpretation of the various trends seen in
the data. To make our physical discussions more convincing, we supplemented the Monte Carlo calculations with suitable analytic calculations, which were helpful to pinpoint the different mechanisms
at play and compare their effects.

In the remaining part of this conclusion, we outline some of our projects for the future. As acknowledged in Section~\ref{sub:beyondDLA}, the current leading-logarithmic picture for the \textit{vacuum-like} evolution has several limitations. Accordingly, the parton shower {\tt JetMed} is currently rather simplistic and we have provided a detailed list of possible improvements in Section~\ref{sec:MCcomp}. We point out however that our step-by-step approach is motivated by the following guideline: in the vacuum, the accuracy of a parton shower means logarithmic accuracy. We believe that an in-medium parton shower should also start with a correct resummation of the large logarithms.
Thus our first step has consisted in resumming all leading-logarithmic contributions associated with vacuum-like emissions in the presence of a dense QCD medium. Then, we have included the dominant (higher-twist) medium-induced emissions through an effective branching process that encompasses the multiple soft branching regime. 

At this stage, the list of physical ingredients that will refine our calculations of jet quenching observables is large and we refer the reader to Table~\ref{Tab:MCs} for a ranked list of such ingredients. Including the medium expansion is the next step of our project, as well as the emissions \`{a} la GLV due to single hard scatterings.

Instead of going through the complex task of completing our Monte
Carlo with a detailed description of the medium and of its
interactions with the jet, one can alternatively think about using our
parton showers as an input for the recently developed
JETSCAPE~\cite{Cao:2017zih,Putschke:2019yrg} framework, which offers various
approaches for treating the interactions between the parton shower and
the medium.

%% file: appendix.tex
\addcontentsline{toc}{part}{Appendices}

\chapter{Propagators and averaged propagators in the background field $\mathcal{A}_m$}
\chaptermark{Propagators and averaged propagators}
\label{app:prop}

\section{Gluon propagators beyond the eikonal approximation}

In this section, we give all the components of the gluon propagator $G^{\mu\nu}{ab}(z,x|\mathcal{A}_m)$ in the background field $\mathcal{A}_m$ defined in \eqref{Gprop} in terms of the scalar propagator $\mathcal{G}$ defined by \eqref{scalar-prop-subei}.
\begin{align}
 G^{ij}_{ab}(z,x|\mathcal{A}_m)&=\int\frac{\dif k^+}{2\pi}\frac{e^{-ik^+(z^--x^-)}}{2k^+}\,\delta^{ij}\mathcal{G}_{ab}(z^+,z_\perp;x^+,x_\perp|k^+)\\
 G^{-i}_{ab}(z,x|\mathcal{A}_m)&=\int\frac{\dif k^+}{2\pi}\frac{e^{-ik^+(z^--x^-)}}{2k^+}\,\frac{-i}{k^+}\partial^i_{z_\perp}\mathcal{G}_{ab}(z^+,z_\perp;x^+,x_\perp|k^+)\\
  G^{i-}_{ab}(z,x|\mathcal{A}_m)&=\int\frac{\dif k^+}{2\pi}\frac{e^{-ik^+(z^--x^-)}}{2k^+}\,\frac{i}{k^+}\partial^i_{x_\perp}\mathcal{G}_{ab}(z^+,z_\perp;x^+,x_\perp|k^+)\\
    G^{--}_{ab}(z,x|\mathcal{A}_m)&=\int\frac{\dif k^+}{2\pi}\frac{e^{-ik^+(z^--x^-)}}{2k^+}\,\left(\frac{1}{{k^+}^2}\partial^i_{z_\perp}\partial^i_{x_\perp}\mathcal{G}_{ab}(z^+,z_\perp;x^+,x_\perp|k^+)\right.\nonumber\\
    &\hspace{4cm}\left.+\frac{2i}{k^+}\delta_{ab}\delta(z^+-x^+)\delta^{(2)}(z_\perp-x_\perp)\right)
\end{align}
with $k^+\rightarrow k^++i\epsilon$ for retarted propagators. We refer to \cite{Iancu:2000hn} or \cite{Iancu:2014kga} for a derivation of these formulas.

\section{The broadening factor $S_{gg}$ for Gaussian correlators}

We calculate now the following average over the background field configurations, often called the ``2-point'' function $S_{z^+,y^+}^{(2)}(\bar{z}_\perp,a_\perp;z_\perp,b_\perp)$
\begin{equation}\label{def-2point}
 S_{z^+,y^+}^{(2)}(\bar{z}_\perp,a_\perp;z_\perp,b_\perp)\equiv \frac{1}{N_c^2-1}\Big\langle \Tr\, \mathcal{G}^{\dagger}(z^+,\bar{z}_\perp;y^+,a_\perp|k^+)\mathcal{G}(z^+,z_\perp;y^+,b_\perp|k^+)\Big\rangle
\end{equation}
We follow the derivation presented in the Appendix of \cite{MehtarTani:2006xq}.
This 2-point function appears in the calculation of the average $\tilde{S}_{gg}$ defined in \eqref{def-Sgg}:
\begin{equation}
 \tilde{S}_{gg}(k_\perp,a_\perp,b_\perp)=\lim\limits_{z^+\rightarrow\infty}\int \dif z_\perp\dif \bar{z}_\perp\,e^{-ik_\perp(a_\perp-b_\perp)} e^{-ik_\perp(z_\perp-\bar{z}_\perp)}S_{z^+,y^+}^{(2)}(\bar{z}_\perp,a_\perp;z_\perp,b_\perp)
\end{equation}
Using the path integral representation \eqref{scalar-prop-subei} and the general result for the average of two Wilson lines \eqref{dipolecross-general} for Gaussian correlators, one finds the following path integral representation of $S^{(2)}$:
\begin{align}\label{Sgg-from2point}
 S_{z^+,y^+}^{(2)}(\bar{z}_\perp,a_\perp;z_\perp,b_\perp)&=\int\mathcal{D}r_\perp(\xi)\mathcal{D}s_\perp(\xi)\exp\left(\int_{y^+}^{z^+}\dif \xi \,\frac{ik^+}{2}(\dot{s}_\perp^2-\dot{r}_\perp^2)\right.\nonumber\\
 &\hspace{5cm}\left.-\frac{g^2}{2}C_A\int_{y^+}^{z^+}\dif \xi\,n(\xi)\sigma_d\big(s_\perp-r_\perp\big)\right)
 \end{align}
It is quite remarkable that this path integral can be calculated exactly for any dipole cross-section $\sigma_d(r_\perp)$.
The calculation starts with a change of variable of determinant one $u_\perp=s_\perp-r_\perp$, $v_\perp=(r_\perp+s_\perp)/2$ to take advantage of the translational invariance of the average of two Wilson lines:
 \begin{align}
S_{z^+,y^+}^{(2)}(\bar{z}_\perp,a_\perp;z_\perp,b_\perp)=\int\mathcal{D}u_\perp(\xi)\mathcal{D}v_\perp(\xi)\exp\left(\int_{y^+}^{z^+}\dif \xi \,ik^+\dot{u}_\perp\dot{v}_\perp-\frac{g^2}{2}C_An(\xi)\sigma_d\big(u_\perp\big)\right)\label{path-1}
 \end{align}
 In terms of these variables, the boundaries of the path integral are $u(y^+)=b_\perp-a_\perp$, $u(z^+)=z_\perp-\bar{z}_\perp$, $v(y^+)=(a_\perp+b_\perp)/2$, $v(z^+)=(z_\perp+\bar{z}_\perp)/2$.
 Integrating by part the $v_\perp$ dependent integral enables to remove the derivative of $v_\perp$:
\begin{equation}
 \int_{y^+}^{z^+}\dif \xi \,\dot{u}_\perp\dot{v}_\perp=\frac{z_\perp+\bar{z}_\perp}{2}\dot{u}_\perp(z^+)-\frac{a_\perp+b_\perp}{2}\dot{u}_\perp(y^+)-\int_{y^+}^{z^+}\dif \xi \,\ddot{u}_\perp v_\perp
\end{equation}
so that, by discretising the path integral over $v_\perp(\xi)$, one shows:
\begin{equation}
\int\mathcal{D}v_\perp(\xi)\exp\left(-ik^+\int_{y^+}^{z^+}\dif \xi \,\ddot{u}_\perp v_\perp\right)=\mathcal{N}\delta(\ddot{u}_\perp(\xi))
\end{equation}
up to a normalisation factor $\mathcal{N}$ to be determined afterwards. The $\delta(\ddot{u}_\perp)$ factor means that $u_\perp$ is fixed to the classical path $u^{\rm cl}_\perp(\xi)$, solution of $\ddot{u}_\perp(\xi)=0$ given the boundaries conditions:
\begin{equation}
 u^{\rm cl}_\perp(\xi)=\frac{1}{z^+-y^+}\Big((z_\perp-\bar{z}_\perp)(\xi-y^+)+(b_\perp-a_\perp)(z^+-\xi)\Big)
\end{equation}
As the $u_\perp$ path is fixed in the remaining $u_\perp$ integral \eqref{path-1} by a delta function of the form $\delta(u_\perp-u^{\rm cl}_\perp)$, the integration is trivial:
 \begin{align}
 S_{z^+,y^+}^{(2)}(\bar{z}_\perp,a_\perp;z_\perp,b_\perp)&=\mathcal{N}\exp\left(\frac{ik^+}{2(z^+-y^+)}\big((\bar{z}_\perp-a_\perp)^2-(z_\perp-b_\perp)^2\big)\right.\nonumber\\
 &\hspace{6cm}\left.-\frac{g^2}{2}C_A\int_{y^+}^{z^+} \dif \xi\, n(\xi)\sigma_d(u^{\rm cl}_\perp)\right)
  \end{align}
with again, a normalisation factor absorbed in $\mathcal{N}$.
This factor is finally inferred from the $g\rightarrow0$ limit where the two-point function should reduce to the product of two free propagators. Using the expression of the free transverse propagator $\mathcal{G}_0$ given in \eqref{G0transverse} plugged in \eqref{def-2point}, one finds:
\begin{equation}
 \mathcal{N}=\left(\frac{k^+}{2\pi(z^+-y^+)}\right)^2
\end{equation}

In order to find the function $\tilde{S}_{gg}$, one must take the Fourier transform as defined in \eqref{Sgg-from2point}. The following change of variable considerably simplifies the calculation:
\begin{equation}
 p_\perp=\bar{z}_\perp-a_\perp-z_\perp+b_\perp\,\qquad q_\perp=\bar{z}_\perp-a_\perp+z_\perp-b_\perp
\end{equation}
since the function $\tilde{S}_{gg}$ now reads:
\begin{align}
 \tilde{S}_{gg}(k_\perp,a_\perp,b_\perp)&=\mathcal{N}\int\dif^2p_\perp\int\dif^2q_\perp e^{-ik_\perp p_\perp}\nonumber\\
 &\hspace{-0.5cm}\times\exp\left(\frac{ik^+}{2(z^+-y^+)}p_\perp q_\perp-\frac{g^2}{2}C_A\int_{y^+}^{z^+} \dif \xi\, n(\xi)\sigma_d\Big(\frac{\xi-y^+}{z^+-y^+}p_\perp+a_\perp-b_\perp\Big)\right)\\
 &=\exp\left(-\frac{g^2}{2}C_A\int_{y^+}^{z^+} \dif \xi\, n(\xi)\sigma_d(a_\perp-b_\perp)\right)
\end{align}
which is precisely \eqref{Sgg} when $z^+\rightarrow\infty$.

\section{The effective propagator $\mathcal{K}$ in the harmonic approximation}

We give first the derivation of formula \eqref{def-K}, valid for Gaussian correlators for the background field. Our starting point is the function $\mathcal{K}_{qg}$ defined by
\begin{equation}
 \mathcal{K}_{qg}(y^+,a_\perp;\bar{y}^+,b_\perp|c_\perp)\equiv\frac{1}{N_c^2-1}\Big\langle \Tr\,\mathcal{G}^{\dagger}(y^+,a_\perp;\bar{y}^+,b_\perp|k^+)\mathcal{W}_{\bar{y}^+}^{y^+}(c_\perp)\Big\rangle
\end{equation}
Using the path integral representation of the gluon propagator \eqref{scalar-prop-subei} and the formula for the average of two Wilson lines \eqref{dipolecross-general}, one finds the following path integral representation of the effective propagator $\mathcal{K}_{qg}$:
\begin{equation}
 \mathcal{K}_{qg}(y^+,a_\perp;\bar{y}^+,b_\perp|c_\perp)=\int_{r_\perp(\bar{y}^+)=b_\perp}^{r_\perp(y^+)=a_\perp}\mathcal{D}r_\perp(\xi)\exp\left(-\int_{\bar{y}^+}^{y^+}\dif\xi\frac{i k^+}{2}\dot{r}_\perp^2(\xi)+\frac{g^2C_A}{2}n(\xi)\sigma(r_\perp-c_\perp)\right)
\end{equation}
After the shift in the path integral $r_\perp\rightarrow r_\perp+c_\perp$, one gets \eqref{kqg-shift} and \eqref{def-K}. In the harmonic approximation, the latter expression reduces to the following Gaussian path integral:
\begin{equation}\label{K-harmonic}
  \mathcal{K}(y^+,a_\perp;\bar{y}^+,b_\perp)=\int_{r_\perp(\bar{y}^+)=b_\perp}^{r_\perp(y^+)=a_\perp}\mathcal{D}r_\perp(\xi)\exp\left(-\int_{\bar{y}^+}^{y^+}\dif\xi\frac{i k^+}{2}\dot{r}_\perp^2(\xi)+\frac{1}{4}\qhat_A(\xi)r_\perp^2\right)
\end{equation}

An explicit formula for $\mathcal{K}$ can be found using the Gel'fand and Yaglom formula \cite{Gelfand:1959nq} for Gaussian path integrals. Given a quadratic action:
\begin{equation}
 S[x(t)]=\frac{1}{2}\int_{t_0}^{t_1}\dif t' \big(m\dot{x}^2-c(t')x^2\big),
\end{equation}
the Gaussian path integral built from it reads:
\begin{equation}\label{Yaglom}
 \int_{x(t_0)=x_0}^{x(t_1)=x_1}\mathcal{D}x(t)\exp(iS[x(t)])=\left(\frac{m}{2\pi if(t_1,t_0)}\right)^{1/2}\exp\Big(iS[x_{\rm cl}(t)]\Big)
\end{equation}
where the function $f(t,t_0)$ satisfies the following linear differential equation:
\begin{equation}
 \frac{\partial^2f}{\partial t^2}+\frac{c(t)}{m}f=0\,,\qquad f(t_0,t_0)=0\,,\qquad \partial_{t=t_0}f(t,t_0)=1
\end{equation}
and $x_{\rm cl}$ is the classical path, extremum of the action $S$, with the boundaries conditions $x_{\rm cl}(t_0)=x_0$, $x_{\rm cl}(t_1)=x_1$.

From \eqref{Yaglom}, one can find an analytic form for the effective propagator. The dictionary that translates the variable used in \eqref{Yaglom} and those used in \eqref{K-harmonic} is:
\begin{equation}
 m\leftrightarrow-k^+\,,\qquad c(t)\leftrightarrow -i\frac{\qhat_A(\xi)}{2}
\end{equation}
The classical path $r^{\rm cl}_\perp$ satisfies the differential equation:
\begin{equation}\label{diff-rcl}
 \frac{\dif^2r^{\rm cl}_\perp}{\dif\xi^2}+\frac{i\qhat_A(\xi)}{2k^+}r^{\rm cl}_\perp=0\,,\qquad r^{\rm cl}_\perp(\bar{y}^+)=b_\perp\,,\qquad r^{\rm cl}_\perp(y^+)=a_\perp
\end{equation}
so that the $S[x^{\rm cl}_\perp(t)]$ reads:
\begin{equation}
 S[x^{\rm cl}_\perp(t)]=\frac{-k^+}{2}\left[r_\perp^{\rm cl}(\xi)\frac{\dif r_\perp^{\rm cl}}{\dif\xi}\right]_{\xi=\bar{y}^+}^{\xi=y^+}
\end{equation}
Combining everything together, without forgetting that the path integral \eqref{K-harmonic} runs over a 2-dimensional path, one finds precisely the formula \eqref{Kqg}:
\begin{equation}
 \mathcal{K}(y^+,a_\perp;\bar{y}^+,b_\perp)=\frac{-k^+}{2\pi iS(y^+,\bar{y}^+)}\exp\left(\frac{-ik^+}{2S(y^+,\bar{y}^+)}\Big(C(\bar{y}^+,y^+)a_\perp^2+C(y^+,\bar{y}^+)b_\perp^2-2a_\perp b_\perp\Big)\right)
\end{equation}
where the functions $S$ and $C$ have been introduced in the main text as independent scalar solutions of the differential equation \eqref{diff-rcl}.

\chapter{Medium-induced spectra for finite jet path length}
\label{app:B}

We provide here the on-shell and off-shell medium induced spectra in the case where the incoming parton has a finite path length $L$ through the medium. Such a finite path length can be implemented with a discontinuous $\qhat(t)$ such that $\qhat(t)=0$ for $t\ge x_0^++L$. However, in this case, one has to deal with the junction at $t=L$ of the solutions $S(t,t_0)$ and $C(t,t_0)$ of the differential equation \eqref{diff-eq-S} in such a way that $S$ and $C$ are continuous and derivable in their first argument.
A way to circumvent this calculation is to divide the amplitude into three parts, with either $y^+\le x_0^+$, or $x_0^+\le y^+\le L+x_0^+$, or $y^+\ge x_0^++L$. Then, the full on-shell cross-section becomes a sum of six terms (the off-shell one is a sum of three terms). For each term, one inserts the Chapman-Kolmogorov relation in order to use the effective propagator $\mathcal{K}$ on open sets where the form \eqref{Kqg} is valid, with $S$ and $C$ the solutions of the differential equation \eqref{diff-eq-S}.

The on-shell and off-shell spectra are given by the following sums:
\begin{equation}
 k^+\frac{\dif^3 N^ {\textrm{on-shell}}}{\dif k^+\dif^2k_\perp}  =  \frac{\alpha_s C_R}{\pi^2}\sum_{\substack{m=b,i \\n=b,i,o}} I^{m/n}\,,\qquad
 k^+\frac{\dif^3 N^ {\textrm{off-shell}}}{\dif k^+\dif^2k_\perp} =  \frac{\alpha_s C_R}{\pi^2}\sum_{\substack{m=i \\n=i,o}} I^{m/n}
\end{equation}
where the indices $b$, $i$ and $o$ stands respectively for ``before'', ``in'' and ``out'' according to the integration domain.

\section{Analytical results for $I^{m/n}$}

The analytical results of the six terms $I^{m/n}$, in the multiple soft scattering approximation are:

\begin{align}\label{I-all-bef}
 I^{b/b}&=\frac{4\pi}{Q_s^2}\,\int\frac{d^2q_\perp}{(2\pi)^2}\frac{1}{q_\perp^2}\exp\left(\frac{-(k_\perp-q_\perp)^2}{Q_s^2}\right)\\
 I^{b/i}&=2\mathfrak{Re}\int_{x_0^+}^{x_0^++L} \dif y^+\frac{\bar{G}(y^+)}{Q_s^2(y^+)+2ik^+\bar{G}(y^+)}\exp\left(\frac{-k_\perp^2}{Q_s^2(y^+)+2ik^+\bar{G}(y^+)}\right)\\
 &-2\mathfrak{Re}\int_{x_0^+}^{x_0^++L} \dif y^+\frac{G(y^+)}{Q_s^2(y^+)+2ik^+G(y^+)}\exp\left(\frac{-k_\perp^2}{Q_s^2(y^+)+2ik^+G(y^+)}\right)\\
 I^{b/o}&=\frac{2}{k_\perp^2}\mathfrak{Re}\left[\exp\left(\frac{-k_\perp^2}{2ik^+\bar{G}(x_0^++L)}\right)-\exp\left(\frac{-k_\perp^2}{2ik^+G(x_0^++L)}\right)\right]
 \end{align}
\begin{align}\label{I-all-in}
 I^{i/i}&=2\mathfrak{Re}\int_{x_0^+}^{x_0^++L} \dif y^+\frac{G(y^+)}{Q_s^2(y^+)+2ik^+G(y^+)}\exp\left(\frac{-k_\perp^2}{Q_s^2(y^+)+2ik^+G(y^+)}\right)\\
 I^{i/o}&=\frac{2}{k_\perp^2}\mathfrak{Re}\left[\exp\left(\frac{-k_\perp^2}{2ik^+G(x_0^++L)}\right)-1\right]\\
 I^{o/o}&=\frac{1}{k_\perp^2}
\end{align}
where $Q_s=Q_s(x_0^+)$.

Note than one cannot recover the formula \eqref{onshell-final} by taking the limit $L\rightarrow\infty$ in these formulas. This is because of the adiabatic switching prescription which is not included in the integrals over compact domains.

\section{Integrated medium-induced spectrum}

In this thesis, we have presented two ways of defining the medium-induced spectrum. The first way, which is standard in the literature is to define:
\begin{align}
 k^+\frac{\dif N^{\mie}}{\dif k^+}&\equiv\int\dif^2k_\perp\left(k^+\frac{\dif^3 N^ {\textrm{off-shell}}}{\dif k^+\dif^2k_\perp}-\frac{\alpha_sC_R}{\pi^2}\frac{1}{k_\perp^2}\right)\\
 &=\frac{\alpha_sC_R}{\pi^2}\int\dif^2k_\perp\,(I^{i/i}+I^{i/o})
\end{align}
Each term in this sum leads to a diverging integral, but the divergence cancels exactly in the sum. To see this, it is convenient to write $I^{i/o}$ as a time integral:
\begin{equation}
 I^{i/o}=\frac{i}{k^+}\int_{x_0^+}^{x_0^+ +L}\dif \bar{y}^+\,\frac{1}{C^2(\bar{y}^+,x_0^+ +L)}\exp\left(\frac{ik_\perp^2S(x_0^++L,\bar{y}^+)}{2k^+ C(\bar{y}^+,x_0^++L)}\right)
\end{equation}
Integrating over $k_\perp$ both $I^{i/i}$ and $I^{i/o}$, one finds after a bit of algebra:
\begin{equation}
 k^+\frac{\dif N^{\mie}}{\dif k^+}=\frac{2\alpha_sC_R}{\pi}\log\Big(|C(x_0^+,x_0^++L)|\Big)
\end{equation}

Another possibility is to isolate in the on-shell spectrum the contributions without explicit $1/k_\perp^2$ dependence. The full on-shell spectrum for finite path length reads:
\begin{align}
 k^+\frac{\dif^3 N^ {\textrm{on-shell}}}{\dif k^+\dif^2k_\perp}  &=  \frac{\alpha_s C_R}{\pi^2}\left(\int\frac{d^2q_\perp}{(2\pi)^2}\frac{1}{q_\perp^2}\exp\left(\frac{-(k_\perp-q_\perp)^2}{Q_s^2}\right)+\frac{1}{k_\perp^2}\left[2\mathfrak{Re}\exp\left(\frac{-k_\perp^2}{2ik^+\bar{G}(x_0^++L)}\right)-1\right]\right.\nonumber\\
 &\left.+2\mathfrak{Re}\int_{x_0^+}^{x_0^++L} \dif y^+\,\frac{\bar{G}(y^+)}{Q_s^2(y^+)+2ik^+\bar{G}(y^+)}\exp\left(\frac{-k_\perp^2}{Q_s^2(y^+)+2ik^+\bar{G}(y^+)}\right)\right)
\end{align}
Thus, defining the medium induced spectrum as the $k_\perp$ integral of the last term, one gets:
\begin{align}
k^+\frac{\dif \tilde{N}^{\mie}}{\dif k^+}&\equiv\frac{2\alpha_sC_R}{\pi^2}\mathfrak{Re}\int\dif^2k_\perp\int_{x_0^+}^{x_0^++L} \dif y^+\,\frac{\bar{G}(y^+)}{Q_s^2(y^+)+2ik^+\bar{G}(y^+)}\exp\left(\frac{-k_\perp^2}{Q_s^2(y^+)+2ik^+\bar{G}(y^+)}\right)\nonumber\\
& =\frac{2\alpha_sC_R}{\pi}\log\Big(|C(x_0^++L,x_0^+)|\Big)
\end{align}
For the brick model, where the size of the medium is taken as large as the jet path length, the function $C$ satisfies $C(x,y)=C(y,x)$ so both definitions agree exactly, as emphasized in Chapter \ref{chapter:DLApic}. In the general case, there is no such symmetry for the function $C$ so $N_{\mie}$ and $\tilde{N}_{\mie}$ give different results. To see why, let us call $\lambda$ the typical time scale for the medium dilution. The difference between these two definitions comes from the fact that in $\tilde{N}^{\mie}$ there are still implicit vacuum-like components included (due to Bremsstrahlung \textit{after} scattering with the medium since the incoming parton was on its mass shell) when $\bar{G}$ starts behaving like $1/t$ for $t\gtrsim\lambda$, as discussed in Section \ref{sub:med-expansion}. If $x_0^++L\gg \lambda$, this scenario occurs and leads to the difference observed between the two spectra. If $x_0^++L\ll\lambda$, the two spectra tend to an agreement.

\section{The junction method}

To conclude this Appendix, we demonstrate that one recovers formulas \eqref{I-all-bef} and \eqref{I-all-in} using the junction method, that is by finding solutions of the differential equation \eqref{diff-eq-S} continuous and derivable in their first argument.

\paragraph{Example: the brick model.} First of all, let us give these solutions in the sinple case of the brick model. The brick model has only two parameters: the constant value $\qhat_0$ of the quenching parameter through the path length $L$ of the jet,
\begin{equation}
 \qhat(t)=\qhat_0\Theta(L-t)\,\qquad\textrm{ for }t\ge0
\end{equation}
The functions $S(t,t_0)$ and $C(t,t_0)$ are defined as the two independent solutions of the differential equation 
\begin{equation}\label{diff2}
 f''+\frac{i\qhat(t)}{2k^+}f=0
\end{equation}
 with initial conditions $S(t_0,t_0)=0$, $\partial_{t=t_0}S(t,t_0)=1$ and 
$C(t_0,t_0)=1$, $\partial_{t=t_0}C(t,t_0)=0$. In particular, $S$ and $C$ must be continuous and derivable as functions of $t$. The method to solve \eqref{diff2} for all $t\in[0,\infty[$ is to solve it on $t\in[0,L[$ and $t\in]L,\infty[$ before adjusting the constants such that $S$ and $C$ have the right initial conditions \textit{and} are continuous and derivable in $t=L$.

One finds, with $\Omega^2\equiv i\qhat_0/(2k^+)$:
\begin{align}
 S(t,t_0)&=\Theta(L-t_0)\Big[\Theta(L-t)\Omega^{-1}\sin\big(\Omega(t-t_0)\big)\nonumber\\
         &+\Theta(t-L)\Big((t-L)\cos\big(\Omega(L-t_0)\big)+\Omega^{-1}\sin\big(\Omega(L-t_0)\big)\Big)\Big]\nonumber\\
         &+\Theta(t_0-L)\Big[\Theta(t-L)(t-t_0)\nonumber\\
         &+\Theta(L-t)\Big((L-t_0)\cos\big(\Omega(L-t)\big)-\Omega^{-1}\sin\big(\Omega(L-t)\big)\Big)\Big]
 \end{align}
and 
\begin{align}
 C(t,t_0)&=\Theta(L-t_0)\Big[\Theta(L-t)\cos\big(\Omega(t-t_0)\big)\nonumber\\
         &+\Theta(t-L)\Big(\cos\big(\Omega(L-t_0)\big)+\Omega(L-t)\sin\big(\Omega(L-t_0)\big)\Big)\Big]\nonumber\\
         &+\Theta(t_0-L)\Big[\Theta(t-L)+\Theta(L-t)\cos\big(\Omega(L-t)\big)\Big]
\end{align}
One notices that $S(t,t_0)$ is also continuous and derivable as a function of $t_0$, whereas $C(t,t_0)$ is only continuous as a function of $t_0$. Moreover, one has the relation $C(t,t_0)=C(t_0,t)$ for all $t\le L$, but \textit{not} for all $t$ values.

\paragraph{Junction in the general case.} To deal with the general case, one notices that if one has the continuous and derivable solutions $\tilde{S}(t,t_0)$ and $\tilde{C}(t,t_0)$, then the functions $\tilde
{G}(t)$ and $\tilde{\bar{G}}(t)$ defined in \eqref{def-G}-\eqref{Gbar-def} are continuous and their derivative are piece-wise continuous. One can then find $\tilde{G}^{-1}(t)$ and $\tilde{\bar{G}}(t)$ as continuous solutions of the differential equations \eqref{riccati-g} and \eqref{riccati-gbar}. 
This means that the function $\tilde{G}^{-1}(t)$ satisfies:
\begin{align}
 \tilde{G}^{-1}(t_0)&=0\label{g-1ini}\\
 \frac{\dif \tilde{G}^{-1}}{\dif t}&=1+\frac{i\qhat(t)}{2k^+}\tilde{G}^{-2}\qquad\textrm{ if }t_0\le t<t_0+L\label{riccatit-L}\\
 \lim\limits_{\epsilon\rightarrow0}\tilde{G}^{-1}(t_0+L-\epsilon)&=\tilde{G}^{-1}(t_0+L+\epsilon)\\
 \frac{\dif \tilde{G}^{-1}}{\dif t}&=1\qquad\textrm{ if }t>t_0+L
\end{align}
The solution of this problem is
\begin{equation}\label{tildeG}
 \tilde{G}^{-1}(t)=\Theta(L+t_0-t)G^{-1}(t)+\Theta(t-L-t_0)\Big(t-t_0-L+G^{-1}(t_0+L)\Big)
\end{equation}
with $G^{-1}$ the solution of \eqref{g-1ini} and \eqref{riccatit-L}. This relation is also true for $\tilde{\bar{G}}$ if one replace $G$ by $\bar{G}$, solution of $\bar{G}'+\bar{G}^2+i\qhat(t)/(2k^+)=0$ and $\bar{G}(t_0)=0$. Using these expressions in the on-shell or off-shell cross-section \eqref{onshell-final}-\eqref{Iii} calculated in Chapter \ref{chapter:emissions}, one finds the expressions \eqref{I-all-bef} and \eqref{I-all-in}.

\chapter{Splitting functions}
\label{app:A}
In this Appendix, we give our conventions for the splitting functions used in this thesis.

\section{Vacuum (DGLAP) splitting functions}

We use the leading order unregularised DGLAP splitting functions. In the notation $\Phi_a^{bc}(z)$, $a$ refers to the parent parton, $b$ to the daughter parton carrying the splitting fraction $z$ and $c$ the daughter parton carrying the splitting fraction $1-z$.

\begin{align}
 \Phi_q^{gq}(z)&=C_F\frac{1+(1-z)^2}{z}\\
 \Phi_q^{qg}(z)&=C_F\frac{1+z^2}{1-z}\\
 \Phi_g^{gg}(z)&=2C_A\left(\frac{1-z}{z}+\frac{z}{1-z}+z(1-z)\right)\\
 \Phi_g^{q\bar{q}}(z)&=n_fT_R\Big(z^2+(1-z)^2\Big)
\end{align}
with obviously $\Phi_g^{q\bar{q}}(z)=\Phi_g^{\bar{q}q}(z)$. These are the building rates for the generating functional \ref{Z-mlla} in Chapter \ref{chapter:jet}.

For $i\in\{q,g\}$, we note $P_i(z)$ the $i$-splitting function summed over all distinct decay channels:
\begin{equation}\label{P-summed-def}
 P_i(z)\equiv\frac{1}{2!}\sum_{(a,b)}\Phi_i^{ab}(z)
\end{equation}
These splitting functions are convenient when one wants to implement the master equation \eqref{Z-mlla} in a Monte Carlo parton shower.
When these splitting functions are integrated between $0$ and $1$, one can exploit the $z\leftrightarrow1-z$ symmetry to write them in such a way that the soft singularity is only at $z=0$:
\begin{align}
 P_q(z)&=C_F\frac{1+(1-z)^2}{z}\\
 P_g(z)&=C_A\left(2\frac{1-z}{z}+z(1-z)+\frac{n_fT_R}{C_A}\big(z^2+(1-z)^2\big)\right)
\end{align}

\section{Medium splitting functions}

When there is no explicit mention of the contrary, $\qhat$ always refers to the adjoint $\qhat_A$. Hence, in \eqref{Z-mie}, there is a $\sqrt{C_A}$ factored out so that the medium-modified splitting kernels $\mathcal{K}_i^{ab}(z)$ read:
\begin{align}
\mathcal{K}_q^{gq}(z)&=C_F\frac{1+(1-z)^2}{z}\sqrt{\frac{C_F}{C_A}z^2+(1-z)}\\
\mathcal{K}_q^{qg}(z)&=C_F\frac{1+z^2}{1-z}\sqrt{\frac{C_F}{C_A}(1-z)^2+z}\\
 \mathcal{K}_g^{gg}(z)&=2C_A\frac{(1-z(1-z))^2}{z(1-z)}\sqrt{(1-z)+z^2}\\
 \mathcal{K}_g^{q\bar{q}}(z)&=n_fT_R\Big(z^2+(1-z)^2\Big)\sqrt{\frac{C_F}{C_A}+z(1-z)}
\end{align}
As for the vacuum splitting functions, we define  $\mathcal{K}_i(z)$ as the splitting function of a parton $i\in\{q,g\}$ summed over all distinct decay channels:
\begin{equation}
\mathcal{K}_i(z)\equiv\frac{1}{2!}\sum_{(a,b)}\mathcal{K}_i^{ab}(z)
\end{equation}
or more explicitly, 
\begin{align}
 \mathcal{K}_q(z)&=\frac{1}{2}\Big(\mathcal{K}_{q}^{qg}(z)+\mathcal{K}_{q}^{gq}(z)\Big)\\
  \mathcal{K}_g(z)&=\frac{1}{2}\Big(\mathcal{K}_{g}^{gg}(z)+2\mathcal{K}_{g}^{q\bar{q}}(z)\Big)
\end{align}

\chapter{Large $x$ jet fragmentation to NLL accuracy}
\label{app:NLL}

Eq.~\eqref{g12} can be deduced from the coherent branching algorithm which resums to all orders
leading and next-to-leading logarithms of the form
$-\alpha_s\log(1-x)$.
Since the fragmentation function is not IRC safe, we introduce a lower transverse momentum cut-off $\ktmin$ for any resolvable splitting. The final result strongly depends on $\ktmin$ so we need to keep track of any $\ktmin$ dependence in the calculation.

\section{Strict NLL result}

We focus on quark-initiated jets and the generalization to gluon-jets
is straightforward. To NLL accuracy, one can neglect the quark/gluon mixing terms. We
discuss this approximation at the end of this appendix.
The MLLA equation for the quark cumulative fragmentation function \eqref{def-sigma} is (integrating \eqref{Dlargex-MLLA} over $x$):
\begin{equation}\label{MLLA}
  Q\frac{\partial \Sigma_q(x,Q)}{\partial Q}=\int_{0}^{1}\dif z\,K_{q}^{q}(z,k_\perp)\left[\Sigma_q\Big(\frac{x}{z},zQ\Big)-\Sigma_q(x,Q)\right]
\end{equation}
where the evolution variable is $Q=p_{T0}\theta$ to account for the ordering in the angle $\th$ of successive emissions and the kernel is
\begin{equation}
 K_q^{q}(z,k_\perp)=\frac{\alpha_s(k_\perp)}{\pi}\Phi_q^{qg}(z)\Theta(k_\perp-\ktmin)
\end{equation}
The initial condition for \eqref{MLLA} is
$\Sigma_q(x,k_\perp=\ktmin)=\Theta(1-x)$. At NLL accuracy,
$k_\perp=z(1-z)Q\simeq (1-z)Q$ and $\Sigma_q(\frac{x}{z},zQ)\simeq
\Sigma_q(\frac{x}{z},Q)$ since the dominant contribution for $x\simeq 1$ comes from $z\simeq 1$. 
In Mellin space, one can solve this equation with our approximations. Anticipating our resummed result, we note
$\lambda_j=\alpha_s\log(j)$ and
$\lambda_0=\alpha_s\log(p_{T0}R/\ktmin)=\alpha_sL_0$, so that the Mellin transform of $\Sigma$ reads:
\begin{align}
 \log(j\tilde{\Sigma}_q(j,p_{T0}R))&=\int_{Q_0}^{p_{T0}R}\frac{\dif Q'}{Q'}\,\int_{0}^{1}\dif z\,(z^{j}-1)K_q^{q}(z,Q')\\
 &=\frac{C_F}{\pi \beta_0}\left[\log(j)\left(1-\log\Big(\frac{1-2\beta_0\lambda_j}{1-2\beta_0\lambda_0}\Big)+\frac{\log(1-2\beta_0\lambda_j)}{2\beta_0\lambda_j}\right)\right.\nonumber \\
 &\hspace{1cm}\left.-\gamma_E\log\Big(\frac{1-2\beta_0\lambda_j}{1-2\beta_0\lambda_0}\Big)+B_q\log(1-2\beta_0\lambda_0)\right]+\mathcal{O}(\alpha_s\lambda_j^n,\alpha_s\lambda_0^n),\label{logDj}
\end{align}
where we used the standard trick
$z^{j}-1\simeq-\Theta(e^{-\gamma_E}/j-z)$ valid at NLL accuracy and we kept only the singular and finite part $B_q=-3/4$ of the quark splitting function when $z\simeq1$.
Eq.~\eqref{logDj} resums to all orders leading and next-to-leading logarithms of the form $\lambda_j$, $\lambda_0$. More explicitly,
\begin{align}
 \log(j\tilde{\Sigma}^{\textrm{NLL}}_q(j,p_{T0}R))&=\log(j)g_1(\lambda_j,\lambda_0)+f_2(\lambda_j,\lambda_0)\\
 g_1(u,v)&=\frac{C_F}{\pi \beta_0}\left[1-\log\Big(\frac{1-2\beta_0u}{1-2\beta_0v}\Big)+\frac{\log(1-2\beta_0u)}{2\beta_0u}\right]\\
 f_{2,q}(u,v)&=\frac{C_F}{\pi \beta_0}\left[-\gamma_E\log\Big(\frac{1-2\beta_0u}{1-2\beta_0v}\Big)+B_q\log(1-2\beta_0v)\right]
\end{align}
The final step is to calculate the inverse Mellin transform of \eqref{logDj}.
\begin{align}
 \Sigma_{q}(x)&=\frac{1}{2\pi i}\int_{\mathcal{C}}\frac{\dif j}{j}\,e^{-j\log(x)}\Big(j\tilde{\Sigma}_{q}(j)\Big)=\frac{1}{2\pi i}\int_{\mathcal{C}}\dif u \,e^{u-\log(u)+G_q[\log(u)-\log(-\log(x))]}
\end{align}
where $\mathcal{C}$ is a contour parallel to the imaginary axis and $G_q[\log(j)]\equiv\log(j\tilde{\Sigma}_{q}(j))$.
For this, we Taylor-expand the function $G_q$ around $L=-\log(-\log(x))\simeq-\log(1-x)$.
\begin{equation}
 G_q[L+\log(u)]=G_q[L]+\log(u)G_q'[L]+\sum_{k=2}^{\infty}\log(u)^k\frac{G_q^{(k)}[L]}{k!}
\end{equation}
For $k\ge2$, $G_q^{(k)}[L]$ is certainly beyond NLL accuracy because the derivatives of $\alpha_s\beta_0 L$ with respect to $L$ bring always at least one extra $\alpha_s$ factor. Thus, we truncate the expansion up to the first derivative. Moreover, the derivative of $f_{2,q}(\alpha_sL,\alpha_sL_0)$ with respect to $L$ is also sub-leading.
Finally, using
\begin{equation}
 \frac{1}{2\pi i}\int_{\mathcal{C}}\dif u\, e^{u+x\log(u)}=\frac{1}{\Gamma(-x)}
\end{equation}
one gets the following result for the cumulative distribution:
\begin{equation}
\label{cumulative}
 \Sigma^{\textrm{NLL}}_{q}(x,p_{T0}R)=\frac{e^{G_{q}[L]}}{\Gamma(1-G_q'[L])}=\frac{\exp\Big(Lg_1(\alpha_sL,\alpha_sL_0)+f_{2,q}(\alpha_sL,\alpha_sL_0)\Big)}{\Gamma\Big(1-\frac{\partial ug_1(u,\alpha_sL_0)}{\partial u}_{|u=\alpha_sL}\Big)}
\end{equation}
which is exactly \eqref{log-expansion}, \eqref{g12}.

\section{Sub-leading $j$ contributions and quark/gluon mixing
  terms.} Besides N$^2$LL contributions, we have neglected terms of
order $\mathcal{O}(\alpha_s^n\log^n(j)/j)$ in formulas \eqref{MLLA} and
\eqref{logDj}. Among such terms, those associated with quark/gluon mixing
give sizeable numerical corrections to the NLL results, especially in
the gluon-jet case.
The main reason for this is that, even though the (power-suppressed)
probability for a gluon to split in a $q\bar q$ pair where the quark
carries most of the momentum ($x\sim 1$) is much
smaller than the probability to find a hard gluon, once such a
splitting occurs, the Sudakov appearing in~\eqref{cumulative} becomes
that of a quark, i.e.\ has a much smaller suppression because of the
colour factor $C_F< C_A$ appearing in the exponential.
In the inclusive fragmentation function, this becomes an increasingly
likely situation~\cite{DeGrand:1978te}.

Including all terms of order $\mathcal{O}(\alpha_s^n\log^n(j)/j)$ is beyond the scope of this simple analysis of the large $x$ behaviour of the fragmentation function. Instead, one can correct Eq.~\eqref{cumulative} for gluon jets with an additional piece $\Sigma_{g,\textrm{mix}}(x,p_{T0}R)$ describing the splitting of the gluon in a $q\bar{q}$ pair, with either the quark or the antiquark carrying a large fraction $x$ of the initial energy:
 \begin{align}
\label{sigmagqq}
 \Sigma_{g,\textrm{mix}}(x,p_{T0}R)&=\int_{0}^{1-x}\dif \xi\, P_g^q(\xi)\int_0^{R}\frac{\dif\theta}{\theta}\frac{\alpha_s(\xi p_{T0}\theta)}{\pi}\Theta(\xi p_{T0}\theta-\ktmin)\nonumber\\ 
 &\times \exp\left(-\frac{2C_A}{\pi}\int_{\xi}^1\frac{\dif z}{z}\int_{\theta}^{R}\frac{\dif\theta'}{\theta'}\alpha_s(zp_{T0}\theta')\Theta(zp_{T0}\theta'-\ktmin)\right)\nonumber\\ 
 &\times \exp\left(-\frac{2C_F}{\pi}\int_{\xi}^1\frac{\dif z}{z}\int_{0}^{\theta}\frac{\dif\theta'}{\theta'}\alpha_s(zp_{T0}\theta')\Theta(zp_{T0}\theta'-\ktmin)\right)
\end{align}
with $P_g^q(\xi)=2n_fT_R(\xi^2+(1-\xi)^2)\simeq 2n_fT_R$ since $\xi\le1-x\ll1$. In Mellin space, this is equivalent to solve the linear differential equation with an inhomogeneous term associated with the non-diagonal elements of the kernel matrix. In Fig.~\ref{Fig:frag-largex}, the analytical ``NLL'' curve for gluon jets is actually $\Sigma^{\textrm{NLL}}_{g}(x)+\Sigma_{g,\textrm{mix}}(x)$.

\chapter{The ISD fragmentation function to NLL accuracy}
\label{app:ISD}

This Appendix deals with the fragmentation function from subjets built from the Iterative Soft Drop procedure as explained in Chapter \ref{chapter:jet}, Section \ref{subsub:frag-ISD}. We aim at calculating this observable to next-to-leading accuracy in the soft sector. The large logarithms in this case are $L=\log(1/z)$ and $L_{\cut}=\log(1/z_{\cut})$.

We start with the evolution equation for the fragmentation function $\vec{D}_{\rm ISD}(z,Q)$ obtained from the master equation \eqref{Z-primary}. The logarithmic counting is actually easier when considering the cumulative fragmentation function $\vec{\Sigma}_{\rm ISD}(z,Q)$ and its logarithmic expansion \eqref{isd-log-expansion}. The validity of our successive approximations can be checked at the level of the evolution equation satisfied by $\vec{\Sigma}_{\rm ISD}$ which is directly deduced from the following equation:

\begin{align}\label{full-Disd-eq}
 \frac{\partial \vec{D}_{\rm ISD}(z,Q)}{\partial \log(Q)}&=\frac{\alpha_s(k_\perp)}{\pi}\Theta_{\cut}(z,Q)\vec{P}_{\sym}(z)+\int_0^{1/2}\dif z'\, \Big([\LL](z',Q)\vec{D}_{\rm ISD}(z',Q)\nonumber\\
 &\hspace{3cm}+[\NLL](z',Q)\big(\vec{D}_{\rm ISD}(z',(1-z')Q)-\vec{D}_{\rm ISD}(z',Q)\big)\Big)
\end{align}
with $\vec{P}_{\sym}(z)$ is the symmetrized splitting function defined as
\begin{equation}
 \vec{P}_{\sym}(z)=\begin{pmatrix}
                    \Phi_q^{gq}(z)+\Phi_q^{qg}(z)\\
                    \Phi_g^{gg}(z)+2\Phi_g^{q\bar{q}}(z)
                   \end{pmatrix}
\end{equation}
The transverse momentum is equal to $k_\perp=z(1-z)Q$ and the matrices $[\LL]$ and $[\NLL]$ are:
\begin{align}
[\LL](z,Q)&=\frac{\alpha_s(k_\perp)}{\pi}\Theta_{\cut}(z,Q)
   \begin{pmatrix}
   -\Phi_q^{qg}(z)        & \Phi_q^{qg}(z)\\
   2 \Phi_g^{q\bar{q}}(z) &- 2\Phi_g^{q\bar{q}}(z) 
   \end{pmatrix}\\
    [\NLL](z,Q)&=\frac{\alpha_s(k_\perp)}{\pi}\Theta_{\cut}(z,Q)
   \begin{pmatrix}
   \Phi_q^{gq}(z) & \Phi_q^{qg}(z)\\
   2\Phi_g^{q\bar{q}}(z)              & \Phi_g^{gg}(z)
   \end{pmatrix}
\end{align}
The step function constrains emissions to satisfy the Soft Drop condition, noting $\bar{Q}=p_TR$:
\begin{equation}
 \Theta_{\cut}(z,Q)=\Theta\big(z-z_{\cut}(Q/\bar{Q})^{\beta}\big)
\end{equation}

\section{LL analytic result}

As explained in the Section \ref{subsub:frag-ISD}, one can neglect the NLL matrix to LL accuracy, since the difference $\vec{D}_{\rm ISD}(z',(1-z')Q)-\vec{D}_{\rm ISD}(z',Q)$ regulates the soft divergence of the splitting functions. In the LL matrix, using $\alpha_s(z(1-z)Q)=\alpha_s(Q)(1-2\alpha_s(Q)\beta_0\log(z(1-z)))$, one sees that one can approximate $\alpha_s(k_\perp)$ by $\alpha_s(Q)$ since the next term generates one extra-power of $\alpha_s$ which is not compensated by a soft logarithmic singularity in the $z$ integration. Finally, in the term proportional to $\vec{P}_{\sym}(z)$, one cannot neglect the $z$ dependence inside $\alpha_s$ since the extra $\alpha_s$ power is compensated by the soft singularity of the splitting function. Therefore, one gets the equation \eqref{diffeq-Disd}:
\begin{equation}\label{diffeq-Disd-app}
 \frac{\partial \vec{D}_{\textrm{ISD}}(z,Q)}{\partial \log(Q)}=\frac{2\vec{C}_R}{\pi}\frac{\alpha_s(zQ)}{z}\Theta_{\textrm{cut}}+\frac{\alpha_s(Q)}{\pi}[F]\vec{D}_{\textrm{ISD}}(z,Q)
\end{equation}
with the matrix $[F]$ defined as:
\begin{equation}
 \begin{pmatrix}
  -f_F & f_F\\
  f_A & -f_A
 \end{pmatrix}
\equiv\int_0^{1/2}\dif z\begin{pmatrix}
      -\Phi_q^{qg}(z) & \Phi_q^{qg}(z) \\
     \Phi_g^{q\bar{q}}(z)  &-\Phi_g^{q\bar{q}}(z)
     \end{pmatrix}=
     \begin{pmatrix}
      -C_F\big(2\log(2)-\frac{5}{8}\big) & C_F\big(2\log(2)-\frac{5}{8}\big)\\
      \frac{2n_f T_R}{3} &-\frac{2n_f T_R}{3}
     \end{pmatrix}
\end{equation}
This equation can be solved using the variation of parameters method. The general solution of the homogeneous equation vanishes because of the initial condition $\vec{D}^{\LL}_{\rm ISD}(z,Q=0)=\vec{0}$, so that only the particular solution remains:
\begin{align}
 \vec{D}^{\LL}_{\rm ISD}(z)=\int_0^{p_TR}\frac{\dif Q'}{Q'}\frac{2}{\pi}\frac{\alpha_s(zQ')}{z}\Theta_{\cut}(z,Q')\exp\left(\int_{Q'}^{p_TR}\frac{\dif Q''}{Q''}\frac{\alpha_s(Q'')}{\pi}
[F]\right)\vec{C}_R
\end{align}
The matrix exponential can be calculated exactly:
\begin{equation}\label{flavor-mix-exp}
 \exp\Big(\lambda(Q')
 \begin{pmatrix}
  -f_F & f_F\\
  f_A  & -f_A
 \end{pmatrix}\Big)= \frac{1}{f_A+f_F}\begin{pmatrix}
  f_A+f_F e^{-\lambda(Q')(f_A+f_F)} & f_F-f_F e^{-\lambda(Q')(f_A+f_F)} \\
  f_A-f_A e^{-\lambda(Q')(f_A+f_F)} & f_F+f_A e^{-\lambda(Q')(f_A+f_F)}
 \end{pmatrix}
\end{equation}
with (noting $\alpha_0=\alpha_s(p_TR)$,)
\begin{equation}
\lambda(Q')\equiv\int_{Q'}^{p_TR}\frac{\dif Q''}{Q''}\alpha_s(Q'')=-\frac{1}{2\pi\beta_0}\log\big(1+2\alpha_0\beta_0\log(Q'/(p_TR))\big)
\end{equation}
After some algebra, the remaining integral over $Q'$ gives the result \eqref{Disd-LL}. In particular, the terms without exponential factors give the term proportional to the function $f_0$ in \eqref{Disd-LL}, whereas the others can be integrated explicitly using the incomplete Gamma function.

\section{NLL equivalent equation}

One can find an equation equivalent to \eqref{full-Disd-eq} at NLL accuracy, with a more suitable form for both analytical and numerical analysis. It is convenient to introduce the following set of matrices:
\begin{align}
 [F_2]&\equiv\int_0^{1/2}\dif z\begin{pmatrix}
      -\Phi_q^{qg}(z)\log(z) & \Phi_q^{qg}(z)\log(z) \\
     \Phi_g^{q\bar{q}}(z)\log(z)  &-\Phi_g^{q\bar{q}}(z)\log(z)
     \end{pmatrix}\\
     &=
     \begin{pmatrix}
      \frac{C_F}{48}\big(8\pi^2+3(1+2\log(2))(-9+8\log(2))\big) & \frac{-C_F}{48}\big(8\pi^2+3(1+2\log(2))(-9+8\log(2))\big)\\
      \frac{-n_fT_R}{72}\big(29+24\log(2)\big) &\frac{n_fT_R}{72}\big(29+24\log(2)\big) 
     \end{pmatrix}\nonumber\\
 [E]&\equiv\int_0^{1/2}\dif z\begin{pmatrix}
   \Phi_q^{gq}(z)\log(1-z) & \Phi_q^{qg}(z)\log(1-z)\\
   2\Phi_g^{q\bar{q}}(z)\log(1-z)              & \Phi_g^{gg}(z)\log(1-z)
     \end{pmatrix}\\
     &=\begin{pmatrix}
        \frac{C_F}{48}\big(33-8\pi^2+6\log(2)(8\log(2)-5)\big)  & \frac{-C_F}{16}\big(-13+2\log(2)(7+8\log(2))\big)\\
        \frac{n_fT_R}{72}\big(24\log(2)-23)  & \frac{-C_A}{72}\big(12\pi^2+132\log(2)-131)
       \end{pmatrix}\nonumber
\end{align}
and to write the differential equation \eqref{full-Disd-eq} in terms of $t\equiv\log(Q/\bar{Q})$, $\bar{Q}=p_TR$. Using the trick 
\begin{equation}
 \vec{D}_{\rm ISD}(z',t+\log(1-z'))-\vec{D}_{\rm ISD}(z',t)=\log(1-z')\frac{\partial \vec{D}_{\rm ISD}(z',t)}{\partial t}+\mathcal{O}\big(\alpha_0^{n+2}\log^n\big)
\end{equation}
one can rewrite \eqref{full-Disd-eq} as:
\begin{align}\label{NLL-Disd-eq1}
 \frac{\partial \vec{D}^{\NLL}_{\rm ISD}(z,t)}{\partial t}&=\frac{\alpha_s(k_\perp)}{\pi}\Theta_{\cut}(z,\bar{Q}e^t)\vec{P}_{\sym}(z)+\left(\frac{\alpha_s(\bar{Q}e^t)}{\pi}[F]-\frac{2\alpha_0^2\beta_0}{\pi}[F_2]\right)\vec{D}^{\NLL}_{\textrm{ISD}}(z,t)\nonumber\\
 &\hspace{8.2cm}+\frac{\alpha_s(\bar{Q}e^t)}{\pi}[E]\frac{\partial \vec{D}^{\NLL}_{\rm ISD}(z,t)}{\partial t}
\end{align}
The last term is a consequence of the fact that at NLL accuracy, one must take into account the energy recoil of the hard branch. The term proportional to $\alpha_0^2[F_2]$ comes from the $z$ dependence of the transverse momentum $k_\perp$ in the matrix $[\LL]$. It must be included as it is of the same logarithmic order as the term proportional to $[E]$.

Equation \eqref{NLL-Disd-eq1} can be further simplified by putting the last term of \eqref{NLL-Disd-eq1} on the left hand side of the equation, and multiplying by the inverse matrix:
\begin{equation}
 \left(\bbone-\frac{\alpha_s(\bar{Q}e^t)}{\pi}[E]\right)^{-1}\simeq\bbone+\frac{\alpha_s(\bar{Q}e^t)}{\pi}[E]+\mathcal{O}\big(\alpha_s^2(Q)\big)
\end{equation}
one gets, using $\alpha_s^2(\bar{Q}e^t)D^{\NLL}_{\rm ISD}=\alpha_0^2D^{\NLL}_{\rm ISD}$ at our accuracy,
\begin{align}\label{NLL-Disd-eq2}
 \frac{\partial \vec{D}^{\NLL}_{\rm ISD}(z,t)}{\partial t}&=\frac{\alpha_s(z\bar{Q}e^t)}{\pi}\Theta_{\cut}(z,\bar{Q}e^t)\vec{P}_{\sym}(z)+\frac{\alpha_s(\bar{Q}e^t)\alpha_s(z\bar{Q}e^t)}{\pi^2}\Theta_{\cut}(z,\bar{Q}e^t)[E]\vec{P}_{\sym}(z)\nonumber\\
 &\hspace{4cm}+\left(\frac{\alpha_s(\bar{Q}e^t)}{\pi}[F]-\frac{2\alpha_0^2\beta_0}{\pi}[F_2]+\frac{\alpha_0^2}{\pi^2}[E][F]\right)\vec{D}^{\NLL}_{\textrm{ISD}}(z,t)
\end{align}
Finally, one notices that $\alpha_0^2D^{\NLL}_{\rm ISD}=\alpha_0^2D^{\LL}_{\rm ISD}+\mathcal{O}(\alpha_0^{n+2}\log^n)$ at NLL precision. This enables to put the terms proportional to $\alpha_0^2$ in the inhomogeneous term of the differential equation, since $D^{\LL}_{\rm ISD}$ is known exactly. Using $\vec{P}_{\sym}(z)\simeq 2\vec{C}_R/z+\vec{B}_R$, where $B_i$ is the finite part of the splitting function, one can put the NLL differential equation into the form \eqref{diffeq-Disd-nll} with the inhomogeneous term $\vec{C}_{\NLL}(z,Q)$ reading:
\begin{align}
 \vec{C}_{\NLL}(z,Q)&=\frac{\alpha_s(zQ)}{\pi}\frac{2}{z}\Theta_{\cut}(z,Q)\left(\bbone+\frac{\alpha_s(Q)}{\pi}[E]\right)\vec{C_R}+\frac{\alpha_s(Q)}{\pi}\Theta\big(Q-p_TR(2z_{\cut})^{-1/\beta}\big)\vec{B}_{R}\nonumber\\
 &\hspace{6cm}+\left(\frac{\alpha_0^2}{\pi^2}[E][F]-\frac{2\alpha_0^2\beta_0}{\pi}[F_2]\right)\vec{D}^{\LL}_{\textrm{ISD}}(z,Q)
\end{align}
This allows for a straightforward evaluation of the solution of \eqref{diffeq-Disd-nll} given the initial conditions:
\begin{equation}\label{Disd-NLLsol}
 \vec{D}^{\NLL}_{\rm ISD}(z)=\int_0^{p_TR}\frac{\dif Q'}{Q'}\exp\left(\lambda(Q')
 \begin{pmatrix}
  -f_F & f_F\\
  f_A  & -f_A
 \end{pmatrix}\right)\vec{C}_{\NLL}(z,Q')
\end{equation}
Since the homogeneous term involves only the matrix $[F]$, its exponential has the simple form \eqref{flavor-mix-exp}.

\paragraph{Other NLL corrections.}

To claim a full NLL answer for $\vec{D}^{\NLL}_{\rm ISD}(z)$, one should also include the two-loop running coupling and splitting functions corrections. These corrections can be incorporated in the function $\vec{C}_{\NLL}(z,Q)$ by adding the standard correction $\delta \vec{C}^{2-\rm loops}(z,Q)$:
\begin{equation}
 \delta \vec{C}^{2-\rm loops}(z,Q)=-\frac{\alpha_0^2\beta_1}{\pi\beta_0}\frac{1}{z}\frac{\log(1+\alpha_0\beta_0\log(zQ)}{(1+\alpha_0\beta_0\log(zQ))^2}\vec{C}_R+\frac{K}{2\pi^2}\frac{\alpha_s^2(zQ)}{z}\vec{C}_R
\end{equation}
where $K$ is the universal two-loop cusp anomalous dimension.

\chapter{Vacuum-like patterns in the off-shell and on-shell gluon spectra in a dense medium}
\chaptermark{Vacuum-like patterns in medium gluon spectra}
\label{app:off-on}

The purpose of this Appendix is twofold:
\begin{enumerate}
 \item we want to show that decomposing the off-shell spectrum according to time integration domains is ambiguous and can lead to unphysical results, such as \eqref{off-y<tf}. In the presence of a dense medium in a finite volume, the quantum formation process is intricate and vacuum formation time arguments can be misleading.
 \item for this reason, we prefer the more physical decomposition given by \eqref{off-new} which is presented in Chapter \ref{chapter:DLApic}. We thus demonstrate formula \eqref{vle-spec-inf} and \eqref{on-shell-physic} as promised. 
\end{enumerate}

\section{Decomposition of the off-shell spectrum from time domains}
\sectionmark{The off-shell spectrum from time domains}
This section aims at explaining where the Bremsstrahlung contributions in the off-shell spectrum \eqref{offshell-final} comes from in the time integral. The results obtained in this section correspond to those obtained in \cite{MehtarTani:2012cy}. As we use analytic expressions valid in the expanding medium case, this analysis can be straightforwardly extended for expanding media using for instance the Taylor expansion method as in Chapter \ref{chapter:DLApic}, Section \ref{sec:veto}.

Let us begin with a rewriting of the result obtained in \eqref{offshell-final} in a physically more transparent way:
\begin{equation}\label{off-shell-v3}
 \om\frac{\dif^3 N^{\textrm{off-shell}}}{\dif\om\dif^2 k_\perp}=\frac{\alpha_s C_F}{\pi^2}\left[2\mathfrak{Re}\frac{1}{2i\om}\int_0^\infty \dif y^+ \int\dif^2 q_\perp\,\mathcal{P}(k_\perp-q_\perp,y^+,\infty)\exp\left(\frac{-q_\perp^2 G^{-1}(y^+)}{2 i\om}\right)-\frac{1}{k_\perp^2}\right]
\end{equation}
where the $\varepsilon$ prescription is implicit. In this form, the final state broadening of the gluon between $y^+$ and $\infty$ is convoluted with the emission spectrum \textit{at formation}. Setting by hand $\mathcal{P}(k_\perp-q_\perp)=\delta(k_\perp-q_\perp)$ to remove the effect of broadening,  all the physics of the medium is encoded in the function $G(y^+)$ defined in \eqref{def-G} (with $x_0^+=0$ in this case.) In the special case of a static medium with constant $\qhat$, the function $G^{-1}(y^+)$ is particularly simple (see the derivation in Appendix \ref{app:B}):
\begin{equation}\label{G-brick}
 G^{-1}(y^+)= \left\{
     \begin{array}{ll}
        j\sqrt{\frac{2\om}{\qhat}}\tanh\Big(\bar{j}\sqrt{\frac{\qhat}{2\om}}y^+\Big) & \mbox{if } y^+\le L \\
y^+-L\Big[1-j\sqrt{\frac{\om}{\omc}}\tanh\Big(\bar{j}\sqrt{\frac{\omc}{\om}}\Big)\Big]\underset{\om\ll\om_c}\approx y^+-L& \mbox{if } y^+\ge L \\
    \end{array}
\right.
\end{equation}
with $j\equiv(1+i)/\sqrt{2}$ and $L$ is the path length in the medium. 

\paragraph{Vacuum formation time.} Emission is a quantum process which is not localized in time. The so-called ``vacuum formation time'' $t_f=2\om/k_\perp^2$ which has been derived before from the Heisenberg inequality comes mathematically from the analysis of the integral giving the Bremsstrahlung spectrum in the vacuum:
\begin{equation}\label{vac-brem-int}
 \om \frac{\dif ^3 N^{\brem}}{\dif \om\dif^2 k_\perp}=\frac{\alpha_sC_F}{\pi^2}\frac{k_\perp^2}{4\om^2}2\mathfrak{Re}\int_0^{\infty}\dif y^+\int_0^\infty \dif \bar{y}^+e^{-\varepsilon(y^++\bar{y}^+)}e^{i k^-(y^+-\bar{y}^+)}
\end{equation}
Hence, the phase in the exponential is controlled by the scale $k^-=k_\perp^2/(2k^+)=k_\perp^2/(2\om)$ so that:
\begin{equation}
 t_f\equiv|y^+-\bar{y}^+|\sim\frac{2\om}{k_\perp^2}
\end{equation}
for Bremsstrahlung emissions in the vacuum. At the level of the amplitude of the process, this formation time controls also the phase of the integral. Consequently, in the vacuum, one has also:
\begin{equation}
 t_f\sim y^+\sim \bar{y}^+
\end{equation}
In our derivation of the BDMPS-Z spectrum, we have used our lemma \eqref{lemma} to integrate over the time $\bar{y}^+$ in the complex conjugate amplitude in order to obtain compact expressions well suited for a numerical analysis. The price to pay is that a spurious $-1/k_\perp^2$ term is generated, spoiling the symmetry of the result all over $k_\perp$. This term comes from a region with $y^+$ and $\bar{y}^+$  much larger than the typical length of the medium. However, it is to be associated with vacuum-like processes with unconstrained formation time $t_f$ since $y^+-\bar{y}^+$ can be as small or as large as desired.


\paragraph{The off-shell spectrum at formation.} Setting by hand $\mathcal{P}(k_\perp-q_\perp)=\delta(k_\perp-q_\perp)$ to remove the effect of broadening, this spectrum at formation becomes simply the emission \textit{rate} integrated over time:
\begin{equation}\label{off-shell-f}
 \om\frac{\dif^3 N^{\textrm{off-shell,f}}}{\dif\om\dif^2 k_\perp}=\frac{\alpha_s C_F}{\pi^2}\left[2\mathfrak{Re}\frac{1}{2i\om}\int_0^\infty \dif y^+\,\exp\left(\frac{-k_\perp^2 G^{-1}(y^+)}{2 i\om}\right)-\frac{1}{k_\perp^2}\right]
\end{equation}

From \eqref{off-shell-v3}, one sees that when the function $G^{-1}(y^+)$ is of the form $G^{-1}(y^+)\sim y^++cste$, the spectrum at formation (without broadening) develops a collinear singularity proportional to $1/k_\perp^2$. More precisely, if $G^{-1}(y^+)\sim y^+$ for $y^+ \le y^+_1$ for instance, the integral in \eqref{off-shell-f} gives:
\begin{align}
 2\mathfrak{Re}\frac{1}{2i\om}\int_{0}^{t_1} \dif y^+\,\exp\left(\frac{-k_\perp^2 G^{-1}(y^+)}{2 i\om}\right)&\simeq2\mathfrak{Re}\frac{1}{2i\om}\int_{0}^{y^+_1}\dif y^+\,\exp\left(\frac{-k_\perp^2 y^+}{2 i\om}\right)\\
 &=\frac{2}{k_\perp^2}\mathfrak{Re}\left(1-e^{\frac{-k_\perp^2 y^+_1}{2 i\om}}\right)
\end{align}
which has the Bremsstrahlung form and vanishes for $k_\perp^2\le 2\om/y^+_1$. 
Note the factor 2 with respect to \eqref{brem}. This is expected as in this calculation, it is as if the quark is created instantaneously at time $t=0$ and vanishes instantaneously at time $t=t_1$, giving twice the genuine spectrum for an accelerated charge.

\paragraph{(1) The domain $t_f\le t_{f,\med}$.} Now, we use in Eq~\eqref{off-shell-v3} the following approximation for the hyperbolic tangent function in \eqref{G-brick}, $\tanh(x)\simeq x$ for $x\le 1$ valid when $y^+\le t_{f,\med}\ll L$ (the second inequality is a consequence of $\om\ll\om_c$). One gets $G^{-1}(y^+)\simeq y^+$. One can neglect the $y^+$ dependence inside the broadening probability $\mathcal{P}$ because the broadening between $0$ and $t_f$ of order $k_{f,\med}$ is negligible in front of $Q_s$. Thus, the $y^+$ integral gives:
\begin{equation}
 \frac{2}{q_\perp^2}\mathfrak{Re}\Big(1-e^{\frac{-q_\perp^2 t_{f,\med}}{2 i\om}}\Big)
\end{equation}
which is suppressed for $q_\perp\le k_{f,\med}=(2\qhat\om)^{1/4}$.
At the level of the full spectrum \eqref{off-shell-v3}, this contributions plus the $-1/k_\perp^2$ term constrained with $t_f\le t_{f,\med}$ (i.e $k_\perp\ge k_{f,\med}$) reads:
\begin{equation}\label{off-y<tf}
 \om\frac{\dif^3 N^{\textrm{off-shell, } t_f<t_{f,\med}}}{\dif\om\dif^2 k_\perp}=\frac{\alpha_s C_F}{\pi^2}\int\dif^2 q_\perp\,\mathcal{P}(k_\perp-q_\perp,0,\infty)\left[ \frac{2}{q_\perp^2}\Theta(q_\perp-k_{f,\med})-\frac{1}{k_\perp^2}\Theta(k_\perp-k_{f,\med})\right]
\end{equation}
The first piece is the convolution between a Bremsstrahlung spectrum with a factor 2 and a Gaussian of width $\sim Q_s$ whereas the second piece (coming from the $\varepsilon$ prescription and lemma \eqref{lemma}) is a genuine Bremsstrahlung spectrum cancelling the factor 2 to restore the expected $1/k_\perp^2$ behaviour for $k_\perp \gg Q_s$.

The latter expression deserves further comments because it is desperately non-symmetrical. The minus term is not convoluted with the broadening probability. Is it an artefact of the way we did the calculation? Actually, it is not as we shall now explain. 
Coming back to the expression \eqref{Iiidouble} where the time in the complex conjugate amplitude is not integrated, it is also possible to factor out the broadening probability as follows:
\begin{align}\label{off-double-v3}
 \om\frac{\dif^3 N^{\textrm{off-shell}}}{\dif\om\dif^2 k_\perp}&=\frac{\alpha_s C_F}{\pi^2}2\mathfrak{Re}\int_0^\infty\dif y^+\,e^{-\varepsilon y^+}\int\dif^2q_\perp \mathcal{P}(k_\perp-q_\perp,y^+,\infty)\nonumber\\ &\hspace{6cm}\times \frac{q_\perp^2}{(2i\om)^2}\int_0^{y^+} \dif\bar{y}^+\,\frac{-e^{-\varepsilon\bar{y}^+}}{C^2(\bar{y}^+,y^+)}e^{\frac{-q_\perp^2S(y^+,\bar{y}^+)}{2i\om  C(\bar{y}^+,y^+)}}
\end{align}
When $t_f=|y^+-\bar{y}^+|\le t_{f,\med}$ one can use $C(\bar{y}^+,y^+)\simeq 1$ and $S(y^+,\bar{y}^+)\simeq y^+-\bar{y}^+$ as explained in Section \ref{sub:veto-anaytic}. The calculation of \eqref{off-double-v2} for $y^+-\bar{y}^+\le t_{f,\med}$ using these approximations for $C$ and $S$ and keeping only the terms with an apparent $1/k_\perp^2$ singularity gives:
\begin{align}
 \om\frac{\dif^3 N^{\textrm{off-shell, }t_f\le t_{f,\med}}}{\dif\om\dif^2 k_\perp}&\simeq
 \frac{\alpha_s C_F}{\pi^2}2\mathfrak{Re}\int_0^{t_{f,\med}}\dif y^+\int\dif^2q_\perp\mathcal{P}(k_\perp-q_\perp,y^+,\infty)\frac{1}{2i\om}\Big(-1+e^{\frac{-q_\perp^2y^+}{2i\om}}\Big)\nonumber\\
 &\hspace{-1cm}+\frac{\alpha_s C_F}{\pi^2}2\mathfrak{Re}\int_0^\infty\dif y^+\,e^{-2\varepsilon y^+}\int\dif^2q_\perp \mathcal{P}(k_\perp-q_\perp,y^+,\infty)\frac{\varepsilon}{q_\perp^2}\Big(-1+e^{\frac{-q_\perp^2t_{f,\med}}{2i\om}}\Big)
\end{align}
Neglecting the broadening during the formation time $t_{f,\med}$, the first term in this expression gives the first term in \eqref{off-y<tf} which has this weird factor 2. In the limit $\varepsilon\rightarrow0$, only large values of $y^+$ contributes to the integral in the second term so the broadening probability becomes $\delta^{(2)}(k_\perp-q_\perp)$. The integral over $y^+$ then gives a factor $1/2$ when the $\varepsilon$ prescription is removed, and one recovers the second term in \eqref{off-y<tf}.

To summarize this short detour, we showed that the result \eqref{off-y<tf} comes not only from the domain $y^+\le t_{f,\med}$ but from the full domain $y^+-\bar{y}^+\le t_{f,\med}$ with unconstrained $y^+$ in agreement with the definition of $t_f=y^+-\bar{y}^+$.
The subsequent broadening treats the short formation times VLEs with $y^+\le L$ differently as those emitted outside the medium, leading to the asymmetrical result \eqref{off-y<tf}. In the next section, we will understand the physical origin of this factor $2$ and the mismatch of \eqref{off-y<tf} in the regime $k_{f,\med}<k_\perp<Q_s$ where one cannot approximate the probability $\mathcal{P}$ by a Dirac delta.

\paragraph{(2) The domain $y^+\ge L$ and $t_f\ge t_{f,\med}$} Let us now come back to formula \eqref{off-shell-v3} and investigate the contribution coming from $y^+\ge L$ and $t_f\ge t_{f,\med}$ (we have already collected the $y^+\ge L$ and $t_f\le t_{f,\med}$ piece in the previous paragraph.) The exact calculation of the $y^+\ge L$ integral in \eqref{off-shell-v3}, where the broadening probability vanishes, reads:
\begin{align}\label{y>L-calc}
 \frac{1}{2i\om}\int_L^\infty \dif y^+\,2\exp\left(\frac{-k_\perp^2 G^{-1}(y^+)}{2 i\om}\right)&=\frac{2}{k_\perp^2}\exp\left(\frac{k_\perp^2L}{2i\om}j\sqrt{\frac{\om}{\om_c}}\tanh\Big(\bar{j}\sqrt{\frac{\om_c}{\om}}\Big)\right)\\
&= \frac{2}{k_\perp^2}\exp\left(-\bar{j}\frac{k_\perp^2}{k^2_{f,\med}}\right)\textrm{ for }\om\ll \om_c\\
 &\approx\frac{2}{k_\perp^2}\Theta(k_{f,\med}-k_\perp)
\end{align}
so that the contribution from $L$ to $\infty$ gives a Bremsstrahlung with support for $k_\perp\le k_{f,\med}$ (and not $k_\perp^2<2\om/L$ !). Combined with the $-1/k_\perp^2$, where the constraint $t_f\ge t_{f,\med}$ is imposed, the contribution from times larger than $L$ and $t_f\ge t_{f,\med}$ to the full spectrum reads:
\begin{equation}
 \om\frac{\dif^3 N^{\textrm{off-shell, }y^+>L,t_f>t_{f,\med}}}{\dif\om\dif^2 k_\perp}=\frac{\alpha_s C_F}{\pi^2}\frac{1}{k_\perp^2}\Theta(k_{f,\med}-k_\perp)
\end{equation}
Note that one would have obtained a wrong result using the approximation $G^{-1}(y^+)\simeq y^+-L$ for $y^+\ge L$, valid when $\om\ll\om_c$ though.

\paragraph{(3) The domain $t_{f,\med}\le y^+\le L$.} In this regime, the emission rate is roughly constant. Indeed, using $\tanh(x)\approx1$ for $x\ge 1$, one gets $G^{-1}(t)\simeq j\sqrt{2\om/\qhat}$. This leads to a purely medium induced component in the full spectrum without $1/k_\perp^2$ behaviour:
\begin{equation}\label{mie-spec-revisited}
 \om\frac{\dif^3 N^{\textrm{off-shell, } t_{f,\med}< y^+ <L}}{\dif\om\dif^2 k_\perp}=\frac{2\alpha_s C_F}{\pi^2}\mathfrak{Re}\frac{1}{2i\om}\int_{0}^L \dif y^+ \int\dif^2 q_\perp\,\mathcal{P}(k_\perp-q_\perp,y^+,\infty)\exp\left(-\bar{j}\frac{q_\perp^2}{k^2_{f,\med}}\right)
\end{equation}
As the broadening probability averaged over time appearing in this formula is not controlled  by small times, one can safely take $y^+=0$ instead of $t_{f,\med}$ in the lower boundary of the time integral. Note that the time integral can be performed exactly in the harmonic approximation and gives the result obtained in \eqref{KCAmultiple} in a different context. With this expression, the medium-induced mechanism is transparent: gluons are created with a constant formation rate and a transverse momentum of order $k_{f,\med}$ (with Gaussian distribution) all along the medium and then undergo transverse momentum broadening so that they end up with a transverse momentum of order $Q_s$. For medium induced emissions, one sees that $t_{f,\med}\le y^+\le L\Leftrightarrow t_f=t_{f,\med}$.

\paragraph{Summary.} Let us summarize the discussion of the one gluon emission spectrum from an off-shell quark. In the presence of the medium, $y^+$ and $\bar{y}^+$ for Bremsstrahlung emissions are not any more solely controlled by $t_f$ contrary to the vacuum calculation \eqref{vac-brem-int}, so the discussion involves both $y^+$ and $\bar{y}^+$ .The full spectrum receives three kinds of contribution from three different integration domains:
\begin{enumerate}
 \item Short formation time vacuum-like emissions from the domain $t_f=|y^+-\bar{y}^+|<t_{f,\med}$ so that $k_\perp\ge k_{f,\med}$.
 \item Medium induced emissions from the domain $t_{f,\med}\le y^+ \le L$ with a constant formation rate. This imposes $|y^+-\bar{y}^+|\simeq t_{f,\med}$. These emissions produce a peak in the spectrum around $k_\perp=Q_s$ but final state broadening aside, these emissions are peaked around $k_\perp=k_{f,\med}$.
 \item Vacuum-like emissions from the domain $y^+\ge L$ and $|y^+-\bar{y}^+|\ge t_{f,\med}$, so that $k_\perp\le k_{f,\med}$.

\end{enumerate}

\section{The on-shell spectrum in a static medium}

To calculate the vacuum-like spectrum defined in \ref{vle-spec-def}, we need to understand the behaviour of the spectrum \eqref{vle-spec-def}, and especially the contributions which have a double-logarithmic structure, and which should therefore be resummed.
To do so, we rely again on the brick model where the analytical calculations are simple. We recall that our aim is solely to extract the physical scales in order to demonstrate the results obtained in Section \ref{sub:qualitative-veto}. For the brick model, the on-shell spectrum reads:
\begin{align}\label{on-shell-recap}
 \om\frac{\dif^3N^{\textrm{on-shell}}}{\dif\om\dif^2 k_\perp}&=\frac{\alpha_sC_F}{\pi^2}\left[\int\frac{\dif^2q_\perp}{q_\perp^2}\mathcal{P}(k_\perp-q_\perp,0,L)\Theta(|q_\perp|-\mu)-\frac{1}{k_\perp^2}\right.\nonumber\\
 &+\frac{2}{k_\perp^2}\mathfrak{Re}\exp\left(-\bar{j}\frac{k_\perp^2}{k_{f,\med}^2}\coth\Big(j\sqrt{\frac{\om_c}{\om}}\Big)\right)\nonumber\\
 &\left.+2\mathfrak{Re}\int_0^L\dif y^+\int\dif^2 q_\perp\mathcal{P}(k_\perp-q_\perp,y^+,L)\frac{1}{2i\om}\exp\left(-\bar{j}\frac{q_\perp^2}{k_{f,\med}^2}\coth\Big(\bar{j}\sqrt{\frac{\om_c}{\om}}y^+\Big)\right)\right]
\end{align}
where $\mu$ is a regularization scale introduced in Chapter \ref{chapter:emissions}-Section \ref{subsub:TMdep-BDMPS}. One easily checks that this expression vanishes in the limit $\qhat\rightarrow0$. 

We now prove formula \eqref{on-shell-physic}. The last term of \eqref{on-shell-recap} is not singular as $k_\perp\rightarrow 0$. Actually, the integral over $k_\perp$ of this term gives precisely the integrated BDMPS-Z spectrum. It is interesting to notice that even if the definition \eqref{off-shell-recap} for the medium-induced spectrum and the alternative definition
\begin{equation}\label{other-bdmpsz}
 \om\frac{\dif^3 \tilde{N}^{\mie}}{\dif^2 \om\dif k_\perp^2}\equiv2\mathfrak{Re}\int_0^L\dif y^+\int\dif^2 q_\perp\mathcal{P}(k_\perp-q_\perp,y^+,L)\frac{1}{2i\om}\exp\left(-\bar{j}\frac{q_\perp^2}{k_{f,\med}^2}\coth\Big(\bar{j}\sqrt{\frac{\om_c}{\om}}y^+\Big)\right)
\end{equation}
give the same result once integrated over $k_\perp$ (we have demonstrated this in Chapter \ref{chapter:emissions}-Section \ref{subsub:integrated-BDMPS} for the brick model, see also Appendix \ref{app:B}) they are not equal and do not have the same $k_\perp$ dependence. This term is denoted as ``BDMPS-Z'' in formula \eqref{on-shell-physic}.

When $\om\ll\om_c$, the third term in \eqref{on-shell-recap} can be approximated by:
\begin{equation}
 \frac{2}{k_\perp^2}\mathfrak{Re}\exp\left(-\bar{j}\frac{k_\perp^2}{k_{f,\med}^2}\coth\Big(j\sqrt{\frac{\om_c}{\om}}\Big)\right)\simeq \frac{2}{k_\perp^2}\Theta(k_{f,\med}-k_\perp)
\end{equation}
Finally, decomposing the $-1/k_\perp^2$ term in \eqref{on-shell-recap} using $1= \Theta(k_{f,\med}-k_\perp)+\Theta(k_\perp-k_{f,\med})$, one gets:
\begin{align}
 \om\frac{\dif^3N^{\textrm{on-shell}}}{\dif\om\dif^2 k_\perp}&\simeq\frac{\alpha_sC_F}{\pi^2}\Big(\int\frac{\dif^2q_\perp}{q_\perp^2}\Big(\mathcal{P}(k_\perp-q_\perp,0,L)-\delta(k_\perp-q_\perp)\Big)\Theta(|q_\perp|-k_{f,\med})\nonumber\\
 &\hspace{2cm}+\frac{1}{k_\perp^2}\Theta(k_{f,\med}-k_\perp)\Big)+\textrm{``BDMPS-Z''}\\
 &\simeq\frac{\alpha_sC_F}{\pi^2} \frac{1}{k_\perp^2}\Theta(k_{f,\med}-k_\perp)+\textrm{``BDMPS-Z''}
\end{align}
for $\mu=k_{f,\med}$.
To get the second line which is precisely formula \eqref{on-shell-physic}, we used the fact that $\mathcal{P}$ is normalised to 1, so that the integral over $k_\perp$ of the first term does not generate any collinear logarithm.

\section{Leading behaviour of the vacuum-like spectrum}

Now, let us explain how this on-shell spectrum combines with the off-shell one. For our purposes here, one can perfectly approximate the cotangent function in the exponential in \eqref{other-bdmpsz} by $1$ and one sees that this third term exactly cancels the medium induced term \eqref{mie-spec-revisited} in the off-shell spectrum.  

The second term, using $\coth(j\sqrt{\om_c/\om})\simeq 1$ for $\om\le\om_c$, cancels the vacuum-like contribution of the off-shell spectrum for $y^+\ge L$ calculated in \eqref{y>L-calc}. The $-1/k_\perp^2$ cancels with the mysterious $-1/k_\perp^2$ in the off-shell spectrum. Thus, for $\om\le\om_c$, the in-medium vacuum-like spectrum as defined in \eqref{vle-spec-def} behaves like:
\begin{align}\label{vle-spec-inf2}
 \om\frac{\dif^3 N^{\textrm{VL}}}{\dif \om\dif^2k_\perp}&\simeq\frac{\alpha_sC_F}{\pi^2}\int\frac{\dif^2q_\perp}{q_\perp^2}\mathcal{P}(k_\perp-q_\perp,0,L)\Big(2\Theta(|q_\perp|-k_{f,\med})-\Theta(|q_\perp|-\mu)\Big)\\
 &\simeq\frac{\alpha_sC_F}{\pi^2}\int\frac{\dif^2q_\perp}{q_\perp^2}\mathcal{P}(k_\perp-q_\perp,0,L)\Theta(|q_\perp|-k_{f,\med})
\end{align}
where we have used $\mu=k_{f,\med}$ in the last line according to our discussion in Chapter \ref{chapter:DLApic}, Section \ref{sec:veto}. This demonstrates formula \eqref{vle-spec-inf}.

\section{Shockwave limit of the gluon spectra}

When the size of the medium is taken into account, via the jet path length $L$, the resulting formula are more complicated and the physics behind may be hidden (see Eqs.~\eqref{off-shell-v3}-\eqref{on-shell-recap}). It is therefore enlightening to consider the ``shockwave limit'' defined as:
\begin{equation}
 L\rightarrow0\,,\quad Q_s=\sqrt{\qhat L}\textrm{ fixed}
\end{equation}
This limit captures the interplay between the virtuality of the initial process and the subsequent multiple soft scatterings via the unique remaining transverse scale $Q_s$.

\paragraph{Shockwave limit of the on-shell spectrum.} In this limit, the medium-induced spectrum \eqref{other-bdmpsz} vanishes, since the integration range scales with $L$. The third term in \eqref{on-shell-recap} has a well-defined limit:
\begin{equation}
 \frac{2}{k_\perp^2}\mathfrak{Re}\exp\left(-\bar{j}\frac{k_\perp^2}{k_{f,\med}^2}\coth\Big(j\sqrt{\frac{\om_c}{\om}}\Big)\right)\underset{\substack{L\rightarrow0\\Q_s=\rm cste}}\longrightarrow\frac{2}{k_\perp^2}\exp\left(-\frac{k_\perp^2}{Q_s^2}\right)
\end{equation}
Combining all the terms of \eqref{on-shell-recap} in this limit, one finds:
\begin{align}
  \om\frac{\dif^3N^{\textrm{on-shell, SW}}}{\dif\om\dif^2 k_\perp}&=\frac{\alpha_sC_F}{\pi^2}\left[\int_{|q_\perp|>\mu}\frac{\dif^2q_\perp}{q_\perp^2}\mathcal{P}(k_\perp-q_\perp,0,L)-\frac{2}{k_\perp^2}\Big(1-e^{-k_\perp^2/Q_s^2}\Big)+\frac{1}{k_\perp^2}\right]\\
&=\frac{\alpha_sC_F}{\pi^2}\left[\frac{4\pi}{Q_s^2}\int_{|q_\perp|>\mu}\frac{\dif^2q_\perp}{(2\pi)^2}\frac{1}{q_\perp^2}e^{-(k_\perp-q_\perp)^2/Q_s^2}-\frac{4\pi}{Q_s^2}\int\frac{\dif^2q_\perp}{(2\pi)^2}\frac{2k_\perp q_\perp}{q_\perp^2k_\perp^2}e^{-(k_\perp-q_\perp)^2/Q_s^2}\right.\nonumber\\
&\hspace{11cm}\left.+\frac{1}{k_\perp^2}\right]\\
  &=\frac{\alpha_s C_F}{\pi^2}\left(\frac{4\pi}{Q_s^2}\int\frac{\dif^2 q_\perp}{(2\pi)^2}\frac{q_\perp^2}{(k_\perp-q_\perp)^2k_\perp^2}e^{-\frac{q_\perp^2}{Q_s^2}}\right)\,,\qquad\textrm{for }\mu\rightarrow0\label{Gunion-Bertsch}
\end{align}
The on-shell spectrum has a Gunion-Bertsch form \cite{Gunion:1981qs}. The exponential corresponds to the Gaussian probability for momentum transfer $q_\perp$. For $k_\perp\ll Q_s$, the spectrum \eqref{Gunion-Bertsch} has a Bremsstrahlung tail in $\frac{\alpha_sC_F}{\pi^2}\frac{1}{k_\perp^2}$. This is precisely why it is interesting to look at the on-shell spectrum: all the Bremsstrahlung components are related to the scattering off the medium. For $k_\perp\gg Q_s$, the spectrum behaves like $\frac{\alpha_sC_F}{\pi^2}\frac{Q_s^2}{k_\perp^4}$.
\paragraph{Shockwave limit of the off-shell spectrum.} The shockwave limit of the off-shell spectrum can be readily deduced from Eq.~\eqref{offshell-final}. One must distinguish two cases: either $x_0^+=x_i^+=0$, meaning that the creation vertex $x_i^+$ of the quark is located at the same position as the shockwave or $x_0^+>0$. In the latter case, $x_0^+$ is the location of the shockwave along the eikonal quark trajectory.
For the former case, one can directly take the shockwave limit of \eqref{off-y<tf}, \eqref{y>L-calc} and \eqref{mie-spec-revisited}:
\begin{equation}
 \om\frac{\dif^3N^{\textrm{off-shell, SW}}}{\dif\om\dif^2 k_\perp}=\frac{\alpha_sC_F}{\pi^2}\frac{1}{k_\perp^2}
\end{equation}
which is exactly the vacuum Bremsstrahlung spectrum. In this scenario, nothing happens. Actually, we have seen in Section \ref{sub:decoherence} that if we start with a virtual onium instead of a virtual quark, the medium plays a non trivial role even in this simple situation, via the decoherence of the colour dipole.

The case $x_0^+>0$ (with $x_i^+=0$) is much more interesting. The function $G$ in \eqref{offshell-final} satisfies:
\begin{align}
G^{-1}(y^+)&=y^+&\textrm{ if }y^+<x_0^+\\
G^{-1}(y^+)&=\frac{1}{\Omega}\frac{\tan\big(\Omega(y^+-x_0^+)\big)+\Omega x_0^+}{1-\Omega x_0^+\tan\big(\Omega(y^+-x_0^+)\big)}&\textrm{ if }x_0^+\le y^+<x_0^++L\\
G^{-1}(y^+)&=y^+-x_0^+-L+\frac{1}{\Omega}\frac{\tan\big(\Omega L\big)+\Omega x_0^+}{1-\Omega x_0^+\tan\big(\Omega L\big)}&\textrm{ if }x_0^++L\le y^+
\end{align}
with $\Omega^2=i\qhat/(2\omega)$. The integral appearing in \eqref{offshell-final} between $x_0^+$ and $x_0^+ +L$ vanishes in the limit $L\rightarrow0$. The integral between $0$ and $x_0^+$ reads
\begin{equation}
 \frac{2\alpha_sC_F}{\pi^2}\int\frac{\dif^2q_\perp}{q_\perp^2}\mathcal{P}(k_\perp-q_\perp,x_0^+,x_0^++L)\Big(1-e^{-\frac{q_\perp^2x_0^+}{2i\om}}\Big)
\end{equation}
This integral is naturally regularised by the finite value of $x_0^+$. Note the presence of the factor $2$ as in \eqref{off-y<tf}. The integral with $y^+\ge x_0^+ +L$ gives:
\begin{equation}
\frac{2\alpha_sC_F}{\pi^2} \frac{1}{k_\perp^2}\,\exp\left(-\frac{k_\perp^2}{2i\om}\frac{1}{\Omega}\frac{\tan\big(\Omega L\big)+\Omega x_0^+}{1-\Omega x_0^+\tan\big(\Omega L\big)}\right)\underset{\substack{L\rightarrow0\\ Q_s=\rm cste}}\longrightarrow\frac{2\alpha_sC_F}{\pi^2}\frac{1}{k_\perp^2}\exp\left(-\frac{k_\perp^2}{Q_s^2}\Big(1+\frac{2i\om}{Q_s^2x_0^+}\Big)^{-1}\right)
\end{equation}
where we emphasize the importance of $x_0^+>0$ to get the shockwave limit. Combining all terms together, one finds:
\begin{align}
 \om\frac{\dif^3N^{\textrm{off-shell, SW}}}{\dif\om\dif^2 k_\perp}&=\frac{\alpha_sC_F}{\pi^2}\mathfrak{Re}\left[2\int\frac{\dif^2q_\perp}{q_\perp^2}\mathcal{P}(k_\perp-q_\perp,x_0^+,x_0^++L)\Big(1-e^{-\frac{q_\perp^2x_0^+}{2i\om}}\Big)\right.\nonumber\\
 &\hspace{3cm}\left.-\frac{2}{k_\perp^2}\left[1-\exp\left(-\frac{k_\perp^2}{Q_s^2}\Big(1+\frac{2i\om}{Q_s^2x_0^+}\Big)^{-1}\right)\right]+\frac{1}{k_\perp^2}\right]\\
 &=\frac{\alpha_sC_F}{\pi^2}\mathfrak{Re}\left[\frac{4\pi}{Q_s^2}\int\frac{\dif^2q_\perp}{(2\pi)^2}\frac{2e^{-(k_\perp-q_\perp)^2/Q_s^2}}{q_\perp^2}\Big(1-e^{-\frac{q_\perp^2x_0^+}{2i\om}}\Big)\right.\nonumber\\
 &\hspace{3cm}\left.-\frac{2}{k_\perp^2}\left[1-\exp\left(-\frac{k_\perp^2}{Q_s^2}\Big(1+\frac{2i\om}{Q_s^2x_0^+}\Big)^{-1}\right)\right]+\frac{1}{k_\perp^2}\right]\label{off-SW}
\end{align}
which looks very much like \eqref{Gunion-Bertsch} except for the factor $2$ in front of the Gaussian broadening convolution. 
To conclude, the spectrum \eqref{off-SW} can be computed fully analytically:
\begin{align}
 \om\frac{\dif^3N^{\textrm{off-shell, SW}}}{\dif\om\dif^2 k_\perp}&=\frac{\alpha_sC_F}{\pi^2}\mathfrak{Re}\left(\frac{2e^{-k_\perp^2/Q_s^2}}{Q_s^2}\left[\textrm{Ei}\left(\frac{k_\perp^2}{Q_s^2}\right)-\textrm{Ei}\left(\frac{k_\perp^2}{Q_s^2}\Big(1+\frac{Q_s^2x_0^+}{2i\om}\Big)^{-1}\right)\right]\right.\nonumber\\
 &\hspace{5cm}\left.+\frac{2}{k_\perp^2}\exp\left(-\frac{k_\perp^2}{Q_s^2}\Big(1+\frac{2i\om}{Q_s^2x_0^+}\Big)^{-1}\right)-\frac{1}{k_\perp^2}\right)
\end{align}
with $\textrm{Ei}(x)$ the exponential integral function. This spectrum behaves like $\frac{\alpha_sC_F}{\pi^2}\frac{1}{k_\perp^2}$ at large and small $k_\perp$. The $k_\perp$-integrated ``medium-induced'' spectrum reads:
\begin{align}
 \om\frac{\dif N^{\textrm{mie, SW}}}{\dif\om}&=\int\dif k_\perp^2\left(\om\frac{\dif^3N^{\textrm{off-shell, SW}}}{\dif\om\dif^2 k_\perp}-\frac{\alpha_sC_F}{\pi^2}\frac{1}{k_\perp^2}\right)\\
 &=\frac{\alpha_sC_F}{\pi}\log\left(1+\frac{x_0^{+2}Q_s^4}{4\om^2}\right)\\
 &\simeq\frac{2\alpha_sC_F}{\pi}\log\left(\frac{x_0^+Q_s^2}{2\om}\right)\,,\qquad\mbox{for }Q_s^2\gg\frac{2\om}{x_0^+} \label{Nmie-SW}
\end{align}
This result is particularly interesting, because it sheds light on the interplay between leading and higher twist effects. As the initial quark is off-shell, it can radiate according to the Bremsstrahlung law: this is a leading twist effect. On top of that, the shockwave can trigger Bremsstrahlung emissions, but these emissions must satisfy $k_\perp\le Q_s$ since $Q_s$ is the transverse momentum transferred by the shockwave. The result \eqref{Nmie-SW} has the following physical content:
\begin{itemize}
\item the leading twist emissions populate all the phase space,
\item whereas the higher twist emissions populate the phase space $k_\perp\le Q_s$.
\item These two mechanisms interfere for $k_\perp^2\le 2\om/x_0^+$, so that a single Bremsstrahlung tail remains at small $k_\perp$.
\end{itemize}
The net result is a combination of two Bremsstrahlung in the phase space $2\om/x_0^+\le k_\perp^2\le Q_s^2$, which gives the logarithmic area of this domain in the $k_\perp$ integrated spectrum \eqref{Nmie-SW}.

\chapter{Saddle-point method for in-medium intrajet multiplicity at DLA}
\chaptermark{Intrajet multiplicity at DLA}
\label{app:DLA}

Our starting point is Eq.~\eqref{master-eq},
assuming~$\omega_L(R)<\ktmin/R$.
For definiteness, we also assume $\th^2\ge\th_c^2$, although it turns out that our
conclusions remain valid for $\th^2\le \th_c^2$.
It is convenient to use logarithmic variables:
$x_1=\log(p_{T0}/\om_1)$, $y_1=\log(R^2/\th_1^2)$, $x_2=\log(\om_2/\om)$,
$y_2=\log(\th_2^2/\th^2)$ and $X\equiv\log(p_{T0}/\om)$,
$Y\equiv\log(R^2/\th^2)$. The energy scales $\omega_0(R)$ and
$\omega_L(R)$, related respectively to the inside and outside domains,
become $x_{0}\equiv\log(p_{T0}/\om_0(R))$ and
$x_L\equiv\log(p_{T0}/\om_L(R))$, and the logarithmic scale associated
with $\th_c^2$ is $y_c\equiv\log(R^2/\th_c^2)=4(x_L-x_{0})/3$.
To get the leading asymptotic behaviour of $T_{i,\textrm{out}}(X,Y)$,
one can neglect the $\delta$ contribution to $T^{\textrm{vac}}$
in~(\ref{Tvac}) since it generates terms with at least one exponential
factor missing. We thus get
\begin{align}\label{master-3}
 T_{i,\textrm{out}}(X,Y)=\abar^3\int_{0}^{\min(X,x_{0})} dx_1\int_{0}^{\min(y_c,\frac{3}{2}(x_{0}-x_1))}dy_1&\int_{0}^{\min(X-x_1,X+Y-x_L)}dx_2\int_{0}^{X+Y-x_L-x_2}dy_2 \nonumber\\
 &\hspace{1.3cm}\textrm{I}_0(2\sqrt{\abar x_1 y_1})\,\textrm{I}_0(2\sqrt{\abar x_2 y_2})
 \end{align}
The integral over $y_1$ and $y_2$ can be performed exactly using the the following relation:
\begin{equation}\label{angular-integral}
  \int_0^s dy\, \textrm{I}_0(2\sqrt{\abar x y})
  =\sqrt{\frac{s}{\abar x}}\textrm{I}_1(2\sqrt{\abar x s})
  \overset{\abar xs\gg 1}{\simeq}\sqrt{\frac{s}{\abar x}}\frac{\exp(2\sqrt{\abar x s})}{\sqrt{4\pi\sqrt{\abar x s}}}.
\end{equation}
Using \eqref{angular-integral}, one gets
\begin{align} \label{master-4}
 T_{i,\textrm{out}}(X,Y)=\abar^3\int_{0}^{\min(X,x_{0})} dx_1&\int_{0}^{\min(X-x_1,X+Y-x_L)}dx_2\,R_1(x_1)R_2(x_2) \nonumber\\
 &\hspace{2.cm} e^{2\sqrt{\abar x_1 \min(y_c,\frac{3}{2}(x_{0}-x_1))}} e^{2\sqrt{\abar x_2 (X+Y-x_L-x_2)}}
\end{align}
with the two non-exponential functions
\begin{equation}
 R_1(x_1)=\frac{1}{\sqrt{4\pi}}\frac{(\min(y_c,\frac{3}{2}(x_{0}-x_1)))^{1/4}}{(\abar x_1)^{3/4}},\qquad
 R_2(x_2)=\frac{1}{\sqrt{4\pi}}\frac{(X+Y-x_L-x_2)^{1/4}}{(\abar x_2)^{3/4}}
\end{equation}

The $x_2$ integrations cannot be performed exactly so we use the
saddle-point approximation:
\begin{equation}
  \int_{x_1}^{x_2} dx\, f(x)e^{Mg(x)}
  \overset{M\to\infty}\simeq\sqrt{ \frac{2\pi}{-Mg''(x^\star)}}f(x^\star)e^{Mg(x^\star)},
\end{equation}
where the saddle point $x^\star$ is the \textit{maximum} of $g(x)$
between $x_1$ and $x_2$. This formula is valid as long as
$x_1<x^\star<x_2$.

Setting $M_2\equiv(X+Y-x_L)=\log(\omega_L(\theta)/\omega)$ and
integrating over $x_2/M_2$, one get
\begin{equation}
 \mathcal{N}_{\textrm{out}} \equiv \abar\int_{0}^{\min(X-x_1,X+Y-x_L)}dx_2\, R_2(x_2)e^{2\sqrt{\abar x_2 (X+Y-x_L-x_2)}}
 \overset{\sqrt{\abar}M_2\rightarrow\infty}{\simeq} \frac{1}{2}e^{\sqrt{\abar}M_2}.
\end{equation}
The corresponding saddle point is
$x_2^\star=M_2/2=\log(\sqrt{\omega_L(\theta)/\omega})$ so that the saddle-point approximation is valid if $x_2^\star<X-x_1$. This gives the condition $x_1\le X-x_2^\star$ in the first integral, in order to ensure energy conservation along the cascade.

Calling $\mathcal{N}_{\textrm{med}}$ the remaining integral over
$x_1$, which is truly a gluon multiplicity \textit{inside} the medium,
we are left with:
\begin{equation}
 \mathcal{N}_{\textrm{med}}\equiv\abar\int_{0}^{\min(x_{0},X-x_2^\star)} dx_1\,R_1(x_1)e^{2\sqrt{\abar x_1 \min(y_c,\frac{3}{2}(x_{0}-x_1))}}.
\end{equation}
Since $\min(X-x_2^\star,x_{0})>x_c\equiv\log(p_{T0}/\om_c)$, the
integral can be split into two pieces: $x_1<x_c$ where
$\min(y_c,3(x_{0}-x_1)/2)=y_c$ and $x_1>x_c$ where
$\min(y_c,3(x_{0}-x_1)/2)=3(x_{0}-x_1)/2$. The first piece is
calculated exactly, and we use again the saddle point method to
evaluate the second piece, assuming
$x_0=\log(p_{T0}/\omega_0(R))\to \infty$. We get (using $x_1'=x_1/x_0$)
\begin{align} 
 \mathcal{N}_{med}&=\int_{0}^{x_c}dx_1\abar \sqrt{\frac{y_c}{\abar x_1}}\textrm{I}_1(2\sqrt{\abar x_1 y_c})+\abar\int_{x_c}^{\min(X-x_2^\star,x_{0})}dx_1\,R(x_1)e^{2\sqrt{\abar x_1\frac{3}{2}(x_{0}-x_1)}}
 \nonumber \\*[0.2cm]
 &=-1+\textrm{I}_0(2\sqrt{\abar x_c y_c})\,+ \abar^{1/4}\sqrt{\frac{x_0}{4\pi}}\int_{x_c/x_0}^{\textrm{min}(1,(X-x_2^\star)/x_0)}\frac{dx_1'}{x_1'^{1/2}}\Big(\frac{3(1-x_1')}{2x_1}\Big)^{1/4} \,
 e^{2x_0\sqrt{\frac{3}{2}\abar x'_1(1-x'_1)}}\nonumber \\*[0.2cm]
 &\overset{\sqrt{\abar}x_0\rightarrow\infty}{\sim} \frac{e^{2\sqrt{\abar x_c y_c}}}{\sqrt{4\pi\sqrt{\abar x_c y_c}}}+\frac{1}{2}e^{\sqrt{\frac{3\abar}{2}}x_{0}}\, \label{Nmed-calc}
\end{align}
The first term in equation \eqref{Nmed-calc} is sub-leading due to the
square root in the argument and in the denominator. Thus, the leading
term for $\mathcal{N}_{med}$ comes from the ``inside-medium'' region
with $\om_1\le\om_c$.\footnote{That is why we can trust our final
  result for $T(\om,\th^2)$ even for $\th^2\le\th_c^2$.}

The saddle point of the integral over $x_1$ is $x_1^\star=x_{0}/2=\log(\sqrt{p_{T0}/\omega_0(R)})$ so our estimation for $\mathcal{N}_{med}$ is valid only if $x_c<x_1^\star<X-x_2^\star$. The condition $x_c<x_1^\star$ leads to the condition~\eqref{eq:cdt-pt0-omc}. The condition $x_1^\star<X-x_2^\star$ leads to the condition~(\ref{ocr}), when $x_2^\star=\log(\sqrt{\omega_L(\theta)/\omega})$ is evaluated at its largest value, that is when $\th=\th_{\textrm{min}}\equiv\ktmin/\omega$.

We have thus demonstrated that when both $\sqrt{\abar}x_0\equiv
\sqrt{\abar}\log(p_{T0}/\omega_0(R))$ and
$\sqrt{\abar}(X+Y-x_L)\equiv\sqrt{\abar}\log(\omega_L(\theta)/\omega)$
are large and $X>x_1^\star+x_2^\star$, i.e. $\omega<\omega_{cr}$, we have
\begin{equation}\label{Tout-asymptot} 
 T_{i,\textrm{out}}(X,Y)\sim\frac{\abar}{4}\exp\left[\sqrt{\abar}\left(X+Y-x_L+\sqrt{\frac{3}{2}}x_{0}\right)\right],
\end{equation}
which is precisely formula \eqref{Tsaddle}.

From~\eqref{Tout-asymptot} and~\eqref{fullT}, one deduces the
asymptotic DLA behaviour of the small-$x$ fragmentation function by
integrating $T_i(\omega,\th^2|p_{T0},R^2)$ over $\th^2$ between
$\ktmin^2/\omega^2$ and $R^2$. The leading contribution comes from the lower limit of this integral or, in logarithmic units, from the upper bound $2(x_{\textrm{max}}-X)$ on the integral on $Y$, 
with $x_{\textrm{max}}=\log(p_{T0}R/\ktmin)$. This reproduces~\eqref{frag-DLA} in logarithmic units:
\begin{equation}
 D^{\textrm{med}}_i(X)
 =\int_0^{2(x_{\textrm{max}}-X)}\dif Y\,T_{i}(X,Y)
\simeq \frac{\sqrt{\abar}C_i}{4C_A}\exp\left[\sqrt{\abar}\Big(-X + 2 x_{\textrm{max}} - x_L+\sqrt{\frac{3}{2}}x_{0}\Big)\right] \label{Dmed-asymptot}
\end{equation}

Finally, the asymptotic form of the ratio $\mathcal{R}_i(X)\equiv\,D^{\textrm{med}}_i(X)/D_i^{\textrm{vac}}(X)$ is obtained from \eqref{Dmed-asymptot} and \eqref{Dg-dla}, using again the asymptotic form of $\textrm{I}_1(x)$ at
 large $x$:
 \begin{align}
 D_i^{\textrm{vac}}(X)
 & \simeq\,\frac{C_i}{\sqrt{4\pi}C_A}\left[\frac{2\abar(x_{\textrm{max}}-X)}{X^3}\right]^{1/4}\exp\Big(2\sqrt{2\abar X(x_{\textrm{max}}-X)}\Big)\label{Dvac-approx}\\
  \mathcal{R}_i(X)& \sim\frac{\sqrt{\abar\pi}}{2}e^{\sqrt{\frac{3}{2}}x_{0}-x_L}\left[\frac{X^3}{2\abar(x_{\textrm{max}}-X)}\right]^{1/4}\exp\left[\sqrt{\abar}\big(\sqrt{X}-\sqrt{2(x_{\textrm{max}}-X)}\big)^2\right].\label{Rfinal}
 \end{align}
 From~\eqref{Dvac-approx}, one can estimate the position of the maximum $x_{\textrm{hump}}$ of $D_i^{\textrm{vac}}(X)$. Neglecting the non-exponential prefactor,  one finds ${\dif D_i^{\textrm{vac}}}/{\dif X}
\propto x_{\textrm{max}}-2X$,
so that the $x_{\textrm{hump}}\simeq x_{\textrm{max}}/2$ and $\om_{\textrm{hump}}\simeq \sqrt{p_{T0}\ktmin/R}$. For $X\ge x_{\textrm{hump}}$ i.e. $\om\le\om_{\textrm{hump}}$, the derivative is negative, hence $D_i^{\textrm{vac}}(\omega)$ decreases when $\omega$ decreases.
Similarly, one can study the variation of $\mathcal{R}_i(X)$ from the exponential factor alone:
\begin{equation}
\frac{\dif \mathcal{R}_i}{\dif X}\simeq\frac{\abar\sqrt{\pi}}{2}e^{\sqrt{\frac{3}{2}}x_{0}-x_L}\frac{\big(\sqrt{2X}+\sqrt{x_{\textrm{max}}-X}\big)\big(\sqrt{X}-\sqrt{2(x_{\textrm{max}}-X)}\big)}{\sqrt{X(x_{\textrm{max}}-X)}}e^{\sqrt{\abar}\big(\sqrt{X}-\sqrt{2(x_{\textrm{max}}-X)}\big)^2}
\end{equation}
The derivative is positive when $\sqrt{X}-\sqrt{2(x_{\textrm{max}}-X)}\ge0$ i.e. when $X\ge 2x_{\textrm{max}}/3$. Hence, for $\om\lesssim(p_{T0}\ktmin^2/R^2)^{1/3}$, the ratio $R_i(\omega)$ increases when $\om$ decreases.

\chapter{Medium-modified Sudakov form factors with running coupling}
\chaptermark{Medium-modified Sudakov factors}
\label{sec:rc-effects}

For the effects of the veto region, the expression corresponding to
Eq.~(\ref{eq:g1-veto}) and including running-coupling effects is
found to be
\begin{equation}\label{eq:g1-veto-rc}
  Lg_{1,i}^{\textrm{veto}}(u,v) = Lg_{1,i}(u,v)+\frac{2C_i}{\pi}\mathcal{A}_{\textrm{veto}}(L)
\end{equation}
where the logarithmic area of the veto region $\mathcal{A}_{\textrm{veto}}(L)$ is defined as:
\begin{equation}\label{eq:aveto-def}
\mathcal{A}_{\textrm{veto}}(L)=\int_{e^{-L}}^1 \frac{\dif z}{z}\int_{0}^{R}\frac{\dif\theta}{\theta}\alpha_s(z p_{T0}\theta)(1-\Theta_{\textrm{veto}})
\end{equation}
and $\Theta_{\textrm{veto}}$ is given by \eqref{step-veto}. Introducing the following function:
\begin{equation}
\mathcal{T}(x,y,z)\equiv \frac{y+zx}{z}\ln(1+\alpha_s\beta_0(y+zx))\,,
\end{equation}
the logarithmic area $\mathcal{A}_{\textrm{veto}}(L)$ reads:

\begin{align*}\label{eq:aveto}
  \mathcal{A}_{\textrm{veto}}(L) \overset{1-x<z_L}=
  &\frac{1}{2\beta_0}\Big[
    \mathcal{T}\big(\ln z_{0},0,2\big)
    -\mathcal{T}\big(\ln z_L,0,2\big)
   +\mathcal{T}\big(\ln z_c ,\tfrac{3}{2}\ln z_{0},\tfrac{1}{2}\big)
    -\mathcal{T}\big(\ln z_{0},\tfrac{3}{2}\ln z_{0},\tfrac{1}{2}\big)\\
  & -\mathcal{T}\big(\ln z_0,\ln z_L,1\big)
    +\mathcal{T}\big(\ln z_L,\ln z_L,1\big)
    -\mathcal{T}\big(\ln z_c,\ln z_L,1\big)
    +\mathcal{T}\big(\ln z_{0},\ln z_L,1\big)\Big]\\
  \overset{z_L<1-x<z_0}=
  &\frac{1}{2\beta_0}\big[
    \mathcal{T}\big(\ln z_{0},0,2\big)
    -\mathcal{T}\big({-L},0,2\big)
    +\mathcal{T}\big(\ln z_c,\tfrac{3}{2}\ln z_{0},\tfrac{1}{2}\big)
    -\mathcal{T}\big(\ln z_0,\tfrac{3}{2}\ln z_0,\tfrac{1}{2}\big)\\
  & -\mathcal{T}\big(\ln z_{0},\ln z_L,1\big)
    +\mathcal{T}\big({-L},\ln z_L,1\big)
    -\mathcal{T}\big(\ln z_c,\ln z_L,1\big)
    +\mathcal{T}\big(\ln z_0,\ln z_L,1\big)\Big]\\
  \overset{z_0<1-x<z_c}=
  &\frac{1}{2\beta_0}\big[
    \mathcal{T}\big(\ln z_c,\tfrac{3}{2}\ln z_0,\tfrac{1}{2}\big)
    -\mathcal{T}\big({-L},\tfrac{3}{2}\ln z_0,\tfrac{1}{2}\big)
    -\mathcal{T}\big(\ln z_c,\ln z_L,1\big)
    +\mathcal{T}\big({-L},\ln z_L,1\big)\Big)\Big]
\end{align*}
with $z_0\equiv\omega_0(R)/p_{T0}=(2\qhat/(p_{T0}^3R^4))^{1/3}$, $z_L\equiv\omega_L(R)/p_{T0}=2/(Lp_{T0}R^2)$ and $z_c\equiv\omega_c/p_{T0}$.

%% file: main.bbl
\providecommand{\href}[2]{#2}\begingroup\raggedright\begin{thebibliography}{100}

\bibitem{Bjorken:1968dy}
J.~Bjorken, ``{Asymptotic Sum Rules at Infinite Momentum},''
  \href{http://dx.doi.org/10.1103/PhysRev.179.1547}{{\em Phys. Rev.} {\bfseries
  179} (1969) 1547--1553}.

\bibitem{Bjorken:1969ja}
J.~Bjorken and E.~A. Paschos, ``{Inelastic Electron Proton and gamma Proton
  Scattering, and the Structure of the Nucleon},''
  \href{http://dx.doi.org/10.1103/PhysRev.185.1975}{{\em Phys. Rev.} {\bfseries
  185} (1969) 1975--1982}.

\bibitem{Feynman:1969ej}
R.~P. Feynman, ``{Very high-energy collisions of hadrons},''
  \href{http://dx.doi.org/10.1103/PhysRevLett.23.1415}{{\em Phys. Rev. Lett.}
  {\bfseries 23} (1969) 1415--1417}.

\bibitem{Yang:1954ek}
C.-N. Yang and R.~L. Mills, ``{Conservation of Isotopic Spin and Isotopic Gauge
  Invariance},'' \href{http://dx.doi.org/10.1103/PhysRev.96.191}{{\em Phys.
  Rev.} {\bfseries 96} (1954) 191--195}.

\bibitem{Gross:1973id}
D.~J. Gross and F.~Wilczek, ``{Ultraviolet Behavior of Nonabelian Gauge
  Theories},'' \href{http://dx.doi.org/10.1103/PhysRevLett.30.1343}{{\em Phys.
  Rev. Lett.} {\bfseries 30} (1973) 1343--1346}.

\bibitem{Politzer:1973fx}
H.~Politzer, ``{Reliable Perturbative Results for Strong Interactions?},''
  \href{http://dx.doi.org/10.1103/PhysRevLett.30.1346}{{\em Phys. Rev. Lett.}
  {\bfseries 30} (1973) 1346--1349}.

\bibitem{tHooft:1973mfk}
G.~'t~Hooft, ``{Dimensional regularization and the renormalization group},''
  \href{http://dx.doi.org/10.1016/0550-3213(73)90376-3}{{\em Nucl. Phys. B}
  {\bfseries 61} (1973) 455--468}.

\bibitem{tHooft:1985mkt}
G.~'t~Hooft, ``{The birth of asymptotic freedom},''
  \href{http://dx.doi.org/10.1016/0550-3213(85)90206-8}{{\em Nucl. Phys. B}
  {\bfseries 254} (1985) 11--18}.

\bibitem{Fischer:2018sdj}
C.~S. Fischer, ``{QCD at finite temperature and chemical potential from
  Dyson--Schwinger equations},''
  \href{http://dx.doi.org/10.1016/j.ppnp.2019.01.002}{{\em Prog. Part. Nucl.
  Phys.} {\bfseries 105} (2019) 1--60},
  \href{http://arxiv.org/abs/1810.12938}{{\ttfamily arXiv:1810.12938
  [hep-ph]}}.

\bibitem{Fukushima:2013rx}
K.~Fukushima and C.~Sasaki, ``{The phase diagram of nuclear and quark matter at
  high baryon density},''
  \href{http://dx.doi.org/10.1016/j.ppnp.2013.05.003}{{\em Prog. Part. Nucl.
  Phys.} {\bfseries 72} (2013) 99--154},
  \href{http://arxiv.org/abs/1301.6377}{{\ttfamily arXiv:1301.6377 [hep-ph]}}.

\bibitem{Hagedorn:1965st}
R.~Hagedorn, ``{Statistical thermodynamics of strong interactions at
  high-energies},'' {\em Nuovo Cim. Suppl.} {\bfseries 3} (1965) 147--186.

\bibitem{Cabibbo:1975ig}
N.~Cabibbo and G.~Parisi, ``{Exponential Hadronic Spectrum and Quark
  Liberation},'' \href{http://dx.doi.org/10.1016/0370-2693(75)90158-6}{{\em
  Phys. Lett. B} {\bfseries 59} (1975) 67--69}.

\bibitem{Bazavov:2014pvz}
{\bfseries HotQCD} Collaboration, A.~Bazavov {\em et al.}, ``{Equation of state
  in ( 2+1 )-flavor QCD},''
  \href{http://dx.doi.org/10.1103/PhysRevD.90.094503}{{\em Phys. Rev. D}
  {\bfseries 90} (2014) 094503},
  \href{http://arxiv.org/abs/1407.6387}{{\ttfamily arXiv:1407.6387 [hep-lat]}}.

\bibitem{Aoki:2006we}
Y.~Aoki, G.~Endrodi, Z.~Fodor, S.~Katz, and K.~Szabo, ``{The Order of the
  quantum chromodynamics transition predicted by the standard model of particle
  physics},'' \href{http://dx.doi.org/10.1038/nature05120}{{\em Nature}
  {\bfseries 443} (2006) 675--678},
  \href{http://arxiv.org/abs/hep-lat/0611014}{{\ttfamily
  arXiv:hep-lat/0611014}}.

\bibitem{deForcrand:2010ys}
P.~de~Forcrand, ``{Simulating QCD at finite density},''
  \href{http://dx.doi.org/10.22323/1.091.0010}{{\em PoS} {\bfseries LAT2009}
  (2009) 010}, \href{http://arxiv.org/abs/1005.0539}{{\ttfamily arXiv:1005.0539
  [hep-lat]}}.

\bibitem{Iancu:2012xa}
E.~Iancu, \href{http://dx.doi.org/10.5170/CERN-2014-003.197}{``{QCD in heavy
  ion collisions},''} in {\em {2011 European School of High-Energy Physics}},
  pp.~197--266.
\newblock 2014.
\newblock \href{http://arxiv.org/abs/1205.0579}{{\ttfamily arXiv:1205.0579
  [hep-ph]}}.

\bibitem{Florkowski:2010zz}
W.~Florkowski, {\em {Phenomenology of Ultra-Relativistic Heavy-Ion
  Collisions}}.
\newblock 3, 2010.

\bibitem{Back:2002wb}
B.~Back {\em et al.}, ``{The Significance of the fragmentation region in
  ultrarelativistic heavy ion collisions},''
  \href{http://dx.doi.org/10.1103/PhysRevLett.91.052303}{{\em Phys. Rev. Lett.}
  {\bfseries 91} (2003) 052303},
  \href{http://arxiv.org/abs/nucl-ex/0210015}{{\ttfamily
  arXiv:nucl-ex/0210015}}.

\bibitem{Aad:2015wga}
{\bfseries ATLAS} Collaboration, G.~Aad {\em et al.}, ``{Measurement of
  charged-particle spectra in Pb+Pb collisions at $\sqrt{{s}_\mathsf{{NN}}} =
  2.76$ TeV with the ATLAS detector at the LHC},''
  \href{http://dx.doi.org/10.1007/JHEP09(2015)050}{{\em JHEP} {\bfseries 09}
  (2015) 050}, \href{http://arxiv.org/abs/1504.04337}{{\ttfamily
  arXiv:1504.04337 [hep-ex]}}.

\bibitem{Bjorken:1982qr}
J.~Bjorken, ``{Highly Relativistic Nucleus-Nucleus Collisions: The Central
  Rapidity Region},'' \href{http://dx.doi.org/10.1103/PhysRevD.27.140}{{\em
  Phys. Rev. D} {\bfseries 27} (1983) 140--151}.

\bibitem{Gardim:2019xjs}
F.~G. Gardim, G.~Giacalone, M.~Luzum, and J.-Y. Ollitrault, ``{Revealing QCD
  thermodynamics in ultrarelativistic nuclear collisions},''
  \href{http://dx.doi.org/10.1038/s41567-020-0846-4}{{\em Nature Phys.}
  {\bfseries 16} no.~6, (2020) 615--619},
  \href{http://arxiv.org/abs/1908.09728}{{\ttfamily arXiv:1908.09728
  [nucl-th]}}.

\bibitem{Aad:2015gqa}
{\bfseries ATLAS} Collaboration, G.~Aad {\em et al.}, ``{Observation of
  Long-Range Elliptic Azimuthal Anisotropies in $\sqrt{s}=$13 and 2.76 TeV $pp$
  Collisions with the ATLAS Detector},''
  \href{http://dx.doi.org/10.1103/PhysRevLett.116.172301}{{\em Phys. Rev.
  Lett.} {\bfseries 116} no.~17, (2016) 172301},
  \href{http://arxiv.org/abs/1509.04776}{{\ttfamily arXiv:1509.04776
  [hep-ex]}}.

\bibitem{ATLAS:2012at}
{\bfseries ATLAS} Collaboration, G.~Aad {\em et al.}, ``{Measurement of the
  azimuthal anisotropy for charged particle production in $\sqrt{s_{NN}}=2.76$
  TeV lead-lead collisions with the ATLAS detector},''
  \href{http://dx.doi.org/10.1103/PhysRevC.86.014907}{{\em Phys. Rev. C}
  {\bfseries 86} (2012) 014907},
  \href{http://arxiv.org/abs/1203.3087}{{\ttfamily arXiv:1203.3087 [hep-ex]}}.

\bibitem{Luzum:2011mm}
M.~Luzum, ``{Flow fluctuations and long-range correlations: elliptic flow and
  beyond},'' \href{http://dx.doi.org/10.1088/0954-3899/38/12/124026}{{\em J.
  Phys. G} {\bfseries 38} (2011) 124026},
  \href{http://arxiv.org/abs/1107.0592}{{\ttfamily arXiv:1107.0592 [nucl-th]}}.

\bibitem{Dusling:2017aot}
K.~Dusling, M.~Mace, and R.~Venugopalan, ``{Parton model description of
  multiparticle azimuthal correlations in $pA$ collisions},''
  \href{http://dx.doi.org/10.1103/PhysRevD.97.016014}{{\em Phys. Rev. D}
  {\bfseries 97} no.~1, (2018) 016014},
  \href{http://arxiv.org/abs/1706.06260}{{\ttfamily arXiv:1706.06260
  [hep-ph]}}.

\bibitem{Borghini:2001vi}
N.~Borghini, P.~M. Dinh, and J.-Y. Ollitrault, ``{Flow analysis from
  multiparticle azimuthal correlations},''
  \href{http://dx.doi.org/10.1103/PhysRevC.64.054901}{{\em Phys. Rev. C}
  {\bfseries 64} (2001) 054901},
  \href{http://arxiv.org/abs/nucl-th/0105040}{{\ttfamily
  arXiv:nucl-th/0105040}}.

\bibitem{Ollitrault:1997vz}
J.-Y. Ollitrault, ``{Flow systematics from SIS to SPS energies},''
  \href{http://dx.doi.org/10.1016/S0375-9474(98)00413-8}{{\em Nucl. Phys. A}
  {\bfseries 638} (1998) 195--206},
  \href{http://arxiv.org/abs/nucl-ex/9802005}{{\ttfamily
  arXiv:nucl-ex/9802005}}.

\bibitem{Alver:2010dn}
B.~H. Alver, C.~Gombeaud, M.~Luzum, and J.-Y. Ollitrault, ``{Triangular flow in
  hydrodynamics and transport theory},''
  \href{http://dx.doi.org/10.1103/PhysRevC.82.034913}{{\em Phys. Rev. C}
  {\bfseries 82} (2010) 034913},
  \href{http://arxiv.org/abs/1007.5469}{{\ttfamily arXiv:1007.5469 [nucl-th]}}.

\bibitem{Gardim:2011xv}
F.~G. Gardim, F.~Grassi, M.~Luzum, and J.-Y. Ollitrault, ``{Mapping the
  hydrodynamic response to the initial geometry in heavy-ion collisions},''
  \href{http://dx.doi.org/10.1103/PhysRevC.85.024908}{{\em Phys. Rev. C}
  {\bfseries 85} (2012) 024908},
  \href{http://arxiv.org/abs/1111.6538}{{\ttfamily arXiv:1111.6538 [nucl-th]}}.

\bibitem{Romatschke:2007mq}
P.~Romatschke and U.~Romatschke, ``{Viscosity Information from Relativistic
  Nuclear Collisions: How Perfect is the Fluid Observed at RHIC?},''
  \href{http://dx.doi.org/10.1103/PhysRevLett.99.172301}{{\em Phys. Rev. Lett.}
  {\bfseries 99} (2007) 172301},
  \href{http://arxiv.org/abs/0706.1522}{{\ttfamily arXiv:0706.1522 [nucl-th]}}.

\bibitem{Heinz:2013th}
U.~Heinz and R.~Snellings, ``{Collective flow and viscosity in relativistic
  heavy-ion collisions},''
  \href{http://dx.doi.org/10.1146/annurev-nucl-102212-170540}{{\em Ann. Rev.
  Nucl. Part. Sci.} {\bfseries 63} (2013) 123--151},
  \href{http://arxiv.org/abs/1301.2826}{{\ttfamily arXiv:1301.2826 [nucl-th]}}.

\bibitem{Alam:1996fd}
J.~Alam, B.~Sinha, and S.~Raha, ``{Electromagnetic probes of quark gluon
  plasma},'' \href{http://dx.doi.org/10.1016/0370-1573(95)00084-4}{{\em Phys.
  Rept.} {\bfseries 273} (1996) 243--362}.

\bibitem{Andronic:2015wma}
A.~Andronic {\em et al.}, ``{Heavy-flavour and quarkonium production in the LHC
  era: from proton--proton to heavy-ion collisions},''
  \href{http://dx.doi.org/10.1140/epjc/s10052-015-3819-5}{{\em Eur. Phys. J. C}
  {\bfseries 76} no.~3, (2016) 107},
  \href{http://arxiv.org/abs/1506.03981}{{\ttfamily arXiv:1506.03981
  [nucl-ex]}}.

\bibitem{Matsui:1986dk}
T.~Matsui and H.~Satz, ``{$J/\psi$ Suppression by Quark-Gluon Plasma
  Formation},'' \href{http://dx.doi.org/10.1016/0370-2693(86)91404-8}{{\em
  Phys. Lett. B} {\bfseries 178} (1986) 416--422}.

\bibitem{Caucal:2020xad}
P.~Caucal, E.~Iancu, A.~Mueller, and G.~Soyez, ``{Nuclear modification factors
  for jet fragmentation},'' \href{http://arxiv.org/abs/2005.05852}{{\ttfamily
  arXiv:2005.05852 [hep-ph]}}.

\bibitem{Caucal:2020vvb}
P.~Caucal, E.~Iancu, A.~Mueller, and G.~Soyez, ``{Jet fragmentation function in
  heavy-ion collisions},'' in {\em {10th International Conference on Hard and
  Electromagnetic Probes of High-Energy Nuclear Collisions}: {Hard Probes
  2020~}}.
\newblock 9, 2020.
\newblock \href{http://arxiv.org/abs/2009.01350}{{\ttfamily arXiv:2009.01350
  [hep-ph]}}.

\bibitem{Caucal:2018dla}
P.~Caucal, E.~Iancu, A.~Mueller, and G.~Soyez, ``{Vacuum-like jet fragmentation
  in a dense QCD medium},''
  \href{http://dx.doi.org/10.1103/PhysRevLett.120.232001}{{\em Phys. Rev.
  Lett.} {\bfseries 120} (2018) 232001},
  \href{http://arxiv.org/abs/1801.09703}{{\ttfamily arXiv:1801.09703
  [hep-ph]}}.

\bibitem{Caucal:2018ofz}
P.~Caucal, E.~Iancu, A.~H. Mueller, and G.~Soyez, ``{A new pQCD based Monte
  Carlo event generator for jets in the quark-gluon plasma},'' {\em PoS}
  {\bfseries HardProbes2018} (2019) 028,
\href{http://arxiv.org/abs/1812.05393}{{\ttfamily arXiv:1812.05393 [hep-ph]}}.

\bibitem{Caucal:2019uvr}
P.~Caucal, E.~Iancu, and G.~Soyez, ``{Deciphering the $z_g$ distribution in
  ultrarelativistic heavy ion collisions},''
  \href{http://dx.doi.org/10.1007/JHEP10(2019)273}{{\em JHEP} {\bfseries 10}
  (2019) 273}, \href{http://arxiv.org/abs/1907.04866}{{\ttfamily
  arXiv:1907.04866 [hep-ph]}}.

\bibitem{Caucal:2020lro}
P.~Caucal, E.~Iancu, and G.~Soyez, ``{Nuclear effects on jet substructure
  observables at the LHC},'' in {\em {28th International Conference on
  Ultrarelativistic Nucleus-Nucleus Collisions}}.
\newblock 1, 2020.
\newblock \href{http://arxiv.org/abs/2001.09071}{{\ttfamily arXiv:2001.09071
  [hep-ph]}}.

\bibitem{Shuryak:2008eq}
E.~Shuryak, ``{Physics of Strongly coupled Quark-Gluon Plasma},''
  \href{http://dx.doi.org/10.1016/j.ppnp.2008.09.001}{{\em Prog. Part. Nucl.
  Phys.} {\bfseries 62} (2009) 48--101},
  \href{http://arxiv.org/abs/0807.3033}{{\ttfamily arXiv:0807.3033 [hep-ph]}}.

\bibitem{Blaizot:2003tw}
J.-P. Blaizot, E.~Iancu, and A.~Rebhan,
  \href{http://dx.doi.org/10.1142/9789812795533\_0002}{``{Thermodynamics of the
  high temperature quark gluon plasma},''} pp.~60--122.
\newblock 3, 2003.
\newblock \href{http://arxiv.org/abs/hep-ph/0303185}{{\ttfamily
  arXiv:hep-ph/0303185}}.

\bibitem{Bjorken:1982tu}
J.~Bjorken, ``{Energy Loss of Energetic Partons in Quark - Gluon Plasma:
  Possible Extinction of High p(t) Jets in Hadron - Hadron Collisions},''.

\bibitem{Braaten:1991we}
E.~Braaten and M.~H. Thoma, ``{Energy loss of a heavy quark in the quark -
  gluon plasma},'' \href{http://dx.doi.org/10.1103/PhysRevD.44.R2625}{{\em
  Phys. Rev. D} {\bfseries 44} no.~9, (1991) 2625}.

\bibitem{Thoma:1990fm}
M.~H. Thoma and M.~Gyulassy, ``{Quark Damping and Energy Loss in the High
  Temperature {QCD}},''
  \href{http://dx.doi.org/10.1016/S0550-3213(05)80031-8}{{\em Nucl. Phys. B}
  {\bfseries 351} (1991) 491--506}.

\bibitem{Mrowczynski:1991da}
S.~Mrowczynski, ``{Energy loss of a high-energy parton in the quark - gluon
  plasma},'' \href{http://dx.doi.org/10.1016/0370-2693(91)90188-V}{{\em Phys.
  Lett. B} {\bfseries 269} (1991) 383--388}.

\bibitem{Djordjevic:2006tw}
M.~Djordjevic, ``{Collisional energy loss in a finite size QCD matter},''
  \href{http://dx.doi.org/10.1103/PhysRevC.74.064907}{{\em Phys. Rev. C}
  {\bfseries 74} (2006) 064907},
  \href{http://arxiv.org/abs/nucl-th/0603066}{{\ttfamily
  arXiv:nucl-th/0603066}}.

\bibitem{Peshier:2006hi}
A.~Peshier, ``{The QCD collisional energy loss revised},''
  \href{http://dx.doi.org/10.1103/PhysRevLett.97.212301}{{\em Phys. Rev. Lett.}
  {\bfseries 97} (2006) 212301},
  \href{http://arxiv.org/abs/hep-ph/0605294}{{\ttfamily arXiv:hep-ph/0605294}}.

\bibitem{Peigne:2008nd}
S.~Peigne and A.~Peshier, ``{Collisional energy loss of a fast heavy quark in a
  quark-gluon plasma},''
  \href{http://dx.doi.org/10.1103/PhysRevD.77.114017}{{\em Phys. Rev. D}
  {\bfseries 77} (2008) 114017},
  \href{http://arxiv.org/abs/0802.4364}{{\ttfamily arXiv:0802.4364 [hep-ph]}}.

\bibitem{PhysRevLett.111.062301}
X.-N. Wang and Y.~Zhu, ``Medium Modification of $\ensuremath{\gamma}$ Jets in
  High-Energy Heavy-Ion Collisions,''
  \href{http://dx.doi.org/10.1103/PhysRevLett.111.062301}{{\em Phys. Rev.
  Lett.} {\bfseries 111} (Aug, 2013) 062301}.

\bibitem{Zakharov:2007pj}
B.~Zakharov, ``{Parton energy loss in an expanding quark-gluon plasma:
  Radiative versus collisional},''
  \href{http://dx.doi.org/10.1134/S0021364007190034}{{\em JETP Lett.}
  {\bfseries 86} (2007) 444--450},
  \href{http://arxiv.org/abs/0708.0816}{{\ttfamily arXiv:0708.0816 [hep-ph]}}.

\bibitem{Baier:2000mf}
R.~Baier, D.~Schiff, and B.~Zakharov, ``{Energy loss in perturbative QCD},''
  \href{http://dx.doi.org/10.1146/annurev.nucl.50.1.37}{{\em Ann. Rev. Nucl.
  Part. Sci.} {\bfseries 50} (2000) 37--69},
  \href{http://arxiv.org/abs/hep-ph/0002198}{{\ttfamily arXiv:hep-ph/0002198}}.

\bibitem{PhysRevC.86.064904}
N.~Armesto, B.~Cole, C.~Gale, W.~A. Horowitz, P.~Jacobs, S.~Jeon, M.~van
  Leeuwen, A.~Majumder, B.~M\"uller, G.-Y. Qin, C.~A. Salgado, B.~Schenke,
  M.~Verweij, X.-N. Wang, and U.~A. Wiedemann, ``Comparison of jet quenching
  formalisms for a quark-gluon plasma ``brick'',''
  \href{http://dx.doi.org/10.1103/PhysRevC.86.064904}{{\em Phys. Rev. C}
  {\bfseries 86} (Dec, 2012) 064904}.

\bibitem{Baier:1994bd}
R.~Baier, Y.~L. Dokshitzer, S.~Peigne, and D.~Schiff, ``{Induced gluon
  radiation in a QCD medium},''
  \href{http://dx.doi.org/10.1016/0370-2693(94)01617-L}{{\em Phys. Lett. B}
  {\bfseries 345} (1995) 277--286},
  \href{http://arxiv.org/abs/hep-ph/9411409}{{\ttfamily arXiv:hep-ph/9411409}}.

\bibitem{Baier:1996vi}
R.~Baier, Y.~L. Dokshitzer, A.~H. Mueller, S.~Peigne, and D.~Schiff, ``{The
  Landau-Pomeranchuk-Migdal effect in QED},''
  \href{http://dx.doi.org/10.1016/0550-3213(96)00426-9}{{\em Nucl. Phys. B}
  {\bfseries 478} (1996) 577--597},
  \href{http://arxiv.org/abs/hep-ph/9604327}{{\ttfamily arXiv:hep-ph/9604327}}.

\bibitem{Baier:1996kr}
R.~Baier, Y.~L. Dokshitzer, A.~H. Mueller, S.~Peigne, and D.~Schiff,
  ``{Radiative energy loss of high-energy quarks and gluons in a finite volume
  quark - gluon plasma},''
  \href{http://dx.doi.org/10.1016/S0550-3213(96)00553-6}{{\em Nucl. Phys. B}
  {\bfseries 483} (1997) 291--320},
  \href{http://arxiv.org/abs/hep-ph/9607355}{{\ttfamily arXiv:hep-ph/9607355}}.

\bibitem{Baier:1996sk}
R.~Baier, Y.~L. Dokshitzer, A.~H. Mueller, S.~Peigne, and D.~Schiff,
  ``{Radiative energy loss and p(T) broadening of high-energy partons in
  nuclei},'' \href{http://dx.doi.org/10.1016/S0550-3213(96)00581-0}{{\em Nucl.
  Phys. B} {\bfseries 484} (1997) 265--282},
  \href{http://arxiv.org/abs/hep-ph/9608322}{{\ttfamily arXiv:hep-ph/9608322}}.

\bibitem{Zakharov:1997uu}
B.~Zakharov, ``{Radiative energy loss of high-energy quarks in finite size
  nuclear matter and quark - gluon plasma},''
  \href{http://dx.doi.org/10.1134/1.567389}{{\em JETP Lett.} {\bfseries 65}
  (1997) 615--620}, \href{http://arxiv.org/abs/hep-ph/9704255}{{\ttfamily
  arXiv:hep-ph/9704255}}.

\bibitem{Zakharov:1998sv}
B.~Zakharov, ``{Light cone path integral approach to the
  Landau-Pomeranchuk-Migdal effect},'' {\em Phys. Atom. Nucl.} {\bfseries 61}
  (1998) 838--854, \href{http://arxiv.org/abs/hep-ph/9807540}{{\ttfamily
  arXiv:hep-ph/9807540}}.

\bibitem{Zakharov:1996fv}
B.~Zakharov, ``{Fully quantum treatment of the Landau-Pomeranchuk-Migdal effect
  in QED and QCD},'' \href{http://dx.doi.org/10.1134/1.567126}{{\em JETP Lett.}
  {\bfseries 63} (1996) 952--957},
  \href{http://arxiv.org/abs/hep-ph/9607440}{{\ttfamily arXiv:hep-ph/9607440}}.

\bibitem{Baier:1998kq}
R.~Baier, Y.~L. Dokshitzer, A.~H. Mueller, and D.~Schiff, ``{Medium induced
  radiative energy loss: Equivalence between the BDMPS and Zakharov
  formalisms},'' \href{http://dx.doi.org/10.1016/S0550-3213(98)00546-X}{{\em
  Nucl. Phys. B} {\bfseries 531} (1998) 403--425},
  \href{http://arxiv.org/abs/hep-ph/9804212}{{\ttfamily arXiv:hep-ph/9804212}}.

\bibitem{Salgado:2003gb}
C.~A. Salgado and U.~A. Wiedemann, ``{Calculating quenching weights},''
  \href{http://dx.doi.org/10.1103/PhysRevD.68.014008}{{\em Phys. Rev. D}
  {\bfseries 68} (2003) 014008},
  \href{http://arxiv.org/abs/hep-ph/0302184}{{\ttfamily arXiv:hep-ph/0302184}}.

\bibitem{Wiedemann:1999fq}
U.~A. Wiedemann and M.~Gyulassy, ``{Transverse momentum dependence of the
  Landau-Pomeranchuk-Migdal effect},''
  \href{http://dx.doi.org/10.1016/S0550-3213(99)00458-7}{{\em Nucl. Phys. B}
  {\bfseries 560} (1999) 345--382},
  \href{http://arxiv.org/abs/hep-ph/9906257}{{\ttfamily arXiv:hep-ph/9906257}}.

\bibitem{Armesto:2003jh}
N.~Armesto, C.~A. Salgado, and U.~A. Wiedemann, ``{Medium induced gluon
  radiation off massive quarks fills the dead cone},''
  \href{http://dx.doi.org/10.1103/PhysRevD.69.114003}{{\em Phys. Rev. D}
  {\bfseries 69} (2004) 114003},
  \href{http://arxiv.org/abs/hep-ph/0312106}{{\ttfamily arXiv:hep-ph/0312106}}.

\bibitem{Baier:1998yf}
R.~Baier, Y.~L. Dokshitzer, A.~H. Mueller, and D.~Schiff, ``{Radiative energy
  loss of high-energy partons traversing an expanding QCD plasma},''
  \href{http://dx.doi.org/10.1103/PhysRevC.58.1706}{{\em Phys. Rev. C}
  {\bfseries 58} (1998) 1706--1713},
  \href{http://arxiv.org/abs/hep-ph/9803473}{{\ttfamily arXiv:hep-ph/9803473}}.

\bibitem{Salgado:2002cd}
C.~A. Salgado and U.~A. Wiedemann, ``{A Dynamical scaling law for jet
  tomography},'' \href{http://dx.doi.org/10.1103/PhysRevLett.89.092303}{{\em
  Phys. Rev. Lett.} {\bfseries 89} (2002) 092303},
  \href{http://arxiv.org/abs/hep-ph/0204221}{{\ttfamily arXiv:hep-ph/0204221}}.

\bibitem{Gyulassy:2000er}
M.~Gyulassy, P.~Levai, and I.~Vitev, ``{Reaction operator approach to
  nonAbelian energy loss},''
  \href{http://dx.doi.org/10.1016/S0550-3213(00)00652-0}{{\em Nucl. Phys. B}
  {\bfseries 594} (2001) 371--419},
  \href{http://arxiv.org/abs/nucl-th/0006010}{{\ttfamily
  arXiv:nucl-th/0006010}}.

\bibitem{Gyulassy:2001nm}
M.~Gyulassy, P.~Levai, and I.~Vitev, ``{Jet tomography of Au+Au reactions
  including multigluon fluctuations},''
  \href{http://dx.doi.org/10.1016/S0370-2693(02)01990-1}{{\em Phys. Lett. B}
  {\bfseries 538} (2002) 282--288},
  \href{http://arxiv.org/abs/nucl-th/0112071}{{\ttfamily
  arXiv:nucl-th/0112071}}.

\bibitem{Wiedemann:2000za}
U.~A. Wiedemann, ``{Gluon radiation off hard quarks in a nuclear environment:
  Opacity expansion},''
  \href{http://dx.doi.org/10.1016/S0550-3213(00)00457-0}{{\em Nucl. Phys. B}
  {\bfseries 588} (2000) 303--344},
  \href{http://arxiv.org/abs/hep-ph/0005129}{{\ttfamily arXiv:hep-ph/0005129}}.

\bibitem{Arnold:2001ms}
P.~B. Arnold, G.~D. Moore, and L.~G. Yaffe, ``{Photon emission from quark gluon
  plasma: Complete leading order results},''
  \href{http://dx.doi.org/10.1088/1126-6708/2001/12/009}{{\em JHEP} {\bfseries
  12} (2001) 009}, \href{http://arxiv.org/abs/hep-ph/0111107}{{\ttfamily
  arXiv:hep-ph/0111107}}.

\bibitem{Arnold:2002ja}
P.~B. Arnold, G.~D. Moore, and L.~G. Yaffe, ``{Photon and gluon emission in
  relativistic plasmas},''
  \href{http://dx.doi.org/10.1088/1126-6708/2002/06/030}{{\em JHEP} {\bfseries
  06} (2002) 030}, \href{http://arxiv.org/abs/hep-ph/0204343}{{\ttfamily
  arXiv:hep-ph/0204343}}.

\bibitem{Mehtar-Tani:2019tvy}
Y.~Mehtar-Tani, ``{Gluon bremsstrahlung in finite media beyond multiple soft
  scattering approximation},''
  \href{http://dx.doi.org/10.1007/JHEP07(2019)057}{{\em JHEP} {\bfseries 07}
  (2019) 057}, \href{http://arxiv.org/abs/1903.00506}{{\ttfamily
  arXiv:1903.00506 [hep-ph]}}.

\bibitem{Mehtar-Tani:2019ygg}
Y.~Mehtar-Tani and K.~Tywoniuk, ``{Improved opacity expansion for
  medium-induced parton splitting},''
  \href{http://arxiv.org/abs/1910.02032}{{\ttfamily arXiv:1910.02032
  [hep-ph]}}.

\bibitem{Gyulassy:1993hr}
M.~Gyulassy and X.-n. Wang, ``{Multiple collisions and induced gluon
  Bremsstrahlung in QCD},''
  \href{http://dx.doi.org/10.1016/0550-3213(94)90079-5}{{\em Nucl. Phys. B}
  {\bfseries 420} (1994) 583--614},
  \href{http://arxiv.org/abs/nucl-th/9306003}{{\ttfamily
  arXiv:nucl-th/9306003}}.

\bibitem{MehtarTani:2006xq}
Y.~Mehtar-Tani, ``{Relating the description of gluon production in pA
  collisions and parton energy loss in AA collisions},''
  \href{http://dx.doi.org/10.1103/PhysRevC.75.034908}{{\em Phys. Rev. C}
  {\bfseries 75} (2007) 034908},
  \href{http://arxiv.org/abs/hep-ph/0606236}{{\ttfamily arXiv:hep-ph/0606236}}.

\bibitem{Barata:2020sav}
J.~Barata and Y.~Mehtar-Tani, ``{Improved opacity expansion at NNLO for medium
  induced gluon radiation},'' \href{http://arxiv.org/abs/2004.02323}{{\ttfamily
  arXiv:2004.02323 [hep-ph]}}.

\bibitem{Barata:2020rdn}
J.~Barata, Y.~Mehtar-Tani, A.~Soto-Ontoso, and K.~Tywoniuk, ``{Revisiting
  transverse momentum broadening in dense QCD media},''
  \href{http://arxiv.org/abs/2009.13667}{{\ttfamily arXiv:2009.13667
  [hep-ph]}}.

\bibitem{Andres:2020vxs}
C.~Andres, L.~Apolinário, and F.~Dominguez, ``{Medium-induced gluon radiation
  with full resummation of multiple scatterings for realistic parton-medium
  interactions},'' \href{http://arxiv.org/abs/2002.01517}{{\ttfamily
  arXiv:2002.01517 [hep-ph]}}.

\bibitem{Arnold:2015qya}
P.~Arnold and S.~Iqbal, ``{The LPM effect in sequential bremsstrahlung},''
  \href{http://dx.doi.org/10.1007/JHEP09(2016)072}{{\em JHEP} {\bfseries 04}
  (2015) 070}, \href{http://arxiv.org/abs/1501.04964}{{\ttfamily
  arXiv:1501.04964 [hep-ph]}}. [Erratum: JHEP 09, 072 (2016)].

\bibitem{Arnold:2016kek}
P.~Arnold, H.-C. Chang, and S.~Iqbal, ``{The LPM effect in sequential
  bremsstrahlung 2: factorization},''
  \href{http://dx.doi.org/10.1007/JHEP09(2016)078}{{\em JHEP} {\bfseries 09}
  (2016) 078}, \href{http://arxiv.org/abs/1605.07624}{{\ttfamily
  arXiv:1605.07624 [hep-ph]}}.

\bibitem{Arnold:2016mth}
P.~Arnold, H.-C. Chang, and S.~Iqbal, ``{The LPM effect in sequential
  bremsstrahlung: dimensional regularization},''
  \href{http://dx.doi.org/10.1007/JHEP10(2016)100}{{\em JHEP} {\bfseries 10}
  (2016) 100}, \href{http://arxiv.org/abs/1606.08853}{{\ttfamily
  arXiv:1606.08853 [hep-ph]}}.

\bibitem{Arnold:2016jnq}
P.~Arnold, H.-C. Chang, and S.~Iqbal, ``{The LPM effect in sequential
  bremsstrahlung: 4-gluon vertices},''
  \href{http://dx.doi.org/10.1007/JHEP10(2016)124}{{\em JHEP} {\bfseries 10}
  (2016) 124}, \href{http://arxiv.org/abs/1608.05718}{{\ttfamily
  arXiv:1608.05718 [hep-ph]}}.

\bibitem{Baier:2000sb}
R.~Baier, A.~H. Mueller, D.~Schiff, and D.~Son, ``{'Bottom up' thermalization
  in heavy ion collisions},''
  \href{http://dx.doi.org/10.1016/S0370-2693(01)00191-5}{{\em Phys. Lett. B}
  {\bfseries 502} (2001) 51--58},
  \href{http://arxiv.org/abs/hep-ph/0009237}{{\ttfamily arXiv:hep-ph/0009237}}.

\bibitem{Baier:2001yt}
R.~Baier, Y.~L. Dokshitzer, A.~H. Mueller, and D.~Schiff, ``{Quenching of
  hadron spectra in media},''
  \href{http://dx.doi.org/10.1088/1126-6708/2001/09/033}{{\em JHEP} {\bfseries
  09} (2001) 033}, \href{http://arxiv.org/abs/hep-ph/0106347}{{\ttfamily
  arXiv:hep-ph/0106347}}.

\bibitem{Wang:2001ifa}
X.-N. Wang and X.-f. Guo, ``{Multiple parton scattering in nuclei: Parton
  energy loss},'' \href{http://dx.doi.org/10.1016/S0375-9474(01)01130-7}{{\em
  Nucl. Phys. A} {\bfseries 696} (2001) 788--832},
  \href{http://arxiv.org/abs/hep-ph/0102230}{{\ttfamily arXiv:hep-ph/0102230}}.

\bibitem{Majumder:2009zu}
A.~Majumder, ``{The In-medium scale evolution in jet modification},''
  \href{http://arxiv.org/abs/0901.4516}{{\ttfamily arXiv:0901.4516 [nucl-th]}}.

\bibitem{Armesto:2009fj}
N.~Armesto, L.~Cunqueiro, and C.~A. Salgado, ``{Q-PYTHIA: A Medium-modified
  implementation of final state radiation},''
  \href{http://dx.doi.org/10.1140/epjc/s10052-009-1133-9}{{\em Eur. Phys. J. C}
  {\bfseries 63} (2009) 679--690},
  \href{http://arxiv.org/abs/0907.1014}{{\ttfamily arXiv:0907.1014 [hep-ph]}}.

\bibitem{Zapp:2013vla}
K.~C. Zapp, ``{JEWEL 2.0.0: directions for use},''
  \href{http://dx.doi.org/10.1140/epjc/s10052-014-2762-1}{{\em Eur. Phys. J. C}
  {\bfseries 74} no.~2, (2014) 2762},
  \href{http://arxiv.org/abs/1311.0048}{{\ttfamily arXiv:1311.0048 [hep-ph]}}.

\bibitem{MehtarTani:2010ma}
Y.~Mehtar-Tani, C.~A. Salgado, and K.~Tywoniuk, ``{Anti-angular ordering of
  gluon radiation in QCD media},''
  \href{http://dx.doi.org/10.1103/PhysRevLett.106.122002}{{\em Phys. Rev.
  Lett.} {\bfseries 106} (2011) 122002},
  \href{http://arxiv.org/abs/1009.2965}{{\ttfamily arXiv:1009.2965 [hep-ph]}}.

\bibitem{MehtarTani:2011tz}
Y.~Mehtar-Tani, C.~Salgado, and K.~Tywoniuk, ``{Jets in QCD Media: From Color
  Coherence to Decoherence},''
  \href{http://dx.doi.org/10.1016/j.physletb.2011.12.042}{{\em Phys. Lett. B}
  {\bfseries 707} (2012) 156--159},
  \href{http://arxiv.org/abs/1102.4317}{{\ttfamily arXiv:1102.4317 [hep-ph]}}.

\bibitem{CasalderreySolana:2011rz}
J.~Casalderrey-Solana and E.~Iancu, ``{Interference effects in medium-induced
  gluon radiation},'' \href{http://dx.doi.org/10.1007/JHEP08(2011)015}{{\em
  JHEP} {\bfseries 08} (2011) 015},
  \href{http://arxiv.org/abs/1105.1760}{{\ttfamily arXiv:1105.1760 [hep-ph]}}.

\bibitem{CasalderreySolana:2012ef}
J.~Casalderrey-Solana, Y.~Mehtar-Tani, C.~A. Salgado, and K.~Tywoniuk, ``{New
  picture of jet quenching dictated by color coherence},''
  \href{http://dx.doi.org/10.1016/j.physletb.2013.07.046}{{\em Phys. Lett. B}
  {\bfseries 725} (2013) 357--360},
  \href{http://arxiv.org/abs/1210.7765}{{\ttfamily arXiv:1210.7765 [hep-ph]}}.

\bibitem{Mehtar-Tani:2014yea}
Y.~Mehtar-Tani and K.~Tywoniuk, ``{Jet (de)coherence in Pb--Pb collisions at
  the LHC},'' \href{http://dx.doi.org/10.1016/j.physletb.2015.03.041}{{\em
  Phys. Lett. B} {\bfseries 744} (2015) 284--287},
  \href{http://arxiv.org/abs/1401.8293}{{\ttfamily arXiv:1401.8293 [hep-ph]}}.

\bibitem{Mehtar-Tani:2017web}
Y.~Mehtar-Tani and K.~Tywoniuk, ``{Sudakov suppression of jets in QCD media},''
  \href{http://dx.doi.org/10.1103/PhysRevD.98.051501}{{\em Phys. Rev. D}
  {\bfseries 98} no.~5, (2018) 051501},
  \href{http://arxiv.org/abs/1707.07361}{{\ttfamily arXiv:1707.07361
  [hep-ph]}}.

\bibitem{Leonidov:2010he}
A.~Leonidov and V.~Nechitailo, ``{Decoherence and energy loss in QCD cascades
  in nuclear collisions},''
  \href{http://dx.doi.org/10.1140/epjc/s10052-011-1537-1}{{\em Eur. Phys. J. C}
  {\bfseries 71} (2011) 1537}, \href{http://arxiv.org/abs/1006.0366}{{\ttfamily
  arXiv:1006.0366 [nucl-th]}}.

\bibitem{Gelis:2010nm}
F.~Gelis, E.~Iancu, J.~Jalilian-Marian, and R.~Venugopalan, ``{The Color Glass
  Condensate},''
  \href{http://dx.doi.org/10.1146/annurev.nucl.010909.083629}{{\em Ann. Rev.
  Nucl. Part. Sci.} {\bfseries 60} (2010) 463--489},
  \href{http://arxiv.org/abs/1002.0333}{{\ttfamily arXiv:1002.0333 [hep-ph]}}.

\bibitem{Catani:2011st}
S.~Catani, D.~de~Florian, and G.~Rodrigo, ``{Space-like (versus time-like)
  collinear limits in QCD: Is factorization violated?},''
  \href{http://dx.doi.org/10.1007/JHEP07(2012)026}{{\em JHEP} {\bfseries 07}
  (2012) 026}, \href{http://arxiv.org/abs/1112.4405}{{\ttfamily arXiv:1112.4405
  [hep-ph]}}.

\bibitem{Bern:1995ix}
Z.~Bern and G.~Chalmers, ``{Factorization in one loop gauge theory},''
  \href{http://dx.doi.org/10.1016/0550-3213(95)00226-I}{{\em Nucl. Phys. B}
  {\bfseries 447} (1995) 465--518},
  \href{http://arxiv.org/abs/hep-ph/9503236}{{\ttfamily arXiv:hep-ph/9503236}}.

\bibitem{Kosower:1999xi}
D.~A. Kosower, ``{All order collinear behavior in gauge theories},''
  \href{http://dx.doi.org/10.1016/S0550-3213(99)00251-5}{{\em Nucl. Phys. B}
  {\bfseries 552} (1999) 319--336},
  \href{http://arxiv.org/abs/hep-ph/9901201}{{\ttfamily arXiv:hep-ph/9901201}}.

\bibitem{Catani:1999ss}
S.~Catani and M.~Grazzini, ``{Infrared factorization of tree level QCD
  amplitudes at the next-to-next-to-leading order and beyond},''
  \href{http://dx.doi.org/10.1016/S0550-3213(99)00778-6}{{\em Nucl. Phys. B}
  {\bfseries 570} (2000) 287--325},
  \href{http://arxiv.org/abs/hep-ph/9908523}{{\ttfamily arXiv:hep-ph/9908523}}.

\bibitem{Ellis:1991qj}
R.~Ellis, W.~Stirling, and B.~Webber, {\em {QCD and collider physics}}, vol.~8.
\newblock Cambridge University Press, 2, 2011.

\bibitem{Dokshitzer:1991wu}
Y.~L. Dokshitzer, V.~A. Khoze, A.~H. Mueller, and S.~Troian, {\em {Basics of
  perturbative QCD}}.
\newblock 1991.

\bibitem{Iancu:2000hn}
E.~Iancu, A.~Leonidov, and L.~D. McLerran, ``{Nonlinear gluon evolution in the
  color glass condensate. 1.},''
  \href{http://dx.doi.org/10.1016/S0375-9474(01)00642-X}{{\em Nucl. Phys. A}
  {\bfseries 692} (2001) 583--645},
  \href{http://arxiv.org/abs/hep-ph/0011241}{{\ttfamily arXiv:hep-ph/0011241}}.

\bibitem{Blaizot:2004wu}
J.~P. Blaizot, F.~Gelis, and R.~Venugopalan, ``{High-energy pA collisions in
  the color glass condensate approach. 1. Gluon production and the Cronin
  effect},'' \href{http://dx.doi.org/10.1016/j.nuclphysa.2004.07.005}{{\em
  Nucl. Phys. A} {\bfseries 743} (2004) 13--56},
  \href{http://arxiv.org/abs/hep-ph/0402256}{{\ttfamily arXiv:hep-ph/0402256}}.

\bibitem{Gelis:2005pt}
F.~Gelis and Y.~Mehtar-Tani, ``{Gluon propagation inside a high-energy
  nucleus},'' \href{http://dx.doi.org/10.1103/PhysRevD.73.034019}{{\em Phys.
  Rev. D} {\bfseries 73} (2006) 034019},
  \href{http://arxiv.org/abs/hep-ph/0512079}{{\ttfamily arXiv:hep-ph/0512079}}.

\bibitem{McLerran:1993ni}
L.~D. McLerran and R.~Venugopalan, ``{Computing quark and gluon distribution
  functions for very large nuclei},''
  \href{http://dx.doi.org/10.1103/PhysRevD.49.2233}{{\em Phys. Rev. D}
  {\bfseries 49} (1994) 2233--2241},
  \href{http://arxiv.org/abs/hep-ph/9309289}{{\ttfamily arXiv:hep-ph/9309289}}.

\bibitem{McLerran:1993ka}
L.~D. McLerran and R.~Venugopalan, ``{Gluon distribution functions for very
  large nuclei at small transverse momentum},''
  \href{http://dx.doi.org/10.1103/PhysRevD.49.3352}{{\em Phys. Rev. D}
  {\bfseries 49} (1994) 3352--3355},
  \href{http://arxiv.org/abs/hep-ph/9311205}{{\ttfamily arXiv:hep-ph/9311205}}.

\bibitem{McLerran:1994vd}
L.~D. McLerran and R.~Venugopalan, ``{Green's functions in the color field of a
  large nucleus},'' \href{http://dx.doi.org/10.1103/PhysRevD.50.2225}{{\em
  Phys. Rev. D} {\bfseries 50} (1994) 2225--2233},
  \href{http://arxiv.org/abs/hep-ph/9402335}{{\ttfamily arXiv:hep-ph/9402335}}.

\bibitem{Aurenche:2002pd}
P.~Aurenche, F.~Gelis, and H.~Zaraket, ``{A Simple sum rule for the thermal
  gluon spectral function and applications},''
  \href{http://dx.doi.org/10.1088/1126-6708/2002/05/043}{{\em JHEP} {\bfseries
  05} (2002) 043}, \href{http://arxiv.org/abs/hep-ph/0204146}{{\ttfamily
  arXiv:hep-ph/0204146}}.

\bibitem{Wang:1991xy}
X.-N. Wang and M.~Gyulassy, ``{Gluon shadowing and jet quenching in A + A
  collisions at s**(1/2) = 200-GeV},''
  \href{http://dx.doi.org/10.1103/PhysRevLett.68.1480}{{\em Phys. Rev. Lett.}
  {\bfseries 68} (1992) 1480--1483}.

\bibitem{Iancu:2018trm}
E.~Iancu, P.~Taels, and B.~Wu, ``{Jet quenching parameter in an expanding QCD
  plasma},'' \href{http://dx.doi.org/10.1016/j.physletb.2018.10.007}{{\em Phys.
  Lett. B} {\bfseries 786} (2018) 288--295},
  \href{http://arxiv.org/abs/1806.07177}{{\ttfamily arXiv:1806.07177
  [hep-ph]}}.

\bibitem{Bjorken:1970ah}
J.~Bjorken, J.~B. Kogut, and D.~E. Soper, ``{Quantum Electrodynamics at
  Infinite Momentum: Scattering from an External Field},''
  \href{http://dx.doi.org/10.1103/PhysRevD.3.1382}{{\em Phys. Rev. D}
  {\bfseries 3} (1971) 1382}.

\bibitem{Feldman:1973ps}
G.~Feldman and P.~Matthews, ``{Light-cone variables and the high energy limit
  of elastic scattering},''
  \href{http://dx.doi.org/10.1088/0305-4470/6/2/013}{{\em J. Phys. A}
  {\bfseries 6} (1973) 236--246}.

\bibitem{CasalderreySolana:2007pr}
J.~Casalderrey-Solana and C.~A. Salgado, ``{Introductory lectures on jet
  quenching in heavy ion collisions},'' {\em Acta Phys. Polon.} {\bfseries B38}
  (2007) 3731--3794,
\href{http://arxiv.org/abs/0712.3443}{{\ttfamily arXiv:0712.3443 [hep-ph]}}.

\bibitem{Hebecker:1999ej}
A.~Hebecker, ``{Diffraction in deep inelastic scattering},''
  \href{http://dx.doi.org/10.1016/S0370-1573(00)00005-3}{{\em Phys. Rept.}
  {\bfseries 331} (2000) 1--115},
  \href{http://arxiv.org/abs/hep-ph/9905226}{{\ttfamily arXiv:hep-ph/9905226}}.

\bibitem{Meggiolaro:2000yf}
E.~Meggiolaro, ``{Eikonal propagators and high-energy parton parton scattering
  in gauge theories},''
  \href{http://dx.doi.org/10.1016/S0550-3213(01)00115-8}{{\em Nucl. Phys.}
  {\bfseries B602} (2001) 261--288},
\href{http://arxiv.org/abs/hep-ph/0009261}{{\ttfamily arXiv:hep-ph/0009261
  [hep-ph]}}.

\bibitem{Iancu:2003xm}
E.~Iancu and R.~Venugopalan, {\em {The Color glass condensate and high-energy
  scattering in QCD}},
  \href{http://dx.doi.org/10.1142/9789812795533\_0005}{pp.~249--3363}.
\newblock 3, 2003.
\newblock \href{http://arxiv.org/abs/hep-ph/0303204}{{\ttfamily
  arXiv:hep-ph/0303204}}.

\bibitem{Iancu:2014kga}
E.~Iancu, ``{The non-linear evolution of jet quenching},''
  \href{http://dx.doi.org/10.1007/JHEP10(2014)095}{{\em JHEP} {\bfseries 10}
  (2014) 095}, \href{http://arxiv.org/abs/1403.1996}{{\ttfamily arXiv:1403.1996
  [hep-ph]}}.

\bibitem{Arnold:2008iy}
P.~B. Arnold, ``{Simple Formula for High-Energy Gluon Bremsstrahlung in a
  Finite, Expanding Medium},''
  \href{http://dx.doi.org/10.1103/PhysRevD.79.065025}{{\em Phys. Rev. D}
  {\bfseries 79} (2009) 065025},
  \href{http://arxiv.org/abs/0808.2767}{{\ttfamily arXiv:0808.2767 [hep-ph]}}.

\bibitem{Landau:1953gr}
L.~Landau and I.~Pomeranchuk, ``{Electron cascade process at very
  high-energies},'' {\em Dokl. Akad. Nauk Ser. Fiz.} {\bfseries 92} (1953)
  735--738.

\bibitem{Migdal:1956tc}
A.~Migdal, ``{Bremsstrahlung and pair production in condensed media at
  high-energies},'' \href{http://dx.doi.org/10.1103/PhysRev.103.1811}{{\em
  Phys. Rev.} {\bfseries 103} (1956) 1811--1820}.

\bibitem{Blaizot:2012fh}
J.-P. Blaizot, F.~Dominguez, E.~Iancu, and Y.~Mehtar-Tani, ``{Medium-induced
  gluon branching},'' \href{http://dx.doi.org/10.1007/JHEP01(2013)143}{{\em
  JHEP} {\bfseries 01} (2013) 143},
\href{http://arxiv.org/abs/1209.4585}{{\ttfamily arXiv:1209.4585 [hep-ph]}}.

\bibitem{Apolinario:2014csa}
L.~Apolinário, N.~Armesto, J.~G. Milhano, and C.~A. Salgado, ``{Medium-induced
  gluon radiation and colour decoherence beyond the soft approximation},''
  \href{http://dx.doi.org/10.1007/JHEP02(2015)119}{{\em JHEP} {\bfseries 02}
  (2015) 119}, \href{http://arxiv.org/abs/1407.0599}{{\ttfamily arXiv:1407.0599
  [hep-ph]}}.

\bibitem{MehtarTani:2012cy}
Y.~Mehtar-Tani, C.~A. Salgado, and K.~Tywoniuk, ``{The Radiation pattern of a
  QCD antenna in a dense medium},''
  \href{http://dx.doi.org/10.1007/JHEP10(2012)197}{{\em JHEP} {\bfseries 10}
  (2012) 197}, \href{http://arxiv.org/abs/1205.5739}{{\ttfamily arXiv:1205.5739
  [hep-ph]}}.

\bibitem{Mehtar-Tani:2017ypq}
Y.~Mehtar-Tani and K.~Tywoniuk, ``{Radiative energy loss of neighboring
  subjets},'' \href{http://dx.doi.org/10.1016/j.nuclphysa.2018.09.041}{{\em
  Nucl. Phys. A} {\bfseries 979} (2018) 165--203},
  \href{http://arxiv.org/abs/1706.06047}{{\ttfamily arXiv:1706.06047
  [hep-ph]}}.

\bibitem{Kinoshita:1962ur}
T.~Kinoshita, ``{Mass singularities of Feynman amplitudes},''
  \href{http://dx.doi.org/10.1063/1.1724268}{{\em J. Math. Phys.} {\bfseries 3}
  (1962) 650--677}.

\bibitem{Lee:1964is}
T.~Lee and M.~Nauenberg, ``{Degenerate Systems and Mass Singularities},''
  \href{http://dx.doi.org/10.1103/PhysRev.133.B1549}{{\em Phys. Rev.}
  {\bfseries 133} (1964) B1549--B1562}.

\bibitem{Sterman:1977wj}
G.~F. Sterman and S.~Weinberg, ``{Jets from Quantum Chromodynamics},''
  \href{http://dx.doi.org/10.1103/PhysRevLett.39.1436}{{\em Phys. Rev. Lett.}
  {\bfseries 39} (1977) 1436}.

\bibitem{Kramer:1986sg}
G.~Kramer and B.~Lampe, ``{Two Jet Cross-Section in e+ e- Annihilation},''
  \href{http://dx.doi.org/10.1007/BF01679868}{{\em Z. Phys. C} {\bfseries 34}
  (1987) 497}. [Erratum: Z.Phys.C 42, 504 (1989)].

\bibitem{Salam:2009jx}
G.~P. Salam, ``{Towards Jetography},''
  \href{http://dx.doi.org/10.1140/epjc/s10052-010-1314-6}{{\em Eur. Phys. J. C}
  {\bfseries 67} (2010) 637--686},
  \href{http://arxiv.org/abs/0906.1833}{{\ttfamily arXiv:0906.1833 [hep-ph]}}.

\bibitem{Dokshitzer:1997in}
Y.~L. Dokshitzer, G.~D. Leder, S.~Moretti, and B.~R. Webber, ``{Better jet
  clustering algorithms},''
  \href{http://dx.doi.org/10.1088/1126-6708/1997/08/001}{{\em JHEP} {\bfseries
  08} (1997) 001},
\href{http://arxiv.org/abs/hep-ph/9707323}{{\ttfamily arXiv:hep-ph/9707323
  [hep-ph]}}.

\bibitem{Wobisch:1998wt}
M.~Wobisch and T.~Wengler, ``{Hadronization corrections to jet cross-sections
  in deep inelastic scattering},'' in {\em {Workshop on Monte Carlo Generators
  for HERA Physics (Plenary Starting Meeting)}}, pp.~270--279.
\newblock 4, 1998.
\newblock \href{http://arxiv.org/abs/hep-ph/9907280}{{\ttfamily
  arXiv:hep-ph/9907280}}.

\bibitem{Catani:1993hr}
S.~Catani, Y.~L. Dokshitzer, M.~Seymour, and B.~Webber, ``{Longitudinally
  invariant $K_t$ clustering algorithms for hadron hadron collisions},''
  \href{http://dx.doi.org/10.1016/0550-3213(93)90166-M}{{\em Nucl. Phys. B}
  {\bfseries 406} (1993) 187--224}.

\bibitem{Cacciari:2008gp}
M.~Cacciari, G.~P. Salam, and G.~Soyez, ``{The anti-$k_t$ jet clustering
  algorithm},'' \href{http://dx.doi.org/10.1088/1126-6708/2008/04/063}{{\em
  JHEP} {\bfseries 04} (2008) 063},
\href{http://arxiv.org/abs/0802.1189}{{\ttfamily arXiv:0802.1189 [hep-ph]}}.

\bibitem{Dasgupta:2014yra}
M.~Dasgupta, F.~Dreyer, G.~P. Salam, and G.~Soyez, ``{Small-radius jets to all
  orders in QCD},'' \href{http://dx.doi.org/10.1007/JHEP04(2015)039}{{\em JHEP}
  {\bfseries 04} (2015) 039}, \href{http://arxiv.org/abs/1411.5182}{{\ttfamily
  arXiv:1411.5182 [hep-ph]}}.

\bibitem{Banfi:2003je}
A.~Banfi, G.~P. Salam, and G.~Zanderighi, ``{Generalized resummation of QCD
  final state observables},''
  \href{http://dx.doi.org/10.1016/j.physletb.2004.01.048}{{\em Phys. Lett. B}
  {\bfseries 584} (2004) 298--305},
  \href{http://arxiv.org/abs/hep-ph/0304148}{{\ttfamily arXiv:hep-ph/0304148}}.

\bibitem{Banfi:2004yd}
A.~Banfi, G.~P. Salam, and G.~Zanderighi, ``{Principles of general final-state
  resummation and automated implementation},''
  \href{http://dx.doi.org/10.1088/1126-6708/2005/03/073}{{\em JHEP} {\bfseries
  03} (2005) 073}, \href{http://arxiv.org/abs/hep-ph/0407286}{{\ttfamily
  arXiv:hep-ph/0407286}}.

\bibitem{Konishi:1979cb}
K.~Konishi, A.~Ukawa, and G.~Veneziano, ``{Jet Calculus: A Simple Algorithm for
  Resolving QCD Jets},''
  \href{http://dx.doi.org/10.1016/0550-3213(79)90053-1}{{\em Nucl.\ Phys.\ B}
  {\bfseries 157} (1979) 45--107}.

\bibitem{Kalinowski:1980ju}
J.~Kalinowski, K.~Konishi, and T.~Taylor, ``{Jet calculus beyond leading
  logarithms},'' \href{http://dx.doi.org/10.1016/0550-3213(81)90351-5}{{\em
  Nucl.\ Phys.\ B} {\bfseries 181} (1981) 221--252}.

\bibitem{Amati:1980ch}
D.~Amati, A.~Bassetto, M.~Ciafaloni, G.~Marchesini, and G.~Veneziano, ``{A
  Treatment of Hard Processes Sensitive to the Infrared Structure of QCD},''
  \href{http://dx.doi.org/10.1016/0550-3213(80)90012-7}{{\em Nucl.\ Phys.\ B}
  {\bfseries 173} (1980) 429--455}.

\bibitem{Catani:1990rr}
S.~Catani, B.~Webber, and G.~Marchesini, ``{QCD coherent branching and
  semiinclusive processes at large x},''
  \href{http://dx.doi.org/10.1016/0550-3213(91)90390-J}{{\em Nucl.\ Phys.\ B}
  {\bfseries 349} (1991) 635--654}.

\bibitem{Catani:1992ua}
S.~Catani, L.~Trentadue, G.~Turnock, and B.~R. Webber, ``{Resummation of large
  logarithms in e+ e- event shape distributions},''
\href{http://dx.doi.org/10.1016/0550-3213(93)90271-P}{{\em Nucl. Phys.}
  {\bfseries B407} (1993) 3--42}.

\bibitem{Khoze:1996dn}
V.~A. Khoze and W.~Ochs, ``{Perturbative QCD approach to multiparticle
  production},'' \href{http://dx.doi.org/10.1142/S0217751X97001638}{{\em Int.
  J. Mod. Phys.} {\bfseries A12} (1997) 2949--3120},
\href{http://arxiv.org/abs/hep-ph/9701421}{{\ttfamily arXiv:hep-ph/9701421
  [hep-ph]}}.

\bibitem{CVITANOVIC1980429}
P.~Cvitanović, P.~Hoyer, and K.~Zalewski, ``Parton evolution as a branching
  process,''
  \href{http://dx.doi.org/https://doi.org/10.1016/0550-3213(80)90461-7}{{\em
  Nuclear Physics B} {\bfseries 176} no.~2, (1980) 429 -- 448}.

\bibitem{Dokshitzer:1992iy}
Y.~L. Dokshitzer and M.~Olsson, ``{Jet cross-sections and multiplicities in the
  modified leading logarithmic approximation},''
  \href{http://dx.doi.org/10.1016/0550-3213(93)90261-M}{{\em Nucl.\ Phys.\ B}
  {\bfseries 396} (1993) 137--160}.

\bibitem{Lupia:1997bs}
S.~Lupia and W.~Ochs, ``{Hadron multiplicity as the limit of jet multiplicity
  at high resolution},''
  \href{http://dx.doi.org/10.1016/S0920-5632(97)01039-6}{{\em Nucl.\ Phys.\ B
  Proc.\ Suppl.} {\bfseries 64} (1998) 74--77},
  \href{http://arxiv.org/abs/hep-ph/9709246}{{\ttfamily arXiv:hep-ph/9709246}}.

\bibitem{Dasgupta:2013ihk}
M.~Dasgupta, A.~Fregoso, S.~Marzani, and G.~P. Salam, ``{Towards an
  understanding of jet substructure},''
  \href{http://dx.doi.org/10.1007/JHEP09(2013)029}{{\em JHEP} {\bfseries 09}
  (2013) 029},
\href{http://arxiv.org/abs/1307.0007}{{\ttfamily arXiv:1307.0007 [hep-ph]}}.

\bibitem{Marzani:2019hun}
S.~Marzani, G.~Soyez, and M.~Spannowsky,
  \href{http://dx.doi.org/10.1007/978-3-030-15709-8}{{\em {Looking inside jets:
  an introduction to jet substructure and boosted-object phenomenology}}},
  vol.~958.
\newblock Springer, 2019.
\newblock \href{http://arxiv.org/abs/1901.10342}{{\ttfamily arXiv:1901.10342
  [hep-ph]}}.

\bibitem{Larkoski:2014wba}
A.~J. Larkoski, S.~Marzani, G.~Soyez, and J.~Thaler, ``{Soft Drop},''
  \href{http://dx.doi.org/10.1007/JHEP05(2014)146}{{\em JHEP} {\bfseries 05}
  (2014) 146}, \href{http://arxiv.org/abs/1402.2657}{{\ttfamily arXiv:1402.2657
  [hep-ph]}}.

\bibitem{Larkoski:2015lea}
A.~J. Larkoski, S.~Marzani, and J.~Thaler, ``{Sudakov Safety in Perturbative
  QCD},'' \href{http://dx.doi.org/10.1103/PhysRevD.91.111501}{{\em Phys. Rev.}
  {\bfseries D91} no.~11, (2015) 111501},
\href{http://arxiv.org/abs/1502.01719}{{\ttfamily arXiv:1502.01719 [hep-ph]}}.

\bibitem{Frye:2017yrw}
C.~Frye, A.~J. Larkoski, J.~Thaler, and K.~Zhou, ``{Casimir Meets Poisson:
  Improved Quark/Gluon Discrimination with Counting Observables},''
  \href{http://dx.doi.org/10.1007/JHEP09(2017)083}{{\em JHEP} {\bfseries 09}
  (2017) 083}, \href{http://arxiv.org/abs/1704.06266}{{\ttfamily
  arXiv:1704.06266 [hep-ph]}}.

\bibitem{Dreyer:2018tjj}
F.~A. Dreyer, L.~Necib, G.~Soyez, and J.~Thaler, ``{Recursive Soft Drop},''
  \href{http://dx.doi.org/10.1007/JHEP06(2018)093}{{\em JHEP} {\bfseries 06}
  (2018) 093}, \href{http://arxiv.org/abs/1804.03657}{{\ttfamily
  arXiv:1804.03657 [hep-ph]}}.

\bibitem{Dreyer:2018nbf}
F.~A. Dreyer, G.~P. Salam, and G.~Soyez, ``{The Lund Jet Plane},''
  \href{http://dx.doi.org/10.1007/JHEP12(2018)064}{{\em JHEP} {\bfseries 12}
  (2018) 064},
\href{http://arxiv.org/abs/1807.04758}{{\ttfamily arXiv:1807.04758 [hep-ph]}}.

\bibitem{Lifson:2020gua}
A.~Lifson, G.~P. Salam, and G.~Soyez, ``{Calculating the primary Lund Jet Plane
  density},'' \href{http://arxiv.org/abs/2007.06578}{{\ttfamily
  arXiv:2007.06578 [hep-ph]}}.

\bibitem{Blaizot:2015lma}
J.-P. Blaizot and Y.~Mehtar-Tani, ``{Jet Structure in Heavy Ion Collisions},''
  \href{http://dx.doi.org/10.1142/S021830131530012X}{{\em Int. J. Mod. Phys.}
  {\bfseries E24} no.~11, (2015) 1530012},
\href{http://arxiv.org/abs/1503.05958}{{\ttfamily arXiv:1503.05958 [hep-ph]}}.

\bibitem{Blaizot:2013vha}
J.-P. Blaizot, F.~Dominguez, E.~Iancu, and Y.~Mehtar-Tani, ``{Probabilistic
  picture for medium-induced jet evolution},''
  \href{http://dx.doi.org/10.1007/JHEP06(2014)075}{{\em JHEP} {\bfseries 06}
  (2014) 075},
\href{http://arxiv.org/abs/1311.5823}{{\ttfamily arXiv:1311.5823 [hep-ph]}}.

\bibitem{Wu:2011kc}
B.~Wu, ``{On p\_T-broadening of high energy partons associated with the LPM
  effect in a finite-volume QCD medium},''
  \href{http://dx.doi.org/10.1007/JHEP10(2011)029}{{\em JHEP} {\bfseries 10}
  (2011) 029}, \href{http://arxiv.org/abs/1102.0388}{{\ttfamily arXiv:1102.0388
  [hep-ph]}}.

\bibitem{Liou:2013qya}
T.~Liou, A.~Mueller, and B.~Wu, ``{Radiative $p_\bot$-broadening of high-energy
  quarks and gluons in QCD matter},''
  \href{http://dx.doi.org/10.1016/j.nuclphysa.2013.08.005}{{\em Nucl. Phys. A}
  {\bfseries 916} (2013) 102--125},
  \href{http://arxiv.org/abs/1304.7677}{{\ttfamily arXiv:1304.7677 [hep-ph]}}.

\bibitem{Blaizot:2019muz}
J.-P. Blaizot and F.~Dominguez, ``{Radiative corrections to the jet quenching
  parameter in dilute and dense media},''
  \href{http://dx.doi.org/10.1103/PhysRevD.99.054005}{{\em Phys. Rev. D}
  {\bfseries 99} no.~5, (2019) 054005},
  \href{http://arxiv.org/abs/1901.01448}{{\ttfamily arXiv:1901.01448
  [hep-ph]}}.

\bibitem{Blaizot:2014bha}
J.-P. Blaizot and Y.~Mehtar-Tani, ``{Renormalization of the jet-quenching
  parameter},'' \href{http://dx.doi.org/10.1016/j.nuclphysa.2014.05.018}{{\em
  Nucl. Phys. A} {\bfseries 929} (2014) 202--229},
  \href{http://arxiv.org/abs/1403.2323}{{\ttfamily arXiv:1403.2323 [hep-ph]}}.

\bibitem{Wu:2014nca}
B.~Wu, ``{Radiative energy loss and radiative $p_{\bot}$-broadening of
  high-energy partons in QCD matter},''
  \href{http://dx.doi.org/10.1007/JHEP12(2014)081}{{\em JHEP} {\bfseries 12}
  (2014) 081}, \href{http://arxiv.org/abs/1408.5459}{{\ttfamily arXiv:1408.5459
  [hep-ph]}}.

\bibitem{Escobedo:2016vba}
M.~A. Escobedo and E.~Iancu, ``{Multi-particle correlations and KNO scaling in
  the medium-induced jet evolution},''
  \href{http://dx.doi.org/10.1007/JHEP12(2016)104}{{\em JHEP} {\bfseries 12}
  (2016) 104},
\href{http://arxiv.org/abs/1609.06104}{{\ttfamily arXiv:1609.06104 [hep-ph]}}.

\bibitem{Blaizot:2013hx}
J.-P. Blaizot, E.~Iancu, and Y.~Mehtar-Tani, ``{Medium-induced QCD cascade:
  democratic branching and wave turbulence},''
  \href{http://dx.doi.org/10.1103/PhysRevLett.111.052001}{{\em Phys. Rev.
  Lett.} {\bfseries 111} (2013) 052001},
\href{http://arxiv.org/abs/1301.6102}{{\ttfamily arXiv:1301.6102 [hep-ph]}}.

\bibitem{Fister:2014zxa}
L.~Fister and E.~Iancu, ``{Medium-induced jet evolution: wave turbulence and
  energy loss},'' \href{http://dx.doi.org/10.1007/JHEP03(2015)082}{{\em JHEP}
  {\bfseries 03} (2015) 082},
\href{http://arxiv.org/abs/1409.2010}{{\ttfamily arXiv:1409.2010 [hep-ph]}}.

\bibitem{Iancu:2015uja}
E.~Iancu and B.~Wu, ``{Thermalization of mini-jets in a quark-gluon plasma},''
  \href{http://dx.doi.org/10.1007/JHEP10(2015)155}{{\em JHEP} {\bfseries 10}
  (2015) 155}, \href{http://arxiv.org/abs/1506.07871}{{\ttfamily
  arXiv:1506.07871 [hep-ph]}}.

\bibitem{Blaizot:2014rla}
J.-P. Blaizot, L.~Fister, and Y.~Mehtar-Tani, ``{Angular distribution of
  medium-induced QCD cascades},''
  \href{http://dx.doi.org/10.1016/j.nuclphysa.2015.03.014}{{\em Nucl. Phys.}
  {\bfseries A940} (2015) 67--88},
\href{http://arxiv.org/abs/1409.6202}{{\ttfamily arXiv:1409.6202 [hep-ph]}}.

\bibitem{Mehtar-Tani:2018zba}
Y.~Mehtar-Tani and S.~Schlichting, ``{Universal quark to gluon ratio in
  medium-induced parton cascade},''
  \href{http://dx.doi.org/10.1007/JHEP09(2018)144}{{\em JHEP} {\bfseries 09}
  (2018) 144},
\href{http://arxiv.org/abs/1807.06181}{{\ttfamily arXiv:1807.06181 [hep-ph]}}.

\bibitem{Mehtar-Tani:2016aco}
Y.~Mehtar-Tani and K.~Tywoniuk, ``{Groomed jets in heavy-ion collisions:
  sensitivity to medium-induced bremsstrahlung},''
  \href{http://dx.doi.org/10.1007/JHEP04(2017)125}{{\em JHEP} {\bfseries 04}
  (2017) 125},
\href{http://arxiv.org/abs/1610.08930}{{\ttfamily arXiv:1610.08930 [hep-ph]}}.

\bibitem{Adhya:2019qse}
S.~P. Adhya, C.~A. Salgado, M.~Spousta, and K.~Tywoniuk, ``{Medium-induced
  cascade in expanding media},''
  \href{http://arxiv.org/abs/1911.12193}{{\ttfamily arXiv:1911.12193
  [hep-ph]}}.

\bibitem{CasalderreySolana:2004qm}
J.~Casalderrey-Solana, E.~Shuryak, and D.~Teaney, ``{Conical flow induced by
  quenched QCD jets},''
  \href{http://dx.doi.org/10.1088/1742-6596/27/1/003}{{\em J. Phys. Conf. Ser.}
  {\bfseries 27} (2005) 22--31},
  \href{http://arxiv.org/abs/hep-ph/0411315}{{\ttfamily arXiv:hep-ph/0411315}}.

\bibitem{Ruppert:2005uz}
J.~Ruppert and B.~Muller, ``{Waking the colored plasma},''
  \href{http://dx.doi.org/10.1016/j.physletb.2005.04.075}{{\em Phys. Lett. B}
  {\bfseries 618} (2005) 123--130},
  \href{http://arxiv.org/abs/hep-ph/0503158}{{\ttfamily arXiv:hep-ph/0503158}}.

\bibitem{Li:2010ts}
H.~Li, F.~Liu, G.-l. Ma, X.-N. Wang, and Y.~Zhu, ``{Mach cone induced by
  $\gamma$-triggered jets in high-energy heavy-ion collisions},''
  \href{http://dx.doi.org/10.1103/PhysRevLett.106.012301}{{\em Phys. Rev.
  Lett.} {\bfseries 106} (2011) 012301},
  \href{http://arxiv.org/abs/1006.2893}{{\ttfamily arXiv:1006.2893 [nucl-th]}}.

\bibitem{Wang:2013cia}
X.-N. Wang and Y.~Zhu, ``{Medium Modification of $\gamma$-jets in High-energy
  Heavy-ion Collisions},''
  \href{http://dx.doi.org/10.1103/PhysRevLett.111.062301}{{\em Phys. Rev.
  Lett.} {\bfseries 111} no.~6, (2013) 062301},
  \href{http://arxiv.org/abs/1302.5874}{{\ttfamily arXiv:1302.5874 [hep-ph]}}.

\bibitem{He:2015pra}
Y.~He, T.~Luo, X.-N. Wang, and Y.~Zhu, ``{Linear Boltzmann Transport for Jet
  Propagation in the Quark-Gluon Plasma: Elastic Processes and Medium
  Recoil},'' \href{http://dx.doi.org/10.1103/PhysRevC.91.054908}{{\em Phys.
  Rev. C} {\bfseries 91} (2015) 054908},
  \href{http://arxiv.org/abs/1503.03313}{{\ttfamily arXiv:1503.03313
  [nucl-th]}}. [Erratum: Phys.Rev.C 97, 019902 (2018)].

\bibitem{Cao:2020wlm}
S.~Cao and X.-N. Wang, ``{Jet quenching and medium response in high-energy
  heavy-ion collisions: a review},''
  \href{http://arxiv.org/abs/2002.04028}{{\ttfamily arXiv:2002.04028
  [hep-ph]}}.

\bibitem{Caucal:2018epd}
P.~Caucal, E.~Iancu, A.~H. Mueller, and G.~Soyez, ``{Jet fragmentation in a
  dense QCD medium},'' \href{http://dx.doi.org/10.22323/1.336.0138}{{\em PoS}
  {\bfseries Confinement2018} (2019) 138},
  \href{http://arxiv.org/abs/1811.12275}{{\ttfamily arXiv:1811.12275
  [hep-ph]}}.

\bibitem{Lupia1998}
S.~Lupia and W.~Ochs, ``Low and high energy limits of particle spectra in QCD
  jets,'' \href{http://dx.doi.org/10.1007/s100520050140}{{\em The European
  Physical Journal C - Particles and Fields} {\bfseries 2} no.~2, (Apr, 1998)
  307--324}.

\bibitem{Gerwick:2012fw}
E.~Gerwick, S.~Schumann, B.~Gripaios, and B.~Webber, ``{QCD Jet Rates with the
  Inclusive Generalized kt Algorithms},''
  \href{http://dx.doi.org/10.1007/JHEP04(2013)089}{{\em JHEP} {\bfseries 04}
  (2013) 089}, \href{http://arxiv.org/abs/1212.5235}{{\ttfamily arXiv:1212.5235
  [hep-ph]}}.

\bibitem{Pablos:2019ngg}
D.~Pablos, ``{Jet Suppression From a Small to Intermediate to Large Radius},''
  \href{http://dx.doi.org/10.1103/PhysRevLett.124.052301}{{\em Phys. Rev.
  Lett.} {\bfseries 124} no.~5, (2020) 052301},
  \href{http://arxiv.org/abs/1907.12301}{{\ttfamily arXiv:1907.12301
  [hep-ph]}}.

\bibitem{Buckley:2014ana}
A.~Buckley, J.~Ferrando, S.~Lloyd, K.~Nordström, B.~Page, M.~Rüfenacht,
  M.~Schönherr, and G.~Watt, ``{LHAPDF6: parton density access in the LHC
  precision era},''
  \href{http://dx.doi.org/10.1140/epjc/s10052-015-3318-8}{{\em Eur. Phys. J. C}
  {\bfseries 75} (2015) 132}, \href{http://arxiv.org/abs/1412.7420}{{\ttfamily
  arXiv:1412.7420 [hep-ph]}}.

\bibitem{Sjostrand:2006za}
T.~Sjostrand, S.~Mrenna, and P.~Z. Skands, ``{PYTHIA 6.4 Physics and Manual},''
  \href{http://dx.doi.org/10.1088/1126-6708/2006/05/026}{{\em JHEP} {\bfseries
  05} (2006) 026}, \href{http://arxiv.org/abs/hep-ph/0603175}{{\ttfamily
  arXiv:hep-ph/0603175}}.

\bibitem{Kutak:2018dim}
K.~Kutak, W.~P\l~aczek, and R.~Straka, ``{Solutions of evolution equations for
  medium-induced QCD cascades},''
  \href{http://dx.doi.org/10.1140/epjc/s10052-019-6838-9}{{\em Eur. Phys. J. C}
  {\bfseries 79} no.~4, (2019) 317},
  \href{http://arxiv.org/abs/1811.06390}{{\ttfamily arXiv:1811.06390
  [hep-ph]}}.

\bibitem{Blanco:2020uzy}
E.~Blanco, K.~Kutak, W.~P\l{}aczek, M.~Rohrmoser, and R.~Straka, ``{Medium
  induced QCD cascades: broadening, entropy and rescattering during
  branching},'' \href{http://arxiv.org/abs/2009.03876}{{\ttfamily
  arXiv:2009.03876 [hep-ph]}}.

\bibitem{Cacciari:2011ma}
M.~Cacciari, G.~P. Salam, and G.~Soyez, ``{FastJet User Manual},''
  \href{http://dx.doi.org/10.1140/epjc/s10052-012-1896-2}{{\em Eur. Phys. J.}
  {\bfseries C72} (2012) 1896},
\href{http://arxiv.org/abs/1111.6097}{{\ttfamily arXiv:1111.6097 [hep-ph]}}.

\bibitem{fastjet-contrib}
M.~Cacciari, G.~P. Salam, and G.~Soyez, {\em FastJet contrib}, 2014 (accessed,
  March 7, 2020).
\newblock \url{https://fastjet.hepforge.org/contrib/}.

\bibitem{Schenke:2009vr}
B.~Schenke, C.~Gale, and S.~Jeon, ``{MARTINI: Monte Carlo simulation of jet
  evolution},'' {\em Acta Phys. Polon. Supp.} {\bfseries 3} (2010) 765--770,
  \href{http://arxiv.org/abs/0911.4470}{{\ttfamily arXiv:0911.4470 [hep-ph]}}.

\bibitem{Majumder:2013re}
A.~Majumder, ``{Incorporating Space-Time Within Medium-Modified Jet Event
  Generators},'' \href{http://dx.doi.org/10.1103/PhysRevC.88.014909}{{\em Phys.
  Rev. C} {\bfseries 88} (2013) 014909},
  \href{http://arxiv.org/abs/1301.5323}{{\ttfamily arXiv:1301.5323 [nucl-th]}}.

\bibitem{Cao:2017qpx}
S.~Cao and A.~Majumder, ``{Nuclear modification of leading hadrons and jets
  within a virtuality ordered parton shower},''
  \href{http://dx.doi.org/10.1103/PhysRevC.101.024903}{{\em Phys. Rev. C}
  {\bfseries 101} no.~2, (2020) 024903},
  \href{http://arxiv.org/abs/1712.10055}{{\ttfamily arXiv:1712.10055
  [nucl-th]}}.

\bibitem{He:2018xjv}
Y.~He, S.~Cao, W.~Chen, T.~Luo, L.-G. Pang, and X.-N. Wang, ``{Interplaying
  mechanisms behind single inclusive jet suppression in heavy-ion
  collisions},'' \href{http://dx.doi.org/10.1103/PhysRevC.99.054911}{{\em Phys.
  Rev. C} {\bfseries 99} no.~5, (2019) 054911},
  \href{http://arxiv.org/abs/1809.02525}{{\ttfamily arXiv:1809.02525
  [nucl-th]}}.

\bibitem{Luo:2018pto}
T.~Luo, S.~Cao, Y.~He, and X.-N. Wang, ``{Multiple jets and $\gamma$-jet
  correlation in high-energy heavy-ion collisions},''
  \href{http://dx.doi.org/10.1016/j.physletb.2018.06.025}{{\em Phys. Lett. B}
  {\bfseries 782} (2018) 707--716},
  \href{http://arxiv.org/abs/1803.06785}{{\ttfamily arXiv:1803.06785
  [hep-ph]}}.

\bibitem{Polosa:2006hb}
A.~D. Polosa and C.~A. Salgado, ``{Jet Shapes in Opaque Media},''
  \href{http://dx.doi.org/10.1103/PhysRevC.75.041901}{{\em Phys. Rev. C}
  {\bfseries 75} (2007) 041901},
  \href{http://arxiv.org/abs/hep-ph/0607295}{{\ttfamily arXiv:hep-ph/0607295}}.

\bibitem{Armesto:2007dt}
N.~Armesto, L.~Cunqueiro, C.~A. Salgado, and W.-C. Xiang, ``{Medium-evolved
  fragmentation functions},''
  \href{http://dx.doi.org/10.1088/1126-6708/2008/02/048}{{\em JHEP} {\bfseries
  02} (2008) 048}, \href{http://arxiv.org/abs/0710.3073}{{\ttfamily
  arXiv:0710.3073 [hep-ph]}}.

\bibitem{Armesto:2009ab}
N.~Armesto, G.~Corcella, L.~Cunqueiro, and C.~A. Salgado, ``{Angular-ordered
  parton showers with medium-modified splitting functions},''
  \href{http://dx.doi.org/10.1088/1126-6708/2009/11/122}{{\em JHEP} {\bfseries
  11} (2009) 122}, \href{http://arxiv.org/abs/0909.5118}{{\ttfamily
  arXiv:0909.5118 [hep-ph]}}.

\bibitem{Dainese:2004te}
A.~Dainese, C.~Loizides, and G.~Paic, ``{Leading-particle suppression in high
  energy nucleus-nucleus collisions},''
  \href{http://dx.doi.org/10.1140/epjc/s2004-02077-x}{{\em Eur. Phys. J. C}
  {\bfseries 38} (2005) 461--474},
  \href{http://arxiv.org/abs/hep-ph/0406201}{{\ttfamily arXiv:hep-ph/0406201}}.

\bibitem{Casalderrey-Solana:2014bpa}
J.~Casalderrey-Solana, D.~C. Gulhan, J.~G. Milhano, D.~Pablos, and
  K.~Rajagopal, ``{A Hybrid Strong/Weak Coupling Approach to Jet Quenching},''
  \href{http://dx.doi.org/10.1007/JHEP09(2015)175}{{\em JHEP} {\bfseries 10}
  (2014) 019}, \href{http://arxiv.org/abs/1405.3864}{{\ttfamily arXiv:1405.3864
  [hep-ph]}}. [Erratum: JHEP 09, 175 (2015)].

\bibitem{Hulcher:2017cpt}
Z.~Hulcher, D.~Pablos, and K.~Rajagopal, ``{Resolution Effects in the Hybrid
  Strong/Weak Coupling Model},''
  \href{http://dx.doi.org/10.1007/JHEP03(2018)010}{{\em JHEP} {\bfseries 03}
  (2018) 010}, \href{http://arxiv.org/abs/1707.05245}{{\ttfamily
  arXiv:1707.05245 [hep-ph]}}.

\bibitem{Ke:2018jem}
W.~Ke, Y.~Xu, and S.~A. Bass, ``{Modified Boltzmann approach for modeling the
  splitting vertices induced by the hot QCD medium in the deep
  Landau-Pomeranchuk-Migdal region},''
  \href{http://dx.doi.org/10.1103/PhysRevC.100.064911}{{\em Phys. Rev. C}
  {\bfseries 100} no.~6, (2019) 064911},
  \href{http://arxiv.org/abs/1810.08177}{{\ttfamily arXiv:1810.08177
  [nucl-th]}}.

\bibitem{vanLeeuwen:2015upu}
M.~van Leeuwen,
  \href{http://dx.doi.org/10.1016/j.nuclphysbps.2016.05.067}{``{Jet
  Fragmentation and Jet Shapes in JEWEL and Q-PYTHIA},''} in {\em {7th
  International Conference on Hard and Electromagnetic Probes of High-Energy
  Nuclear Collisions}}.
\newblock 11, 2015.
\newblock \href{http://arxiv.org/abs/1511.06108}{{\ttfamily arXiv:1511.06108
  [hep-ph]}}.

\bibitem{Adcox:2001jp}
{\bfseries PHENIX} Collaboration, K.~Adcox {\em et al.}, ``{Suppression of
  hadrons with large transverse momentum in central Au+Au collisions at
  $\sqrt{s_{NN}}$ = 130-GeV},''
  \href{http://dx.doi.org/10.1103/PhysRevLett.88.022301}{{\em Phys. Rev. Lett.}
  {\bfseries 88} (2002) 022301},
  \href{http://arxiv.org/abs/nucl-ex/0109003}{{\ttfamily
  arXiv:nucl-ex/0109003}}.

\bibitem{Adler:2003qi}
{\bfseries PHENIX} Collaboration, S.~Adler {\em et al.}, ``{Suppressed pi0
  production at large transverse momentum in central Au+ Au collisions at
  S(NN)**1/2 = 200 GeV},''
  \href{http://dx.doi.org/10.1103/PhysRevLett.91.072301}{{\em Phys. Rev. Lett.}
  {\bfseries 91} (2003) 072301},
  \href{http://arxiv.org/abs/nucl-ex/0304022}{{\ttfamily
  arXiv:nucl-ex/0304022}}.

\bibitem{Adler:2003cb}
{\bfseries PHENIX} Collaboration, S.~Adler {\em et al.}, ``{Identified charged
  particle spectra and yields in Au+Au collisions at S(NN)**1/2 = 200-GeV},''
  \href{http://dx.doi.org/10.1103/PhysRevC.69.034909}{{\em Phys. Rev. C}
  {\bfseries 69} (2004) 034909},
  \href{http://arxiv.org/abs/nucl-ex/0307022}{{\ttfamily
  arXiv:nucl-ex/0307022}}.

\bibitem{Adler:2003au}
{\bfseries PHENIX} Collaboration, S.~Adler {\em et al.}, ``{High $p_{T}$
  charged hadron suppression in Au + Au collisions at $\sqrt{s}_{NN} = 200$
  GeV},'' \href{http://dx.doi.org/10.1103/PhysRevC.69.034910}{{\em Phys. Rev.
  C} {\bfseries 69} (2004) 034910},
  \href{http://arxiv.org/abs/nucl-ex/0308006}{{\ttfamily
  arXiv:nucl-ex/0308006}}.

\bibitem{Adler:2005ee}
{\bfseries PHENIX} Collaboration, S.~Adler {\em et al.}, ``{Dense-Medium
  Modifications to Jet-Induced Hadron Pair Distributions in Au+Au Collisions at
  s(NN)**(1/2) = 200-GeV},''
  \href{http://dx.doi.org/10.1103/PhysRevLett.97.052301}{{\em Phys. Rev. Lett.}
  {\bfseries 97} (2006) 052301},
  \href{http://arxiv.org/abs/nucl-ex/0507004}{{\ttfamily
  arXiv:nucl-ex/0507004}}.

\bibitem{Adare:2008qa}
{\bfseries PHENIX} Collaboration, A.~Adare {\em et al.}, ``{Suppression pattern
  of neutral pions at high transverse momentum in Au$+$Au collisions at
  $\sqrt{s_{NN}}=$ 200 GeV and constraints on medium transport coefficients},''
  \href{http://dx.doi.org/10.1103/PhysRevLett.101.232301}{{\em Phys. Rev.
  Lett.} {\bfseries 101} (2008) 232301},
  \href{http://arxiv.org/abs/0801.4020}{{\ttfamily arXiv:0801.4020 [nucl-ex]}}.

\bibitem{Adare:2008ae}
{\bfseries PHENIX} Collaboration, A.~Adare {\em et al.}, ``{Dihadron azimuthal
  correlations in Au$+$Au collisions at $\sqrt{s_{NN}}=$ 200 GeV},''
  \href{http://dx.doi.org/10.1103/PhysRevC.78.014901}{{\em Phys. Rev. C}
  {\bfseries 78} (2008) 014901},
  \href{http://arxiv.org/abs/0801.4545}{{\ttfamily arXiv:0801.4545 [nucl-ex]}}.

\bibitem{Adler:2002xw}
{\bfseries STAR} Collaboration, C.~Adler {\em et al.}, ``{Centrality dependence
  of high $p_{T}$ hadron suppression in Au+Au collisions at $\sqrt{s}_{NN}$ =
  130-GeV},'' \href{http://dx.doi.org/10.1103/PhysRevLett.89.202301}{{\em Phys.
  Rev. Lett.} {\bfseries 89} (2002) 202301},
  \href{http://arxiv.org/abs/nucl-ex/0206011}{{\ttfamily
  arXiv:nucl-ex/0206011}}.

\bibitem{Adler:2002tq}
{\bfseries STAR} Collaboration, C.~Adler {\em et al.}, ``{Disappearance of
  back-to-back high $p_{T}$ hadron correlations in central Au+Au collisions at
  $\sqrt{s_{NN}}$ = 200-GeV},''
  \href{http://dx.doi.org/10.1103/PhysRevLett.90.082302}{{\em Phys. Rev. Lett.}
  {\bfseries 90} (2003) 082302},
  \href{http://arxiv.org/abs/nucl-ex/0210033}{{\ttfamily
  arXiv:nucl-ex/0210033}}.

\bibitem{Adams:2003kv}
{\bfseries STAR} Collaboration, J.~Adams {\em et al.}, ``{Transverse momentum
  and collision energy dependence of high p(T) hadron suppression in Au+Au
  collisions at ultrarelativistic energies},''
  \href{http://dx.doi.org/10.1103/PhysRevLett.91.172302}{{\em Phys. Rev. Lett.}
  {\bfseries 91} (2003) 172302},
  \href{http://arxiv.org/abs/nucl-ex/0305015}{{\ttfamily
  arXiv:nucl-ex/0305015}}.

\bibitem{Adams:2003am}
{\bfseries STAR} Collaboration, J.~Adams {\em et al.}, ``{Particle type
  dependence of azimuthal anisotropy and nuclear modification of particle
  production in Au + Au collisions at s(NN)**(1/2) = 200-GeV},''
  \href{http://dx.doi.org/10.1103/PhysRevLett.92.052302}{{\em Phys. Rev. Lett.}
  {\bfseries 92} (2004) 052302},
  \href{http://arxiv.org/abs/nucl-ex/0306007}{{\ttfamily
  arXiv:nucl-ex/0306007}}.

\bibitem{Adams:2003im}
{\bfseries STAR} Collaboration, J.~Adams {\em et al.}, ``{Evidence from d + Au
  measurements for final state suppression of high p(T) hadrons in Au+Au
  collisions at RHIC},''
  \href{http://dx.doi.org/10.1103/PhysRevLett.91.072304}{{\em Phys. Rev. Lett.}
  {\bfseries 91} (2003) 072304},
  \href{http://arxiv.org/abs/nucl-ex/0306024}{{\ttfamily
  arXiv:nucl-ex/0306024}}.

\bibitem{Adams:2005ph}
{\bfseries STAR} Collaboration, J.~Adams {\em et al.}, ``{Distributions of
  charged hadrons associated with high transverse momentum particles in pp and
  Au + Au collisions at s(NN)**(1/2) = 200-GeV},''
  \href{http://dx.doi.org/10.1103/PhysRevLett.95.152301}{{\em Phys. Rev. Lett.}
  {\bfseries 95} (2005) 152301},
  \href{http://arxiv.org/abs/nucl-ex/0501016}{{\ttfamily
  arXiv:nucl-ex/0501016}}.

\bibitem{Adams:2006yt}
{\bfseries STAR} Collaboration, J.~Adams {\em et al.}, ``{Direct observation of
  dijets in central Au+Au collisions at s(NN)**(1/2) = 200-GeV},''
  \href{http://dx.doi.org/10.1103/PhysRevLett.97.162301}{{\em Phys. Rev. Lett.}
  {\bfseries 97} (2006) 162301},
  \href{http://arxiv.org/abs/nucl-ex/0604018}{{\ttfamily
  arXiv:nucl-ex/0604018}}.

\bibitem{Adamczyk:2016fqm}
{\bfseries STAR} Collaboration, L.~Adamczyk {\em et al.}, ``{Dijet imbalance
  measurements in $Au+Au$ and $pp$ collisions at $\sqrt{s_{NN}} = 200$ GeV at
  STAR},'' \href{http://dx.doi.org/10.1103/PhysRevLett.119.062301}{{\em Phys.
  Rev. Lett.} {\bfseries 119} no.~6, (2017) 062301},
  \href{http://arxiv.org/abs/1609.03878}{{\ttfamily arXiv:1609.03878
  [nucl-ex]}}.

\bibitem{Adamczyk:2017yhe}
{\bfseries STAR} Collaboration, L.~Adamczyk {\em et al.}, ``{Measurements of
  jet quenching with semi-inclusive hadron+jet distributions in Au+Au
  collisions at $\sqrt{s_{NN}}$ = 200 GeV},''
  \href{http://dx.doi.org/10.1103/PhysRevC.96.024905}{{\em Phys. Rev. C}
  {\bfseries 96} no.~2, (2017) 024905},
  \href{http://arxiv.org/abs/1702.01108}{{\ttfamily arXiv:1702.01108
  [nucl-ex]}}.

\bibitem{Adam:2020wen}
{\bfseries STAR} Collaboration, J.~Adam {\em et al.}, ``{Measurement of
  inclusive charged-particle jet production in Au+Au collisions at
  $\sqrt{s_{NN}}$=200 GeV},'' \href{http://arxiv.org/abs/2006.00582}{{\ttfamily
  arXiv:2006.00582 [nucl-ex]}}.

\bibitem{Aad:2014wha}
{\bfseries ATLAS} Collaboration, G.~Aad {\em et al.}, ``{Measurement of
  inclusive jet charged-particle fragmentation functions in Pb+Pb collisions at
  $\sqrt{s_{NN}}=2.76$ TeV with the ATLAS detector},''
  \href{http://dx.doi.org/10.1016/j.physletb.2014.10.065}{{\em Phys. Lett. B}
  {\bfseries 739} (2014) 320--342},
  \href{http://arxiv.org/abs/1406.2979}{{\ttfamily arXiv:1406.2979 [hep-ex]}}.

\bibitem{ATLAS:2017rmz}
{\bfseries ATLAS} Collaboration, ``{Measurement of nuclear modification factor
  $R_\mathrm{AA}$ in Pb+Pb collisions at $\sqrt{s_{NN}}=5.02$TeV with the ATLAS
  detector at the LHC},''.

\bibitem{Chatrchyan:2011sx}
{\bfseries CMS} Collaboration, S.~Chatrchyan {\em et al.}, ``{Observation and
  studies of jet quenching in PbPb collisions at nucleon-nucleon center-of-mass
  energy = 2.76 TeV},''
  \href{http://dx.doi.org/10.1103/PhysRevC.84.024906}{{\em Phys. Rev. C}
  {\bfseries 84} (2011) 024906},
  \href{http://arxiv.org/abs/1102.1957}{{\ttfamily arXiv:1102.1957 [nucl-ex]}}.

\bibitem{Chatrchyan:2011pb}
{\bfseries CMS} Collaboration, S.~Chatrchyan {\em et al.}, ``{Dependence on
  pseudorapidity and centrality of charged hadron production in PbPb collisions
  at a nucleon-nucleon centre-of-mass energy of 2.76 TeV},''
  \href{http://dx.doi.org/10.1007/JHEP08(2011)141}{{\em JHEP} {\bfseries 08}
  (2011) 141}, \href{http://arxiv.org/abs/1107.4800}{{\ttfamily arXiv:1107.4800
  [nucl-ex]}}.

\bibitem{CMS:2012aa}
{\bfseries CMS} Collaboration, S.~Chatrchyan {\em et al.}, ``{Study of high-pT
  charged particle suppression in PbPb compared to $pp$ collisions at
  $\sqrt{s_{NN}}=2.76$ TeV},''
  \href{http://dx.doi.org/10.1140/epjc/s10052-012-1945-x}{{\em Eur. Phys. J. C}
  {\bfseries 72} (2012) 1945}, \href{http://arxiv.org/abs/1202.2554}{{\ttfamily
  arXiv:1202.2554 [nucl-ex]}}.

\bibitem{Chatrchyan:2012gw}
{\bfseries CMS} Collaboration, S.~Chatrchyan {\em et al.}, ``{Measurement of
  jet fragmentation into charged particles in $pp$ and PbPb collisions at
  $\sqrt{s_{NN}}=2.76$ TeV},''
  \href{http://dx.doi.org/10.1007/JHEP10(2012)087}{{\em JHEP} {\bfseries 10}
  (2012) 087}, \href{http://arxiv.org/abs/1205.5872}{{\ttfamily arXiv:1205.5872
  [nucl-ex]}}.

\bibitem{Chatrchyan:2012nia}
{\bfseries CMS} Collaboration, S.~Chatrchyan {\em et al.}, ``{Jet momentum
  dependence of jet quenching in PbPb collisions at $\sqrt{s_{NN}}=2.76$
  TeV},'' \href{http://dx.doi.org/10.1016/j.physletb.2012.04.058}{{\em Phys.
  Lett. B} {\bfseries 712} (2012) 176--197},
  \href{http://arxiv.org/abs/1202.5022}{{\ttfamily arXiv:1202.5022 [nucl-ex]}}.

\bibitem{Khachatryan:2016odn}
{\bfseries CMS} Collaboration, V.~Khachatryan {\em et al.}, ``{Charged-particle
  nuclear modification factors in PbPb and pPb collisions at $
  \sqrt{s_{\mathrm{N}\;\mathrm{N}}}=5.02 $ TeV},''
  \href{http://dx.doi.org/10.1007/JHEP04(2017)039}{{\em JHEP} {\bfseries 04}
  (2017) 039}, \href{http://arxiv.org/abs/1611.01664}{{\ttfamily
  arXiv:1611.01664 [nucl-ex]}}.

\bibitem{Abelev:2012hxa}
{\bfseries ALICE} Collaboration, B.~Abelev {\em et al.}, ``{Centrality
  Dependence of Charged Particle Production at Large Transverse Momentum in
  Pb--Pb Collisions at $\sqrt{s_{\rm{NN}}} = 2.76$ TeV},''
  \href{http://dx.doi.org/10.1016/j.physletb.2013.01.051}{{\em Phys. Lett. B}
  {\bfseries 720} (2013) 52--62},
  \href{http://arxiv.org/abs/1208.2711}{{\ttfamily arXiv:1208.2711 [hep-ex]}}.

\bibitem{Abelev:2013kqa}
{\bfseries ALICE} Collaboration, B.~Abelev {\em et al.}, ``{Measurement of
  charged jet suppression in Pb-Pb collisions at $\sqrt{s_{NN}}$ = 2.76 TeV},''
  \href{http://dx.doi.org/10.1007/JHEP03(2014)013}{{\em JHEP} {\bfseries 03}
  (2014) 013}, \href{http://arxiv.org/abs/1311.0633}{{\ttfamily arXiv:1311.0633
  [nucl-ex]}}.

\bibitem{Abelev:2014laa}
{\bfseries ALICE} Collaboration, B.~B. Abelev {\em et al.}, ``{Production of
  charged pions, kaons and protons at large transverse momenta in pp and Pb--Pb
  collisions at $\sqrt{s_{\rm NN}}$ =2.76 TeV},''
  \href{http://dx.doi.org/10.1016/j.physletb.2014.07.011}{{\em Phys. Lett. B}
  {\bfseries 736} (2014) 196--207},
  \href{http://arxiv.org/abs/1401.1250}{{\ttfamily arXiv:1401.1250 [nucl-ex]}}.

\bibitem{Adam:2015kca}
{\bfseries ALICE} Collaboration, J.~Adam {\em et al.}, ``{Centrality dependence
  of the nuclear modification factor of charged pions, kaons, and protons in
  Pb-Pb collisions at $\sqrt{s_{\rm NN}}=2.76$ TeV},''
  \href{http://dx.doi.org/10.1103/PhysRevC.93.034913}{{\em Phys. Rev. C}
  {\bfseries 93} no.~3, (2016) 034913},
  \href{http://arxiv.org/abs/1506.07287}{{\ttfamily arXiv:1506.07287
  [nucl-ex]}}.

\bibitem{Adam:2015ewa}
{\bfseries ALICE} Collaboration, J.~Adam {\em et al.}, ``{Measurement of jet
  suppression in central Pb-Pb collisions at $\sqrt{s_{\rm NN}}$ = 2.76 TeV},''
  \href{http://dx.doi.org/10.1016/j.physletb.2015.04.039}{{\em Phys. Lett. B}
  {\bfseries 746} (2015) 1--14},
  \href{http://arxiv.org/abs/1502.01689}{{\ttfamily arXiv:1502.01689
  [nucl-ex]}}.

\bibitem{Acharya:2017goa}
{\bfseries ALICE} Collaboration, S.~Acharya {\em et al.}, ``{First measurement
  of jet mass in Pb--Pb and p--Pb collisions at the LHC},''
  \href{http://dx.doi.org/10.1016/j.physletb.2017.11.044}{{\em Phys. Lett. B}
  {\bfseries 776} (2018) 249--264},
  \href{http://arxiv.org/abs/1702.00804}{{\ttfamily arXiv:1702.00804
  [nucl-ex]}}.

\bibitem{Acharya:2018qsh}
{\bfseries ALICE} Collaboration, S.~Acharya {\em et al.}, ``{Transverse
  momentum spectra and nuclear modification factors of charged particles in pp,
  p-Pb and Pb-Pb collisions at the LHC},''
  \href{http://dx.doi.org/10.1007/JHEP11(2018)013}{{\em JHEP} {\bfseries 11}
  (2018) 013}, \href{http://arxiv.org/abs/1802.09145}{{\ttfamily
  arXiv:1802.09145 [nucl-ex]}}.

\bibitem{Miller:2007ri}
M.~L. Miller, K.~Reygers, S.~J. Sanders, and P.~Steinberg, ``{Glauber modeling
  in high energy nuclear collisions},''
  \href{http://dx.doi.org/10.1146/annurev.nucl.57.090506.123020}{{\em Ann. Rev.
  Nucl. Part. Sci.} {\bfseries 57} (2007) 205--243},
  \href{http://arxiv.org/abs/nucl-ex/0701025}{{\ttfamily
  arXiv:nucl-ex/0701025}}.

\bibitem{Aaboud:2018twu}
{\bfseries ATLAS} Collaboration, M.~Aaboud {\em et al.}, ``{Measurement of the
  nuclear modification factor for inclusive jets in Pb+Pb collisions at
  $\sqrt{s_\mathrm{NN}}=5.02$ TeV with the ATLAS detector},''
  \href{http://dx.doi.org/10.1016/j.physletb.2018.10.076}{{\em Phys. Lett. B}
  {\bfseries 790} (2019) 108--128},
  \href{http://arxiv.org/abs/1805.05635}{{\ttfamily arXiv:1805.05635
  [nucl-ex]}}.

\bibitem{Acharya:2019jyg}
{\bfseries ALICE} Collaboration, S.~Acharya {\em et al.}, ``{Measurements of
  inclusive jet spectra in pp and central Pb-Pb collisions at
  $\sqrt{s_{\rm{NN}}}$ = 5.02 TeV},''
  \href{http://dx.doi.org/10.1103/PhysRevC.101.034911}{{\em Phys. Rev. C}
  {\bfseries 101} no.~3, (2020) 034911},
  \href{http://arxiv.org/abs/1909.09718}{{\ttfamily arXiv:1909.09718
  [nucl-ex]}}.

\bibitem{Aad:2014bxa}
{\bfseries ATLAS} Collaboration, G.~Aad {\em et al.}, ``{Measurements of the
  Nuclear Modification Factor for Jets in Pb+Pb Collisions at
  $\sqrt{s_{\mathrm{NN}}}=2.76$ TeV with the ATLAS Detector},''
  \href{http://dx.doi.org/10.1103/PhysRevLett.114.072302}{{\em Phys. Rev.
  Lett.} {\bfseries 114} no.~7, (2015) 072302},
  \href{http://arxiv.org/abs/1411.2357}{{\ttfamily arXiv:1411.2357 [hep-ex]}}.

\bibitem{Khachatryan:2016jfl}
{\bfseries CMS} Collaboration, V.~Khachatryan {\em et al.}, ``{Measurement of
  inclusive jet cross sections in $pp$ and PbPb collisions at $\sqrt{s_{NN}}=$
  2.76 TeV},'' \href{http://dx.doi.org/10.1103/PhysRevC.96.015202}{{\em Phys.
  Rev. C} {\bfseries 96} no.~1, (2017) 015202},
  \href{http://arxiv.org/abs/1609.05383}{{\ttfamily arXiv:1609.05383
  [nucl-ex]}}.

\bibitem{Aad:2010bu}
{\bfseries ATLAS} Collaboration, G.~Aad {\em et al.}, ``{Observation of a
  Centrality-Dependent Dijet Asymmetry in Lead-Lead Collisions at
  $\sqrt{s_{NN}}=2.77$ TeV with the ATLAS Detector at the LHC},''
  \href{http://dx.doi.org/10.1103/PhysRevLett.105.252303}{{\em Phys. Rev.
  Lett.} {\bfseries 105} (2010) 252303},
  \href{http://arxiv.org/abs/1011.6182}{{\ttfamily arXiv:1011.6182 [hep-ex]}}.

\bibitem{Aaboud:2017eww}
{\bfseries ATLAS} Collaboration, M.~Aaboud {\em et al.}, ``{Measurement of jet
  $p_{\mathrm{T}}$ correlations in Pb+Pb and $pp$ collisions at
  $\sqrt{s_{\mathrm{NN}}}=$ 2.76 TeV with the ATLAS detector},''
  \href{http://dx.doi.org/10.1016/j.physletb.2017.09.078}{{\em Phys. Lett. B}
  {\bfseries 774} (2017) 379--402},
  \href{http://arxiv.org/abs/1706.09363}{{\ttfamily arXiv:1706.09363
  [hep-ex]}}.

\bibitem{Aaboud:2018anc}
{\bfseries ATLAS} Collaboration, M.~Aaboud {\em et al.}, ``{Measurement of
  photon--jet transverse momentum correlations in 5.02 TeV Pb + Pb and $pp$
  collisions with ATLAS},''
  \href{http://dx.doi.org/10.1016/j.physletb.2018.12.023}{{\em Phys. Lett. B}
  {\bfseries 789} (2019) 167--190},
  \href{http://arxiv.org/abs/1809.07280}{{\ttfamily arXiv:1809.07280
  [nucl-ex]}}.

\bibitem{Chatrchyan:2012gt}
{\bfseries CMS} Collaboration, S.~Chatrchyan {\em et al.}, ``{Studies of jet
  quenching using isolated-photon+jet correlations in PbPb and $pp$ collisions
  at $\sqrt{s_{NN}}=2.76$ TeV},''
  \href{http://dx.doi.org/10.1016/j.physletb.2012.11.003}{{\em Phys. Lett. B}
  {\bfseries 718} (2013) 773--794},
  \href{http://arxiv.org/abs/1205.0206}{{\ttfamily arXiv:1205.0206 [nucl-ex]}}.

\bibitem{Sirunyan:2017qhf}
{\bfseries CMS} Collaboration, A.~M. Sirunyan {\em et al.}, ``{Study of jet
  quenching with isolated-photon+jet correlations in PbPb and pp collisions at
  $\sqrt{s_{_{\mathrm{NN}}}} =$ 5.02 TeV},''
  \href{http://dx.doi.org/10.1016/j.physletb.2018.07.061}{{\em Phys. Lett. B}
  {\bfseries 785} (2018) 14--39},
  \href{http://arxiv.org/abs/1711.09738}{{\ttfamily arXiv:1711.09738
  [nucl-ex]}}.

\bibitem{Khachatryan:2015lha}
{\bfseries CMS} Collaboration, V.~Khachatryan {\em et al.}, ``{Measurement of
  transverse momentum relative to dijet systems in PbPb and pp collisions at $
  \sqrt{s_{\mathrm{NN}}}=2.76 $ TeV},''
  \href{http://dx.doi.org/10.1007/JHEP01(2016)006}{{\em JHEP} {\bfseries 01}
  (2016) 006}, \href{http://arxiv.org/abs/1509.09029}{{\ttfamily
  arXiv:1509.09029 [nucl-ex]}}.

\bibitem{Milhano:2015mng}
J.~G. Milhano and K.~C. Zapp, ``{Origins of the di-jet asymmetry in heavy ion
  collisions},'' \href{http://dx.doi.org/10.1140/epjc/s10052-016-4130-9}{{\em
  Eur. Phys. J. C} {\bfseries 76} no.~5, (2016) 288},
  \href{http://arxiv.org/abs/1512.08107}{{\ttfamily arXiv:1512.08107
  [hep-ph]}}.

\bibitem{Brewer:2018mpk}
J.~Brewer, A.~Sadofyev, and W.~van~der Schee, ``{Jet shape modifications in
  holographic dijet systems},''
  \href{http://arxiv.org/abs/1809.10695}{{\ttfamily arXiv:1809.10695
  [hep-ph]}}.

\bibitem{Aaboud:2018hpb}
{\bfseries ATLAS} Collaboration, M.~Aaboud {\em et al.}, ``{Measurement of jet
  fragmentation in Pb+Pb and $pp$ collisions at $\sqrt{s_{NN}} = 5.02$ TeV with
  the ATLAS detector},''
  \href{http://dx.doi.org/10.1103/PhysRevC.98.024908}{{\em Phys. Rev.}
  {\bfseries C98} no.~2, (2018) 024908},
\href{http://arxiv.org/abs/1805.05424}{{\ttfamily arXiv:1805.05424 [nucl-ex]}}.

\bibitem{Sirunyan:2018jqr}
{\bfseries CMS} Collaboration, A.~M. Sirunyan {\em et al.}, ``{Jet properties
  in PbPb and pp collisions at $ \sqrt{s_{\mathrm{N}\;\mathrm{N}}}=5.02 $
  TeV},'' \href{http://dx.doi.org/10.1007/JHEP05(2018)006}{{\em JHEP}
  {\bfseries 05} (2018) 006}, \href{http://arxiv.org/abs/1803.00042}{{\ttfamily
  arXiv:1803.00042 [nucl-ex]}}.

\bibitem{Aaboud:2019oac}
{\bfseries ATLAS} Collaboration, M.~Aaboud {\em et al.}, ``{Comparison of
  Fragmentation Functions for Jets Dominated by Light Quarks and Gluons from
  $pp$ and Pb+Pb Collisions in ATLAS},''
  \href{http://dx.doi.org/10.1103/PhysRevLett.123.042001}{{\em Phys. Rev.
  Lett.} {\bfseries 123} no.~4, (2019) 042001},
  \href{http://arxiv.org/abs/1902.10007}{{\ttfamily arXiv:1902.10007
  [nucl-ex]}}.

\bibitem{Sirunyan:2018qec}
{\bfseries CMS} Collaboration, A.~M. Sirunyan {\em et al.}, ``{Observation of
  Medium-Induced Modifications of Jet Fragmentation in Pb-Pb Collisions at
  $\sqrt{s_{NN}}=$ 5.02 TeV Using Isolated Photon-Tagged Jets},''
  \href{http://dx.doi.org/10.1103/PhysRevLett.121.242301}{{\em Phys. Rev.
  Lett.} {\bfseries 121} no.~24, (2018) 242301},
  \href{http://arxiv.org/abs/1801.04895}{{\ttfamily arXiv:1801.04895
  [hep-ex]}}.

\bibitem{Chatrchyan:2014ava}
{\bfseries CMS} Collaboration, S.~Chatrchyan {\em et al.}, ``{Measurement of
  Jet Fragmentation in PbPb and pp Collisions at $\sqrt{s_{NN}}= 2.76$ TeV},''
  \href{http://dx.doi.org/10.1103/PhysRevC.90.024908}{{\em Phys. Rev. C}
  {\bfseries 90} no.~2, (2014) 024908},
  \href{http://arxiv.org/abs/1406.0932}{{\ttfamily arXiv:1406.0932 [nucl-ex]}}.

\bibitem{Aaboud:2017bzv}
{\bfseries ATLAS} Collaboration, M.~Aaboud {\em et al.}, ``{Measurement of jet
  fragmentation in Pb+Pb and $pp$ collisions at $\sqrt{{s_\mathrm{NN}}} = 2.76$
  TeV with the ATLAS detector at the LHC},''
  \href{http://dx.doi.org/10.1140/epjc/s10052-017-4915-5}{{\em Eur. Phys. J. C}
  {\bfseries 77} no.~6, (2017) 379},
  \href{http://arxiv.org/abs/1702.00674}{{\ttfamily arXiv:1702.00674
  [hep-ex]}}.

\bibitem{Chatrchyan:2013kwa}
{\bfseries CMS} Collaboration, S.~Chatrchyan {\em et al.}, ``{Modification of
  Jet Shapes in PbPb Collisions at $\sqrt {s_{NN}} = 2.76$ TeV},''
  \href{http://dx.doi.org/10.1016/j.physletb.2014.01.042}{{\em Phys. Lett. B}
  {\bfseries 730} (2014) 243--263},
  \href{http://arxiv.org/abs/1310.0878}{{\ttfamily arXiv:1310.0878 [nucl-ex]}}.

\bibitem{Sirunyan:2018ncy}
{\bfseries CMS} Collaboration, A.~M. Sirunyan {\em et al.}, ``{Jet Shapes of
  Isolated Photon-Tagged Jets in Pb-Pb and pp Collisions at
  $\sqrt{s_\mathrm{NN}} =$ 5.02 TeV},''
  \href{http://dx.doi.org/10.1103/PhysRevLett.122.152001}{{\em Phys. Rev.
  Lett.} {\bfseries 122} no.~15, (2019) 152001},
  \href{http://arxiv.org/abs/1809.08602}{{\ttfamily arXiv:1809.08602
  [hep-ex]}}.

\bibitem{Acharya:2018uvf}
{\bfseries ALICE} Collaboration, S.~Acharya {\em et al.}, ``{Medium
  modification of the shape of small-radius jets in central Pb-Pb collisions at
  $\sqrt{s_{\mathrm {NN}}} = 2.76\,\rm{TeV}$},''
  \href{http://dx.doi.org/10.1007/JHEP10(2018)139}{{\em JHEP} {\bfseries 10}
  (2018) 139}, \href{http://arxiv.org/abs/1807.06854}{{\ttfamily
  arXiv:1807.06854 [nucl-ex]}}.

\bibitem{Acharya:2019ssy}
{\bfseries ALICE} Collaboration, S.~Acharya {\em et al.}, ``{Measurement of jet
  radial profiles in Pb$–$Pb collisions at $\sqrt{s_{\rm NN}}=$ 2.76 TeV},''
  \href{http://dx.doi.org/10.1016/j.physletb.2019.07.020}{{\em Phys. Lett. B}
  {\bfseries 796} (2019) 204--219},
  \href{http://arxiv.org/abs/1904.13118}{{\ttfamily arXiv:1904.13118
  [nucl-ex]}}.

\bibitem{Cal:2019hjc}
P.~Cal, F.~Ringer, and W.~J. Waalewijn, ``{The jet shape at NLL},''
  \href{http://dx.doi.org/10.1007/JHEP05(2019)143}{{\em JHEP} {\bfseries 05}
  (2019) 143}, \href{http://arxiv.org/abs/1901.06389}{{\ttfamily
  arXiv:1901.06389 [hep-ph]}}.

\bibitem{Sirunyan:2017bsd}
{\bfseries CMS} Collaboration, A.~M. Sirunyan {\em et al.}, ``{Measurement of
  the Splitting Function in $pp$ and Pb-Pb Collisions at
  $\sqrt{s_{_{\mathrm{NN}}}} =$ 5.02 TeV},''
  \href{http://dx.doi.org/10.1103/PhysRevLett.120.142302}{{\em Phys. Rev.
  Lett.} {\bfseries 120} no.~14, (2018) 142302},
\href{http://arxiv.org/abs/1708.09429}{{\ttfamily arXiv:1708.09429 [nucl-ex]}}.

\bibitem{Acharya:2019djg}
{\bfseries ALICE} Collaboration, S.~Acharya {\em et al.}, ``{Exploration of jet
  substructure using iterative declustering in pp and Pb-Pb collisions at LHC
  energies},''
\href{http://arxiv.org/abs/1905.02512}{{\ttfamily arXiv:1905.02512 [nucl-ex]}}.

\bibitem{Sirunyan:2018gct}
{\bfseries CMS} Collaboration, A.~M. Sirunyan {\em et al.}, ``{Measurement of
  the groomed jet mass in PbPb and pp collisions at $
  \sqrt{s_{\mathrm{NN}}}=5.02 $ TeV},''
  \href{http://dx.doi.org/10.1007/JHEP10(2018)161}{{\em JHEP} {\bfseries 10}
  (2018) 161}, \href{http://arxiv.org/abs/1805.05145}{{\ttfamily
  arXiv:1805.05145 [hep-ex]}}.

\bibitem{Adam:2020kug}
{\bfseries STAR} Collaboration, J.~Adam {\em et al.}, ``{Measurement of Groomed
  Jet Substructure Observables in $pp$ Collisions at $\sqrt{s} = 200$ GeV with
  STAR},'' \href{http://arxiv.org/abs/2003.02114}{{\ttfamily arXiv:2003.02114
  [hep-ex]}}.

\bibitem{ATLAS:2019rmd}
{\bfseries ATLAS} Collaboration, ``{Measurement of suppression of large-radius
  jets and its dependence on substructure in Pb+Pb at 5.02 TeV by ATLAS
  detector},''.

\bibitem{Khachatryan:2016tfj}
{\bfseries CMS} Collaboration, V.~Khachatryan {\em et al.}, ``{Decomposing
  transverse momentum balance contributions for quenched jets in PbPb
  collisions at $ \sqrt{s_{\mathrm{N}\;\mathrm{N}}}=2.76 $ TeV},''
  \href{http://dx.doi.org/10.1007/JHEP11(2016)055}{{\em JHEP} {\bfseries 11}
  (2016) 055}, \href{http://arxiv.org/abs/1609.02466}{{\ttfamily
  arXiv:1609.02466 [nucl-ex]}}.

\bibitem{Andrews:2018jcm}
H.~A. Andrews {\em et al.}, ``{Novel tools and observables for jet physics in
  heavy-ion collisions},''
\href{http://arxiv.org/abs/1808.03689}{{\ttfamily arXiv:1808.03689 [hep-ph]}}.

\bibitem{Cacciari:2008gn}
M.~Cacciari, G.~P. Salam, and G.~Soyez, ``{The Catchment Area of Jets},''
  \href{http://dx.doi.org/10.1088/1126-6708/2008/04/005}{{\em JHEP} {\bfseries
  04} (2008) 005},
\href{http://arxiv.org/abs/0802.1188}{{\ttfamily arXiv:0802.1188 [hep-ph]}}.

\bibitem{Spousta:2015fca}
M.~Spousta and B.~Cole, ``{Interpreting single jet measurements in Pb $+$ Pb
  collisions at the LHC},''
  \href{http://dx.doi.org/10.1140/epjc/s10052-016-3896-0}{{\em Eur. Phys. J. C}
  {\bfseries 76} no.~2, (2016) 50},
  \href{http://arxiv.org/abs/1504.05169}{{\ttfamily arXiv:1504.05169
  [hep-ph]}}.

\bibitem{Casalderrey-Solana:2018wrw}
J.~Casalderrey-Solana, Z.~Hulcher, G.~Milhano, D.~Pablos, and K.~Rajagopal,
  ``{Simultaneous description of hadron and jet suppression in heavy-ion
  collisions},'' \href{http://dx.doi.org/10.1103/PhysRevC.99.051901}{{\em Phys.
  Rev.} {\bfseries C99} no.~5, (2019) 051901},
\href{http://arxiv.org/abs/1808.07386}{{\ttfamily arXiv:1808.07386 [hep-ph]}}.

\bibitem{KunnawalkamElayavalli:2017hxo}
R.~Kunnawalkam~Elayavalli and K.~C. Zapp, ``{Medium response in JEWEL and its
  impact on jet shape observables in heavy ion collisions},''
  \href{http://dx.doi.org/10.1007/JHEP07(2017)141}{{\em JHEP} {\bfseries 07}
  (2017) 141}, \href{http://arxiv.org/abs/1707.01539}{{\ttfamily
  arXiv:1707.01539 [hep-ph]}}.

\bibitem{Chesler:2015nqz}
P.~M. Chesler and K.~Rajagopal, ``{On the Evolution of Jet Energy and Opening
  Angle in Strongly Coupled Plasma},''
  \href{http://dx.doi.org/10.1007/JHEP05(2016)098}{{\em JHEP} {\bfseries 05}
  (2016) 098}, \href{http://arxiv.org/abs/1511.07567}{{\ttfamily
  arXiv:1511.07567 [hep-th]}}.

\bibitem{Rajagopal:2016uip}
K.~Rajagopal, A.~V. Sadofyev, and W.~van~der Schee, ``{Evolution of the jet
  opening angle distribution in holographic plasma},''
  \href{http://dx.doi.org/10.1103/PhysRevLett.116.211603}{{\em Phys. Rev.
  Lett.} {\bfseries 116} no.~21, (2016) 211603},
  \href{http://arxiv.org/abs/1602.04187}{{\ttfamily arXiv:1602.04187
  [nucl-th]}}.

\bibitem{Casalderrey-Solana:2016jvj}
J.~Casalderrey-Solana, D.~Gulhan, G.~Milhano, D.~Pablos, and K.~Rajagopal,
  ``{Angular Structure of Jet Quenching Within a Hybrid Strong/Weak Coupling
  Model},'' \href{http://dx.doi.org/10.1007/JHEP03(2017)135}{{\em JHEP}
  {\bfseries 03} (2017) 135}, \href{http://arxiv.org/abs/1609.05842}{{\ttfamily
  arXiv:1609.05842 [hep-ph]}}.

\bibitem{Casalderrey-Solana:2019ubu}
J.~Casalderrey-Solana, G.~Milhano, D.~Pablos, and K.~Rajagopal, ``{Modification
  of Jet Substructure in Heavy Ion Collisions as a Probe of the Resolution
  Length of Quark-Gluon Plasma},''
  \href{http://dx.doi.org/10.1007/JHEP01(2020)044}{{\em JHEP} {\bfseries 01}
  (2020) 044}, \href{http://arxiv.org/abs/1907.11248}{{\ttfamily
  arXiv:1907.11248 [hep-ph]}}.

\bibitem{Tachibana:2017syd}
Y.~Tachibana, N.-B. Chang, and G.-Y. Qin, ``{Full jet in quark-gluon plasma
  with hydrodynamic medium response},''
  \href{http://dx.doi.org/10.1103/PhysRevC.95.044909}{{\em Phys. Rev. C}
  {\bfseries 95} no.~4, (2017) 044909},
  \href{http://arxiv.org/abs/1701.07951}{{\ttfamily arXiv:1701.07951
  [nucl-th]}}.

\bibitem{Chen:2017zte}
W.~Chen, S.~Cao, T.~Luo, L.-G. Pang, and X.-N. Wang, ``{Effects of jet-induced
  medium excitation in $\gamma$-hadron correlation in A+A collisions},''
  \href{http://dx.doi.org/10.1016/j.physletb.2017.12.015}{{\em Phys. Lett. B}
  {\bfseries 777} (2018) 86--90},
  \href{http://arxiv.org/abs/1704.03648}{{\ttfamily arXiv:1704.03648
  [nucl-th]}}.

\bibitem{Chien:2015hda}
Y.-T. Chien and I.~Vitev, ``{Towards the understanding of jet shapes and cross
  sections in heavy ion collisions using soft-collinear effective theory},''
  \href{http://dx.doi.org/10.1007/JHEP05(2016)023}{{\em JHEP} {\bfseries 05}
  (2016) 023}, \href{http://arxiv.org/abs/1509.07257}{{\ttfamily
  arXiv:1509.07257 [hep-ph]}}.

\bibitem{Cao:2017zih}
{\bfseries JETSCAPE} Collaboration, S.~Cao {\em et al.}, ``{Multistage
  Monte-Carlo simulation of jet modification in a static medium},''
  \href{http://dx.doi.org/10.1103/PhysRevC.96.024909}{{\em Phys. Rev.}
  {\bfseries C96} no.~2, (2017) 024909},
\href{http://arxiv.org/abs/1705.00050}{{\ttfamily arXiv:1705.00050 [nucl-th]}}.

\bibitem{Putschke:2019yrg}
J.~H. Putschke {\em et al.}, ``{The JETSCAPE framework},''
\href{http://arxiv.org/abs/1903.07706}{{\ttfamily arXiv:1903.07706 [nucl-th]}}.

\bibitem{Gelfand:1959nq}
I.~Gelfand and A.~Yaglom, ``{Integration in functional spaces and it
  applications in quantum physics},''
  \href{http://dx.doi.org/10.1063/1.1703636}{{\em J. Math. Phys.} {\bfseries 1}
  (1960) 48}.

\bibitem{DeGrand:1978te}
T.~A. DeGrand, ``{Structure Functions of Quarks, Gluons, and Hadrons in Quantum
  Chromodynamics},'' \href{http://dx.doi.org/10.1016/0550-3213(79)90452-8}{{\em
  Nucl.\ Phys.\ B} {\bfseries 151} (1979) 485--517}.

\bibitem{Gunion:1981qs}
J.~Gunion and G.~Bertsch, ``Hadronization by color Bremsstrahlung,''
  \href{http://dx.doi.org/10.1103/PhysRevD.25.746}{{\em Phys. Rev. D}
  {\bfseries 25} (1982) 746}.

\end{thebibliography}\endgroup
